\documentclass[10pt]{iopart}

\usepackage[a4paper, margin=1in]{geometry}

\usepackage{dsfont}

\usepackage{fancyhdr}  
\usepackage[utf8]{inputenc} 	
\usepackage[T1]{fontenc} 	
\usepackage{soul} 
\usepackage[perpage,bottom]{footmisc}  
\usepackage[english]{babel}  
\usepackage{bbm, amsthm, placeins, graphicx, wrapfig, subfigure, sidecap, float, verbatim, color, multirow, scalerel, makecell, iopams, mathrsfs, enumerate}

\expandafter\let\csname equation*\endcsname\relax
\expandafter\let\csname endequation*\endcsname\relax
\usepackage{amsmath}

\usepackage{etoolbox}

\makeatletter
\newcommand{\mainmatter}{%
  \setcounter{footnote}{0}%
  \patchcmd{\@makefntext}{\fnsymbol}{\arabic}{}{}%
  \patchcmd{\@thefnmark}{\fnsymbol}{\arabic}{}{}%
  \def\@makefnmark{\textsuperscript{\arabic{footnote}}}%
}
\makeatother

\usepackage{caption}
\usepackage[sort&compress,numbers,square,comma]{natbib}

\usepackage{tikz}
\usetikzlibrary{svg.path}

\definecolor{orcidlogocol}{HTML}{A6CE39}
\tikzset{
  orcidlogo/.pic={
    \fill[orcidlogocol] svg{M256,128c0,70.7-57.3,128-128,128C57.3,256,0,198.7,0,128C0,57.3,57.3,0,128,0C198.7,0,256,57.3,256,128z};
    \fill[white] svg{M86.3,186.2H70.9V79.1h15.4v48.4V186.2z}
                 svg{M108.9,79.1h41.6c39.6,0,57,28.3,57,53.6c0,27.5-21.5,53.6-56.8,53.6h-41.8V79.1z M124.3,172.4h24.5c34.9,0,42.9-26.5,42.9-39.7c0-21.5-13.7-39.7-43.7-39.7h-23.7V172.4z}
                 svg{M88.7,56.8c0,5.5-4.5,10.1-10.1,10.1c-5.6,0-10.1-4.6-10.1-10.1c0-5.6,4.5-10.1,10.1-10.1C84.2,46.7,88.7,51.3,88.7,56.8z};
  }
}

\usepackage[unicode=true, breaklinks=false, pdfborder={0 0 1}, backref=false, colorlinks=true, linkcolor=blue, citecolor=blue]{hyperref}

\newcommand\orcidicon[1]{\href{https://orcid.org/#1}{\mbox{\scalerel*{
\begin{tikzpicture}[yscale=-1,transform shape]
\pic{orcidlogo};
\end{tikzpicture}
}{|}}}}

\allowdisplaybreaks

\numberwithin{equation}{subsection}

\usepackage{xparse} 
\usepackage[most]{tcolorbox} 

\newtcbtheorem[auto counter, number within=section]{Definition}{}{
    enhanced,
    sharp corners,
    attach boxed title to top left={
        xshift=-1mm,
        yshift=-5mm,
        yshifttext=-1mm
    },
    top=1.5em,
    colback=white,
    colframe=blue!75!black,
    fonttitle=\bfseries,
    boxed title style={
        sharp corners,
        size=small,
        colback=blue!75!black,
        colframe=blue!75!black,
    }
}{def}

\newcounter{definition}[section]
\renewcommand{\thedefinition}{\thesection.\arabic{definition}}

\NewDocumentEnvironment{myDefinition}{ o m o }{%
  \begin{Definition}[title=#2]%
    \IfValueT{#3}{\refstepcounter{definition}\label{#3}} 
    \IfValueT{#1}{\textbf{Definition \thedefinition. } #1}{\textbf{Definition \thedefinition. }}%
}{%
  \end{Definition}
}

\newenvironment{myDefinition*}[2]{ \begin{Definition}[adjusted title=#1]{}{#2}
  }{\end{Definition}}
  
\usepackage{titlesec}

\titleformat{\section}
       {\normalfont\large\bfseries\fontsize{12}{16}}{\thesection}{1em}{}

\titleformat{\subsection}
       {\normalfont\bfseries\fontsize{10}{12}}{\thesubsection}{1em}{}

\graphicspath{ {Images/} }

\newcommand{\Acal}{\mathcal{A}}

\newcommand{\Ccal}{\mathcal{C}}
\newcommand{\Dcal}{\mathcal{D}}
\newcommand{\Ecal}{\mathcal{E}}
\newcommand{\Fcal}{\mathcal{F}}
\newcommand{\Hcal}{\mathcal{H}}
\newcommand{\Ical}{\mathcal{I}}
\newcommand{\Jcal}{\mathcal{J}}
\newcommand{\Mcal}{\mathcal{M}}
\newcommand{\Wcal}{\mathcal{W}}
\newcommand{\Scal}{\mathcal{S}}

\newcommand{\Tcal}{\mathcal{T}}
\newcommand{\Ucal}{\mathcal{U}}
\newcommand{\Pprob}{\mathds{P}}
\newcommand{\ident}{\mathds{1}}
\newcommand{\Lcal}{\mathcal{L}}
\newcommand{\Rcal}{\mathcal{R}}

\newcommand{\Hscr}{\mathscr{H}}
\newcommand{\Lscr}{\mathscr{L}}

\newcommand{\inp}{\texttt{i}}
\newcommand{\out}{\texttt{o}}
\newcommand{\aux}{\texttt{a}}
\newcommand{\bux}{\texttt{b}}
\newcommand{\xt}{\texttt{x}}
\newcommand{\yt}{\texttt{y}}

\definecolor{darkorange}{RGB}{255,100,0}

\def\BraVert{\egroup\,\mid\,\bgroup}

\def\ketbra#1#2{\ket{#1\vphantom{#2}}\!\bra{#2\vphantom{#1}}}

\def\bra#1{\mathinner{\langle{#1}|}}
\def\bbra#1{\mathinner{\langle \! \langle{#1}|}}
\def\ket#1{\mathinner{|{#1}\rangle}}
\def\kket#1{\mathinner{|{#1}\rangle \! \rangle}}
\def\braket#1{\mathinner{\langle{#1}\rangle}}

\def\bbrakket#1{\mathinner{\langle \! \langle{#1}\rangle \! \rangle}}

\def\kketbra#1#2{\kket{#1\vphantom{#2}}\!\bbra{#2\vphantom{#1}}}

\def\tr#1{\mbox{tr}\left[{#1}\right]}
\newcommand{\ptr}[2]{\mbox{tr}_{#1}\left[ #2 \right]}

\makeatletter
\theoremstyle{plain}

\theoremstyle{plain}

\theoremstyle{plain}

\theoremstyle{remark}
\newtheorem*{rem*}{\protect\remarkname}
\theoremstyle{plain}

\theoremstyle{plain}

\theoremstyle{definition}

\theoremstyle{plain}
\newtheorem*{thm*}{\protect\theoremname}
\theoremstyle{plain}
\newtheorem*{lem*}{\protect\lemmaname}
\theoremstyle{definition}
\newtheorem{example}{\protect\examplename}
 
\providecommand{\propositionname}{Proposition}
\providecommand{\theoremname}{Theorem}
\providecommand{\lemmaname}{Lemma}
\providecommand{\remarkname}{Remark}
\providecommand{\conjecturename}{Conjecture}
\providecommand{\definitionname}{Definition}
\providecommand{\corollaryname}{Corollary}
\providecommand{\examplename}{Example}

\makeatletter
\renewcommand\@dotsep{500}   
\makeatother

\begin{document}

\review[Higher-Order Quantum Operations]{Higher-Order Quantum Operations}

\author{Philip Taranto \orcidicon{0000-0002-4247-3901}}
\address{Department of Physics \& Astronomy, University of Manchester, Manchester M13 9PL, \\ United Kingdom}\vspace{0.25em}
\address{Department of Physics, Graduate School of Science, The University of Tokyo, 7-3-1 Hongo, \\ Bunkyo-ku, Tokyo 113-0033, Japan}
\ead{philip.taranto@manchester.ac.uk}

\author{Simon Milz \orcidicon{0000-0002-6987-5513}}
\address{Institute of Photonics and Quantum Sciences, School of Engineering and Physical Sciences, Heriot-Watt University, Edinburgh EH14 4AS, United Kingdom}\vspace{0.25em}
\address{School of Physics, Trinity College Dublin, Dublin 2, Ireland}\vspace{0.25em}
\address{Trinity Quantum Alliance, Unit 16, Trinity Technology and Enterprise Centre, Pearse Street, \\ Dublin 2, D02YN67, Ireland}
\ead{s.milz@hw.ac.uk}

\author{Mio Murao \orcidicon{0000-0001-7861-1774}}
\address{Department of Physics, Graduate School of Science, The University of Tokyo, 7-3-1 Hongo, \\ Bunkyo-ku, Tokyo 113-0033, Japan}\vspace{0.25em}
\address{Trans-scale Quantum Science Institute, The University of Tokyo, Hongo 7-3-1, Bunkyo-ku, \\ Tokyo 113-0033, Japan}
\ead{murao@phys.s.u-tokyo.ac.jp}

\author{Marco T\'ulio Quintino \orcidicon{0000-0003-1332-3477}}
\address{Sorbonne Universit{\' e}, CNRS, LIP6, F-75005 Paris, France} 
\ead{marco.quintino@lip6.fr}

\author{Kavan Modi \orcidicon{0000-0002-2054-9901}}
\address{Science, Mathematics and Technology Cluster, Singapore University of Technology and Design, \\8 Somapah Road, 487372 Singapore}\vspace{0.25em}
\address{School of Physics and Astronomy, Monash University, Clayton, Victoria 3800, Australia}
\ead{kavan@quantumlah.org}

\vspace{10pt}

\begin{indented}
\item[] \today
\end{indented}

\begin{abstract}
An operational description of quantum phenomena concerns developing models that describe experimentally observed behaviour. \textit{Higher-order quantum operations}---quantum operations that transform quantum operations---are fundamental to modern quantum theory, extending beyond basic state preparations, evolutions, and measurements described by the Born rule. These operations naturally emerge in quantum circuit architectures, correlated open dynamics, and investigations of quantum causality, to name but a few fields of application. This Review Article provides both a pedagogical introduction to the framework of higher-order quantum operations and a comprehensive survey of current literature, illustrated through physical examples. We conclude by identifying open problems and future research directions in this rapidly evolving field.
\end{abstract}

\vspace{2pc}

\maketitle 
\newpage
\tableofcontents


\mainmatter

\section{Introduction}
\label{pt::introduction}

The advent of quantum information theory and its applications for building quantum technologies has led to a paradigm shift regarding the connection between \textit{physics} and \textit{information}. On the one hand, the laws of physics limit one's ability to store, process, and retrieve information; indeed, ``information is physical''~\cite{Landauer_1961}. In other words, information processing protocols make use of physical systems and, therefore, must abide by the laws of physics. On the other hand, information-theoretic principles have proven instrumental in axiomatically deriving quantum theory (see, e.g., Refs.~\cite{Hardy_2001,Clifton_2003,Chiribella_2011,D'ariano_Chiribella_Perinotti_2017}). This symbiosis manifests across diverse domains, from quantum computation~\cite{Preskill_1997, Preskill_2018,Preskill_2025} and cryptography~\cite{Gisin_2002, Pirandola_2020} to precision metrology~\cite{Giovannetti_2011,Toth_2014}, quantum thermodynamics~\cite{Goold_2016,Binder_2018}, and open quantum dynamics~\cite{Devega_2017,Weimer_2021}, establishing a fundamental principle: Physics defines the boundaries of information processing, while information processing shapes our understanding of physical reality.

This interplay becomes particularly crucial in quantum technologies. The fundamental challenge across quantum computation, communication, and cryptography lies in efficiently and reliably encoding, processing, transmitting, and reading quantum information---a task complicated by the inherent fragility of quantum systems. Short coherence times and thermal fluctuations in quantum devices, whether in trapped ions~\cite{Bruzewicz_2019} or superconducting systems~\cite{Siddiqi_2021,Chatterjee_2021}, as well as spurious noise harming optical setups~\cite{Kok_2007,OBrien_2007}, necessitate optimal resource utilisation to achieve meaningful quantum advantages~\cite{Acin_2018,awschalom2022roadmap}. Both static, \textit{spatial} properties (e.g., coherence, entanglement, and non-local correlations) as well as dynamic, \textit{temporal} ones (e.g., the generation, transmission, and/or detection of quantum features) are of utmost importance to understand, quantify, mitigate, and control for next-generation quantum applications. 

However, traditional formulations of quantum theory often treat space and time on an uneven footing. This asymmetry manifests itself in several ways: State preparations are typically deterministic, while measurements are inherently probabilistic. Moreover, averaging over measurement outcomes performed on part of a multi-partite state leads to a correct reduced description of the local statistics, whereas similar temporal averaging generally fails due to measurement invasiveness~\cite{Milz_2020_Quantum}. These apparent inconsistencies point to a fundamental limitation in describing quantum information processing across space and time.

\textbf{Higher-order quantum operations (HOQOs)} offer a solution through their operational approach to quantum theory. HOQOs describe transformations between standard quantum objects---states, channels, measurements, and instruments---while ensuring valid probability distributions for \textit{any} experimental intervention. While these operations often encode spatiotemporal structures of an envisaged scenario---as is the case for multi-time quantum processes~\cite{Chiribella_2008_PRL, Chiribella_2009, Pollock_2018_PRA, Pollock_2018_PRL}---they can also describe more general causally indefinite processes~\cite{Oreshkov_2012, Chiribella_2013, Shrapnel_2017}.

Within this framework, all quantum objects become operations that can be considered as \textit{linear maps}: State preparations map probability distributions to quantum states, measurements transform states into outcome probabilities, and channels convert input states to output states. Truly higher-order operations include operations that take quantum channels to quantum channels via encoding/decoding procedures and mappings from quantum instruments to probability distributions. 

The framework's power lies in its unification of spatial and temporal aspects of quantum theory and its broad realm of applicability: it both characterises spatiotemporal structures in quantum experiments and enables enhanced information processing through temporal feed-forward, circuit optimisation, and causal structure verification. For instance, feeding forward information in time is critical for benchmarking and correcting errors that are correlated in both space and time in quantum computers~\cite{White_2020,White_2021_Many,Figueroa-Romero_2021_PRXQ,Figueroa-Romero_2022,Morris_2022}. Quantum circuits can be optimised via pre- and post-processing to transform certain operations into other more desirable ones~\cite{Chiribella_2008_PRL,Chiribella_2008,Quintino_2019_PRL,Quintino_2019_PRA,Quintino_2022_Quantum}. In the fields of quantum cryptography, computational complexity, and distributed quantum computation, HOQOs can be employed to model multi-round games and protocols where correlations are shared and/or communication is possible~\cite{Kitaev2000Parallelization,ambainis_quantum_2000, Gutoski_2007,Portmann_2015} and to analyse complexity classes of quantum computation with indefinite causal structures~\cite{Baumeler2017noncausal_computation,araujo_quantum_2017,Baumeler_2016}. Lastly, spatiotemporal structures, such as the causal connection and/or causal ordering between laboratories, can be verified via operational procedures~\cite{Araujo_2015, Bavaresco_2019,Milz_2022}. In short, the quantum information community broadly understands the resourcefulness of \textit{spatial} correlations in quantum states, and the framework of HOQOs allows one to unambiguously quantify and rigorously understand \textit{spatiotemporal} resources such as non-Markovianity (memory)~\cite{Pollock_2018_PRL, Pollock_2018_PRA, Costa_2016, Taranto_2019_PRL, Taranto_2019_PRA, Taranto_2021_npj} and causal non-separability~\cite{Oreshkov_2012,Araujo_2015,Milz_2022}. 

Understandably, such operational descriptions of experimental scenarios have been (re)discovered within a variety of different contexts: analysing memory in open quantum dynamics~\cite{Kretschmann_2005,Modi_2012_SciRep,Modi_2012_PRA,Ringbauer_2015,Pollock_2018_PRL,Pollock_2018_PRA, Costa_2016}, developing quantum circuit architectures~\cite{Chiribella_2008_PRL,Chiribella_2008}, characterising correlations in complex networks~\cite{Chiribella_2009}, determining optimal quantum strategies for games and information-processing tasks~\cite{Gutoski_2007}, and describing situations with indefinite causal relations~\cite{Oreshkov_2012,Chiribella_2013,Araujo_2015,Oreshkov_2016}, to name but a few. 

In recent years, it has become increasingly apparent that these fields---while seemingly disparate---overlap significantly through their shared foundation in HOQOs. This Review Article synthesises these developments, beginning with a Tutorial section on fundamental HOQO concepts, followed by a comprehensive literature Review summarising the key applications of HOQOs in higher-order quantum information tasks (Sec.~\ref{subsec::higherorderquantumsubroutines}), open system dynamics \& memory effects (Sec.~\ref{subsec::opensystemdynamicsquantummemory}), many-time quantum physics (Sec.~\ref{subsec::manytimequantumphysics}), causality \& quantum foundations (Sec.~\ref{subsec::causalityquantumfoundations}), and finally characterisation \& experimental demonstrations (Sec.~\ref{subsec::characterisationexperimentaldemonstrations}). We conclude by outlining promising future directions in this rapidly evolving field. 


\vspace{0.25cm}\noindent
\textbf{\textul{Reader's Guide.}} This Review Article aims to provide both a pedagogical introduction and a comprehensive overview of HOQOs. We begin with an extensive Tutorial section that systematically develops the fundamental concepts, carefully deriving key results to help newcomers enter this exciting field. This part is mostly self-contained, and we only assume familiarity with the basic tenets of quantum mechanics and linear algebra (see, e.g., Ref.~\cite{Nielsen_Chuang_2010} for a basic introduction). The Tutorial is followed by a Review section that surveys major developments and relevant applications. Given that HOQOs have emerged in various contexts under different nomenclature, we have attempted to strike a balance between maintaining consistent notation throughout while respecting established conventions within different subfields. Our coverage is necessarily selective---we explore certain topics in considerable depth while presenting others more broadly, particularly in areas where detailed reviews already exist. While we have endeavoured to be thorough, the expansive nature of the field means that some relevant works may not be included. Nevertheless, we believe this structure provides both an accessible entry point for newcomers and a valuable resource for researchers already working in the field.


\section{Tutorial: Higher-Order Quantum Operations}
\label{pt::tutorial}

This Tutorial introduces the fundamental concepts and applications of higher-order quantum operations (HOQOs). We begin with three motivating examples (Secs.~\ref{subsubsec::me-opensystemdynamicswithinitialcorrelations}---\ref{subsubsec::me-causalityquantumtheory}) that build intuition. The rigorous mathematical framework is systematically developed starting in Sec.~\ref{subsec::theoreticalframework}, where we first review the basic elements of quantum theory---states, measurements, channels, and instruments---before showing how these primitives can be combined to form HOQOs. We then examine temporally-ordered multi-time quantum processes (Sec.~\ref{subsec::timeorderedquantumprocesses}) both from a constructive and axiomatic perspective, and present a foundational approach to HOQOs (Sec.~\ref{subsec::axiomatichoqos}) that reveals exotic spacetime structures beyond conventional causally ordered processes; Sec.~\ref{subsec::indefinitecausalorder} explores such causally indefinite quantum processes in detail. Throughout, our goal is to equip the Reader with both conceptual understanding and technical prowess, enabling them to apply and extend this framework across quantum theory.

\FloatBarrier


\subsection{Motivating Examples}\label{subsec::motivatingexamples}

\subsubsection{Open System Dynamics with Initial Correlations}\hfill\\
\label{subsubsec::me-opensystemdynamicswithinitialcorrelations}

\noindent Let us begin with an illuminating example from open quantum system dynamics that naturally introduces higher-order quantum operations (HOQOs) and highlights their key properties. Broadly speaking, the main goal of the study of open quantum system dynamics is to describe the evolution of an experimentally accessible system that is coupled to an experimentally inaccessible environment. Concretely, one could, for example, consider a qubit in a quantum experiment that is evolving in time and subject to noise due to unwanted interactions with its surroundings. 

\begin{figure}[t]
\centering
\subfigure[\textbf{Channel from Open System Dynamics.}]
{
\includegraphics[scale=0.5]{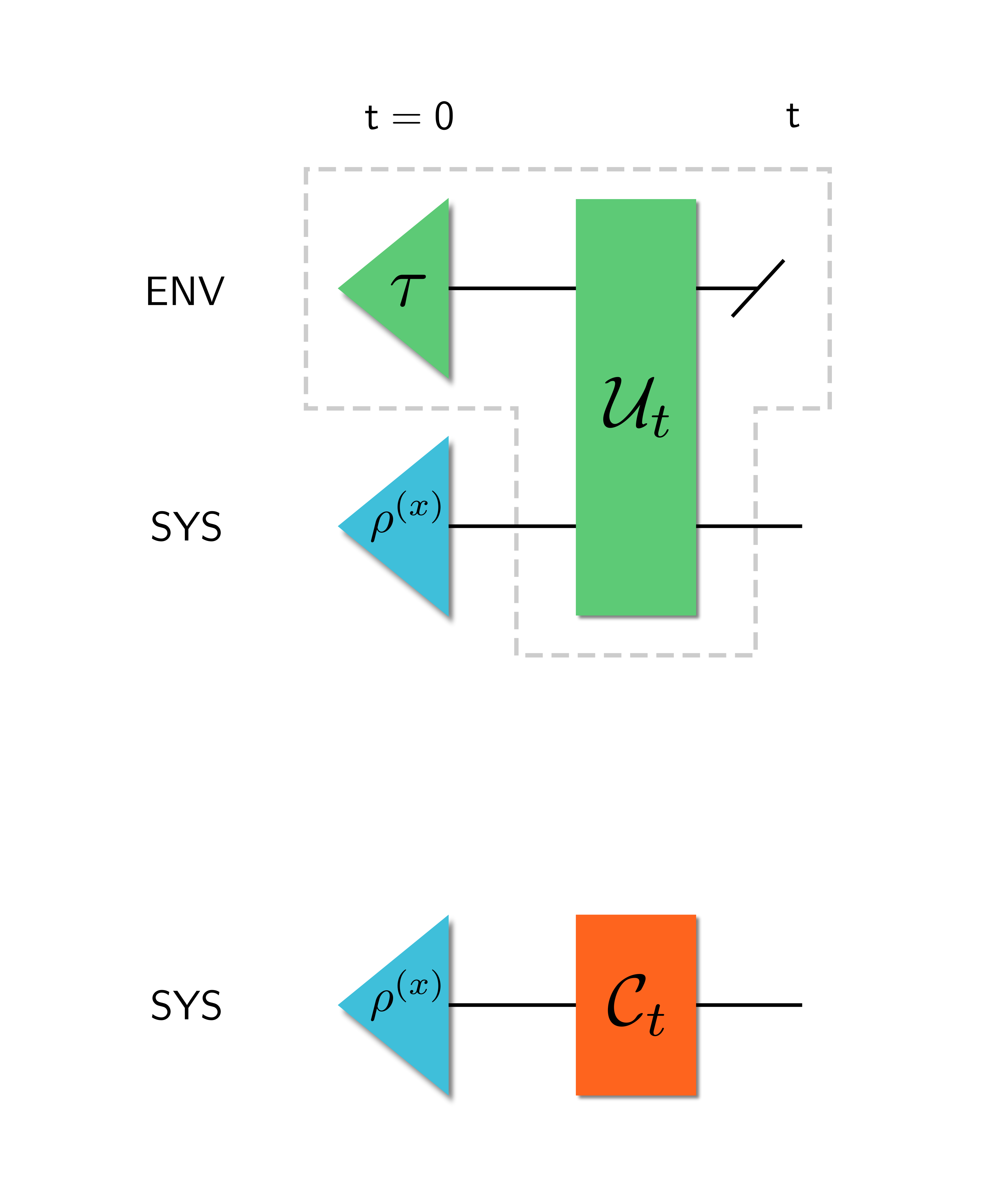}
\label{fig::me-osdic-channel}
}\hspace{1.5cm}
\subfigure[\textbf{Superchannel from Open System Dynamics.}]
{
\includegraphics[scale=0.5]{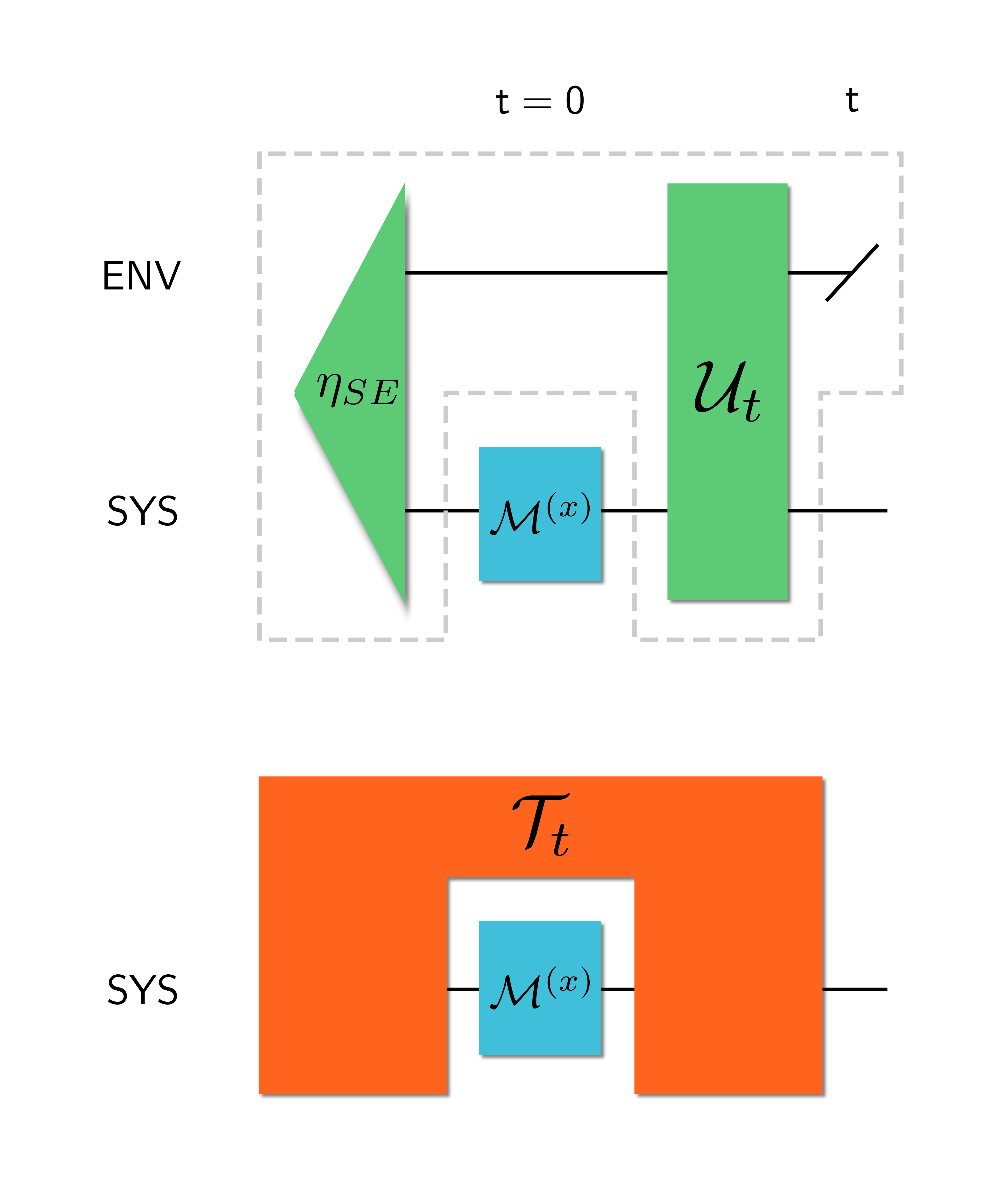}
\label{fig::me-osdic-superchannel}
}\hfill
\caption{\textbf{Open System Dynamics with Initial Correlations.} Here and throughout this Review Article, information along wires `flows' from left to right and we denote states by triangles with `output wires', operations by boxes with `input' \textit{and} `output' wires, and the trace (i.e., discarding of degrees of freedom) by a short diagonal wire. \textbf{(a)} If the system and environment begin uncorrelated, the influence of the environment (ENV) on the system (SYS) is characterised by a quantum channel (i.e., a completely positive and trace-preserving map) $\Ccal_t$ that acts on the space of the system alone, taking any input state $\rho^{(x)}$ to the corresponding output state $\rho_t^{(x)} = \mathcal{C}_t[\rho^{(x)}]$. \textbf{(b)} If the system and environment are initially correlated, the dynamical situation is operationally described by a superchannel $\Tcal_t$ that takes preparation maps (labelled by a choice of preparation $x$) $\mathcal{M}^{(x)}$ to (subnormalised) output states $\rho_t^{(x)} = \mathcal{T}_t[\mathcal{M}^{(x)}]$. This example depicts a special case of a superchannel (i.e., a HOQO with one open slot) with a trivial initial input space (i.e., there is no wire `entering' the superchannel from the left). We distinguish between the overall (higher-order) quantum operations (orange), enclosed in the grey dashed outline in the upper panels and comprising primitive elements (green), and the `input' objects upon which said operation acts (blue). While it is difficult to adhere to an unambiguous colour coding throughout---e.g., in many cases `input' objects can themselves be HOQOs---we strive to follow these conventions as much as possible.}
\end{figure}

\vspace{0.25cm}\noindent
\textbf{\textul{Uncorrelated Case.}} If the system and environment begin \textit{uncorrelated}, then any such evolution can be formally described by the mapping
\begin{align}\label{eq::me-osdic-uncorrelatedcase}
    \rho_t^{(x)} = \mathrm{tr}_E \big( \Ucal_t[\rho^{(x)} \otimes \tau] \big) =:\Ccal_t[\rho^{(x)}] \, ,
\end{align}
where $\tau$ is the initial state of the environment, $x$ denotes the choice of input state, $\mathcal{U}_t[\, \bullet\, ] := U_t \bullet U_t^\dagger$ is the unitary dynamics pertaining to the closed system-environment evolution from time $t=0$ to $t$, and $\text{tr}_E$ corresponds to a trace over (or discarding of) the environmental degrees of freedom [see Fig.~\ref{fig::me-osdic-channel}]. 


This evolution defines a \textit{quantum channel} $\Ccal_t$---a linear, \textbf{completely positive and trace-preserving (CPTP)} map acting solely on the system's Hilbert space. A quantum channel fully describes the evolution of \textit{any} input state $\rho^{(x)}$ of interest under the influence of the environment between times $t=0$ and $t$. One could reconstruct such a channel experimentally via tomography by preparing a complete basis of states $\{\hat \rho^{(x)}\}_{x=1}^{d_S^2}$ and measuring their time-evolved counterparts $\{\hat \rho_t^{(x)}\}_{x=1}^{d_S^2}$; applying linear inversion techniques to this set of data yields the full description of the channel (see Sec.~\ref{subsec::characterisationexperimentaldemonstrations}).

Noticeably, the quantum channel $\Ccal_t$ is an example of a \textit{(first-order) quantum operation}: it maps input quantum states $\rho^{(x)}$ (at time $t=0$) to output quantum states $\rho_t^{(x)}$ (at time $t$). As we will see throughout the remainder of this Tutorial, CPTP maps such as $\Ccal_t$ constitute a lower rung of an infinite ladder of conceivable \textit{higher-order} quantum operations, i.e., quantum operations that act on quantum operations, and so on. 

\vspace{0.25cm}\noindent
\textbf{\textul{Correlated Case.}} The necessity of truly \textit{higher}-order quantum operations becomes apparent when slightly altering the above example. The situation changes fundamentally when system and environment start in a \textit{correlated} state $\eta_{SE}$ at time $t=0$; this is a rather natural case to consider, since generically a system will have built up correlations with its environment due to previous interactions. Nonetheless, an experimenter may wish to probe the system dynamics by preparing a set of input states $\{\hat \rho^{(x)}\}$ as discussed previously. However, in contrast to the uncorrelated case, preparing different input states now inevitably affects the environment due to the system-environment correlations. 

For example, the experimenter might prepare an initial state by measuring the system in the computational basis and post-selecting on a particular outcome; or they might do so by discarding the system and replacing it with a fresh state (which is, by construction, uncorrelated to the environment). Let us assume that any such preparation procedure is labelled by $x$ and corresponds to a map $\Mcal^{(x)}$ that acts (linearly) on the system alone. Then, Eq.~\eqref{eq::me-osdic-uncorrelatedcase} would read [see Fig.~\ref{fig::me-osdic-superchannel}]
\begin{align}\label{eq::me-osdic-correlatedcase}
    \rho_t^{(x)} = \mathrm{tr}_E\bigg(\Ucal_t\big[(\Mcal^{(x)} \otimes \Ical_E)[\eta_{SE}]\big]\bigg) =: \Tcal_t[\Mcal^{(x)}]\, ,
\end{align}
where $\Ical_E$ is the identity map on the environment, encoding the assumption that the experimenter cannot directly influence it. 

However, crucially, the experimenter could prepare \textit{the same} initial state of the system in different ways. For instance, the state $\rho = \ketbra{0}{0}$ could be prepared by measuring in the computational basis and feeding it forward upon recording outcome $0$, or by applying a \verb!NOT! gate whenever recording outcome $1$. Generally, both preparations leave the environment in different states, thus leading to different dynamics of the system. Consequently, in contradistinction to Eq.~\eqref{eq::me-osdic-uncorrelatedcase}, Eq.~\eqref{eq::me-osdic-correlatedcase} cannot be understood as a linear map acting on initial system states as inputs as soon as $\eta_{SE}$ is correlated. Nonetheless, the equation remains linear with respect to the \textit{preparation map} $\Mcal^{(x)}$. This gives rise to a mapping $\Tcal_t$ called a \textit{superchannel}: a higher-order quantum operation mapping preparation procedures $\Mcal^{(x)}$ (applied at time $t=0$) to output states $\rho_t^{(x)}$ (at time $t$). 

We have thus encountered the superchannel $\Tcal_t$~\cite{Modi_2012_SciRep, Ringbauer_2015}, the first non-trivial example of a HOQO. That is, a map that takes operations $\Mcal^{(x)}$ to (subnormalised) states $\rho_t^{(x)}$ [cf. $\Mcal^{(x)}$ is a map that takes states to (subnormalised) states]. Although here we have a map that takes input operations to output states, we will also consider similar maps with one open slot that take input quantum channels to output channels~\cite{Chiribella_2008}; we will refer to both cases as `superchannels' throughout.\footnote{The latter are often called `supermaps' throughout the literature.} We discuss the physical relevance of this shift of perspective in more detail in Sec.~\ref{subsec::theoreticalframework} and review the usage of HOQOs in the field of open quantum system dynamics to understand complex quantum processes in Secs.~\ref{subsec::opensystemdynamicsquantummemory} and~\ref{subsec::manytimequantumphysics}. Here, we simply present some pertinent properties of the superchannel $\Tcal_t$: 
\begin{enumerate}[(i)]
    \item While it is clear how a quantum channel $\Ccal_t$ is made up of elementary building blocks---namely the joint unitary dynamics $\Ucal_t$, the initial environment state $\tau_E$, and the partial trace $\text{tr}_E$ leading to the Stinespring representation of quantum channels~\cite{Stinespring_1955}---it is \textit{a priori} unclear how to obtain $\Tcal_t$ in a similar manner from the basic elements $U_t, \rho_{SE}$ and $\text{tr}_E$. 
    \item As we mentioned, $\Ccal_t$ is CPTP, i.e., $\mathrm{tr}\left[\Ccal_t[\rho]\right] = \tr{\rho}$ and $\Ccal_t \otimes \Ical_\texttt{a}$ is a positive map for any arbitrarily-sized auxiliary system \texttt{a}. \textit{A priori}, neither of these properties seems to apply to---or even be meaningful for---a superchannel $\Tcal_t$, begging the question of what properties define a `proper' or `valid' superchannel in quantum theory. 
    \item Finally, $\Ccal_t$ and $\Tcal_t$ appear to be fundamentally different objects. While the former acts on states (i.e., matrices), the latter acts on preparations (i.e., transformations of matrices). As we will see, though, they can be treated on exactly the same mathematical footing. 
\end{enumerate}
These questions motivate our systematic development of the HOQO framework throughout this Tutorial. Before doing so, let us first provide a second example where HOQOs crop up naturally, namely in designing and optimising quantum circuit architectures.

\FloatBarrier


\subsubsection{Quantum Circuit Architecture}\hfill\\
\label{subsubsec::me-quantumcircuitarchitecture}

\begin{figure}[t]
\centering
\subfigure[\textbf{Complex Conjugation of a Qubit Unitary.}]
{
\includegraphics[width=0.47\linewidth]{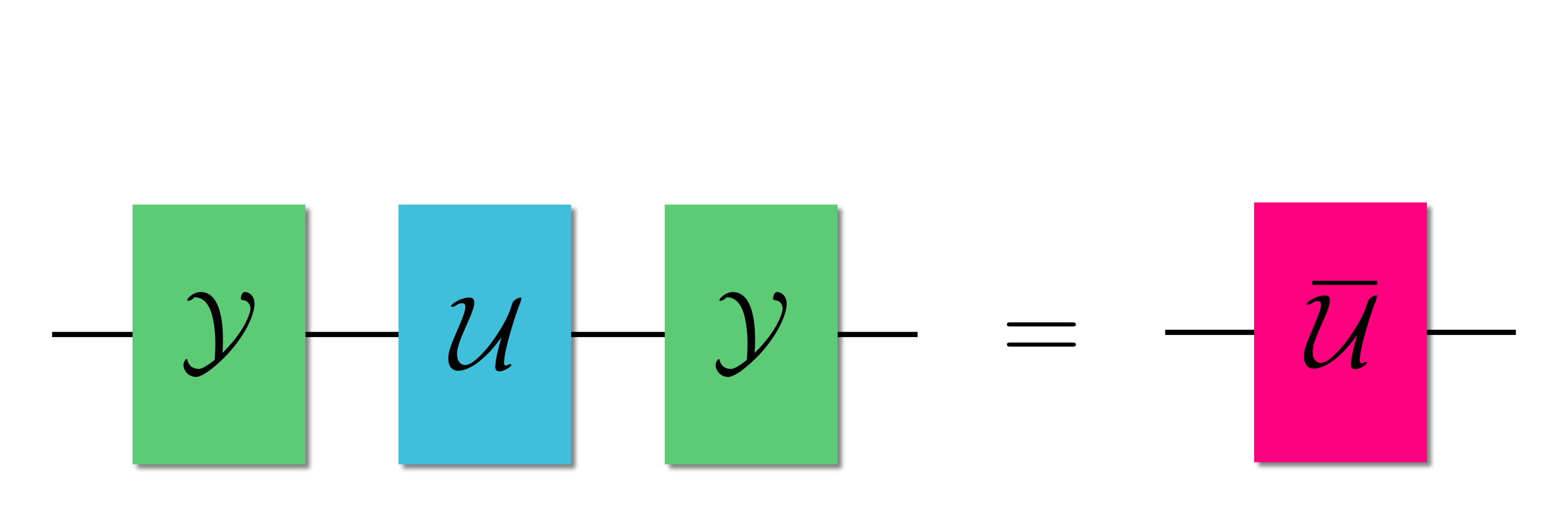}
\label{fig::me-qca-qubitunitaryconjugation}
}\hfill
\subfigure[\textbf{General Transformation of a Quantum Channel.}]
{
\includegraphics[width=0.47\linewidth]{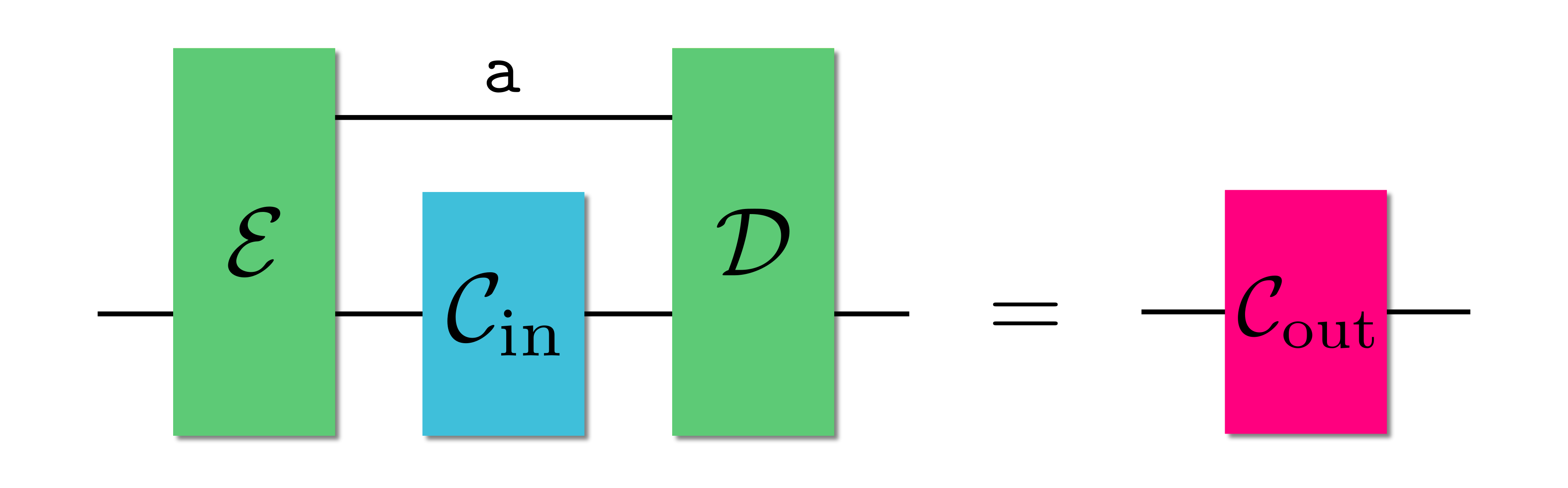}
\label{fig::me-qca-encoderdecoder}
}\hfill \vspace{0.2cm}

\subfigure[\textbf{Sequential Transformation of Multiple Channels.}]
{
\includegraphics[width = 0.7\linewidth]{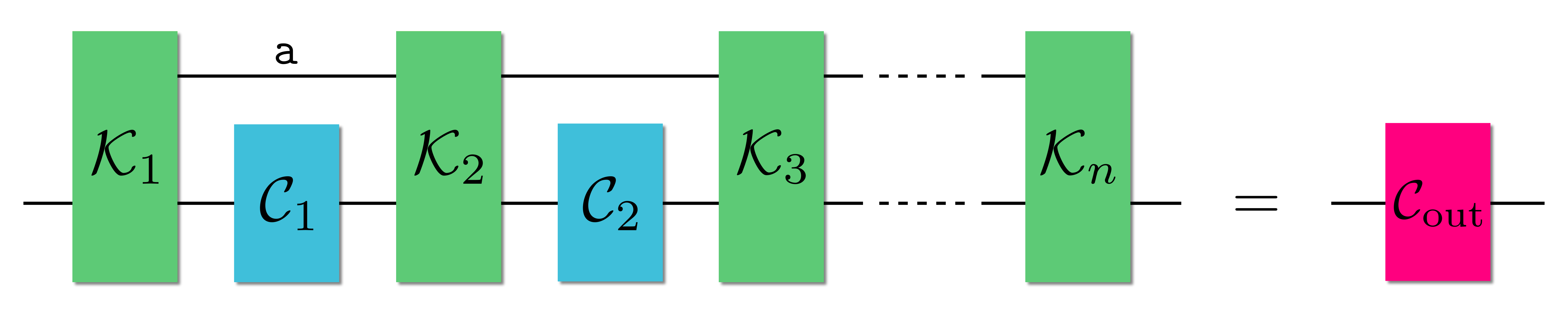}
\label{fig::me-qca-sequentialtransformationmultiplechannels}
}
\hfill
\caption{\textbf{Circuits Transforming Quantum Operations.} \textbf{(a)} The Pauli operation $\mathcal{Y}$ can be applied before and after an arbitrary qubit unitary $\Ucal$ to obtain its complex conjugate $\overline{\Ucal}$. \textbf{(b)} More generally, an encoder channel $\mathcal{E}$ and a decoder channel $\mathcal{D}$ can be used---potentially with an auxiliary system (\texttt{a})---to transform an input quantum channel $\mathcal{C}_\text{in}$ into some other channel $\mathcal{C}_\text{out}$. Such a transformation from channels to channels constitutes the more general case of a superchannel (compared with that with a trivial input space described in the previous section). \textbf{(c)} More generally still, one can concatenate a sequence of intermediate system-auxiliary channels $\{ \mathcal{K}_i \}$ to transform an entire sequence of quantum channels $\{ \mathcal{C}_{i} \}$ into another channel $\mathcal{C}_\text{out}$. In turn, each $\mathcal{K}_i$ could be further decomposed into an encoder-decoder pair, e.g., representing a measurement on the auxiliary system followed by a measurement-dependent encoder applied to the system. Here, and throughout, we depict objects in pink if we wish to emphasise that they are the output/result of the action of a HOQO applied to another object, independent of what specific type of object they represent.}
\end{figure}


\noindent Higher-order quantum operations (HOQOs) naturally emerge not only in open quantum dynamics with initial correlations but also in quantum computation paradigms that transcend the conventional prepare-evolve-measure framework. This extended setting enables us to move beyond merely enhancing the efficiency of problems with classical inputs and outputs, allowing us to explore a fully quantum information-processing paradigm where both inputs and outputs can be intrinsically quantum mechanical, without classical analogues.

From a computational perspective, standard quantum operations can be viewed as black-box devices that transform an input state $\rho_\text{in}$ into an output state $\rho_\text{out}$ (which may subsequently be measured to yield a probability distribution). While such operations act as gates---or boxes---that transform quantum states, higher-order quantum operations work on a higher level: they transform quantum operations themselves into other quantum operations. Such operations can be achieved through appropriate pre- and post-processing, potentially enabling the optimisation of entire quantum circuits for specific tasks.

The key distinction here is that the operations themselves serve as `inputs' to a higher-order `super-operation' that produces a valid `output' operation; thus, the input operations play an analogous role to that of an initial state in the standard setting. A straightforward way to transform a quantum operation is to insert it as an input into a larger quantum circuit and treat the resulting concatenated circuit as the overall output operation.
	
\vspace{0.25cm}\noindent
\textbf{\textul{Qubit Unitary Complex Conjugation.}} To illustrate this concept, let us examine a concrete example where we transform a qubit unitary gate into its complex conjugate~\cite{Chiribella_2016,Miyazaki_2019,Ebler_2022}. Consider a unitary operator $U$ representing an arbitrary unitary qubit operation $\mathcal{U}(\rho)=U\, \rho\,  U^\dagger$ and the Pauli-$y$ operator $\sigma_y := i(\ketbra{1}{0}-\ketbra{0}{1})$. Up to an irrelevant global phase,\footnote{Similarly to pure quantum states, unitary operators differing only by global phases represent identical physical operations, as verified by the identity $\mathcal{U}(\rho)=U \, \rho \, U^\dagger = (e^{i\varphi}U) \, \rho \, (e^{i\varphi}U)^\dagger$ for all real $\varphi$.} all qubit unitary operators $U$ satisfy respect
\begin{align}\label{eq::me-qca-complexconjugation}
    \sigma_y U \sigma_y= \overline{U}, 
\end{align}
where $\overline{U}$ is the complex conjugate of $U$ in the computational basis. This relationship constitutes a HOQO on the level of the operator $U$ as opposed to that of the corresponding linear map $\mathcal{U}$, i.e., left and right multiplication by $\sigma_y^\dagger$ and $\sigma_y$ respectively define the Kraus form for the HOQO implementing qubit unitary conjugation. Denoting the qubit state Pauli channel $\mathcal{Y}[\bullet] := \sigma_y \bullet \sigma_y^\dagger$, we can express the HOQO $\mathcal{T}_\ast$ achieving complex conjugation for any qubit unitary map $\mathcal{U}$ as follows. Let $\mathcal{T}_\ast(\mathcal{U})[\rho] $ denote the overall map $\mathcal{T}_\ast(\mathcal{U})$ applied to an arbitrary initial state $\rho$. Direct calculation shows that
\begin{align}\label{eq::me-qca-complexconjugationhoqo}
    \mathcal{T}_\ast(\mathcal{U})[\rho] = \mathcal{Y} \circ \mathcal{U} \circ \mathcal{Y}[\rho] = \sigma_y U \sigma_y \rho \sigma_y^\dagger U^\dagger \sigma_y^\dagger = \overline{U} \rho U^{\textup{T}} = \overline{\mathcal{U}}[\rho].
\end{align}
This holds for all $\rho$, implying that $\mathcal{T}_\ast(\mathcal{U}) = \overline{\mathcal{U}}$. One interpretation of Eq.~\eqref{eq::me-qca-complexconjugationhoqo} is: \textit{If one plugs an arbitrary qubit unitary map} $\mathcal{U}$ \textit{as an input to the circuit between two Pauli channels} $\mathcal{Y}$, \textit{then the output map is its complex conjugate} $\mathcal{U}^\prime =\overline{\mathcal{U}}$, as illustrated in Fig.~\ref{fig::me-qca-qubitunitaryconjugation}. We therefore have a constructive, \textit{universal} method to transform any unitary qubit operation into its complex conjugate by simply performing another unitary before and after it (namely $\mathcal{Y}$)---all without \textit{any} knowledge of the input unitary to be transformed (beyond is dimension). 

\vspace{0.25cm}\noindent
\textbf{\textul{Key Insights and Generalisations.}} While seemingly straightforward, this example highlights several important concepts. The pre- and post-processing operations that constitute a channel-to-channel HOQO are referred to respectively as the \textit{encoder} and \textit{decoder}. Here, the encoder and decoder are equal to each other and given by simple unitary qubit Pauli rotations $\mathcal{Y}$, chosen specifically to implement qubit unitary conjugation and applied before and after the input unitary in an uncorrelated manner. More generally, one could choose encoder-decoder pairs to achieve other desired tasks (e.g., unitary inversion, implementing linear functions, etc.). Furthermore, they could be unrelated to each other and could incorporate noise, therefore possibly corresponding to (non-unitary) quantum channels. Thus, a generic \textit{uncorrelated} (uc) encoder-decoder circuit to transform quantum operations is formed by pair of quantum channels $\mathcal{E}$ and $\mathcal{D}$ that transforms an arbitrary input quantum channel $\mathcal{C}_\text{in}$ (again, not necessarily unitary) into an output one $\Ccal_\text{out}$ by sequentially composing the operations:
\begin{align}\label{eq::me-qca-encoderdecodermarkov}\mathcal{C}_\text{out}=\Dcal\circ\Ccal_\text{in}\circ\Ecal =: \Tcal_{\textup{uc}}(\Ccal_\text{in}). 
\end{align}
Here, the fact that the encoder and decoders are themselves quantum channels ensure that the output is a valid quantum channel for any input quantum channel, i.e., $\Tcal_{\textup{uc}}$ is a valid HOQO. In terms of applicability, suppose that one can implement some channel $\Ccal_\text{in}$, but desires to achieve a certain task (e.g., quantum state discrimination) for which it is known that some other channel, $\Ccal_\text{out}$ performs better. Then, one possible way to improve performance would be to construct an appropriate encoder $\Ecal$ and decoder $\Dcal$, i.e., construct $\Tcal_{\textup{uc}}$, such that Eq.~\eqref{eq::me-qca-encoderdecodermarkov} yields the desired channel.

Moreover, the encoder and decoder channels may additionally be \textit{correlated} via an auxiliary quantum system that is independent of the Hilbert space upon which the input operation $\Ccal_\text{in}$ acts, as shown in Fig.~\ref{fig::me-qca-encoderdecoder}. More precisely, a general encoder-decoder scheme transforms an input operation $\Ccal_\text{in}$ into the output operation $\Ccal_\text{out}$ according to
	\begin{align}\label{eq::me-qca-encoderdecodergeneral}
		\mathcal{C}_\text{out} =\Dcal\circ(\Ical_\aux\otimes \Ccal_\text{in})\circ\Ecal =:\Tcal_{\text{SC}}(\Ccal_\text{in}),
	\end{align}
where $\Ical$ denotes the identity map (here, the maps $\Ecal$ and $\Dcal$ act on the principle system $S$ and auxiliary labelled by $\aux$; cf. the uncorrelated scheme of Eq.~\eqref{eq::me-qca-encoderdecodermarkov} where they both only act on $S$). The subscript on $\Tcal_{\text{SC}}$ stands for \textit{superchannel}, which here has a global past and thus take quantum channels to quantum channels (in contrast to that of the form in Eq.~\eqref{eq::me-osdic-correlatedcase}, which takes input channels to output states). As discussed later in this Tutorial, under reasonable physical assumptions, every transformation that takes any single input quantum channel into an output quantum channel can be realised in this way, i.e., by composing an encoder and decoder channel (which may make use of a sufficiently large auxiliary space to correlate them)~\cite{Chiribella_2008,Chiribella_2009}. Hence, this constructive encoder-decoder circuit approach to transform quantum operations provides a completely general characterisation of superchannels.

Following this circuit-based approach, another relevant paradigm consists of transformations where an input quantum channel can be used several times sequentially, as represented in Fig.~\ref{fig::me-qca-sequentialtransformationmultiplechannels}. In general, the input channel queried at each time need not be the same. To describe such a setting, one can concatenate sequences of (potentially different) system-auxiliary channels $\{\mathcal{K}_i\}$ over the auxiliary system to build what is called a \textit{quantum comb}~\cite{Chiribella_2008_PRL}. Again, the quantum comb could help one transform sequences of sub-optimal channels for a given task into a better one.


\begin{figure}[t]
\centering
\includegraphics[scale=.4]{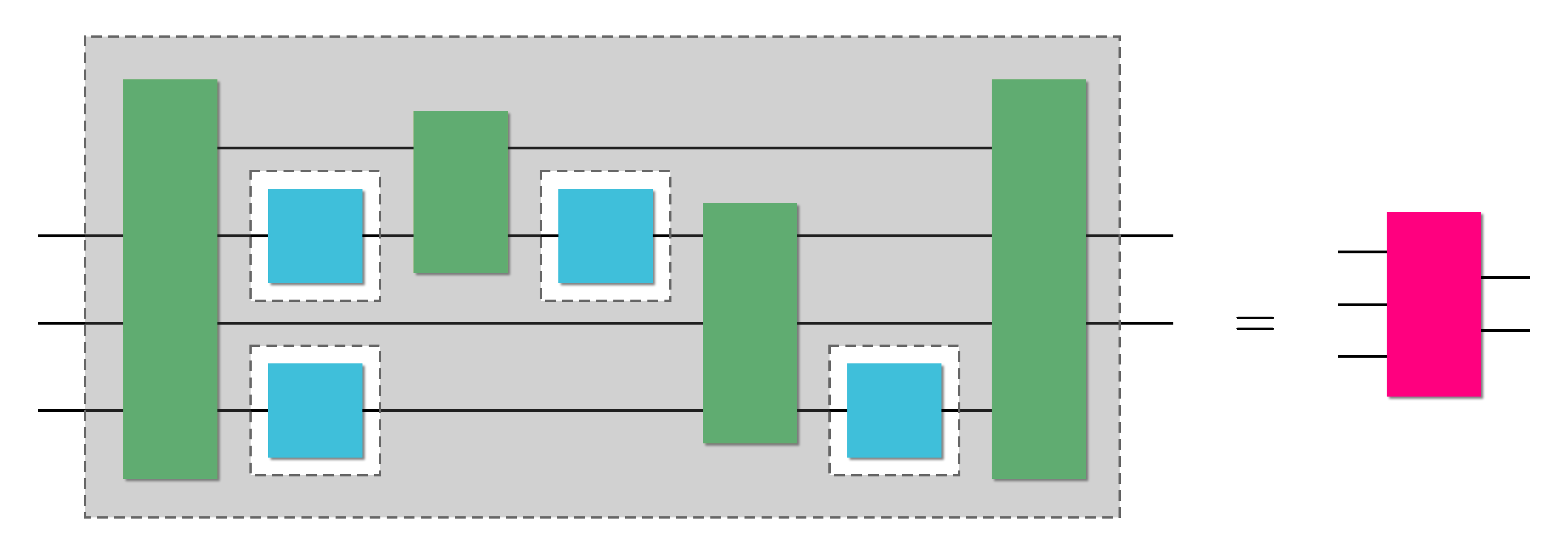}
\caption{\textbf{Customisable Quantum Circuit Architecture.} A quantum circuit where some parts (green, within the shaded area) are fixed, but other parts (blue) may be `plugged in' at the desire of the experimenter in order to customise the overall circuit to achieve a particular goal, i.e., the channel (pink) on the r.h.s.}
\label{fig::me-qca-customisablequantumcircuitarchitecture}
\end{figure}


Lastly, the scope of HOQOs extends beyond simple operations to encompass modifications of entire quantum circuits, complex networks, and probabilistic circuits. A compelling future application lies in modular quantum computing: imagine pre-fabricated quantum circuits where certain components remain fixed while others serve as customisable `plug-and-play' elements~\cite{Thompson_2018}, allowing users to tailor circuits for specific applications (see Fig. \ref{fig::me-qca-customisablequantumcircuitarchitecture}). Similar, albeit distinct, settings that we will discuss below concerns the ability for one to \textit{verify} properties of quantum devices, \textit{estimate} quantum operations, and \textit{discriminate} between them. Notably, these tasks can be accomplished \textit{without} requiring direct access to the internal mechanisms of the devices---a significant practical advantage. The mathematical framework of HOQOs provides the tools necessary for analysing all these scenarios, supporting various quantum information processing tasks such as transforming~\cite{Chiribella_2016,Miyazaki_2019,Quintino_2019_PRL,Quintino_2019_PRA,Quintino_2022_Quantum,Ebler_2022,Yoshida_2022,Yoshida_2023,Yoshida_2024,Chen_2024,Mo_2024,Zhu_2024_Reversing}, discriminating~\cite{Bavaresco_2021,Bavaresco_2022}, and estimating/learning operations~\cite{Bisio_2009,Bisio_2010,Zhao_2024,Raza_2024}, amongst others~\cite{Sedlak_2019,Chen_2024_Hypothesis,Zhu_2024_Optimal}.

Throughout this Tutorial, we will explore the fundamental properties of HOQOs within the context of quantum circuit architecture design. Like their quantum channel counterparts, HOQOs must satisfy specific physical constraints to remain valid---constraints we will carefully detail. By controlling the parameters that characterise these operations, we can design optimal circuits (or at minimum, effective black-box implementations) to achieve desired quantum processes, which is an important application in quantum information processing.

Furthermore, just as quantum state tomography allows us to characterise quantum channels through input-output relationships, one can probe the structure of HOQOs by analysing how they transform different input operations. This method proves particularly powerful for verifying spatiotemporal structures in which said operation is embedded, providing a mechanism to test the principles of quantum foundations. We will review developments in line with this perspective throughout Sec.~\ref{subsec::characterisationexperimentaldemonstrations}. Before moving on to present the general formalism and tackle such questions, we consider one last motivating example.

\FloatBarrier


\subsubsection{Causality in Quantum Theory}\hfill\\
\label{subsubsec::me-causalityquantumtheory}

\begin{figure}[t]
    \centering
    \hspace{-1em}\subfigure[\textbf{Sequential Process.}]
    {
    \includegraphics[scale =0.56]{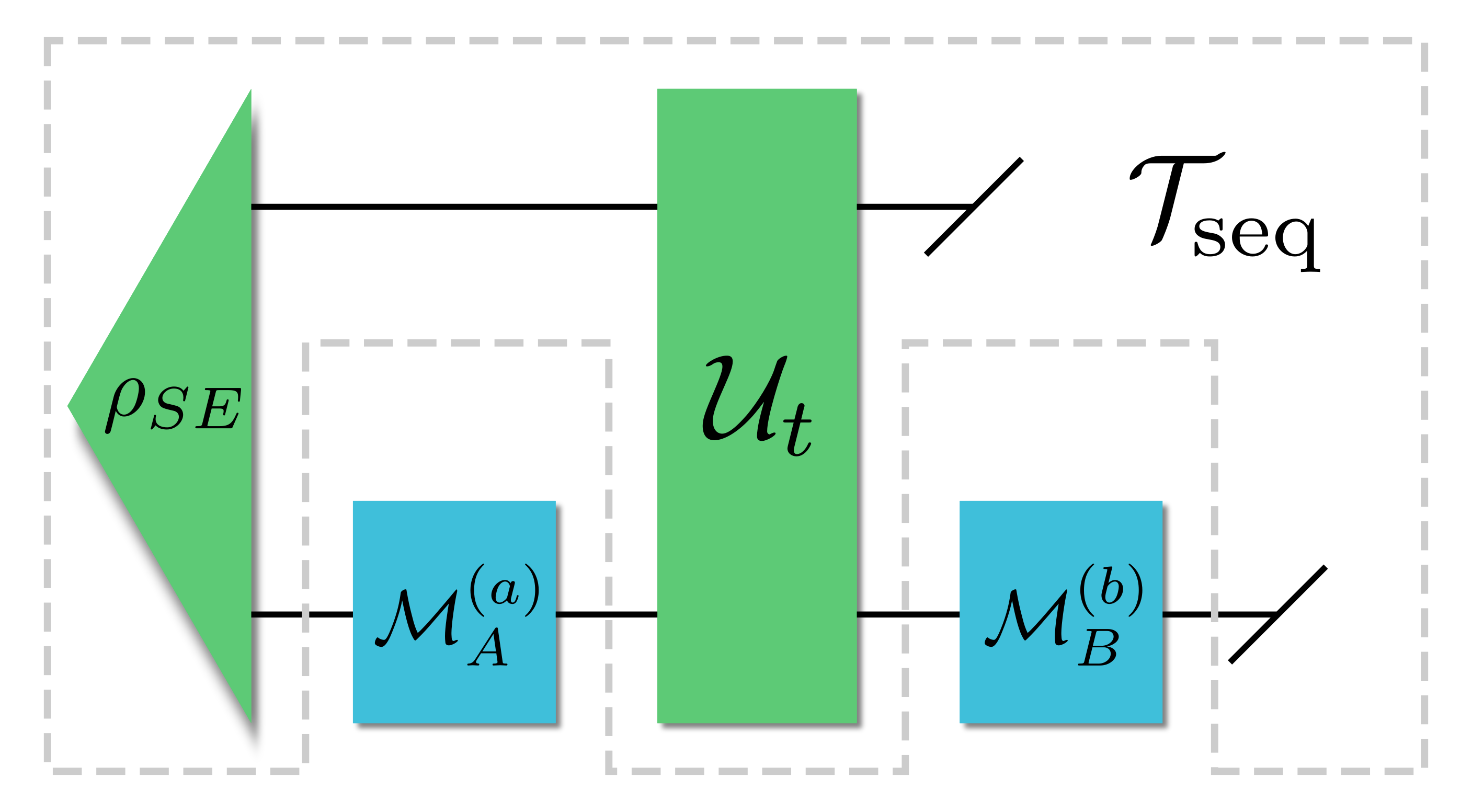}
    \label{fig::me-cqt-sequentialprocess}
    }\hspace{0.1cm}
    \subfigure[\textbf{Parallel Process.}]
    {
    \includegraphics[scale =0.56]{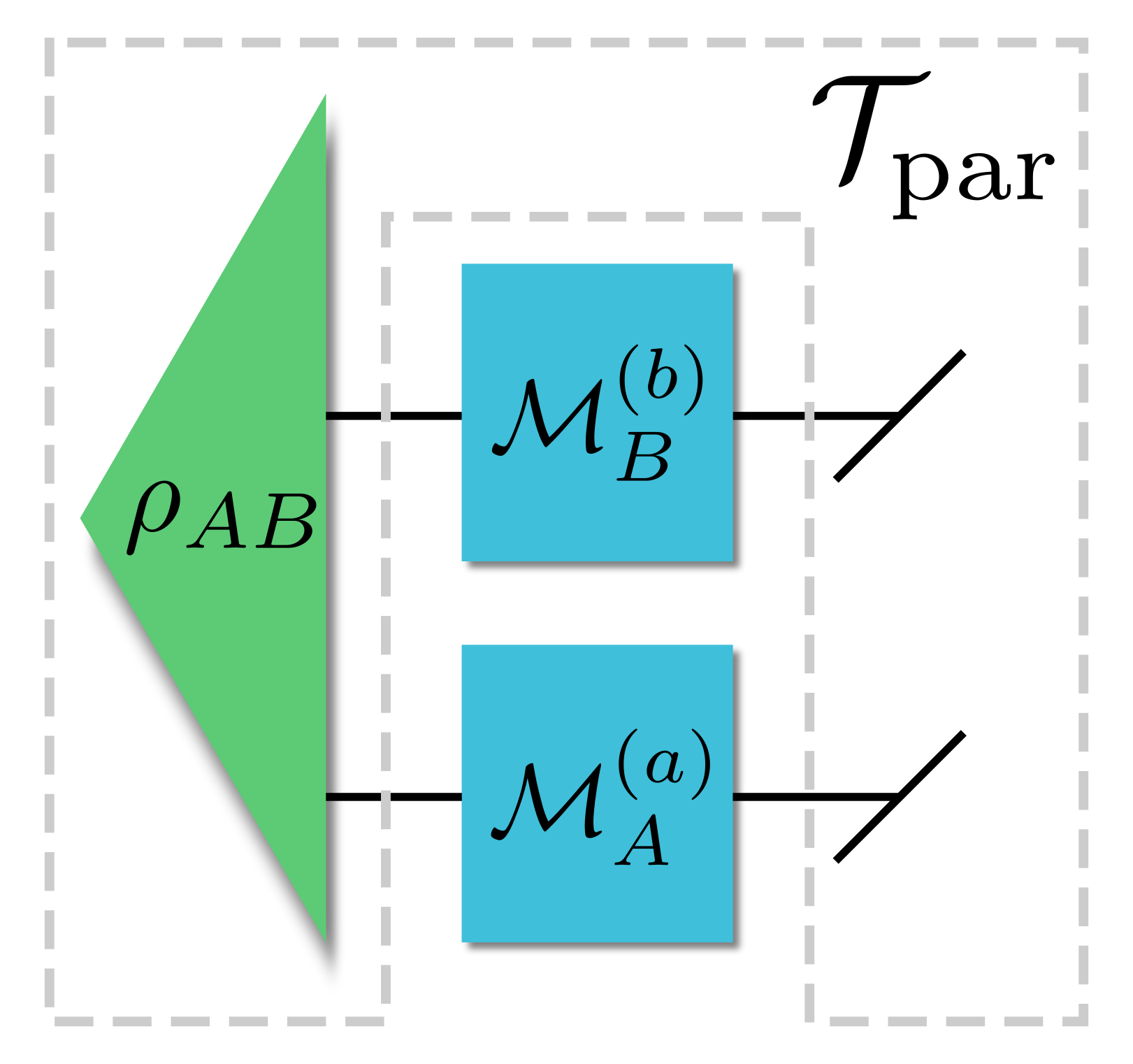}
    \label{fig::me-cqt-parallelprocess}
    }\hspace{0.cm}
    \subfigure[\textbf{Causally Indefinite Process.}]
    {
    \includegraphics[scale=0.56]{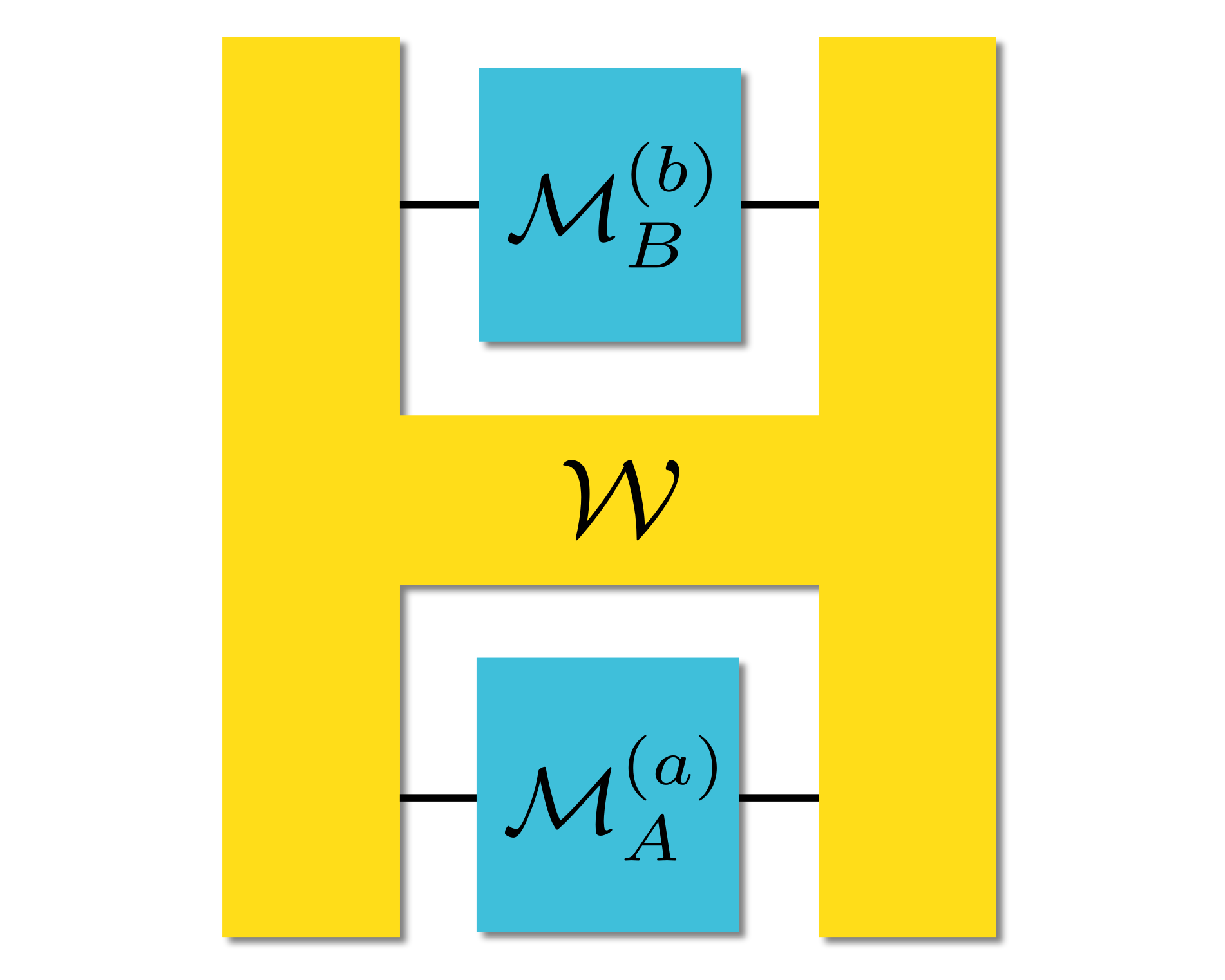}
    \label{fig::me-cqt-causallyindefiniteprocess}
    }\hfill
    \caption{\textbf{Causal Structures of Quantum Processes.} \textbf{(a)} Alice and Bob interrogate a quantum system that enters their respective laboratories. Here, the state that Alice feeds forward enters Bob's laboratory, while the state that Bob feeds forward is discarded. Most generally, the state that enters Alice's laboratory is correlated with some inaccessible environment and undergoes some unitary evolution between Alice and Bob (green). The resulting causally ordered HOQO $\mathcal{T}_{\text{seq}}$ is depicted by the dotted grey boundary [see Eq.~\eqref{eq::me-cqt-sequential}]. \textbf{(b)} In another scenario, Alice and Bob could be causally disconnected. Here, the most general setting is that they each make measurements on two independent subsystems of a joint initial state (green). Again, the resulting HOQO $\mathcal{T}_{\text{par}}$ is depicted by the dotted grey outline, but in this case of causal independence, the map simply boils down to the initial joint state that they share [since the post-measurement states are irrelevant; see Eq.~\eqref{eq::me-cqt-parallel}]. \textbf{(c)} In principle, the global causal order between Alice and Bob could be indefinite and therefore be described by a HOQO $\mathcal{W}$ that does not allow for a fixed causally ordered representation [like, for example, the ones in (a) and (b)], depicted by the yellow object. It is not necessarily composed of primitive elements such as states and unitaries, but nonetheless leads to a valid probability distribution for any choices of instruments applied by Alice and Bob [see Eq.~\eqref{eq::me-cqt-causallyindefinite}].}
\end{figure}


\noindent While we have explored HOQOs in open quantum systems and circuit architectures---where they emerge from clear physical pictures involving quantum circuits or global unitary dynamics---their applications extend into more `exotic' territory: quantum processes that may not possess a pre-defined global causal order. This final motivating example explores how HOQOs provide a natural framework for analysing such scenarios.

\vspace{0.25cm}\noindent
\textbf{\textul{Three Causal Scenarios.}} Consider two parties, Alice and Bob, performing experiments in separate laboratories. In each experimental run, these two players perform measurements on a quantum system that enters their laboratory, record measurement outcomes, and send forward the resulting quantum state. More precisely, Alice implements an \textit{instrument} $\Jcal_A = \{ \Mcal_A^{(a)}\}_{a=1}^{n_A}$ on the incoming state $\rho_{A^\inp}$, obtaining outcomes $\{a\}$ and sending forward states $\{ \Mcal_A^{(a)}(\rho_{A^\inp})\}$. Similarly, Bob applies the instrument $\Jcal_B = \{\Mcal_B^{(b)}\}_{b=1}^{n_B}$. (We will formally define instruments in Sec.~\ref{subsubsec::tf-statesmeasurementschannelsinstruments}.) This setup allows one to explore three distinct causal scenarios---sequential, parallel, and causally indefinite---and compute the associated joint probability distribution over outcomes $\{a,b\}$ that the two players might observe. 

\textit{Sequential.---}Consider first a scenario where the laboratories of Alice and Bob are connected in a causally ordered way---say, with Alice's output feeding into Bob's laboratory. The most general quantum mechanical description involves: Alice first probing the system part of a global state $\rho_{SE}$, the post-measurement state feeding forward and interacting again with the environment via some global unitary $\Ucal$, before Bob finally accessing the system [see Fig. \ref{fig::me-cqt-sequentialprocess}]. In this case, the probabilities for Alice and Bob to respectively observe outcomes $a$ and $b$, given that they employed the instruments $\Jcal_A$ and $\Jcal_B$, are computed by combining the underlying `building blocks' into a single HOQO, such that 
\begin{align}\label{eq::me-cqt-sequential}
    \Pprob(a,b|\Jcal_A, \Jcal_B) = \tr{(\Mcal_B^{(b)}\otimes \Ical_E)\circ \Ucal \circ (\Mcal_A^{(a)}\otimes \Ical_E) [\rho_{SE}]} =: \Tcal_{\text{seq}}[\Mcal_A^{(a)}, \Mcal_B^{(b)}]\, ,
\end{align}
where $\Tcal_{\text{seq}}$ is a linear map acting on the quantum operations $\Mcal_A^{(a)}$ and $\Mcal_B^{(b)}$. This is analogous to the superchannel $\Tcal_t$ of Eq.~\eqref{eq::me-osdic-correlatedcase}, with the slight difference being that, here, two quantum operations are mapped to probabilities, whereas in the previous example, $\Tcal_t$ mapped a quantum operation to a (potentially unnormalised) quantum state. Note finally that one could envisage a situation where Bob first has access to the system and Alice comes second, in which case a similar expression to above would hold, but with the order of the local interrogations switched. 

\textit{Parallel.---}Next, imagine a scenario where Alice and Bob have no influence on each other, i.e., they are causally independent and their instruments act upon distinct subsystems of some (potentially correlated) initial global state $\rho_{AB}$ [see Fig.~\ref{fig::me-cqt-parallelprocess}]. In this case, the initial state $\rho_{AB}$ itself acts as the HOQO, mapping measurement pairs to probabilities via the Born rule
\begin{align}\label{eq::me-cqt-parallel}
    \Pprob(a,b|\Jcal_A, \Jcal_B) = \tr{ (\Mcal_A^{(a)}\otimes \Mcal_B^{(b)}) [\rho_{AB}]} =: \Tcal_{\text{par}}[\Mcal_A^{(a)},\Mcal_B^{(b)}].
\end{align}
This \textit{common cause} scenario attributes correlations to shared past interactions between the subsystems~\cite{Costa_2016,Feix_2017,Allen_2017,Guo_2021,Nery_2021}.

The HOQO in Eqs.~\eqref{eq::me-cqt-sequential} and~\eqref{eq::me-cqt-parallel} possesses the information about the particular structural information about the underlying physical situation at hand; in particular, the map in Eq.~\eqref{eq::me-cqt-sequential} accounts for the unitary dynamics through which Alice's outcome can influence Bob's, whereas that in Eq.~\eqref{eq::me-cqt-parallel}---since it simply comprises a trace with the joint state, $\Tcal_{\textup{par}}[\,\bullet\, , \,\circ \,] = \tr{(\bullet \otimes \circ) \,\rho_{AB}}$---encodes the fact that Alice and Bob cannot exert any causal influence over each other.

\textit{Causally Indefinite.---}Moving beyond scenarios with well-defined causal order established by some underlying circuit, one might ask: \textit{What is the most general rule for computing joint probabilities that agrees with quantum physics and respects causality locally} (i.e., inside the respective laboratories) \textit{but not necessarily globally} (i.e., there is no assumption of an underlying circuit)? The answer comes in the form of a linear mapping $\Wcal$ (see, e.g., Refs.~\cite{Oreshkov_2012,Chiribella_2013,Shrapnel_2017} and Sec.~\ref{subsec::causalityquantumfoundations}) which yields these joint probabilities via
\begin{align}\label{eq::me-cqt-causallyindefinite}
    \Pprob(a,b|\Jcal_A, \Jcal_B) = \Wcal[\Mcal_A^{(a)}, \Mcal_B^{(b)}], \quad \text{with} \quad \Wcal[\Mcal_A, \Mcal_B] = 1 \quad \forall \, \Mcal_A, \Mcal_B\, \in \mathsf{CPTP} ,
\end{align}
where $\mathsf{CPTP}$ is the set of all CPTP maps [see Fig.~\ref{fig::me-cqt-causallyindefiniteprocess}]. Specifically, Eq.~\eqref{eq::me-cqt-causallyindefinite} guarantees that CPTP maps are implemented deterministically, i.e., with unit probability, but imposes no further structure beyond that. While all HOQOs $\Tcal$ that emerge from some underlying dynamics (e.g., an initial state and some global unitary) satisfy the properties of a map $\Wcal$ as defined above, there exist HOQOs that satisfy Eq.~\eqref{eq::me-cqt-causallyindefinite} but \textit{cannot} be represented by any quantum circuit (or convex combinations thereof)~\cite{Oreshkov_2012,Araujo_2015,Oreshkov_2016}. Employing HOQOs thus enables the investigation of \textit{causally indefinite} processes. Although the physical reality of such processes remains debated~\cite{Araujo_2017,Oreshkov_2019,Wechs_2021,wechs_existence_2023}, their mathematical description closely parallels that of causally ordered processes, differing only in specific structural properties (detailed in Sec.~\ref{subsec::causalityquantumfoundations}). Most importantly, this framework allows us to treat all HOQOs---whether causally ordered or not---within a unified, versatile formalism that makes their respective properties transparent.

\vspace{0.25cm}\noindent
\textbf{\textul{Summary.}} The three motivating examples presented above demonstrate the remarkable versatility of HOQOs across quantum physics. Their application spans open quantum system dynamics, quantum circuit architectures, and the investigation of causal order. Other examples include---but are not limited to---the study of quantum games~\cite{Gutoski_2007}, within which distributed parties can maximise their success probability of a game including measurements and guesses by optimising a corresponding strategy (which corresponds to the HOQO they plug their measurements into); the analysis of entropy accumulation, used in the security analysis of device-independent quantum key distribution~\cite{Dupuis_2020,Metger_2022}; and the field of quantum causal modelling~\cite{Costa_2016, Allen_2017,Cotler_2019,Barrett_2019}, where active interventions are used to determine causal relations between different events (the former being fixed by the HOQO that describes the causal connections and signalling possibilities encoded in the experimental situation at hand). While these various applications may appear distinct, with different perspectives and underlying assumptions, they share a common mathematical foundation: HOQOs. The primary aim of this Review Article is to present a unified framework that not only encompasses these diverse applications but also illuminates their similarities and differences.

\FloatBarrier


\subsection{Theoretical Framework}
\label{subsec::theoreticalframework}

We will now move to present the theoretical framework that encompasses the aforementioned scenarios, following much of the discussion of Refs.~\cite{Chiribella_2009, Bisio_2011}. As we have seen, there are two natural ways of motivating the theory of HOQOs: a \textit{constructive} one and an \textit{axiomatic} one. Regarding the former approach, HOQOs have emerged as natural descriptors of general quantum circuits~\cite{Chiribella_2008_PRL,Chiribella_2009}, quantum games~\cite{Gutoski_2007}, non-Markovian open dynamics~\cite{Pollock_2018_PRA,Pollock_2018_PRL}, or non-anticipatory quantum channels (and transformations thereof)~\cite{Kretschmann_2005,Caruso_2014,Portmann_2015}. We have seen this explicitly above in the examples of open quantum system dynamics in the presence of initial correlations as well as general circuit architectures, where HOQOs followed naturally from the respective underlying dynamical building blocks. 

On the other hand, HOQOs can be motivated in purely axiomatic terms as the set of valid (in quantum mechanics) transformations of quantum operations, transformations of transformations of quantum operations, and so on~\cite{Hardy_2001,Hardy_2009,Hardy_2012,Hardy_2015,D'ariano_Chiribella_Perinotti_2017,Perinotti_2017,Bisio_2019}. An example for this vantage point was the discussion of causal order in quantum mechanics in the previous section. While the constructive approach leads---by construction---to HOQOs that encode causal ordering (stemming from the spatiotemporal relations of the underlying circuit), the axiomatic approach offers greater flexibility. When causality is not explicitly imposed---either directly or through other requirements (see Sec.~\ref{subsubsec::axiomatichoqos-quantumcomb})---it can describe more general scenarios where causality holds only locally, without requiring a global causal order~\cite{Oreshkov_2012,Chiribella_2013,Araujo_2015,Oreshkov_2016}. Here, we first lay out the constructive approach to HOQOs and subsequently circle back to the more general axiomatic one in order to provide a well-rounded picture.

\FloatBarrier


\subsubsection{Notation and Linear Transformations}\hfill\\
\label{subsubsec::tf-notationlineartransformations}

\noindent Throughout this Review Article, we only consider finite dimensional complex linear spaces $\mathscr{H}$; hence $\mathscr{H}$ is always isomorphic to $\mathbb{C}^d$ for some dimension $d\in\mathbb{N}$, and we use the words `operator' and `matrix' interchangeably. With respect to a Hilbert space $\mathscr{H}$, we denote the set of linear operators thereupon as $\mathscr{L}(\mathscr{H})$, the set of quantum states by $\mathsf{St}(\mathscr{H})$, and the set of positive semidefinite matrices by $\mathsf{Pos}(\mathscr{H})$; all of these form convex sets, meaning that any element $X$ can be constructed via a convex mixture of other elements in the set, $\sum_i p_i X^i$, where $\{ p_i \}$ forms a probability distribution (i.e., a collection of non-negative real numbers summing to unity). We will also typically restrict ourselves to discrete measurement outcomes $x \in \mathds{N}$, although with careful analysis most concepts can be extended to the continuous outcome setting.

Regarding notation, we will often label the system that a state corresponds with by capital Latin letters $A, B, ...$ as subscripts, e.g., $\rho_A \in \mathscr{L}(\mathscr{H}_A)$. On the other hand, we will typically denote summation with $i,j,...$, time with $t$, discrete times with $k$, and outcome labels as $(a), (b), \ldots$ superscripts. We will occasionally omit either type of label wherever no confusion can arise, and in very few cases (i.e., whenever the notation becomes too dense) we will make exceptions. We will furthermore often require spaces that are isomorphic to one of interest $X$, in which case we will label these with a prime, i.e., $X \cong X^\prime$ is such that $\text{dim}(X) = \text{dim}(X^\prime)$. Whenever clear from context, we write $X \cong X^\prime$ instead of $\mathscr{H}_{X}\cong \mathscr{H}_{X'}$ and $d_X = d_{X^\prime}$ instead of $\text{dim}(\mathscr{H}_{X})=\text{dim}(\mathscr{H}_{X'})$. Lastly, we will denote reduced states of composite systems by removing the label of the ignored system, i.e., $\rho_A := \ptr{B}{\rho_{AB}}$.

For better bookkeeping, throughout this Review Article we will often distinguish between the input (\inp) and output (\out) space of a map (and also use them as subscripts to denote spaces), even if their dimensions coincide, such that a map $\Fcal: \mathscr{L}(\mathscr{H}_\inp) \rightarrow \mathscr{L}(\mathscr{H}_\out)$ takes linear operators acting on an input Hilbert space $\mathscr{H}_\inp$ to those on an output Hilbert space $\mathscr{H}_\out$. We will, however, often drop the Hilbert space labels where there is no risk of confusion. 

In order to intuitively present HOQOs, we begin by defining \textit{linear maps}, which make up the primitive building blocks of HOQOs. 
\begin{myDefinition}{Linear Maps}{}
A linear map is a function $\Fcal: \mathscr{L}(\mathscr{H}_\inp) \rightarrow \mathscr{L}(\mathscr{H}_\out)$ that satisfies 
\begin{align}
\label{eq::tf-def-lineartransformation}
    \Fcal(\alpha A + \beta B) = \alpha \Fcal(A) + \beta \Fcal(B)
\end{align}
for all complex numbers $\alpha, \beta \in \mathds{C}$ and all operators/matrices $A, B \in \mathscr{L}(\mathscr{H}_\inp)$.
Any linear map $\Fcal$ acting on an operator/matrix $A$ can be written as
\begin{align}\label{eq::tf-def-lineartransformationaction}
    \Fcal(A)= \sum_k L_k \, A \, R_k^\dagger = A',
\end{align}
where $L_k, R_k: \mathscr{H}_\inp \to \mathscr{H}_\out$ are linear operators.
\end{myDefinition}

\noindent Above, the matrix $A$ belongs to an \textit{input} space $\mathscr{L}(\mathscr{H}_\inp)$, which is mapped to a matrix $A'$ belonging to the \textit{output} space $\mathscr{L}(\mathscr{H}_\out)$. Here, $L_k$ ($R_k$) are the left (right) operators such that $L_k, R_k: \mathscr{H}_\inp \to \mathscr{H}_\out$. When $d_\inp \ne d_\out$, $L_k$ and $R_k$ are rectangular matrices.

The fact that physical quantum operations must be linear (or at least convex linear) in their arguments follows from the linearity of the mixing principle, which is essential to any reasonable probabilistic theory~\cite{Barrett_2005,Masanes_2011,Hardy_2015}. We will show below that both quantum measurement and quantum dynamics can be cast as linear maps acting on quantum states, thereby placing fundamentally different objects on a similar footing. Viewing all quantum objects as linear maps in this way also lends itself to considering higher-order quantum operations, which themselves are nothing but linear maps acting on `larger' spaces; thus, this definition will provide a natural foundation for understanding HOQOs.\footnote{In this Review Article, we reserve the term `HOQOs' to refer to \textit{linear} maps, thus excluding analogous models such as closed time-like curves~\cite{Deutsch_1991, Bennett_2005, Lloyd_2011_PRL, Lloyd_2011} and multi-time states~\cite{Aharonov_1964, Aharonov_1990,silva_connecting_2017}, which generally exhibit non-linear features.} 


\subsubsection{Quantum States, Measurements, Channels, and Instruments}\hfill\\
\label{subsubsec::tf-statesmeasurementschannelsinstruments}

\noindent\textbf{\textul{Quantum States.}} We are now in a position to present the lowest order quantum linear operation, i.e., a quantum operator. In particular, we will focus on \textit{quantum states}, as described by unit-trace, positive semidefinite matrices acting on a Hilbert space $\mathscr{H}$. We have the following: 
\begin{myDefinition}{States}{}
A quantum state is an operator $\rho \in \mathscr{L}(\mathscr{H})$ satisfying
\begin{align}\label{eq::tf-smci-def-quantumstates}
    \rho \geq 0 \quad \text{and} \quad \tr{\rho} = 1.
\end{align}
\end{myDefinition}
\noindent Throughout this Review Article, we use the term `state' to mean any positive semidefinite operator with bounded trace; the normalisation condition (unit trace) may be relaxed when clear from context. Apart from quantum states, there are two other major facets to quantum theory: \textit{measurements} and \textit{dynamics}. Below we will show that each of these is a linear transformation acting on quantum states. The former is a mapping from states (operators) to probabilities (real positive numbers); the latter is a mapping from one state to another state. Thus, these seemingly distinct concepts can be unified through the lens of linear maps, providing a cohesive mathematical structure. We begin with quantum measurements.

\vspace{0.25cm}\noindent
\textbf{\textul{Quantum Measurements.}} Quantum states represent the static component of quantum mechanics, containing all information that can be extracted through measurements. These \textit{measurements} are mathematically described by \textbf{positive-operator valued measures} \textbf{(POVMs)}: collections $\Jcal = \{\xi^{(x)}\}$ of positive semidefinite matrices summing to the identity matrix, encoding that fact that \textit{some} measurement outcome $x$ must occur. 
\begin{myDefinition}{Measurements/POVMs}{}
A measurement/POVM is a collection of matrices $\Jcal = \{\xi^{(x)}\}$ satisfying
\begin{align}\label{eq::tf-smci-def-measurements}
    \xi^{(x)} \geq 0 \quad \text{and} \quad \sum_x \xi^{(x)} = \ident.
\end{align}
\end{myDefinition}

\noindent Each POVM element $\xi^{(x)}$ corresponds to a possible outcome $x$ that could be recorded by the measurement device $\Jcal$. For a given quantum state $\rho$, the probability $\Pprob(x|\Jcal)$ of observing outcome $x$ when using measurement device $\Jcal = \{ \xi^{(x)} \}$ is given by the \textit{Born rule}:
\begin{align}\label{eq::tf-smci-bornrule}
    \Pprob(x|\rho,\Jcal) = \tr{\rho \, \xi^{(x)}}. 
\end{align}
The Born rule manifests a linear map commonly known as the \textit{effect map}~\cite{Kraus_1983,D'ariano_Chiribella_Perinotti_2017}:
{\hypersetup{citecolor=white}
\begin{myDefinition}{Effects}{}
An effect is a collection of linear maps $\{ \mathcal{E}^{(x)} : \mathscr{L}(\mathscr{H}) \to \mathds{C} \}$ that transforms states into probability distributions via the Born rule
\begin{align}\label{eq::tf-smci-def-effects}
    \mathcal{E}^{(x)}[\bullet] := \tr{\bullet \, \xi^{(x)}}.
\end{align}
\end{myDefinition}
}

\noindent This mapping aligns with the general form of linear transformations [see Eq.~\eqref{eq::tf-def-lineartransformationaction}] when expressed with $L_k^{(x)} = R_k^{(x)} = \bra{k} \sqrt{\xi^{(x)}}$, where $\{\ket{k}\}_{k = 1}^{d}$ forms an orthogonal basis of $\Hscr$. Here, the relation $L_k^{(x)} = R_k^{(x)}$ ensures real-valued measurement outcomes and complete positivity (see below) of  $\mathcal{E}^{(x)}$.\footnote{Eq.~\eqref{eq::tf-smci-def-measurements} ensures that each outcome of an effect is positive and together they sum to unity, i.e., the Born rule leads to a valid probability distribution.} Nevertheless, we have denoted the output space as $\mathbb{C}$ to emphasise that the effect map takes an input quantum state to the trivial output Hilbert space of complex numbers.

\vspace{0.25cm}\noindent
\textbf{\textul{Quantum Evolution.}} Beyond measurement, physical quantum systems evolve in time. In its simplest form, noiseless quantum evolution follows unitary dynamics
\begin{align}\label{eq::tf-smci-unitaryevolution}
   \rho \to U \, \rho \, U^\dag = \rho',
\end{align}
where $U:\mathscr{H}_{\inp}\to\mathscr{H}_{\out}$ is a unitary matrix. This represents a linear transformation as per Eq.~\eqref{eq::tf-def-lineartransformationaction} with $L_k = R_k = U$. This operation maps operators $\rho \in \mathscr{L}(\mathscr{H}_{\inp})$ to operators $\rho' \in \mathscr{L}(\mathscr{H}_{\out})$ such that the input and output spaces are the same, i.e., $\mathscr{H}_{\inp}\cong \mathscr{H}_{\out}$. More generally, evolution may occur between different input and output spaces (as in isometries) or even be non-unitary due to noise. Any such evolution that takes arbitrary \textit{input} states $\rho \in \mathsf{St}(\mathscr{H}_{\inp})$ to \textit{output} states $\rho^\prime \in \mathsf{St}(\mathscr{H}_{\out})$ follow the general form of linear transformations, with properties determined by the underlying physical theory. 

Quantum mechanics imposes three fundamental requirements on open quantum evolution. First, the operation must preserve Hermiticity, which is accomplished by setting $L_k = \pm R_k$ for all $k$. Second, the operation must ensure the positivity of the output, even when acting in the presence of any trivial extension, which requires it to be \textbf{completely positive (CP)}. Complete positivity condition ensures that any input state, belonging to a larger system, is mapped to a valid output state.\footnote{When we say that a operation is CP, we automatically mean that it is also Hermiticity preserving.} Third, to conserve probability (encoded in the state's trace), the operation must be \textbf{trace preserving (TP)}. The first two properties guarantee physical realisability, permitting a physical implementation via dilated system-environment dynamics. While complete positivity is universally required, trace preservation applies only to deterministic operations; probabilistic processes like measurements correspond to trace non-increasing CP maps. 

Linear transformations satisfying both complete positivity and trace preservation are called \textit{CPTP maps} or \textit{quantum channels}:
\begin{myDefinition}{Quantum Channels}{}\label{def::channels}
 A quantum channel is a linear map $\Ccal: \mathscr{L}(\mathscr{H}_\inp) \rightarrow \mathscr{L}(\mathscr{H}_\out)$ satisfying
\begin{alignat}{3}   
    &\text{1. Complete positivity: } &&\Ccal \otimes \Ical [\eta] \geq 0 \quad \quad &&\forall \; \eta \in \mathsf{Pos}(\mathscr{H}_\inp \otimes \mathscr{H}_\aux) \label{eq::tf-smci-def-quantumchannels-cp} \\
    &\text{2. Trace preservation: } &&\tr{\Ccal[\rho]} = \tr{\rho} \quad \quad &&\forall \; \rho \in \mathsf{St}(\mathscr{H}_\inp). \label{eq::tf-smci-def-quantumchannels-tp}
\end{alignat}
\end{myDefinition}
\noindent Above, $\mathscr{H}_\aux$ can be of arbitrary dimension; however, for complete positivity, it is sufficient if Eq.~\eqref{eq::tf-smci-def-quantumchannels-cp} holds for auxiliary systems up to dimension $d_\aux = d_\inp$~\cite{Choi_1975}. Stinespring's dilation theorem~\cite{Stinespring_1955} (see also Sec.~\ref{subsec::axiomatichoqos}) ensures that every CPTP map can be represented through an initially uncorrelated environment state $\tau_\aux$, unitary system-environment dynamics $\Ucal(\bullet)=U\,\bullet\, U^\dagger$ for some unitary operator $U\in\mathscr{H}_{\inp \aux}\to\mathscr{H}_{\out \aux}$, and a discarding of the environmental degrees of freedom (represented by a partial trace)
\begin{align}\label{eq::tf-smci-channeldilation}
    \Ccal[\, \bullet\,] = \mathrm{tr}_\aux\left(\Ucal[ \,\bullet \otimes \tau_\aux]\right).
\end{align}
Furthermore, a linear map is CPTP \textbf{if and only if (iff)} it admits a Kraus representation~\cite{PhysRev.121.920, Kraus_1983} (see also Sec.~\ref{subsec::axiomatichoqos}):
\begin{align}\label{eq::tf-smci-channelkraus}
    \Ccal[\, \bullet\,] = \sum_{k} C_k \bullet C_k^\dagger \quad \text{with} \quad  \sum_k C_k^\dagger C_k = \ident\, ,
\end{align}
where $\{C_k: \mathscr{H}_\inp \rightarrow \mathscr{H}_\out\}$ are arbitrary matrices satisfying $\sum_k C_k^\dagger C_k = \ident$. We note that the above form of a CP map $\Ccal$ corresponds to the representation of a general linear map [given in Eq.~\eqref{eq::tf-def-lineartransformation}] with $L_k = R_k =: C_k$. The additional completeness relation $\sum_k C_k^\dagger C_k = \ident$ is equivalent to trace preservation of $\Ccal$.

\vspace{0.25cm}\noindent
\textbf{\textul{Quantum Instruments.}} Effect maps and quantum channels represent two extremes of quantum operations: Effects extract classical information from quantum states, yielding outcome $x$ with probability $\mathbb{P}(x|\mathcal{\rho, J})$, while channels convert one quantum state into another without extracting any information. A \textit{quantum instrument} unifies these concepts, encapsulating both operations simultaneously.

To construct instruments, one must first relax the trace preservation condition for quantum channels to trace non-increasing. While trace preservation ensures a deterministic operation (preserving total probability), trace non-increasing maps describe probabilistic operations, such as measurement with post-selection. Intuitively, complete positivity captures the physicality of the map (i.e., all physical dynamics in quantum theory \textit{must} be completely positive), and trace preservation concerns whether or not any such operation occurs with overall certainty or only with some (non-unit) probability. Maps describing such probabilistic operations, called CP maps (as opposed to CPTP), retain the Kraus representation but with a modified normalisation condition $\sum_k C^\dagger_k C_k \leq \ident$. We denote the sets of CPTP and CP (trace non-increasing) maps as $\mathsf{CPTP}[\mathscr{L}(\mathscr{H}_\inp),\mathscr{L}(\mathscr{H}_\out)]$ and $\mathsf{CP}[\mathscr{L}(\mathscr{H}_\inp),\mathscr{L}(\mathscr{H}_\out)]$, respectively, and drop the arguments whenever clear from context.

Such non-deterministic quantum operations can be physically implemented by coupling the system to an auxiliary one through unitary interaction and performing selective measurements on the auxiliary system (see Sec.~\ref{subsec::axiomatichoqos}). Recording specific outcomes $x$ leads to probabilistic state transformations (CP maps $\mathcal{C}^{(x)}$), while averaging over all outcomes (i.e., tracing the final auxiliary system out) recovers deterministic CPTP dynamics.

{\hypersetup{citecolor=white}
\begin{myDefinition}{Instruments}{}
An instrument $\Jcal$ is a collection $\{\Ccal^{(x)}\}$ of trace non-increasing CP maps summing to a CPTP map, i.e., $\Ccal = \sum_x \Ccal^{(x)} \in \mathsf{CPTP}$. The instrument $\Jcal: \mathscr{L}(\mathscr{H}_\inp) \to \{\mathbb{N}, \mathscr{L}(\mathscr{H}_\out)\}$ maps operators to operators plus an outcome label $x\in \mathbb{N}$, such that for all $\rho \geq 0$
\begin{align}\label{eq::tf-smci-def-instruments}
\Ccal^{(x)}[\rho] = \rho^{(x)} \geq 0 \qquad \text{with} \qquad \Pprob(x|\rho,\Jcal)=\tr{\rho^{(x)}}.
\end{align}
\end{myDefinition}
}

\noindent Instruments naturally generalise POVMs (or, more precisely, effects)---hence our decision to denote them by the same label $\Jcal$---addressing not only outcome probabilities but also state transformations under measurement, which is important in the context of sequentially probed systems. For a given instrument $\mathcal{J} = \{ \Ccal^{(x)}\}$, each element $\Ccal^{(x)}$ corresponds to observing the outcome $x$, with the system state being transformed to $\rho^{(x)} = \Ccal^{(x)}[\rho]$. The probability of observing the outcome $x$ is given by the trace of the output state.

This probability rule is reminiscent of the Born rule of Eq.~\eqref{eq::tf-smci-bornrule}, which provides the probability of observing a certain outcome when a POVM is applied. Indeed, for any instrument, one can write
\begin{align}
 \tr{\Ccal^{(x)}[\rho]}= \tr{\Ccal^{(x)}[\rho]\ident}= \tr{\rho\,{\Ccal^{(x)}}^\dagger[\ident]},
\end{align}
where the adjoint map $\Acal^\dagger$ is the unique linear map such that $\tr{\mathcal{A}(X) Y^\dag } = \tr{X \mathcal{A}^\dag(Y^\dag) }$ for all linear operators $X,Y$; then $\Acal(\bullet)= \sum_k L_k \, \bullet \, R_k^\dagger $ implies $\mathcal{A}^\dag(\bullet)= \sum_k R_k^\dag \bullet L_k$. Evidently, the set $\{{\Ccal^{(x)}}^\dagger[\ident]\}$ forms a POVM yielding identical probabilities to those of the instrument $\{\Ccal^{(x)}\}$. However, a POVM does not determine the post-measurement state, leaving room for various non-equivalent instrument implementations. In contrast, an instrument uniquely encompasses the dynamics of the system due to measurement, providing the observed measurement outcome, the probability of observing that outcome, and the post-measurement output state, all in one object.

This distinction can be illustrated by considering two different instrument implementations of a POVM $\{\xi^{(x)}\}$:
\begin{enumerate}
    \item L{\" u}ders instrument: $\Ccal_\text{L}^{(x)}(\rho):=\sqrt{\xi^{(x)}}\rho\sqrt{\xi^{(x)}}  = \Pprob(x|\rho,\Jcal_{L}) \, \rho_L$, 
    \item Measure-and-prepare instrument: $\Ccal_\text{MP}^{(x)}(\rho):=\tr{\rho\, \xi^{(x)}}  \sigma^{(x)} = \Pprob(x|\rho,\Jcal_{MP}) \ \sigma^{(x)}$.
\end{enumerate}
The first is the L{\"u}ders instrument~\cite{Lueders_2006}, where $\sqrt{\xi^{(x)}}$ is the unique positive semidefinite square root of $\xi^{(x)}$ and $\Pprob(x|\rho,\Jcal_{L}) := \tr{\Ccal_\text{L}^{(x)}(\rho)}$. The second is the `measure-and-prepare' instrument, where after the measurement a quantum state $\sigma^{(x)}$ is prepared, which can depend upon the observed outcome $x$. While these instruments yield identical outcome probabilities [$\Pprob(x|\rho,\Jcal_{L}) = \Pprob(x|\rho,\Jcal_{MP})$], they produce distinct post-measurement states, demonstrating how instruments provide a more complete description of quantum measurement processes. 

\vspace{0.25cm}\noindent
\textbf{\textul{Revisiting Motivating Examples \& Summary.}} Several key observations emerge from our discussion. While channels, measurements, effects, and instruments are all linear maps acting on quantum states (density operators), the operations presented in our three motivating examples operate upon quantum operations themselves---specifically on instruments or CPTP channels---hence their designation as \textit{higher-order} quantum operations. 

Another important point is that the operational description of quantum theory distinguishes two fundamental classes of objects. The first class comprises those that occur \textit{deterministically}, such as quantum states or quantum channels, as encoded respectively by unit trace and trace preservation conditions. These conditions, though manifesting at different levels in the hierarchy of quantum objects, encode the same underlying property. Importantly, this notion of determinism does not mean that the state is pure or known, or that a dynamics does not `mix' the spectrum of the system, or any such concept tied to notions of non-randomness; rather, it signifies that the state preparation occurs with unit probability, or that the fixed underlying circuit describing the dynamics can be implemented deterministically.

The second class encompasses objects that occur \textit{probabilistically}, such as measurements or instruments with specific outcomes. While these operations are necessarily CP, they need not be TP, reflecting their probabilistic nature. As previously noted, any CP (but not necessarily TP) dynamics can be physically implemented through a Stinespring dilation where the environment undergoes measurement rather than being traced out (see Sec.~\ref{subsec::axiomatichoqos}). The fact that POVMs and instruments comprise collections of CP operations summing to a CPTP operation captures an essential principle: While \textit{some} measurement device or instrument certainly interrogates the system, individual outcomes manifest probabilistically.

This deterministic-probabilistic dichotomy extends naturally to HOQOs. While all of the objects and the properties that we have discussed here so far are found in `standard' quantum theory, we recall them both in order to introduce notation as well as to show how they are all instances of linear maps, which will allow us to naturally generalise our analysis to the case of HOQOs. By conceptualising states/measurements and channels/instruments as linear transformations on appropriate spaces, we recognise them as the first two layers in the HOQO hierarchy. To develop a unified framework for the entire hierarchy of valid quantum operations, we next introduce the Choi-Jamio{\l}kowski isomorphism. This representation treats all linear maps on equal footing and is endowed with many appealing properties, particularly in simplifying the verification of CP and TP conditions that can be challenging to determine for general maps. Finally, we will introduce the link product, which provides a systematic method for composing HOQOs in this representation, enabling the construction of complex higher-order circuits from simpler primitives.

\FloatBarrier


\subsubsection{Choi-Jamio{\l}kowski Isomorphism: Representation of Linear Maps}\label{subsubsec::tf-cji}\hfill\\

\noindent While all statements about HOQOs are, quite naturally, representation-independent, many properties and relations find particularly elegant expression through the \textbf{Choi-Jamio{\l}kowski isomorphism} \textbf{(CJI)}~\cite{Jamiolkowski_1972,Choi_1975}. The CJI enables us to represent linear maps as \textit{operators/matrices}, placing all spatiotemporal quantum objects on equal mathematical footing. This representation proves especially advantageous for higher-order operations---which, while acting on linear maps themselves, remain linear maps in their own right. Additionally, the relevant properties for quantum operations, e.g., CP and TP and their higher-order analogues, adopt particularly clear forms in this representation. This operator formalism also facilitates the application of established quantum state techniques, including semidefinite programming, matrix product operator representations, and tools from entanglement and quantum resource theories, amongst others. Following our discussion of the CJI, we will subsequently introduce the \textit{link product}, which provides the appropriate composition rule in this representation, enabling both the definition of a spatiotemporal Born rule and the construction of complex higher-order objects like superchannels from elementary components.

\begin{figure}[t]
\centering
\vspace{0.5em}
\includegraphics[scale=0.6]{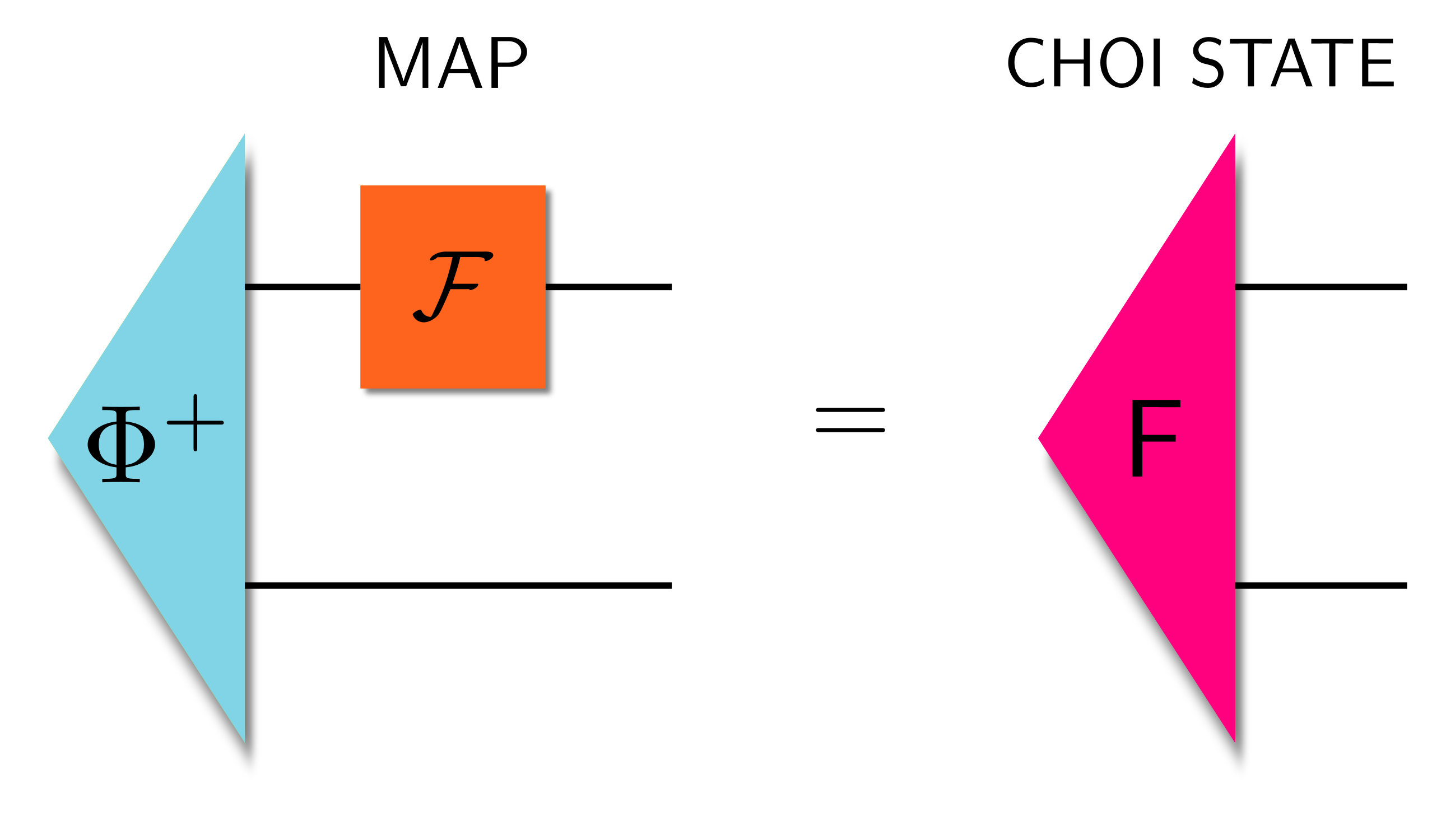}
\caption{\textbf{Choi-Jamio{\l}kowski Isomorphism.} A linear map $\mathcal{F}$ can be expressed as a matrix $\mathsf{F}$ via its action on one half of an unnormalised maximally entangled state $\Phi^+$.} \label{fig::tf-cji-def}
\end{figure}

The CJI establishes a correspondence between any linear map $\Fcal: \mathscr{L}(\mathscr{H}_\inp) \rightarrow \mathscr{L}(\mathscr{H}_\out)$ and a bipartite matrix $\mathsf{F} \in \mathscr{L}(\mathscr{H}_\out\otimes \mathscr{H}_\inp)$ (see Fig.~\ref{fig::tf-cji-def}).\footnote{Since by definition $\Phi^+ \in \mathscr{L}(\mathscr{H}_\inp \otimes \mathscr{H}_{\inp'})$, technically, the resulting matrix $\mathsf{F}$ is an element of $\mathscr{L}(\mathscr{H}_\out\otimes \mathscr{H}_{\inp^\prime})$ and an appropriate relabelling of spaces is tacitly assumed; whenever more diligent bookkeeping is required, we will be explicit.} Formally:

{\hypersetup{citecolor=white}
\begin{myDefinition}{Choi-Jamio{\l}kowski Isomorphism~\cite{Jamiolkowski_1972,Choi_1975}}{}
A linear map $\Fcal: \mathscr{L}(\mathscr{H}_\inp) \rightarrow \mathscr{L}(\mathscr{H}_\out)$ is isomorphic to the matrix $\mathsf{F} \in \mathscr{L}(\mathscr{H}_\out\otimes \mathscr{H}_\inp)$ via
\begin{align}
\label{eq::tf-cji-def}
    \mathsf{F} := \mathsf{Choi}(\Fcal) = (\Fcal \otimes \Ical) [\Phi^+] = \sum_{ij} \mathcal{F}(\ket{i}\!\bra{j}) \otimes \ket{i}\!\bra{j} ,
\end{align}
where $\Phi^+ := \sum_{i,j=1}^{d_\inp} \ketbra{ii}{jj} \in \mathscr{L}(\mathscr{H}_\inp \otimes \mathscr{H}_{\inp'})$ denotes the \textit{unnormalised} maximally entangled state.
\end{myDefinition}
}

\noindent Throughout this Review Article, we denote abstract maps by calligraphic letters and their corresponding Choi matrices by sans-serif versions (or occasionally Greek letters following established conventions) and will interchangeably employ the terms `Choi operator', `Choi matrix' and `Choi state'.\footnote{Note that the Choi operator is technically not a quantum state (since it does not have unit trace), but rather a \textit{super}normalised one.} Thus, $\mathsf{F}$ represents the Choi matrix of the map $\Fcal$, and when unambiguous, we omit explicit distinction between maps and their Choi representations. While various versions of the CJI exist in the literature---differing in map ordering ($\Ical \otimes \Fcal$ versus $\Fcal \otimes \Ical$), transposition conventions, or normalisation---these differences, while affecting specific expressions, do not impact the fundamental results presented here. Lastly, note that in principle, the CJI could alternatively be defined through the action of the map on half of any pure state of full Schmidt rank (indeed, such a generalisation proves necessary to properly define the CJI in infinite dimensions~\cite{Holevo_2010}).  


With our definition of the CJI at hand, the action of a map $\Fcal:\mathscr{L}(\mathscr{H}_\inp) \rightarrow \mathscr{L}(\mathscr{H}_\out)$ on an arbitrary input operator $\rho_\inp \in \mathscr{L}(\mathscr{H}_\inp)$ can be expressed in terms of the Choi matrix $\mathsf{F}_{\out \inp} \in \mathscr{L}(\mathscr{H}_\out \otimes \mathscr{H}_\inp)$ as
\begin{align}
\label{eq::tf-cji-choiaction}
    \Fcal[\rho_\inp] = \ptr{\inp}{\mathsf{F}_{\out \inp}(\ident_\out \otimes \rho_\inp^\mathrm{T})} ,
\end{align}
where $\text{tr}_\inp$ denotes the trace over $\mathscr{H}_\inp$. Direct substitution verifies this expression.

\vspace{0.25cm}\noindent
\textbf{\textul{Standard Quantum Objects Revisited with the CJI.}} While the CJI establishes a general correspondence between linear maps and matrices, quantum theory imposes additional constraints on physically valid Choi matrices. Any CPTP map $\Ccal \in \mathsf{CPTP}$ has a corresponding Choi matrix that equivalently satisfies  
\begin{align}\label{eq::tf-cji-choicptp}
    \mathsf{C}_{\out \inp} \geq 0 \ \ \text{(CP)}   \quad \text{and} \quad \ptr{\out}{\mathsf{C}_{\out \inp}} = \ident_\inp \ \ \text{(TP)}\, ,
\end{align}
as verified by direct insertion and comparison to Eqs.~\eqref{eq::tf-smci-def-quantumchannels-cp},~\eqref{eq::tf-smci-def-quantumchannels-tp} and~\eqref{eq::tf-smci-channelkraus}.\footnote{We use $\Fcal$ for generic linear maps and reserve $\mathcal{C}$ specifically for quantum channels (or instruments when labelled with an outcome $x$).} These conditions imply that the Choi matrix of a CPTP map is a supernormalised quantum state---a positive semidefinite matrix with trace equal to $d_\inp$. For trace non-increasing CP maps $\mathcal{C} \in \mathsf{CP}$, the conditions modify to $\mathsf{C}_{\out \inp} \geq 0$ and $\ptr{\out}{\mathsf{C}_{\out \inp}} \leq \ident_\inp$. An instrument in the Choi representation thus comprises a collection $\{\mathsf{C}_{\out \inp}^{(x)}\}$ of Choi matrices of trace non-increasing CP maps such that $\ptr{\out}{\sum_x \mathsf{C}_{\out \inp}^{(x)}} = \ident_\inp$, ensuring their sum constitutes a CPTP map $\mathsf{C}_{\out \inp} := \sum_x \mathsf{C}_{\out \inp}^{(x)} \in \mathsf{CPTP}$. 

If the output space of an instrument is trivial (i.e., post-measurement states are disregarded), the instrument reduces to an effect map $\{ \Ecal^{(x)} \}: \mathscr{L}(\mathscr{H}) \rightarrow \mathds{C}$. Normalisation of the instrument ensures that we have $\sum_x \mathsf{E}^{(x)} = \ident$, confirming that $\{\mathsf{E}^{(x)}\}$ constitutes a POVM. In this case, Eq.~\eqref{eq::tf-cji-choiaction} coincides with the Born rule Eq.~\eqref{eq::tf-smci-bornrule}. For an effect map $\mathcal{E}^{(x)}[\,\bullet\,] := \tr{\bullet \, \xi^{(x)}}$ with POVM elements $ \{ \xi^{(x)} \}$, the Choi matrix is simply the transpose of the relevant POVM element, i.e., $\mathsf{E}^{(x)} = \mathsf{Choi}(\mathcal{E}^{(x)}) =  \xi^{(x) \textup{T}}$. From Eq.~\eqref{eq::tf-cji-choiaction}, we obtain
\begin{align}\label{eq::tf-cji-choieffectmap}
    \mathcal{E}^{(x)}[\rho] = \tr{\mathsf{E}^{(x)} \rho^{\textup{T}}} = \tr{\mathsf{E}^{(x) \textup{T}} \rho} = \tr{\xi^{(x)} \rho} = \mathds{P}(x | \rho, \mathcal{J}),
\end{align}
using the self-duality of transposition $\tr{A^{\textup{T}} B} = \tr{A B^{\textup{T}}}$ and expressing the Choi matrix in terms of the original POVM element.\footnote{Recall that this additional transpose arises from the definition we take for the CJI; therefore it is entirely conventional (we choose it so that the standard Born rule applies) and has no physical consequence.} 

More generally, computing probabilities for an instrument $\mathcal{J} = \{ \mathsf{C}_{\out \inp}^{(x)}\}$ via Eq.~\eqref{eq::tf-cji-choiaction} yields
\begin{align}\label{eq::tf-cji-choiinstrument}
    \Pprob(x|\rho, \Jcal) = \tr{\Ccal^{(x)}[\rho]} = \ptr{\out\inp}{\mathsf{C}_{\out \inp}^{(x)}(\ident_\out \otimes \rho_\inp^\mathrm{T})} = \tr{\left(\ptr{\out}{\mathsf{C}_{\out \inp}^{(x)}}\right)^\mathrm{T}\rho_\inp}.
\end{align}
Consequently, every instrument $\mathcal{J} = \{\mathsf{C}_{\out \inp}^{(x)}\}$ defines a unique POVM $\{\left(\ptr{\out}{\mathsf{C}_{\out \inp}^{(x)}}\right)^\mathrm{T}\}$ by tracing out the output degrees of freedom [see Eq.~\eqref{eq::tf-smci-def-instruments}]. Conversely, a given POVM may be compatible with multiple instruments, as it contains no information regarding the post-measurement state. 

Two particularly important Choi matrices are those of the identity map $\Ical_X$ and the partial trace $\text{tr}_X$. Direct computation shows $\mathsf{Choi}(\Ical_X) = \Phi^+_{XX'}$ and $\mathsf{Choi}(\text{tr}_X) = \ident_X$. Notably, this latter relation identifies the trace map with the trivial single-element POVM; indeed, the trace (or equivalently, the identity matrix $\ident$) represents the unique deterministic effect~\cite{Dariano_2018}. Additionally, quantum states $\rho$ themselves emerge as Choi matrices of state preparation maps with trivial input spaces $\Rcal:\mathds{C} \rightarrow \mathscr{L}(\mathscr{H})$. For a map $\mathcal{R}$ preparing state $\rho$ with unit probability, $\rho = \mathsf{Choi}(\mathcal{R})$, with complete positivity and trace preservation translating to the familiar conditions $\rho \geq 0$ and $\tr{\rho} = 1$ for all $\rho \in \mathsf{St}(\mathscr{H})$.\footnote{While our notation convention suggests that quantum states should be denoted by sans-serif letters, i.e., $\mathsf{R} = \mathsf{Choi}(\Rcal)$, we maintain the traditional Greek letter notation, i.e., $\rho = \mathsf{Choi}(\Rcal)$.} 

Finally, an alternative construction of the CJI involves first \textit{vectorising} an arbitrary linear operator $A: \Hscr_{\inp} \to \Hscr_{\out}$ via
\begin{align}\label{eq::tf-cji-choivector}
    \kket{\mathsf{A}} := (A \otimes \mathds{1}) \ket{\Phi^+} \in \mathscr{H}_{\out} \otimes \mathscr{H}_{\inp}. 
\end{align}
With this, if $\mathcal{F}$ is an arbitrary linear map given by $\mathcal{F}(\bullet) = \sum_k L_k \bullet R_k^\dagger$, its Choi matrix is given by $\mathsf{F} = \sum_k \kketbra{L_k}{R_k}$. Analogously, if $\Ccal$ is CP, such that $\Ccal[\bullet] = \sum_k C_k \bullet C_k^\dagger$, then its Choi matrix is $\mathsf{C} = \sum_k \kketbra{C_k}{C_k}$. In particular, if $\mathcal{U}$ is a unitary channel given by $\mathcal{U}(\bullet )=U \bullet  U^\dagger$, its Choi matrix is $\mathsf{U}=\kketbra{U}{U}$.


\begin{figure}[t]
    \centering
    \subfigure[\textbf{Superchannel}]
    {
    \includegraphics[scale =0.55]{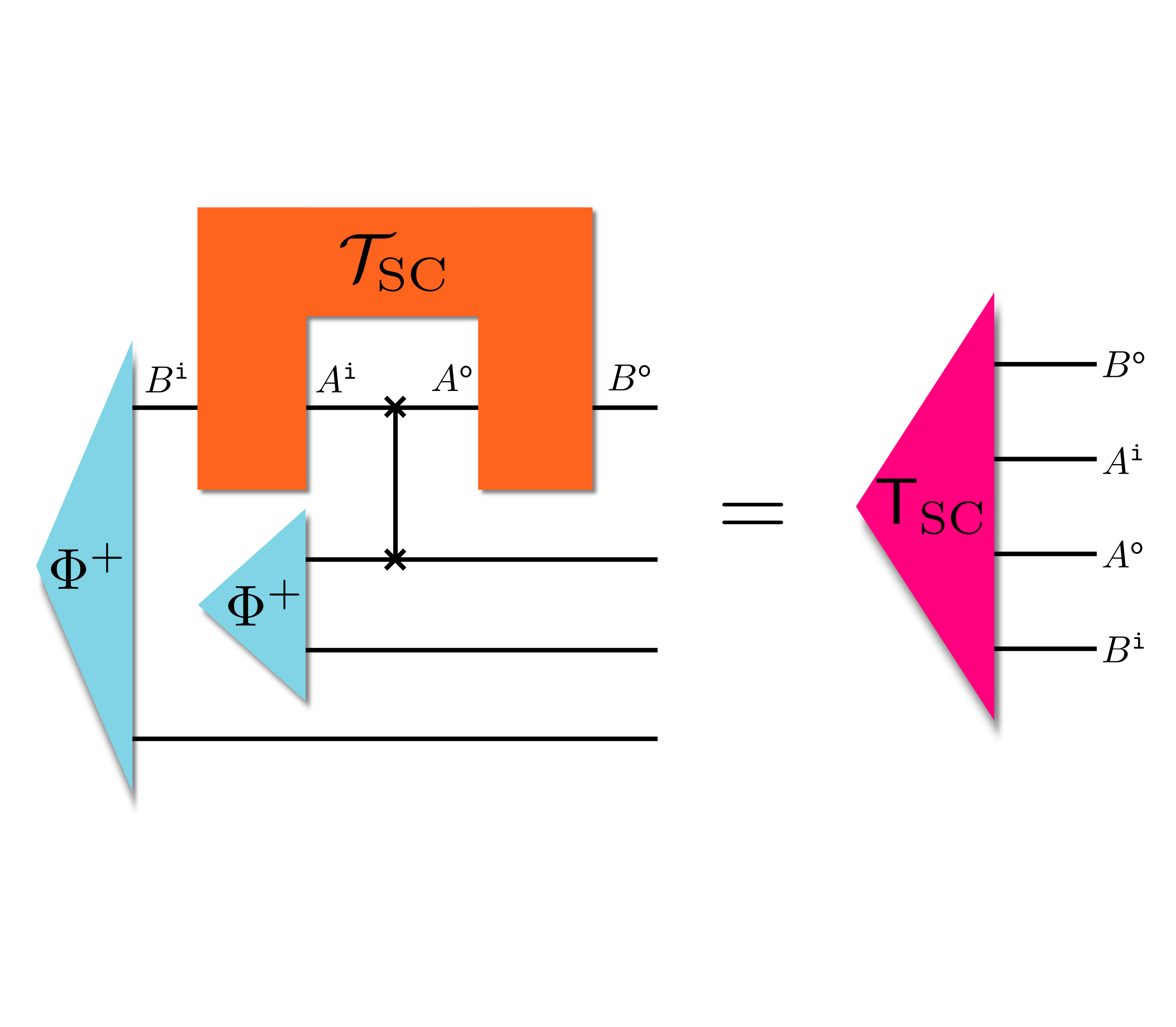}
    \label{fig::tf-cji-choisuperchannel}
    } \hspace{1.2cm}
    \subfigure[\textbf{Process Matrix.}]
    {
    \includegraphics[scale=0.55]{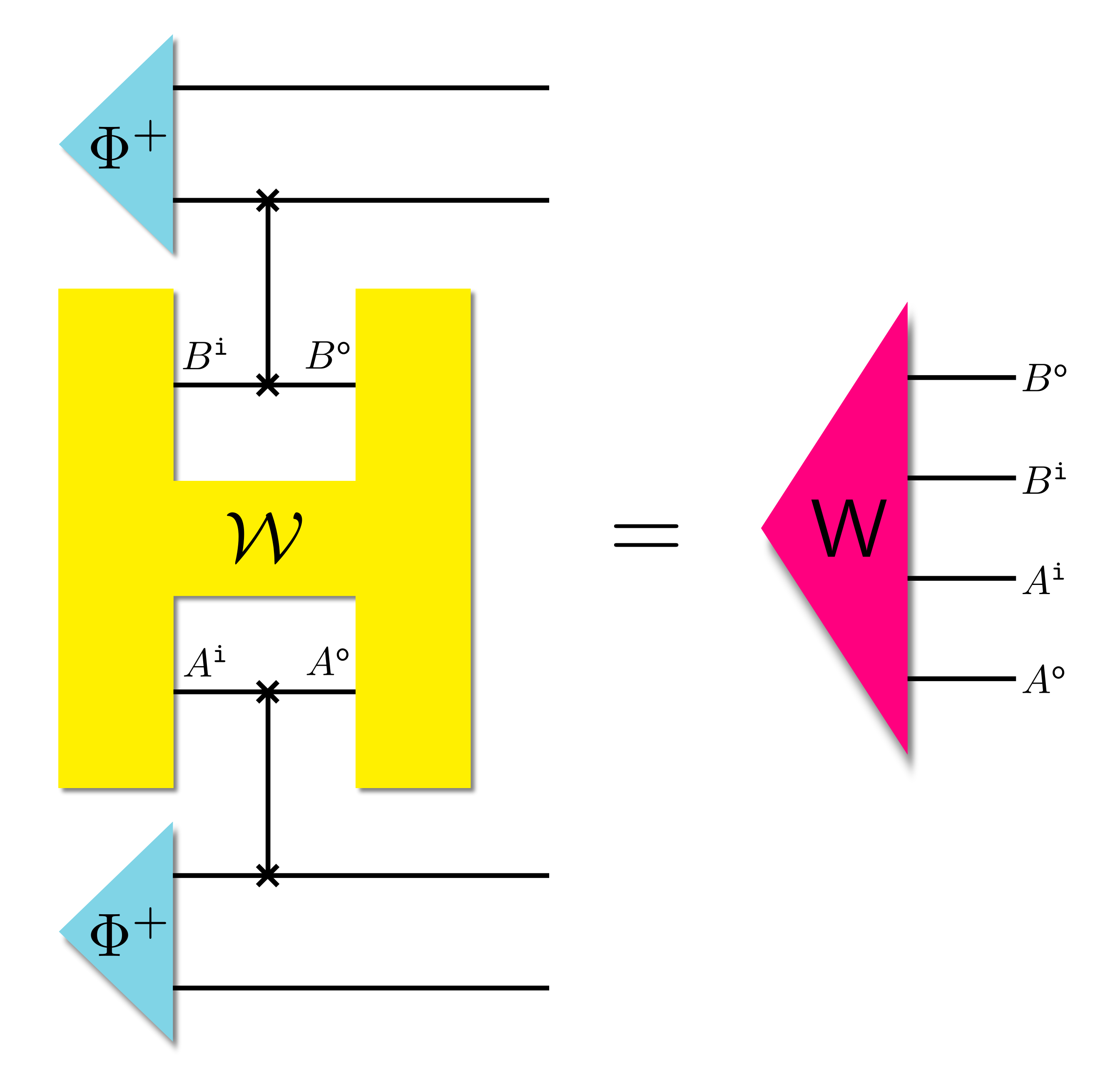}
    \label{fig::tf-cji-choiprocessmatrix}
    }\hfill
    \caption{\textbf{Choi States for HOQOs.} Just as the Choi state of a channel can be constructed by feeding in half of a maximally entangled state, the Choi state of any HOQO can be constructed by swapping in half of a maximally entangled state $\Phi^+$ at each open slot (denoted by the vertical line with an \textsf{x} on both ends). \textbf{(a)} The superchannel $\mathcal{T}_{\textup{SC}}$ gets mapped to a four-partite Choi state $\mathsf{T}_{\textup{SC}}$; \textbf{(b)} The process matrix $\mathcal{W}$ gets mapped to a four-partite Choi state $\mathsf{W}$. Throughout, we will reserve orange and yellow to depict processes with a fixed causal order (see Sec.~\ref{subsec::timeorderedquantumprocesses}) and processes with indefinite causal order (see Sec.~\ref{subsec::indefinitecausalorder}), respectively. \label{fig::hoqt-choi-examples}}
\end{figure}


\vspace{0.25cm}\noindent \textbf{\textul{CJI of Motivating Examples.}} The Choi-Jamio{\l}kowski isomorphism (CJI) extends beyond its familiar application to quantum channels: It fundamentally provides an isomorphism between linear maps acting on vector spaces and linear operators acting on joint vector spaces. This correspondence naturally applies to HOQOs, as these are themselves linear maps acting on spaces of input maps. We can thus represent HOQOs as multi-partite quantum states through the CJI by inserting halves of maximally entangled pairs into each open slot of the HOQO, as depicted in Fig.~\ref{fig::hoqt-choi-examples} and as we will discuss in detail in Sec.~\ref{subsec::timeorderedquantumprocesses}. Just as with quantum channels, the resulting Choi states of valid HOQOs exhibit specific structural properties that generalise complete positivity and trace preservation.

Let us revisit our three motivating examples of Sec.~\ref{subsec::motivatingexamples} in light of the CJI. In our first example of open quantum system dynamics with initial correlations (Sec.~\ref{subsubsec::me-opensystemdynamicswithinitialcorrelations}), we encountered the superchannel $\mathcal{T}$ that transforms any preparation procedure $\mathcal{M}^{(x)}$ at the initial time $t=0$ to the corresponding output state\footnote{We emphasise that the space the `output' state is defined on is labelled by $1^{\inp}$ here, i.e., an `input' ($\inp$) label. This is owed to the fact that we label spaces from the perspective of the experimenter, not the superchannel, and the roles of inputs and outputs are exchanged for both. This notational ambiguity is unavoidable and we adhere to the convention that when we denote something as an input/output of an object, we refer to what is inserted into it/comes out of it, independent of the labelling of the respective spaces.} [see Eq.~\eqref{eq::me-osdic-correlatedcase}]
\begin{align}\label{eq::superCh-example-CJI}
\Tcal: [\Lscr(\Hscr_{0^{\inp}}) \to \Lscr(\Hscr_{0^{\out}})] \to \Lscr(\Hscr_{1^{\inp}}) \quad \mbox{with} \quad
 \mathcal{M}^{(x)}: \Lscr(\Hscr_{0^{\inp}}) \to \Lscr(\Hscr_{0^{\out}}),
\end{align}
where we relabel the final time as $t=1$ for notational simplicity. 

Both objects admit Choi matrix representations. The input preparation element $\mathcal{M}^{(x)}$ is represented by a bipartite matrix $\mathsf{M}^{(x)}_{0^{\out} 0^{\inp}} \in \Lscr(\Hscr_{0^{\out}} \otimes \Hscr_{0^{\inp}})$ that is positive semidefinite $(\mathsf{M}^{(x)}_{0^{\out} 0^{\inp}} \geq 0)$ and satisfies $\ptr{\out}{\sum_{x} \mathsf{M}^{(x)}_{0^{\out} 0^{\inp}}} = \mathds{1}_{0^{\inp}}$, which follow from the fact that $\{ \mathcal{M}^{(x)} \}$ forms an instrument. Similarly, by swapping in half of a maximally entangled state into the `input' space at time $t_0$, the superchannel $\Tcal$ maps to a tripartite matrix $\mathsf{T}_{1^{\inp} 0^{\out} 0^{\inp}} \in \Lscr(\Hscr_{1^{\inp}} \otimes \Hscr_{0^{\out}} \otimes \Hscr_{0^{\inp}})$. This Choi state must be positive semidefinite $(\mathsf{T}_{1^{\inp} 0^{\out} 0^{\inp}} \geq 0)$ and satisfy $\ptr{1^{\inp}}{\mathsf{T}_{1^{\inp} 0^{\out} 0^{\inp}}} = \mathds{1}_{0^{\out}} \otimes \rho_{0^{\inp}}$ (for some state $\rho_{0^{\inp}}$), generalising the concepts of CP and TP to HOQOs (see Sec.~\ref{subsec::timeorderedquantumprocesses}).

Our second example [Sec.~\ref{subsubsec::me-quantumcircuitarchitecture}, see Fig.~\ref{fig::me-qca-encoderdecoder}] featured a quantum superchannel $\mathcal{T}_{\textup{SC}}$ that transforms operations between laboratories\footnote{Throughout, we frequently change the labelling of spaces depending on whether we want to emphasise the time---leading to labels $0^\inp, 0^\out, 1^\inp, \dots$---or the laboratory---leading to labels $A^\inp, A^\out, B^\inp, \dots$---of the respective spaces.}  [see Eq.~\eqref{eq::me-qca-encoderdecodergeneral}] 
\begin{align}\label{eq::Encoder-decoder-example-CJI}
\mathcal{T}_{\textup{SC}}: [\Lscr(\mathscr{H}_{A^{\inp}}) \to \Lscr(\mathscr{H}_{A^{\out}})] \to [\Lscr(\mathscr{H}_{B^{\inp}}) \to \Lscr(\mathscr{H}_{B^{\out}})].
\end{align}

\noindent In words, this superchannel takes an input operation $\mathcal{C}_{A}: \Lscr(\mathscr{H}_{A^{\inp}}) \to \Lscr(\mathscr{H}_{A^{\out}})$ to an output operation $\mathcal{C}_{B} := \mathcal{T}_{\textup{SC}}[\Ccal_A]: \Lscr(\mathscr{H}_{B^{\inp}}) \to \Lscr(\mathscr{H}_{B^{\out}})$ (here, we have relabelled the spaces `in' and `out' in the example to `A' and `B' respectively for convenience). This superchannel has 4 `wires'---two that will eventually contract with those of the input map $\Ccal_A$, leaving two to correspond to the output map $\Ccal_B$. The four-partite Choi state $\mathsf{T}_{B^{\out} A^{\out} A^{\inp} B^{\inp}} \in \mathscr{L}(\mathscr{H}_{B^{\out}}\otimes\mathscr{H}_{A^{\out}}\otimes\mathscr{H}_{A^{\inp}}\otimes\mathscr{H}_{B^{\inp}})$ is constructed by feeding in half of a maximally entangled state to the input spaces of both labs [see Fig.~\ref{fig::tf-cji-choisuperchannel}]. Similarly to the tripartite superchannel above, this four-partite superchannel is completely positive $(\mathsf{T}_{B^{\out} A^{\out} A^{\inp} B^{\inp}} \geq 0)$. Moreover, as the overall operation takes place with a fixed causal ordering, the superchannel satisfies $\ptr{B^{\out}}{\mathsf{T}_{B^{\out} A^{\out} A^{\inp} B^{\inp}}}= \mathds{1}_{A^{\out}} \otimes \mathsf{T}_{A^{\inp} B^{\inp}}$, where $\mathsf{T}_{A^{\inp} B^{\inp}}$ itself is a quantum channel, i.e., it in turn satisfies $\mathsf{T}_{A^{\inp} B^{\inp}} \geq 0$ and $\ptr{A^{\inp}}{\mathsf{T}_{A^{\inp} B^{\inp}}} = \mathds{1}_{B^{\inp}}$. We will later see that this cascading structure of trace conditions characterises HOQOs that occur overall deterministically with a fixed temporal ordering.

Finally, we examined process matrices $\Wcal$ that go beyond fixed causal order (Sec.~\ref{subsubsec::me-causalityquantumtheory}). These take two instruments as inputs
\begin{align}\label{eq::instruments-example-CJI}
\Jcal_A \ni \Mcal_A^{(a)} : \mathscr{L}(\Hscr_{A^{\inp}}) \to \Lscr(\Hscr_{A^{\out}}),\quad \Jcal_B \ni \Mcal_B^{(b)} : \mathscr{L}(\Hscr_{B^{\inp}}) \to \Lscr(\Hscr_{B^{\out}})    
\end{align}
and output a probability distribution $\mathds{P}(a,b|\Jcal_A,\Jcal_B)=\mathcal{W}(\mathcal{M}_A^{(a)},\mathcal{M}_B^{(b)})$, i.e.,
\begin{align}\label{eq::processmatrix-example-CJI}
\Wcal: \bigg([\mathscr{L}(\Hscr_{A^{\inp}}) \to \Lscr(\Hscr_{A^{\out}})], \,[\mathscr{L}(\Hscr_{B^{\inp}}) \to \Lscr(\Hscr_{B^{\out}})]\bigg) \to \mathds{C}.    
\end{align} 
Thus, the process matrix has four `wires' and is represented by a four-partite Choi state $\mathsf{W}_{A^{\inp} A^{\out} B^{\inp} B^{\out}} \in \mathscr{L}(\Hscr_{A^{\inp}}\otimes\Hscr_{A^{\out}}\otimes\Hscr_{B^{\inp}}\otimes\Hscr_{B^{\out}})$ by feeding in half of a maximally entangled state into the input spaces of both labs [see Fig.~\ref{fig::tf-cji-choiprocessmatrix}], as was the case for the four-partite superchannel $\mathcal{T}_{\text{SC}}$ above. The Choi state must be positive semidefinite to ensure valid probability distributions for any pair of independent instruments, and to guarantee that the process matrix maps completely positive maps onto completely positive maps, even when only acting non-trivially on a part of them (see Sec.~\ref{subsubsec::axiomatichoqos-qc-transtranstrans} for a more detailed discussion). Notably, the absence of fixed causal order means there are fewer additional constraints on the Choi state of $\mathcal{W}$ than there were for the case of superchannels, which abide by a fixed causal order (see Sec.~\ref{subsubsec::axiomatichoqos-qc-transtranstrans}). 

\vspace{0.25cm}\noindent
\textbf{\textul{Summary.}} With these foundational elements---the basic quantum formalism and the CJI representation of linear maps (see Tab.~\ref{tab::tf-cji-basicquantumobjects})---we have almost all the pieces needed to derive HOQO theory from operational principles. The final crucial element required is the \textit{link product}, which provides a unified rule for composing quantum objects through their Choi matrices. This encompasses everything from fundamental operations (like channels acting on states and the Born rule) to higher-order quantum operations (such as superchannels acting on maps and process matrices acting on instruments). The link product ultimately gives rise to the spatiotemporal Born rule, enabling one to calculate observed probability distributions for quantum experiments with arbitrary causal structure and establishing a comprehensive framework for transformations between higher-order quantum objects.


\begin{table}
\caption{\label{tab::tf-cji-basicquantumobjects}
\textbf{Basic Quantum Objects as Linear Maps.} All fundamental elements of quantum theory can be understood as maps between appropriate spaces. Lower-order quantum operations include states (mapping probability distributions to density operators) and measurements (mapping density operators to probability distributions). Their Choi matrices have simple representations: The Choi matrix of a preparation map $\mathcal{R}$ is simply the prepared density operator $\rho$ itself; while a measurement map $\mathcal{E}^{(x)}$ with outcome $x$, defined by the effect $\mathcal{E}^{(x)}[\bullet] := \tr{ \bullet \, \xi^{(x)} }$, has Choi matrix $\mathsf{E}^{(x)} = \xi^{(x) \textup{T}}$. The next tier comprises dynamical transformations between quantum states, including deterministic quantum channels and probabilistic instruments (which capture both measurement outcomes and post-measurement states). These are all special cases of HOQOs, with their properties generalising naturally to the higher-order framework.} 
\footnotesize\rm
\begin{tabular*}{\textwidth}{llllll}
\br
Object & Map & Action & Choi & CP & TP \\
\mr
Quantum State & $\mathcal{R}: \mathds{C} \to \mathscr{L}(\mathscr{H})$ & $\mathcal{R}[1] = \rho$ & $\mathsf{R} = \rho$ & $\rho \geq 0$ & $\tr{\rho} = 1$ \\
Measurement/Effect & $ \{\mathcal{E}^{(x)}\}: \mathscr{L}(\mathscr{H}) \to \mathds{C}$ & $ \mathcal{E}^{(x)}[\rho] = \tr{\rho\,\xi^{(x)}}$ & $\mathsf{E}^{(x)} = \xi^{(x)\textup{T}}$ & $\mathsf{E}^{(x)} \geq 0$ & $\sum_x \mathsf{E}^{(x)} = \mathds{1}$ \\
Quantum Channel & $\mathcal{C} : \mathscr{L}(\mathscr{H}_\inp) \to \mathscr{L}(\mathscr{H}_\out)$ & $\rho^\prime = \mathcal{C}[\rho]$ & $\mathsf{C}$ & $\mathsf{C} \geq 0$ & $\ptr{\out}{\mathsf{C}} = \mathds{1}_\inp$ \\
Instrument & $\{ \mathcal{C}^{(x)} \} : \mathscr{L}(\mathscr{H}_\inp) \to \mathscr{L}(\mathscr{H}_\out)$ & $\rho^{(x)} = \mathcal{C}^{(x)}[\rho]$ & $\mathsf{C}^{(x)}$ & $\mathsf{C}^{(x)} \geq 0$ & $\ptr{\out}{\sum_x \mathsf{C}^{(x)}} = \mathds{1}_\inp$ \\
\br
\end{tabular*}
\end{table}


\FloatBarrier


\subsubsection{Link Product: Composing Linear Maps in the Choi Representation}\label{subsubsec::tf-linkproduct}\hfill\\

\noindent The \textit{link product} is the mathematical operation that enables the composition of linear maps in their Choi representation. Consider the simplest case: a channel acting on a quantum state, expressed as $\rho^\prime = \mathcal{C}[\rho]$. When the map $\mathcal{C}: \mathscr{L}(\mathscr{H}_\inp) \to \mathscr{L}(\mathscr{H}_\out)$ is represented by its Choi matrix $\mathsf{C}_{\out \inp} \in \mathscr{L}(\mathscr{H}_\out\otimes\mathscr{H}_\inp)$, its action on an input state $\rho_\inp \in \mathscr{L}(\mathscr{H}_\inp)$ can be written as
\begin{align}\label{eq::tf-lp-channelaction}
    \rho^\prime_\out = \ptr{\inp}{(\mathds{1}_\out \otimes \rho_\inp^\textup{T}) \, \mathsf{C}_{\out \inp}} =: \rho_\inp \star \mathsf{C}_{\out\inp} \in \mathscr{L}(\mathscr{H}_\out).
\end{align}
This represents the most basic application of the link product $\star$, as the input quantum state here can itself be considered the Choi matrix of a preparation map (i.e., a map with trivial input). The key message is that the resulting Choi state (here, $\rho^\prime_\out$) comes from the link product of the constituent Choi matrices. 

Similarly, when composing sequential channels $\Mcal : \Lscr(\Hscr_1) \to \Lscr(\Hscr_2)$ and $\mathcal{N} : \Lscr(\Hscr_2) \to \Lscr(\Hscr_3)$, the Choi matrix of their composition $\mathcal{K} := \mathcal{N} \circ \Mcal: \Lscr(\Hscr_1) \to \Lscr(\Hscr_3)$ is given by
\begin{align}\label{eq::tf-lp-channelcomposition}
    \mathsf{K}_{31} = \ptr{2}{(\mathds{1}_3 \otimes \mathsf{M}_{21}^{\textup{T}_2}) \, (\mathsf{N}_{32} \otimes \mathds{1}_1)} =: \mathsf{M}_{21}\star \mathsf{N}_{32} \in \mathscr{L}(\mathscr{H}_3 \otimes \mathscr{H}_1).
\end{align}
Here, $\bullet^{\textup{T}_{X}}$ represents a partial transposition with respect to the Hilbert space $\mathscr{H}_{X}$. The validity of this expression can be checked by direct insertion. 

This formalism extends naturally to build HOQOs by contracting certain spaces of maps while leaving others open. Since all quantum objects can be described as linear maps with associated Choi matrices, the link product provides a unified rule for describing all sorts of complicated scenarios, from complex quantum circuits to causally indefinite process matrices. It accommodates various operations: composing independent quantum objects (in which case it reduces to the tensor product), computing measurement outcome probabilities (yielding the Born rule), determining the output state of a quantum channel for a given input, and concatenating sequential quantum operations.

As we have discussed, we can compose many such elemental objects in various ways to yield complex quantum circuit architectures, potentially with remaining `open' slots. When constructing such complex networks by linking elementary components, it is important to keep track of the spaces upon which each object/map acts in order to ensure sensible results. For instance, input states must match channel dimensions, output wires cannot link to their own inputs (avoiding causal loops), and input wires of different channels cannot be linked together. We now present the formal definition of the link product:

{\hypersetup{citecolor=white}
\begin{myDefinition}{Link Product~\cite{Chiribella_2009}}[def::linkproduct]{}
Consider two sets, $\alpha$ and $\beta$ with intersection $\alpha \cap \beta$ and set differences $\alpha \backslash \beta$ and $\beta \backslash \alpha$. For matrices $\mathsf{A} \in \mathscr{L}(\otimes_{\alpha_j \in \alpha} \mathscr{H}_{\alpha_j})$ and $\mathsf{B} \in \mathscr{L}(\otimes_{\beta_j \in \beta} \mathscr{H}_{\beta_j})$, the link product $\mathsf{A} \star \mathsf{B}$ is
\begin{align}
\label{eq::tf-lp-def-linkproduct}
    \mathsf{A}\star\mathsf{B} := \ptr{\alpha\cap \beta}{(\mathds{1}_{\beta\backslash \alpha} \otimes \mathsf{A}^{\textup{T}_{\alpha \cap \beta}})(\mathsf{B} \otimes \mathds{1}_{\alpha\backslash \beta})} \in \mathscr{L}(\mathscr{H}_{\beta \backslash \alpha}\otimes\mathscr{H}_{\alpha \backslash \beta}).
\end{align}
Here, $\mathds{1}_{x\backslash y}$ is the identity matrix on $\mathscr{H}_{x\backslash y}:=\otimes_{j\in x\backslash y} \mathscr{H}_j$ and $\bullet^{\textup{T}_{x \cap y}}$ denotes transposition on $\mathscr{H}_{x\cap y}:=\otimes_{j \in x\cap y} \mathscr{H}_j$.
\end{myDefinition}
}

\noindent As we will discuss in detail below, Eqs.~\eqref{eq::tf-cji-choiaction},~\eqref{eq::tf-cji-choieffectmap}, and~\eqref{eq::tf-cji-choiinstrument} can all be seen as examples of the link product being applied to certain quantum objects. Intuitively, the link product between two matrices $A$ and $B$ consists of: i) `padding' both of them out with identity matrices (on the spaces labelled by the sets $\beta\setminus \alpha$ and $\alpha \setminus \beta$, respectively) so that they are defined on the same space overall; ii) partially transposing one of the matrices with respect to the spaces that both of them act non-trivially upon (i.e., the spaces labelled by the set $\alpha\cap \beta$); and iii) multiplying the resulting matrices and taking the partial trace over the spaces that both of them act non-trivially upon.\footnote{Since $\mathscr{H}_{\alpha\cap \beta}$ is traced over, the partial transposition (which acts only on this space) can be applied to either matrix $\mathsf{A}$ or $\mathsf{B}$ due to the duality of transposition within said trace; we simply choose $\mathsf{A}$ as a matter of convention.} 

For pure/unitary operators, the link product can also be expressed in terms of Choi vectors $\kket{\mathsf{U}}_{12} \in \Hscr_1 \otimes \Hscr_2$ and $\kket{\mathsf{V}}_{23} \in \Hscr_2 \otimes \Hscr_3$ [see Eq.~\eqref{eq::tf-cji-choivector}] as
\begin{align}\label{eq::tf-lp-linkproductvector}
    \kket{\mathsf{U}}_{12} \star \kket{\mathsf{V}}_{23} 
    :=& \sum_i (\mathds{1}_1 \otimes \bra{i}_2 ) \kket{\mathsf{U}}_{12} \otimes (\bra{i}_2 \otimes \mathds{1}_3) \kket{\mathsf{V}}_{23} \notag \\ =& \left(\kket{\mathsf{U}}_{12}^{T_2} \otimes \ident_3\right) \Big(\ident_1\otimes \kket{\mathsf{V}}_{23} \Big) \in \Hscr_{1} \otimes \Hscr_{3} .
\end{align}
With this, it holds that $\kket{\mathsf{U}}_{12} \star \kket{\mathsf{V}}_{23} =  \kket{\mathsf{VU}}_{13} $ and for $\ket{\psi} \in \Hscr_{1}$ we have $ \kket{\mathsf{U}}_{12} \star \ket{\psi}_1 = U\ket{\psi} $. Moreover,
the resulting Choi matrix $\kket{\mathsf{U}}\bbra{\mathsf{U}} \star \kket{\mathsf{V}}\bbra{\mathsf{V}}$ can be calculated via $(\kket{\mathsf{U}}\star\kket{\mathsf{V}}) (\kket{\mathsf{U}}\star\kket{\mathsf{V}})^\dagger$~\cite{Wechs_2021}.

\newpage \noindent The link product possesses several important properties:

{\hypersetup{citecolor=white}
\begin{myDefinition*}{Link Product Properties~\cite{Chiribella_2009}}{def::linkproductproperties}
\begin{enumerate}
    \item \textit{Hermiticity Preservation:} The link product of Hermitian matrices is Hermitian. 
    \item \textit{Positivity Preservation:} The link product of positive semidefinite matrices is positive semidefinite.
    \item \textit{Associativity:} For matrices $\mathsf{A}$, $\mathsf{B}$, $\mathsf{C}$ acting on Hilbert spaces labelled by the sets $\alpha, \beta, \gamma$ respectively, with $\alpha \cap \beta \cap \gamma = \emptyset$, we have $\mathsf{A} \star (\mathsf{B} \star \mathsf{C}) = (\mathsf{A} \star \mathsf{B}) \star \mathsf{C}$. 
    \item \textit{Commutativity:} $\mathsf{A} \star \mathsf{B} = \$ (\mathsf{B} \star \mathsf{A}) \$$, where $\$$ is the unitary swap operator on $\mathscr{H}_{\beta \backslash \alpha}\otimes\mathscr{H}_{\alpha \backslash \beta}$.
\end{enumerate}
\end{myDefinition*}
}

\noindent The first two properties ensure that concatenating physically valid quantum operations yield valid CP maps, therefore remaining in the realm of physically realisable objects. The latter two properties relate to proper bookkeeping: The condition $\alpha\cap \beta\cap \gamma=\emptyset$ is always satisfied in practice (by labelling spaces appropriately) and prohibits ill-defined linkages, while the commutativity property (up to appropriate space relabelling) ensures consistent results regardless of calculation order. Again, with proper bookkeeping this property holds automatically, as the `position' of the operators in an expression is rendered irrelevant due to the labels, e.g., for two independent states, we clearly have $\rho_A \star \rho_B = \rho_B \otimes \rho_A = \rho_A \otimes \rho_B = \rho_B \star \rho_A$; this property holds true for the link product more generally and the swap operator becomes superfluous. Of course, in practical situations it is sometimes necessary to explicitly account for the swap operator, e.g., when calculating a link product in a computer program. Importantly, this commutativity property does \textit{not} imply that sequential operations commute ($\mathcal{M} \circ \mathcal{N} \neq \mathcal{N} \circ \mathcal{M}$ in general), but rather that properly labelled Choi matrices yield consistent results regardless of their position in the link product formula. Naturally, properties (iii) and (iv) equally apply to the link product of Choi \textit{vectors} [see Eq.~\eqref{eq::tf-lp-linkproductvector}].


\begin{figure}[t]
    \centering
    \subfigure[\textbf{Example~\ref{ex::tf-lp-linkproductindependent}: Independent Objects.} ]
    {
    \includegraphics[scale =0.6]{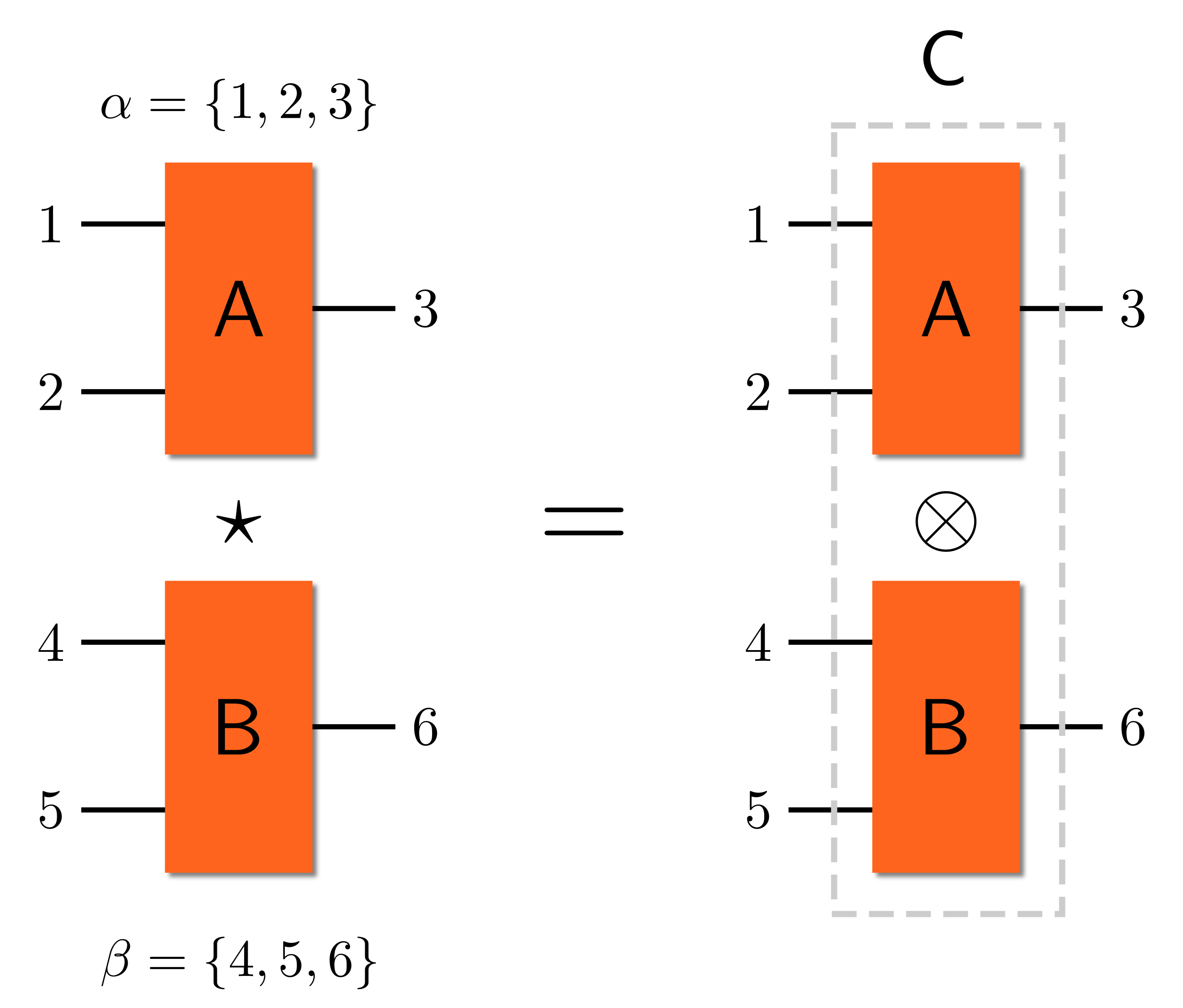}
    \label{fig::tf-lp-linkproductindependent}
    }\hspace{1cm}
    \subfigure[\textbf{Example~\ref{ex::tf-lp-linkproducttrace}: Link Product on the Same Space.}]
    {
    \includegraphics[scale=0.6]{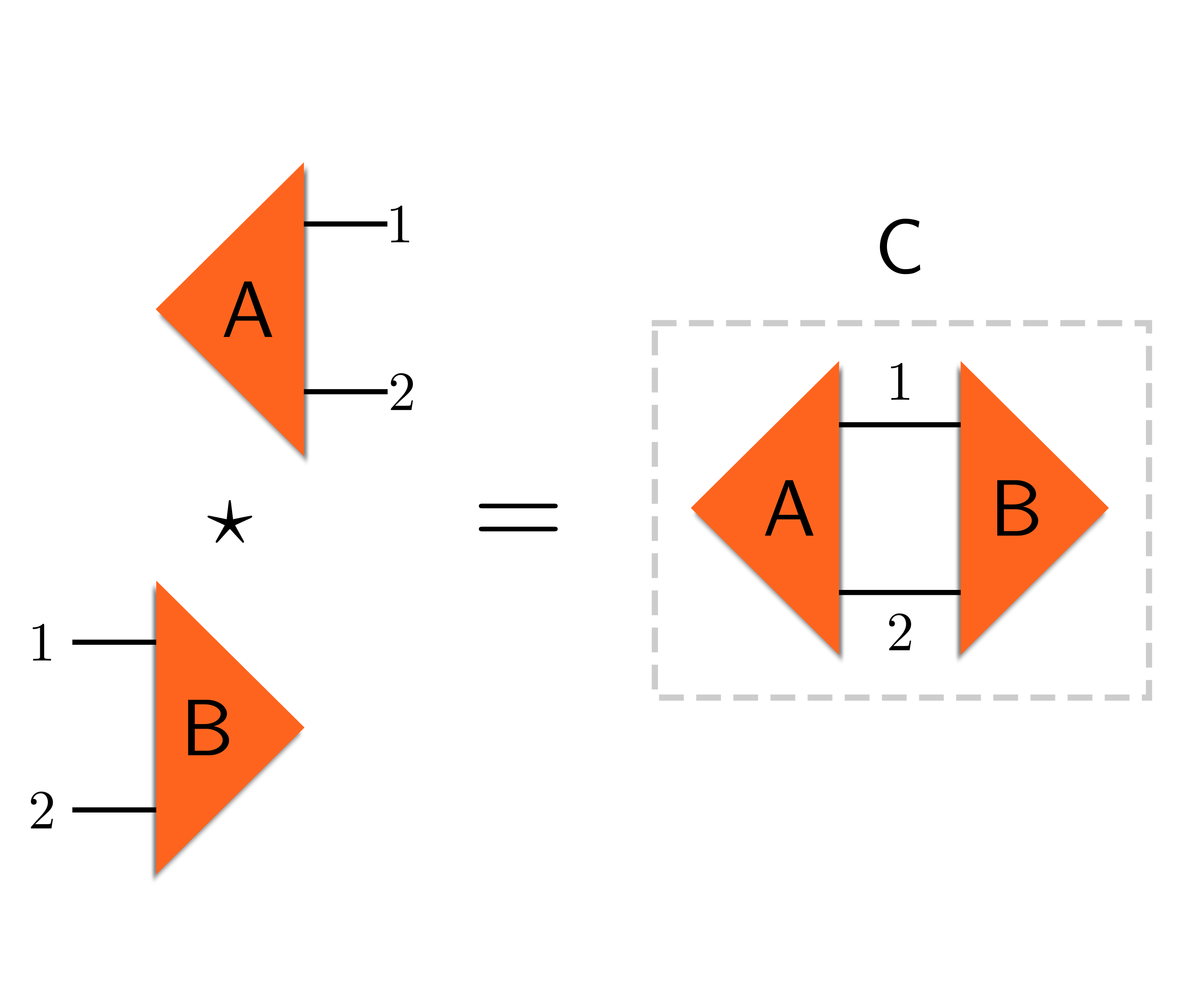}
    \label{fig::tf-lp-linkproducttrace}
    }\hfill
    \caption{\textbf{Link Product Examples (1).} In each panel, $\mathsf{C} = \mathsf{A} \star \mathsf{B}$ is depicted in the grey dashed outline. \textbf{(a)} When the Choi matrices $\mathsf{A}$ and $\mathsf{B}$ are defined upon mutually exclusive spaces, i.e., $\alpha$ and $\beta$ such that $\alpha \cap \beta = \emptyset$, the link product reduces to the tensor product. \textbf{(b)} When the Choi matrices $\mathsf{A}$ and $\mathsf{B}$ are defined upon the same space, i.e., $\alpha$ and $\beta$ such that $\alpha = \beta$, the link product reduces to the trace, yielding a scalar. Note that the above depiction `mixes' the representation of HOQOs in terms of maps (boxes with inputs and outputs) and in terms of Choi operators (which, since they are states, should technically be depicted as triangles with wires to the right) by dropping the explicit distinction between them. This mixed representation has the advantage of clarifying the physical meaning of the link product and we will opt for it whenever there is no risk of confusion.}
\end{figure}


\vspace{0.25cm}\noindent
\textbf{\textul{Simple Link Product Examples.}} The link product is particularly useful when considering (partial) compositions of quantum objects, as we now highlight through some simple examples. In what follows, we often drop the explicit distinction between a map and its corresponding Choi state; in particular, in the subsequent figures we depict the action of the link product between Choi states in terms of the corresponding action on the respective maps. Although somewhat of an abuse of notation, this choice of representation has the advantage of clarifying the physical meaning of the link product. \vspace{0.25cm}

\begin{example}[Link product of independent objects]\label{ex::tf-lp-linkproductindependent}
When one independently prepares two quantum states, $\rho_A \in \mathscr{L}(\mathscr{H}_A)$ and $\sigma_B \in  \mathscr{L}(\mathscr{H}_B)$, the link product reduces to the tensor product: $\rho_A \star \sigma_B = \sigma_B \otimes \rho_A \in \mathscr{L}(\mathscr{H}_B \otimes \mathscr{H}_A)$. This is the standard way of composing independent quantum objects. As an aside, note that when keeping track of Hilbert spaces through appropriate labelling, the order of operators in the link product becomes irrelevant, as the operation commutes up to relabelling: $\rho_A \otimes \sigma_B = \sigma_B \otimes \rho_A$; throughout, we will adhere to this convention of denoting the space that an operator belongs to by labelling, rather than its position in the tensor product. More generally, for any two Choi matrices $\mathsf{A}, \mathsf{B}$ defined on Hilbert spaces labelled by disjoint sets $\alpha, \beta$ (i.e., $\alpha\cap \beta=\emptyset$), we have [see Fig.~\ref{fig::tf-lp-linkproductindependent}]
\begin{align}\label{eq::tf-lp-linkproductindependent}
    \mathsf{A}\star\mathsf{B} = \mathsf{B} \otimes \mathsf{A} \quad \textup{whenever} \quad \alpha\cap \beta=\emptyset.
\end{align}
\end{example}\hfill$\blacksquare$

\begin{example}[Link product on the same space]\label{ex::tf-lp-linkproducttrace}
Consider measuring a quantum system using POVM elements $\{ \xi^{(x)} \}$. When one takes the link product between two Choi matrices defined on the same space---say, $\rho \in \mathscr{L}(\mathscr{H}_A)$ and $\{ \mathsf{E}^{(x)} = \xi^{(x) \textup{T}} \} \in \mathscr{L}(\mathscr{H}_A)$---it reduces to the trace: $\rho \star \mathsf{E}^{(x)} = \tr{\rho \, \mathsf{E}^{(x) \textup{T}}} = \tr{\rho \, \xi^{(x)}} \in \mathds{C}$. This yields the standard Born rule. More generally, for any two Choi matrices $\mathsf{A}, \mathsf{B}$ defined on Hilbert spaces labelled by the same set (i.e., $\alpha = \beta$), we have [see Fig.~\ref{fig::tf-lp-linkproducttrace}]
\begin{align}\label{eq::tf-lp-linkproducttrace}
    \mathsf{A}\star\mathsf{B} = \tr{\mathsf{A}^{\textup{T}}\,\mathsf{B}} \quad \textup{whenever} \quad \alpha = \beta.
\end{align}
\end{example}\hfill$\blacksquare$

\begin{figure}[t]
    \centering
    \subfigure[\textbf{Example~\ref{ex::tf-lp-linkproductchannels}: Link Product of Quantum Channels.}]
    {
    \includegraphics[scale =0.5]{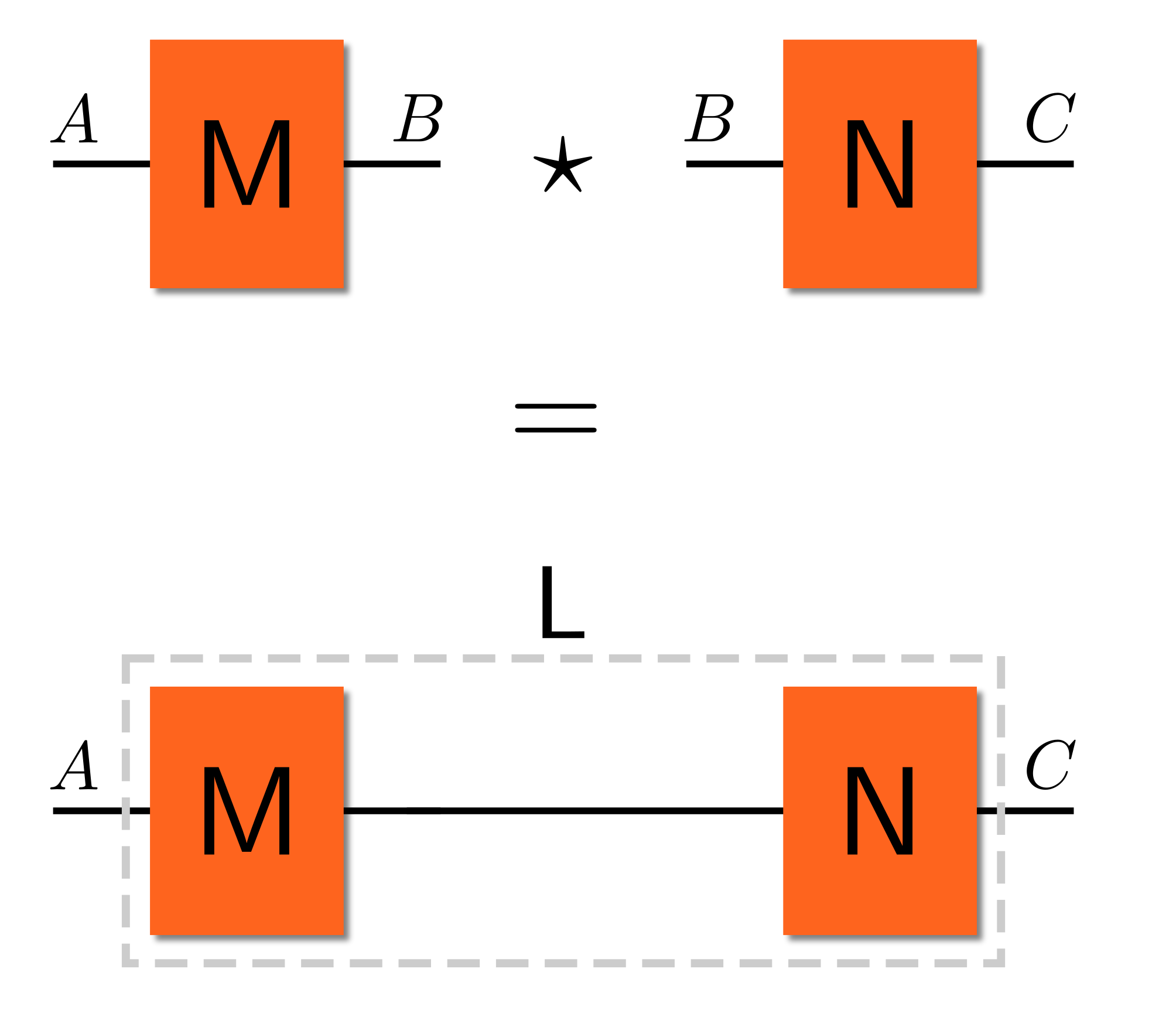}
    \label{fig::tf-lp-linkproductchannels}
    }\hspace{2cm}
    \subfigure[\textbf{Example~\ref{ex::tf-lp-linkproductconditionalstates}: Conditional States.} ]
    {
    \includegraphics[scale=0.5]{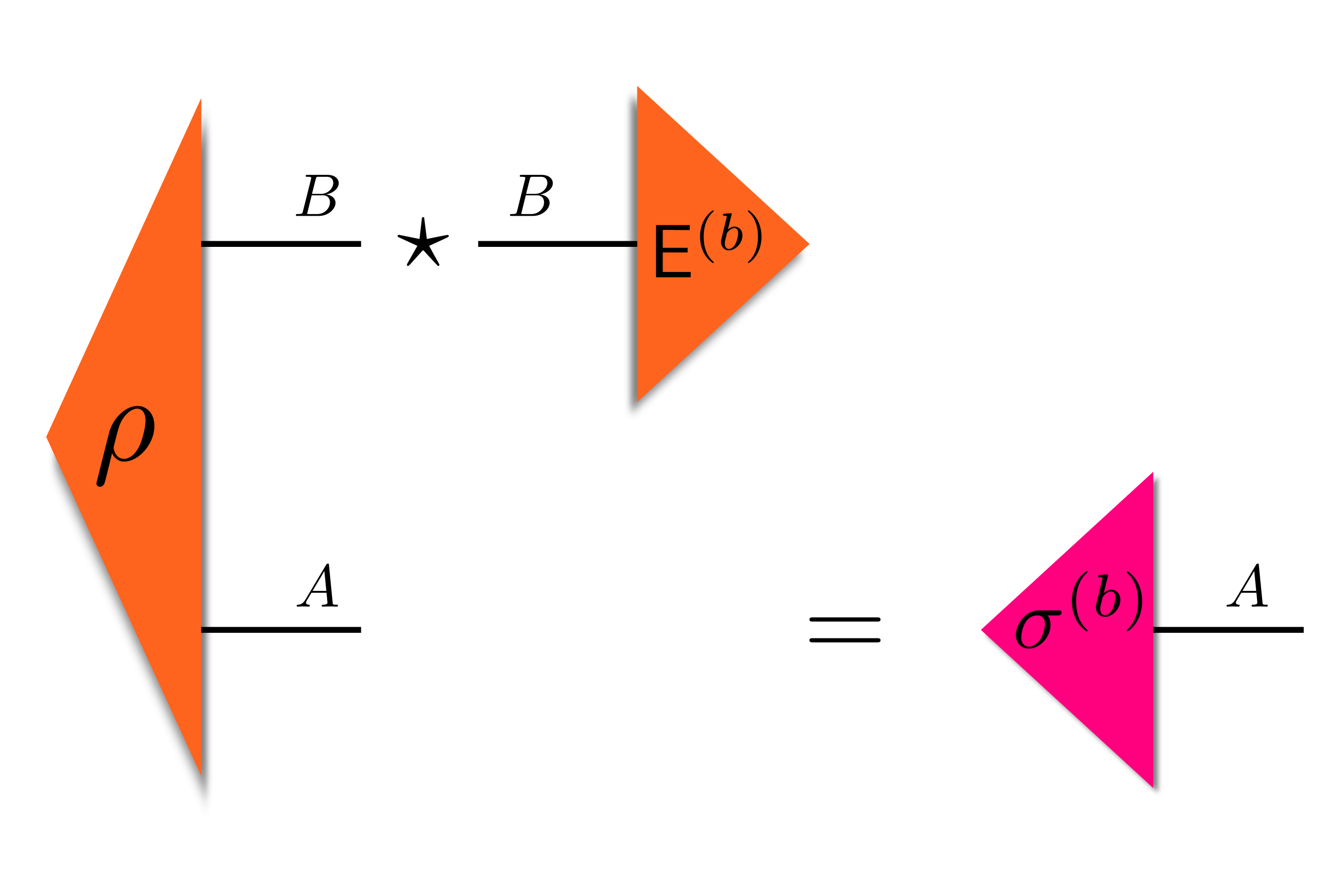}
    \label{fig::tf-lp-linkproductconditionalstates}
    }
    \caption{\textbf{Link Product Examples (2).} \textbf{(a)} The concatenation of two quantum channels $\mathsf{M}$ and $\mathsf{N}$ yields the quantum channel $\mathsf{L} = \mathsf{M} \star \mathsf{N}$ (within the grey dashed outline). \textbf{(b)} Consider a bipartite state $\rho_{AB}$ upon which Bob makes a measurement $\{ \mathsf{E}_B^{(b)} \}$; the conditional (subnormalised) state of Alice for each outcome $b$ is $\sigma_{A}^{(b)} = \rho_{AB} \star \mathsf{E}_B^{(b)}$ (pink).}
\end{figure}


\begin{example}[Concatenating quantum channels]\label{ex::tf-lp-linkproductchannels}
Consider two quantum channels, $\mathcal{M}: \mathscr{L}(\mathscr{H}_A) \to \mathscr{L}(\mathscr{H}_B)$ and $\mathcal{N}: \mathscr{L}(\mathscr{H}_B) \to \mathscr{L}(\mathscr{H}_C)$, described by the Choi matrices $\mathsf{M}_{BA} \in \mathscr{L}(\mathscr{H}_B \otimes \mathscr{H}_A)$ and $\mathsf{N}_{CB} \in \mathscr{L}(\mathscr{H}_C \otimes \mathscr{H}_B)$ respectively. Their concatenation $\mathcal{L} := \mathcal{N} \circ \mathcal{M} : \mathscr{L}(\mathscr{H}_A) \to \mathscr{L}(\mathscr{H}_C)$ has Choi matrix [see Fig.~\ref{fig::tf-lp-linkproductchannels}]
\begin{align}\label{eq::tf-lp-linkproductchannels}
    \mathsf{L}_{CA} = \mathsf{M}_{BA} \star \mathsf{N}_{CB} = \ptr{B}{(\mathds{1}_C \otimes \mathsf{M}_{BA}^{\textup{T}_B})(\mathsf{N}_{CB}\otimes\mathds{1}_A)} \in \mathscr{L}(\mathscr{H}_C \otimes \mathscr{H}_A).
\end{align}
\end{example}\hfill$\blacksquare$

\noindent The link product can equally be employed to denote the action of trace non-increasing CP maps, as well as the action of maps on only a subset of wires in order to, e.g., compute quantum states conditioned on measurement outcomes.


\begin{figure}[t!]
    \centering
    \subfigure[\textbf{Example~\ref{ex::tf-lp-linkproductpostmeasurementstates}: Post-measurement States.}]
    {
    \includegraphics[scale=0.7]{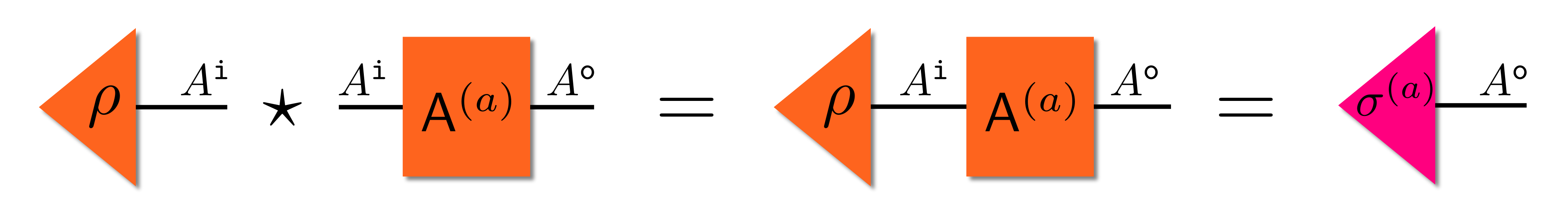}
    \label{fig::tf-lp-linkproductpostmeasurementstates}
    }
    \subfigure[\textbf{Example~\ref{ex::tf-lp-linkproductpartialchannels}: Partial Concatenation.}]
    {
    \includegraphics[scale =0.6]{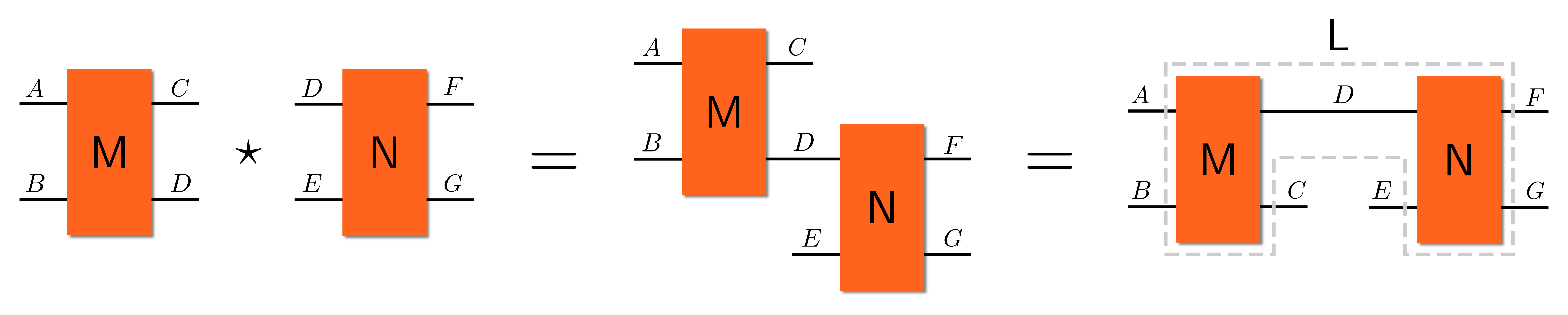}
    \label{fig::tf-lp-linkproductpartialchannels}
    }
    \hfill
    \caption{\textbf{Link Product Examples (3).} \textbf{(a)} Suppose Alice performs an instrument $\{ \mathsf{A}_{A^\out A^\inp}^{(a)} \}$ on a quantum state $\rho_{A^\inp}$; the post-measurement (subnormalised) state for each outcome is $\sigma_{A^\out}^{(a)} = \rho_{A^\inp} \star \mathsf{A}_{A^\out A^\inp}^{(a)}$ (pink). \textbf{(b)} The partial concatenation of two channels $\mathsf{M}$ and $\mathsf{N}$ over subsystem $D$ leads to a channel $\mathsf{L}$ that takes states in $\mathscr{L}(\mathscr{H}_A \otimes \mathscr{H}_B \otimes \mathscr{H}_E)$ to those in $\mathscr{L}(\mathscr{H}_C \otimes \mathscr{H}_F \otimes \mathscr{H}_G)$. Equivalently, it can be understood as a superchannel $\mathsf{L}$ with one slot (outlined in grey, rightmost), into which one could insert any channel $\mathcal{Q}: \Lscr(\Hscr_C) \rightarrow \Lscr(\Hscr_E)$ (not shown) to yield the correct resulting channel from $\Lscr(\Hscr_A \otimes \Hscr_B)$ to $\Lscr(\Hscr_F \otimes \Hscr_G)$ [see Eq.~\eqref{eqn::actionLsuperchann}].}
\end{figure}


\begin{example}[Computing conditional states]\label{ex::tf-lp-linkproductconditionalstates}
Consider a bipartite quantum state $\rho_{AB} \in \mathscr{L}(\mathscr{H}_A \otimes \mathscr{H}_B)$. Suppose that Bob performs a local measurement described by $\{ \mathsf{E}_B^{(b)} \} \in \mathscr{L}(\mathscr{H}_B)$. The conditional (subnormalised) state of Alice's system is [see Fig.~\ref{fig::tf-lp-linkproductconditionalstates}]
\begin{align}\label{eq::tf-lp-linkproductconditionalstates}
    \sigma_A^{(b)} = \rho_{AB} \star \mathsf{E}^{(b)}_B = \ptr{B}{\rho_{AB}(\mathds{1}_A \otimes \mathsf{E}_B^{(b) \textup{T}})} \in \mathscr{L}(\mathscr{H}_A).
\end{align}
\end{example}\hfill$\blacksquare$

\noindent Similarly, whenever a state is measured with an instrument but is not discarded, the resulting (sub-normalised) state can be computed via the link product.

\begin{example}[Post-measurement states and sequential statistics]\label{ex::tf-lp-linkproductpostmeasurementstates}
For an initial state $\rho\in\mathscr{L}(\mathscr{H}_{A^\inp})$ measured by an instrument $\mathcal{J}_{\mathcal{A}} = \{ \mathcal{A}^{(a)} \}:\mathscr{L}(\mathscr{H}_{A^\inp})\to\mathscr{L}(\mathscr{H}_{A^\out})$ described by Choi matrices $\mathsf{A}_{A^\out A^\inp}^{(a)} = \mathsf{Choi}(\mathcal{A}^{(a)})$, the post-measurement state is [see Fig.~\ref{fig::tf-lp-linkproductpostmeasurementstates}]
\begin{align}\label{eq::tf-lp-linkproductpostmeasurementstates}
    \sigma_{A^\out}^{(a)} = \rho_{A^\inp} \star \mathsf{A}_{A^\out A^\inp}^{(a)} = \ptr{A^\inp}{(\mathds{1}_{A^\out} \otimes \rho_{A^\inp}^{\textup{T}})\mathsf{A}_{A^\out A^\inp}^{(a)}}  \in \mathscr{L}(\mathscr{H}_{A^\out}).
\end{align}
The post-measurement state encodes the probability to record outcome $a$ in its trace: $\mathds{P}(a|\rho,\mathcal{J}_{\mathcal{A}})=\tr{\mathcal{A}^{(a)}[\rho]} = \tr{\sigma_{A^\out}^{(a)}}$. 

One can extend this example to a more general scenario of calculating the probability distribution over a pair of sequential measurements. Continuing from above, suppose that after the interrogation $\mathcal{J}_{\mathcal{A}}$, the system evolves according to a quantum channel $\mathcal{C}:\mathscr{L}(\mathscr{H}_{A^\out})\to\mathscr{L}(\mathscr{H}_{B^\inp}) \in \mathsf{CPTP}$, before being subject to a final measurement $\mathcal{J}_{\mathcal{B}} = \{ B^{(b)} \} \in\mathscr{L}(\mathscr{H}_{B^\inp})$. The probability of sequentially obtaining the outcomes $a$ and then $b$ in this experiment is given by the standard Born rule
\begin{align}\label{eq::tf-lp-linkproductsequentialprobability}
	\mathds{P}(a,b|\mathcal{\Jcal}_{\Acal}, \mathcal{\Jcal}_{\mathcal{B}}) = \tr{ B^{(b)} \left( \mathcal{C} \left( \mathcal{A}^{(a)}[\rho] \right) \right) }.
\end{align}
When all maps are described in terms of Choi matrices, i.e., $\mathsf{A}^{(a)}_{A^\out A^\inp} = \mathsf{Choi}(\mathcal{A}^{(a)}), \mathsf{C}_{B^\inp A^\out} = \mathsf{Choi}(\mathcal{C}),  \mathsf{B}^{(b)}_{B^\inp} = \mathsf{Choi}(\mathcal{B}^{(b)})$ (where $\mathcal{B}^{(b)}[\bullet] := \tr{\bullet \, B^{(b)}}$ represents Bob's effect), the expression can be written as
\begin{align}
    	\mathds{P}(a,b|\mathcal{\Jcal}_{\Acal}, \mathcal{\Jcal}_{\mathcal{B}}) = \mathsf{B}^{(b)}_{B^\inp} \star \mathsf{C}_{B^\inp A^\out} \star \mathsf{A}^{(a)}_{A^\out A^\inp} \star \rho_{A^\inp}.
\end{align}
\end{example}\hfill$\blacksquare$

\noindent As we can see, the link product generalises both spatial (i.e., tensor product) and temporal (i.e., trace or concatenation) composition rules for quantum objects, and it can be readily employed when operations only act on a subset of wires. This allows one to build up more complex HOQOs by linking together elementary building blocks. Here, we demonstrate how the (partial) concatenation of two quantum channels leads to a superchannel; this construction readily generalises to arbitrary numbers of quantum channels (see Sec.~\ref{subsubsec::toqp-deterministichoqos} for more details).

\begin{example}[Concatenating parts of channels]\label{ex::tf-lp-linkproductpartialchannels}
Consider two quantum channels, $\mathcal{M}: \mathscr{L}(\mathscr{H}_A\otimes\mathscr{H}_B) \to \mathscr{L}(\mathscr{H}_C\otimes\mathscr{H}_D)$ and $\mathcal{N}: \mathscr{L}(\mathscr{H}_D\otimes \mathscr{H}_E) \to \mathscr{L}(\mathscr{H}_F \otimes \mathscr{H}_G)$, described by the Choi matrices $\mathsf{M}_{DCBA} \in \mathscr{L}(\mathscr{H}_D \otimes \mathscr{H}_C \otimes \mathscr{H}_B \otimes \mathscr{H}_A)$ and $\mathsf{N}_{GFED} \in \mathscr{L}(\mathscr{H}_G \otimes \mathscr{H}_F \otimes \mathscr{H}_E \otimes \mathscr{H}_D)$ respectively. Concatenating these channels over the system $D$ leads to the dynamics $\mathcal{L} := \mathcal{N} \circ \mathcal{M} : \mathscr{L}(\mathscr{H}_A \otimes \mathscr{H}_B \otimes \mathscr{H}_E) \to \mathscr{L}(\mathscr{H}_C \otimes \mathscr{H}_F \otimes \mathscr{H}_G)$ described by the Choi matrix $\mathsf{L}_{GFECBA} \in \mathscr{L}(\mathscr{H}_G \otimes \mathscr{H}_F \otimes \mathscr{H}_E \otimes \mathscr{H}_C \otimes \mathscr{H}_B \otimes \mathscr{H}_A)$ 
\begin{align}\label{eq::tf-lp-linkproductpartialchannels}
    \mathsf{L}_{GFECBA} = \mathsf{M}_{DCBA} \star \mathsf{N}_{GFED} = \ptr{D}{(\mathds{1}_{GFE} \otimes \mathsf{M}^{\textup{T}_D}_{DCBA})(\mathsf{N}_{GFED}\otimes\mathds{1}_{CBA})}.
\end{align}
As depicted in Fig.~\ref{fig::tf-lp-linkproductpartialchannels}, the resulting object $\mathsf{L}_{GFECBA}$ can be understood as a superchannel with one open slot. Indeed, it correctly reproduces the channel resulting from `plugging' an arbitrary CPTP map $\mathcal{Q}:\Lscr(\Hscr_C) \rightarrow \Lscr(\Hscr_E)$ with corresponding Choi matrix $\mathsf{Q}_{CE}$ into said slot, since 
\begin{gather}
\label{eqn::actionLsuperchann}
    \mathsf{L}_{GFECBA} \star \mathsf{Q}_{CE} = \mathsf{M}_{DCBA} \star \mathsf{N}_{GFED}  \star \mathsf{Q}_{CE} = \mathsf{M}_{DCBA} \star \mathsf{Q}_{CE} \star \mathsf{N}_{GFED}   = \mathsf{Choi}[\mathcal{N} \circ \mathcal{Q} \circ \mathcal{M}].
\end{gather}
Building HOQOs with more open slots follows in a similar vein to Eq.~\eqref{eq::tf-lp-linkproductpartialchannels} by linking together more than two CPTP maps.
\end{example}\hfill$\blacksquare$

\vspace{0.25cm}\noindent
\textbf{\textul{Link Product for Motivating Examples.}} The power of the link product becomes particularly evident when applying it to HOQOs. Two key properties---commutativity and associativity---allow us to contract specific spaces while leaving others open, enabling the construction of HOQOs with open slots as we will formalise in Sec.~\ref{subsec::timeorderedquantumprocesses}. We first return to see this applied in practice to the motivating examples.

\begin{enumerate}[1.]
    \item \textit{Superchannels on Preparation Procedures:} The action of a superchannel $\Tcal$ on a preparation procedure $\mathcal{M}^{(x)}$ [see Eqs.~\eqref{eq::me-osdic-correlatedcase} and~\eqref{eq::superCh-example-CJI}] can be expressed as $\rho_{1^\inp}^{(x)} = \mathsf{T}_{1^{\inp} 0^{\out} 0^{\inp}} \star \mathsf{M}^{(x)}_{0^{\out} 0^{\inp}}$.
    \item \textit{Encoder-Decoder Superchannels:} The action of a superchannel $\mathcal{T}_{\textup{SC}}$ on an input operation $\mathcal{C}_{A}$ [see Eqs.~\eqref{eq::me-qca-encoderdecodergeneral} and~\eqref{eq::Encoder-decoder-example-CJI}] becomes $\mathsf{C}'_{B^{\out} B^{\inp}} = \mathsf{T}_{B^{\out} A^{\out} A^{\inp} B^{\inp}} \star \mathsf{C}_{A^{\out} A^{\inp}}$. Moreover, the superchannel itself can be constructed in terms of the encoder-decoder circuit elements as $\mathsf{T}_{B^{\out} A^{\out} A^{\inp} B^{\inp}} = \mathsf{E}_{\textup{aux} A^{\inp} B^{\inp}} \star \mathsf{D}_{B^{\out} A^{\out} \textup{aux}}$.
    \item \textit{Process Matrices:} For a process matrix $\Wcal$ applied to instrument elements $\Mcal_A^{(a)} \in \mathcal{J}_A,\Mcal_B^{(b)} \in \mathcal{J}_B$ [see Eqs.~\eqref{eq::me-cqt-causallyindefinite},~\eqref{eq::instruments-example-CJI} and~\eqref{eq::processmatrix-example-CJI}], we have $\mathds{P}(a,b|\Jcal_A,\Jcal_B) = \mathsf{W}_{A^{\inp} A^{\out} B^{\inp} B^{\out}} \star \mathsf{M}_{A^{\inp} A^{\out}}^{(a)} \star \mathsf{M}_{B^{\inp} B^{\out}}^{(b)}$.
\end{enumerate}

\noindent It is straightforward to verify the above expressions via the definitions of the Choi state and the link product. We hope that it is now clear to the Reader why this representation is so favourable: All concatenations of quantum objects---be they simple tensor products for independent composition, the standard Born rule, quantum channel dynamics, or more intricate higher-order scenarios---are manifestations of the same rule, namely the link product. Thus, working in this setting, one need not concern oneself too much with worrying about complicated expressions involving different kinds of objects (maps, operators, etc.) but rather can simply cast everything into the Choi representation and apply the link product appropriately. As we shall see in more detail below (see Secs.~\ref{subsubsec::axiomatichoqos-qc-transtranstrans} and~\ref{subsec::indefinitecausalorder}) this ease of framework even goes beyond the concatenation of objects familiar in standard quantum mechanics and also applies to \textit{all} kinds of HOQOs, e.g., those displaying indefinite causal order.

\vspace{0.25cm}\noindent
\textbf{\textul{Link Product `Rules of Thumb'.}} The properties of the link product imply several `rules of thumb' that generalise naturally to the higher-order setting: 
\begin{enumerate}
    \item \textit{Independent Composition}: Independent objects are composed via the tensor product.
    \item \textit{Spatiotemporal Born Rule}: When all wires in an expression are contracted, the result is a probability.
    \item \textit{Preservation of Overall Determinism}: Linking deterministic objects always yields another deterministic object. Linking probabilistic objects yields a probabilistic object (i.e., no supernormalised `probability distributions' can arise).
\end{enumerate}

\vspace{0.25cm}\noindent
\textbf{\textul{Summary.}} We have explored a favourable representation of linear maps---as matrices via the CJI---that exhibits particularly nice properties for the operations relevant in quantum theory. We demonstrated how simple objects such as states, measurements and channels, can be concatenated via the link product when represented in this way. The link product generalises both spatial (tensor product) and temporal (trace) compositions; all concatenations---from simple tensor products to complex higher-order scenarios---follow this same fundamental rule. Working in the Choi representation thus eliminates the need to handle different types of objects (maps, operators, etc.) separately, placing them all on equivalent mathematical footing. While this is not the only meaningful way of representing HOQOs, given these advantages, we shall opt for the CJI representation throughout whenever appropriate.

So far, our discussion has mostly centred on `lower-order' quantum operations that are familiar from standard quantum theory, such as state preparations, channels and measurements. However, we also provided glimpses on how these concepts can naturally be applied to `higher-order' settings by way of the motivating examples. The Choi representation naturally extends to the higher-order realm by linking elementary components across difference spaces. This formalism sets the stage for constructing more sophisticated quantum objects in a systematic way, as we now move to discuss.

\FloatBarrier


\subsection{Time-Ordered Quantum Processes}\label{subsec::timeorderedquantumprocesses}

\begin{figure}[t]
\centering
\vspace{0.5em}
\includegraphics[scale=0.6]{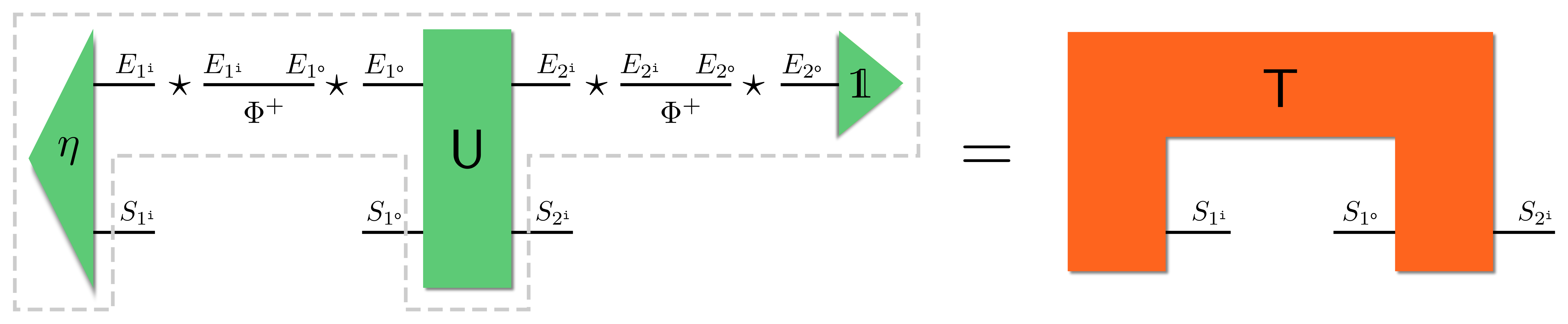}
\caption{\textbf{Superchannel Construction.} The superchannel $\mathsf{T}$ (orange, right), which acts only on the space of the system, can be constructed explicitly by linking the Choi matrices of the initial joint state $\eta$, the joint unitary $\mathsf{U}$ and the trace $\mathds{1}$ (green, left) over the environment degrees of freedom (explicitly denoted by linking identity channels $\Phi^+$ over the environment between times).} \label{fig::toqp-linkproductsuperchannel}
\end{figure}

So far, we have explored how the link product can be used to concatenate `standard' objects in quantum theory---such as states, measurements, channels and instruments---to yield other such objects or probability distributions. More generally, one can concatenate all sorts of complicated spatiotemporal quantum objects using the link product, e.g., parts of quantum circuits, as shown in Fig.~\ref{fig::toqp-linkproductsuperchannel}, in order to construct HOQOs from such elementary building blocks. We begin with a simple such construction, namely that of the superchannel in terms of its underlying dilation, before generalising to multi-time quantum processes.


\subsubsection{Quantum Superchannels: 1-Slot Quantum Combs}
\label{subsubsec::toqp-superchannels}\hfill\\

\noindent Quantum superchannels represent the simplest case of HOQOs, characterised by having a single `open slot' where an operation can be inserted. These mathematical objects are particularly useful for describing open quantum dynamics where a system interacts with an environment, especially in scenarios with initial system-environment correlations (as we considered in Sec.~\ref{subsubsec::me-opensystemdynamicswithinitialcorrelations}).

A superchannel combines several fundamental elements of quantum mechanics: an initial system-environment state $\eta \in \mathscr{L}(\mathscr{H}_{S_{1^\inp}} \otimes \mathscr{H}_{E_{1^\inp}})$, a unitary evolution $\Ucal: \mathscr{L}(\mathscr{H}_{S_{1^\out}} \otimes \mathscr{H}_{E_{1^\out}}) \rightarrow \mathscr{L}(\mathscr{H}_{S_{2^\inp}} \otimes \mathscr{H}_{E_{2^\inp}})$ acting on both system and environment, and a final partial trace over the environment $\text{tr}_{E_{2^\out}}: \mathscr{L}(\mathscr{H}_{E_{2^\out}}) \rightarrow \mathds{C}$. Additionally, we include identity maps representing the evolution of the environment that remains inaccessible between times, i.e., $\mathcal{I}: \mathscr{L}(\mathscr{H}_{E_{1^\inp}}) \to \mathscr{L}(\mathscr{H}_{E_{1^\out}})$ and $\mathcal{I}: \mathscr{L}(\mathscr{H}_{E_{2^\inp}}) \to \mathscr{L}(\mathscr{H}_{E_{2^\out}})$. Here, we focus on superchannels with one open slot and a future output wire (with no global past) which conveys the crucial points; the case with a global past can be derived similarly including an additional space labelled by $S_{0^\out}$ and replacing the initial system-environment state with an isometry map $\mathcal{V} : \mathscr{L}(\mathscr{H}_{S_{0^\out}}) \to  \mathscr{L}(\mathscr{H}_{E_{1^\inp}} \otimes \mathscr{H}_{S_{1^\inp}})$.

As mentioned previously, we distinguish between input and output spaces (in the definition of $\Ucal$ and $\Ical$) and do not identify the systems $S_{1^\inp}$ and $S_{1^\out}$. This distinction between input and output spaces is crucial as it denotes where an experimenter can intervene throughout the process: Any map $\Mcal:\mathscr{L}(\mathscr{H}_{S_{1^\inp}}) \rightarrow \mathscr{L}(\mathscr{H}_{S_{1^\out}})$ acts on the system in between $\eta$ and $\Ucal$. In contrast, the environment part of the initial state `feeds forward' without intervention to become an input to the global unitary evolution, represented by the identity map; similarly for the environment degrees of freedom output by the unitary that eventually get traced over. 

The mathematical construction of the superchannel in terms of its primitives can be described via the link product. Expressing all of the constituent maps by their Choi matrices, i.e., $\mathsf{U}_{S_{2^\inp} E_{2^\inp} S_{1^\out} E_{1^\out}} = \mathsf{Choi}(\Ucal) \in \mathscr{L}(\mathscr{H}_{S_{2^\inp}} \otimes \mathscr{H}_{E_{2^\inp}} \otimes \mathscr{H}_{S_{1^\out}} \otimes \mathscr{H}_{E_{1^\out}})$, $\Phi^+_{E_{X^\out} E_{X^\inp}} = \mathsf{Choi}(\Ical) \in \mathscr{L}(\mathscr{H}_{E_{X^\out}} \otimes \mathscr{H}_{E_{X^\inp}})$ and $\mathds{1}_{E_{2^\out}} = \mathsf{Choi}(\text{tr}_{E_{2^\out}}) \in \mathscr{L}(\mathscr{H}_{E_{2^\out}})$, the Choi matrix $\mathsf{T}_{S_{2^\inp} S_{1^\out} S_{1^\inp}} \in \mathscr{L}(\mathscr{H}_{S_{2^\inp}} \otimes \mathscr{H}_{S_{1^\out}} \otimes \mathscr{H}_{S_{1^\inp}})$ of the superchannel $\Tcal: \mathscr{L}[\mathscr{L}(\mathscr{H}_{S_{1^\inp}}) \to \mathscr{L}(\mathscr{H}_{S_{1^\out}})] \to \mathscr{L}(\mathscr{H}_{S_{2^\inp}})$ is given by (see Fig.~\ref{fig::toqp-linkproductsuperchannel})
\begin{align}\label{eq::toqp-superchannelconstruction}
    \mathsf{T}_{S_{2^\inp} S_{1^\out} S_{1^\inp}} &= \mathds{1}_{E_{2^\out}} \star \Phi^+_{E_{2^\inp} E_{2^\out}} \star \mathsf{U}_{S_{2^\inp} E_{2^\inp} S_{1^\out} E_{1^\out}} \star \Phi^+_{E_{1^\inp} E_{1^\out}} \star \eta_{S_{1^\inp} E_{1^\inp}} \notag \\
    &= \mathds{1}_{E_{2^\out}} \star \mathsf{U}_{S_{2^\inp} E_{2^\out} S_{1^\out} E_{1^\inp}} \star \eta_{S_{1^\inp} E_{1^\inp}},
\end{align}
where in the final line we used the property $X_{AB} \,\star\, \Phi^+_{BC} = X_{AC}$, which encodes the fact that linking an identity map with some object simply relabels the wire appropriately. Although we explicitly included the identity maps in the top line of the formula above, in most cases we will only include it implicitly by identifying and relabelling the wires that get connected. Finally, from this point on, we will often drop the labels $S$ and $E$ wherever unambiguous: For instance, we understand the superchannel above $\mathsf{T}$ to act on the Hilbert space associated to the system $S$ at times $t_1$ (both input and output spaces) and $t_2$ (only input space), and therefore will write the condensed $\mathsf{T}_{2^\inp 1^{\out} 1^\inp}$ instead of $\mathsf{T}_{S_{2^\inp} S_{1^\out} S_{1^\inp}}$; on the other hand, the labelling of objects such as $\mathsf{U}_{S_{2^\inp} E_{2^\out} S_{1^\out} E_{1^\inp}}$ above (where system and environment spaces must remain distinguishable) will be condensed to $\mathsf{U}_{S_{2^\inp 1^\out} E_{2^\out 1^\inp}}$. 

The construction above ensures two essential physical properties for the superchannel: generalised complete positivity and generalised trace preservation. Generalised complete positivity ensures that the superchannel maintains positivity when acting on any (part of a) physical operation, mathematically expressed in the Choi representation as $\mathsf{T}_{{2^\inp} {1^\out} {1^\inp}} \geq 0$. This property guarantees that the output of a superchannel remains positive semidefinite (even when only acting on parts of objects). Generalised trace preservation ensures overall deterministic evolution and enforces causality by preventing any future measurements from influencing past statistics (on average). This is expressed in the Choi representation as $\ptr{{2^\inp}}{\mathsf{T}_{{2^\inp} {1^\out} {1^\inp}}} = \mathds{1}_{1^\out} \otimes\rho_{{1^\inp}}$ for some quantum state $\rho_{{1^\inp}}$. Conversely, any matrix satisfying said conditions represents some fixed underlying system-environment dynamics (see Sec.~\ref{subsec::axiomatichoqos}). Thus, we have:

\begin{myDefinition}{Superchannels}{}
A superchannel $\mathsf{T}_{{2^\inp} {1^\out} {1^\inp}} \in \mathscr{L}(\mathscr{H}_{S_{2^\inp}} \otimes \mathscr{H}_{S_{1^\out}} \otimes \mathscr{H}_{S_{1^\inp}})$ is characterised by
\begin{alignat}{2} 
    &\text{1. (Generalised) Complete positivity: } &&\mathsf{T}_{{2^\inp} {1^\out} {1^\inp}} \geq 0 \label{eq::toqp-def-superchannelcp} \\
    &\text{2. (Generalised) Trace preservation: } &&\ptr{{2^\inp}}{\mathsf{T}_{{2^\inp} {1^\out} {1^\inp}}} = \mathds{1}_{1^\out} \otimes\rho_{{1^\inp}} \label{eq::toqp-def-superchanneltp} \\ & &&\text{for } \rho_{{1^\inp}} \in \mathsf{St}(\mathscr{H}_{{1^\inp}}). \notag
\end{alignat}
\end{myDefinition}

\begin{figure}[t]
\centering
\vspace{0.5em}
\includegraphics[scale=0.6]{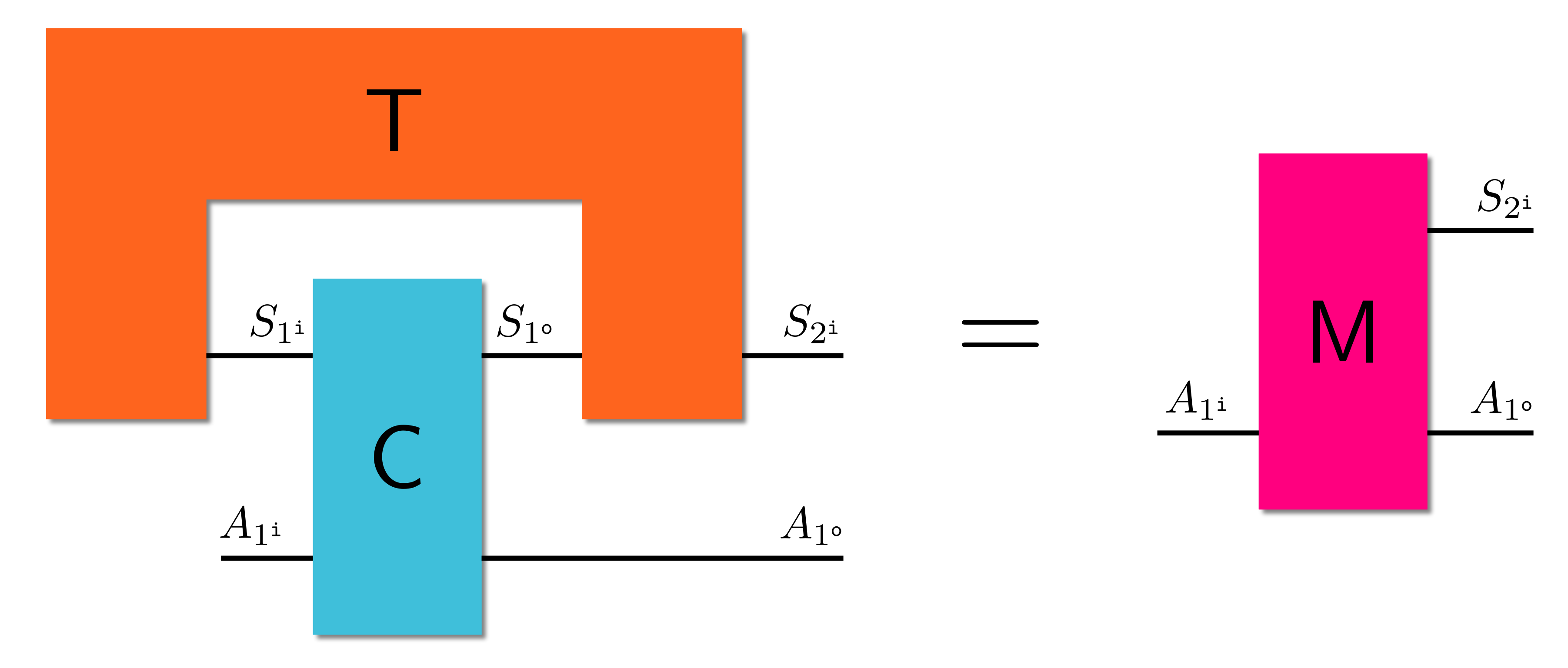}
\caption{\textbf{Complete Positivity of a Superchannel.} The natural extension of the notion of complete positivity to a superchannel $\mathsf{T}$ means that its action upon any CP map $\mathsf{C}$ which acts on the system and an arbitrarily large auxiliary system yields a map $\mathsf{M}$ that is CP.} \label{fig::toqp-superchannelcp}
\end{figure}

\noindent The fact that these conditions hold for superchannels arising from an underlying system-environment dynamics can easily be verified by applying the properties of the link product and the fact that $\mathcal{U} \in \mathsf{CPTP}$ to Eq.~\eqref{eq::toqp-superchannelconstruction}. In particular, since $\mathsf{T}_{S_{2^\inp} S_{1^\out} S_{1^\inp}} = \mathds{1}_{E_{2^\out}} \star \mathsf{U}_{S_{2^\inp} E_{2^\out} S_{1^\out} E_{1^\inp}} \star \eta_{S_{1^\inp} E_{1^\inp}}$, the fact that $\mathsf{T}_{S_{2^\inp} S_{1^\out} S_{1^\inp}} \geq 0$ follows immediately from property (ii) of the link product; moreover, we can directly calculate
\begin{align}\label{eq::toqp-superchannelinherittraceconditions}
    \ptr{{2^\inp}}{\mathsf{T}_{{2^\inp} {1^\out} {1^\inp}}} &= \ptr{S_{2^\inp}}{\mathds{1}_{E_{2^\out}} \star \mathsf{U}_{S_{2^\inp} E_{2^\out} S_{1^\out} E_{1^\inp}} \star \eta_{S_{1^\inp} E_{1^\inp}}} = \ptr{S_{2^\inp} E_{2^\out}}{\mathsf{U}_{S_{2^\inp} E_{2^\out} S_{1^\out} E_{1^\inp}} \star \eta_{S_{1^\inp} E_{1^\inp}}} \notag \\
    &= \mathds{1}_{S_{1^\out} E_{1^\inp}} \star \eta_{S_{1^\inp} E_{1^\inp}} = \mathds{1}_{S_{1^\out}} \otimes \ptr{E_{1^\inp}}{\eta_{S_{1^\inp} E_{1^\inp}}} = \mathds{1}_{1^\out} \otimes\rho_{{1^\inp}},
\end{align}
where $\rho_{{1^\inp}}:= \ptr{E_{1^\inp}}{\eta_{S_{1^\inp} E_{1^\inp}}}$. As in the case of quantum channels, these two properties refer to the physicality and the overall deterministic occurrence of the dynamics. Firstly, the generalised notion of complete positivity means that for any physical operation $\Ccal_{SA} \in \mathsf{CP}$ that an experimenter might implement on both the system $S$ and some additional auxiliary degrees of freedom $A$ (of arbitrarily large dimension), the resulting object $\Tcal_S \otimes \Ical_A[\Ccal_{SA}]$ is a CP map~\cite{Modi_2012_SciRep} (see Fig.~\ref{fig::toqp-superchannelcp}); this also implies Hermiticity preservation and that for any local trace non-increasing CP map $\Ccal_S$ applied, the resulting output (potentially subnormalised) state is positive semidefinite. 


Secondly, the generalised notion of trace preservation ensures that whenever the superchannel acts on an operation $\Ccal_S$ that is TP, i.e., overall deterministic, then the resulting output state $\eta_{2^\inp}$ is of unit trace. Indeed, if $\mathsf{C}^S_{1^\out 1^\inp} = \textsf{Choi}[\Ccal_S]$, then
\begin{gather}
    \tr{\eta_{2^\inp}} = \tr{\mathsf{T}_{2^\inp 1^\out 1^\inp} \star \mathsf{C}^S_{1^\out 1^\inp}} = \tr{\ptr{2^\inp}{\mathsf{T}_{2^\inp 1^\out 1^\inp}} \star \mathsf{C}^S_{1^\out 1^\inp}} = \tr{(\ident_{1^\out} \otimes \rho_{1^\inp}) \star \mathsf{C}^S_{1^\out 1^\inp}} = 1,
\end{gather}
where we have explicitly used $\ptr{1^\out}{\mathsf{C}^S_{1^\out 1^\inp}} = \ident_{1^\inp}$ in the last equality.

In addition, generalised trace preservation implies that the Choi matrix of the superchannel is a supernormalised quantum state with trace equal to the dimension of the $S_{1^\out}$ output Hilbert space. As we will discuss in more detail later, for HOQOs, this generalised notion of trace preservation means that the superchannel itself is an overall deterministic object, which equivalently encodes a \textit{causality constraint}: On average, no information can be sent from the future to the past. This can be seen by considering an experimenter that aims to send information from $1^\out$ back in time to $1^\inp$ by feeding states $\{\mu^{x}_{1^\out}\}$ into the process. On $2^\inp$, they can perform measurements, corresponding to a POVM $\{ \mathsf{E}_{2^\inp}^{(y)} \}$, which overall adds up to the unique deterministic effect $\sum_y \mathsf{E}_{2^\inp}^{(y)} = \mathds{1}_{2^\inp}$. Thus, on average, when aiming to send the state $\mu^{x}_{1^\out}$, the experimenter would perform an operation that corresponds to the Choi state $\ident_{2^\inp} \otimes \mu^{x}_{1^\out}$. Plugging this operation into the superchannel yields 
\begin{gather}
(\ident_{2^\inp} \otimes \mu^{x}_{1^\out}) \star \mathsf{T}_{{2^\inp} {1^\out} {1^\inp}} = \mu^{x}_{1^\out} \star \ptr{{2^\inp}}{\mathsf{T}_{{2^\inp} {1^\out} {1^\inp}}} = \mu^{x}_{1^\out} \star (\ident_{1^\out} \otimes \rho_{1^\inp}) = \tr{\mu^{x}_{1^\out}} \rho_{1^\inp} = \rho_{1^\inp},
\end{gather}
which yields the initial state $\rho_{1^\inp}$ of the system, independent of the choice of $\mu^{x}_{1^\out}$, i.e., no information was transmitted. This impossibility of sending information back in time (and more general variants thereof) is sometimes referred to as \textit{no backwards-in-time signalling}~\cite{chiribella_probabilistic_2010,Brukner_2014_NBTS,Guryanova_2019}.

Conversely, for any matrix satisfying the (generalised) complete positivity [Eq.~\eqref{eq::toqp-def-superchannelcp}] and trace preservation [Eq.~\eqref{eq::toqp-def-superchanneltp}] properties, there exists an underlying dilation in terms of an initial system-environment state, a joint unitary evolution and a final partial trace over the environment~\cite{Chiribella_2008,Chiribella_2008_PRL,Chiribella_2009,Pollock_2018_PRA} (as per the l.h.s.\ in Fig.~\ref{fig::toqp-linkproductsuperchannel}); thus, said matrix is the Choi representation of some quantum dynamics. This representation theorem is the higher-order analogue of Stinespring's dilation theorem for quantum channels~\cite{Stinespring_1955} (see Sec.~\ref{subsec::axiomatichoqos}). 

\begin{figure}[t]
\centering
\vspace{0.5em}
\includegraphics[scale=0.5]{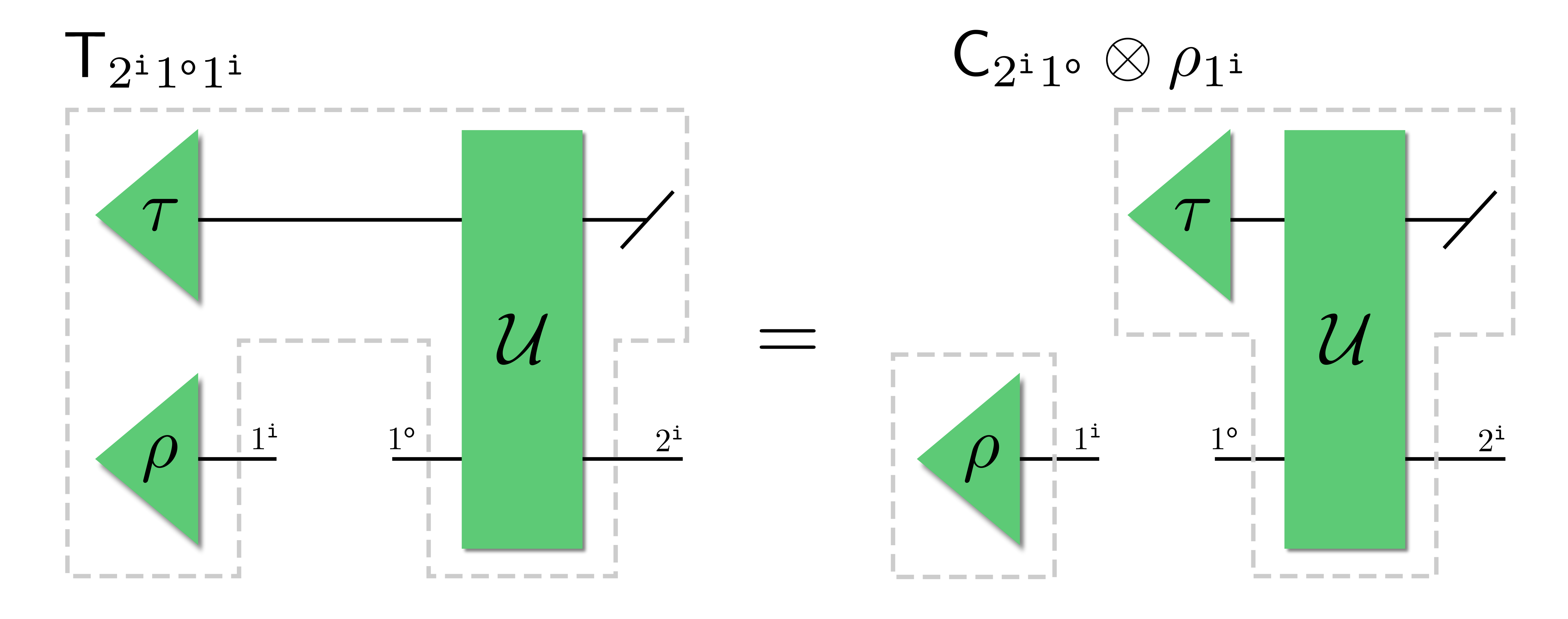}
\caption{\textbf{Superchannel with no Initial Correlations.} If the initial system-environment state $\eta_{SE} = \rho_S \otimes \tau_E$ is uncorrelated, the superchannel $\mathsf{T}_{2^\inp 1^\out 1^\inp}$ (grey, dashed, left) reduces to a tensor product form between the initial system state and an independent channel $\mathsf{C}_{2^\inp 1^\out} \otimes \rho_{1^\inp}$ (grey, dashed, right).} \label{fig::toqp-superchannelinitialuncorrelated}
\end{figure}

A particularly interesting case arises when the initial system-environment state is uncorrelated, $\eta_{S_{1^\inp} E_{1^\inp}} = \rho_{S_{1^\inp}} \otimes \tau_{E_{1^\inp}}$. In this scenario, the superchannel simplifies to $\mathsf{T}_{2^\inp 1^\out 1^\inp} = \mathsf{C}_{2^\inp 1^\out} \otimes \rho_{1^\inp}$, where $\rho_{1^\inp} := \rho_{S_{1^\inp}}$ is the initial system state and $\mathsf{C}_{2^\inp 1^\out} := \mathsf{C}_{S_{2^\inp 1^\out}} = \mathds{1}_{E_{2^\inp}} \star \mathsf{U}_{S_{2^\inp 1^\out} E_{2^\inp 1^\inp} } \star \tau_{E_{1^\inp}}$ is the quantum channel describing the open system dynamics between times due to interactions with the (initially uncorrelated) environment (see Fig.~\ref{fig::toqp-superchannelinitialuncorrelated}). This can be seen directly by inserting $\eta_{S_{1^\inp} E_{1^\inp}} = \rho_{S_{1^\inp}} \otimes \tau_{E_{1^\inp}}$ into Eq.~\eqref{eq::toqp-superchannelconstruction}. This special case represents \textit{Markovian} (memoryless) quantum dynamics, where the system's future state depends only on its current state, not on its history. That both the initial state $\rho_{1^\inp}$ and the channel $\mathsf{C}_{2^\inp 1^\out}$ here act on the Hilbert space of the system alone reflects the fact that memoryless dynamics can be fully understood in terms of two-point dynamical state transformations of the system of interest.  


\begin{figure}[t]
\centering
\vspace{0.5em}
\includegraphics[scale=0.6]{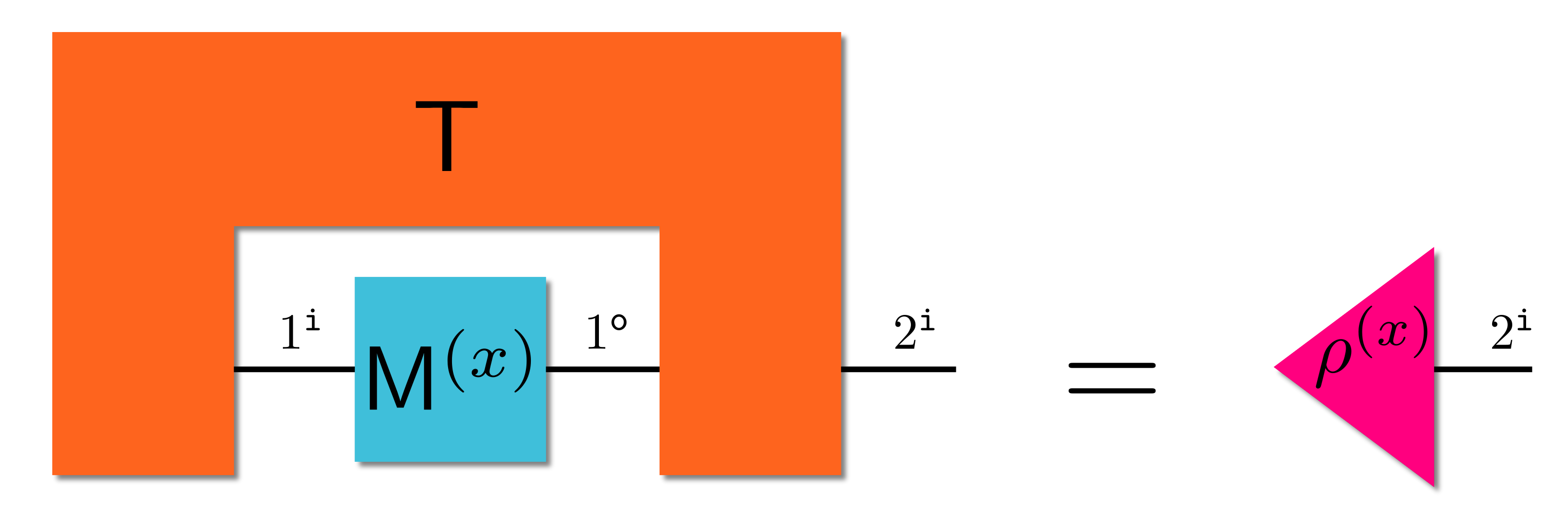}
\caption{\textbf{Superchannel Action.} The superchannel $\mathsf{T}$ (orange) acts upon an instrument applied at the first time $\{ \mathsf{M}^{(x)}\}$ (blue) to yield the (subnormalised) state $\rho^{(x)}$ at the second time.} \label{fig::toqp-superchannelaction}
\end{figure}

In experimental settings, superchannels prove invaluable for computing output states and analysing sequential measurements. When an instrument $\mathcal{J}_1 = \{ \mathsf{M}_{1^\out 1^\inp }^{(x)}\} \in \mathscr{L}(\mathscr{H}_{{1^\out}} \otimes \mathscr{H}_{{1^\inp}})$ is applied to the system, the output state is given by (see Fig.~\ref{fig::toqp-superchannelaction})
\begin{align}\label{eq::toqp-superchannelaction}
    \rho_{{2^\inp}}^{(x)} = \mathcal{T}[\mathcal{M}^{(x)}] = \mathsf{T}_{{2^\inp} {1^\out} {1^\inp}} \star \mathsf{M}^{(x)}_{{1^\out} {1^\inp}}.
\end{align}
Here, the (subnormalised) state $\rho_{{2^\inp}}^{(x)}$ output by the superchannel has trace equal to the probability of observing outcome $x$ when the system is probed with the instrument $\mathcal{J}_1$. It is labelled with $2^\inp$ as it could be considered as the `input' state to a final measurement of the experimenter's choice at time $t_2$. Supposing that such a measurement $\Jcal_2 = \{\mathsf{E}^{(y)}_{2^\inp}\} \in \mathscr{L}(\mathscr{H}_{{2^\inp}})$ is made on this state, then the probability of sequentially observing outcome $x$ and then $y$ is computed as
\begin{align}\label{eq::toqp-superchannelsequentialprobability}
    \mathds{P}(x,y|\Jcal_1,\Jcal_2) = \mathsf{T}_{{2^\inp} {1^\out} {1^\inp}} \star \mathsf{M}^{(x)}_{{1^\out} {1^\inp}} \star \mathsf{E}^{(y)}_{2^\inp}.
\end{align}

\begin{figure}[t]
\centering
\vspace{0.5em}
\includegraphics[scale=0.6]{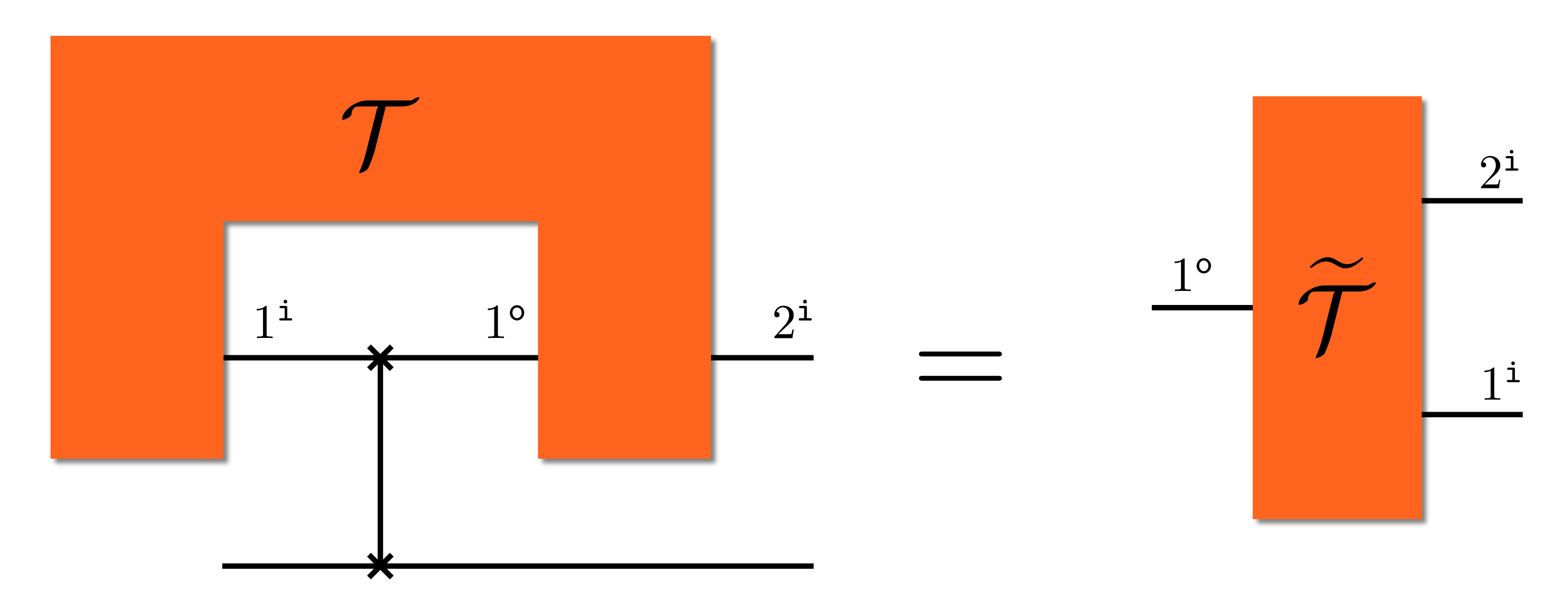}
\caption{\textbf{Alternative Representation of a Superchannel.} A superchannel $\mathcal{T}$ can be represented as a particular type of 1-to-2 channel $\widetilde{\mathcal{T}}$ by pulling all wires corresponding to input spaces to the left and those corresponding to output spaces to the right.} \label{fig::toqp-superchannelalternativemap}
\end{figure}


The superchannel is a deterministically occurring object and is the two-time version of the more general \textit{multi-time quantum processes} that we will analyse in Sec.~\ref{subsubsec::toqp-deterministichoqos}. In contrast, the sequence of instrument and measurement here together constitutes a probabilistically occurring object; more generally, the experimenter could correlate the instrument applied at $t_1$ with the measurement performed at $t_2$, leading to a generalised (multi-time) instrument known as a \textit{superinstrument}, which we will explore in Sec.~\ref{subsubsec::toqp-superinstruments}. Both of these objects possess a representation theorem analogous to Stinespring's dilation theorem~\cite{Stinespring_1955} (see Sec.~\ref{subsec::axiomatichoqos}). 

So far, we have phrased all of the properties of the superchannel in terms of its Choi matrix; however, a superchannel is fundamentally a linear map $\mathcal{T}$ from CP maps, which themselves act as $\mathscr{L}(\mathscr{H}_{1^{\inp}}) \to \mathscr{L}(\mathscr{H}_{1^\out})$, to output states in $\mathscr{L}(\mathscr{H}_{2^\inp})$, i.e., $\mathcal{T} : \mathscr{L}[\mathscr{L}(\mathscr{H}_{1^\inp}) \to \mathscr{L}(\mathscr{H}_{1^\out})] \to \mathscr{L}(\mathscr{H}_{2^\inp})$. The Choi representation is constructed by swapping in half of a maximally entangled state at time $t_1$ [see Fig.~\ref{fig::tf-cji-choisuperchannel}]. However, instead of representing such a map via the CJI, one could represent it by pulling all `input' wires to the right and all `output' wires to the left (see Fig. \ref{fig::toqp-superchannelalternativemap}). This leads to an equivalent description of the superchannel in terms of a linear map from outputs to inputs; by construction, this linear map is CPTP in the regular sense~\cite{Milz_2017}. However, it is not a general one-to-two type quantum channel, as the causality constraint imposes additional structure. 

Superchannels find applications across various areas of quantum physics, from modelling open quantum systems with initial correlations to analysing sequential quantum measurements. They provide a powerful framework for studying quantum memory effects and characterising non-Markovian quantum dynamics, making them essential tools in modern quantum physics research. The framework of superchannels exemplifies how HOQOs can capture complex quantum dynamics while maintaining mathematical rigour and physical interpretability. Their ability to describe correlated quantum processes while respecting causality makes them fundamental to our understanding of temporal quantum processes and their applications in quantum information science.

\FloatBarrier


\subsubsection{Deterministic Higher-Order Quantum Operations: Quantum Combs / Process Tensors} \label{subsubsec::toqp-deterministichoqos}\hfill\\

\noindent The superchannel described above is the first example of a deterministic HOQO. As mentioned previously, by `deterministic', we mean that it describes some fixed, underlying quantum circuit in which all elements occur with unit probability; equivalently, its action upon any valid `probabilistic' object such as a sequence of measurements (or, more generally, a superinstrument; see below) is guaranteed to yield a probability distribution [see Eq.~\eqref{eq::toqp-superchannelsequentialprobability}]. It is thus the natural generalisation of other deterministic quantum objects, such as states and channels, to the case of describing two-time open quantum dynamics with initial correlations. 

\begin{figure}[t]
\centering
\vspace{0.5em}
\includegraphics[scale=0.6]{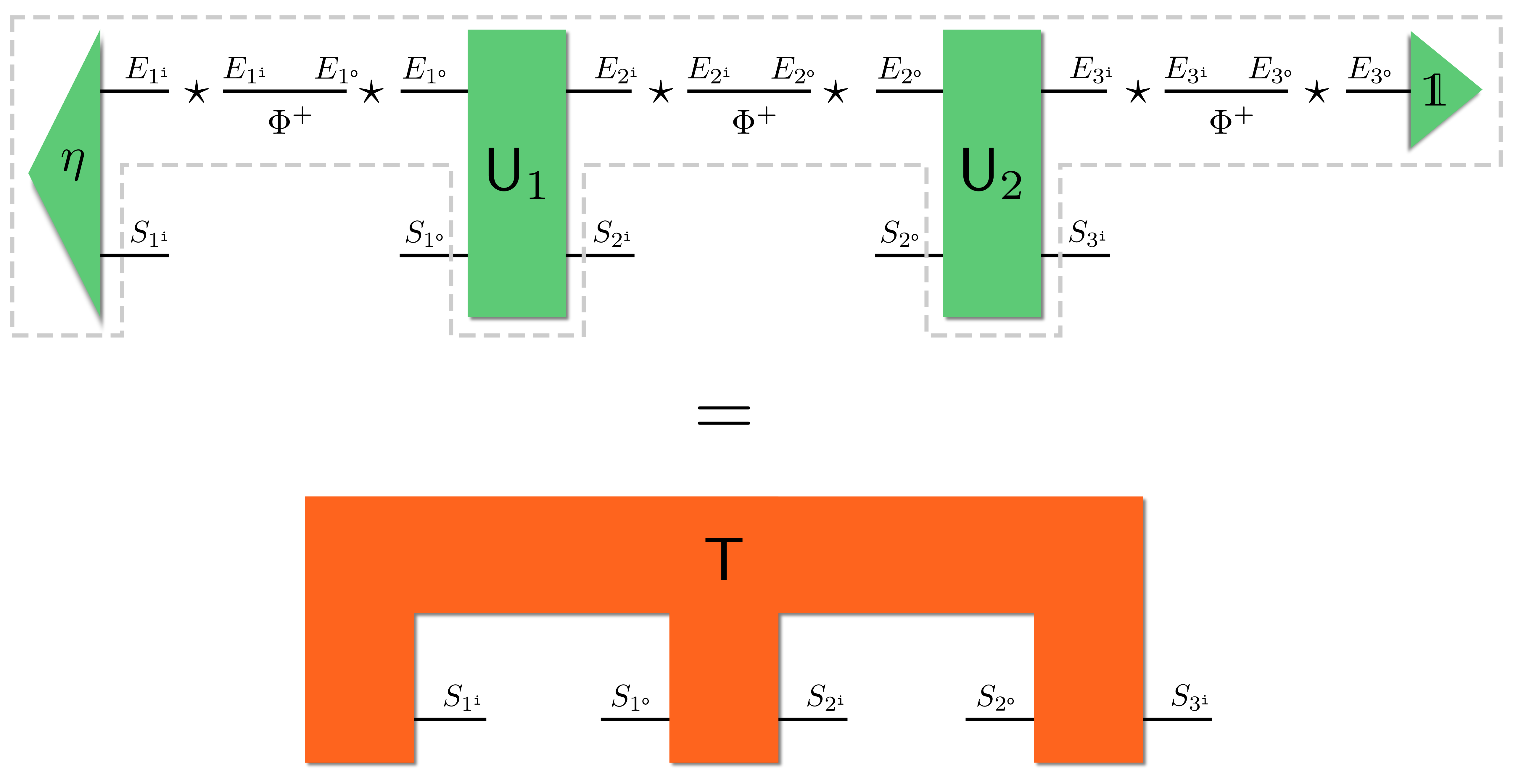}
\caption{\textbf{Quantum Comb Construction.} The quantum comb $\mathsf{T}$ (orange, lower) can be constructed by linking the Choi matrices of the initial joint state $\eta$, the global unitaries $\mathsf{U}_{i}$, and the final partial trace $\mathds{1}$ (green, upper) over the environment degrees of freedom (here we explicitly link identity channels $\Phi^+$ on the environment between times and show the construction for $n=3$). } \label{fig::toqp-dethoqo-linkquantumcomb}
\end{figure}

Quite naturally, one can extend the superchannel to allow for experimental interventions on the system of interest at multiple points in time, yielding a \textit{multi-time quantum process} known as a \textit{quantum comb} or \textit{process tensor}. Such processes provide the most general description of time-ordered open quantum dynamics probed at any finite number of times. Just as we constructed the superchannel, we can build a quantum comb by connecting several basic elements: an initial joint system-environment state, a sequence of system-environment unitaries, and a final partial trace over the environment. This construction yields a $(2n-1)$-partite Choi matrix $\mathsf{T}_{S_{n^\inp} \hdots S_{2^\inp} S_{1^\out} S_{1^\inp}} \in \mathscr{L}(\mathscr{H}_{S_{n^\inp}} \otimes \hdots \otimes \mathscr{H}_{S_{1^\inp}})$, where $n$ represents the number of times at which one can intervene (see Fig.~\ref{fig::toqp-dethoqo-linkquantumcomb})
\begin{align}\label{eq::toqp-dethoqo-quantumcombconstruction}
    \mathsf{T}_{S_{n^\inp} \hdots S_{2^\inp} S_{1^\out} S_{1^\inp}} 
    &= \mathds{1}_{E_{n}} \star \mathsf{U}_{S_{n^\inp} E_{n} S_{n-1^\out} E_{n-1}} \star \hdots \star \mathsf{U}_{S_{2^\inp} E_{2} S_{1^\out} E_{1}} \star \eta_{S_{1^\inp} E_{1}} 
    =: \mathsf{T}_{{n^\inp} : {1^\inp}}.
\end{align}
Although we have been explicit with all of the labels in the above equation out of necessity, from now on we will understand such a multi-time quantum process to always act on the tensor product of Hilbert spaces associated to some system of interest and compress the notation to $\mathsf{T}_{n^\inp : 1^\inp}$, with the understanding that all input and output spaces between $n^\inp$ and $1^\inp$ (inclusive) are accounted for (unless specified otherwise). Lastly, note that this structure does not always have to correspond to the same system being probed at many times, or the same environment carrying the memory; all that is necessary is that there is \textit{some} fixed underlying dynamics with respect to which the interventions occur in a fixed temporal order. Such processes can therefore naturally describe multi-time generalisations of both the superchannels considered in Secs.~\ref{subsubsec::me-opensystemdynamicswithinitialcorrelations} and~\ref{subsubsec::me-quantumcircuitarchitecture}.


\begin{figure}[t]
\centering
\vspace{0.5em}
\includegraphics[scale=0.5]{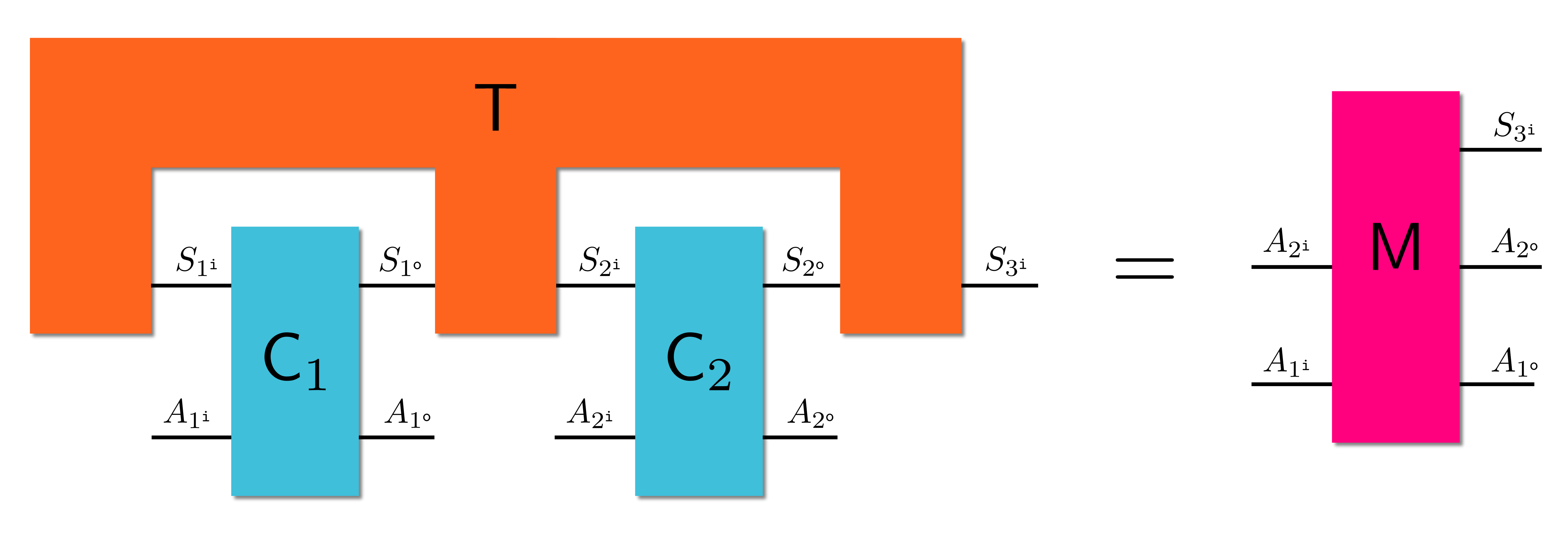}
\caption{\textbf{Complete Positivity of a Quantum Comb.} For two slots, complete positivity of quantum combs implies that \textit{any} pair of CP maps---here, $\mathsf{C}_1$ and $\mathsf{C}_2$ (blue)---is mapped onto a CP map (here, $\mathsf{M}$), i.e., $\mathsf{M}=T_{3^\inp:1^\inp}\star\mathsf{C}_1 \star \mathsf{C}_2\geq 0$ for all $\mathsf{C}_1,\mathsf{C}_2\geq 0$.  This may be viewed as the natural higher-order generalisation of complete positivity for linear maps. Since this must hold for arbitrary auxiliary spaces $\Hscr_{A_{1^\inp}}, \Hscr_{A_{1^\out}}, \Hscr_{A_{2^\inp}}$ and $\Hscr_{A_{2^\out}}$, complete positivity of the comb is equivalent to $T_{3^\inp:1^\inp} \geq 0$.}
\label{fig::toqp-dethoqo-quantumcombcp}
\end{figure}

The construction of a quantum comb $\mathsf{T}_{{n^\inp} : {1^\inp}}$ above follows directly from linking together various components over the environment degrees of freedom. Just like was the case for superchannels [see the discussion around Eq.~\eqref{eq::toqp-superchannelinherittraceconditions}], this systematic building-up of processes ensures that quantum combs inherit two crucial properties: generalised complete positivity and generalised trace preservation, summarised in Eqs.~\eqref{eq::toqp-dethoqo-def-quantumcombcp} and~\eqref{eq::toqp-dethoqo-def-quantumcombtp} below. In the multi-time setting, generalised complete positivity means that when the comb acts on any sequence of completely positive maps (which might act on an additional arbitrarily large auxiliary system), it results in another completely positive map (see Fig.~\ref{fig::toqp-dethoqo-quantumcombcp}). This property ensures that the description of the physics remains valid even when we consider interventions that might be correlated or entangled with additional systems. 

The trace preservation property manifests as a hierarchy of trace conditions that enforce causality in the form of no backwards-in-time signalling~\cite{chiribella_probabilistic_2010,Brukner_2014_NBTS, Guryanova_2019}. This means that no deterministic operation performed at a later time can influence the statistics of measurements made at earlier times (for any choice of instruments). Although this causality constraint emerges automatically from the constructive approach to multi-time HOQOs [i.e., by building them from elementary primitives as per Eq.~\eqref{eq::toqp-dethoqo-quantumcombconstruction}], it is worth noting that there exist more exotic deterministic HOQOs (such as the process matrix considered in Sec.~\ref{subsubsec::me-causalityquantumtheory}) that do not obey such temporal ordering (see Sec.~\ref{subsec::axiomatichoqos}). Both generalised complete positivity and trace preservation follow automatically from the construction of the quantum comb in terms of system-environment circuit primitives [see Eq.~\eqref{eq::toqp-dethoqo-quantumcombconstruction}].

In the converse direction, any matrix satisfying said conditions represents some fixed underlying system-environment dynamics~\cite{Chiribella_2008,Chiribella_2008_PRL,Chiribella_2009,Bisio_2011,Pollock_2018_PRA} (see also Sec.~\ref{subsec::axiomatichoqos}). Thus, we have the following: 
\begin{myDefinition}{Quantum Combs / Process Tensors}[def::toqp-def-quantumcombs]
A quantum comb / process tensor $\mathsf{T}_{{n^\inp}:{1^\inp}} \in \mathscr{L}(\mathscr{H}_{{n^\inp}} \otimes \hdots \otimes \mathscr{H}_{{1^\inp}})$ is characterised by
\begin{alignat}{2}
    &\text{1. (Generalised) Complete positivity: } &&\mathsf{T}_{{n^\inp} : {1^\inp}} \geq 0 \label{eq::toqp-dethoqo-def-quantumcombcp} \\
    &\text{2. (Generalised) Trace preservation: } &&\ptr{{n^\inp}}{\mathsf{T}_{{n^\inp} : {1^\inp}}} = \mathds{1}_{{n-1^\out}}\otimes\mathsf{T}_{{n-1^\inp : {1^\inp}}} \label{eq::toqp-dethoqo-def-quantumcombtp} \\ 
    & &&\ptr{{n-1^\inp}}{\mathsf{T}_{{n-1^\inp : {1^\inp}}}} = \mathds{1}_{{n-2^\out}}\otimes\mathsf{T}_{{n-2^\inp : {1^\inp}}} \notag \\ & && \quad \quad \quad \quad \quad  \vdots \notag \\
    & &&\ptr{{2^\inp}}{\mathsf{T}_{{2^\inp} : {1^\inp}}} = \mathds{1}_{{1^\out}}\otimes\rho_{{1^\inp}} \notag \\
    & &&\text{where } \mathsf{T}_{{j^\inp} : {1^\inp}} \geq 0 \;\; \forall \, j \text{ and } \rho_{{1^\inp}} \in \mathsf{St}(\mathscr{H}_{{1^\inp}}). \notag
\end{alignat} 
\end{myDefinition}


\noindent One of the most elegant aspects of quantum combs is their ability to unify various quantum objects within a single framework. This definition supersedes all previously defined deterministic quantum objects since any of the Hilbert spaces considered above could be trivial. By considering such special cases, we can recover simpler quantum objects: When all but one input space is trivial, we have a (single-party) quantum state; when we have just one input and one output space, we have a (one-to-one) quantum channel; and when we have three or four non-trivial spaces arranged appropriately, we recover the superchannels discussed earlier.\footnote{As mentioned previously, throughout we reserve the term `superchannel' for quantum combs / process tensors with one slot, both for trivial and non-trivial initial input space $0^\out$.} Again, just like for superchannels, quantum combs can also start on a space labelled by $0^\out$, representing situations in which an initially uncorrelated system state can be fed in to the dynamics; this merely represents a slight notational change to the above expressions without impacting any results. 

We emphasise that the above trace conditions can be endowed with physical meaning in terms of \textit{signalling constraints} that reflect the fixed causal order of the comb. To see this, consider the first trace condition, $\ptr{{n^\inp}}{\mathsf{T}_{{n^\inp} : {1^\inp}}} = \mathds{1}_{{n-1^\out}}\otimes\mathsf{T}_{{n-1^\inp : {1^\inp}}}$. This implies that discarding the final output of the comb---denoted by the operation $\text{tr}_{n^\inp}$ on the l.h.s.---amounts to also discarding the final input of the comb---denoted by $\mathds{1}_{{n-1^\out}}$ on the r.h.s., which is the Choi state of $\text{tr}_{n-1^\out}$. As a consequence, any operation applied to the final input at time $t_{n-1}$ cannot influence/signal information to \textit{any} previous times. Since the same condition holds for $\mathsf{T}_{{n-1^\inp : {1^\inp}}}$ and so on further down the hierarchy, we see that the above trace conditions nicely encode the fact that there is no signalling backwards in time for quantum combs. Since this is a reflection of the inherent causal order of quantum combs / process tensors, we will often refer to these trace conditions as \textit{causality constraints} (see Sec.~\ref{subsec::indefinitecausalorder} for a more detailed discussion on causal order in HOQOs).

\begin{figure}[t]
\centering
\vspace{0.5em}
\includegraphics[scale=0.6]{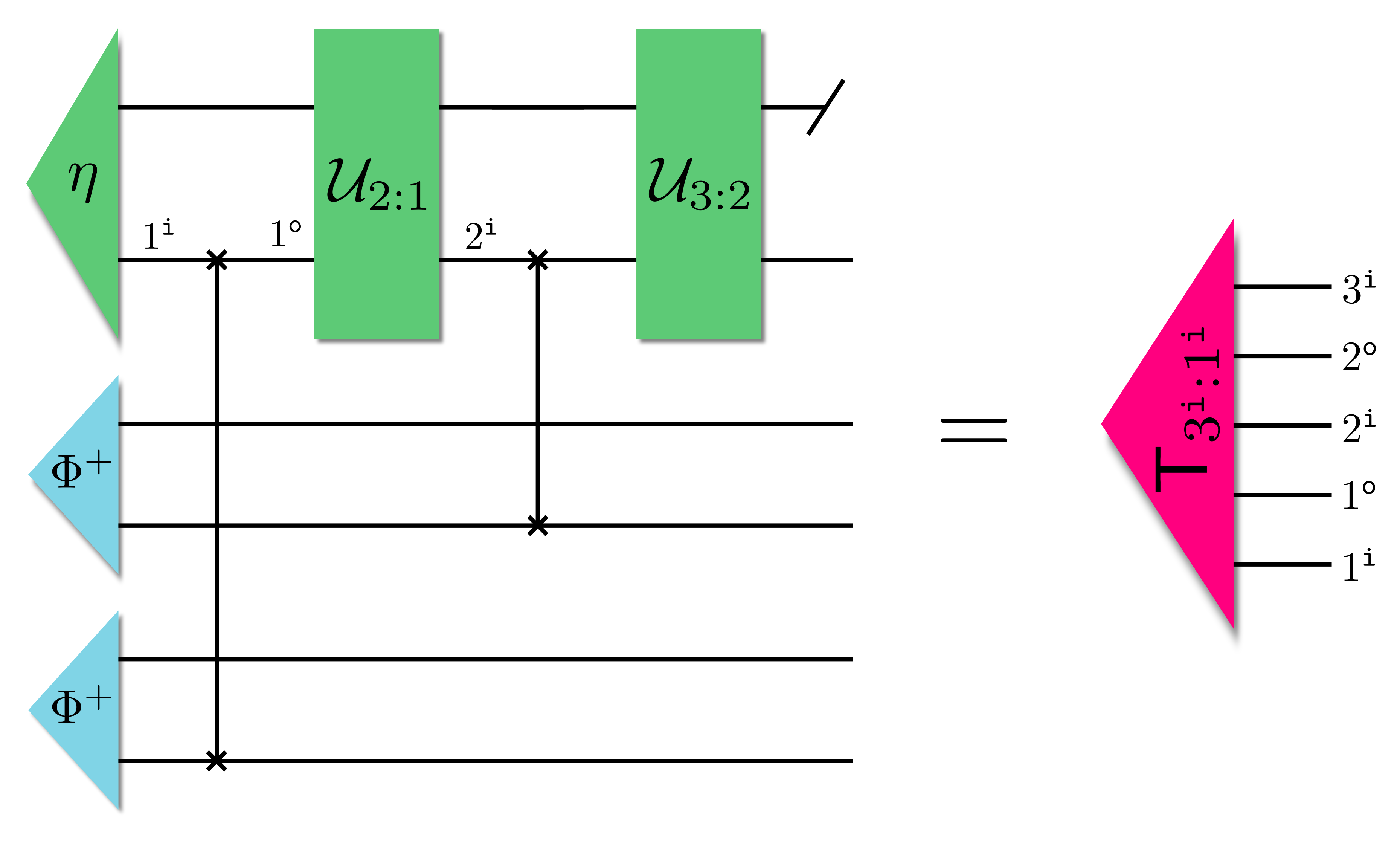}
\caption{\textbf{Choi State of a Quantum Comb.} Here we depict the Choi state of an $n=3$-step quantum comb. As previously, this is constructed by swapping in half of a maximally entangled state $\Phi^+$ at every open slot.} \label{fig::toqp-dethoqo-quantumcombchoi}
\end{figure}

Just like the superchannel, the Choi matrix of a quantum comb can be operationally prepared (up to normalisation) by swapping the system with one half of a maximally entangled state at each intervention time (see Fig.~\ref{fig::toqp-dethoqo-quantumcombchoi}). More precisely, begin with the system-environment dynamics shown in Fig.~\ref{fig::toqp-dethoqo-linkquantumcomb} and denote the initial system-environment state by $\eta$ and the unitary maps describing the joint evolution between times $t_{j-1}$ and $t_j$ by $\mathcal{U}_{j:j-1}$. Now consider $n-1$ additional maximally entangled pairs, $\Phi^+_{j^\out j^{\out\prime}}$ associated to auxiliary systems $A_{j^\out}\simeq A_{j^{\out\prime}}\simeq S$, collectively described as $\Phi^{+}_{n-1:1} := \bigotimes_{j=1}^{n-1}  \Phi^+_{j^\out j^{\out \prime}}$. Letting the unitary maps between times act on the environment and one half of the appropriately swapped in auxiliary systems, i.e., $\mathcal{U}_{j:j-1}: \mathscr{L}(\mathscr{H}_{A_{j-1^\out}} \otimes \mathscr{H}_{E_{j-1^\out}}) \to \mathscr{L}(\mathscr{H}_{S_{j^\inp}} \otimes \mathscr{H}_{E_{j^\inp}})$ yields the Choi matrix of the multi-time quantum process
\begin{align}\label{eq::toqp-dethoqo-quantumcombchoimaprelation}
    \mathsf{T}_{{n^\inp} : {1^\inp}}  =&\, \mathrm{tr}_E[\mathcal{U}_{n:n-1}  \hdots \mathcal{U}_{2:1}(\eta \otimes \Phi^{+}_{n-1:1})].
\end{align}


\noindent What makes quantum combs particularly powerful is their ability to describe complex quantum dynamics where the system of interest might change between interventions, or where different environments might carry the quantum memory. The only requirement is the existence of some fixed underlying dynamics with respect to which the interventions occur in a definite temporal order.

\FloatBarrier


\subsubsection{Probabilistic Higher-Order Quantum Operations: Superinstruments}\label{subsubsec::toqp-superinstruments}\hfill\\

\noindent Having explored deterministic quantum combs, we now turn to their probabilistic counterparts: \textit{superinstruments}. These objects naturally extend the concept of quantum measurements and instruments to multiple time steps, providing a framework for describing complex measurement sequences in quantum mechanics. 

A quantum superinstrument can be thought of as a collection of measurement operators that act across multiple times while preserving the fundamental requirements of quantum mechanics; as such, they have been referred to as `measuring co-strategies', `process POVMs', `testers', or `superinstruments' throughout the literature~\cite{Gutoski_2007,Ziman_2008,Quintino_2019_PRA,Quintino_2019_PRL}. Typically, the former two terms have been used to describe situations in which \textit{only} classical information is extracted from a quantum process: Process POVMs or testers are higher-order analogues of POVMs/effects. We will use the term `superinstrument' to refer to the more general case where a quantum object can also be output (alongside classical information): In this sense, superinstruments are to quantum combs what instruments are to quantum channels. As such, they comprise multi-time objects that individually satisfy generalised complete positivity and sum to a proper quantum comb. These requirements ensure that whenever we perform multi-time interrogations described by a superinstrument, we always obtain valid probability distributions and quantum outputs.

\begin{figure}[t]
\centering
\vspace{0.5em}
\includegraphics[scale=0.7]{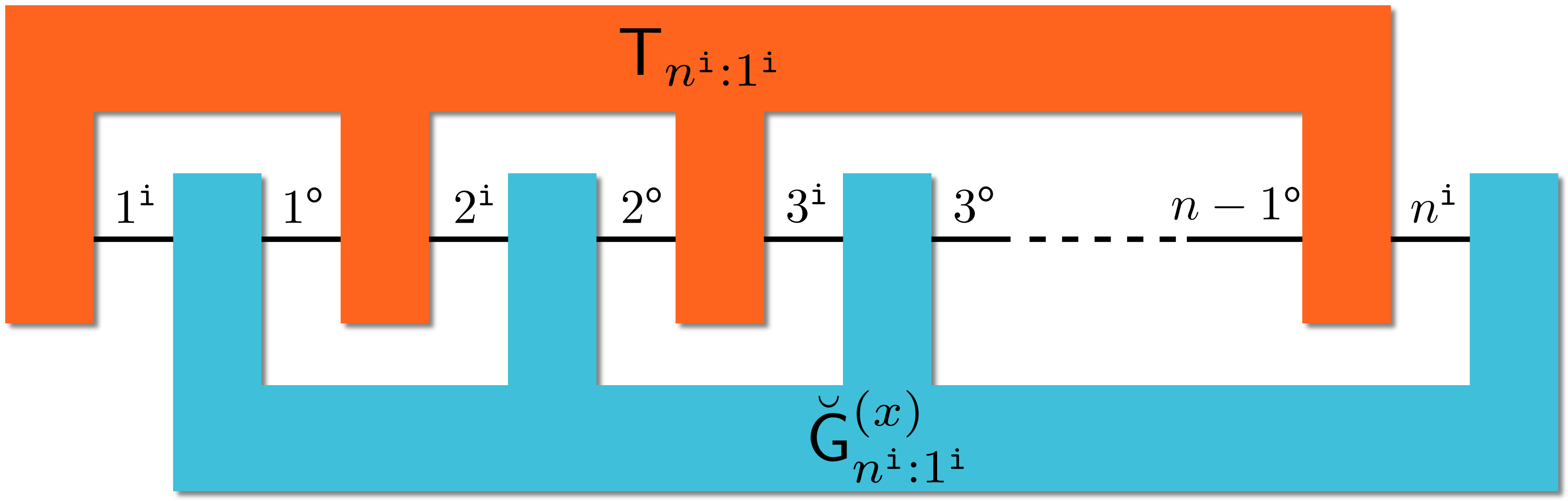}
\caption{\textbf{Spatiotemporal Born Rule.} The probability of observing outcome $x$ when a quantum comb $\mathsf{T}_{n^\inp:1^\inp}$ is interrogated with a superinstrument $\breve{\mathsf{G}}_{n^\inp:1^\inp}$ is given by $\mathds{P}(x| \mathcal{J}) = \breve{\mathsf{G}}_{n^\inp : 1^\inp}^{(x)} \star \mathsf{T}_{n^\inp : 1^\inp}$. Throughout, we use blue when we want to signify that an object is considered as an `input' to a HOQO; naturally, such input operations can be HOQOs themselves, as is the case here. We emphasise that the role of input and output spaces are reversed for combs and superinstruments that can meaningfully be applied to them; inputs to the comb are outputs of the superinstrument.} \label{fig::toqp-testers-spatiotemporalbornrule}
\end{figure}

The simplest way to understand superinstruments is to first consider how a quantum comb $\mathsf{T}_{n^\inp:1^\inp}$ interacts with a sequence of independent instruments $\{ \breve{\mathsf{G}}_{1^\out 1^\inp}^{(x_1)}, \hdots, \breve{\mathsf{E}}_{n^\inp}^{(x_n)}\}$. Here, and in what follows, we add additional `breves' on operators to stress that their role of inputs and outputs is reversed compared to the quantum comb that they act on, i.e., their inputs are outputs of the comb, and \textit{vice versa}. When the sequence of instruments is applied, a sequence of outcomes $x_1, \hdots x_n$ is observed. The probability of obtaining any such sequence is calculated via a particular case of the generalised spatiotemporal Born rule (see below)
\begin{align}\label{eq::toqp-testers-uncorrelatedspatiotemporalbornrule}
    \mathds{P}(x_n, \hdots, x_1 | \mathcal{J}_n, \hdots, \mathcal{J}_1) = \breve{\mathsf{E}}_{n^\inp}^{(x_n)} \star \hdots \star \breve{\mathsf{G}}_{1^\out 1^\inp}^{(x_1)} \star \mathsf{T}_{n^\inp : 1^\inp} = \tr{(\breve{\mathsf{E}}_{n^\inp}^{(x_n)} \otimes \hdots \otimes \breve{\mathsf{G}}_{1^\out 1^\inp}^{(x_1)} )^\textup{T} \, \mathsf{T}_{n^\inp : 1^\inp} }.
\end{align}
\noindent The above expression constitutes a special case of how HOQOs act on each other: Here, we are contracting \textit{all} open slots of the deterministic quantum comb with a probabilistic operation associated to each time in order to yield a probability distribution (thus, one might refer to the superinstrument above as a `tester'). However, measurements need not be independent across different times. Consider, for instance, a scenario where we feed forward some auxiliary system between measurements, creating temporal correlations in our measurement strategy. This is where superinstruments truly shine---they can capture such correlated measurement sequences in a mathematically elegant way. A superinstrument consists of any collection of positive semidefinite operators that sum to a quantum comb, formalised as follows:

\begin{myDefinition}{Superinstruments}[def::toqp-def-superinstruments]
A (causally ordered) superinstrument $\mathcal{J}$ that can be applied to a quantum comb $\mathsf{T}_{n^\inp:1^\inp}$ is a collection $\{ \breve{\mathsf{G}}^{(x)}_{{n^\inp}:{1^\inp}}\} \in \mathscr{L}(\mathscr{H}_{{n^\inp}} \otimes \hdots \otimes \mathscr{H}_{{1^\inp}})$ of positive semidefinite operators $\breve{\mathsf{G}}^{(x)}_{{n^\inp}:{1^\inp}} \geq 0$ such that overall $\breve{\mathsf{G}}_{n^\inp : 1^\inp} := \sum_x \breve{\mathsf{G}}_{n^\inp : 1^\inp}^{(x)}$ is a quantum comb, i.e., it satisfies the hierarchy of trace conditions in Eq.~(\ref{eq::toqp-dethoqo-def-quantumcombtp}), but with input and output labels reversed (see Fig.~\ref{fig::toqp-testers-spatiotemporalbornrule}), i.e.,
\begin{gather}
    \breve{\mathsf{G}}_{n^\inp : 1^\inp} = \ident_{n^\inp} \otimes \breve{\mathsf{G}}_{n-1^\out : 1^\inp}, \quad \ptr{n-1^\out}{\breve{\mathsf{G}}_{n-1^\out : 1^\inp}} = \ident_{n-1^\inp} \otimes  \breve{\mathsf{G}}_{n-2^\out : 1^\inp}, \dots, \ptr{1^\out}{\breve{\mathsf{G}}_{1^\out : 1^\inp}} = \ident_{1^\inp}.
\end{gather}
\end{myDefinition}
\noindent Similar to how the definition of multi-time quantum processes supersedes that of all previous deterministic objects (such as states and channels), the definition of a superinstrument indeed encapsulates instruments and POVMs by allowing certain Hilbert spaces to be trivial. Just as an instrument can always be thought of as arising from a system-environment dynamics where the environment is finally measured (instead of traced out), so too can a superinstrument be considered a multi-time open system dynamics where the environment is finally measured (see Sec.~\ref{subsubsec::axiomatichoqos-quantumcomb}). Moreover, just as the space of POVMs forms the dual cone to the space of quantum states, here the space of superinstruments forms the dual cone to that of multi-time quantum combs. 

\noindent Thus, intuitively, superinstruments are the most general objects for which \textit{any} deterministic multi-time quantum process is guaranteed to yield valid output quantum objects (e.g., probability distributions, quantum states, channels, etc.) when acted upon. Whenever a superinstrument is applied to a quantum comb, the probability of observing outcome $x$ is given by the higher-order contraction, i.e., the \textit{spatiotemporal Born rule}~\cite{Gutoski_2007,Chiribella_2009} (see Fig.~\ref{fig::toqp-testers-spatiotemporalbornrule})
\begin{align}\label{eq::toqp-testers-spatiotemporalbornrule}
    \mathds{P}(x| \mathcal{J}) &= \breve{\mathsf{G}}_{n^\inp : 1^\inp}^{(x)} \star \mathsf{T}_{n^\inp : 1^\inp} = \tr{\breve{\mathsf{G}}_{n^\inp : 1^\inp}^{(x) \textup{T}} \mathsf{T}_{n^\inp : 1^\inp} }.
\end{align} 
Quite naturally, as stated in the above definition, it is important to keep track of the labelling of Hilbert spaces (or timesteps) when contracting multi-time quantum combs with superinstruments: Sensible results only ensue when objects with the same temporal order are contracted. When contracting a superinstrument with a deterministic comb, the input and output labels on the superinstrument essentially function alternatingly to those of the comb with which it is contracted, meaning that the hierarchy of trace conditions apply to the opposite labels (i.e., inputs interchange with outputs); this can be seen by explicitly connecting identity map wires during the contraction, which serves to swap the input and output labels appropriately. 


One particularly powerful aspect of superinstruments is their ability to describe partial measurements. When we apply a superinstrument to only some of the time slots in a quantum comb, we obtain a `conditional process' for the remaining times that describes the correct behaviour of the concatenated dynamics~\cite{Taranto_2021_npj}. In other words, the resulting object contains all of the information required to compute the correct probability distribution for \textit{any} instruments applied to the remaining times. While these conditional processes might not themselves be proper quantum combs (due to the probabilistic nature of measurements), their average over all possible measurement outcomes always yields a valid quantum comb. 

\newpage

\begin{figure}[t]
\centering
\vspace{0.5em}
\includegraphics[scale=0.6]{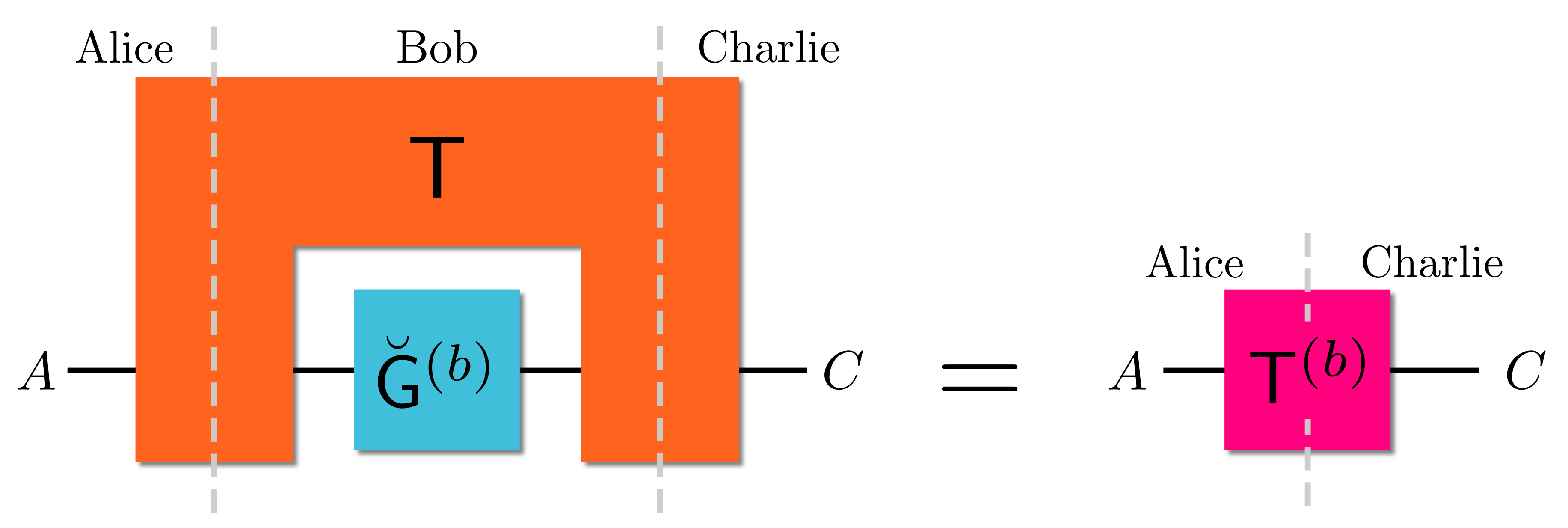}
\caption{\textbf{Conditional Quantum Processes.} Here the HOQO $\mathsf{T}$ (orange) connects Alice, Bob, and Charlie in a temporally ordered way. If Bob applies the (super)instrument $\Jcal_B = \{ \breve{\mathsf{G}}_B^{(b)} \}$ (here depicted as a single step instrument, blue), then the conditional process that Alice and Charlie `sees' (pink), for each outcome $b$, is $\mathsf{T}_{AC}^{(b)} = \mathsf{T}_{ABC} \star \breve{\mathsf{G}}_B^{(b)}$.} \label{fig::toqp-testers-conditionalprocesses}
\end{figure}

\begin{example}
    To illustrate this, consider a simple scenario with three parties---Alice, Bob, and Charlie---arranged in a temporal sequence that is encoded in the quantum comb $\mathsf{T}_{ABC}$ (see Fig.~\ref{fig::toqp-testers-conditionalprocesses}). If Bob applies the (super)instrument $\Jcal_B = \{ \breve{\mathsf{G}}_B^{(b)} \}$, Alice and Charlie are left with a conditional process $\mathsf{T}_{AC}^{(b)} = \mathsf{T}_{ABC} \star \breve{\mathsf{G}}_B^{(b)}$ that depends on Bob's measurement outcome $b$. For this simple one-slot scenario, the condition on $\Jcal_B$ to be a superinstrument boils down to the `normal' instrument case, i.e.,  $\sum_b \breve{\mathsf{G}}_B^{(b)}$ corresponds to a CPTP map.    
    
    The probability of Bob recording outcome $b$, which depends on state $\rho_{A^\out}$ input by Alice, is encoded in $\tr{\mathsf{T}_{AC}^{(b)}}$ via $\mathds{P}(b|\Jcal_B, \rho_{A^\out}) = \rho_{A^\out} \star \tr{\mathsf{T}_{AC}^{(b)}}$. Since this procedure amounts to Bob post-selecting an outcome after Alice, this conditional process $\mathsf{T}_{AC}^{(b)}$ is not necessarily a proper comb/channel for any specific outcome (although it is positive semidefinite); nonetheless, by averaging over all of Bob's possible outcomes, one recovers a valid quantum comb describing the effective/average dynamics between Alice and Charlie. More precisely, $\mathsf{T}_{AC}^{\mathcal{J}_B} := \sum_b \mathsf{T}_{AC}^{(b)}$ satisfies Eqs.~(\ref{eq::toqp-dethoqo-def-quantumcombcp},\,\ref{eq::toqp-dethoqo-def-quantumcombtp}) for any (super)instrument $\mathcal{J}_B$, which follows from the fact that Bob's (super)instrument sums to a proper quantum comb (in this case, a quantum channel).\hfill$\blacksquare$
\end{example}


The notion of superinstruments provides the most general description of quantum measurements across multiple times while respecting causality. Just as quantum combs unify various deterministic quantum objects, superinstruments unify probabilistic quantum objects like POVMs and instruments under a single mathematical framework. They form the dual cone to quantum combs, meaning that they are the most general objects that yield valid probability distributions when contracted with any quantum comb of compatible temporal order. As we saw above, care is necessary when considering their structural properties, since inputs of superinstruments are outputs of the comb they act on, and vice versa. Finally, having seen how HOQOs (both quantum combs and superinstruments) arise from underlying system-environment dynamics and/or quantum circuits, we mow take a complimentary approach to the \textit{constructive} one to deriving HOQOs that we have so far focused on, namely an \textit{axiomatic} one.


\subsection{Axiomatic Approach to Higher-Order Quantum Operations}
\label{subsec::axiomatichoqos}

Up to this point, most of our discussion on HOQOs focused on building intuition and was based on a constructive approach, i.e., higher-order processes were built up from quantum circuits by combining lower-order elementary quantum objects such as states, channels, measurements, and instruments. From this perspective, we derived mathematical properties of multi-time quantum processes and superinstruments based on linear/affine and positive semidefinite constraints of Choi matrices (see Defs.~\ref{def::toqp-def-quantumcombs} and~\ref{def::toqp-def-superinstruments}). In this section, building upon the developed intuition, we will consider HOQOs in a more formal way and consider them from an abstract point of view, namely as transformations of quantum objects, and analyse whether or not this leads to the same set of HOQOs. 

For example, while it is clear that every superchannel maps quantum channels to quantum channels, it is, \textit{a priori}, unclear if \textit{every} map that maps quantum channels to quantum channels (and preserves complete positivity) is indeed a superchannel in the \textit{constructive} sense, i.e., it can be represented as a concatenation of states, channels and measurements. Answering this question splits into two parts: i) showing that the Choi state of a general mapping from quantum channels to quantum channels satisfies the properties laid out in Eqs.~\eqref{eq::toqp-def-superchannelcp} and~\eqref{eq::toqp-def-superchanneltp}; and ii) demonstrating that any object that satisfies these properties can be represented as a concatenation of quantum states and channels, i.e., admits a (generalised) Stinespring dilation.

\begin{figure}[t]
    \centering
    \subfigure[\textbf{Superchannels.}]
    {
    \includegraphics[scale =0.5]{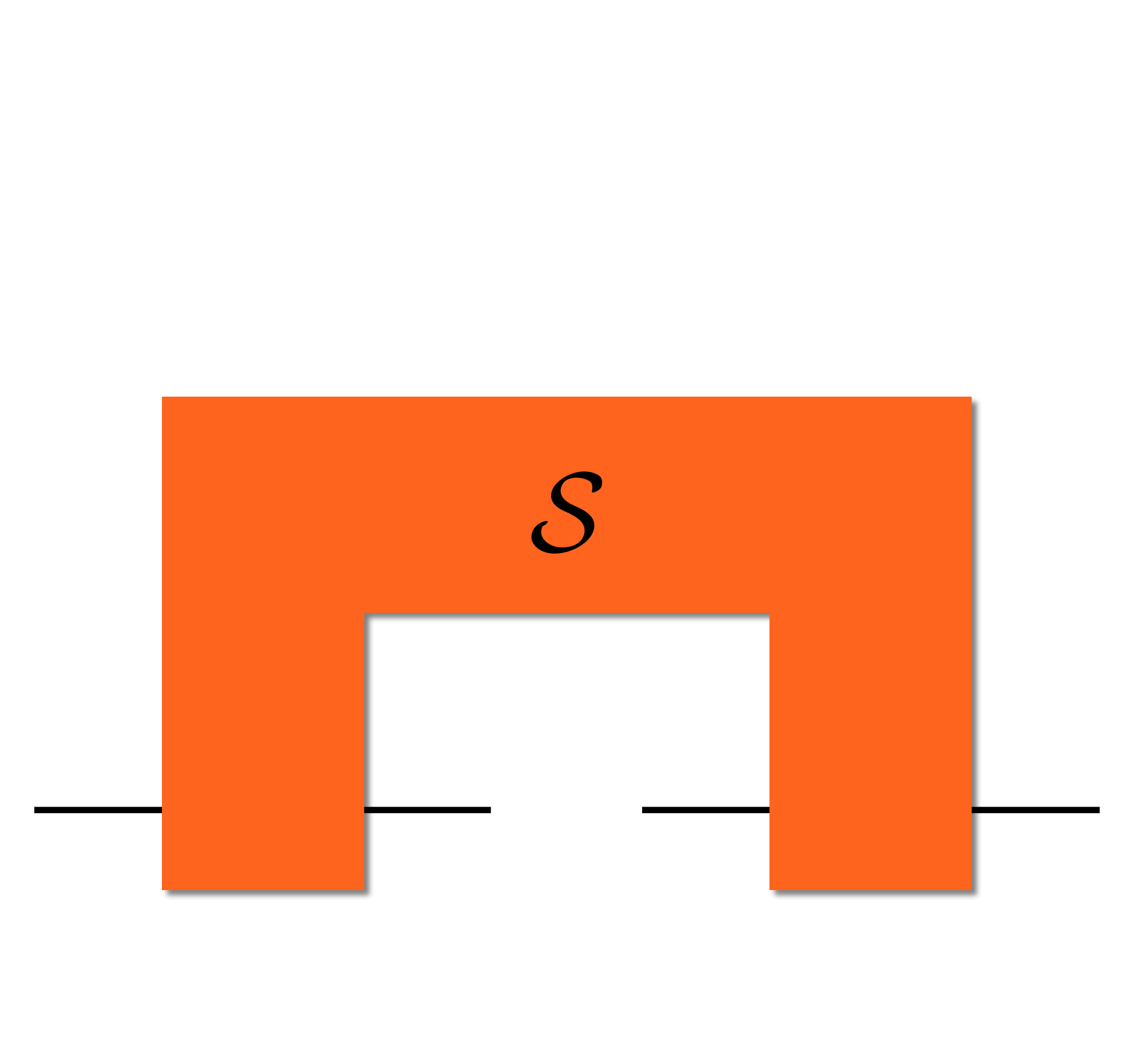}
    \label{fig::axiomatichoqos-superchannels}
    }\hspace{2cm}
    \subfigure[\textbf{Super-superchannels.} ]
    {
    \includegraphics[scale=0.5]{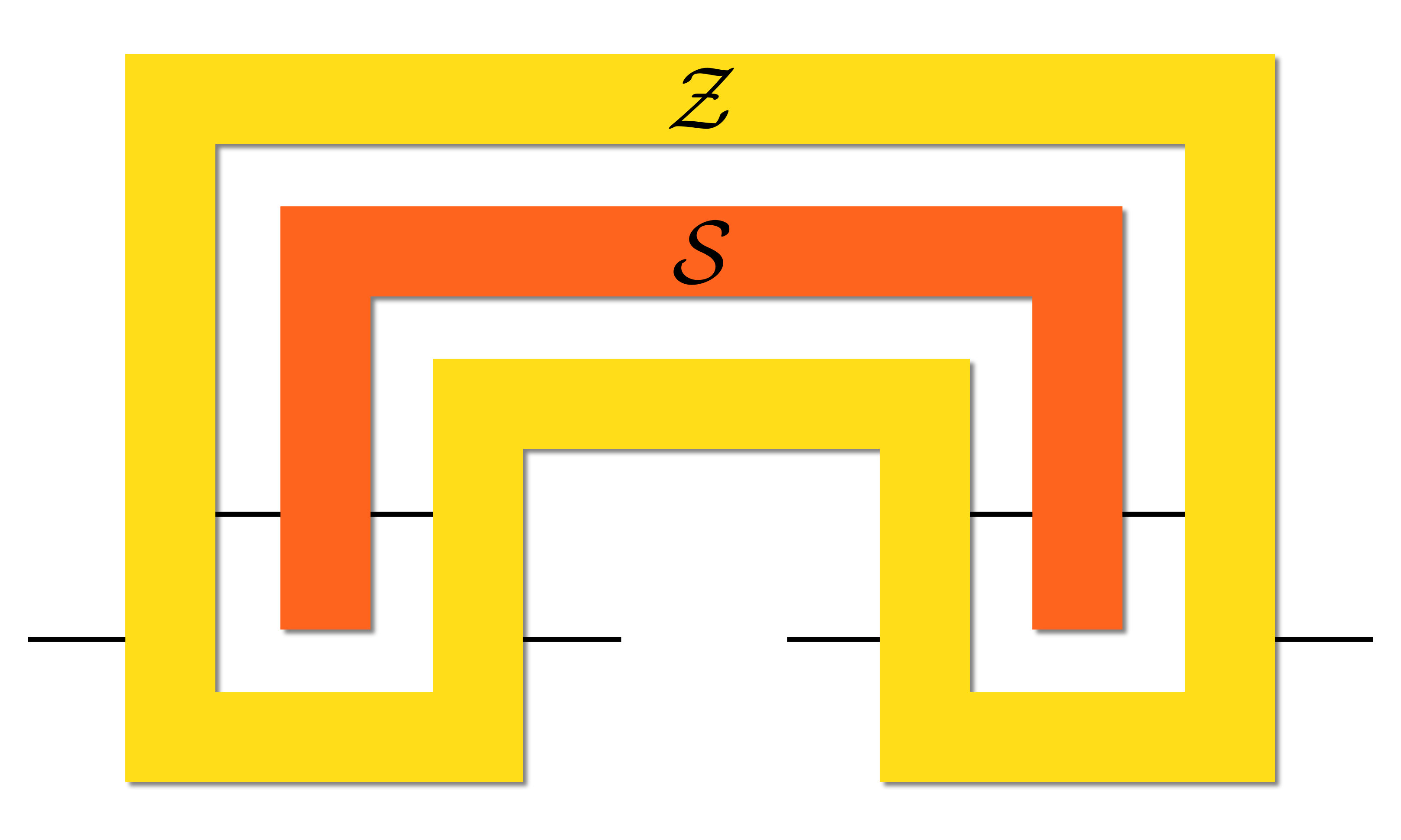}
    \label{fig::axiomatichoqos-supersuperchannels}
    }\hspace{1cm}\\
    \hfill
    \caption{\textbf{HOQOs With and Without a Stinespring Dilation.} \textbf{(a)} Transformations $\mathcal{S}$ of channels to channels can always be represented as a quantum circuit (see Sec.~\ref{subsubsec::axiomatichoqos-qc-onetwoslotcombs}). \textbf{(b)} Transformations $\mathcal{Z}$ that map superchannels to superchannels cannot necessarily be represented by a quantum circuit~\cite{milz_characterising_2024}. \label{fig::axiomatichoqos-stinespringornot}}
\end{figure}

These considerations arise in the same vein for multi-time quantum combs, i.e., objects that satisfy Eqs.~\eqref{eq::toqp-dethoqo-def-quantumcombcp} and~\eqref{eq::toqp-dethoqo-def-quantumcombtp}, as well as more general HOQOs. In the following, we will see that the constructive and abstract/axiomatic approach to HOQOs indeed coincide for a specific, well-justified choice of HOQOs, and we will derive a (generalised) Stinespring dilation for such HOQOs. Importantly though, the ensuing equivalence between quantum circuits and HOQOs, does not hold for \textit{all} conceivable HOQOs; e.g., not all possible, physically motivated mappings from superchannels to superchannels admit a representation as a quantum circuit~\cite{milz_characterising_2024} (see Fig.~\ref{fig::axiomatichoqos-stinespringornot}). The axiomatic approach to HOQOs will thus be useful to identify the appropriate scope of the aforementioned definitions of combs and superchannels, and to tackle HOQOs that go beyond the paradigm of global causal order (and thus do not admit a generalised Stinespring dilation). 


\begin{figure}[t]
\centering
\vspace{0.5em}
\includegraphics[width=0.9\linewidth]{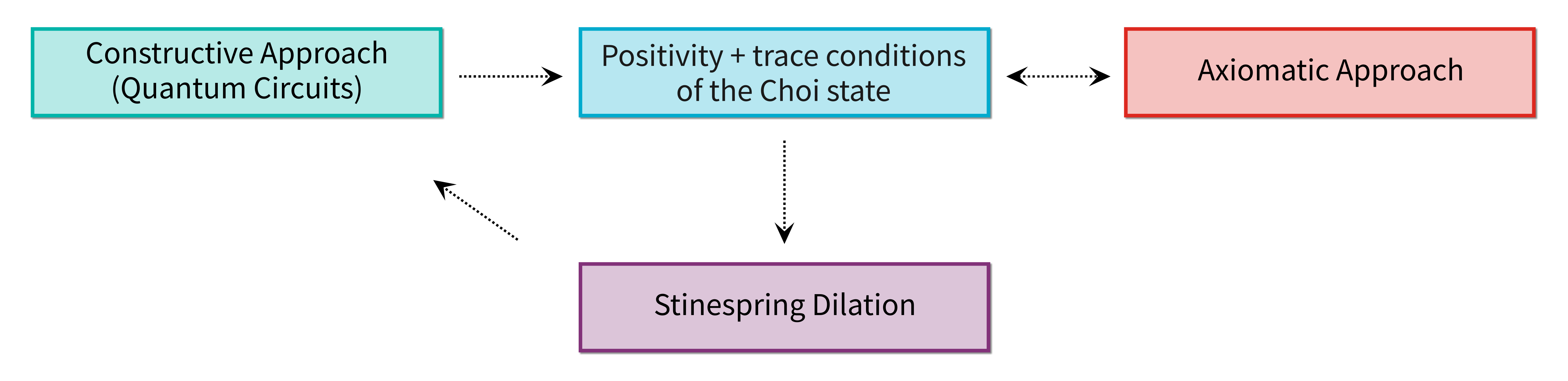}
\caption{\textbf{Constructive and Axiomatic Approaches to HOQOs.} The Choi state of any quantum circuit is positive semidefinite and satisfies the trace conditions of Def.~\ref{def::toqp-def-quantumcombs}. These conditions coincide exactly with the axiomatic approach to quantum combs/HOQOs (see Secs.~\ref{subsubsec::axiomatichoqos-quantumchannels} and~\ref{subsubsec::axiomatichoqos-quantumcomb}). On the other hand, any positive semidefinite Choi matrix that satisfies the trace conditions of Def.~\ref{def::toqp-def-quantumcombs} admits a Stinespring dilation, i.e., it can be understood as stemming from a quantum circuit only containing isometries and partial traces. Together, this diagram implies the equivalence of the axiomatic and the constructive approach to HOQOs. We emphasise that this equivalence only holds true if the axiomatic requirements of HOQOs are rather strict (see Def.~\ref{def::axiomatichoqos-qc-def-axiomaticcombs}), and there are many meaningfully HOQOs that satisfy weaker requirements and do \textit{not} admit a circuit representation (see Sec.~\ref{subsubsec::axiomatichoqos-qc-transtranstrans}). }  \label{fig::axiomatichoqos-flowchart}
\end{figure}

In what follows, we will first reiterate the axiomatic considerations that lead to quantum channels and more generally quantum combs~\cite{Chiribella_2008_PRL, Bisio_2011}, and then see how the axiomatic approach readily leads to quantum operations that lie outside the quantum comb formalism when the axiomatic considerations are weakened~\cite{Oreshkov_2012, Chiribella_2012, Chiribella_2013, Bisio_2019,  Milz_2022, Hoffreumon_2021, simmons_higher-order_2022, milz_characterising_2024}. As a warm-up, before considering a truly `higher-order' case, we begin with the axiomatic approach to quantum channels in order to build intuition. For a simple flow chart of the logic that leads to proving the equivalence of the constructive and axiomatic approach to HOQOs, see Fig.~\ref{fig::axiomatichoqos-flowchart}.


\subsubsection{Axiomatic Derivation of Quantum Channels}\hfill\\
\label{subsubsec::axiomatichoqos-quantumchannels}

\begin{figure}[t]
\centering
\subfigure[\textbf{CPTP Map}]
{
\includegraphics[scale=0.7]{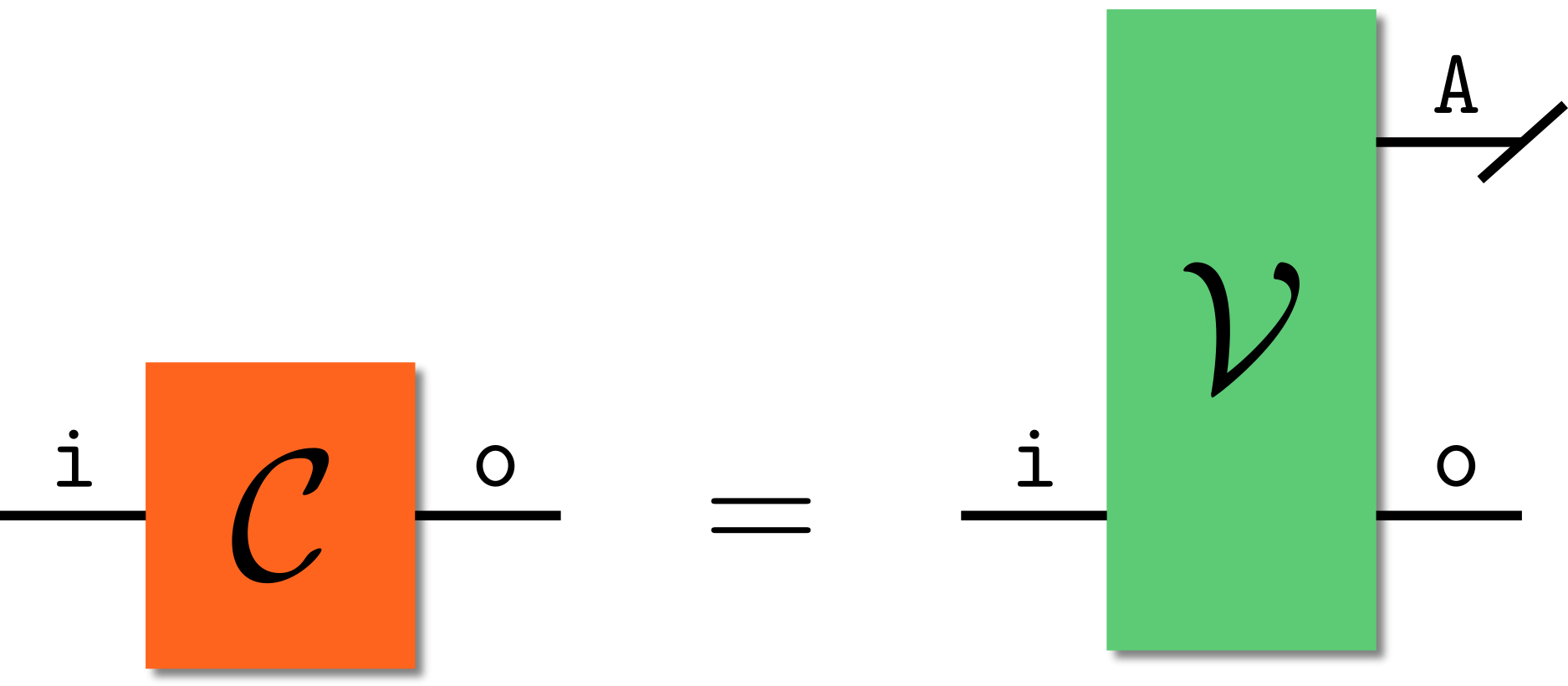}
\label{fig::axiomatichoqos-channelstinespring_a}
}
\hspace{1.5cm}
\subfigure[\textbf{CP Trace Non-increasing Map}]
{
\includegraphics[scale=0.7]{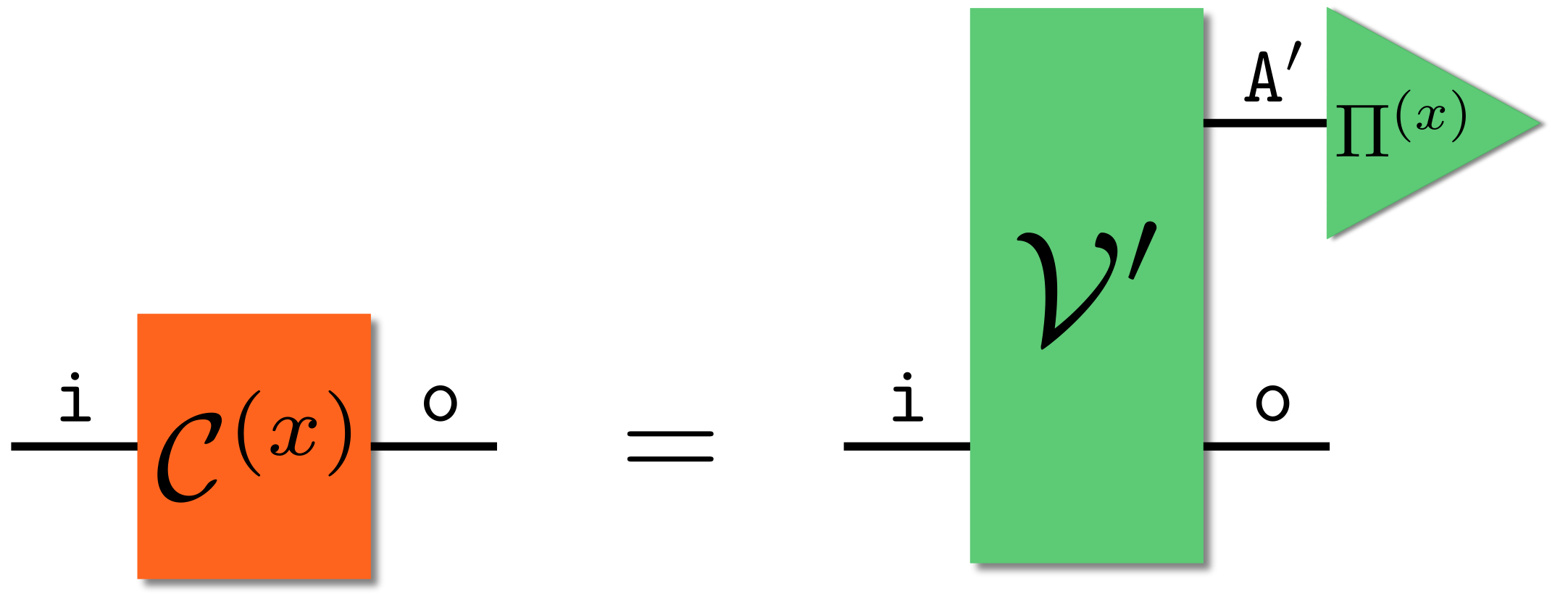}
\label{fig::axiomatichoqos-channelstinespring_b}
}\hfill
\caption{\textbf{Stinespring Dilation.} \textbf{(a)} Any quantum channel $\Ccal:\Lscr(\Hscr_\inp) \rightarrow \Lscr(\Hscr_\out)$ can be understood as arising from an isometry $V:\Hscr_\inp \rightarrow \Hscr_\out \otimes \Hscr_\texttt{A}$, with $\mathcal{V}[\bullet] = V\bullet V^\dagger$, and a partial trace over $\Hscr_\texttt{A}$. \textbf{(b)} Similarly, any CP trace non-increasing map can be understood as arising from an isometry and a projective measurement on the auxiliary space (with corresponding POVM element $\Pi^{(x)}_{\texttt{A}'}$).}  
\end{figure}

\noindent As discussed in Sec.~\ref{subsubsec::me-opensystemdynamicswithinitialcorrelations}, a constructive approach to general deterministic transformations between quantum states is to define a quantum channel via [see Fig.~\ref{fig::axiomatichoqos-channelstinespring_a}]
\begin{align} \label{eq::axiomatichoqos-constructivechannelstinespring}
	\mathcal{C}[\rho_{\inp}] := \ptr{\texttt{A}}{U\left(\rho_\inp\otimes \ketbra{0}{0} _\texttt{A}\right) U^\dagger} =: \ptr{\texttt{A}}{V\rho V^\dagger},
\end{align}
where $\ketbra{0}{0}_\texttt{A}$ is an arbitrary (pure) state in some auxiliary Hilbert space $\mathscr{H}_\texttt{A}$, $U:\Hscr_\texttt{i} \otimes \Hscr_\texttt{A}\to \Hscr_\out\otimes\Hscr_{\texttt{A}}$ is a unitary, and $V: \Hscr_\inp \to \Hscr_\out \otimes \Hscr_\texttt{A}$ is an isometry.\footnote{For simplicity, here, we assume $\Hscr_\out \cong \Hscr_\inp$. The more general case follows by replacing the trace over the auxiliary space \texttt{A} in Eq.~\eqref{eq::axiomatichoqos-constructivechannelstinespring} by a trace over a space $\texttt{A}'$, such that $\text{dim}(\Hscr_\inp)\times \text{dim}(\Hscr_\texttt{A}) = \text{dim}(\Hscr_\out)\times \text{dim}(\Hscr_{\texttt{A}'})$. We predominantly discuss dilations in terms of isometries $V$ such that $V^\dagger V = \mathds{1}_{\texttt{o}}$ instead of unitaries. Since isometries can always be completed to unitaries, this distinction is purely notational.} From the circuit representation of Eq.~\eqref{eq::axiomatichoqos-constructivechannelstinespring}, it is straightforward to see that $\mathcal{C}: \Lscr(\Hscr_\inp) \rightarrow \Lscr(\Hscr_\out)$ inherits the properties of being a linear CPTP map and thus that its Choi state $\mathsf{C}_{\inp\out}\in \Lscr(\Hscr_{\out} \otimes \Hscr_\inp)$ satisfies $\mathsf{C} \geq 0$ and $\ptr{\out}{\mathsf{C}} = \ident_\inp$ [see Eq.~\eqref{eq::tf-cji-choicptp}].\footnote{In this and the following section, we simplify notation as compared to the previous sections, denoting Hilbert spaces by $\Hscr_\inp, \Hscr_\out, \dots$ instead of $\Hscr_{S^\inp}, \Hscr_{S^\out}, \dots$ and labelling auxiliary spaces by $\aux, \texttt{A}, \dots$. Additionally, since it is more intuitive for the channel case, we switch the role of input and input and output and denote them by $\inp$ and $\out$ instead of $1^\out$ and $2^\inp$ respectively as would follow from the convention in the previous section. We will return to the latter convention when we discuss genuinely multi-time quantum combs.}

\vspace{0.25cm}\noindent
\textbf{\textul{Axiomatic Approach: Desiderata.}} In contrast, an axiomatic approach to quantum channels starts by listing their basic requirements, and then accepting any transformation that satisfies them. We have already encountered these basic requirements in Sec.~\ref{subsubsec::tf-notationlineartransformations}, but we reiterate them here in order to make the subsequent discussion of HOQOs more transparent. To respect the linear structure of quantum theory (or, more generally, any probabilistic theory), it is natural to impose that any valid channel must be a linear map $\mathcal{C} : \Lscr(\Hscr_\inp)\to\Lscr(\Hscr_\out)$ from the considered input space [here, $\Lscr(\Hscr_\inp)$] to the output space [here, $\Lscr(\Hscr_\out)$]. Focusing further on deterministic operations, one imposes that for any deterministic input object, the output is also a deterministic object within the considered theory. For transformations of quantum states, this corresponds to demanding that any unit trace input state $\rho$ is mapped to a unit trace output state $\rho' = \mathcal{C}[\rho]$, i.e., the map $\mathcal{C}$ should be trace preserving (TP). Additionally, the map $\mathcal{C}$ must preserve positivity, i.e., for any $\rho \geq 0$, we should have $\mathcal{C}[\rho] \geq 0$. Moreover, it is reasonable to impose that trivial extensions $\Ccal \otimes \Ical$ of $\mathcal{C}$ should also lead to valid output states; in other words, $\mathcal{C}$ should be completely positive (CP), i.e., $\mathcal{C}\otimes\mathcal{I}(\rho_{\inp\texttt{A}})\geq0$ for every auxiliary space $\mathscr{H}_\texttt{A}$ and every operator $\rho_{\inp\texttt{A}}\geq0$. Hence, we can argue: 

\begin{myDefinition*}{Quantum Channels---Axiomatic Approach}{}
For a quantum channel $\mathcal{C}:\Hscr_\inp \rightarrow \Hscr_\out$ to be valid, it must be linear, completely positive (CP), and trace preserving (TP). 
\end{myDefinition*}


\noindent\textbf{\textul{Stinespring Dilation.}} As mentioned above, the Stinespring dilation~\cite{Stinespring_1955} ensures that for every CPTP linear map $\mathcal{C}$, there exists an auxiliary space $\mathscr{H}_\texttt{A}$ and an isometry $V_{\inp \rightarrow \out\texttt{A}}:{\Hscr_\texttt{i}} \to \Hscr_\texttt{\out}\otimes\Hscr_{\texttt{A}}$ such that $\mathcal{C}[\rho_\texttt{i}] = \ptr{\texttt{A}}{V\rho_\texttt{i} V^\dagger}$ [see Fig.~\ref{fig::axiomatichoqos-channelstinespring_a}]. This can be proven directly: Given any Kraus decomposition $\{K_\alpha\}_{\alpha = 1}^{N}$ of $\mathcal{C}$, i.e., such that $\mathcal{C}[\rho_\inp] = \sum_{\alpha=1}^N K_\alpha \rho_\inp K_\alpha^\dagger$, it is straightforward to see that for any orthonormal set of vectors $\{\ket{v_\alpha}\}_{\alpha = 1}^N$, an isometry of the form $V_{\inp \rightarrow \out \texttt{A}} = \sum_{\alpha=1}^N K_\alpha \otimes \ket{v_\alpha}: \Hscr_\inp \rightarrow \Hscr_\out \otimes \Hscr_\texttt{A}$, with $\Hscr_\texttt{A} = \text{span}(\{\ket{v_\alpha}\}_{\alpha = 1}^N)$ and $\text{dim}(\Hscr_\texttt{A}) = N$ indeed provides a dilation for $\mathcal{C}$ with an $N$-dimensional dilation space $\Hscr_\texttt{A}$, since $V^\dagger_{\inp \rightarrow \out \texttt{A}} V_{\inp \rightarrow \out \texttt{A}} = \ident_\inp$ (i.e., it is an isometry) and
\begin{align}\label{eq::axiomatichoqos-constructivechannelstinespringkraus}
 \ptr{\texttt{A}}{V_{\inp \rightarrow \out \texttt{A}} \rho_\inp V_{\inp \rightarrow \out \texttt{A}}^\dagger} = \sum_{\alpha, \beta}\ptr{\texttt{A}}{(K_\alpha \otimes \ket{v_\alpha})\rho_\inp (K^\dagger_\beta \otimes \bra{v_\beta})} = \sum_{\alpha} K_\alpha \rho_\inp K^\dagger_\alpha = \mathcal{C}[\rho_\inp], 
\end{align}
where we have used $\braket{v_\alpha|{v_\beta}} = \delta_{\alpha\beta}$. In the same way, any CP \textit{trace non-increasing} map $\Ccal^{(x)}$ with Kraus operators $\{L_\alpha\}_{\alpha=1}^N$ and $\sum_{\alpha=1}^N L_\alpha^\dagger L_\alpha =: \mathsf{F} \leq \ident$ can be implemented by means of an isometry and a projective measurement on the environment. Concretely, set $L_{N+1} := \sqrt{\ident - \mathsf{F}} \geq 0$, such that $V'_{\inp \rightarrow \out \texttt{A}'} = \sum_{\alpha=1}^{N+1} L_\alpha \otimes \ket{u_\alpha}$ is an isometry and $\text{dim}(\Hscr_{\texttt{A}'}) = N+1$. Then 
\begin{gather}
\Ccal^{(x)} = \ptr{\texttt{A}'}{V'_{\inp \rightarrow \out \texttt{A}'} \rho_\inp V_{\inp \rightarrow \out \texttt{A}'}^\dagger \Pi^{(x)}_{\texttt{A}'}}, 
\end{gather}
where $\Pi^{(x)}_{\texttt{A}'} := \sum_{\alpha=1}^N \ketbra{u_\alpha}{u_\alpha}$ is a POVM element on $\texttt{A}'$ [see Fig.~\ref{fig::axiomatichoqos-channelstinespring_b}]. That is, every trace non-increasing CP map can be understood as stemming from an isometry and a selective, projective measurement on the auxiliary degrees of freedom.

However, the thusly constructed isometries manifestly depend upon the choice of the set of Kraus operators, and might not be minimal. That is, both $\Ccal$ and $\Ccal^{(x)}$ might be dilatable by means of smaller dilation spaces $\Hscr_\aux$ and $\Hscr_{\aux'}$, respectively.\footnote{We denote general dilation spaces by $\Hscr_\texttt{A}$ and reserve $\Hscr_\aux$ for \textit{minimal} dilation spaces.} In what follows, we will focus on the trace preserving case (the trace non-increasing case follows in a similar manner~\cite{Bisio_2011}), for which we have the following Proposition:

\begin{myDefinition*}{Quantum Channels---Minimal Dilation}{}
For a quantum channel $\mathcal{C}: \Lscr(\Hscr_\inp) \rightarrow \Lscr(\Hscr_\out)$ with Choi state $\mathsf{C}_{\inp\out} = \sum_\beta^R \lambda_\beta \ketbra{\beta}{\beta} \in \Lscr(\Hscr_\inp \otimes \Hscr_\out)$ in its spectral decomposition, the dimension of the minimal required dilation space is given by $\text{dim}(\Hscr_\aux) = \text{rank}(\mathsf{C}_{\inp\out}) =:R$. A possible choice for the corresponding isometry is 
\begin{align}
\label{eq::axiomatichoqos-minimalstinespringchannels}
    V_{\inp \rightarrow \out\aux} := \sum_{\beta=1}^R K_\beta \otimes \ket{v_\beta}, \quad \text{where} \ K_\beta := \sqrt{\lambda_\beta} \sum_{k} \braket{k_\inp|\beta}\bra{k_\inp}\, ,
\end{align}
with $\braket{v_\beta|v_\gamma} = \delta_{\beta\gamma}$. We then have $\Ccal[\rho_\inp] = \sum_\beta K_\beta \rho_\inp K_\beta^\dagger$, and this choice of Kraus operators is \textit{canonical} in the sense that it satisfies $\tr{K_\beta^\dagger K_\gamma} = \delta_{\beta\gamma} \, \tr{K_\beta^\dagger K_\beta} = \delta_{\beta\gamma}\lambda_\beta$.
\end{myDefinition*}
\noindent 
We emphasise that, as mentioned, the Kraus decomposition itself is non-unique and only fixed up to an isometric freedom. In particular, let $\{K_\beta\}$ be a minimal set of Kraus operators for a map $\Ccal$. If $\left\{L_\gamma := \sum_\beta w_{\gamma \beta} K_\beta\right\}$, where $w_{\gamma\beta} \in \mathds{C}$ are the matrix elements of an isometry, i.e., $\sum_{\beta} w_{\gamma \beta}^ * w_{\gamma \alpha}  = \delta_{\alpha \beta}$, then $\{L_\gamma\}$ is also a (generally non-minimal) set of Kraus operators for $\mathcal{C}$~\cite{Nielsen_Chuang_2010}.

\newpage To prove that the above is a minimal dilation, we first check that the isometry of Eq.~\eqref{eq::axiomatichoqos-minimalstinespringchannels} indeed dilates $\Ccal$. By insertion, we see that
\begin{align}
    V^\dagger_{\inp \rightarrow \out\aux}V_{\inp \rightarrow \out\aux} &= \sum_{\beta,k,k'} \lambda_\beta \ket{k_\inp}\braket{\beta|k_\inp}\braket{k'_\inp|\beta}\bra{k'_\inp} = \sum_{\beta,k,k',\ell} \lambda_\beta \ket{k_\inp}\braket{\beta|k_\inp \ell_\out}\braket{k'_\inp\ell_\out|\beta}\bra{k'_\inp} \notag \\*
    &= \sum_{k,k'}\ketbra{k_\inp}{k'_\inp}{\ptr{\out}{\mathsf{C}_{\inp\out}}}\ketbra{k_\inp}{k'_\inp} = \ptr{\out}{\mathsf{C}_{\inp\out}}^\mathrm{T} = \ident_\inp, 
\end{align} 
where $\{\ket{\ell_\out}\}$ is an orthonormal basis of $\Hscr_\out$ and we have used $\ptr{\out}{\mathsf{C}_{\inp\out}} = \ident_\inp$. We emphasise that here and in what follows, for the sake of precise bookkeeping, we denote the spaces that bras and kets act on by additional subscripts in typewriter font. In contrast, summation indices will be denoted in normal font. Additionally, note that $\{\ket{\beta} \in \Hscr_\inp \otimes \Hscr_\out\}$ corresponds to the eigenbasis of $\mathsf{C}_{\inp\out}$ and $\{\ket{k_\inp} \in \Hscr_\inp\}$ denotes the computational basis of $\Hscr_\inp$. We do not assume a fixed relationship between them, e.g., $\ket{\beta = 0}$ is generally not equal to $\ket{0_\inp 0_\out}$, etc.

In a similar way as above, we can also verify the action of the channel by computing
\begin{align}
    \ptr{\aux}{V_{\inp \rightarrow \out\aux}\rho_\inp V_{\inp \rightarrow \out\aux}^\dagger} &= \sum_{\beta,\gamma,k,k'} \sqrt{\lambda_\beta\lambda_\gamma} \, \ptr{\aux}{(\braket{k_\inp|\beta}\bra{k_\inp} \otimes \ket{v_\beta})\rho_\inp (\ket{k'_\inp}\braket{\gamma|k'_\inp} \otimes \bra{v_\gamma})} \notag \\
    &= \sum_{\beta,k,k'} \lambda_\beta \braket{k_\inp|\beta}\bra{k_\inp} \rho_\inp \ket{k'_\inp}\braket{\beta|k'_\inp} = \sum_\beta K_\beta \rho_\inp K_\beta^\dagger \notag \\
    & =\ptr{\inp}{\mathsf{C}_{\inp\out} (\rho_\inp^\mathrm{T} \otimes \ident_\out)} = \mathcal{C}[\rho_\inp].
\end{align}
The fact that the chosen Kraus decomposition is canonical follows directly from its definition, since 
\begin{align}
 \tr{K_\beta^\dagger K_\gamma} = \sqrt{\lambda_\beta\lambda_\gamma} \sum_{k,k'} \tr{\ket{k_\inp}\braket{\beta|k_\inp} \braket{k'_\inp|\gamma}\bra{k'_\inp}} = \delta_{\beta\gamma} \lambda_{\beta} = \delta_{\beta\gamma}\, \tr{K_\beta^\dagger K_\beta}.
\end{align} 
Finally, let us assume for the sake of contradiction that there exists a dilation with a smaller auxiliary space $\Hscr_{\bux}$ that satisfies $R':=\text{dim}(\Hscr_{\bux}) < \text{dim}(\Hscr_{\aux}) = R$. Denoting the corresponding isometry by $V'_{\inp \rightarrow \out\bux}$, we would obtain for the Choi state $\mathsf{C}_{\inp\out}$ of $\mathcal{C}$
\begin{align}
 \mathsf{C}_{\inp\out} = (\Ccal \otimes \Ical)[\ketbra{\Phi^+_{\inp\inp'}}{\Phi^+_{\inp\inp'}}] = \ptr{\bux}{V'_{\inp \rightarrow \out\bux} \ketbra{\Phi^+_{\inp\inp'}}{\Phi^+_{\inp\inp'}} V^{\prime\dagger}_{\inp \rightarrow \out\bux}} =: \sum_{\mu =1}^{R'} \kketbra{K_\mu'}{K_\mu'},
\end{align}
where $\kket{K_\mu'} = \braket{\mu_{\bux}|V'_{\inp \rightarrow \out\bux} | \Phi^+_{\inp\inp'}}$. This implies that $\text{rank}(\mathsf{C}_{\inp\out}) \leq R'$, which contradicts the assumption that $\text{rank}(\mathsf{C}_{\inp\out}) = R > R'$. \hfill $\square$

\vspace{0.25cm}\noindent
\textbf{\textul{Stinespring Dilation from the Choi Matrix.}} The above provides a minimal Stinespring dilation of $\mathcal{C}$, but the choice of   $V_{\inp \rightarrow \out\aux}$ is manifestly dependent on the choice of vectors $\{\ket{v_\beta}\}$ spanning $\Hscr_\aux$. Additionally, it is not phrased in terms of the Choi matrix $\mathsf{C}_{\inp\out}$ which---given that we predominantly use the Choi representation for HOQOs throughout this Tutorial---makes it difficult to generalise the explicit form of required isometry to the multi-slot case. Consequently, in anticipation of the following sections, we now re-derive said minimal Stinespring dilation manifestly in terms of its Choi matrix $\mathsf{C}_{\inp\out}$. This, in turn, provides us with a blueprint for the generalisation to the multi-slot case, and yields a canonical choice for the vectors $\{\ket{v}_\beta\}$ in the definition of $V_{\inp\rightarrow \out \aux}$. 

While rather straightforward, the corresponding derivation requires careful bookkeeping of the involved spaces; we recall the convention that `primed' spaces are isomorphic to `unprimed' ones, e.g., $\Hscr_{\inp'} \cong \Hscr_{\inp}$, and analogously for matrices defined on `primed' and `unprimed' spaces, e.g., $\mathsf{C}_{\inp'\out} \cong \mathsf{C}_{\inp\out}$, etc.

Now, in order to derive the Stinespring dilation, firstly, let $\mathsf{C}_{\inp\out} \in \Lscr(\Hscr_\out \otimes \Hscr_\inp)$ be the Choi matrix of a CPTP map $\mathcal{C}$, whose action is given by 
\begin{align}
    \label{eq::axiomatichoqos-choichannelaction}
    \mathcal{C}[\rho] = \ptr{\inp}{\mathsf{C}_{\inp\out}(\rho_\inp^\mathrm{T} \otimes \ident_\out)} \quad \forall \, \rho \in \Lscr(\Hscr_\inp).
\end{align} As already discussed around Eq.~\eqref{eq::tf-cji-choicptp}, the Choi matrix of any CPTP map satisfies 
\begin{align}
\label{eq::axiomatichoqos-choichannelcp}
 \mathsf{C}_{\inp\out}\geq 0 \quad \text{and} \quad \ptr{\out}{\mathsf{C}_{\inp\out}} = \ident_\inp.
\end{align}
To show that there exists a Stinespring dilation for \textit{any} $\mathsf{C}$ that satisfies Eq.~\eqref{eq::axiomatichoqos-choichannelcp}, let us introduce a \textit{purification} of $\mathsf{C}_{\inp\out}$ as follows
\begin{align}
\label{eq::axiomatichoqos-choichannelpurification}
    \ket{\tilde{\mathsf{C}}}_{\inp\out\aux} := (\sqrt{\mathsf{C}}^\dagger_{\inp\out} \otimes \mathds{1}_{\inp'\out'}) \ket{\Phi^+_{\inp\inp'}} \ket{\Phi^+_{\out\out'}}  = 
    \sum_
    {\beta = 1}^{R} \lambda_\beta \ket{\beta}\ket{\beta^*} \in \Hscr_{\inp} \otimes \Hscr_{\out} \otimes \Hscr_{\aux},
\end{align}
where $\ket{\beta^*}:=\ket{\beta}^*$ is the complex conjugate of $\ket{\beta}$ with respect to the product basis $\{\ket{k_\inp \ell_\out}\}$ of $\Hscr_{\inp}\otimes \Hscr_\out$.\footnote{Since $\mathsf{C}_{\inp\out}$ is positive semidefinite, the additional dagger on $\sqrt{\mathsf{C}}^\dagger$ in the definition of the purification of $\mathsf{C}_{\inp\out}$ is somewhat superfluous. Here, we explicitly include it to adhere to the convention chosen in Refs.~\cite{Chiribella_2009,Bisio_2011}.} Setting $\widetilde{\mathsf{C}}_{\inp\out\aux}:=  \ketbra{\widetilde{\mathsf{C}}_{\inp\out\aux}}{\widetilde{\mathsf{C}}_{\inp\out\aux}}$, it is straightforward to see by insertion that $\mathsf{C}_{\inp\out} = \ptr{\aux}{\widetilde{\mathsf{C}}_{\inp\out\aux}}$ holds. Additionally, from Eq.~\eqref{eq::axiomatichoqos-choichannelpurification}, we can deduce that the required purification space $\Hscr_\aux \subseteq \Hscr_{\out'}\otimes \Hscr_{\inp'}$---which turns out to be the dilation space for the Stinespring dilation of $\Ccal$---satisfies $\text{dim}(\Hscr_\aux) = \text{rank}(\mathsf{C}_{\inp\out})$, coinciding with the minimal possible dimension for the purification. 

From Eq.~\eqref{eq::axiomatichoqos-choichannelcp}, we further have that $\ptr{\out\texttt{a}}{\widetilde{\mathsf{C}}_{\inp\out\texttt{a}}} = \ident_\inp$ holds, making $\widetilde{\mathsf{C}}_{\inp\out\texttt{a}}$ a purification of $\ident_\inp$. On the other hand, a \textit{minimal} purification of $\ident_\inp$ is given by $\Phi^+_{\inp \inp''} \in \Lscr(\Hscr_\inp \otimes \Hscr_{\inp''})$. Now, any purification $\ket{\widetilde{\mathsf{C}}_{\inp\out\aux}}$ of $\mathds{1}_\inp$ is isometrically related to this minimal one via 
\begin{align}
\label{eq::axiomatichoqos-choichannelpurificationisometry}
 \ket{\widetilde{\mathsf{C}}}_{\inp\out\aux} = V_{\inp'' \to \out\aux}  \ket{\Phi^+_{\inp \inp''}},
\end{align}
where $V_{\inp'' \to \out \aux}: \Hscr_{\inp''} \to \Hscr_{\out} \otimes \Hscr_{\aux}$ is an isometry that acts only on the purification spaces. As one would expect, this is exactly the isometry required for the Stinespring dilation of $\Ccal$; we derive its explicit form below [see Eq.~\eqref{eq::axiomatichoqos-choichannelisometrystinespring}]. First, we insert the above into Eq.~\eqref{eq::axiomatichoqos-choichannelaction} to show that the CPTP map $\Ccal$ can indeed be understood as stemming from this isometry. Specifically, we have 
\begin{align}
\mathcal{C}[\rho] &= \ptr{\inp\aux}{\widetilde{\mathsf{C}}_{\inp\out\aux}(\rho_\inp^\mathrm{T} \otimes \ident_{\out\aux})} = \ptr{\inp\aux}{ V_{\inp'' \to \out\aux}  \ketbra{\Phi^+_{\inp \inp''}}{\Phi^+_{\inp \inp''}} V_{\inp'' \to \out\aux}^\dagger (\rho_\inp^\mathrm{T} \otimes \ident_{\out\aux})} \notag \\
&=\sum_{k,\ell}\ptr{\aux}{V_{\inp'' \to \out\aux}  \ketbra{k_{\inp''}}{\ell_{\inp''}} V_{\inp'' \to \out\aux}^\dagger \braket{k_\inp|\rho_\inp|\ell_\inp}}  = \ptr{\aux}{V_{\inp'' \to \out\aux}  \rho_{\inp''}  V_{\inp'' \to \out\aux}^\dagger} \notag \\
&= \ptr{\aux}{V_{\inp \to \out\aux}  \rho_{\inp}  V_{\inp \to \out\aux}^\dagger}, 
\end{align}
where we made the replacement $\inp'' \mapsto \inp$ in the last line. 

\begin{figure}[t]
\centering
\vspace{0.5em}
\includegraphics[width=0.5\linewidth]{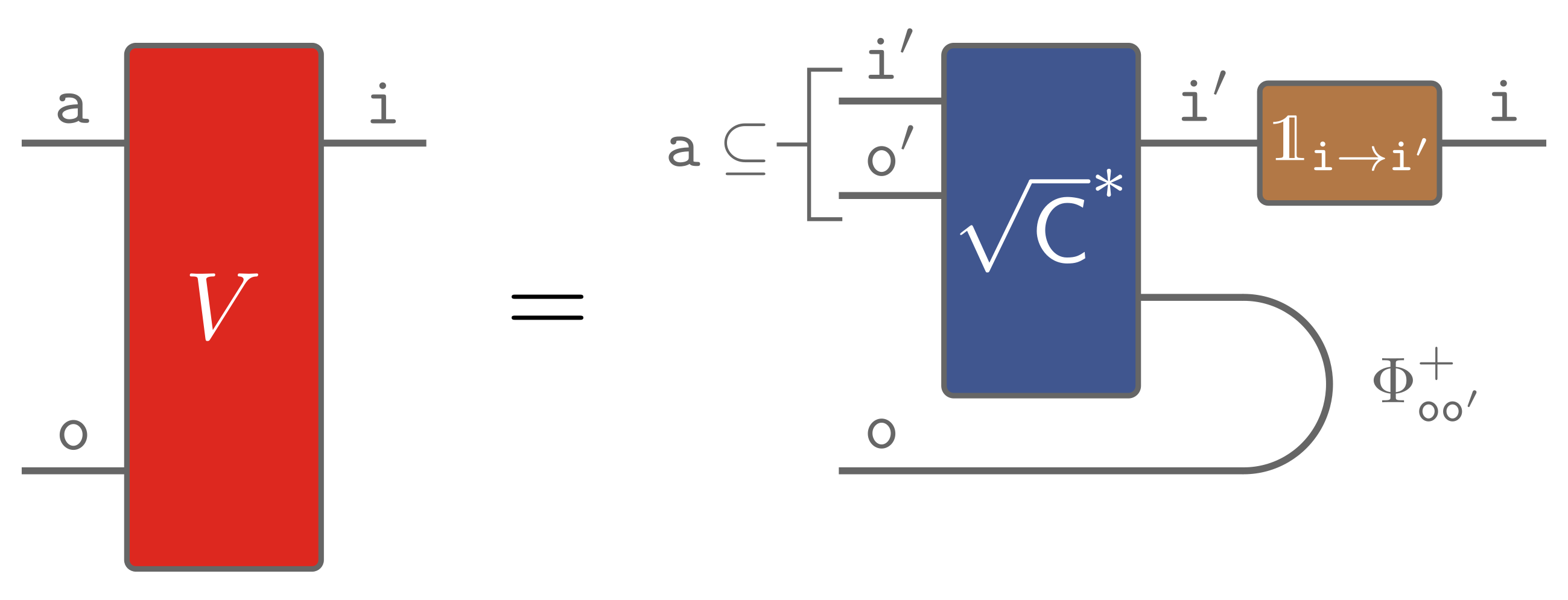}
\caption{\textbf{Isometry of a Quantum Channel from its Choi Matrix.} The isometry $V_{\inp \rightarrow \out\aux}: \Hscr_{\inp} \rightarrow \Hscr_{\out} \otimes \Hscr_{\aux}$ leading to a channel $\Ccal:\Lscr(\Hscr_\inp) \rightarrow \Lscr(\Hscr_\out)$ via  $\Ccal[\bullet] = \ptr{\aux}{V_{\inp \rightarrow \out\aux}\bullet V_{\inp \rightarrow \out\aux}^\dagger}$ coincides with $\sqrt{\mathsf{C}}_{\inp\out}^*$---where $\mathsf{C}_{\inp\out}$ is the Choi matrix of $\mathcal{C}$---up to a `bending of a wire', corresponding to multiplication with $\ket{\Phi_{\out\out'}}$, and the relabelling of some spaces. We re-iterate that $\Hscr_\aux \subseteq \Hscr_{\inp'} \otimes \Hscr_{\out'}$ and we tacitly assume the combination of the wires labelled by $\inp'$ and $\out'$ into the wire labelled by $\aux$ in the above figure. Note that here, the depicted objects correspond to \textit{matrices}, not maps, like in the circuit diagrams above. To emphasise this difference, matrices are depicted with a grey, rounded frame and labelled in white, and `wires' are depicted in grey. } \label{fig::axiomatichoqos-choichannelisometry}
\end{figure}

It remains to derive the isometry $V_{\inp \to \out\aux}$ explicitly. This can be done by using the fact that 
\begin{align}
    \braket{\Phi^+_{\inp \inp''}|V_{\inp'' \to \out\aux} |\Phi^+_{\inp \inp''}} = V_{\inp'' \to \out\aux}. 
\end{align}
With this, one can directly read off $V_{\inp'' \to \out\aux}$ from Eq.~\eqref{eq::axiomatichoqos-choichannelpurificationisometry} as 
\begin{align}
\label{eq::axiomatichoqos-choichannelisometrystinespring}
    V_{\inp'' \to \out\aux} = \braket{\Phi^+_{\inp \inp''}|\widetilde{\mathsf{C}}_{\inp\out\aux}} \quad \text{and thus} \quad V_{\inp \to \out\aux} = \braket{\Phi^+_{\inp'' \inp}|\widetilde{\mathsf{C}}_{\inp''\out\aux}}.
\end{align}
This indeed defines an isometry, since 
\begin{align}
    V_{\inp \to \out\aux}^\dagger V_{\inp \to \out\aux} &= \braket{\widetilde{\mathsf{C}}_{\inp''\out\aux}|\Phi^+_{\inp'' \inp}}\braket{\Phi^+_{\inp'' \inp}|\widetilde{\mathsf{C}}_{\inp''\out\aux}} \notag \\
    &= \bra{\Phi^+_{\inp''\inp'}} \bra{\Phi^+_{\out\out'}} \sqrt{\mathsf{C}}_{\inp''\out} \ketbra{\Phi^+_{\inp'' \inp}}{\Phi^+_{\inp'' \inp}} \sqrt{\mathsf{C}}^\dagger_{\inp''\out}  \ket{\Phi^+_{\inp''\inp'}} \ket{\Phi^+_{\out\out'}} \notag  \\
    &= \sum_{k,\ell,m, n, p, q} \braket{k_{\inp'} k_{\inp''}\ell_{\out}\ell_{\out'}| \sqrt{\mathsf{C}}_{\inp''\out}|m_{\inp''} m_\inp} \braket{n_{\inp''}n_\inp|\sqrt{\mathsf{C}}_{\inp''\out}^\dagger|p_{\inp''}p_{\inp'}q_\out q_{\out'}} \notag  \\
    &=\sum_{k,\ell,m, n, p, q} \delta_{kp} \delta_{\ell q} \braket{k_{\inp''}\ell_\out|  \sqrt{\mathsf{C}}_{\inp''\out}|  m_{\inp''}} \braket{n_{\inp''} |\sqrt{\mathsf{C}}_{\inp''\out}^\dagger | q_\out p_{\inp''}} \ketbra{m_\inp}{n_\inp} \notag  \\
    &= \sum_{m,n} \tr{\sqrt{\mathsf{C}}_{\inp''\out} \ketbra{m_{\inp''}}{n_{\inp''}} \sqrt{\mathsf{C}}_{\inp''\out}^\dagger}\ketbra{m_\inp}{n_\inp} \notag  \\
    &= \sum_{m,n} \braket{n_{\inp''}|\ptr{\out}{\mathsf{C}_{\inp''\out}} |m_{\inp''}}\ketbra{m_\inp}{n_\inp} = \sum_{m,n} \delta_{mn} \ketbra{m_\inp}{n_\inp} = \ident_\inp,
\end{align}
where we have used $\ptr{\out}{\mathsf{C}_{\inp''\out}} = \ident_{\inp''}$. 
While Eq.~\eqref{eq::axiomatichoqos-choichannelisometrystinespring} expresses the isometry $V_{\inp\rightarrow \out\aux}$ in terms of the Choi state $\mathsf{C}_{\inp\out}$, in this form it does not lend itself nicely to generalisation to the multi-slot quantum comb case. In anticipation of this generalisation, following Refs.~\cite{ChiribellaReal2009, Bisio_2011}, it proves advantageous to re-write Eq.~\eqref{eq::axiomatichoqos-choichannelisometrystinespring} as
\begin{align}\label{eq::axiomatichoqos-choichannelisometrystinespringfinal}
    V_{\inp \to \out\aux} &= \braket{\Phi^+_{\inp'' \inp}|\widetilde{\mathsf{C}}_{\inp''\out\aux}} = \bra{\Phi^+_{\inp'' \inp}}\sqrt{\mathsf{C}}^\dagger_{\inp''\out} \ket{\Phi^+_{\inp'' \inp'}}\ket{\Phi^+_{\out' \out}} \notag \\
    &= \sum_{k,\ell,m} \bra{k_{\inp''}k_\inp}\sqrt{\mathsf{C}} ^\dagger_{\inp''\out} \ket{\ell_{\inp''}\ell_{\inp'}}\ket{m_{\out'}m_\out} = \sum_{k,\ell,m,n} \bra{k_{\inp''}n_\out}\sqrt{\mathsf{C}} ^\dagger_{\inp''\out} \ket{\ell_{\inp''}m_\out} \ketbra{n_\out \ell_{\inp'} m_{\out'}}{k_\inp} \notag \\
    &= \sum_{k,\ell,m,n} \bra{\ell_{\inp'}m_{\out'}}\sqrt{\mathsf{C}} ^*_{\inp'\out'} \ket{k_{\inp'}n_{\out'}} \ketbra{n_\out \ell_{\inp'} m_{\out'}}{k_\inp} \notag \\
    &= \sum_{k,\ell,m,n,r} \left[\bra{\ell_{\inp'}m_{\out'}}\sqrt{\mathsf{C}} ^*_{\inp'\out'} \ket{k_{\inp'}n_{\out'}} \ketbra{\ell_{\inp'} m_{\out'}}{k_{\inp'} n_{\out'}}\right]\ket{\Phi^+_{\out \out'}}\ketbra{r_{\inp'}}{r_\inp} \notag \\
    &= \sqrt{\mathsf{C}} ^*_{\inp'\out'} \ket{\Phi^+_{\out \out'}} \ident_{\inp \rightarrow \inp'}\, 
\end{align}
where we have used that $\bra{\ell_{\inp'}m_{\out'}}\sqrt{\mathsf{C}} ^*_{\inp'\out'} \ket{k_{\inp'}n_{\out'}} = \bra{k_{\inp''}k_\inp}\sqrt{\mathsf{C}} ^\dagger_{\inp''\out} \ket{\ell_{\inp''}\ell_{\inp'}}$ together with $\ket{n_\out} = \braket{n_{\out'}|\Phi^+_{\out\out'}}$ and $\ket{k_\inp} = \bra{k_{\inp'}} \sum_r\ketbra{r_{\inp'}}{r_\inp} =: \bra{k_{\inp'}} \ident_{\inp \rightarrow \inp'}$. Intuitively, Eq.~\eqref{eq::axiomatichoqos-choichannelisometrystinespringfinal} expresses the fact that $V_{\inp\rightarrow \out\aux}$ coincides with $\sqrt{\mathsf{C}}^*_{\inp'\out'}$ up to a transformation of an `input space' into an `output space'---given by the multiplication with $\ket{\Phi^+_{\out\out'}}$---and a relabelling of one of the involved spaces---given by the matrix $\ident_{\inp \rightarrow \inp'}$ (see Fig.~\ref{fig::axiomatichoqos-choichannelisometry}). 


As an aside, if we did not already know the size of the required dilation space $\Hscr_{\aux}$, it could be deduced from the above form of $V_{\inp \rightarrow \out\aux}$ as $\text{dim}(\Hscr_\aux) = \text{dim}(\text{rng}(\sqrt{\mathsf{C}}^*_{\inp'\out'}))$, where $\text{rng}(\mathsf{X}) := \{\ket{\varphi}: \exists \ket{\xi} \text{s.t.} \ket{\varphi} = \mathsf{X}\ket{\xi}\}$ is the range of the operator $\mathsf{X}$. Since we have that $\dim(\text{rng}(\sqrt{\mathsf{C}}^*_{\inp'\out'})) = \text{dim}(\text{rank}(\mathsf{C}^*_{\inp'\out'})) = \text{dim}(\text{supp}(\sqrt{\mathsf{C}}^*_{\inp'\out'}))$, where $\text{supp}(\mathsf{X})$ is the support of $\mathsf{X}$, this coincides with the dimension of the dilation space given above (and also the one provided in Refs.~\cite{ChiribellaReal2009,Bisio_2011}).

Let us finally summarise these results on the Stinespring dilation of quantum channels: 

{\hypersetup{citecolor=white}
\begin{myDefinition*}{Quantum Channels---Stinespring Dilation~\cite{ChiribellaReal2009,Bisio_2011}}{}
Let $\Ccal:\Lscr(\Hscr_\inp) \rightarrow \Lscr(\Hscr_\out)$ be a quantum channel with corresponding Choi operator $\mathsf{C}_{\inp \out} \in \Lscr (\Hscr_{\inp} \otimes \Hscr_\out)$. A minimal dilation $\Ccal[\rho_\inp] = \ptr{\aux}{V_{\inp \rightarrow \out \aux} \rho_\inp V_{\inp \rightarrow \out \aux}^\dagger}$ of $\Ccal$ in terms of an isometry $V_{\inp \rightarrow \out \aux} : \Hscr_\inp \rightarrow \Hscr_\out \otimes \Hscr_\aux $, where $\text{dim}(\Hscr_\aux) = \text{rank}(\sqrt{\mathsf{C}}^*_{\inp'\out'})$, is given by $V_{\inp \rightarrow \out \aux} = \sqrt{\mathsf{C}} ^*_{\inp'\out'} \ket{\Phi^+_{\out \out'}} \ident_{\inp \rightarrow \inp'}$.
\end{myDefinition*}
}
\noindent For completeness, it remains to show how this form of isometry $V_{\inp\rightarrow \out\aux}$ corresponds to the representation $V_{\inp\rightarrow \out\aux} = \sum_\beta K_\beta \otimes \ket{v_\beta}$ provided in Eq.~\eqref{eq::axiomatichoqos-minimalstinespringchannels}. To this end, we can re-write Eq.~\eqref{eq::axiomatichoqos-choichannelisometrystinespringfinal} as
\begin{align}
 V_{\inp\rightarrow \out\aux} &= \sqrt{\mathsf{C}}^*_{\inp'\out'} \ket{\Phi^+_{\out \out'}} \ident_{\inp \rightarrow \inp'} 
 = \sum_{\beta,k,\ell}  \sqrt{\lambda_\beta} \ket{\beta^*_{\inp'\out'}} \braket{\beta^*_{\inp'\out'}|k_{\inp'} \ell_{\out'}}\ketbra{\ell_\out}{k_{\inp}} \notag \\
 &= \sum_{\beta,k,\ell}  \sqrt{\lambda_\beta} \ket{\beta^*_{\inp'\out'}} \ket{\ell_\out} \braket{k_{\inp'}\ell_{\out'}|\beta_{\inp'\out'}}	\bra{k_{\inp}}
 = \sum_{\beta,k}  \sqrt{\lambda_\beta} \ket{\beta^*_{\inp'\out'}} \braket{k_{\inp}|\beta_{\inp\out}}\bra{k_{\inp}} \notag \\
 &= \sum_\beta K_\beta \otimes \ket{\beta^*_{\inp'\out'}} 
 = \sum_\beta K_\beta \otimes \frac{\ket{K_\beta^*}}{\sqrt{\lambda_\beta}} = \sum_\beta K_\beta \otimes \frac{\ket{K_\beta^*}}{\sqrt{\tr{K_\beta^\dagger K_\beta}}}\, 
\end{align}
where we have used $K_\beta = \sqrt{\lambda_\beta}\sum_k \braket{k_\inp|\beta}\bra{k_\inp}$ and $\ket{K^*_\beta} :=  K_\beta^*\ket{\Phi^+_{\inp\inp'}} = \sqrt{\lambda_\beta} \ket{\beta^*_{\inp'\out'}}$. By setting $\ket{v_\beta} = \ket{K_\beta^*}/\sqrt{\tr{K_\beta^\dagger K_\beta}}$ this (canonical) form of $V_{\inp \rightarrow \out\aux}$ coincides with that given in Eq.~\eqref{eq::axiomatichoqos-minimalstinespringchannels}.

This concludes our axiomatic consideration of quantum channels. We have demonstrated that every axiomatically admissible channel $\mathcal{C}$ can be obtained from a quantum circuit only containing isometries (and partial traces) and \textit{vice versa}. In addition, we have explicitly constructed the corresponding isometry $V_{\inp \rightarrow \out\aux}$ based on the Choi matrix of $\mathcal{C}$ and massaged it into a form that can straightforwardly be generalised to the multi-slot case. Now, we will follow a similar logic in order to first provide an axiomatically motivated derivation of quantum combs and show that it indeed coincides with the constructive approach taken in Sec.~\ref{subsubsec::toqp-deterministichoqos}.  


\subsubsection{Representation Theorem for Quantum Combs}
\label{subsubsec::axiomatichoqos-quantumcomb}\hfill\\

\noindent \textbf{\textul{Axiomatic Approach to Quantum Combs.}} While the axiomatic requirements for a quantum channel discussed in the previous section are immediately evident, as we shall see, the situation presents itself somewhat murkier in the multi-slot scenario. To elucidate this point, let us assume that $\mathsf{T}_{n^\inp:0^\out} \in \Lscr(\Hscr_{{n^{\inp}}} \otimes \cdots \Hscr_{{0^{\out}}})$ is the Choi matrix of a multi-slot HOQO that we wish to characterise axiomatically. By the end of this section, by generalising the considerations that led to quantum channels, we will obtain a  set of valid HOQOs that coincides exactly with that derived constructively in Sec.~\ref{subsubsec::toqp-deterministichoqos}, i.e., it yields the set of quantum combs / process tensors (see Def.~\ref{def::toqp-def-quantumcombs}).\footnote{In this section, we often let HOQOs start on an output space $\Hscr_{0^\out}$. This amounts merely to making a specific choice and has no bearing on the following arguments; the case where combs start on an input space $1^\inp$ can easily be recovered by setting $\Hscr_{0^\out} \cong \mathds{C}$.} Following the arguments of the previous section, we require that $\mathsf{T}_{n^\inp:0^\out}$ maps positive objects to positive objects, even when only acting non-trivially on a part of them, i.e.,
\begin{align}
    \mathsf{T}_{n^\inp:0^\out} \star \mathsf{X} \geq 0  \qquad \forall \ \mathsf{X} \geq 0.
\end{align}
Since this has to hold for \textit{any} positive $\mathsf{X}$---no matter what space $\mathsf{X}$ is defined on---this implies $\mathsf{T}_{n^\inp:0^\out} \geq 0$, just like for the case of quantum channels where we argued for the positivity of $\mathsf{C}$ on the same axiomatic grounds. To derive the remaining conditions on $\mathsf{T}_{n^\inp:0^\out}$ na\"ively, we would require that $\mathsf{T}_{n^\inp:0^\out}$ maps deterministic objects to deterministic objects---where `deterministic' is yet to be properly defined. Consequently, a one-slot comb/superchannel should map a channel to a channel; the next levels of HOQOs should map superchannels to channels, and then superchannels to superchannels [see Fig.~\ref{fig::axiomatichoqos-supersuperchannels}], and so on. In general, we should then be able to build up the hierarchy of admissible quantum combs by starting from states and channels (the most basic deterministic objects), and then demanding that all admissible higher-level HOQOs map `lower rungs' of the hierarchy to admissible objects. 

\begin{figure}[t]
\centering
\subfigure[\textbf{Process Matrix: Definition.}]
{
\includegraphics[scale=0.6]{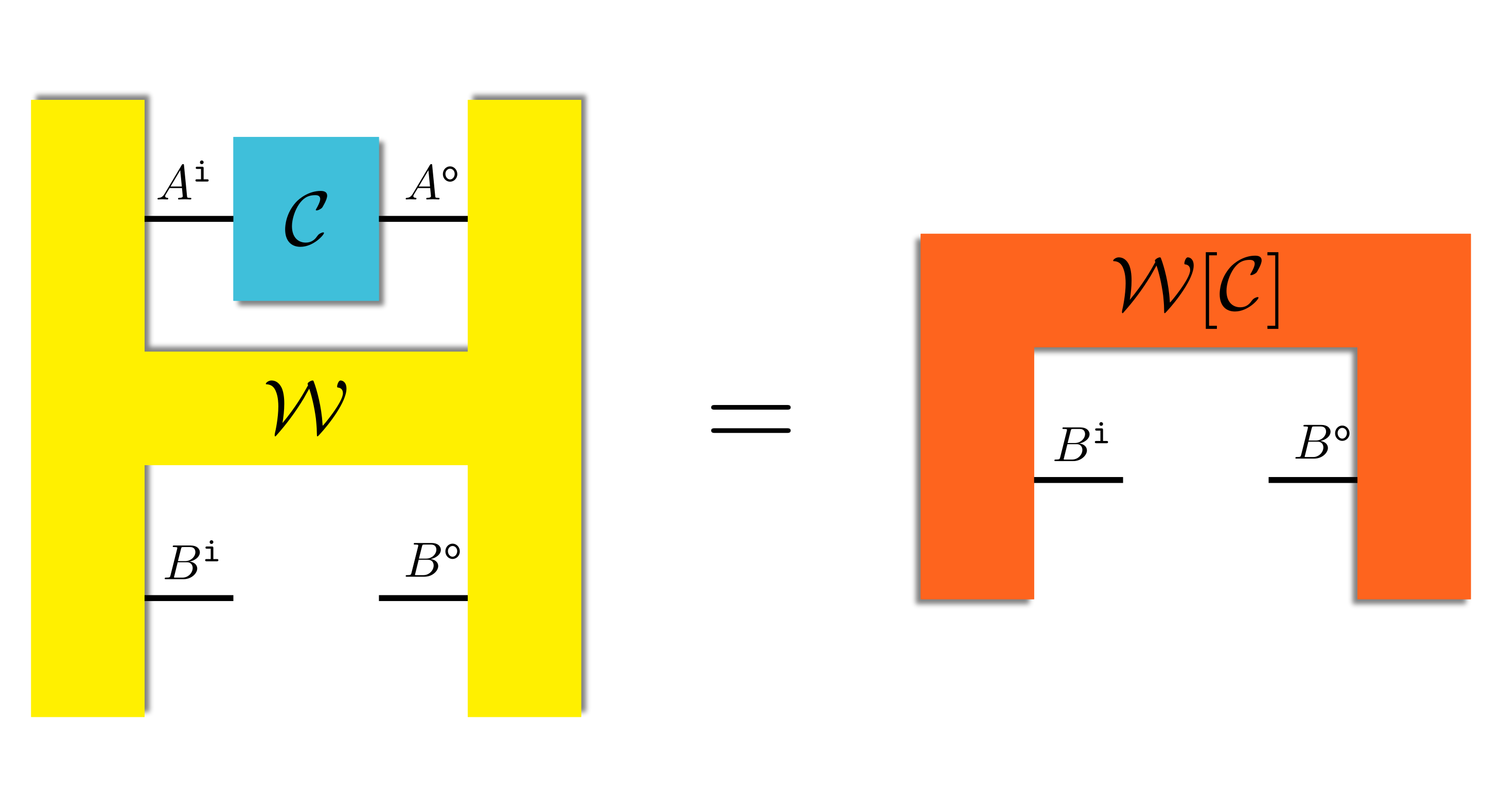}
\label{fig::axiomatichoqos-qc-processmatrix}
}\hspace{1.5cm}
\subfigure[\textbf{Process Matrix Action on a Superchannel.}]
{
\includegraphics[scale=.6]{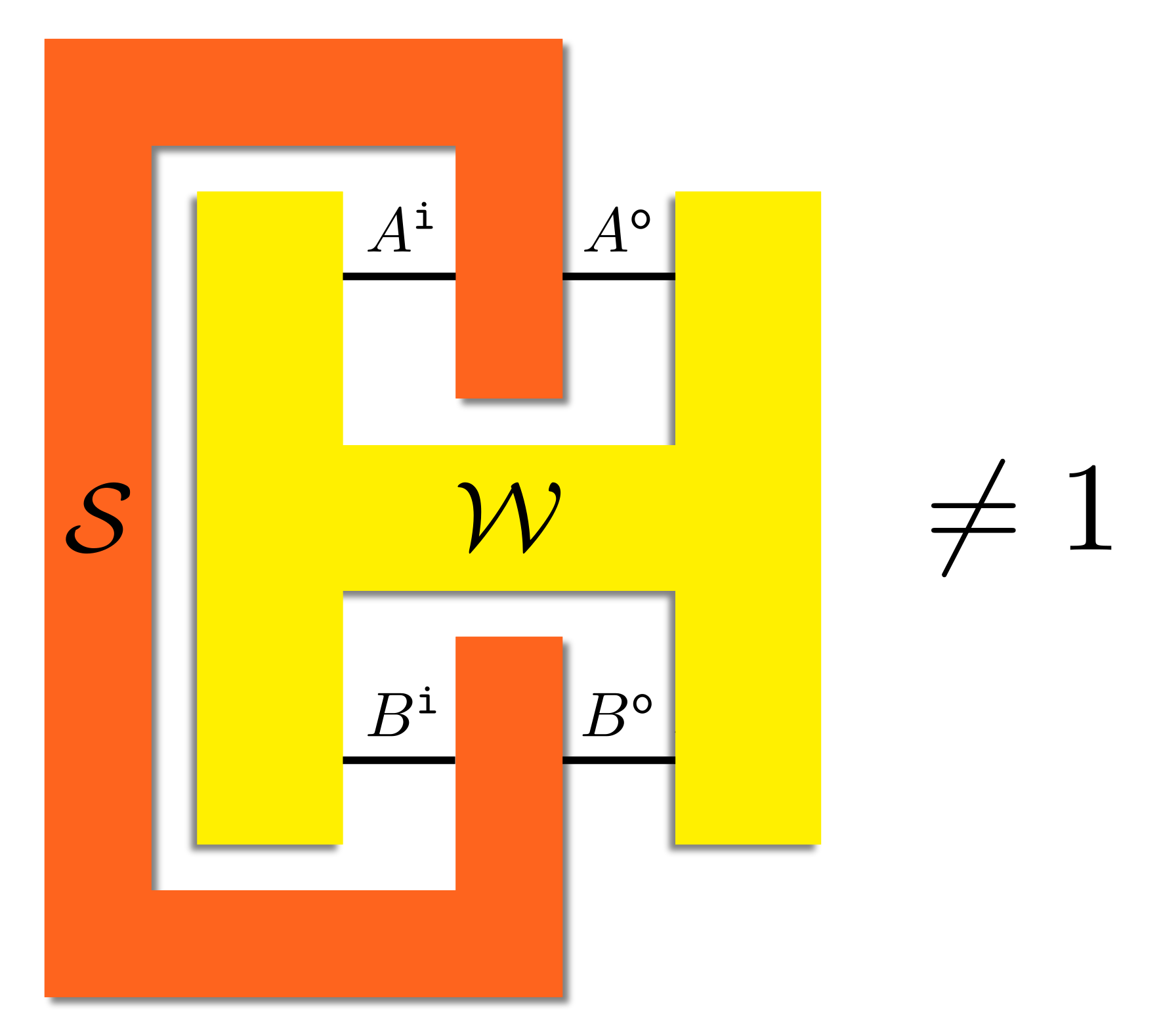}
\label{fig::axiomatichoqos-qc-processmatrixsuperchannel}
}\hfill
\caption{\textbf{Process Matrices. (a)} Process matrices comprise the set of all quantum operations $\mathcal{M}$ that map channels to proper superchannels without initial input and final output space. \textbf{(b)} When acting on a superchannel---depicted vertically here, to `fit' into the process matrix---a process matrix does generally \textit{not} yield unit probability.  For example, the causal ordering of $\Scal$ could be $A_\inp \prec A_\out \prec B_\inp \prec B_\out$ such that its Choi state satisfies $\ptr{B_\out}{S_{A_\inp A_\out B_\inp B_\out}} = \ident_{B_\inp} \otimes S_{A_\inp A_\out}$ and $\ptr{A_\out}{S_{A_\inp A_\out}} = \ident_{A_\inp}$. We note that since process matrices do not necessarily abide by a fixed causal order, spaces are not labelled in increasing order, but each slot corresponds to a laboratory $A, B, \dots$ with respective input and output spaces labelled accordingly (see Sec.~\ref{subsec::indefinitecausalorder}).}
\end{figure}

However, as it turns out, this recipe is too broad~\cite{Chiribella_2009} and only demanding that \textit{particular} deterministic objects are mapped to other deterministic objects (say, superchannels to superchannels) would lead to HOQOs that do not abide by a global causal order and can thus not be represented by a quantum circuit with only isometries and partial traces. The simplest example of this phenomenon is the (bipartite) process matrix case~\cite{Oreshkov_2012}, i.e., the set of transformations that map pairs of independent CPTP maps to unit probability (or, equivalently, the set of transformations that map quantum channels to one-slot combs with trivial input and output space; see Sec.~\ref{subsec::indefinitecausalorder}) [see Figs.~\ref{fig::me-cqt-causallyindefiniteprocess} and~\ref{fig::axiomatichoqos-qc-processmatrix}]. 


As shown in Ref.~\cite{Oreshkov_2012}, such HOQOs can be used to violate so-called \textit{causal inequalities}---a feat impossible under the assumption of a global causal order---and thus lie outside the set of processes that can be obtained from a quantum circuit (or convex combinations thereof; see Sec.~\ref{subsec::indefinitecausalorder}). To remedy this issue, i.e., to make the sets of constructively and axiomatically motivated quantum combs coincide, we must enforce stricter axiomatic requirements (for a more detailed analysis, see Ref.~\cite{Chiribella_2009}). 

\begin{figure}[t]
\centering
\vspace{0.5em}
\includegraphics[width=0.85\linewidth]{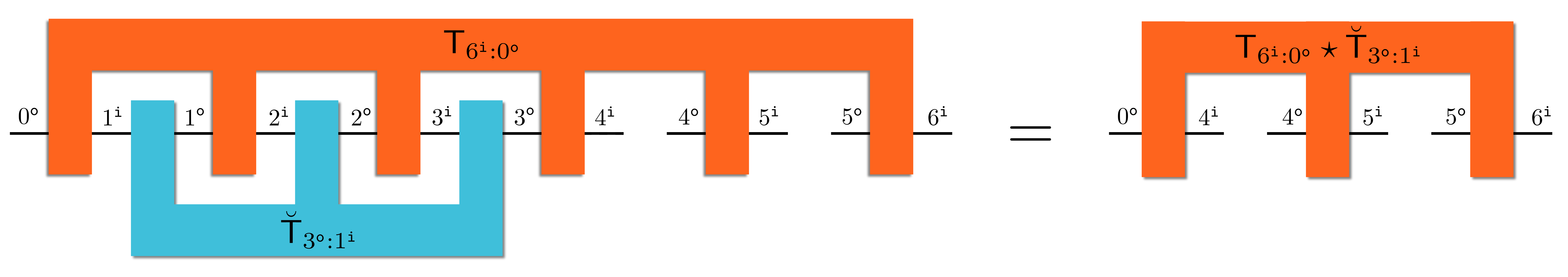}
\caption{\textbf{Action of Quantum Combs on Quantum Combs.} A valid $n$-slot comb should map any valid $k$-slot comb to a valid $(n-k)$-slot comb. Here, this situation is depicted for $n=5$ and $k=2$. } \label{fig::axiomatichoqos-qc-def-axiomaticcombsaction} 
\end{figure}

Intuitively, an admissible quantum comb should not only map certain deterministic objects to deterministic objects, but rather, it should map \textit{all} the `lower rungs' of the hierarchy onto admissible objects. That is a matrix $\mathsf{T}_{n+1^\inp:0^\out} \geq 0$---corresponding to an $n$-slot comb---would be valid iff 
\begin{align}
\label{eq::axiomatichoqos-qc-def-axiomaticcombs}
\mathsf{T}_{n^\inp:0^\out} \star \breve{\mathsf{T}}_{k+1^\out:1^\inp} \quad \text{is a valid } (n-k)\text{-slot quantum comb}
\end{align}
for all $k<n$ and all valid $k$-slot combs $\breve{\mathsf{T}}_{k+1^\out:1^\inp}$.\footnote{Here, for simplicity, we assume that $\mathsf{T}_{n+1^\inp:0^\out}$ and $\breve{\mathsf{T}}_{k+1^\out:1^\inp}$ `start' on fixed spaces that are labelled by $0$ and $1$, respectively, and that they are defined on all spaces between $0$ $(1)$ and $n+1$ $(k+1)$. The arguments can be extended without any added difficulty to the case where these starting times are chosen differently, and where there are `holes' in the set of spaces they are defined on.} We emphasise that the roles of `input' and `output' spaces for $\mathsf{T}_{n+1^\inp:0^\out}$ and $\breve{\mathsf{T}}_{k+1^\out:1^\inp}$ are interchanged---inputs of $\mathsf{T}_{n+1^\inp:0^\out}$ are outputs of $\breve{\mathsf{T}}_{k+1^\out:1^\inp}$ and \textit{vice versa} (see Fig.~\ref{fig::axiomatichoqos-qc-def-axiomaticcombsaction}). As in Sec.~\ref{subsubsec::toqp-deterministichoqos}, this difference in the role of spaces is indicated through the breve on $\breve{\mathsf{T}}_{k+1^\out:1^\inp}$. 


Requiring Eq.~\eqref{eq::axiomatichoqos-qc-def-axiomaticcombs} to hold for \textit{all} $k<n$ is a much stronger requirement than only demanding it for a \textit{fixed} $k$. In contrast, for example, process matrices only require that channels are mapped to $1$-slot combs (without initial input and final output space), but \textit{not} that all superchannels are mapped to the value $1$, the only valid `comb' on $\Lscr(\mathds{C})$ [see Fig.~\ref{fig::axiomatichoqos-qc-processmatrixsuperchannel}]. As a result, process matrices can display causal indefiniteness (see Sec.~\ref{subsec::indefinitecausalorder}), while quantum combs can always be represented by a causally ordered quantum circuit (see below). 

\begin{figure}[t]
\centering
\vspace{0.5em}
\includegraphics[width=0.95\linewidth]{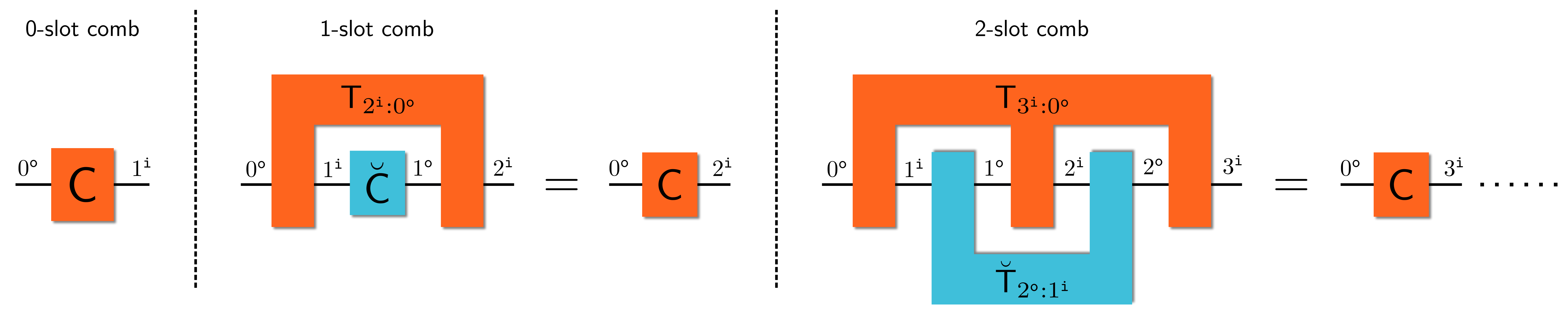}
\caption{\textbf{Axiomatic Definition of $\boldsymbol{n}$-Slot Quantum Combs.} Starting from quantum channels ($0$-slot combs), every $n$-slot quantum comb must map $(n-1)$-slot combs to channels. Note that in this definition spaces can be trivial, i.e., isomorphic to $\mathds{C}$, e.g., combs can start on an output space such that we have $\Hscr_{0^\out} \cong \mathds{C}$.}\label{fig::axiomatichoqos-qc-def-axiomaticcombs}
\end{figure}

Notably, we have not yet defined in an axiomatic way what we actually mean by an $n$-slot comb, but rather, we have borrowed intuition from the previous sections for the short discussion above. A proper, axiomatic definition that `builds up' the set of valid quantum combs can be given recursively:

{\hypersetup{citecolor=white, linkcolor=white}
\begin{myDefinition}{$\boldsymbol{n}$-Slot Quantum Combs---Axiomatic Approach [see Fig.~\ref{fig::axiomatichoqos-qc-def-axiomaticcombs}]}[def::axiomatichoqos-qc-def-axiomaticcombs]
A $0$-slot quantum comb is a quantum channel. For arbitrary $n \in \mathds{N}$, an $n$-slot quantum comb $\mathsf{T}_{n+1^\inp:0^\out}\geq 0$ is recursively defined as a HOQO that maps $(n-1)$-slot quantum combs onto quantum channels, i.e., for all valid $(n-1)$-slot quantum combs $\breve{\mathsf{T}}_{n^\out:1^\inp}$, we have that $\mathsf{T}_{n+1^\inp:0^\out} \star \breve{\mathsf{T}}_{n^\out:1^\inp} =: \mathsf{C}_{n+1^\inp 0^\out}$ is the Choi matrix of a quantum channel $\mathcal{C}:\Lscr(\Hscr_{0^\out}) \rightarrow \Lscr(\Hscr_{n+1^\inp})$. 
\end{myDefinition}}


\noindent This definition establishes a natural axiomatic framework for characterising \textit{deterministic} quantum combs: \textit{Any} valid comb within the hierarchy defined by the above criteria is regarded as deterministic, as it is constructed from quantum channels, which are themselves deterministic objects. Conversely, any HOQO that lies outside this hierarchy—i.e., outside the set of objects specified by the definition—is considered non-deterministic (or may result in supernormalised `probabilities’). This concept of `deterministic’ naturally extends to less constrained scenarios, such as the process matrix framework, where all objects that map pairs of CPTP maps to unit probability are classified as deterministic. 

While \textit{a priori} seemingly different in spirit, this definition agrees well with that of Eq.~\eqref{eq::axiomatichoqos-qc-def-axiomaticcombs}: Naturally, if a HOQO $\mathsf{T}_{n+1^\inp:0^\out}$ satisfies Eq.~\eqref{eq::axiomatichoqos-qc-def-axiomaticcombs} for all $k<n$, then it also satisfies the above definition [set $k=n-1$ in Eq.~\eqref{eq::axiomatichoqos-qc-def-axiomaticcombs}]. For the converse direction, let us assume for the moment that a tensor product of quantum combs remains a valid comb, i.e., $\mathsf{T}_{n^\out:1^\inp} = \breve{\mathsf{T}}^{(1)}_{n^\out:k^\out} \otimes \breve{\mathsf{T}}^{(2)}_{k^\inp:1^\inp}$ is a valid $(n-1)$-slot comb whenever $\breve{\mathsf{T}}^{(1)}_{n^\out:k^\out}$ and $\breve{\mathsf{T}}^{(2)}_{k^\inp:1^\inp}$ are valid $n-k-1$ and $(k-1)$-slot combs, respectively.\footnote{When taking the tensor product of two quantum combs, subtleties with respect to the temporal ordering of the resulting comb arise (see Ref.~\cite{Chiribella_2009}). Here, we do not discuss these issues in detail and assume that the times the resulting comb is defined on are given by the labels of the involved Hilbert spaces.} It then follows that $\breve{\mathsf{T}}_{n+1^\inp:0^\out} \star \breve{\mathsf{T}}^{(1)}_{n-1^\inp:k^\out}$ maps any $(k-1)$-slot comb to a quantum channel and is thus a valid comb itself
\begin{align}
    \mathsf{C}_{n+1^\inp 0^\out} = \mathsf{T}_{n+1^\inp:0^\out} \star \breve{\mathsf{T}}_{n^\out:1^\inp} = \mathsf{T}_{n^\inp:0^\out} \star ( \breve{\mathsf{T}}^{(1)}_{n^\out:k^\out} \otimes \breve{\mathsf{T}}^{(2)}_{k^\inp:1^\inp}) = (\mathsf{T}_{n^\inp:0^\out} \star \breve{\mathsf{T}}^{(1)}_{n^\out:k^\out}) \star \breve{\mathsf{T}}^{(2)}_{k^\inp:1^\inp}.
\end{align}
Since this holds for all $k$ and $(n-k-1)$-slot combs, Def.~\ref{def::axiomatichoqos-qc-def-axiomaticcombs} of $n$-slot combs coincides with Eq.~\eqref{eq::axiomatichoqos-qc-def-axiomaticcombs}. Importantly, this equivalence hinges on the assumption that the space of valid quantum combs (defined according to Def.~\ref{def::axiomatichoqos-qc-def-axiomaticcombs}) is stable under tensor product composition. As we will see below, \textit{all} $n$-slot combs that satisfy Def.~\ref{def::axiomatichoqos-qc-def-axiomaticcombs} are representable by a quantum circuit, and tensor products of quantum combs are thus still valid quantum combs, justifying this assumption. 

\vspace{0.25cm}\noindent\textbf{\textul{Hierarchy of Causality Constraints on Quantum Combs.}} With this axiomatic motivation of quantum combs in hand, we can now analyse whether or not it coincides with the constructive derivation of quantum combs presented in Sec.~\ref{subsec::timeorderedquantumprocesses}. This will be done in two steps: First, we will show that a matrix $\mathsf{T}_{n+1^\inp:0^\out}$ is a quantum comb according to Def.~\ref{def::axiomatichoqos-qc-def-axiomaticcombs} iff it satisfies the hierarchy of trace conditions of Def.~\ref{def::toqp-def-quantumcombs}, i.e., the trace conditions that follow from the constructive approach to quantum combs. Subsequently---similarly to the case of channels---we will show that every comb satisfying the trace conditions of Def.~\ref{def::toqp-def-quantumcombs} can be understood as stemming from a quantum circuit that only contains isometries and partial traces, i.e., it possesses a Stinespring dilation. Let us phrase the first step as a Proposition:

{\hypersetup{citecolor=white}
\begin{myDefinition*}{Properties of $\boldsymbol{n}$-Slot Quantum Combs~\cite{Chiribella_2009}}{}
A matrix $\mathsf{T}_{n+1^\inp:0^\out}\geq 0$ is the Choi state of an $n$-slot quantum comb according to Def.~\ref{def::axiomatichoqos-qc-def-axiomaticcombs} iff it satisfies the hierarchy of trace conditions in Def.~\ref{def::toqp-def-quantumcombs}.
\end{myDefinition*}
}

\noindent For the proof, first assume that $\mathsf{T}_{n+1^\inp:0^\out}\geq 0$ and $\breve{\mathsf{T}}_{n^\out:1^\inp}\geq 0$ are two matrices satisfying the trace conditions of Def.~\ref{def::toqp-def-quantumcombs}, i.e.,  
\begin{align}
\label{eq::axiomatichoqos-qc-traceproof1}
&\ptr{n+1^{\inp}}{\mathsf{T}_{n+1^\inp:0^\out}} = \ident_{n^\out} \otimes \mathsf{T}_{n^\inp:0^\out}, \dots, \ptr{2^\inp}{\mathsf{T}_{2^\inp:0^\out}} = \ident_{1^\out} \otimes \mathsf{T}_{1^\inp:0^\out}, \ptr{1^\inp}{\mathsf{T}_{1^\inp:0^\out}} = \ident_{0^\out},\\
\label{eq::axiomatichoqos-qc-traceproof2}
\text{and} \quad & \ptr{n^{\out}}{\breve{\mathsf{T}}_{n^\out:1^\inp}} = \ident_{n^\inp} \otimes \breve{\mathsf{T}}_{n-1^\out:1^\inp}, \dots, \ptr{2^\out}{\breve{\mathsf{T}}_{2^\out:1^\inp}} = \ident_{2^\inp} \otimes \breve{\mathsf{T}}_{1^\out:1^\inp}, \ptr{1^\out}{\breve{\mathsf{T}}_{1^\out:1^\inp}} = \ident_{1^\inp}.
\end{align}
We reiterate that the role of spaces is exchanged for the two combs $\mathsf{T}_{n+1^\inp:0^\out}$ and $\breve{\mathsf{T}}_{n^\out:1^\inp}$, i.e., inputs of $\mathsf{T}_{n+1^\inp:0^\out}$ are identified with outputs of $\breve{\mathsf{T}}_{n^\out:1^\inp}\geq 0$ and vice versa; hence the different trace conditions on the two combs. By invoking the conditions~\eqref{eq::axiomatichoqos-qc-traceproof1} and~\eqref{eq::axiomatichoqos-qc-traceproof2} alternatingly, we can now show that $\mathsf{T}_{n+1^\inp:0^\out} \star \breve{\mathsf{T}}_{n^\out:1^\inp} =: \mathsf{C}_{n+1^\inp0^\out}$ is indeed the Choi matrix of a channel, i.e, it satisfies $\ptr{n+1^\inp}{\mathsf{C}_{n+1^\inp0^\out}} = \ident_{0^\out}$, as follows
\begin{align}
\ptr{n+1^\inp}{\mathsf{C}_{n+1^\inp0^\out}} &= \ptr{n+1^\inp}{\mathsf{T}_{n+1^\inp:0^\out}} \star \breve{\mathsf{T}}_{n^\out:1^\inp} = (\ident_{n^\out} \otimes \mathsf{T}_{n^\inp:0^\out}) \star \breve{\mathsf{T}}_{n^\out:1^\inp} = \mathsf{T}_{n^\inp:0^\out} \star \ptr{n^\out}{\breve{\mathsf{T}}_{n^\out:1^\inp}} \notag\\
&= \cdots = (\ident_{1^\out} \otimes \mathsf{T}_{1^\inp:0^\out}) \star \breve{\mathsf{T}}_{1^\out:1^\inp} = \mathsf{T}_{1^\inp:0^\out} \star \ptr{1^\out}{\breve{\mathsf{T}}_{1^\out:1^\inp}} = \mathsf{T}_{1^\inp:0^\out} \star \ident_{1^\inp} = \ident_{0^\out}.
\end{align}
That is, if $n$-combs satisfy the trace conditions of Def.~\ref{def::toqp-def-quantumcombs} for all $n\in \mathds{N}_0$, then $n$-combs map $(n-1)$-slot combs to quantum channels, thereby satisfying Def.~\ref{def::axiomatichoqos-qc-def-axiomaticcombs}. 

The converse direction is shown by induction. To this end, let us assume that $(k-1)$-slot combs coincide exactly with the set of matrices $\breve{\mathsf{T}}_{k^\out:1^\inp} \geq 0$ that satisfy the trace conditions of Def.~\ref{def::toqp-def-quantumcombs}, i.e., the conditions of Eq.~\eqref{eq::axiomatichoqos-qc-traceproof2}. This implies that $\breve{\mathsf{T}}_{k^\out:1^\inp} = \tfrac{1}{d_{\mathbf{k^\out}}} \ident_{k^\out:1^\inp} + \varepsilon \,\sigma_{k^\out}^\mathrm{T} \otimes \mathsf{X}^\mathrm{T}_{k^\inp:1^\inp}$---where $\sigma_{k^\out}$ is some traceless matrix, $\mathsf{X}_{k^\inp:1^\inp}$ is arbitrary, and $d_{\mathbf{k^\out}}$ is the total dimension of the spaces labelled by $\out$---is a proper $(k-1)$-slot comb for sufficiently small $\varepsilon$ (such that $\breve{\mathsf{T}}_{k^\out:1^\inp}\geq 0$). If a $k$-slot quantum comb $\mathsf{T}_{k+1^\inp:0^\out}$ satisfies Def.~\ref{def::axiomatichoqos-qc-def-axiomaticcombs}, then
\begin{align}
\ptr{k+1^\inp}{\mathsf{T}_{k+1^\inp:0^\out} \star \breve{\mathsf{T}}_{k^\out:1^\inp}} = \frac{1}{d_{\mathbf{k^\out}}} \ptr{k+1^{\inp}:1^\inp}{\mathsf{T}_{k+1^\inp:0^\out}} +  \varepsilon \, \ptr{k^{\out}:1^\inp}{\ptr{k+1^{\inp}:1^\inp}{\mathsf{T}_{k+1^\inp:0^\out}} (\sigma_{k^\out} \otimes \mathsf{X}_{k^\inp:1^\inp})} = \ident_{0^\out}.
\end{align}
This must hold for all traceless matrices $\sigma_{k^\out}$, all $\mathsf{X}_{k^\inp:1^\inp}$, and all $\varepsilon$ sufficiently small. Since the r.h.s.\ of the above equation is independent of all of these terms, this implies that $\ptr{k+1^{\inp}:1^\inp}{\mathsf{T}_{k+1^\inp:0^\out}} = \ident_{k^\out} \otimes \mathsf{T}_{k^\inp:0^\out}$. To show the full hierarchy of trace conditions on $\mathsf{T}_{k+1^\inp:0^\out}$, we employ the fact that---due to Eq.~\eqref{eq::axiomatichoqos-qc-traceproof2}---we have that $\breve{\mathsf{T}}_{k^\out:1^\inp} = \tfrac{1}{d_{\mathbf{k^\out}}} \ident_{k^\out:1^\inp} + \varepsilon \, \ident_{k^\out:k^\inp} \otimes \sigma_{k-1^\out}^\mathrm{T} \otimes \mathsf{X}^\mathrm{T}_{k-1^\inp:1^\inp}$ is also a proper $(k-1)$-slot comb for all traceless $\sigma_{k-1^\out}$, all $\mathsf{X}^\mathrm{T}_{k-1^\inp:1^\inp}$, and sufficiently small $\varepsilon$. Repeating the above argument with this choice of $\breve{\mathsf{T}}_{k^\out:1^\inp}$, we obtain $\ptr{k^\inp}{\mathsf{T}_{k^\inp:0^\out}} = \ident_{k-1^\out} \otimes {\mathsf{T}_{k-^\inp:0^\out}}$. Continuing this procedure then yields the entire hierarchy of trace conditions on $\mathsf{T}_{k+1^\inp:0^\out}$. Finally, we remark that $0$-slot combs---quantum channels---satisfy the trace conditions of Def.~\ref{def::toqp-def-quantumcombs}, thus providing a suitable starting point for the induction. 

\vspace{0.25cm} \noindent \textbf{\textul{Stinespring Dilation of Quantum Combs.}} To show that the axiomatic and constructive approach to quantum combs coincide, it finally remains to demonstrate that every comb $\mathsf{T}_{n
+1^\inp:0^\out}$ that satisfies the trace conditions of Eq.~\eqref{eq::axiomatichoqos-qc-traceproof1} can be understood as stemming from a quantum circuit; in other words, we must derive a Stinespring dilation for quantum combs~\cite{Chiribella_2009, Bisio_2011}. 

\begin{figure}[t]
\centering
\vspace{0.5em}
\includegraphics[width=0.8\linewidth]{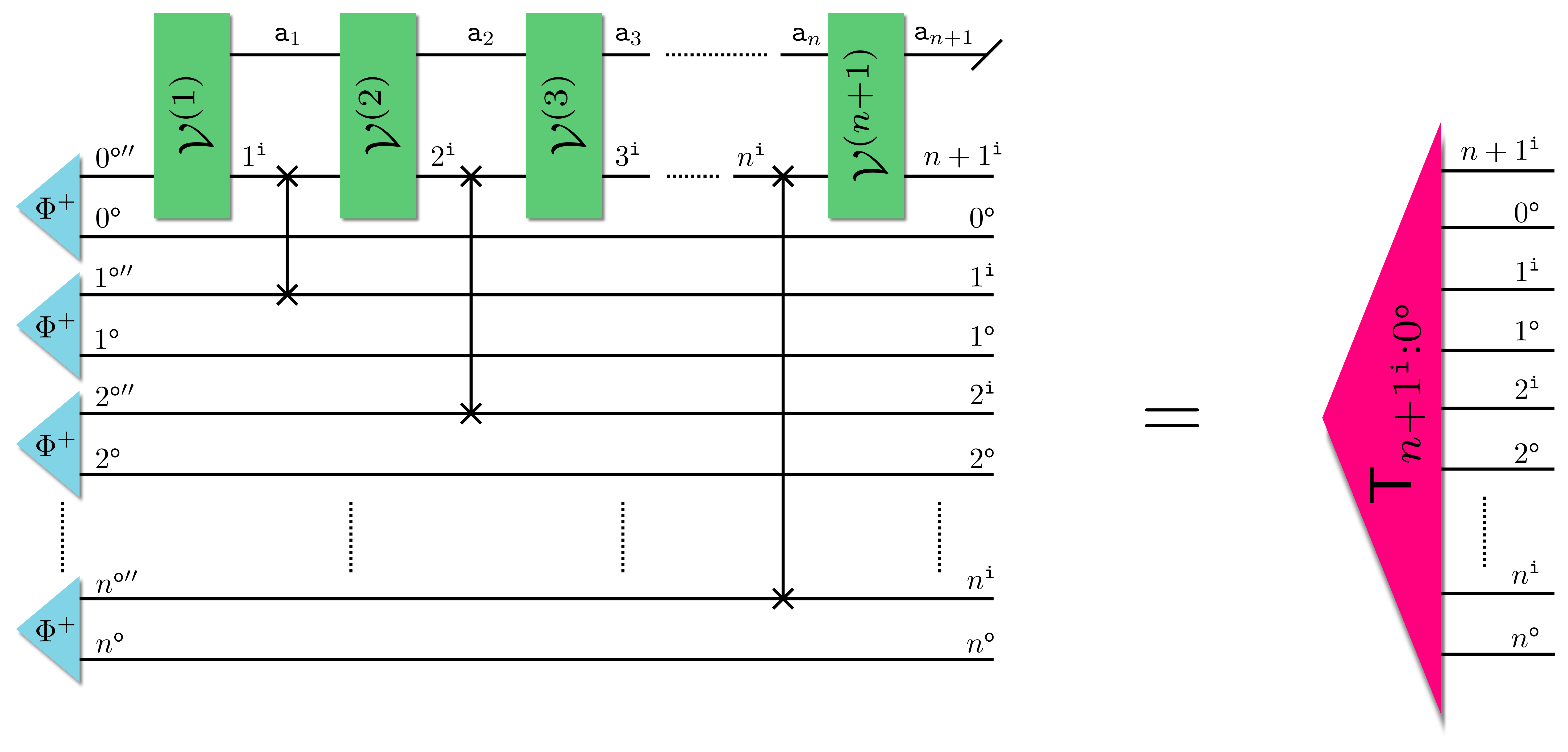}
\caption{\textbf{Stinespring Dilation of $\boldsymbol{n}$-Slot Quantum Combs.} The matrix $\mathsf{T}_{n+1^\inp:0^\out}$ corresponding to an $n$-slot comb can be understood as the Choi matrix of a circuit only consisting of isometries $\mathcal{V}^{(k+1)}[\bullet] = V^{(k+1)} \bullet V^{(k+1)\dagger}$ [see Eq.~\eqref{eq::axiomatichoqos-qc-combisometryconstruction}].} 
\label{fig::axiomatichoqos-qc-combstinespring}
\end{figure}

To arrive at such a representation theorem of a quantum comb $\mathsf{T}_{n+1^\inp:0^\out}$, i.e., to derive the $n$ isometries that make up the circuit that leads to $\mathsf{T}_{n+1^\inp:0^\out}$, we follow similar logic as for the case of quantum channels discussed in the previous section, albeit with a slightly more cumbersome notation. Let $\mathsf{T}_{n+1^\inp:0^\out} \in \Lscr(\Hscr_{{n+1^\inp}} \otimes \Hscr_{{{n}^\out}} \otimes \cdots \otimes \Hscr_{{0^\out}})$ be a quantum comb that satisfies the trace conditions of Eq~\eqref{eq::axiomatichoqos-qc-traceproof1}. Similar to the channel case, a purification of $\mathsf{T}_{n^\inp:1^\inp}$ can be constructed as follows
\begin{align}
\label{eq::axiomatichoqos-qc-combpurification}
    \ket{\widetilde{\mathsf{T}}_{n+1^\inp:0^\out\aux_{n+1}}} := \sqrt{\mathsf{T}}^\dagger_{n+1^\inp:0^\out} \ket{\Phi^+_{n+1^\inp n+1^{\inp\prime}}} \ket{\Phi^+_{{{n}^\out} {n}^{\out\prime}}} \cdots \ket{\Phi^+_{0^\out 0^{\out\prime}}} \in \Hscr_{n+1^\inp} \otimes \cdots \otimes \Hscr_{{0}^\out} \otimes \Hscr_{\aux_{n+1}}, 
\end{align}
where $\Hscr_{\aux_{n+1}} = \text{supp}(\mathsf{T}_{n+1^\inp:0^\out}) \subseteq \Hscr_{n+1^{\inp\prime}} \otimes \Hscr_{{n}^{\out\prime}} \otimes \cdots \otimes \Hscr_{0^{\out \prime}}$. From the trace conditions of Eq.~\eqref{eq::axiomatichoqos-qc-traceproof1}, have that $\ptr{n+1^\inp}{\mathsf{T}_{n+1^\inp:0^\out}} = \ident_{{n}^\out} \otimes \mathsf{T}_{{n}^\inp:0^\out}$, such that $\widetilde{\mathsf{T}}_{n+1^\inp:0^\out\aux_{n+1}} := \ketbra{\widetilde{\mathsf{T}}_{n+1^\inp:0^\out\aux_{n+1}}}{\widetilde{\mathsf{T}}_{n+1^\inp:0^\out\aux_{n+1}}}$ is a purification of $\ident_{{n}^\out} \otimes \mathsf{T}_{{n}^\inp:0^\out}$. On the other hand, $\ket{\Phi^+_{{n}^\out {n}^{\out\prime\prime}}} \otimes \ket{\widetilde{\mathsf{T}}_{{n}^\inp:0^\out \aux_{n}}}$ with 
\begin{align}
\ket{\widetilde{\mathsf{T}}_{{n}^\inp:0^\out \aux_{n}}} := \sqrt{\mathsf{T}}^\dagger_{{n}^\inp:0^\out} \ket{\Phi^+_{{n}^\inp {n}^{\inp\prime\prime}}} \ket{\Phi^+_{{n-1}^\out {n-1}^{\out\prime\prime}}} \cdots \ket{\Phi^+_{0^\out 0^{\out\prime\prime}}} \in \Hscr_{{n}^\inp} \otimes \cdots \otimes \Hscr_{{0}^\out} \otimes \Hscr_{\aux_{n}} 
\end{align}
provides another (minimal) purification of $\ident_{{n}^\out} \otimes \mathsf{T}_{{n}^\inp:0^\out}$ with $\Hscr_{\aux_{n}} = \text{supp}(\mathsf{T}_{n^\inp:0^\out}) \subseteq \Hscr_{{n}^{\inp\prime\prime}} \otimes \Hscr_{{n-1}^{\out\prime\prime}} \otimes \cdots \otimes \Hscr_{{0}^{\out \prime \prime}}$. Consequently, there exists an isometry $V^{(n+1)}: \Hscr_{{n}^{\out\prime\prime}} \otimes \Hscr_{\aux_{n}} \rightarrow \Hscr_{{n+1}^{\inp}} \otimes \Hscr_{\aux_{n+1}}$ that acts only on the purification spaces and connects both purifications via
\begin{align}
\label{eq::axiomatichoqos-qc-combpurificationisometry}
    \ket{\widetilde{\mathsf{T}}_{n+1^\inp:0^\out\aux_{n+1}}} = V^{(n+1)}\ket{\Phi^+_{{n}^\out {n}^{\out\prime\prime}}}  \ket{\widetilde{\mathsf{T}}_{{n}^\inp:0^\out \aux_{n}}}       
\end{align}
We can now run this argument again for $\ket{\widetilde{\mathsf{T}}_{{n}^\inp:0^\out \aux_{n}}}$, $\ket{\widetilde{\mathsf{T}}_{{n-1}^\inp:0^\out \aux_{n-1}}}, \hdots ,$ yielding a set of $n+1$ isometries $\{V^{(k+1)}: \Hscr_{{k}^{\out \prime\prime}} \otimes \Hscr_{\aux_{k}} \rightarrow \Hscr_{k+1^\inp} \otimes \Hscr_{\aux_{k+1}}\}_{k=0}^{n}$ that we can use to express $\ket{\widetilde{\mathsf{T}}_{n+1^\inp:0^\out\aux_{n+1}}}$ as
\begin{align}
\label{eq::axiomatichoqos-qc-combisometryconstruction}
\ket{\widetilde{\mathsf{T}}_{n+1^\inp:0^\out\aux_{n+1}}} = V^{(n+1)} \ket{\Phi^+_{{n}^\out {n}^{\out\prime\prime}}} V^{(n)} \ket{\Phi^+_{{n-1}^\out {n-1}^{\out\prime\prime}}} \cdots V^{(2)}\ket{\Phi^+_{{1}^\out {1}^{\out\prime\prime}}}V^{(1)}\ket{\Phi^+_{{0}^\out {0}^{\out\prime\prime}}}.
\end{align}
Since $\mathsf{T}_{n+1^\inp:0^\out} = \ptr{\aux_{n+1}}{\widetilde{\mathsf{T}}_{n+1^\inp:0^\out\aux_{n+1}}}$, we see that the $n$-comb $\mathsf{T}_{n+1^\inp:0^\out}$ indeed stems from a sequence of isometries acting on halves of maximally entangled states followed by a partial trace over the ancillary space $\Hscr_{\aux_{n+1}}$ [see Fig.~\ref{fig::axiomatichoqos-qc-combstinespring}], i.e., there exists a quantum circuit that yields $\mathsf{T}_{n+1^\inp:0^\out}$.


\begin{figure}
    \centering
    \includegraphics[width=0.8\linewidth]{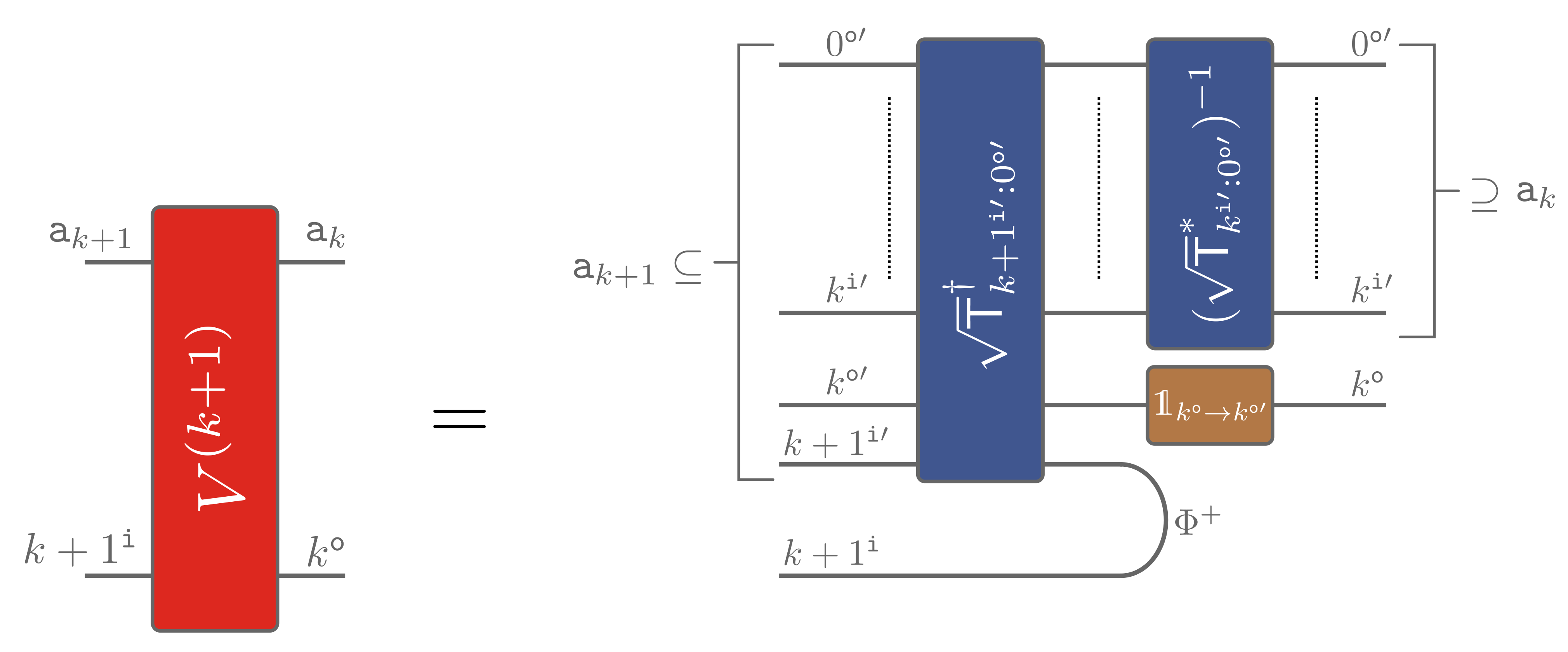}
    \caption{\textbf{Isometries for Constructing Quantum Combs.} The isometries $V^{(k+1)}: \Hscr_{k^\out} \otimes \Hscr_{\aux_k} \rightarrow \Hscr_{k+1^\inp} \otimes \Hscr_{\aux_{k+1}} $ making up the circuit corresponding to an $n$-slot comb $\mathsf{T}_{n+1^\inp:0^\out}$ can be built from the `reduced' combs $\{\mathsf{T}_{k^\inp:0^\out}\}_{k=1}^{n+1}$ [see Eq.~\eqref{eq::axiomatichoqos-qc-isometrycombsimple}]. }
    \label{fig::axiomatichoqos-qc-isometrycombsimple}
\end{figure}

As for the case of quantum channels, it remains to derive the explicit form of each $V^{(k+1)}$ and show that they are indeed isometries. To this end, we first note that Eq.~\eqref{eq::axiomatichoqos-qc-combpurificationisometry} implies 
\begin{align}
\ket{\widetilde{\mathsf{T}}_{k+1^\inp:0^\out\aux_{k+1}}} = V^{(k+1)} \ket{\Phi^+_{{k}^\out {k}^{\out\prime\prime}}}  \sqrt{\mathsf{T}}^\dagger_{{k}^\inp:0^\out} \ket{\Phi^+_{{k}^\inp {n}^{\inp\prime \prime}}} \ket{\Phi^+_{{k-1}^\out {k-1}^{\out\prime \prime}}} \cdots \ket{\Phi^+_{0^\out 0^{\out\prime\prime}}}.
\end{align}
As a result, we have 
\begin{align}
     V^{(k+1)}\ket{\Phi^+_{{k}^\out {k}^{\out\prime\prime}}} \ket{\Phi^+_{{k}^\inp {k}^{\inp\prime\prime}}} \ket{\Phi^+_{{k-1}^\out {k-1}^{\out\prime\prime}}} \cdots \ket{\Phi^+_{0^\out 0^{\out\prime\prime}}} = (\sqrt{\mathsf{T}}^{\dagger}_{{k}^\inp:0^\out})^{-1} \ket{\widetilde{\mathsf{T}}_{k+1^\inp:0^\out\aux_{n+1}}} ,
\end{align}
where $(\sqrt{\mathsf{T}}^{\dagger}_{{k}^\inp:1^\inp})^{-1} $ is the Moore-Penrose pseudoinverse of $\sqrt{\mathsf{T}}^\dagger_{{k}^\inp:1^\inp}$ and we have used that $ V^{(k+1)} \ket{\Phi^+_{{k}^\out {k}^{\out\prime\prime}}} $ and $\sqrt{\mathsf{T}}^\dagger_{{k}^\inp:0^\out} $ are defined on disjoint Hilbert spaces. Similar to the case of quantum channels [see Eq.~\eqref{eq::axiomatichoqos-choichannelisometrystinespring}], we can now directly read off the isometry $V^{(k+1)}$ by using the fact that $\braket{\Phi^+_{j^\xt j^{\xt\prime\prime}}|V^{(k+1)}|\Phi^+_{j^\xt j^{\xt\prime\prime}}} = V^{(k+1)}$ for all $j\leq k$ and $\xt \in \{\inp,\out\}$, which yields
\begin{align}
\label{eq::axiomatichoqos-qc-isometrycombexplicit}
    V^{(k+1)} = \bra{\Phi^+_{{k}^\out {k}^{\out\prime\prime}}} \bra{\Phi^+_{{k}^\inp {k}^{\inp\prime\prime}}} \bra{\Phi^+_{{k-1}^\out {n-1}^{\out\prime\prime}}} \cdots \bra{\Phi^+_{0^\out 0^{\out\prime\prime}}} (\sqrt{\mathsf{T}}^\dagger_{{k}^\inp:0^\out})^{-1} \ket{\widetilde{\mathsf{T}}_{k+1^\inp:0^\out\aux_{k+1}}}. 
\end{align}
To show that the set  $\{V^{(k+1)}\}$ comprises only isometries, we bring $V^{(k+1)}$ into a more manageable form, along the lines of Eq.~\eqref{eq::axiomatichoqos-choichannelisometrystinespringfinal} for the case of quantum channels. We first note that for any matrix $\mathsf{X}$, its partial transpose can be expressed as  
\begin{align}
\label{eq::axiomatichoqos-qc-transposeidentity}
    \braket{\Phi^+_{\ell^\texttt{x} \ell^{\texttt{x}\prime\prime}}|\mathsf{X}| \Phi^+_{\ell^\texttt{x} \ell^{\texttt{x}\prime}}} = \ident_{\ell^\texttt{x} \rightarrow \ell^{\texttt{x}\prime}} \mathsf{X}^{\mathrm{T}_{\ell^\texttt{x}}} \ident_{\ell^{\texttt{x}\prime \prime} \rightarrow \ell^{\texttt{x}}}, 
\end{align} 
where the extra matrices $\ident_{\ell^\texttt{x} \rightarrow \ell^{\texttt{x}\prime}}$ and $\ident_{\ell^{\texttt{x}\prime \prime} \rightarrow \ell^{\texttt{x}}}$ are merely required for proper labelling. Inserting this identity into Eq.~\eqref{eq::axiomatichoqos-qc-isometrycombexplicit} and employing the definition of $\ket{\widetilde{\mathsf{T}}_{k+1^\inp:0^\out\aux_{n+1}}}$ from Eq.~\eqref{eq::axiomatichoqos-qc-combpurification}, we obtain
\begin{align}\label{eqn::isomety_ncomb_final1}
    V^{(k+1)} &= \ident_{\mathbf{k^\out}\rightarrow \mathbf{k^{\out\prime}}} \left[(\sqrt{\mathsf{T}}^\dagger_{k^\inp:0^\out})^{-1} \sqrt{\mathsf{T}}^\dagger_{k+1^\inp:0^\out} \right]^{\mathrm{T}_{\mathbf{k^\out}}}\ket{\Phi^+_{k+1^\inp k+1^{\inp\prime}}} \ident_{\mathbf{k^{\out\prime\prime}} \rightarrow \mathbf{k^\out} } \notag \\
    &= \ident_{\mathbf{k^\out}\rightarrow \mathbf{k^{\out\prime}}} \braket{\Phi^+_{k+1^\inp k+1^{\inp\prime \prime}}|\left[(\sqrt{\mathsf{T}}^\dagger_{k^\inp:0^\out})^{-1} \sqrt{\mathsf{T}}^\dagger_{k+1^\inp:0^\out} \right]^{\mathrm{T}_{\mathbf{k^\out}}} |\Phi^+_{k+1^\inp k+1^{\inp\prime}}} \ket{\Phi^+_{k+1^\inp k+1^{\inp \prime\prime}}}\ident_{\mathbf{k^{\out\prime\prime}} \rightarrow \mathbf{k^\out} } \notag \\
    &=  \left[(\sqrt{\mathsf{T}}^\dagger_{k^{\inp\prime}:0^{\out\prime}})^{-1} \sqrt{\mathsf{T}}^\dagger_{k+1^{\inp\prime}:0^{\out\prime}} \right]^{\mathrm{T}}  \ket{\Phi^+_{k+1^{\inp} k+1^{\inp \prime}}}\ident_{\mathbf{k^{\out\prime\prime}} \rightarrow \mathbf{k^{\out\prime}} },
\end{align}
where we have set $\mathbf{k^\texttt{o}} := k^\texttt{o}:0^\out$ and have used that $\braket{\Phi^+_{k^\inp k^{\inp\inp''}}|\mathsf{X}|\Phi^+_{k^\inp k^{\inp\inp''}}} = \mathsf{X}$ in the second line; the last line follows from Eq.~\eqref{eq::axiomatichoqos-qc-transposeidentity} and relabelling terms appropriately. For improved clarity, we relabel the input spaces of $V^{(k+1)}$ such that they are labelled in the same way as the outputs of $V^{(k)}$; this notational change amounts to replacing `double primes' by `single primes' in Eq.~\eqref{eqn::isomety_ncomb_final1}. Finally, we strip spaces that are exclusive to $V^{(k+1)}$ of their primes, such that $V^{(k+1)}: \Hscr_{k^\out} \otimes \Hscr_{\aux_{k}} \rightarrow \Hscr_{k+1^{\inp}} \otimes \Hscr_{\aux_{k+1}}$, with $\Hscr_{\aux_{k}} \subseteq \Hscr_{k^{\inp\prime}} \otimes \cdots \otimes \Hscr_{0^{\out\prime}}$ and $\Hscr_{\aux_{k+1}} \subseteq \Hscr_{k+1^{\inp\prime}} \otimes \cdots \otimes \Hscr_{0^{\out\prime}}$. Performing these notational changes on Eq.~\eqref{eqn::isomety_ncomb_final1} yields (see Fig.~\ref{fig::axiomatichoqos-qc-isometrycombsimple})
\begin{align}
\label{eq::axiomatichoqos-qc-isometrycombsimple}
    V^{(k+1)} =  \left[\sqrt{\mathsf{T}}^*_{k+1^{\inp\prime}:0^{\out\prime}} (\sqrt{\mathsf{T}}^*_{k^{\inp\prime}:0^{\out\prime}})^{-1}  \right]  \ket{\Phi^+_{k+1^{\inp} k+1^{\inp \prime}}}\ident_{k^\out \rightarrow k^{\out\prime} }.
\end{align}


\noindent In this form, it is easy to see that $V^{(k+1)}$ is an isometry since
\begin{align}
  V^{(k+1)\dagger} V^{(k+1)} &=  \ident_{k^{\out\prime} \rightarrow k^{\out}}  \bra{\Phi^+_{k+1^\inp k+1^{\inp\prime}}} \left[ (\sqrt{\mathsf{T}}^*_{k^{\inp\prime}:0^{\out\prime}})^{-1} \mathsf{T}^*_{k+1^{\inp\prime}:0^{\out\prime}} (\sqrt{\mathsf{T}}^*_{k^{\inp\prime}:0^{\out\prime}})^{-1}  \right]  \ket{\Phi^+_{k+1^{\inp} k+1^{\inp \prime}}}\ident_{k^\out \rightarrow k^{\out\prime} } \notag \\
  &= \ident_{k^{\out\prime} \rightarrow k^{\out}}   \left[ (\sqrt{\mathsf{T}}^*_{k^{\inp\prime}:0^{\out\prime}})^{-1} \ptr{k+1^{\inp\prime}}{\mathsf{T}^*_{k+1^{\inp\prime}:0^{\out\prime}}} (\sqrt{\mathsf{T}}^*_{k^{\inp\prime}:0^{\out\prime}})^{-1}  \right]  \ident_{k^\out \rightarrow k^{\out\prime} } = \ident_{k^\out} \otimes \ident_{\aux_{k}},
\end{align}
where we have made use of $\ptr{k+1^{\inp\prime}}{\mathsf{T}^*_{k+1^{\inp\prime}:0^{\out\prime}}} = \ident_{k^{\out\prime}} \otimes \mathsf{T}^*_{k^{\inp\prime}:0^{\out\prime}}$ and $(\sqrt{\mathsf{T}}^*_{k^{\inp\prime}:0^{\out\prime}})^{-1} \sqrt{\mathsf{T}}^*_{k^{\inp\prime}:0^{\out\prime}} = \ident_{\aux_{k}}$. 

Similarly to the channel case, the required dilation space $\Hcal_{\aux_{k+1}}$ for the constructed isometries $V^{(k+1)}: \Hscr_{k^\out} \otimes \Hscr_{\aux_{k}} \rightarrow \Hscr_{k+1^{\inp}} \otimes \Hscr_{\aux_{k+1}}$, with $\Hscr_{\aux_{k+1}} = \text{supp}(\mathsf{T}_{{k+1}^\inp:0^\out}) = \text{supp}(\sqrt{\mathsf{T}}^*_{{k+1}^\inp:0^\out})$, is \textit{minimal}~\cite{Chiribella_2009, Bisio_2011}: \textit{Any} isometry $V^{(k+1)\prime}: \Hscr_{k^{\out\prime}} \otimes \Hscr_{\bux_{k}} \rightarrow \Hscr_{k+1^\inp} \otimes \Hscr_{\bux_{k+1}}$ with potentially smaller dilation space $\bux_{k+1}$ in the circuit that leads to $\mathsf{T}_{k+1^\inp:0^\out}$ connects purifications of $\mathsf{T}_{k+1^\inp:0^\out}$ and $\mathsf{T}_{k^\inp:0^\out}$ via
\begin{align}
    \ket{\widetilde{\mathsf{T}}_{k+1^\inp:0^\out\bux_{k+1}}} = V^{(k+1)\prime} \ket{\Phi^+_{k^\out k^{\out\prime}}}\ket{\widetilde{\mathsf{T}}_{k^\inp:0^\out \bux_{k}}}.
\end{align}
Since the purification $\ket{\widetilde{\mathsf{T}}_{k+1^\inp:0^\out\aux_{k+1}}}$ that we used in our construction was minimal, we see that $\text{dim}(\Hscr_{\bux_{k+1}}) < \text{dim}(\Hscr_{\aux_{k+1}})$ is not possible. The above argument culminates in the following:

{\hypersetup{citecolor=white}
\begin{myDefinition*}{Stinespring Dilation of Quantum Combs~\cite{Chiribella_2009, Bisio_2011}}{}
A positive semidefinite matrix $\mathsf{T}_{n+1^\inp:0^\out} \in \Lscr(\Hscr_{n+1^\inp} \otimes \cdots \Hscr_{0^\out})$ is the Choi state of a quantum circuit with isometries $\{V^{(k+1)}: \Hscr_{k^\out} \otimes \Hscr_{\aux_k} \rightarrow \Hscr_{k+1^\inp} \otimes \Hscr_{\aux_{k+1}}\}_{k=1}^{n}$ iff it satisfies the hierarchy of trace conditions of Def.~\ref{def::toqp-def-quantumcombs}. A minimal choice of isometries is given by 
\begin{align}
\label{eq::axiomatichoqos-qc-def-stinespringdilationcomb}
V^{(k+1)} =  \left[\sqrt{\mathsf{T}}^*_{k+1^{\inp\prime}:0^{\out\prime}} (\sqrt{\mathsf{T}}^*_{k^{\inp\prime}:0^{\out\prime}})^{-1}  \right]  \ket{\Phi^+_{k+1^{\inp} k+1^{\inp \prime}}}\ident_{k^\out \rightarrow k^{\out\prime} }.
\end{align}
\end{myDefinition*}}

\noindent As a result, every matrix $\mathsf{T}_{n+1^\inp:0^\out}\geq 0$ satisfying the trace conditions of Def.~\ref{def::toqp-def-quantumcombs} is the Choi matrix of a quantum circuit. Put differently, the axiomatic and constructive approach to quantum combs yield exactly the same set of valid processes.

We emphasise that this equivalence of the set of axiomatically defined HOQOs and the set of \mbox{HOQOs} stemming from quantum circuits \textit{only} holds if we demand that a proper HOQO has to map $\textit{all}$ HOQOs on `lower rungs' of the hierarchy to deterministic objects.\footnote{Alternatively, one can employ the requirement of `compatibility with remote connection'~\cite{Chiribella_2009}, which yields an axiomatic definition of HOQOs that coincides with that of Def.~\ref{def::axiomatichoqos-qc-def-axiomaticcombs}.} If, however, one only requests that a HOQO of interest maps \textit{particular} sets of HOQOs to other HOQOs---like, e.g., superchannels to superchannels---then quantum combs do \textit{not} constitute the maximum set of admissible quantum operations, and one readily obtains HOQOs that do cannot be represented by quantum circuits (or convex combinations therof). The discussion of such more general HOQOs is the subject of he following sections.


\subsubsection{Example: 1- and 2-Slot Quantum Combs}\hfill\\
\label{subsubsec::axiomatichoqos-qc-onetwoslotcombs}

\begin{figure}[t]
\centering
\subfigure[\textbf{$\boldsymbol{1}$-Slot Quantum Comb and Corresponding Circuit.}]
{
\includegraphics[scale=0.7]{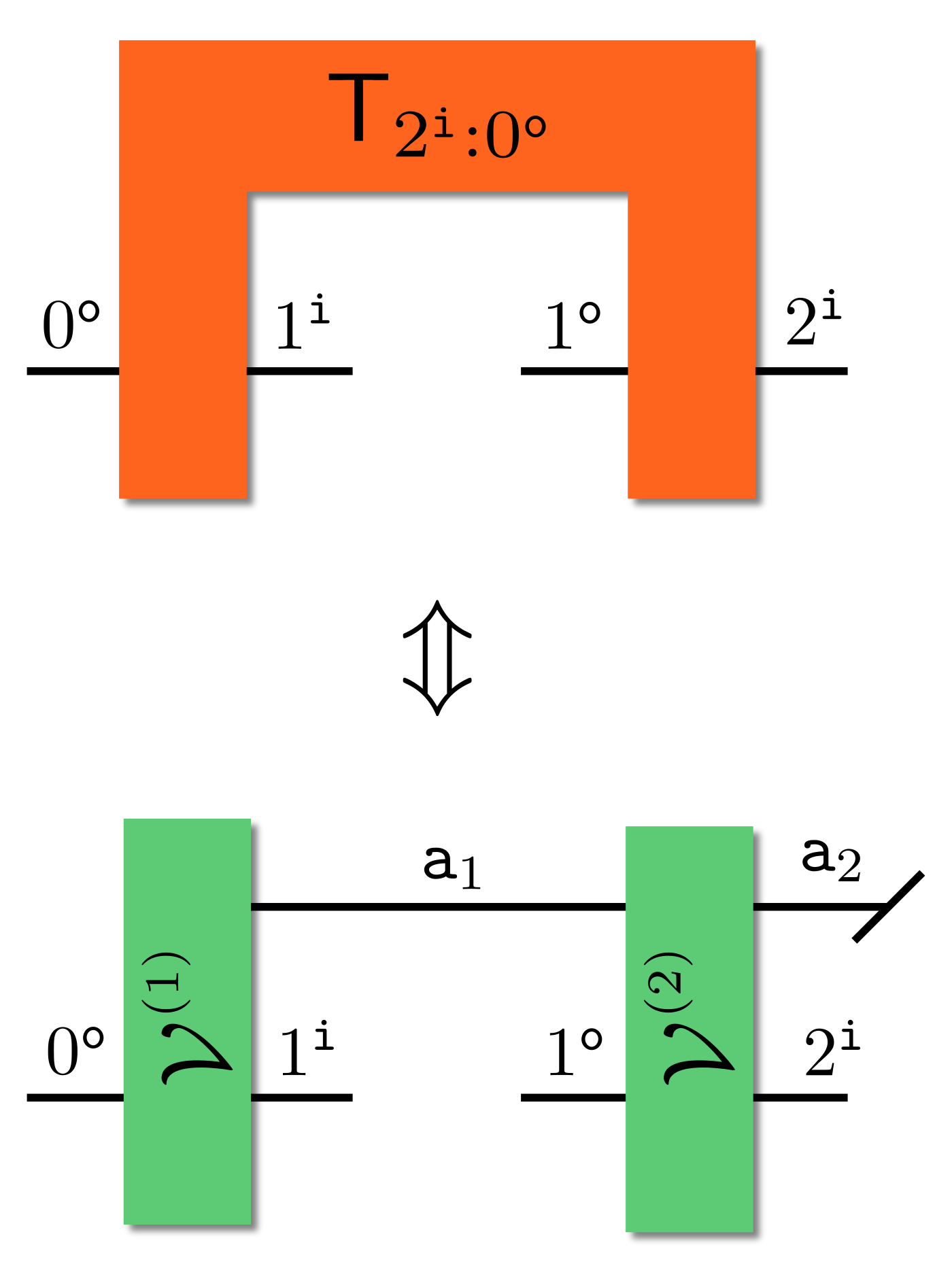}
\label{fig::axiomatichoqos-qc-1slot}
}\hspace{2.2cm}
\subfigure[\textbf{$\boldsymbol{2}$-Slot Quantum Comb and Corresponding Circuit.}]
{
\includegraphics[scale=0.7]{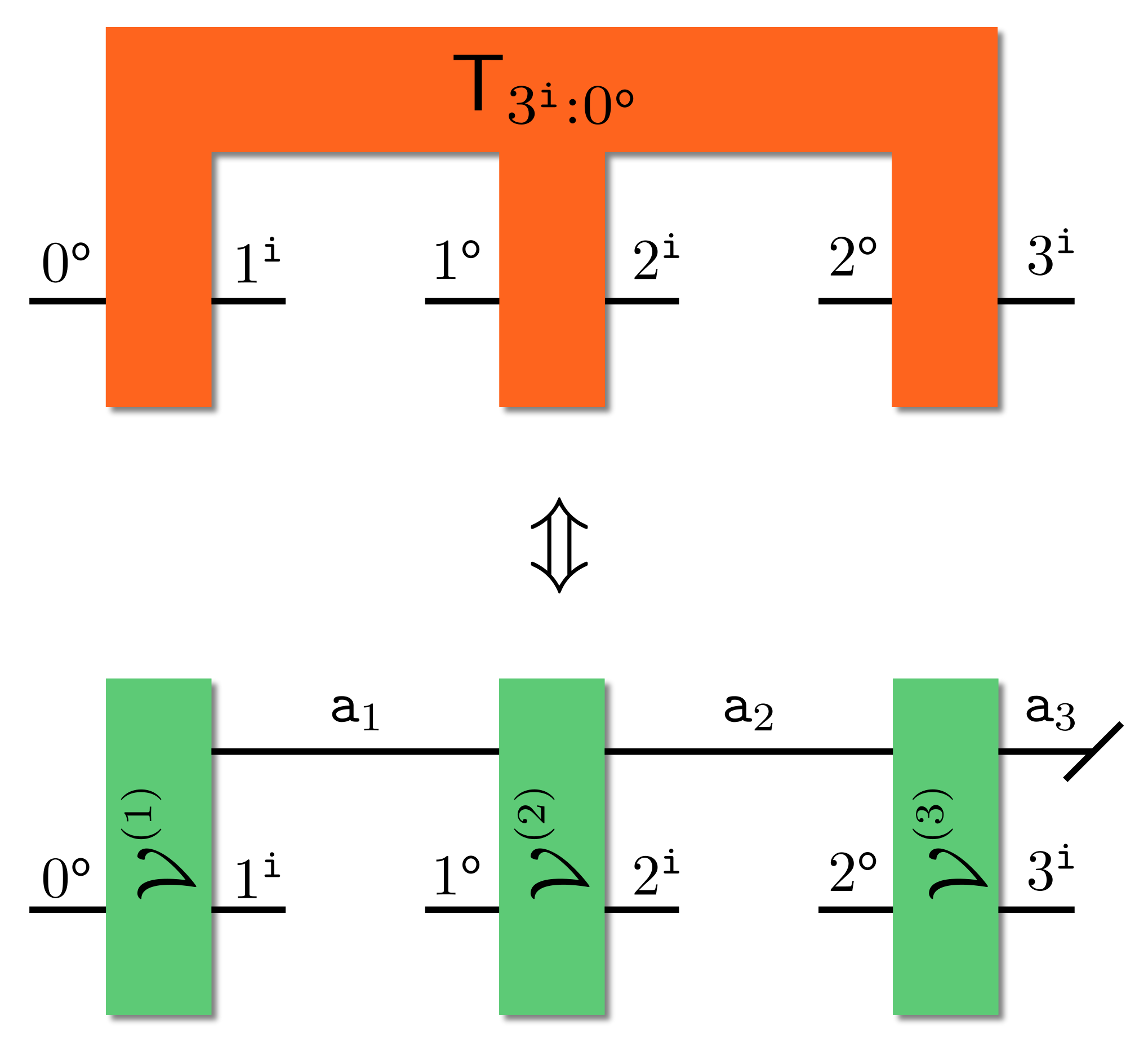}
\label{fig::axiomatichoqos-qc-2slot}
}
\caption{\textbf{Circuit Representation of $\boldsymbol{1}-$ and $\boldsymbol{2}$-Slot Quantum Combs.} \textbf{(a)} Any $1$-slot quantum comb can be obtained via two isometries $\mathcal{V}^{(1)}$ (called the `encoder') and $\mathcal{V}^{(2)}$ followed by a partial trace over the ancillary space (together, this final isometry and the partial trace form the `decoder'). \textbf{(b)} Analogously, every $2$-slot quantum comb can be represented by three isometries and a final partial trace.}
\label{fig::axiomatichoqos-qc-12slot}
\end{figure}

\noindent We now demonstrate the general considerations of the previous section for the simplest cases: $1$- and $2$-slot quantum combs. Respectively, these are HOQOs that map channels to channels and $1$-slot combs to channels (see Fig.~\ref{fig::axiomatichoqos-qc-12slot}). 


\newpage \vspace{0.25cm} \noindent \textbf{\textul{$\mathbf{1}$-Slot Quantum Combs.}} We begin by deriving the building blocks, i.e., the isometries $V^{(1)}: \Hscr_{0^\out} \rightarrow \Hscr_{1^\inp} \otimes \Hscr_{\aux_1}$ and $V^{(2)}: \Hscr_{1^\out} \otimes \Hscr_{\aux_1} \rightarrow \Hscr_{1^\inp} \otimes \Hscr_{\aux_2}$, for a $1$-slot comb $\mathsf{T}_{2^\inp:0^\out}$ that satisfies
\begin{align}
\ptr{2^\inp}{\mathsf{T}_{2^\inp:0^\out}} = \ident_{1^\out} \otimes \mathsf{T}_{1^\inp:0^\out} \quad \text{and} \quad \ptr{1^\inp}{\mathsf{T}_{1^\inp:0^\out}} = \ident_{0^\out}.
\end{align}
Employing Eqs.~\eqref{eq::axiomatichoqos-qc-def-stinespringdilationcomb}, we can directly read off $V^{(1)}$ and $V^{(2)}$ as 
\begin{align}
\label{eq::axiomatichoqos-qc-isometry1slot}
    V^{(2)} = \left[\sqrt{\mathsf{T}}^*_{2^{\inp\prime}:0^{\out\prime}} (\sqrt{\mathsf{T}}^*_{1^{\inp\prime}:0^{\out\prime}})^{-1}  \right]  \ket{\Phi^+_{2^{\inp} 2^{\inp \prime}}}\ident_{1^\out \rightarrow 1^{\out\prime}} \quad \text{and} \quad V^{(1)} = \left[\sqrt{\mathsf{T}}^*_{1^{\inp\prime}:0^{\out\prime}} \right]  \ket{\Phi^+_{1^{\inp} 1^{\inp \prime}}}\ident_{0^\out \rightarrow 0^{\out\prime}}. 
\end{align}
It is often useful to not only construct the isometries, but also their Choi matrices $\mathsf{V}^{(1)}_{0^\out1^\inp\aux_1} \in \Lscr(\Hscr_{0^\out} \otimes \Hscr_{1^\inp} \otimes \Hscr_{\aux_1})$ and $\mathsf{V}^{(2)}_{1^\out\aux_12^\inp\aux_2} \in \Lscr(\Hscr_{1^\out} \otimes \Hscr_{\aux_1} \otimes \Hscr_{2^\inp} \otimes \Hscr_{\aux_2})$, such that
\begin{align}
\label{eq::axiomatichoqos-qc-isometry1slotchoi}
    \mathsf{T}_{2^\inp:0^\out} = \ptr{\aux_2}{\mathsf{V}^{(1)}_{0^\out1^\inp\aux_1} \star \mathsf{V}^{(2)}_{1^\out\aux_12^\inp\aux_2}} = \mathsf{V}^{(1)}_{0^\out1^\inp\aux_1}\star \ptr{\aux_2}{\mathsf{V}^{(2)}_{1^\out\aux_12^\inp\aux_2}} =: \mathsf{E}_{0^\out 1^\inp \aux_1} \star \mathsf{D}_{\aux_1 1^\out 2^\inp}, 
\end{align}
where we have introduced the `encoder' $\mathsf{E}_{0^\out 1^\inp \aux_1}$ and `decoder' $\mathsf{D}_{\aux_1 1^\out 2^\inp}$ channels. Concretely, we have\footnote{In what follows, to avoid an additional relabelling step in the derivation, we tacitly suppress the terms $\ident_{1^\out \rightarrow 1^{\inp\prime}}$ and $\ident_{0^\out \rightarrow 0^{\inp\prime}}$ that respectively occur in the definitions of $V^{(2)}$ and $V^{(1)}$; this is merely done for notational convenience and not a conceptual step.} 
\begin{align}
    \mathsf{E}_{0^\out 1^\inp \aux_1} = \mathsf{Choi}[\mathcal{V}^{(1)}] = V^{(1)} \Phi^+_{0^\out 0^{\out\prime}} V^{(1)\dagger} = \sqrt{\mathsf{T}}^*_{1^{\inp\prime}:0^{\out\prime}} (\Phi^+_{0^\out 0^{\out\prime}} \otimes \Phi^+_{1^{\inp} 1^{\inp \prime}}) \sqrt{\mathsf{T}}^\mathrm{T}_{1^{\inp\prime}:0^{\out\prime}}.
\end{align}
We emphasise that $\ptr{\aux_1}{\mathsf{E}_{0^\out 1^\inp \aux_1}} =  \mathsf{T}_{1^{\inp}:0^{\out}}$ holds, i.e., $\mathsf{E}_{0^\out 1^\inp \aux_1} = \ketbra{\mathsf{T}'_{0^\out 1^\inp \aux_1}}{\mathsf{T}'_{0^\out 1^\inp \aux_1}}$ is a purification of $\mathsf{T}_{1^{\inp}:0^{\out}}$. 

The Choi matrix $\mathsf{V}^{(2)}$ follows analogously, with a slightly longer calculation to bring it into a nicer form. First, we note that 
\begin{align}
\label{eq::axiomatichoqos-qc-isometry1slotchoi-2}
    \mathsf{Choi}[\mathcal{V}^{(2)}] = V^{(2)} \left(\Phi^+_{0^{\out\prime} 0^{\out\prime \prime}} \otimes \Phi^+_{1^{\inp \prime} 1^{\inp\prime \prime}} \otimes \Phi^+_{1^{\out} 1^{\out\prime}}\right) V^{(2)\dagger} = \kketbra{V^{(2)}}{V^{(2)}},
\end{align}
where we require the double primes on the maximally entangled states for proper bookkeeping, such that $\kket{V^{(2)}} \in \Hscr_{1^\out}\otimes \Hscr_{\aux_1} \otimes \Hscr_{1^\inp} \otimes \Hscr_{\aux_2}$ with $\Hscr_{\aux_2} \subseteq \Hscr_{0^{\out\prime}} \otimes  \Hscr_{1^{\inp\prime}} \otimes  \Hscr_{1^{\out\prime}} \otimes  \Hscr_{2^{\inp\prime}}$ and $\Hscr_{\aux_1} \subseteq \Hscr_{0^{\out\prime\prime}} \otimes  \Hscr_{1^{\inp\prime\prime}}$. Since in Eq.~\eqref{eq::axiomatichoqos-qc-isometry1slotchoi} we trace over the final auxiliary space $\aux_2$, for the $1$-slot case we are not necessarily interested in the full Choi matrix $\mathsf{V}^{(2)}$ but rather $\ptr{\aux_2}{\mathsf{V}^{(2)}}$: \newpage
\begin{align}
    \ptr{\aux_2}{\mathsf{V}^{(2)}_{1^\out\aux_12^\inp\aux_2}} &= \ptr{{0^{\out\prime}} 1^{\inp\prime} 1^{\out\prime}2^{\inp\prime}}{\mathsf{V}^{(2)}_{1^\out\aux_12^\inp\aux_2}} \notag \\
    &=  \ptr{\aux_2}{\left[\sqrt{\mathsf{T}}^*_{2^{\inp\prime}:0^{\out\prime}} (\sqrt{\mathsf{T}}^*_{1^{\inp\prime}:0^{\out\prime}})^{-1}  \right]  \Phi^+_{0^{\out\prime} 0^{\out\prime \prime}} \otimes \Phi^+_{1^{\inp \prime} 1^{\inp\prime \prime}} \otimes \Phi^+_{1^{\out} 1^{\out\prime}}\otimes \Phi^+_{2^{\inp} 2^{\inp \prime}} \left[\sqrt{\mathsf{T}}^*_{2^{\inp\prime}:0^{\out\prime}} (\sqrt{\mathsf{T}}^*_{1^{\inp\prime}:0^{\out\prime}})^{-1}  \right]^\dagger} \notag \\
    &= \ptr{\aux_2}{\left[(\sqrt{\mathsf{T}}_{1^{\inp\prime}:0^{\out\prime}})^{-1} \mathsf{T}_{2^{\inp\prime}:0^{\out\prime}} (\sqrt{\mathsf{T}}_{1^{\inp\prime}:0^{\out\prime}})^{-1}  \right]  \$_{0^{\out\prime} 0^{\out\prime \prime}} \otimes \$_{1^{\inp \prime} 1^{\inp\prime \prime}} \otimes \$_{1^{\out} 1^{\out\prime}}\otimes \$_{2^{\inp} 2^{\inp \prime}} }.
\end{align}
Here, $\$_{k^{\xt} k^{\yt}} = (\Phi^+_{k^{\xt} k^{\yt}})^{\mathrm{T}_{k^{\xt}}} = (\Phi^+_{k^{\xt} k^{\yt}})^{\mathrm{T}_{k^{\yt}}}$ denotes the swap operator between the spaces $\Hscr_{k^{\xt}}$ and $\Hscr_{k^{\yt}}$ and we have used both the cyclicity and invariance under transposition of the trace. It is easy to see that `tracing against swap operators' amounts to a simple relabelling of spaces, i.e., 
\begin{align}
    \ptr{0^{\out\prime} 1^{\inp\prime} 1^{\out\prime}2^{\inp\prime}}{\mathsf{X}_{2^{\inp\prime}:0^{\out\prime}}(\$_{0^{\out\prime} 0^{\out\prime \prime}} \otimes \$_{1^{\inp \prime} 1^{\inp\prime \prime}} \otimes \$_{1^{\out} 1^{\out\prime}}\otimes \$_{2^{\inp} 2^{\inp \prime}})} = \ident_{ 1^{\inp} 0^\out \rightarrow 1^{\inp\prime \prime} 0^{\out\prime \prime}} \mathsf{X}_{2^{\inp}:0^{\out}} \ident_{1^{\inp\prime \prime} 0^{\out\prime \prime}\rightarrow 1^{\inp} 0^\out}.
\end{align}
Consequently, we obtain 
\begin{align}
    \mathsf{D}_{\aux_1 1^\out 2^\inp} = \ptr{\aux_2}{\mathsf{V}_{1^\out\aux_12^\inp\aux_2}} = \ident_{ 1^{\inp} 0^\out \rightarrow 1^{\inp\prime \prime} 0^{\out\prime \prime}}(\sqrt{\mathsf{T}}_{1^{\inp}:0^{\out}})^{-1} \mathsf{T}_{2^{\inp}:0^{\out}} (\sqrt{\mathsf{T}}_{1^{\inp}:0^{\out}})^{-1} \ident_{1^{\inp\prime \prime} 0^{\out\prime \prime}\rightarrow 1^{\inp} 0^\out}. 
\end{align}
Finally, we replace the double primes in the above equation by single primes so that the auxiliary spaces in $\mathsf{E}_{0^\out 1^\inp \aux_1}$ and $\mathsf{D}_{\aux_1 1^\out 2^\inp}$ are labelled appropriately, yielding:

\begin{myDefinition*}{Encoder and Decoder for $\boldsymbol{1}$-Slot Quantum Combs}{}
Any $1$-slot quantum comb $\mathsf{T}_{2^\inp:0^\out}$ can be obtained as a concatenation of two CPTP maps $\mathsf{E}_{0^\out 1^\inp \aux}$ (the encoder) and $\mathsf{D}_{\aux 1^\out 2^\inp}$ (the decoder), such that 
\begin{align}
\label{eq::axiomatichoqos-qc-1slotencoderdecoder}
   &\phantom{asdfasdfasdfasdf} \mathsf{T}_{2^\inp:0^\out} = \mathsf{E}_{0^\out 1^\inp \aux_1} \star \mathsf{D}_{\aux_1 1^\out 2^\inp} \\ 
 \text{with} \quad &\mathsf{E}_{0^\out 1^\inp \aux}  =  \sqrt{\mathsf{T}}^*_{1^{\inp\prime}:0^{\out\prime}} (\Phi^+_{0^\out 0^{\out\prime}} \otimes \Phi^+_{1^{\inp} 1^{\inp \prime}}) \sqrt{\mathsf{T}}^\mathrm{T}_{1^{\inp\prime}:0^{\out\prime}} \\
     \text{and} \quad & \mathsf{D}_{\aux 1^\out 2^\inp} =  \ident_{ 1^{\inp} 0^\out \rightarrow 1^{\inp\prime} 0^{\out\prime }}(\sqrt{\mathsf{T}}_{1^{\inp}:0^{\out}})^{-1} \mathsf{T}_{2^{\inp}:0^{\out}} (\sqrt{\mathsf{T}}_{1^{\inp}:0^{\out}})^{-1} \ident_{1^{\inp\prime} 0^{\out\prime}\rightarrow 1^{\inp} 0^\out} \, ,
\end{align}
where $\Hscr_{\aux} \subseteq \Hscr_{1^{\inp\prime}} \otimes \Hscr_{0^{\out\prime}}$.
\end{myDefinition*}

\noindent For easier manipulation, it is advantageous to incorporate the relabelling  $\ident_{1^{\inp\prime} 0^{\out\prime}\rightarrow 1^{\inp} 0^\out}$ into the notation, which can be done by making  all the spaces $\mathsf{T}_{2^{\inp}:0^{\out}} = \mathsf{T}_{2^{\inp}1^\out 1^\inp 0^{\out}}$ is defined on explicit. With this, we have 
\begin{align}
\label{eq::axiomatichoqos-qc-1slotencdecsimple}
\mathsf{E}_{0^\out 1^\inp \aux_1} = \sqrt{\mathsf{T}}^*_{1^{\inp\prime}0^{\out\prime}} (\Phi^+_{0^\out 0^{\out\prime}} \otimes \Phi^+_{1^{\inp} 1^{\inp \prime}}) \sqrt{\mathsf{T}}^\mathrm{T}_{1^{\inp\prime}0^{\out\prime}}  \quad \text{and} \quad \mathsf{D}_{\aux_1 1^\out 2^\inp} =  (\sqrt{\mathsf{T}}_{1^{\inp\prime}0^{\out\prime}})^{-1} \mathsf{T}_{2^{\inp}1^\out 1^{\inp\prime} 0^{\out\prime}} (\sqrt{\mathsf{T}}_{1^{\inp\prime}0^{\out \prime}})^{-1}.
\end{align}
It is now easy to prove Eq.~\eqref{eq::axiomatichoqos-qc-1slotencoderdecoder}---which holds by construction---explicitly, by direct insertion:
\begin{align}
\mathsf{E}_{0^\out 1^\inp \aux_1} \star \mathsf{D}_{\aux_1 1^\out 2^\inp} &= \ptr{\aux_1}{\mathsf{E}_{0^\out 1^\inp \aux_1} \mathsf{D}_{\aux_1 1^\out 2^\inp}^{\mathrm{T}_{\aux_1}} } \notag \\
&= \ptr{0^{\out\prime} 1^{\inp\prime}}{\sqrt{\mathsf{T}}^*_{1^{\inp\prime}0^{\out\prime}} (\Phi^+_{0^\out 0^{\out\prime}} \otimes \Phi^+_{1^{\inp} 1^{\inp \prime}}) \sqrt{\mathsf{T}}^\mathrm{T}_{1^{\inp\prime}0^{\out\prime}}(\sqrt{\mathsf{T}}^\mathrm{T}_{1^{\inp\prime}0^{\out\prime}})^{-1} \mathsf{T}^{\mathrm{T}_{0^{\out\prime}1^{\inp\prime}}}_{2^{\inp}1^\out 1^{\inp\prime} 0^{\out\prime}} (\sqrt{\mathsf{T}}^\mathrm{T}_{1^{\inp\prime}0^{\out \prime}})^{-1}} \notag \\
&=\ptr{0^{\out\prime} 1^{\inp\prime}}{(\Phi^+_{0^\out 0^{\out\prime}} \otimes \Phi^+_{1^{\inp} 1^{\inp \prime}}) \mathsf{T}^{\mathrm{T}_{0^{\out\prime}1^{\inp\prime}}}_{2^{\inp}1^\out 1^{\inp\prime} 0^{\out\prime}} } = \mathsf{T}_{2^{\inp}1^\out 1^{\inp} 0^{\out}}, 
\end{align}
where we have used the cyclicity of the trace and Hermiticity of $\mathsf{T}$, as well as the identity $\ptr{0^{\out\prime} 1^{\inp\prime}}{(\Phi^+_{0^\out 0^{\out\prime}} \otimes \Phi^+_{1^{\inp} 1^{\inp \prime}}) \mathsf{X}_{2^{\inp}1^\out 1^{\inp\prime} 0^{\out\prime}}} = \mathsf{X}_{2^{\inp}1^\out 1^{\inp} 0^{\out}}^{\mathrm{T}_{0^\out 1^\inp}}$. 

Importantly, for the $n$-slot case, the construction of the initial encoder $\mathsf{E}_{0^\out 1^\inp \aux_1}$ does not change. On the other hand, the final decoder $\mathsf{D}_{\aux_n n^\out n+1^\inp}$ follows from a simple generalisation of Eq.~\eqref{eq::axiomatichoqos-qc-1slotencdecsimple} as $\mathsf{D}_{\aux_n n^\out n+1^\inp} =  (\sqrt{\mathsf{T}}_{n^{\inp\prime}n-1^{\out\prime} \cdots 0^{\out\prime}})^{-1} \mathsf{T}_{n+1^{\inp}n^\out n-1^{\inp\prime} \cdots 0^{\out\prime}} (\sqrt{\mathsf{T}}_{n^{\inp\prime}n-1^{\out\prime} \cdots 0^{\out\prime}})^{-1}$. The only building block we have not yet discussed in detail in the Choi formalism are the `intermediate' isometries $\mathcal{V}^{(k)}$ for $k\neq 1$ and $k\neq {n+1}$. We will now address this point explicitly when discussing the $2$-slot case. 

\vspace{0.25cm} \noindent \textbf{\textul{$\mathbf{2}$-Slot Quantum Combs.}} For a comprehensive discussion of the $2$-slot case [see Fig.~\ref{fig::axiomatichoqos-qc-2slot}] with Choi matrix $\mathsf{T}_{3^\inp:0^\out} = \mathsf{T}_{3^\inp 2^\out 2^\inp 1^\out 1^\inp 0^\out}$, it remains to provide a concise form of the Choi matrix $\mathsf{V}^{(2)}_{1^\out\aux_1 2^\inp\aux_2}$ of $\mathcal{V}^{(2)}$. As mentioned, the respective encoder $\mathsf{E}_{0^\out 1^\inp \aux_1} = \sqrt{\mathsf{T}}^*_{1^{\inp\prime}0^{\out\prime}} (\Phi^+_{0^\out 0^{\out\prime}} \otimes \Phi^+_{1^{\inp} 1^{\inp \prime}}) \sqrt{\mathsf{T}}^\mathrm{T}_{1^{\inp\prime}0^{\out\prime}}$ and decoder $\mathsf{D}_{\aux_2 2^\out 3^\inp} = (\sqrt{\mathsf{T}}_{2^{\inp\prime}1^{\out\prime} 1^{\inp\prime} 0^{\out\prime}})^{-1} \mathsf{T}_{3^\inp 2^\out 2^{\inp\prime} 1^{\out\prime} 1^{\inp\prime} 0^{\out\prime}} (\sqrt{\mathsf{T}}_{2^{\inp\prime}1^{\out\prime} 1^{\inp\prime} 0^{\out\prime}})^{-1}$ immediately follow from the $1$-slot case. 

\noindent We can obtain a concise form of $\mathsf{V}^{(2)}_{1^\out \aux_1 2^\inp \aux_2} = \kketbra{V^{(2)}}{V^{(2)}}$ by inserting the definition of $V^{(2)}$ [see Eq.~\eqref{eq::axiomatichoqos-qc-isometry1slot}] into Eq.~\eqref{eq::axiomatichoqos-qc-isometry1slotchoi-2}:\footnote{We again suppress the term $\ident_{1^\out\rightarrow 1^{\out\prime}}$ to avoid additional relabelling.}
\begin{align}
 \kket{V^{(2)}} &=  \left[\sqrt{\mathsf{T}}^*_{2^{\inp\prime}1^{\out\prime} 1^{\inp\prime}0^{\out\prime}} (\sqrt{\mathsf{T}}^*_{1^{\inp\prime}0^{\out\prime}})^{-1}  \right]  \ket{\Phi^+_{0^{\out\prime} 0^{\out\prime \prime}}} \ket{\Phi^+_{1^{\inp \prime} 1^{\inp\prime \prime}}} \ket{\Phi^+_{1^{\out} 1^{\out\prime}}} \ket{\Phi^+_{2^{\inp} 2^{\inp \prime}}} \notag \\
 &= \left[(\sqrt{\mathsf{T}}_{1^{\inp\prime\prime}0^{\out\prime\prime}})^{-1} \sqrt{\mathsf{T}}^*_{2^{\inp\prime}1^{\out\prime} 1^{\inp\prime}0^{\out\prime}} \right]  \ket{\Phi^+_{0^{\out\prime} 0^{\out\prime \prime}}} \ket{\Phi^+_{1^{\inp \prime} 1^{\inp\prime \prime}}} \ket{\Phi^+_{1^{\out} 1^{\out\prime}}} \ket{\Phi^+_{2^{\inp} 2^{\inp \prime}}}, 
\end{align}
such that we obtain
\begin{align}
\notag
&\mathsf{V}^{(2)}_{1^\out \aux_1 2^\inp \aux_2} \\
\label{eq::axiomatichoqos-qc-2slotisometrychoi2}
&= \left[(\sqrt{\mathsf{T}}_{1^{\inp\prime\prime}0^{\out\prime\prime}})^{-1} \sqrt{\mathsf{T}}^*_{2^{\inp\prime}1^{\out\prime} 1^{\inp\prime}0^{\out\prime}} \right]  \Phi^+_{0^{\out\prime} 0^{\out\prime \prime}} \otimes \Phi^+_{1^{\inp \prime} 1^{\inp\prime \prime}} \otimes \Phi^+_{1^{\out} 1^{\out\prime}} \otimes \Phi^+_{2^{\inp}2^{\inp\prime}} \left[\sqrt{\mathsf{T}}^\mathrm{T}_{2^{\inp\prime}1^{\out\prime} 1^{\inp\prime}0^{\out\prime}} (\sqrt{\mathsf{T}}_{1^{\inp\prime\prime}0^{\out\prime\prime}})^{-1} \right],  
\end{align}
where $\Hscr_{\aux_1} \subseteq \Hscr_{0^{\out\prime\prime}} \otimes  \Hscr_{1^{\inp\prime\prime}}$ and $\Hscr_{\aux_2} \subseteq \Hscr_{0^{\out\prime}} \otimes  \Hscr_{1^{\inp\prime}} \otimes \Hscr_{1^{\out\prime}} \otimes \Hscr_{2^{\inp\prime}}$, and we have used $\mathsf{X}_{1^{\inp\prime}0^{\out\prime}} \ket{\Phi^+_{0^{\out\prime} 0^{\out\prime \prime}}} \ket{\Phi^+_{1^{\inp \prime} 1^{\inp\prime \prime}}} = \mathsf{X}_{1^{\inp\prime\prime}0^{\out\prime\prime}}^\mathrm{T} \ket{\Phi^+_{0^{\out\prime} 0^{\out\prime \prime}}} \ket{\Phi^+_{1^{\inp \prime} 1^{\inp\prime \prime}}}$ as well as the Hermiticity of $\mathsf{T}_{1^{\inp\prime\prime}0^{\out\prime\prime}}$. We emphasise the proximity to the form of the encoder $\mathsf{E}_{0^\out1^\inp\aux_1}$ in Eq.~\eqref{eq::axiomatichoqos-qc-1slotencoderdecoder}, which is a purification of $\mathsf{T}_{1^\inp0^\out}$. Here, re-writing Eq.~\eqref{eq::axiomatichoqos-qc-2slotisometrychoi2} we have 
\begin{align}
  \mathsf{V}^{(2)}_{1^\out \aux_1 2^\inp \aux_2} =  (\sqrt{\mathsf{T}}_{1^{\inp\prime\prime}0^{\out\prime\prime}})^{-1}  \ketbra{\mathsf{T}'_{0^{\out\prime\prime} 1^{\inp\prime\prime} 1^\out 2^\inp \aux_2}}{\mathsf{T}'_{0^{\out\prime\prime} 1^{\inp\prime\prime} 1^\out 2^\inp \aux_2}} (\sqrt{\mathsf{T}}_{1^{\inp\prime\prime}0^{\out\prime\prime}})^{-1},
\end{align}
where $\ptr{\aux_2}{\mathsf{T}'_{0^{\out\prime\prime} 1^{\inp\prime\prime} 1^\out 2^\inp \aux_2}} = \mathsf{T}_{0^{\out\prime\prime} 1^{\inp\prime\prime} 1^\out 2^\inp}$. If the spaces $\Hscr_{0^{\out\prime\prime}}$ and $\Hscr_{1^{\inp\prime\prime}}$ are trivial, then this form of $\mathsf{V}^{(2)}_{1^\out \aux_1 2^\inp \aux_2}$ coincides---up to the labels of the involved spaces---with that of $\mathsf{E}_{0^\out1^\inp\aux_1}$ in Eq.~\eqref{eq::axiomatichoqos-qc-1slotencoderdecoder}. In summary, we have:

\begin{myDefinition*}{Building Blocks for $\boldsymbol{2}$-Slot Quantum Combs}{}
Any $2$-slot quantum comb $\mathsf{T}_{3^\inp:0^\out} = \mathsf{T}_{3^\inp 2^\out 2^\inp 1^\out 1^\inp 0^\out}$ can be obtained as a concatenation of three CPTP maps $\mathsf{E}_{0^\out 1^\inp \aux_1}$, $\mathsf{V}^{(2)}_{1^\out \aux_1 2^\inp \aux_2}$ and $\mathsf{D}_{\aux_2 2^\out 3^\inp}$, such that 
\begin{align}
\label{eq::axiomatichoqos-qc-2slotencoderdecoder}
   &\phantom{asdfasdfasdfasdf}\mathsf{T}_{3^\inp 2^\out 2^\inp 1^\out 1^\inp 0^\out} = \mathsf{E}_{0^\out 1^\inp \aux_1} \star \mathsf{V}^{(2)}_{1^\out \aux_1 2^\inp \aux_2} \star \mathsf{D}_{\aux_2 2^\out 3^\inp}, \\ 
   \label{eq::axiomatichoqos-qc-2slotencdecmid}
 \text{with} \quad &\mathsf{E}_{0^\out 1^\inp \aux_1} = \sqrt{\mathsf{T}}^*_{1^{\inp\prime\prime}0^{\out\prime\prime}} (\Phi^+_{0^\out 0^{\out\prime\prime}} \otimes \Phi^+_{1^{\inp} 1^{\inp \prime\prime}}) \sqrt{\mathsf{T}}^\mathrm{T}_{1^{\inp\prime\prime}0^{\out\prime\prime}}, \\
 & \mathsf{D}_{\aux_2 2^\out 3^\inp} = (\sqrt{\mathsf{T}}_{2^{\inp\prime}1^{\out\prime} 1^{\inp\prime} 0^{\out\prime}})^{-1} \mathsf{T}_{3^\inp 2^\out 2^{\inp\prime} 1^{\out\prime} 1^{\inp\prime} 0^{\out\prime}} (\sqrt{\mathsf{T}}_{2^{\inp\prime}1^{\out\prime} 1^{\inp\prime} 0^{\out\prime}})^{-1},\\
\text{and} \quad &\mathsf{V}^{(2)}_{1^\out \aux_1 2^\inp \aux_2} = \left[(\sqrt{\mathsf{T}}_{1^{\inp\prime\prime}0^{\out\prime\prime}})^{-1} \sqrt{\mathsf{T}}^*_{2^{\inp\prime}1^{\out\prime} 1^{\inp\prime}0^{\out\prime}} \right]  \Phi^+_{0^{\out\prime} 0^{\out\prime \prime}} \otimes \Phi^+_{1^{\inp \prime} 1^{\inp\prime \prime}} \otimes \Phi^+_{1^{\out} 1^{\out\prime}} \otimes \Phi^+_{2^{\inp}2^{\inp\prime}} \times \notag \\
\phantom{\text{and} \quad} &\phantom{\mathsf{V}^{(2)}_{1^\out \aux_1 2^\inp \aux_2} =} \times \left[\sqrt{\mathsf{T}}^\mathrm{T}_{2^{\inp\prime}1^{\out\prime} 1^{\inp\prime}0^{\out\prime}} (\sqrt{\mathsf{T}}_{1^{\inp\prime\prime}0^{\out\prime\prime}})^{-1} \right],
\end{align}
where $\Hscr_{\aux_1} \subseteq \Hscr_{1^{\inp\prime\prime}} \otimes \Hscr_{0^{\out\prime\prime}}$ and $\Hscr_{\aux_2} \subseteq \Hscr_{0^{\out\prime}} \otimes \Hscr_{1^{\inp\prime}} \otimes \Hscr_{1^{\out\prime}} \otimes \Hscr_{2^{\inp\prime}}$.
\end{myDefinition*}

\noindent Note that, in order to make the auxiliary space $\aux_1$ match in the definition of $\mathsf{E}_{0^\out 1^\inp \aux_1}$ and $\mathsf{V}^{(2)}_{1^\out \aux_1 2^\inp \aux_2}$, we have changed single primes to double primes in Eq.~\eqref{eq::axiomatichoqos-qc-2slotencdecmid}. Similarly to the $1$-slot case, we can show Eq.~\eqref{eq::axiomatichoqos-qc-2slotencoderdecoder}---which holds by construction---explicitly:
\begin{align}
&\mathsf{E}_{0^\out 1^\inp \aux_1} \star \mathsf{V}^{(2)}_{1^\out \aux_1 2^\inp \aux_2} \star \mathsf{D}_{\aux_2 2^\out 3^\inp} \notag \\
&= \ptr{1^{\inp\prime\prime}0^{\out\prime\prime}}{(\$_{0^\out 0^{\out\prime\prime}} \otimes \$_{1^{\inp} 1^{\inp \prime\prime}}) \sqrt{\mathsf{T}}^*_{2^{\inp\prime}1^{\out\prime} 1^{\inp\prime}0^{\out\prime}}  ( \Phi^+_{0^{\out\prime} 0^{\out\prime \prime}} \otimes \Phi^+_{1^{\inp \prime} 1^{\inp\prime \prime}} \otimes \Phi^+_{1^{\out} 1^{\out\prime}} \otimes \Phi^+_{2^{\inp}2^{\inp\prime}} )\sqrt{\mathsf{T}}^\mathrm{T}_{2^{\inp\prime}1^{\out\prime} 1^{\inp\prime}0^{\out\prime}}} \star \mathsf{D}_{\aux_2 2^\out 3^\inp} \notag \\
&= \ptr{2^{\inp\prime}1^{\out\prime} 1^{\inp\prime}0^{\out\prime}}{(\$_{0^{\out\prime} 0^{\out\prime \prime}} \otimes \$_{1^{\inp \prime} 1^{\inp\prime \prime}} \otimes \$_{1^{\out} 1^{\out\prime}} \otimes \$_{2^{\inp}2^{\inp\prime}} )\mathsf{T}_{3^\inp 2^\out 2^{\inp\prime} 1^{\out\prime} 1^{\inp\prime} 0^{\out\prime}}} = \mathsf{T}_{3^\inp 2^\out 2^{\inp} 1^{\out} 1^{\inp} 0^{\out}},
 \end{align}
where we have again used the fact that `tracing with the swap operator' amounts to a relabelling. Naturally, the above definition of $\mathsf{V}^{(2)}_{1^\out \aux_1 2^\inp \aux_2} $ directly generalises to $\mathsf{V}^{(k)}_{k-1^\out \aux_{k-1} k^\inp \aux_k}$ and the corresponding multi-slot case. 

This concludes our discussion of the axiomatic and constructive approach to causally ordered HOQOs. As we have seen, starting from quantum circuits, we always end up with HOQOs that satisfy the causality constraints of Def.~\ref{def::toqp-def-quantumcombs} and \textit{vice versa}: Any HOQO that satisfies said trace conditions can be understood as the Choi matrix of a quantum circuit. Finally, we have seen that, if we impose strict axiomatic constraints, then we recover the same set of HOQOs when following only axiomatic considerations. This begs the question what happens to the hierarchy of HOQOs when these axiomatic requirements are loosened. The ensuing HOQOs, which lie outside of the set of causally ordered combs, as well as the notion of causal indefiniteness of in HOQOs is the subject of the following section, after briefly discussing axiomatic considerations for superinstruments. 

\FloatBarrier


\subsubsection{Axiomatic Considerations for Superinstruments} \hfill \\

\noindent Having considered the axiomatic foundations of $n$-combs, it remains to briefly discuss the analogous motivation for Quantum superinstruments, which we have already encountered in Sec.~\ref{subsubsec::toqp-superinstruments}. There, we introduced them as objects that extract classical information from quantum channels, quantum combs, and other higher-order quantum objects, while at the same time potentially outputting a valid quantum object. A special case are the so-called \textit{testers} or \textit{process POVMs}~\cite{Ziman_2008}, which typically describe higher-order quantum objects that are fully contracted with the object on which they act, thereby leading to only a classical output (in the form of a probability distribution). In other words, superinstruments are the higher-order analogue of instruments, while testers / process POVMs are the higher order analogues of effects/POVMs. 

In the previous section, we have seen that quantum combs---i.e., deterministic (causally ordered) HOQOs---can be derived from axiomatic considerations; here, we will extend this analysis to superinstruments---i.e., their probabilistic counterparts. As per Def.~\ref{def::toqp-def-superinstruments}, a (causally ordered) superinstrument $\mathcal{J}$ is a collection $\{ \mathsf{G}^{(x)}_{{n^\inp}:{1^\inp}}\} \in \mathscr{L}(\mathscr{H}_{{n^\inp}} \otimes \hdots \otimes \mathscr{H}_{{1^\inp}})$ of positive semidefinite operators $\mathsf{G}^{(x)}_{{n^\inp}:{1^\inp}} \geq 0$ such that overall $\mathsf{G}_{n^\inp : 1^\inp} := \sum_x \mathsf{G}_{n^\inp : 1^\inp}^{(x)}$ is a (deterministic) quantum comb.\footnote{Since in this section we only consider superinstruments---and not the quantum combs that they act on---we omit the `breves' we added to the added to the notation of superinstruments in Sec.~\ref{subsubsec::toqp-superinstruments}.} First, as is the case with quantum combs, the elements of a superinstrument must be positive semidefinite; otherwise, they would potentially map positive semidefinite operators to non-positive operators, which would yield negative `probabilities'---an unphysical scenario. Moreover, when all elements of a superinstrument are summed over, one must obtain a deterministic transformation overall since this represents an event that occurs with certainty. Within a causally ordered setting, such a sum of superinstrument elements must therefore form a quantum comb (see below for a generalisation beyond the causally ordered case).

\begin{figure}[t!]
    \centering
    \includegraphics[width=1.\linewidth]{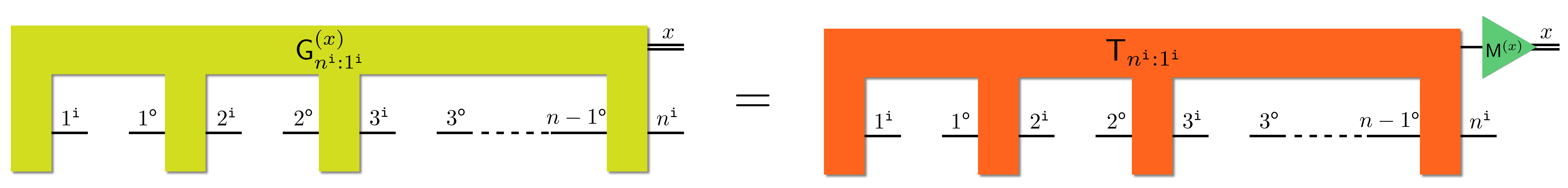}
    \caption{\textbf{Realisation of Superinstruments.} Any (causally ordered) superinstrument $\{ \mathsf{G}^{(x)}_{{n^\inp : 1^\inp}}\}$ can be realised by a quantum comb $\mathsf{T}:=\sum_x \mathsf{G}_{n^\inp : 1^\inp}^{(x)} \otimes \ketbra{x}{x}_\aux$ followed by a measurement $\{ \mathsf{M}^{(x)}:=\ketbra{x}{x} \}$ on an auxiliary system. 
    The classical outcome $x$ is explicitly represented by double lines. While the depicted superinstrument begins and ends on input wires---$1^\inp$ and $n^\inp$, respectively---superinstruments can also begin and/or end on output wires.}
    \label{fig::axiomatic_instrument}
\end{figure}

Next, we observe that \textit{all} (causally ordered) superinstruments can indeed be realised by a quantum comb followed by a measurement on an auxiliary system. That is, all quantum superinstruments can be implemented by means of a quantum circuit and an additional (projective) measurement. To see this, consider a superinstrument $\{ {\mathsf{G}^{(x)}_{n^\inp:1^\inp}} \}_{x=1}^N$. We can construct a quantum comb
\begin{align}
    \mathsf{T}:=\sum_x \mathsf{G}_{n^\inp : 1^\inp}^{(x)} \otimes \ketbra{x}{x}_\aux,
\end{align}
where $\{\ket{x}\}_{x=1}^N$ forms the computational basis for the auxiliary system ${\mathscr{H}_{\aux}} := \mathbb{C}^N$. Direct calculation shows that $\mathsf{T}$ is indeed a valid comb. Intuitively, this comb makes use of an auxiliary system to output a quantum `flag' state $\ket{x}$ that indicates which superinstrument element was applied on the system, and we have $\mathsf{T} \star \ketbra{x}{x}_\aux = \mathsf{G}_{n^\inp : 1^\inp}^{(x)}$ for all $x$. More precisely, by performing a measurement (in the computational basis) on the auxiliary system, we obtain a realisation for any superinstrument by combining a quantum comb with a final measurement (see Fig.~\ref{fig::axiomatic_instrument}). 

In light of discussions to come regarding the application of HOQOs to scenarios that go beyond a definite causal structure, i.e., HOQOs where the respective slots are not necessarily temporally ordered in a fixed way (see Secs.~\ref{subsec::indefinitecausalorder} and~\ref{subsec::causalityquantumfoundations}), we emphasise here that this relationship between deterministic HOQOs (e.g., quantum combs) and their probabilistic counterparts (e.g., causally ordered superinstruments) holds true more generally, in particular for the case where the admissible set of deterministic HOQOs can comprise HOQOs that are causally indefinite. In this case, in complete analogy to the causally ordered case, we can then consider higher-order analogues to instruments that are not necessarily causally ordered in time. More precisely, if $\{\mathsf{D}^{(x)}_{{n^\inp}:{1^\inp}}\}$ is a set of positive semidefinite operators that add up to some general deterministic $k$-slot HOQO---which may not have a definite causal order---the operator $\mathsf{W}:=\sum_x \mathsf{D}_{n^\inp : 1^\inp}^{(x)} \otimes \ketbra{x}{x}_\aux$ will also be a general deterministic $k$-slot HOQO of the same type, providing a method to `implement' the probabilistic object from its deterministic counterpart. We will refer to such objects as \textit{general superintruments}.\footnote{Such general superinstruments are, e.g., used to demonstrate the advantage of causally indefinite HOQOs over all causally ordered ones for the task of channel discrimination~\cite{Bavaresco_2021, Bavaresco_2022}.}


Returning to the causally ordered setting, since superinstruments can be implemented in terms of quantum combs plus a final measurement on an auxiliary system, their axiomatic derivation follows that for quantum combs described above. However, for the sake of concreteness, below we present a brief alternative axiomatic approach to one-slot superinstruments without a global future, i.e., objects that transform  quantum channels to probability distributions.

Consider an arbitrary quantum channel in its Choi representation $\mathsf{C}_{\inp \out} \in \Lscr(\Hscr_\inp \otimes \Hscr_\out)$. We seek the set of maps $\mathcal{T}^{(x)}:\Lscr(\Hscr_\inp \otimes \Hscr_\out)\to\mathbb{C}$ that take any such input channel into a probability distribution $\mathds{P}(x)$, thereby playing the role of a higher-order effect. Concretely, we require that $\mathcal{T}^{(x)}[\mathsf{C}_{\inp\out}]\geq0$, and $\sum_x \mathcal{T}^{(x)}[\mathsf{C}_{\inp\out}] = 1$ for all $\mathsf{C}_{\inp\out} \in \mathsf{CPTP}$. In order to be compatible with quantum theory, these superinstruments must comprise only linear maps; hence, by the Riez representation lemma, for any constituent element, there exists an associated unique linear operator $\mathsf{T}^{(x)}_{\inp \out}$ such that\footnote{Note that for this definition of $\mathsf{T}_{\inp\out}$, we have $\mathcal{T}^{(x)}[\mathsf{C}_{\inp \out}]= \mathsf{T}_{\inp\out}^{(x)\mathrm{T}} \star \mathsf{C}_{\inp \out}$, due to the definition of the link product.} $\mathcal{T}^{(x)}[\mathsf{C}_{\inp \out}]=\tr{\mathsf{T}^{(x)}_{\inp \out} \, \mathsf{C}_{\inp \out} }$ for any self-adjoint linear operator $\mathsf{C}_{\inp \out} \in \Lscr(\Hscr_\inp \otimes \Hscr_\out)$ (such as any quantum channel in its Choi representation). In order to ensure that any superinstrument will lead to non-negative probabilities (see also Ref.~\cite{Burniston_2020})---even when applied to part of a bipartite channel---we require the elements $\mathcal{T}^{(x)}$ to be completely positive; hence, we must have that $\mathsf{T}_{\inp \out}^{(x)}\geq 0$. The normalisation condition imposes that
\begin{align} \label{eq::axiomatichoqos-qc-tester_universal}
    \sum_x \tr{\mathsf{T}^{(x)}_{\inp \out}\, \mathsf{C}_{\inp \out} } = 1.
\end{align}
As we require testers to be universal, the above expression must hold for arbitrary quantum channels $\mathsf{C}_{\inp \out} \in \mathsf{CPTP}$. Now, we define the matrix $\mathsf{T}_{\inp \out}:=\sum_x \mathsf{T}_{\inp \out}^{(x)}$. Since we must have $\tr{\mathsf{T}_{\inp \out} \mathsf{\mathsf{C}_{\inp \out}}}=1$ for every quantum channel $\mathsf{C}_{\inp \out}$, which in particular satisfy $\ptr{\out}{\mathsf{C}_{\inp \out}} = \mathds{1}_{\inp}$, it is straightforward to see that there must exist a quantum state $\sigma_\inp \in\Lscr(\Hscr_\inp)$ such that $\mathsf{T}_{\inp \out}=\sigma_\inp\otimes\ident_\out$. Lastly, as proven in Ref.~\cite{Chiribella_2009,Ziman_2008} (and using purification methods that are analogous to the ones from the previous section), for any set of operators  $\mathsf{T}_{\inp \out}^{(x)}\geq 0$ respecting $\sum_x \mathsf{T}_{\inp \out}^{(x)}=\sigma_\inp\otimes\ident_\out $, there exists a bipartite quantum state $\rho_{\inp \aux} \in   \Lscr(\Hscr_\inp \otimes \Hscr_\aux) $, and a POVM with elements $\mathsf{E}^{(x)}_{\aux \out} \in \Lscr(\Hscr_\aux \otimes \Hscr_\out)$ such that $\mathsf{T}^{(x)}=\rho_{\inp \aux} \star \mathsf{E}^{(x)}_{\aux \out}$, providing a `Stinespring/Naimark dilation' for any such one-slot superinstrument.


\subsubsection{Transformations of Transformations of Transformations...}\hfill\\
\label{subsubsec::axiomatichoqos-qc-transtranstrans}

\begin{figure}[t!]
    \centering
    \includegraphics[width=0.95\linewidth]{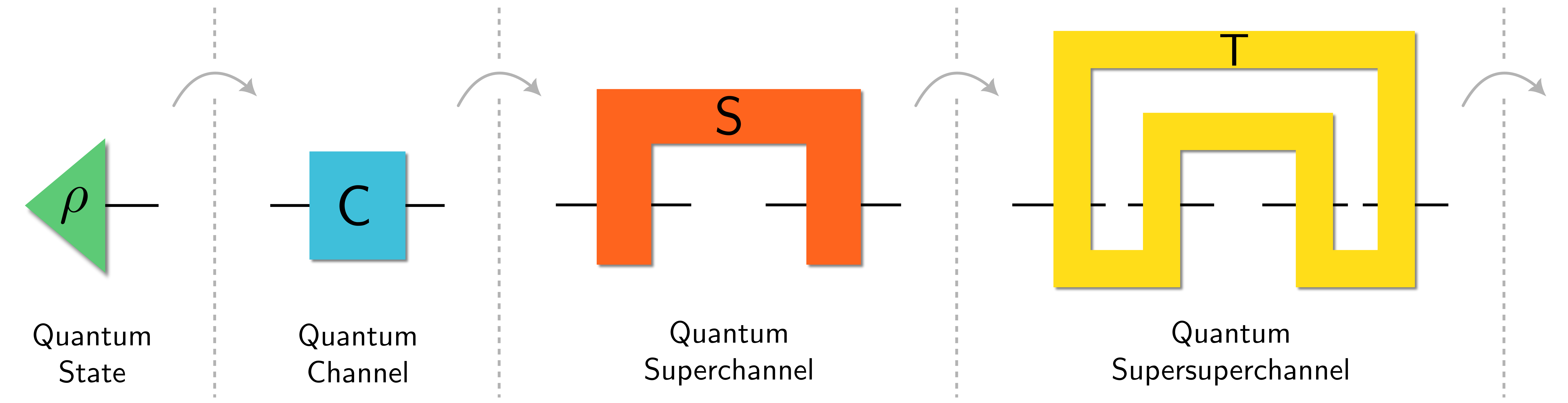}
    \caption{\textbf{Sequence of Transformations Between HOQOs.} A possible requirement on HOQOs is to map \textit{certain} types of maps onto other types of maps. Here, e.g., channels map states to states, superchannels map channels to channels, super-superchannels map superchannels to superchannels, and so on. One could continue this reasoning accordingly, with maps beyond the quantum superchannel possibly lying outside of what can be described by $n$-slot quantum combs. }
    \label{fig::axiomatichoqos-qc-ttt-sequence}
\end{figure}

\noindent Up to this point, our axiomatic considerations for HOQOs have been rather strict; so strict that they led to $n$-slot quantum combs being the only admissible transformations. Both from a foundational as well as a practical point of view, it is often reasonable to weaken the requirements on valid HOQOs. For example, one might be interested in the properties of particular types of transformations---such as mappings from superchannels onto superchannels (see Fig.~\ref{fig::axiomatichoqos-qc-ttt-sequence})---and the study of when or under which circumstances such mappings can go beyond the quantum comb formalism~\cite{lmcs:4426, Bisio_2019, hoffreumon_projective_2022, simmons_higher-order_2022, apadula2024, milz_characterising_2024}.


In addition, the considerations leading to the set of quantum combs often do not adequately capture a physical situation at hand. As an example, consider the (two-party) process matrix case that we have already encountered in previous sections. There, one considers two distinct laboratories---corresponding to the two slots of the process matrix, in each of which Alice and Bob can freely and independently perform operations---and aims to find the most general signalling/correlation structure between them that is locally (i.e., in the respective laboratories) compatible with quantum mechanics and does not lead to logical inconsistencies. Consequently, the process matrix $\mathsf{W}_{A^\inp A^\out B^\inp B^\out}\geq 0$ describing the correlation and signalling structure between the laboratories must satisfy $\mathsf{W}_{A^\inp A^\out B^\inp B^\out} \star (\mathsf{M}_{A^\inp A^\out} \otimes \mathsf{N}_{B^\inp B^\out}) = 1$ for all CPTP maps $\mathsf{M}_{A^\inp A^\out} \in \Lscr(\Hscr_{A^\inp} \otimes \Hscr_{A^\out})$ and $\mathsf{N}_{B^\inp B^\out} \in \Lscr(\Hscr_{B^\inp} \otimes \Hscr_{B^\out})$---to ensure proper normalisation of probabilities---but does not need to satisfy particular additional constraints when acting, e.g., on a superchannel, since such operations are not considered possible in this setting [see Fig.~\ref{fig::axiomatichoqos-qc-processmatrixsuperchannel}]. The weakened axiomatic considerations thus mirror the physical situation that is investigated. 

Finally, it is often insightful to even further restrict the set of maps one acts on/onto. For instance, in the study of quantum operations with indefinite time direction~\cite{chiribella_quantum_2022}, the set of HOQOs that map \textit{unital} channels to \textit{unital} channel---instead of channels to channels---emerges naturally as an interesting set of transformations to consider (and as it turns out this set contains HOQOs that go beyond the quantum comb formalism, e.g., the quantum time flip~\cite{chiribella_quantum_2022}). Similar situations occur in the study of higher-order resource theories, i.e., resource theories where channels~\cite{theurer_quantifying_2019, liu_resource_2019, gour_dynamical_2020, gour_dynamical_2020a, gour_entanglement_2021, gour_inevitable_2024} or HOQOs~\cite{Araujo_2015, Berk_2021, Milz_2022} are considered as resources, and one studies their interconvertibility under designated \textit{free} transformations. Such free transformations are HOQOs themselves, with the restriction that they map free objects to free objects, i.e., they cannot create resources. Requesting such properties instead of (or in addition to) the axiomatic constraints imposed on HOQOs in the previous sections can lead to sets of HOQOs that do not necessarily coincide with the set of $n$-slot quantum combs. As a result, the hierarchy of $n$-slot combs derived in the previous sections is not the end of the story and does not comprise all conceivable HOQOs of interest. Rather, there is a large variety of HOQOs both in the `vertical' direction---i.e., going up the ladder of HOQOs---as well as `laterally'---i.e., when restricting the sets of input and output elements. Discussing the complete structural characterisation of all possible HOQOs would go beyond the scope of this work (see Refs.~\cite{Perinotti_2017, Bisio_2019, hoffreumon_projective_2022} for a detailed analysis). Here, we content ourselves with a brief discussion of how the Choi states of such more general HOQOs can be characterised, and how causal indefiniteness and indefinite time direction naturally emerge when relaxing the axiomatic considerations of HOQOs. 

\vspace{0.25cm} \noindent \textbf{\textul{Types of HOQOs.}} Up until this point, we have---deliberately---not touched upon the fact that HOQOs come in different \textit{types}. For instance, a channel $\Ccal: \Lscr(\Hscr_\inp) \rightarrow \Lscr(\Hscr_\out)$ maps matrices onto matrices, a superchannel $\Scal: [\Lscr(\Hscr_\inp) \rightarrow \Lscr(\Hscr_\out)] \rightarrow [\Lscr(\Hscr_{\inp'}) \rightarrow \Lscr(\Hscr_{\out'})]$ maps linear mappings $\Ccal: \Lscr(\Hscr_\inp) \rightarrow \Lscr(\Hscr_\out)$ to linear mappings $\Ccal': \Lscr(\Hscr_{\inp'}) \rightarrow \Lscr(\Hscr_{\out'})$, while a super-superchannel $\mathcal{T}$ would map superchannels (i.e., mappings of mappings) to superchannels (i.e., mappings of mappings). At first glance, all of these objects (and those further up the hierarchy) not only have differing input and output spaces, but also act on fundamentally different types of objects. Consequently, the hierarchy of admissible HOQOs and their causal properties are most rigorously and comprehensively discussed in the language of type/category theory~\cite{Perinotti_2017, Bisio_2019, simmons_higher-order_2022, hoffreumon_projective_2022, Wilson_2022_Mathematical, Wilson_2022_Locality, Wilson_2022_Unitary, Hefford_2023, Hefford_2024, apadula2024, jencova_structure_2024}. Here, we have circumvented the need to address the varying types of objects we consider by fully relying on the Choi-Jamio{\l}kowski isomorphism: \textit{All} types of HOQOs that we encounter in this Review Article can isomorphically be mapped onto matrices, putting them all on the same mathematical footing and allowing for flexible notation. This simplicity, however, comes at a cost: Without additional context, a Choi matrix alone does not specify the type of transformation it represents. 

For instance [see Fig.~\ref{fig::axiomatichoqos-qc-1slot}], a (positive semidefinite) Choi matrix $\mathsf{T}_{2^\inp:0^\out}\in \Lscr(\Hscr_{0^\out} \otimes \Hscr_{1^\inp} \otimes\Hscr_{1^\out} \otimes\Hscr_{2^\inp})$ satisfying $\ptr{2^\inp}{\mathsf{T}_{2^\inp:0^\out}} =\ident_{1^\out} \otimes \mathsf{T}_{1^\inp:0^\out}$ and $\ptr{1^\inp}{\mathsf{T}_{1^\inp:0^\out}} =\ident_{0^\out}$ could pertain to both a quantum channel $\mathcal{C}: \Lscr(\Hscr_{0^\out} \otimes \Hscr_{1^\out}) \rightarrow \Lscr(\Hscr_{1^\inp} \otimes \Hscr_{2^\inp})$ or a superchannel $\mathcal{S}: [\Lscr(\Hscr_{1^\inp}) \rightarrow \Lscr(\Hscr_{1^\out})] \rightarrow [\Lscr(\Hscr_{0^\out}) \rightarrow \Lscr(\Hscr_{2^\inp})]$---two fundamentally different types of operations. Throughout this Review Article, the type of all encountered operations will be clear from context---even if given only in terms of their Choi matrices---and we will thus keep notation simple and refrain from discussing HOQOs in type theoretic terms. This being said, we emphasise that type theory does not merely add a `nice-to-have' labelling of the type of transformation one deals with, but provides a powerful tool to systematically analyse the properties of the entire hierarchy of admissible HOQOs under different axioms~\cite{Perinotti_2017, Bisio_2019, hoffreumon_projective_2022}. In contrast, we will stay more modest in scope and only discuss the means to characterise the Choi states of HOQOs. 

\vspace{0.25cm} \noindent \textbf{\textul{Characterising HOQOs via Choi States.}} A first step in discussing any type of HOQO is a proper characterisation of the mathematical properties that uniquely define them. Naturally, this can be achieved in any chosen representation. Here, leveraging the insights of the previous sections, we focus on a characterisation of the respective Choi states. We have already encountered such a characterisation in Sec.~\ref{subsubsec::axiomatichoqos-quantumcomb}, where we derived the hierarchy of trace conditions on $n$-slot quantum combs from axiomatic requirements. For more general HOQOs, it proves insightful to provide a more systematic recipe for the derivation of their properties. 

This approach could be developed in an entirely abstract way, but for the sake of concreteness, let us focus on a specific case. Consider the set of HOQOs $\mathsf{W}_{A^\inp A^\out B^\inp B^\out}$ that map quantum channels $\mathsf{C}_{A^\inp A^\out}$ onto $1$-slot combs $\mathsf{S}_{B^\inp B^\out}$ without a global past or future [see Fig.~\ref{fig::axiomatichoqos-qc-processmatrix}]. One could consider this definition of the set of process matrices $\mathsf{W}_{A^\inp A^\out B^\inp B^\out}$ as a purely mathematical example that will lead us outside the set of (causally ordered) $n$-slot quantum combs. More conceptually though, this situation reflects the requirement of \textit{local causality}: Whatever deterministic operation (i.e., a quantum channel $ \mathsf{C}_{A^\inp A^\out}$) Alice performs, Bob will always `see' a causally ordered comb (represented by $\mathsf{S}_{B^\inp B^\out}$) with which they can interact (see below for a more detailed discussion). Independent of the respective motivation, we are tasked with deriving the set of matrices $\mathsf{W}_{A^\inp A^\out B^\inp B^\out}$ such that 
\begin{gather}
\label{eq::ahoqos-qc-ttt-def-processmatrix}
\mathsf{S}_{B^\inp B^\out} = \mathsf{W}_{A^\inp A^\out B^\inp B^\out}\star \mathsf{C}_{A^\inp A^\out}
\end{gather}
is a valid $1$-slot comb without global future and past for all quantum channels $\mathsf{C}_{A^\inp A^\out} \in \mathsf{CPTP}$. We emphasise that $\mathsf{S}_{B^\inp B^\out} \geq 0$ corresponds to a quantum superchannel without global future and past iff it satisfies $\mathsf{S}_{B^\inp B^\out} =  \rho_{B^\inp} \otimes \ident_{B^\out}$ for some quantum state $\rho_{B^\inp}$ (see Def.~\ref{def::toqp-def-quantumcombs}).

\begin{figure}[t!]
    \centering
    \includegraphics[scale=0.65]{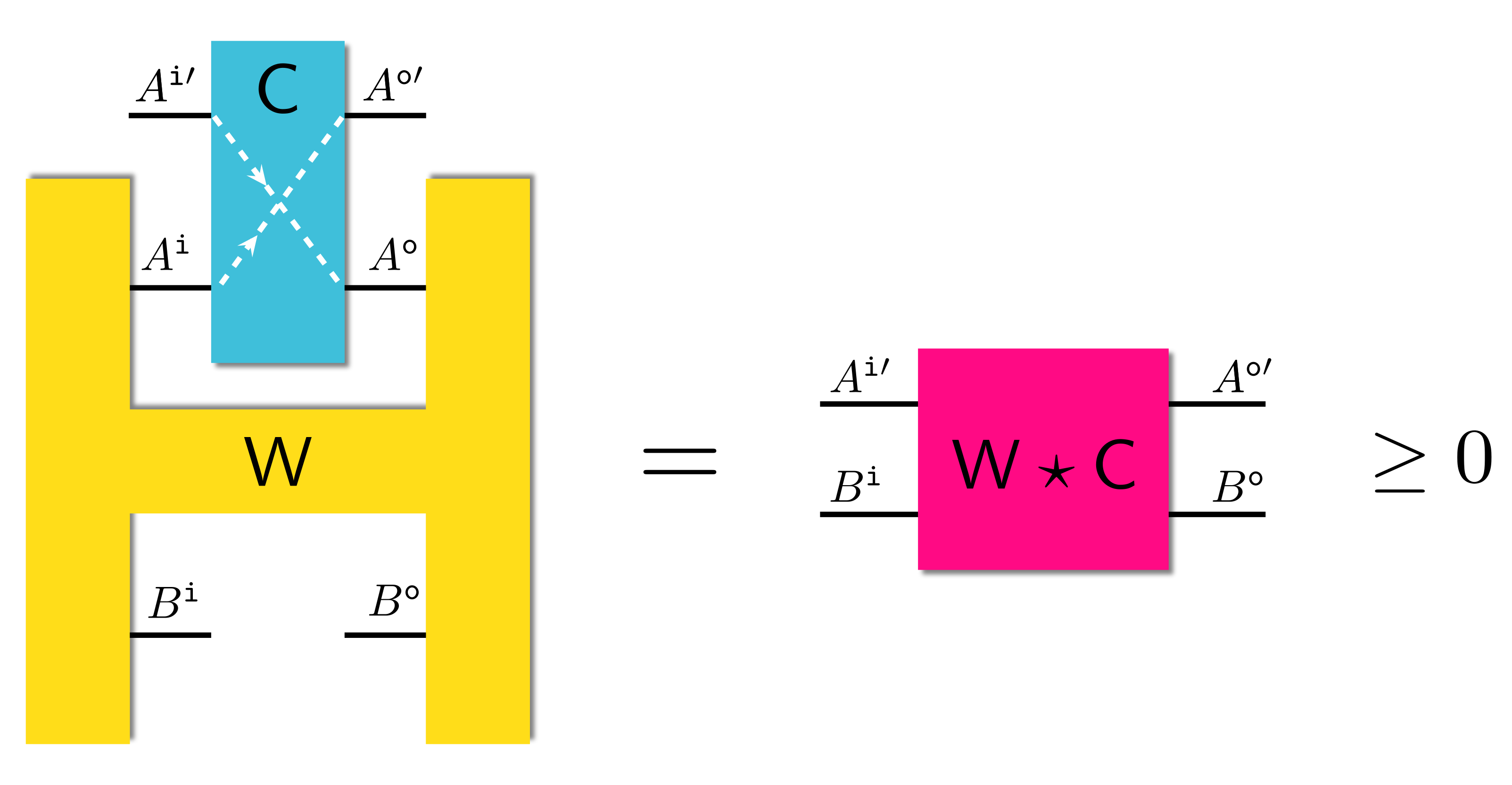}
    \caption{\textbf{Positivity of the Process Matrix.} Demanding that a process matrix yields a positive output whenever it acts on part of a CPTP map implies that its Choi operator $\mathsf{W}$ is necessarily positive semidefinite. This can, e.g., be seen by by letting $\mathsf{W}$ act on part of a swap channel, depicted by the white dotted lines (left).}
    \label{fig::w_is_positive}
\end{figure}

Here, positivity of the Choi operator $\mathsf{W}_{A^\inp A^\out B^\inp B^\out} \geq 0$ can be argued for by demanding that the process matrix is completely positive in the correct sense, i.e., by requiring that $\mathsf{W}_{A^\inp A^\out B^\inp B^\out} \star \mathsf{C}_{A^\inp A^{\inp\prime} A^\out A^{\out \prime}} \geq 0$ for all channels $\mathsf{C}_{A^\inp A^{\inp\prime} A^\out A^{\out \prime}}$ and all auxiliary systems $A^{\inp\prime}$ and $A^{\out\prime}$ (see Fig.~\ref{fig::w_is_positive}). This situation includes the case of inserting part of a swap channel, which leads to $\mathsf{W}_{A^\inp A^\out B^\inp B^\out} \star \Phi^+_{A^\inp A^{\inp\prime}} \star \Phi^+_{A^\inp A^{\inp\prime}} = \mathsf{W}_{A^{\inp\prime} A^{\out\prime} B^\inp B^\out} \cong \mathsf{W}_{A^\inp A^\out B^\inp B^\out}$; demanding that the latter is positive semidefinite then yields $\mathsf{W}_{A^\inp A^\out B^\inp B^\out} \geq 0$ as required.\footnote{In more general situations, depending on the sets of objects between which the HOQO of interest maps, the additional demand of `completeness' may be insufficient to guarantee positivity and lead to structurally different sets of HOQOs~\cite{milz_characterising_2024}. Consequently, in all following considerations, we will require positivity of the considered (Choi matrices of) HOQOs as an explicit assumption.}


 Deriving the remaining characterising properties of $\mathsf{W}_{A^\inp A^\out B^\inp B^\out}$ now takes two steps. First, we note that $\tr{\mathsf{C}_{A^\inp A^\out}} = d_{A^\inp}$ and $\tr{\mathsf{S}_{B^\inp B^\out}} = d_{B^\out}$ holds. Second, let us denote the linear spans of the set of channels and the set of $1$-slot combs without global future and past respectively by
\begin{align}
    \texttt{SP} := \text{span}(\{\mathsf{C}_{A^\inp A^\out}\})\quad \text{and} \quad \texttt{SP}' := \text{span}(\{\mathsf{S}_{B^\inp B^\out}\}).
\end{align} 
Since both $\texttt{SP}$ and $\texttt{SP}'$ are vector spaces, they are characterised by projectors $\mathcal{P}_A: \Lscr(\Hscr_{A^\inp} \otimes \Hscr_{A^\out}) \rightarrow \Lscr(\Hscr_{A^\inp} \otimes \Hscr_{A^\out})$ and $\mathcal{P}_B: \Lscr(\Hscr_{B^\inp} \otimes \Hscr_{B^\out}) \rightarrow \Lscr(\Hscr_{B^\inp} \otimes \Hscr_{B^\out})$ with $\mathcal{P}_A \circ \mathcal{P}_A = \mathcal{P}_A$ and $\mathcal{P}_B \circ \mathcal{P}_B = \mathcal{P}_B$, such that $\mathcal{P}_A[\mathsf{C}] = \mathsf{C}$ for all $\mathsf{C} \in \texttt{SP}$ and $\mathcal{P}_B[\mathsf{S}] = \mathsf{S}$ for all $\mathsf{S} \in \texttt{SP}$. In particular, we have $\ptr{A^\out}{\mathsf{C}} \propto \ident_{A^\inp}$ for all $\mathsf{C} \in \texttt{SP}$ and $\mathsf{S} = \mathsf{R}_{B^\inp} \otimes \ident_{B^\out}$ for all $\mathsf{C} \in \texttt{SP}'$, where $ \mathsf{R}_{B^\inp}$ is some Hermitian matrix. Employing the operator ${}_X\bullet := \frac{\ident_X}{d_X} \otimes \ptr{X}{\bullet}$, we can compactly write the projectors $\mathcal{P}_A$ and $\mathcal{P}_B$ as 
\begin{align}\label{eq::ico-projectorchannelandoneslotcomb}
   \mathsf{C} \in \texttt{SP} \quad \Leftrightarrow \quad  \mathcal{P}_A[\mathsf{C}] := \mathsf{C} - {}_{A^\out}\mathsf{C} + {}_{A^\inp A^\out}\mathsf{C} = \mathsf{C} \quad 
   \text{and} \quad \mathsf{S} \in \texttt{SP}' \quad \Leftrightarrow \quad  \mathcal{P}_B[\mathsf{S}] := {}_{B^\out}\mathsf{S} = \mathsf{S}.
\end{align}
Both projectors $\mathcal{P}_A$ and $\mathcal{P}_B$ are self-adjoint, unital, and commute with transposition, i.e., we have $\mathcal{P}_X^\dagger =  \mathcal{P}_X$, $\mathcal{P}_X[\ident_X] =  \ident_X$, and $\mathcal{P}_X[\bullet^\mathrm{T}]=  (\mathcal{P}_X[\bullet])^\mathrm{T}$ for all $X \in \{A, B\}$, respectively.

With these projectors at hand, we can rephrase the characterising criteria of Eq.~\eqref{eq::ahoqos-qc-ttt-def-processmatrix} in terms of the following two conditions
\begin{align}
\label{eq::ahoqos-qc-ttt-projector_cond_proc_mat}
 \mathcal{P}_B[\mathsf{W}_{A^\inp A^\out B^\inp B^\out}\star \mathsf{C}_{A^\inp A^\out}] &= \mathsf{W}_{A^\inp A^\out B^\inp B^\out}\star \mathsf{C}_{A^\inp A^\out} \\
 \quad \tr{\mathsf{W}_{A^\inp A^\out B^\inp B^\out}\star \mathsf{C}_{A^\inp A^\out}} &= d_{B^\out}, \label{eq::ahoqos-qc-ttt-projector_cond_proc_mat_trace}
\end{align}
which must hold for all quantum channels $\mathsf{C}_{A^\inp A^\out} \in \mathsf{CPTP}$.\footnote{Since $\mathsf{W}_{A^\inp A^\out B^\inp B^\out} \geq 0$, it is already ensured that $\mathsf{W}_{A^\inp A^\out B^\inp B^\out}$ yields a positive matrix when acting on a positive object, so that we do not have to demand this property as an additional requirement here.}  Importantly, since both of these equations are linear, the first one holds on \textit{all} elements of $\texttt{SP}$ (not only for quantum channels), while the second one holds on all elements of $\texttt{SP}$ up to a normalisation factor accounting for the fact that $\tr{\mathsf{X}_{A^\inp A^\out}} \neq d_{A^\inp}$ for some elements $\mathsf{X}_{A^\inp A^\out} \in \texttt{SP}$.

We can now use the fact that any self-adjoint operation $\mathcal{P}$ that commutes with transposition `can be moved around' inside the link product, i.e., $\mathsf{Y} \star \mathcal{P}[\mathsf{X}] = \mathcal{P}[\mathsf{Y}] \star \mathsf{X}$ (this can be seen by direct insertion into the definition of the link product; see Def.~\ref{def::linkproduct}). In addition, we have that $\mathcal{P}_A[\mathsf{X}_{A^\inp A^\out}] \in \texttt{SP}$ for all $\mathsf{X}_{A^\inp A^\out} \in \Lscr(\Hscr_{A^\inp} \otimes \Hscr_{A^\out})$. These two facts imply that Eq.~\eqref{eq::ahoqos-qc-ttt-projector_cond_proc_mat} can be expressed as
\begin{align}
    \mathcal{P}_B [\mathsf{W}_{A^\inp A^\out B^\inp B^\out}] \star \mathcal{P}_A[\mathsf{X}_{A^\inp A^\out}] = \mathsf{W}_{A^\inp A^\out B^\inp B^\out} \star \mathcal{P}_A[\mathsf{X}_{A^\inp A^\out}] \quad \forall \ \mathsf{X}_{A^\inp A^\out},
\end{align}
which further implies that
\begin{align}
    (\mathcal{P}_A \otimes \mathcal{P}_B) [\mathsf{W}_{A^\inp A^\out B^\inp B^\out}] \star \mathsf{X}_{A^\inp A^\out} = \mathcal{P}_A[\mathsf{W}_{A^\inp A^\out B^\inp B^\out}] \star \mathsf{X}_{A^\inp A^\out} \quad \forall \ \mathsf{X}_{A^\inp A^\out}.
\end{align}
Since this holds for all $\mathsf{X}_{A^\inp A^\out}$, we obtain a first property that characterises $\mathsf{W}_{A^\inp A^\out B^\inp B^\out}$, namely that $(\mathcal{P}_A \otimes \mathcal{P}_B) [\mathsf{W}_{A^\inp A^\out B^\inp B^\out}] = \mathcal{P}_A[\mathsf{W}_{A^\inp A^\out B^\inp B^\out}]$, which is equivalent to 
\begin{align}
    \label{eqn::projProp1}
    \mathsf{W}_{A^\inp A^\out B^\inp B^\out} = \mathsf{W}_{A^\inp A^\out B^\inp B^\out} - \mathcal{P}_A[\mathsf{W}_{A^\inp A^\out B^\inp B^\out}] + (\mathcal{P}_A \otimes \mathcal{P}_B) [\mathsf{W}_{A^\inp A^\out B^\inp B^\out}] =: \mathcal{P}^{(1)}_{AB}[\mathsf{W}_{A^\inp A^\out B^\inp B^\out}].
\end{align}
Here, we have introduced the projector $\mathcal{P}^{(1)}_{AB}$, which satisfies $\mathcal{P}^{(1)}_{AB} \circ \mathcal{P}^{(1)}_{AB} = \mathcal{P}^{(1)}_{AB}$.

Returning to the trace condition of Eq.~\eqref{eq::ahoqos-qc-ttt-projector_cond_proc_mat_trace}, we obtain
\begin{align}
 \mathrm{tr}\left(\mathsf{W}_{A^\inp A^\out B^\inp B^\out} \star \mathcal{P}_A[\mathsf{X}_{A^\inp A^\out}] \right) = \tr{\mathcal{P}_A[
 \mathrm{tr}_{B^\inp B^\out}\left(\mathsf{W}_{A^\inp A^\out B^\inp B^\out}\right)] \, \mathsf{X}^\mathrm{T}_{A^\inp A^\out}} = \frac{d_{B^\out}}{d_{A^\inp}}\tr{\mathsf{X}^\mathrm{T}_{A^\inp A^\out}}, 
\end{align}
where the extra factor on the r.h.s.\ accounts for the fact that $\tr{\mathsf{X}_{A^\inp A^\out}}$ is not necessarily equal to $d_{A^\inp}$. Since the above must hold for all $\mathsf{X}_{A^\inp A^\out}$, we obtain our second characterising constraint on $\mathsf{W}_{A^\inp A^\out B^\inp B^\out}$, namely 
\begin{align}
    \mathcal{P}_A[
 \mathrm{tr}_{B^\inp B^\out}\left(\mathsf{W}_{A^\inp A^\out B^\inp B^\out}\right)] = \frac{d_{B^\out}}{d_{A^\inp}} \ident_{A^\inp A^\out}.
\end{align}
Since $\mathcal{P}_A$ is unital and self-adjoint, it is also trace preserving, which means we can re-write this equation as 
\begin{align}
    \label{eq::ahoqos-qc-ttt-projector2}
    \mathsf{W} = \mathsf{W}- \mathcal{P}_A[{}_{B^\inp B^\out} \mathsf{W}] + {}_{A^\inp A^\out B^\inp B^\out}\mathsf{W} =: \mathcal{P}^{(2)}_{AB}[\mathsf{W}]  \quad \text{and} \quad \tr{\mathsf{W}} = d_{A^\out}d_{B^\out}, 
\end{align}
where we have introduced the projector $\mathcal{P}^{(2)}_{AB}$, which satisfies $\mathcal{P}^{(2)}_{AB} \circ \mathcal{P}^{(2)}_{AB} = \mathcal{P}^{(2)}_{AB}$, and omitted the subscripts on $\mathsf{W}_{A^\inp A^\out B^\inp B^\out}$ for improved readability. Importantly, $\mathcal{P}_{AB}^{(1)}$ and $\mathcal{P}_{AB}^{(2)}$ commute, such that they can be \textit{uniquely} combined into a single joint projector $\mathcal{P}_{AB} := \mathcal{P}^{(1)}_{AB} \circ \mathcal{P}^{(2)}_{AB} = \mathcal{P}^{(2)}_{AB} \circ \mathcal{P}^{(1)}_{AB}$, which yields  
\begin{align}
\label{eq::ahoqos-qc-ttt-projectorfull}
    \mathcal{P}_{AB}[\mathsf{W}] =  \mathsf{W} - \mathcal{P}_A[\mathsf{W}] + (\mathcal{P}_A \otimes \mathcal{P}_B)[\mathsf{W}] - \mathcal{P}_A[{}_{B^\inp B^\out}\mathsf{W}] + {}_{A^\inp A^\out B^\inp B^\out}\mathsf{W}, 
\end{align}
where we have used ${}_{B^\inp B^\out}\mathcal{P}_B[\mathsf{Y}] = \mathcal{P}_B[{}_{B^\inp B^\out}\mathsf{Y}] = {}_{B^\inp B^\out}\mathsf{Y}$ due to the unitality of $\mathcal{P}_B$. One can readily verify that $\mathcal{P}_{AB}[\mathsf{W}] = \mathsf{W}$ implies $\mathcal{P}^{(1)}_{AB}[\mathsf{W}] = \mathsf{W}$, since 
\begin{align}
    \mathcal{P}^{(1)}_{AB}[\mathsf{W}] = \mathcal{P}^{(1)}_{AB}[\mathcal{P}_{AB}[\mathsf{W}]] = (\mathcal{P}^{(1)}_{AB} \circ \mathcal{P}^{(1)}_{AB} \circ \mathcal{P}^{(2)}_{AB})[\mathsf{W}] = (\mathcal{P}^{(1)}_{AB} \circ \mathcal{P}^{(2)}_{AB})[\mathsf{W}] = \mathcal{P}_{AB}[\mathsf{W}] = \mathsf{W},
\end{align}
and analogously $\mathcal{P}^{(2)}_{AB}[\mathsf{W}] = \mathsf{W}$. As a consequence, the projector $\mathcal{P}_{AB}$ together with the trace condition of Eq.~\eqref{eq::ahoqos-qc-ttt-projector2} characterises the set of valid transformations from CPTP maps to $1$-slot quantum combs without global future and past. Inserting the definitions of $\mathcal{P}^{(1)}_{AB}$ and $\mathcal{P}^{(2)}_{AB}$ into Eq.~\eqref{eq::ahoqos-qc-ttt-projectorfull} then yields:
\begin{myDefinition*}{Characterisation of Process Matrices}{}
A process matrix $\mathsf{W}_{A^\inp A^\out B^\inp B^\out} \geq 0$ maps quantum channels $\mathsf{C}_{A^\inp A^\out}$ to $1$-slot quantum combs $\mathsf{S}_{B^\inp B^\out}$ without global past and future iff 
\begin{align}
\label{eq::ico-proj_proc_mat}
    \mathcal{P}_{AB}[\mathsf{W}] :=& {}_{A^\out}\mathsf{W} + {}_{B^\out}\mathsf{W} - {}_{A^\out B^\out}\mathsf{W} - {}_{A^\inp A^\out}\mathsf{W} - {}_{B^\inp B^\out}\mathsf{W} + {}_{A^\out B^\inp B^\out}\mathsf{W} + {}_{A^\inp A^\out B^\out}\mathsf{W} = \mathsf{W} \\
    \label{eq::ico-proj_proc_mat2}
    \text{and} \quad \tr{\mathsf{W}}=& d_{A^\out}d_{B^\out}.
\end{align}
\end{myDefinition*}
\noindent This characterisation of process matrices coincides exactly with that provided in Ref.~\cite{Araujo_2015}, where the characterisation of HOQOs by means of projectors was first introduced. 

It is insightful to take stock of what we have shown here. We started from the set of quantum channels and saw that this set---let us denote it $\Omega$---is defined by a positivity constraint, a projector, and a trace constraint, i.e., 
\begin{align}
    \mathsf{C}_{A^\inp A^\out} \in \Omega \quad \Leftrightarrow \quad \mathsf{C}_{A^\inp A^\out} \geq 0, \ \ \mathcal{P}_{A}[\mathsf{C}_{A^\inp A^\out}] = \mathsf{C}_{A^\inp A^\out}, \ \ \text{and} \ \ \tr{\mathsf{C}_{A^\inp A^\out}} = d_{A^\inp}.
\end{align}
Similarly, the set $\Omega^\prime$ of $1$-slot quantum combs without global future and past is given by 
\begin{align}
    \mathsf{S}_{B^\inp B^\out} \in \Omega^\prime \quad \Leftrightarrow \quad \mathsf{S}_{B^\inp B^\out} \geq 0, \ \ \mathcal{P}_{B}[\mathsf{S}_{B^\inp B^\out}] = \mathsf{S}_{B^\inp B^\out}, \ \ \text{and} \ \ \tr{\mathsf{S}_{B^\inp B^\out}} = d_{B^\out}.
\end{align}
Finally, the set $\Theta$ of (completely positive) mappings \textit{between} the sets $\Omega$ and $\Omega'$ is also characterised by a positivity constraint, a projector, and a trace constraint:
\begin{align}
    \mathsf{W}_{A^\inp A^\out B^\inp B^\out} \in \Theta \quad \Leftrightarrow \quad \mathsf{W}_{A^\inp A^\out B^\inp B^\out} \geq 0, \ \ \mathcal{P}_{AB}[\mathsf{W}] = \mathsf{W}, \ \ \text{and} \ \ \tr{\mathsf{W}}= d_{A^\out}d_{B^\out}, 
\end{align}
where the `new' projector $\mathcal{P}_{AB}$ is constructed from the two initial ones $\mathcal{P}_{A}, \mathcal{P}_{B}$ via Eq.~\eqref{eq::ahoqos-qc-ttt-projectorfull}. 

From this construction, the process to `go further up the ladder' in the hierarchy of HOQOs becomes clear. For instance, if one wished to derive the set of HOQOs that map process matrices onto process matrices~\cite{Castro-Ruiz_2018, Milz_2022}, the same recipe applies: both input and output sets (i.e., those of valid process matrices) are defined by a positivity constraint, a projector, and a trace condition, from which the projector onto the (span of the) set of mappings between them can be readily deduced (as too can the corresponding trace condition). Since this procedure only makes use of the structural properties of $\mathcal{P}_A$ and $\mathcal{P}_B$ (rather than their explicit forms), we can directly obtain a general characterisation of HOQOs that map between other sets of HOQOs (see Fig.~\ref{fig::ahoqos-qc-ttt-transformationsoftransformations}):


\begin{figure}[t!]
    \centering
    \includegraphics[width=0.7\linewidth]{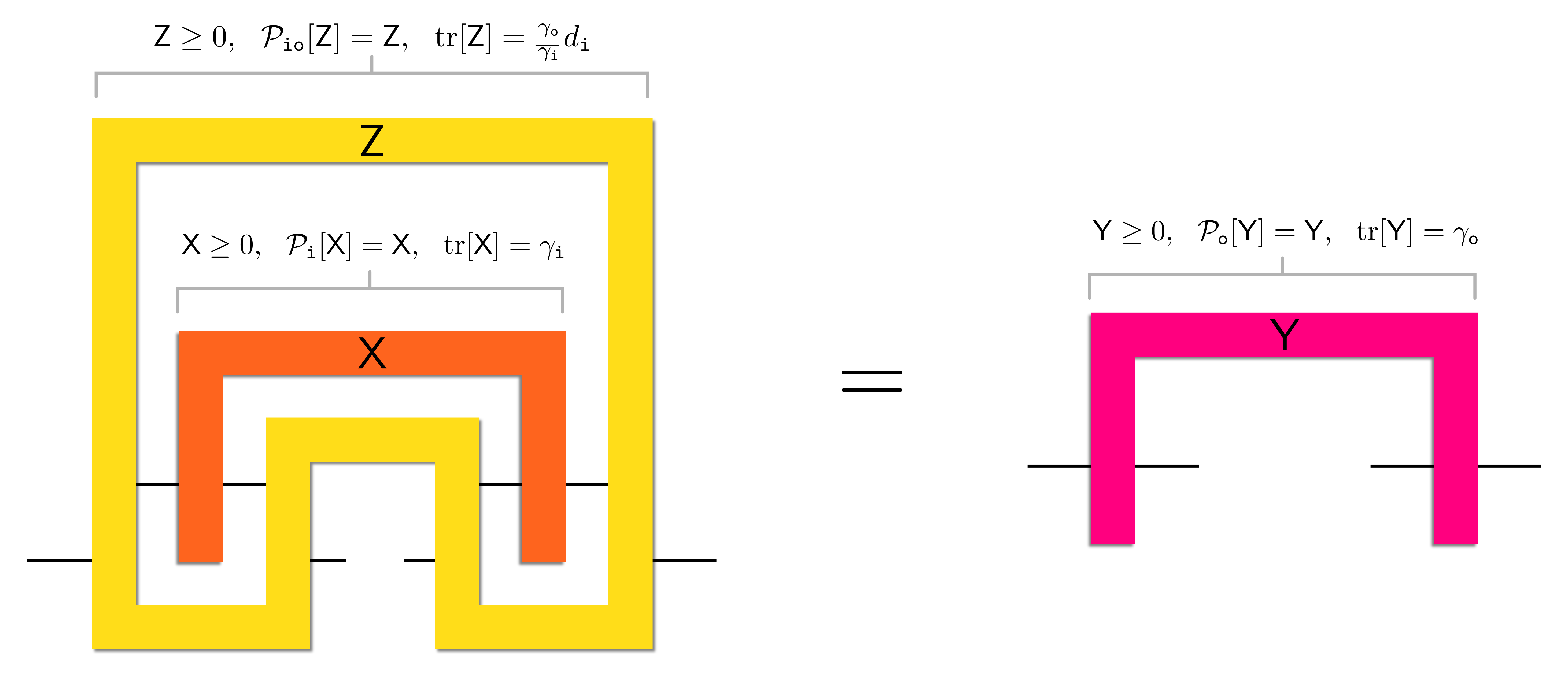}
    \caption{\textbf{Transformations of Transformations of Transformations.} Characterising a specific set  of HOQOs---depicted here is the set of transformations $\mathsf{Z}$ that map superchannels to superchannels---amounts to finding an appropriate projector and trace conditions, which can be derived from the corresponding projector and trace conditions that characterise the input and output objects.}
    \label{fig::ahoqos-qc-ttt-transformationsoftransformations}
\end{figure}


{\hypersetup{citecolor=white}
\begin{myDefinition*}{General Characterisation of HOQOs~\cite{Castro-Ruiz_2018, simmons_higher-order_2022, hoffreumon_projective_2022, milz_characterising_2024}}{}
Let two sets of HOQOs be given by 
\begin{align}
    &\Omega = \{\mathsf{X}_\inp|\ \mathsf{X}_\inp \in \Lscr(\Hscr_\inp), \ \mathsf{X}_\inp \geq 0, \ \mathcal{P}_\inp[\mathsf{X_\inp}] = \mathsf{X}_\inp,\ \tr{\mathsf{X}_\inp} = \gamma_\inp \} \\
    \text{and} \quad &\Omega' = \{\mathsf{Y}_\out|\ \mathsf{Y}_\out \in \Lscr(\Hscr_\out), \ \mathsf{X}_\out \geq 0, \ \mathcal{P}_\out[\mathsf{Y_\out}] = \mathsf{Y}_\out,\ \tr{\mathsf{Y}_\out} = \gamma_\out\},
\end{align}
where both $\mathcal{P}_\inp$ and $\mathcal{P}_\out$ are self-adjoint and unital projectors that commute with transposition and $\gamma_\inp, \gamma_\out \neq 0$. The set $\Theta$ of transformations $\mathsf{Z}_{\inp\out} \geq 0$ between them, i.e., such that $\mathsf{Z}_{\inp\out} \star \mathsf{X}_\inp \in \Omega'$ for all $\mathsf{X}_\inp \in \Omega$, is given by
\begin{align}\label{eq::ice-gen_char_hoqos}
    \mathsf{Z}_{\inp\out} \in \Theta \quad \Leftrightarrow \quad \mathsf{Z}_{\inp\out} \geq 0, \ \mathcal{P}_{\inp\out}[\mathsf{Z}_{\inp\out}] = \mathsf{Z}_{\inp\out}, \ \text{and} \ \tr{\mathsf{Z}_{\inp\out}} = \frac{\gamma_\out}{\gamma_\inp}d_\inp, 
\end{align}
where $d_\inp := \text{dim}(\Hscr_\inp)$ and 
\begin{align}\label{eq::ice-gen_char_hoqos2}
    \mathcal{P}_{\inp\out}[\mathsf{Z}] :=  \mathsf{Z} - \mathcal{P}_\inp[\mathsf{Z}] + (\mathcal{P}_\inp \otimes \mathcal{P}_\out)[\mathsf{Z}] - \mathcal{P}_\inp[{}_{\inp}\mathsf{Z}] + {}_{\inp\out}\mathsf{Z}
\end{align}
is a self-adjoint, unital projector that commutes with transposition.
\end{myDefinition*}
}
\noindent We emphasise that the properties of $n$-slot quantum combs that we worked out in Sec.~\ref{subsubsec::axiomatichoqos-quantumcomb} could have equivalently been derived in this manner: Begin with quantum channels $\mathsf{C}_{1^\out2^\inp}\geq 0$, which are defined by $\mathcal{P}_\inp[\mathsf{C}_{1^\out2^\inp}] = \mathsf{C}_{1^\out2^\inp} - {}_{2^\inp}\mathsf{C}_{1^\out2^\inp} + {}_{1^\out2^\inp}\mathsf{C}_{1^\out2^\inp} = \mathsf{C}_{1^\out2^\inp}$ and $\tr{\mathsf{C}_{1^\out2^\inp}} = d_{1^\out}$. From this, one readily obtains $1$-slot combs as mappings from channels $\mathsf{C}_{1^\out2^\inp}$ to channels $\mathsf{C}'_{1^\inp 2^\out}$ via the above theorem. Next, $2$-slot combs are obtained as mappings from $1$-slot combs---for which the characterising projector and trace conditions are known from the previous step---to channels; and so forth. 

Similarly, one can readily deduce mappings where the input and output spaces contain only a subset of all possible deterministic objects, which is often the case when considering free operations in higher-order quantum resource theories. For example, when investigating quantum operations with indefinite time direction~\cite{chiribella_quantum_2022}, the set of transformations that map unital channels onto unital channels, emerges naturally as a set of HOQOs of interest. Any unital channel $\mathcal{C}:\Lscr(\Hscr_1) \rightarrow \Lscr(\Hscr_2)$  with $\mathcal{C}[\ident] = \ident$ satisfies $\tr{\mathsf{C}_{12}} = d_{1}$ as well as both
\begin{enumerate}
    \item \textit{Trace Preservation:} $\ptr{2}{\mathsf{C}_{12}} = \ident_{1}$, with corresponding projector $\mathcal{P}^{(\text{tp})}[\mathsf{C}] := \mathsf{C} - {}_2 \mathsf{C} + {}_{12} \mathsf{C}$; and 
    \item \textit{Unitality:} $\ptr{1}{\mathsf{C}_{12}} = \ident_{2}$, with corresponding projector $\mathcal{P}^{(\text{ut})}[\mathsf{C}] := \mathsf{C} - {}_1 \mathsf{C} + {}_{12} \mathsf{C}$.
\end{enumerate}
The combined projector on the set $\Omega$ of unital channels is thus of the form
\begin{gather}
\label{eq::unital_proj}
    \mathcal{P}_\inp[\mathsf{C}] = (\mathcal{P}^{(\text{tp})} \circ \mathcal{P}^{(\text{ut})})[\mathsf{C}] = (\mathcal{P}^{(\text{ut})} \circ \mathcal{P}^{(\text{tp})})[\mathsf{C}] = \mathsf{C} - {}_1\mathsf{C} - {}_2\mathsf{C} + 2({}_{12}\mathsf{C}).
\end{gather}
The output set $\Omega^\prime$ of unital channels $\mathcal{C}':\Lscr(\Hscr_0) \rightarrow \Lscr(\Hscr_3)$ can be described similarly. Characterising the set $\Theta$ of superchannels between the sets $\Omega$ and $\Omega'$ of unital channels is then merely a matter of applying the above Theorem to derive the projector $\mathcal{P}_{\inp\out}$ onto the (span of the) set of HOQOs that map unital channels to unital channels. Interestingly, the resulting set of HOQOs already fails to coincide with the set of $1$-slot combs and contains HOQOs that do not abide by a fixed time direction~\cite{chiribella_quantum_2022} (i.e., that cannot be understood as stemming from a quantum circuit). Finally, we note that while here we have only considered the case where all involved projectors are `well-behaved' (self-adjoint, unital, and commute with transposition)---which is the case almost exclusively found in the literature---similar results can be derived for more general situations where the individual projectors satisfy none of these properties~\cite{milz_characterising_2024}.

To briefly summarise, we have seen that weakening the axiomatic requirements on HOQOs reveals a whole host of transformations that lie outside of the standard quantum comb formalism. Their description in terms of Choi matrices places them all on the same footing and permits the systematic derivation of their properties (at the expense at losing their type information) in terms of projectors and trace conditions~\cite{Araujo_2015, Castro-Ruiz_2018, Milz_2022, simmons_higher-order_2022, hoffreumon_projective_2022, milz_characterising_2024}. As we have seen already for the process matrix case, in many scenarios, HOQOs that go beyond the quantum comb formalism are more than just objects of mathematical interest, but rather used to describe the most general spatiotemporal correlations between distinct laboratories that quantum mechanics allows for. We now conclude this Tutorial with a brief discussion of this aspect of HOQOs and their application for the investigation of causal order in quantum theory.

\FloatBarrier


\subsection{Quantum Processes with Indefinite Causal Order \& Indefinite Time Direction}
\label{subsec::indefinitecausalorder}

Although a crucial part of our everyday experience, the assumption that events occur in a definite causal order or time direction is not explicitly enshrined in the axioms of quantum mechanics. Previously, we saw that if one begins from constructive considerations, i.e., on a dilated space, then one naturally arrives at quantum combs / process tensors, where individual slots---corresponding to events---are sorted in a clear temporal order. However, we have also seen that there is no fundamental problem with defining meaningful HOQOs that lie outside of the set of processes with a fixed causal order. This immediately begs the question: \textit{what types of causal order can quantum mechanics---in principle---allow for?} 

Investigations to this end can be motivated from (at least) two directions. First, it is possible to imagine that the laws of causality in nature might be more general than our intuition would permit---particularly in the regime of quantum gravity---necessitating an understanding of the HOQOs that could describe such scenarios. Second, such analysis allows one to map out the fundamental limits of quantum processes (say, for information theoretic tasks) and enables one to pinpoint the consequences of restricting events to occur in a fixed causal order. In turn, this provides a deeper insight into the nature of causality itself, independent of whether or not processes with `exotic' causal structures truly exist in nature or not. 

Here, we aim to provide a brief introduction to the analysis of causality in quantum mechanics. We begin by discussing how the notion of causal order is encapsulated by signalling properties of the HOQO at hand and analysing some key examples that lie outside the framework of (causally ordered) quantum combs / process tensors. We then discuss three paradigmatic examples of HOQOs that lie outside the quantum comb formalism: the process matrix and the \textit{quantum switch}---which are the standard examples of quantum process with indefinite causal order---as well as the \textit{quantum time flip}, which exhibits indefinite time direction.

\subsubsection{Process Matrices \& Spatiotemporal Signalling Structures}\hfill\\
\label{subsubsec::ico-processmatrices}


\begin{figure}[t]
    \centering
    \includegraphics[width = 0.99\linewidth]{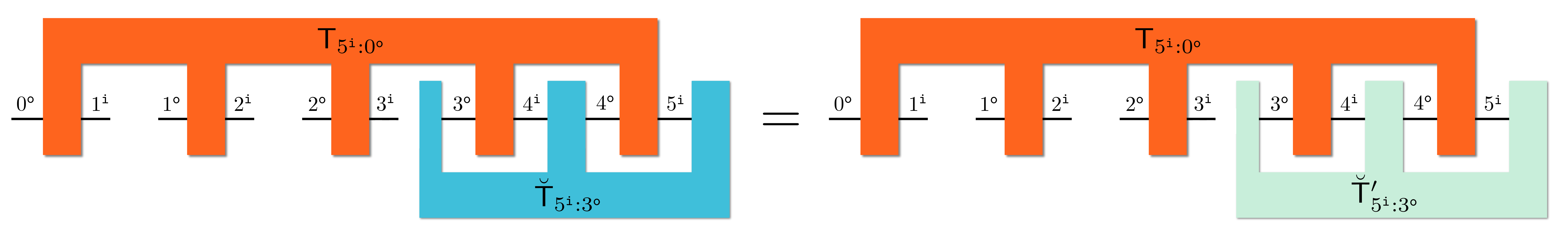}
    \caption{\textbf{Quantum Combs Cannot Signal Backwards in Time.} Whenever a quantum comb $\mathsf{T}_{n+1^\inp:0^\out}$ is contracted with another comb $\breve{\mathsf{T}}_{n+1^\inp:n-k^\out}$ that acts on all of the `final' degrees of freedom of the former, the resulting comb on the `earlier' $k$ times is independent of the latter comb [see Eq.~\eqref{eq::ico-causality-nbts}]. Here, this is shown for $n=4$ and $k=2$; contracting $\mathsf{T}_{n+1^\inp:0^\out}$ with two different combs $\breve{\mathsf{T}}_{5^\inp:3^\out}$ and $\breve{\mathsf{T}}'_{5^\inp:3^\out}$ on the final degrees of freedom yields the same resulting comb $\mathsf{T}_{3^\inp:0^\out} = \mathsf{T}_{5^\inp:0^\out} \star \breve{\mathsf{T}}_{5^\inp:3^\inp} = \mathsf{T}_{5^\inp:0^\out} \star \breve{\mathsf{T}}'_{5^\inp:3^\inp}$.}
    \label{fig::ico-causality_pic}
\end{figure}


\noindent
\textbf{\textul{Quantum Processes with Fixed Causal Order.}} As we briefly discussed below~\ref{def::toqp-def-quantumcombs}---and as is evident from their temporally ordered implementation---quantum circuits and consequently $n$-slot combs naturally display a fixed causal ordering: Any input at a `later' time---say, $t_\ell$---\textit{cannot} influence the resulting comb defined on any earlier times---say, $t_k$ for $k<\ell$. More concretely, contracting an $n$-slot comb $\mathsf{T}_{n+1^\inp:0^\out}$ with any $k$-slot comb\footnote{Here, again,  we employ the breve to signify that the `input' spaces of one comb correspond to the `output' spaces of the other.} $\breve{\mathsf{T}}_{n+1^\inp:n-k^\out}$ that acts on all of the `final' degrees of freedom of the former (see Fig.~\ref{fig::ico-causality_pic}) yields a comb that is \textit{independent} of $\breve{\mathsf{T}}_{n+1^\inp:n-k^\out}$, i.e., 
\begin{align}\label{eq::ico-causality-nbts}
    \mathsf{T}_{n+1^\inp:0^\out} \star \breve{\mathsf{T}}_{n+1^\inp:n-k^\out} \quad \text{is independent of } \breve{\mathsf{T}}_{n+1^\inp:n-k^\out}
\end{align}
as long as $\breve{\mathsf{T}}_{n+1^\inp:n-k^\out}$ is a proper (i.e., deterministic) quantum comb. This can be seen by direct insertion: Since $\breve{\mathsf{T}}_{n+1^\inp:n-k^\out}$ ends on an input space, it must be of the form $\breve{\mathsf{T}}_{n+1^\inp:n-k^\out} = \ident_{n+1^\inp} \otimes \breve{\mathsf{T}}_{n^\out:n-k^\out}$, and so we have 
\begin{align}
 \mathsf{T}_{n+1^\inp:0^\out} \star \breve{\mathsf{T}}_{n+1^\inp:n-k^\out} &= \ptr{n+1^\inp}{\mathsf{T}_{n+1^\inp:0^\out}} \star \breve{\mathsf{T}}_{n^\out:n-k^\out} = (\ident_{n^\out} \otimes \mathsf{T}_{n^\inp:0^\out}) \star \breve{\mathsf{T}}_{n^\out:n-k^\out} \notag \\
 &= \mathsf{T}_{n^\inp:0^\out} \star \ptr{n^\out}{\breve{\mathsf{T}}_{n^\out:n-k^\out}} = \dots = \mathsf{T}_{n-k^\inp:0^\out}, 
\end{align}
where we have alternatingly employed the trace conditions satisfied by quantum combs. Since this expression is independent of $\breve{\mathsf{T}}_{n+1^\inp:n-k^\out}$, we see that indeed any `later' choice of input to the $n$-comb $\mathsf{T}_{n+1^\inp:0^\out}$ cannot influence what is `seen' at the preceding times. Put differently, quantum combs do not allow for signalling from the future to the past, making causal ordering imply a signalling constraint on quantum combs. Notationally, we will denote such signalling structures by either $0^\out \prec 1^\inp \prec \cdots \prec n^\out \prec n+1^\inp$ or $A\prec B \prec C \prec \cdots$, depending on whether we wish to emphasise the individual spaces (corresponding to a single degree of freedom) or the parties (corresponding to an overall slot) that can signal to each other.


Crucially, the above reasoning only holds true if $\breve{\mathsf{T}}_{n+1^\inp:n-k^\out}$ is a \textit{deterministic} $k$-slot comb. In contrast, if it were merely a tester element $\breve{\mathsf{G}}^{(x)}_{n+1^\inp:n-k^\out} \geq 0$, then $\mathsf{T}_{n+1^\inp:0^\out} \star \breve{\mathsf{G}}^{(x)}_{n+1^\inp:n-k^\out}$ would generically depend upon $\breve{\mathsf{G}}^{(x)}_{n+1^\inp:n-k^\out}$. However, since a tester element cannot be implemented deterministically, this situation does \textit{not} amount to signalling backwards in time, which pertains to overall rather than post-selected phenomena. Nonetheless, setups with tester elements can be used to probabilistically \textit{simulate} processes with indefinite causal order~\cite{genkina_optimal_2012, oreshkov_operational_2016, silva_connecting_2017, Milz_2017_NJP}. 

On an abstract level, by identifying, each slot of a quantum comb---let us denote them by $A, B, C, \dots$ for simple notation---as a distinct laboratory with an input and output space, we can think of a quantum comb as the object containing all spatio-temporal quantum correlations between these distinct laboratories. That is, if we do not allow for any additional communication between the parties $A, B, C, \dots$ besides the comb, it fully specifies in what way they can signal to each other, and how correlated their respective measurement outcomes can be. Naturally, in an experiment, different slots of a comb would not necessarily have to correspond to physically distinct laboratories; but rather to, say, measurements made in a single laboratory, but at different points in time. Since each slot of a comb comes equipped with its own Hilbert spaces, mathematically there is no problem with this separation into distinct laboratories. Adopting the viewpoint of distinct laboratories proves fruitful for considering causally disordered processes as the most general admissible HOQOs between separated parties. Here, we focus on the two-party case, i.e., two slots only, since for this case the notion of causal indefiniteness/causal disorder is rather unambiguous and easy to define~\cite{Oreshkov_2012, Araujo_2015}. The multi-party case presents itself much more layered, since it allows for dynamical notions of causal order---i.e., the causal order is determined on-the-fly as the parties act, and their respective actions can potentially influence the causal ordering of the others---leading to a more involved concept of `causally ordered'~\cite{Baumeler_2014_PefectSignalling,Oreshkov_2016, abbott_multipartite_2016,Wechs2019MultipartiteCausality}.

\begin{figure}
    \centering
    \includegraphics[width=0.5\linewidth]{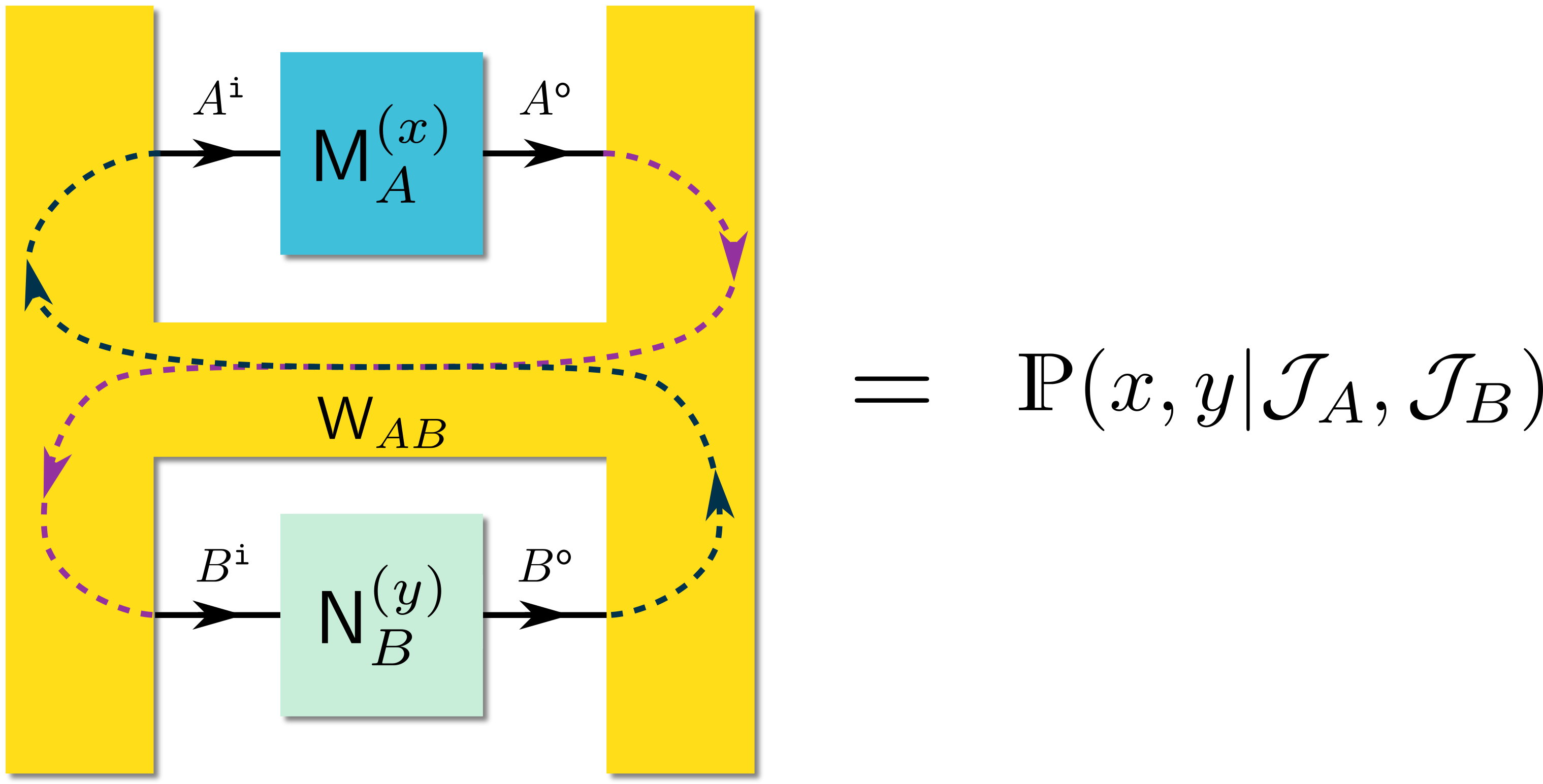}
    \caption{\textbf{Process Matrix Action.} A process matrix $\mathsf{W}_{AB}$ maps pairs of instruments $\mathcal{J}_A = \{ \mathsf{M}_A^{(x)} \}$ and $\mathcal{J}_B = \{ \mathsf{N}_B^{(y)}\}$ to probability distributions $\mathds{P}(x,y|\mathcal{J}_A, \mathcal{J}_B)$. Unlike in the causally ordered case, here there can be both signalling from Alice to Bob (depicted by the navy dotted line) and from Bob to Alice (violet dotted line). Since, overall, probabilities are guaranteed to be normalised whenever $\mathsf{W}_{AB}$ acts on any pair of CPTP maps, such `two-way' signalling does not lead to logical paradoxes.}
    \label{fig::ico-processmatrixaction}
\end{figure}

\vspace{0.25cm} \noindent
\textbf{\textul{Quantum Processes with Indefinite Causal Order.}} While first conceived in the context of computational advantages without definite causal order~\cite{Hardy_2009, Chiribella_2012, Chiribella_2013}, we will here begin with a discussion of causally indefinite quantum processes that is closer in spirit to our previous considerations. Such a discussion was provided in Ref.~\cite{Oreshkov_2012}, in which the \textit{process matrix} formalism---which we have already encountered in Secs.~\ref{subsubsec::me-causalityquantumtheory} and~\ref{subsubsec::axiomatichoqos-qc-transtranstrans}---was introduced. To motivate it, consider two parties, Alice ($A$) and Bob ($B$), who are spatially separated (i.e., they are in distinct laboratories) and assumed to be able to act freely. In their laboratories, they can receive quantum states, manipulate them, and send them forward. Assuming that quantum mechanics holds in both laboratories, the most general (independent) procedures that they could perform consist of local instruments $\Jcal_A = \{\mathcal{M}_{A}^{(a)}: \Lscr(\Hscr_{A^\inp}) \rightarrow \Lscr(\Hscr_{A^\out})\}_{a=1}^{n_A}$ and $\Jcal_B= \{\mathcal{N}_B^{(b)}: \Lscr(\Hscr_{B^\inp}) \rightarrow \Lscr(\Hscr_{B^\out})\}_{b=1}^{n_B}$ on the quantum systems that they receive and subsequently send forward (see Fig.~\ref{fig::ico-processmatrixaction}). Now, we might wonder what the most general spatiotemporal structure that Alice and Bob can be embedded in looks like, i.e., what is the most general set of admissible HOQOs that connect them. We emphasise that `admissible' is a rather important qualifier here; naturally, when going from quantum combs to causally indefinite processes, one should not allow for \textit{arbitrary} processes which might lead to, e.g., unnormalised probability distributions. We will use the term `admissible' to mean that the HOQOs we consider lead to \textit{local causality} being fulfilled, or equivalently, that all possible deducible probability distributions are normalised, i.e., they are positive and sum to unity. 


To introduce these requirements, we note that---as was the case previously---linearity forces one to compute joint outcome probabilities via the generalised spatiotemporal Born rule
\begin{align}\label{eq::ico-processmatrixprobabilityrule}
    \Pprob(a,b|\Jcal_A,\Jcal_B) = \tr{(\mathsf{M}_{A^\inp A^\out}^{(a)} \otimes \mathsf{N}_{B^\inp B^\out}^{(b)})^\mathrm{T} \,\mathsf{W}_{A^\inp A^\out B^\inp B^\out}} = \mathsf{W}_{A^\inp A^\out B^\inp B^\out} \star \mathsf{M}_{A^\inp A^\out}^{(a)} \star \mathsf{N}_{B^\inp B^\out}^{(b)}, 
\end{align}
where the process matrix $\mathsf{W}_{A^\inp A^\out B^\inp B^\out}\in \mathscr{L}(\mathscr{H}_{A^{\inp}}\otimes \mathscr{H}_{A^{\out}} \otimes \mathscr{H}_{B^{\inp}} \otimes \mathscr{H}_{B^{\out}})$ encapsulates all possible spatiotemporal correlations between Alice and Bob. Naturally, one requires that $\Pprob(a,b|\Jcal_A,\Jcal_B) \geq 0$ holds for all CP maps $\mathsf{M}_{A^\inp A^\out}^{(a)} \star \mathsf{N}_{B^\inp B^\out}^{(b)}$. As discussed in Sec.~\ref{subsubsec::axiomatichoqos-qc-transtranstrans}, this condition by itself does \textit{not} suffice to guarantee overall positivity of $\mathsf{W}_{A^\inp A^\out B^\inp B^\out}$, but only when restricted to acting on \textit{pure tensors}~\cite{klay_tensor_1987, Barnum_2005}, i.e., positivity on tensor products of operations that Alice and Bob perform without additional auxiliary spaces. Positivity of $\mathsf{W}_{A^\inp A^\out B^\inp B^\out}$ is enforced by the additional demand that $\mathsf{W}$ yields positive outcomes even when only partially acting on CPTP maps.

Evidently, the process matrix plays a similar role to a quantum comb $\mathsf{T}$ in the causally ordered case, in the sense that it provides a single object from which all observable behaviour [i.e., the probability distribution $\Pprob(a,b|\Jcal_A,\Jcal_B)$] can be computed for any choice of instruments $\Jcal_A,\Jcal_B$. However, a quantum comb further assumes a \textit{global} causal order: Either Alice can send signals to Bob ($A \prec B$) or Bob can send signals to Alice ($B\prec A$), but not both. Here, instead, we merely demand that causal ordering holds \textit{locally}, i.e., in each laboratory individually. Concretely, this means that for whatever deterministic operation either party performs, the resulting HOQO describing the situation in the \textit{other} laboratory is causally ordered. Mathematically, this implies that
\begin{align}
\label{eq::ico-localcausality}
\mathsf{W}_{AB} \star \mathsf{M}_{A} = \rho_{B^{\inp}}^{(\mathsf{M}_{A})} \otimes \ident_{B^{\out}} \ \quad \mathrm{and} \quad \mathsf{W}_{AB}\star \mathsf{N}_B = \eta_{A^{\inp}}^{(\mathsf{N}_B)}  \otimes \ident_{A^{\out}} 
\end{align}
for all CPTP maps $\{\mathsf{M}_A,\mathsf{N}_B\}$, where $\rho_{B^{\inp}}^{(\mathsf{M}_A)}$ and $\eta_{A^{\inp}}^{(\mathsf{M}_B)}$ are quantum states that depend on $\mathsf{M}_A$ and $\mathsf{M}_B$, respectively, and we have set $A := A^\inp A^\out$ and $B := B^\inp B^\out$. We note that it is unnecessary to demand \textit{both} requirements in Eq.~\eqref{eq::ico-localcausality}, since they are equivalent to each other, as can be seen by direct insertion. Moreover, while a $2$-slot quantum comb without global past or future would satisfy similar structural constraints, only one of the states $\rho_{B^{\inp}}^{(\mathsf{M}_{A})}$ and $\eta_{A^{\inp}}^{(\mathsf{M}_B)}$ could depend on the CPTP map applied by the other party. For instance, given a comb $\mathsf{T}_{AB}$ (without global past or future) with causal ordering $A\prec B$, then $\mathsf{T}_{AB} \star \mathsf{M}_A = \rho_{B^{\inp}}^{(\mathsf{M}_{A})} \otimes \ident_{B^\out}$ generally depends on $\mathsf{M}_A$, but $\mathsf{T}_{AB} \star \mathsf{N}_B = \eta_{A^{\inp}} \otimes \ident_{A^\out}$ is independent of the choice of CPTP map $\mathsf{N}_B$, reflecting the fact that Bob cannot signal to Alice in this case.

\vspace{0.25cm} \noindent
\textbf{\textul{Process Matrices and Non-Signalling Channels.}}
The requirements of Eq.~\eqref{eq::ico-localcausality} can equivalently be expressed more concisely as 
\begin{align}
\label{eq::ico-normalised_prob}
    \mathsf{W}_{AB} \star \mathsf{M}_A \star \mathsf{N}_B = 1 \quad \forall \, \mathsf{M}_A,\mathsf{N}_B  \in \mathsf{CPTP}.
\end{align}
This implies that \textit{all} pairs of trace preserving maps are mapped to unit probability by the process matrix $\mathsf{W}_AB$ according to Eq.~\eqref{eq::ico-processmatrixprobabilityrule}. The equivalence between Eq.~\eqref{eq::ico-localcausality} and Eq.~\eqref{eq::ico-normalised_prob} follows by direct insertion: If $\mathsf{W}_{AB}$ satisfies Eq.~\eqref{eq::ico-localcausality}, then $\mathsf{W}_{AB} \star \mathsf{M}_A \star \mathsf{N}_B = (\mathsf{W}_{AB} \star \mathsf{M}_A) \star \mathsf{N}_B = (\rho_{B^{\inp}}^{(\mathsf{M}_{A})} \otimes \ident_{B^{\out}}) \star \mathsf{N}_B = 1$ for all $\mathsf{M}_A, \mathsf{M}_B \in \mathsf{CPTP}$. Conversely, if $\mathsf{W}_{AB}$ satisfies Eq.~\eqref{eq::ico-normalised_prob}, then $(\mathsf{W}_{AB} \star \mathsf{M}_A) \star \mathsf{N}_B = 1$ for all $\mathsf{M}_A, \mathsf{M}_B \in \mathsf{CPTP}$. This can only hold if $(\mathsf{W}_{AB} \star \mathsf{M}_A)$ is of the form $\rho_{B^\inp} \otimes \ident_{B^\out}$ for some unit trace positive semidefinite matrix $\rho_{B^\inp}$; analogous reasoning leads to the second identity in Eq.~\eqref{eq::ico-localcausality}.

Recalling the discussion of HOQOs as transformations between sets of HOQOs, we thus see that the more foundational motivation for process matrices presented in the current section chimes well with the abstract considerations of Sec.~\ref{subsubsec::axiomatichoqos-qc-transtranstrans}. Process matrices $\mathsf{W}_{A^\inp A^\out B^\inp B^\out} \geq 0$ are precisely those operations that map quantum channels $\mathsf{M}_{A^\inp A^\out}$ to $1$-slot combs $\mathsf{S}_{B^\inp B^\out} = \rho_{B^{\inp}}^{(\mathsf{M}_{A})} \otimes \ident_{B^{\out}}$ without global past or future [see Eq.~\eqref{eq::ico-localcausality}], or equivalently, pairs of channels $\mathsf{M}_{A} \otimes \mathsf{N}_{B}$ to unit probability [see Eq.~\eqref{eq::ico-normalised_prob}]. While the former definition agrees exactly with the one discussed in Sec.~\ref{subsubsec::axiomatichoqos-qc-transtranstrans}---thus leading to the characterisation of process matrices given in Eqs.~\eqref{eq::ico-proj_proc_mat} and~\eqref{eq::ico-proj_proc_mat2}---the latter seems slightly at odds with it, since the set of product channels $\mathsf{M}_{A} \otimes \mathsf{N}_{B}$ is \textit{not} defined by a positivity constraint, a projector, and a trace constraint (which can be seen by noting that the set of product maps is not convex). 

Evidently, though, if Eq.~\eqref{eq::ico-normalised_prob} holds for all pairs of CPTP maps, then it follows that
\begin{align}
\label{eq::ico-non_signalling}
    \mathsf{W}_{AB} \star \mathsf{R}_{AB} = 1, \ \ \text{for} \ \ \mathsf{R}_{AB} = \sum_\lambda r_\lambda \mathsf{M}^{(\lambda)}_A \otimes \mathsf{N}^{(\lambda)}_B, \ \ \text{with} \ \ \mathsf{M}^{(\lambda)}_A, \mathsf{N}^{(\lambda)}_B \in \mathsf{CPTP} \quad \text{and} \ \sum_\lambda r_\lambda = 1.
\end{align}
Any quantum channel $\mathsf{R}_{AB}$ that can be written in this way is \textit{non-signalling}, i.e., it cannot be used to send signals from Alice to Bob and \textit{vice versa}. For example, suppose that Alice feeds in an arbitrary quantum state $\rho_{A^\inp}$. The resulting channel that Bob `sees' is given by
\begin{align}
    (\rho_{A^\inp} \otimes \ident_{A^\out}) \star \mathsf{R}_{AB} = \sum_\lambda r_\lambda [(\rho_{A^\inp} \otimes \ident_{A^\out}) \star \mathsf{M}^{(\lambda)}_A] \otimes \mathsf{N}^{(\lambda)}_B = \sum_\lambda r_\lambda  \mathsf{N}^{(\lambda)}_B\, ,
\end{align}
where we have used the fact that each $\mathsf{M}^{(\lambda)}_A$ itself is a quantum channel. Since the above expression is independent of $\rho_{A^\inp}$, the channel that Bob `sees' locally is independent of any state that Alice feeds into the overall channel $\mathsf{R}_{AB}$. Analogous reasoning for Bob then shows that such a channel cannot be used for any signalling between the two parties. 

As it turns out, \textit{all} non-signalling channels $\mathsf{R}_{AB} \geq 0$ between Alice and Bob can be written as an affine combination (i.e., with prefactors that sum up to unity but are not necessarily all positive) of product channels~\cite{Gutoski09, Chiribella_2013}, as in Eq.~\eqref{eq::ico-non_signalling}. Consequently, we obtain yet a third, equivalent definition of (bipartite) process matrices as those HOQOs $\mathsf{W}_{AB} \geq 0$ that map non-signalling channels $\mathsf{R}_{AB}$ to unit probability. In contrast to the set of product channels, the set of non-signalling channels is characterised by a positivity constraint, i.e., $\mathsf{R}_{AB} \geq 0$, a projector $\mathcal{P}_A \otimes \mathcal{P}_B$, where $\mathcal{P}_X$ is the projector onto the (span of the) set of CPTP maps from $\Lscr(\Hscr_{X^{\inp}})$ to $\Lscr(\Hscr_{X^{\out}})$ for $X\in \{A,B\}$ [see the first condition of Eq.~\eqref{eq::ico-projectorchannelandoneslotcomb}], and a trace constraint $\tr{\mathsf{R_{AB}}} = d_{A^\inp}d_{B^\inp}$. Consequently, the properties of $\mathsf{W}_{AB}$ can again be deduced by using the methods presented in Sec.~\ref{subsubsec::axiomatichoqos-qc-transtranstrans}.\footnote{Since the output of $\mathsf{W}_{AB}$ when acting on non-signalling maps is equal to $1$, the projector on the output space is the trivial identity projector $\Ical$, and we have $\tr{1} = 1$  as the (trivial) trace condition on the output space.} We also see quite clearly that process matrices form a strict superset of the set of $2$-slot quantum combs without global past and global future: Such $2$-slot combs $\mathsf{T}$ (with order $A \prec B$ w.l.o.g.) must map \textit{any} $1$-slot comb $\breve{\mathsf{T}}$ with order $A\prec B$ to $1$ (put differently, they must be compatible with `remote connection'~\cite{Chiribella_2009}, i.e., yield valid probability distributions even if the respective parties can send signals to each other outside the comb $\mathsf{T}$). Since such $1$-slot quantum combs $\breve{\mathsf{T}}$ clearly form a strict superset of the set of non-signalling channels (as they can signal one way), the corresponding set of admissible $2$-slot combs $\mathsf{T}$ must be strictly smaller than that of process matrices. 

Finally, the understanding of process matrices as the set of positive matrices that map non-signalling maps to the value $1$---i.e., process matrices form the \textit{affine dual} of the set of non-signalling channels---also allows for a straightforward extension to the multipartite process matrix case~\cite{Araujo_2015}. In particular, a multipartite quantum channel $\mathsf{R}_{ABC\cdots} \geq 0$ that does not allow for any signalling between the involved parties is characterised by the projector $\mathcal{P}_A \otimes \mathcal{P}_B \otimes \mathcal{P}_C \otimes \cdots $, where $\mathcal{P}_X$ is the projector onto the (span of the) set of CPTP maps from $\Lscr(\Hscr_{X^{\inp}})$ to $\Lscr(\Hscr_{X^{\out}})$ for $X\in \{A,B, C, \dots\}$ [see the first condition of Eq.~\eqref{eq::ico-projectorchannelandoneslotcomb}], and the trace condition $\tr{\mathsf{R}_{ABC\cdots}} = d_{A^\inp}d_{B^\inp}d_{C^\inp}\cdots$. The set of multipartite process matrices is then the set of all matrices $\mathsf{W}_{ABC\cdots}$ that satisfy~\cite{Araujo_2015}
\begin{align}
    \mathsf{W}_{ABC\cdots} \geq 0 \quad \text{and} \quad \mathsf{W}_{ABC\cdots} \star \mathsf{R}_{ABC\cdots} = 1 \quad \text{for all non-signalling channels} \ \mathsf{R}_{ABC\cdots}\, .
\end{align}
While we will not discuss the multipartite case in detail, we note that the properties of such multipartite process matrices could again be straightforwardly derived by means of the projector based approach presented in Sec.~\ref{subsubsec::axiomatichoqos-qc-transtranstrans}~\cite{Araujo_2015, milz_characterising_2024}. 


\subsubsection{Witnessing Causal Indefiniteness}\hfill\\
\label{subsubsec::ico-witnessingcausalindefiniteness}


\noindent Up to this point, we have emphasised that process matrices can lie outside of the set of $2$-slot combs without a global past and future due to the weaker axiomatic requirements that the former must satisfy. Such process matrices that go beyond the quantum comb formalism do not abide by a \textit{fixed} causal order and cannot be represented by an underlying quantum circuit. However, this fact alone does not render them causally indefinite \textit{per se}. To see this, consider a process that is sometimes (with probability $p$) given by a quantum comb $\mathsf{T}^{A\prec B}$ with causal ordering $A \prec B$, and the rest of the times (with probability $1-p$) by a quantum comb $\mathsf{T}^{B\prec A}$ with causal ordering $B\prec A$. This happens, for instance, in situations where one flips a coin (with probability $p$ for heads and probability $1-p$ for tails) in each run of an experiment to choose between two causally ordered circuits: one in which Alice comes before Bob, and the other in which Bob comes before Alice. The overall HOQO describing this scenario is
\begin{align}
\label{eqn::ico-caus_sep}
    \mathsf{W}^{\text{cs}}_{AB} = p\mathsf{T}^{A\prec B} + (1-p)\mathsf{T}^{B\prec A}. 
\end{align}

HOQOs $\mathsf{W}^{\text{cs}}_{AB}$ that can be written as such a convex combination of processes with a fixed causal order are called \textbf{causally separable (cs)}. These processes typically lie outside of the set of quantum combs and allow for signalling in two directions. In other words, both $\mathsf{W}^{\text{cs}}_{AB} \star \mathsf{M}_A$ and $\mathsf{W}^{\text{cs}}_{AB} \star \mathsf{N}_B$ generally depend on the CPTP maps $\mathsf{M}_A$ and $\mathsf{N}_B$. Nonetheless, there is nothing particularly exotic about their causal structure, since they can be implemented by means of standard quantum circuits and a coin flip that decides which of the circuits is used in each run of the experiment; crucially, in each run, the process has a clear causal ordering. Consequently, we shall only call process matrices that lie outside of the convex hull of causally ordered processes \textbf{causally indefinite / non-separable (cns)}:
{\hypersetup{citecolor=white}
\begin{myDefinition}{Causally Indefinite / Non-Separable Process Matrices~\cite{Oreshkov_2012}}{}
A (bipartite) process matrix $\mathsf{W}_{AB}$ is causally indefinite / non-separable iff it cannot be written as a convex combination of causally ordered quantum combs, i.e., 
\begin{align}
    \mathsf{W}_{AB} \neq p\mathsf{T}^{A\prec B} + (1-p)\mathsf{T}^{B\prec A}\, ,
\end{align}
where $\mathsf{T}^{X\prec Y} \geq 0$, with $\mathsf{T}^{X\prec Y} = \ident_{Y^\out} \otimes \mathsf{T}_{X^\inp X^\out Y^\inp}$ and $\ptr{Y^\inp}{\mathsf{T}_{X^\inp X^\out Y^\inp}} = \rho_{X^\inp} \otimes \ident_{X^\out}$ for $X,Y \in \{A,B\}$, with $\rho_{X^\inp}$ a quantum state.
\end{myDefinition}
}
\noindent Interestingly, there indeed exist causally non-separable process matrices~\cite{Oreshkov_2012}; \textit{a priori}, quantum mechanics therefore allows for processes that lie completely outside of the paradigm of causal order (even when probabilistic mixtures of causal orders are permitted). 

An example of such a causally indefinite process matrix on qubit systems is $\mathsf{W}^{\text{OCB}}_{A^\inp A^\out B^\inp B^\out}$ (named after the authors of the paper where it was introduced---Oreshkov, Costa, and Brukner~\cite{Oreshkov_2012}), which is of the form
\begin{align}
\label{eq::WOCB}
    \mathsf{W}^{\text{OCB}}_{A^\inp A^\out B^\inp B^\out} = \frac{1}{4}\left[\ident_{A^\inp A^\out B^\inp B^\out} + \frac{1}{\sqrt{2}} \left( \ident_{A^\inp} \otimes \sigma^z_{A^\out} \otimes \sigma^z_{B^\inp} + \sigma^z_{A^\inp}\otimes \ident_{A^\out} \otimes \sigma^x_{B^\inp} \otimes  \sigma^z_{B^\out}\right)\right]\, ,
\end{align}
    where $\sigma^x := \ketbra{0}{1} + \ketbra{1}{0}$ and $\sigma^z := \ketbra{0}{0} - \ketbra{1}{1}$ are respectively the Pauli-$x$ and Pauli-$z$ matrices. It can be shown by direct insertion [e.g., into Eqs~\eqref{eq::ico-proj_proc_mat} and~\eqref{eq::ico-proj_proc_mat2}] that $\mathsf{W}^{\text{OCB}}_{A^\inp A^\out B^\inp B^\out}$ is indeed a valid process matrix. To prove that it is causally non-separable, there are two possible ways. 
    
    For the first one, note that the set $\Theta^{\text{cs}}$ of causally separable process matrices is, by construction, convex. Consequently, via the hyperplane separation theorem, for any causally non-separable $\mathsf{W}^{\text{cns}}_{A B} \notin \Theta^{\text{cs}}$ there exists a \textit{causal witness} $\mathsf{D}_{AB}$ such that $\tr{\mathsf{W}^{\text{cns}}_{A B} \mathsf{D}_{AB}} < 0$, while $\tr{\mathsf{W}^{\text{cs}}_{A B} \mathsf{D}_{AB}} \geq 0$ for all causally separable $\mathsf{W}^{\text{cs}}_{A B} \in \Theta^{\text{cs}}$. As such, if minimising the quantity $\tr{\mathsf{W}_{A B} \mathsf{D}_{AB}}$ over all valid witnesses---i.e., all $\mathsf{D}_{AB}$ that satisfy $\tr{\mathsf{W}^{\text{cs}}_{A B} \mathsf{D}_{AB}} \geq 0$ for all $\mathsf{W}^{\text{cs}}_{A B} \in \Theta^{\text{cs}}$---yields a negative value, one can be certain that $\mathsf{W}_{A B}$ is causally non-separable. Since, in the bipartite case, the definition of causal separability is relatively simple [see Eq.~\eqref{eqn::ico-caus_sep}], the set of causal witnesses can be characterised in terms of conic constraints and the minimisation of $\tr{\mathsf{W}_{A B} \mathsf{D}_{AB}}$ can be phrased as a \textbf{semidefinite program (SDP)}~\cite{Skrzypczyk_2023} that can readily be numerically solved to high precision~\cite{Araujo_2015,Branciard_2016}: 
\begin{align}
\begin{split}
\textbf{given} \ \ & \mathsf{W}_{AB} \\ 
\textbf{minimise} \ \ & \tr{\mathsf{W}_{AB} \mathsf{D}_{AB}},  \\
\textbf{subject to} \ \ & \mathsf{D}_{AB} = \mathcal{P}_{AB}[\mathsf{G}_{AB}], \ \ {}_{A^\out}\mathsf{G}_{AB} \geq 0, \ \ {}_{B^\out}\mathsf{G}_{AB} \geq 0, \\
& \frac{\ident_{AB}}{d_{A^\out}d_{B^\out}} - \mathsf{D}_{AB} = \mathcal{P}_{AB}[\mathsf{J}_{AB}], \ \mathsf{J}_{AB} \geq 0, 
\end{split}
\end{align}
where $\mathcal{P}_{AB}$ is the projector onto (the span of) the set of process matrices given in Eq.~\eqref{eq::ico-proj_proc_mat}. We forego the derivation of this SDP here (the full derivation can be found in Ref.~\cite{Araujo_2015}) but note that it is easy to see that any such witness $\mathsf{D}_{AB}$ satisfies $\tr{\mathsf{D}_{AB} \mathsf{W}^{\text{cs}}_{AB}} \geq 0$ for all $\mathsf{W}^{\text{cs}}_{AB} \in \Theta^{\text{cs}}$, since 
\begin{align}
    \tr{\mathsf{D}_{AB} \mathsf{W}^{\text{cs}}_{AB}} &= \tr{\mathcal{P}_{AB}[\mathsf{G}_{AB}] \mathsf{W}^{\text{cs}}_{AB}} = \tr{\mathsf{G}_{AB} \mathcal{P}_{AB}[\mathsf{W}^{\text{cs}}_{AB}}] \notag \\
    &= p\, \tr{\mathsf{G}_{AB} \mathsf{T}^{A\prec B}} + (1-p) \, \tr{\mathsf{G}_{AB} \mathsf{T}^{B\prec A}} \notag \\
    &=  p \, \tr{\mathsf{G}_{AB} ({}_{B^\out}\mathsf{T}^{A\prec B})} + (1-p) \, \tr{\mathsf{G}_{AB} ({}_{A^\out}\mathsf{T}^{B\prec A})} \notag \\
    &= p \, \tr{{}_{B^\out}\mathsf{G}_{AB} \mathsf{T}^{A\prec B}} + (1-p) \, \tr{{}_{A^\out}\mathsf{G}_{AB} \mathsf{T}^{B\prec A}} \geq 0, 
\end{align}
where we have used the invariance of $\mathsf{W}^{\text{cs}}_{AB}$ under the projector $\mathcal{P}_{AB}$, the fact that $\mathcal{P}_{AB}$ and ${}_{X^\out}\bullet$ can be moved around freely within the trace, as well as that $\mathsf{T}^{X\prec Y} ={}_{Y^\out} \mathsf{T}^{X\prec Y}$ and ${}_{X^\out}\mathsf{G}_{AB} \geq 0$ for $X,Y \in \{X,Y\}$. The remaining restrictions on $\mathsf{D}_{AB}$, i.e., the last line in the above SDP, that we did not use in this reasoning, ensure that the value of $\tr{\mathsf{D}_{AB}\mathsf{W}_{AB}}$ is bounded from below~\cite{Araujo_2015}. The above program can be solved analytically and yields a value of $\text{min}\left(\tr{\mathsf{D}_{AB} \mathsf{W}_{AB}^{\text{OCB}}}\right) = 1 -\sqrt{2} < 0$, demonstrating that $\mathsf{W}_{AB}^{\text{OCB}}$ is indeed causally non-separable~\cite{Araujo_2015, Branciard_2016}. 

The notion of causal non-separability is fundamentally device-dependent; the fact that $\mathsf{W}_{AB}^{\text{OCB}}$ is causally non-separable means that there exist no two causally ordered processes $\mathsf{T}^{A\prec B}$ and $\mathsf{T}^{B\prec A}$ on qubits such that $\mathsf{W}_{AB}^{\text{OCB}}$ can be written as a convex combination of them. This does however not necessarily mean that the statistics $\Pprob(a,b|\Jcal_A,\Jcal_B)$ that can be obtained from $\mathsf{W}_{AB}^{\text{OCB}}$ cannot be reproduced in some causally ordered way. This distinction is akin to the difference between entanglement and the violation of Bell inequalities: A (bipartite) quantum state is entangled iff it cannot be written as a convex combination of product states, but this does not imply that there does not exist a local hidden-variable model that reproduces the measurement statistics exhibited~\cite{werner_quantum_1989}. Put differently, any quantum state that does not admit a local hidden-variable model is necessarily entangled, but the converse is not true. Similarly, the violation of~\textit{causal inequalities}---i.e., inequalities that are satisfied by \textit{any} causally definite process---provides a stronger, device-independent notion of causal indefiniteness~\cite{Oreshkov_2012, baumeler_maximal_2014, branciard_simplest_2015}. Indeed,  although all processes that violate a causal inequality must be causally non-separable, there exist causally non-separable processes that do not violate any causal inequalities~\cite{Araujo_2015, feix_causally_2016, purves_quantum_2021,Wechs_2021}, such as the quantum switch (see below). We will not go into the details of the derivation of causal inequalities here (see Ref.~\cite{branciard_simplest_2015} for a collation of the simplest causal inequalities), but remark that $\mathsf{W}_{AB}^\text{OCB}$ is not only causally non-separable, but also violates a causal inequality (which we will call the \textit{OCB inequality} below)~\cite{Oreshkov_2012}, thereby demonstrating that its statistics are impossible to be reproduced in a causally definite manner. Additionally, relaxing the underlying assumptions that lead to causal inequalities, one can derive broader classes of device-independent and semi-device independent certifications~\cite{Bavaresco_2019,Dourdent_2022,Dourdent_2024,Lugt_2023,vanderLugt2024possibilistic} of causal indefiniteness; we discuss some of these ideas in more detail in Sec.~\ref{subsubsec::cqf-quantumswitch}.

While `highly causally indefinite' processes like $\mathsf{W}^\text{OCB}_{AB}$ are admissible from a mathematical and foundational perspective, it remains \textit{a priori} unclear how they arise in nature and whether they can be faithfully implemented experimentally. Furthermore, their inherently strong causal indefiniteness begs the question whether or not they are potentially excluded by means of introducing reasonable additional postulates to quantum mechanics~\cite{Araujo_2017}. We finish this exposition on causally indefinite processes with a discussion of the \textit{quantum switch}~\cite{Chiribella_2013}, an HOQO with two slots---similar to process matrices---but with an additional global past and future.


\subsubsection{Indefinite Causal Order: Quantum Switch}\hfill\\
\label{subsubsec::ico-quantumswitch}

\begin{figure}
    \centering
    \includegraphics[width=0.4\linewidth]{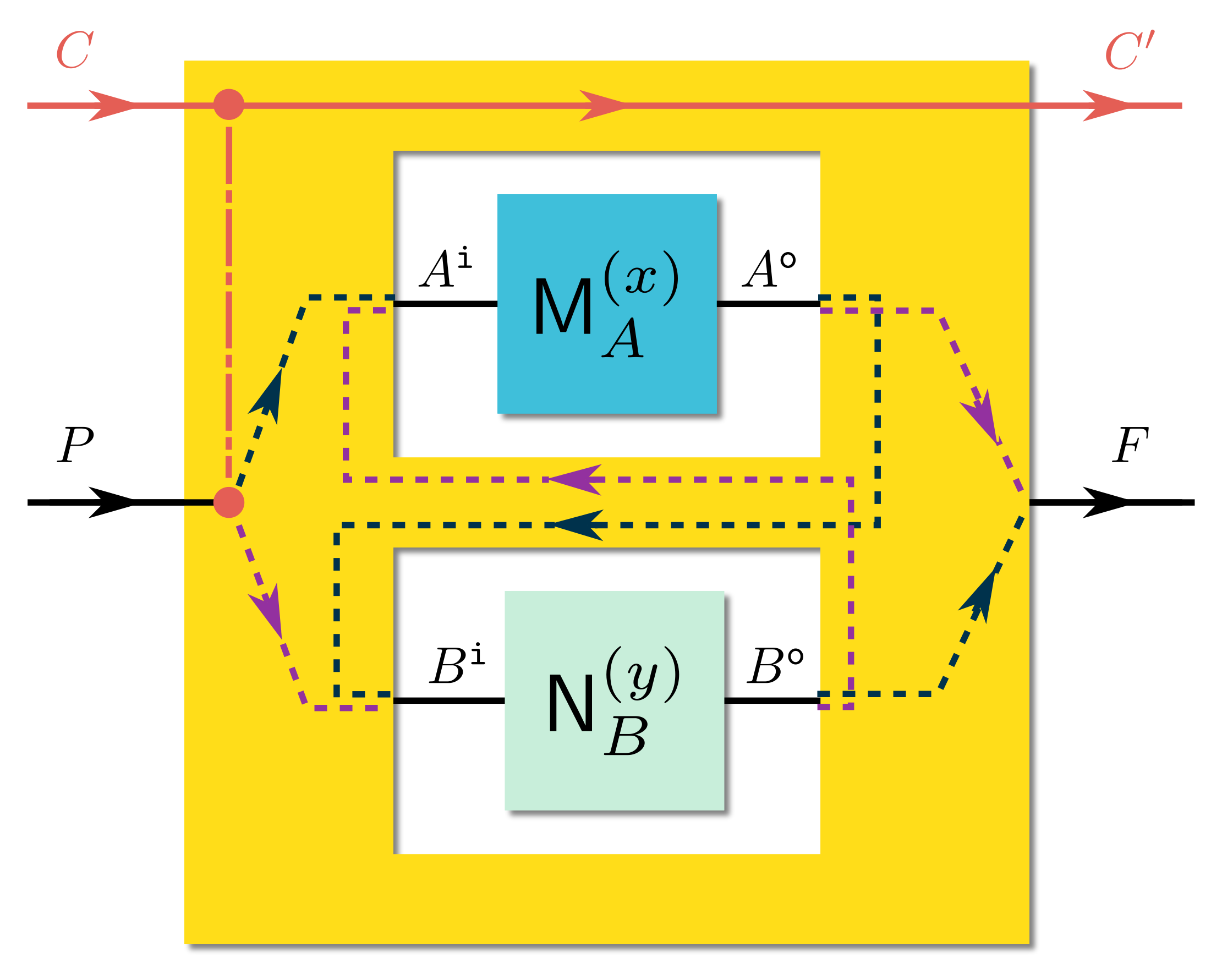}
    \caption{\textbf{Quantum Switch.} Depending on the state of the control $C$, any state that is inserted on system $P$ either goes first through Alice's laboratory and then through Bob's (if the control is in state $\ket{0_C}$; navy dotted path) or vice versa (if the control is in state $\ket{1_C}$; violet dotted path). The influence of the control on the causal path is depicted by the red dotted line: If the control is in a superposition of $\ket{0_C}$ and $\ket{1_C}$, then the two paths are taken in superposition and the final state of the control system $C'$ is generally entangled with that of the system $F$.}
    \label{fig::ico-quantumswitch}
\end{figure}

\noindent
\textbf{\textul{Causally Indefinite HOQOs with Global Past and Future.}} Thus far, the HOQOs we have considered for the analysis of causal order (or lack thereof) had two slots, but no global past or future. We will now go beyond this scenario and allow for an additional input that lies in the global past of both slots, and an additional output that lies in the future of both slots (see Fig.~\ref{fig::ico-quantumswitch}). For reasons that will become clear shortly, we split the global past in two systems $P$ and $C$ (for past and control) and the global future in two systems $F$ and $C'$ (for future and control). We denote the corresponding HOQOs on the two slots plus past and future as $\mathsf{W}_{PCA^\inp A^\out B^\inp B^\out F C'}$ and will shorten or omit the subscript whenever there is no risk of confusion. Axiomatically, in a similar vein to the process matrix case, we make the minimal demand that $\mathsf{W} \geq 0$ maps all pairs of CPTP maps $\mathsf{M}_A \otimes \mathsf{N}_B$ onto a CPTP map from the global past to the global future. Put differently, 
\begin{align}
\label{eq::ico-Switchlike}
    \mathsf{W}_{PCA^\inp A^\out B^\inp B^\out F C'} \star \mathsf{M}_A \star \mathsf{N}_B =: \mathsf{K}_{PCFC'}
\end{align} 
is the Choi state of a CPTP map $\mathcal{K}: \Lscr(\Hscr_{P} \otimes \Hscr_{C}) \rightarrow \Lscr(\Hscr_{F} \otimes \Hscr_{C'})$. For the case where global past and global future are trivial, i.e., $\Hscr_{P} = \Hscr_{C} = \Hscr_{F} = \Hscr_{C'} = \mathds{C}$, this definition coincides with that of process matrices [see Eq.~\eqref{eq::ico-normalised_prob}]. In addition, it ensures that both Alice and Bob locally `see' a causally ordered process. For instance, if Alice implements a CPTP map $\mathsf{M}_A$, the remaining HOQO is given by $\mathsf{S}_{PCA^\inp B^\out F C'} := \mathsf{W}_{PCA^\inp A^\out B^\inp B^\out F C'} \star \mathsf{M}_A$. From Eq.~\eqref{eq::ico-Switchlike}, it follows that $\mathsf{S}_{PCA^\inp B^\out F C'}$ maps \textit{every} CPTP map $\mathsf{N}_B$ onto a CPTP map $\mathsf{K}_{PCFC'}$ and is thus a valid $1$-slot comb, i.e., a HOQO with a fixed causal order (see Def.~\ref{def::axiomatichoqos-qc-def-axiomaticcombs}). 


Analogous to the case of process matrices, Eq.~\eqref{eq::ico-Switchlike} can be equivalently phrased as a mapping from non-signalling channels $\mathsf{R}_{AB}$ to (general) channels $\mathsf{K}_{PCFC'}$, making it amenable to a characterisation of the HOQOs $\mathsf{W}_{PCA^\inp A^\out B^\inp B^\out F C'}$ in terms of a positivity constraint, a projector, and a trace constraint (see Sec.~\ref{subsubsec::axiomatichoqos-qc-transtranstrans}). In particular, if $\mathsf{W}\geq 0$ is a valid HOQO [according to Eq.~\eqref{eqn::def_switchlike}], then
\begin{gather}
\label{eqn::def_switchlike}
    \mathsf{K}_{PCFC'} := \mathsf{W}_{PCA^\inp A^\out B^\inp B^\out F C'} \star \mathsf{R}_{AB}
\end{gather}
is a quantum channel for all non-signalling channels $\mathsf{R}_{AB}$. Since we have already encountered the projectors onto the (span of the) set of non-signalling channels $\mathsf{R}_{AB}$, as well as the (span of the) set of CPTP maps $\mathsf{K}_{PCFC'}$, we could readily deduce the characterisation of all such admissible HOQOs $\mathsf{W}$ using the procedure outlined in Sec.~\ref{subsubsec::axiomatichoqos-qc-transtranstrans}. However, rather than providing this general derivation, we leave the characterisation of all admissible HOQOs $\mathsf{W}_{PCA^\inp A^\out B^\inp B^\out F C'}$ as an exercise to the Reader and instead turn our attention towards a particular HOQO: the \textit{quantum switch}.

\vspace{0.25cm}\noindent
\textbf{\textul{Quantum Switch.}} The quantum switch $\mathsf{S}_{PCA^\inp A^\out B^\inp B^\out F C'}$ is a particularly insightful example of a HOQO with two slots and a global past and global future. It was introduced in Ref.~\cite{Chiribella_2013} in the context of quantum computation without causal order and its advantages for information theoretic tasks have been investigated both theoretically~\cite{Chiribella_2012,araujo_computational_2014} as well as experimentally~\cite{procopio_experimental_2015,goswami_indefinite_2018,rubino2017experimental,Goswami_2020,Rozema_2024}. We refer the Reader to Sec.~\ref{subsubsec::cqf-quantumswitch} for more applications of the quantum switch and Sec.~\ref{subsubsec::ced-experimentaldemonstrations} for more on the experimental realisations of the quantum switch. 

In the two-party, the quantum switch permits the coherent superposition of causal orders. In particular, the quantum switch $\mathsf{S} := \kketbra{\mathsf{S}}{\mathsf{S}}$ is of the form
\begin{align}
    \kket{\mathsf{S}} := \ket{0_C}\ket{\Phi^+_{PA^\inp}} \ket{\Phi^+_{A^\out B^\inp}} \ket{\Phi^+_{B^\out F}} \ket{0_{C'}} + \ket{1_C}\ket{\Phi^+_{PB^\inp}} \ket{\Phi^+_{B^\out A^\inp}} \ket{\Phi^+_{A^\out F}} \ket{1_{C'}},  
\end{align}
where all systems are qubits. By direct (albeit somewhat lengthy) calculation one can check that this HOQO indeed maps pairs of pairs of CPTP maps to a CPTP map (we provide a simpler proof below). In addition, since $\mathsf{S} = \kketbra{\mathsf{S}}{\mathsf{S}}$ is rank-$1$ and does not abide by a \textit{fixed} causal order (it does not satisfy the trace conditions of Def.~\ref{def::toqp-def-quantumcombs}), it cannot be represented as a convex combination of causally ordered processes and is therefore causally non-separable (as for the case of process matrices, the causal non-separability of the quantum switch can also be shown via causal witnesses~\cite{Chiribella_2013, Araujo_2015, rubino2017experimental, goswami_indefinite_2018}). 

To investigate the behaviour of the quantum switch, we note that the process displays two different causal orderings, depending on the initial state of the control. In particular, we have 
\begin{align}
    \mathsf{S} \star \ketbra{0_C}{0_C} =  \Phi^+_{PA^\inp} \otimes \Phi^+_{A^\out B^\inp} \otimes \Phi^+_{B^\out F} \otimes \ketbra{0_{C'}}{0_{C'}}.
\end{align}
Since $\Phi^+$ is the Choi state of the identity channel, the above expression corresponds to an identity channel from $P$ to $A^\inp$, followed by an identity channel from $A^\out$ to $B^\inp$ and finally an identity channel from $B^\out$ to $F$, while the control qubit remains unchanged. That is, the process exhibits a causal ordering $P\prec A \prec B \prec FC'$ if the control qubit is initialised as $\ket{0_C}$. On the other hand, if the control qubit is initialised as $\ket{1_C}$, we yield
\begin{align}
    \mathsf{S} \star \ketbra{1_C}{1_C} =  \Phi^+_{PB^\inp} \otimes \Phi^+_{B^\out A^\inp} \otimes \Phi^+_{A^\out F} \otimes \ketbra{1_{C'}}{1_{C'}}.
\end{align}
That is, we have the same process (i.e., identity channels between the laboratories), but in this case the causal ordering is $P\prec B \prec A \prec FC'$. 

This phenomenon can be succinctly expressed in terms of Kraus operators. If $\mathsf{M}_A = \sum_\alpha \kketbra{K^{(\alpha)}_A}{K^{(\alpha)}_A}$ and $\mathsf{N}_B = \sum_\beta \kketbra{L^{(\beta)}_B}{L^{(\beta)}_B}$ are CPTP maps with respective Kraus operators $\{K_A^{(\alpha)}: \Hscr_{A^\inp} \rightarrow \Hscr_{A^\out}\}$ and $\{L_B^{(\beta)}: \Hscr_{B^\inp} \rightarrow \Hscr_{B^\out}\}$, then
\begin{align}
    \kketbra{\mathsf{S}}{\mathsf{S}} \star \mathsf{M}_A \star \mathsf{N}_B &= \sum_{\alpha, \beta}  \kketbra{\mathsf{S}}{\mathsf{S}} \star \kketbra{K^{(\alpha)}_A}{K^{(\alpha)}_A} \star \kketbra{L^{(\beta)}_B}{L^{(\beta)}_B} \notag \\
    &= \sum_{\alpha, \beta}  \left(\bbra{K^{(\alpha)*}_A}\bbrakket{L^{(\beta)*}_B | \mathsf{S}} \right) \left(\bbrakket{\mathsf{S}|L^{(\beta)*}_B } \kket{K_A^{(\alpha)*}}\right) =: \sum_{\alpha, \beta}  \kketbra{S^{(\alpha\beta)}}{S^{(\alpha\beta)}}. 
\end{align}
The resulting Kraus operators $\{S^{(\alpha\beta)}: \Hscr_{P} \otimes \Hscr_C \rightarrow \Hscr_{F} \otimes \Hscr_{C'}\}$ of $\mathsf{S} \star \mathsf{M}_A \star \mathsf{N}_B$ can be derived by inserting the definition of $\kket{\mathcal{S}}$ into the above equation, which gives
\begin{align}
    S^{(\alpha\beta)} = \ketbra{0_{C'}}{0_C} \otimes L^{(\beta)} K^{(\alpha)} + \ketbra{1_{C'}}{1_C} \otimes K^{(\alpha)} L^{(\beta)}, 
\end{align}
with the understanding that $K^{(\alpha)}$ maps from $\Hscr_{P}$ to $\Hscr_{B^\inp}$ in the first term, and from $\Hscr_{A^\inp}$ to $\Hscr_F$ in the second, while $L^{(\beta)}$ maps from  $\Hscr_{P}$ to $\Hscr_{A^\inp}$ in the second term, and from $\Hscr_{B^\inp}$ to $\Hscr_F$ in the first. With this, it can easily be checked that $\sum_{\alpha \beta}S^{(\alpha\beta)\dagger} S^{(\alpha\beta)} = \ident$ holds, i.e., the quantum switch indeed maps pairs of quantum channels to quantum channels. 

Perhaps the most interesting behaviour of the switch occurs when, instead of $\ket{0_C}$ or $\ket{1_C}$, one feeds a superposition such as $\ket{+}_C := \tfrac{1}{\sqrt{2}} (\ket{0_C}+\ket{1_C})$ into the control. In this case, the two different causal orders from above occur in superposition, enabling the aforementioned information processing advantages that the quantum switch permits (see Sec.~\ref{subsubsec::cqf-quantumswitch}). We finish with a brief analysis of what happens when the final control system $C'$ is discarded (i.e., traced out). In this case, we obtain
\begin{align}
\label{eq::conv_mix_switch}
    \ptr{C'}{\mathsf{S}} =   \ketbra{0_{C}}{0_{C}} \otimes \Phi^+_{PA^\inp} \otimes \Phi^+_{A^\out B^\inp} \otimes \Phi^+_{B^\out F} + \ketbra{1_{C}}{1_{C}} \otimes \Phi^+_{PB^\inp} \otimes \Phi^+_{B^\out A^\inp} \otimes \Phi^+_{A^\out F}, 
\end{align}
i.e., a \textit{classical switch} between two causally ordered processes. For \textit{any} input state $\rho_C$ of the control, the classical switch yields
\begin{gather}
\ptr{C'}{\mathsf{S}} \star \rho_C = \braket{0|\rho_C|0} \cdot \Phi^+_{PA^\inp} \otimes \Phi^+_{A^\out B^\inp} \otimes \Phi^+_{B^\out F} + \braket{1|\rho_C|1} \cdot \Phi^+_{PB^\inp} \otimes \Phi^+_{B^\out A^\inp} \otimes \Phi^+_{A^\out F},
\end{gather}
which is a convex mixture of the two causally ordered processes $\Phi^+_{PA^\inp} \otimes \Phi^+_{A^\out B^\inp} \otimes \Phi^+_{B^\out F}$ (with order $P\prec A \prec B \prec F$) and $\Phi^+_{PB^\inp} \otimes \Phi^+_{B^\out A^\inp} \otimes \Phi^+_{A^\out F}$ (with order $P \prec B \prec A \prec F$). Consequently, as soon as the final control qubit is discarded, the causal non-separability of the quantum switch is lost---and with it, many of the advantages that the switch provides over causally ordered processes~\cite{Chiribella_2013}. This also means that the quantum switch is the purification of a rather benign process, the mixture of an identity process in one direction with an identity process in the other. This is in contrast to the process matrix $\mathsf{W}^{\text{OCB}}$ we encountered above [see Eq.~\eqref{eq::WOCB}], which is not the purification of a causally separable process. Since in quantum mechanics, purifications of admissible objects are generally themselves admissible, this fact provides yet another justification for the quantum switch as an interesting and physically relevant object. 

\subsubsection{Indefinite Time Direction: Quantum Time Flip}\hfill\\
\label{subsubsec::ico-quantumtimeflip}

\noindent Both process matrices as well as mappings from (bipartite) non-signalling channels to channels (like the quantum switch) are examples of HOQOs with two slots that can display causal indefiniteness. This $2$-slot case is the simplest possible example of causal indefiniteness---since at least two parties are required for any kind of causal `ordering'. Additionally, as we have seen, \textit{any} HOQO with only one slot that maps quantum channels to quantum channels admits a Stinespring dilation and can thus be represented by a causally ordered quantum circuit (see Sec.~\ref{subsubsec::axiomatichoqos-qc-onetwoslotcombs}), which seemingly implies that at least two slots are required for HOQOs to possibly lie outside of the quantum comb formalism. However, even in the $1$-slot scenario, meaningful HOQOs beyond the comb formalism exist if we \textit{restrict} the respective input and output spaces, e.g., to subsets of quantum channels. In turn, this restriction can lead to HOQOs with indefinite \textit{time direction}~\cite{chiribella_quantum_2022}---as opposed to indefinite \textit{causal order}, as we saw for process matrices and the quantum switch---which we now discuss in more detail to conclude this section.

\begin{figure}[t]
\centering
\hspace{0.5cm}
\subfigure[\textbf{HOQOs that Map Unital Channels to Unital Channels.}]
{
\includegraphics[scale=0.75]{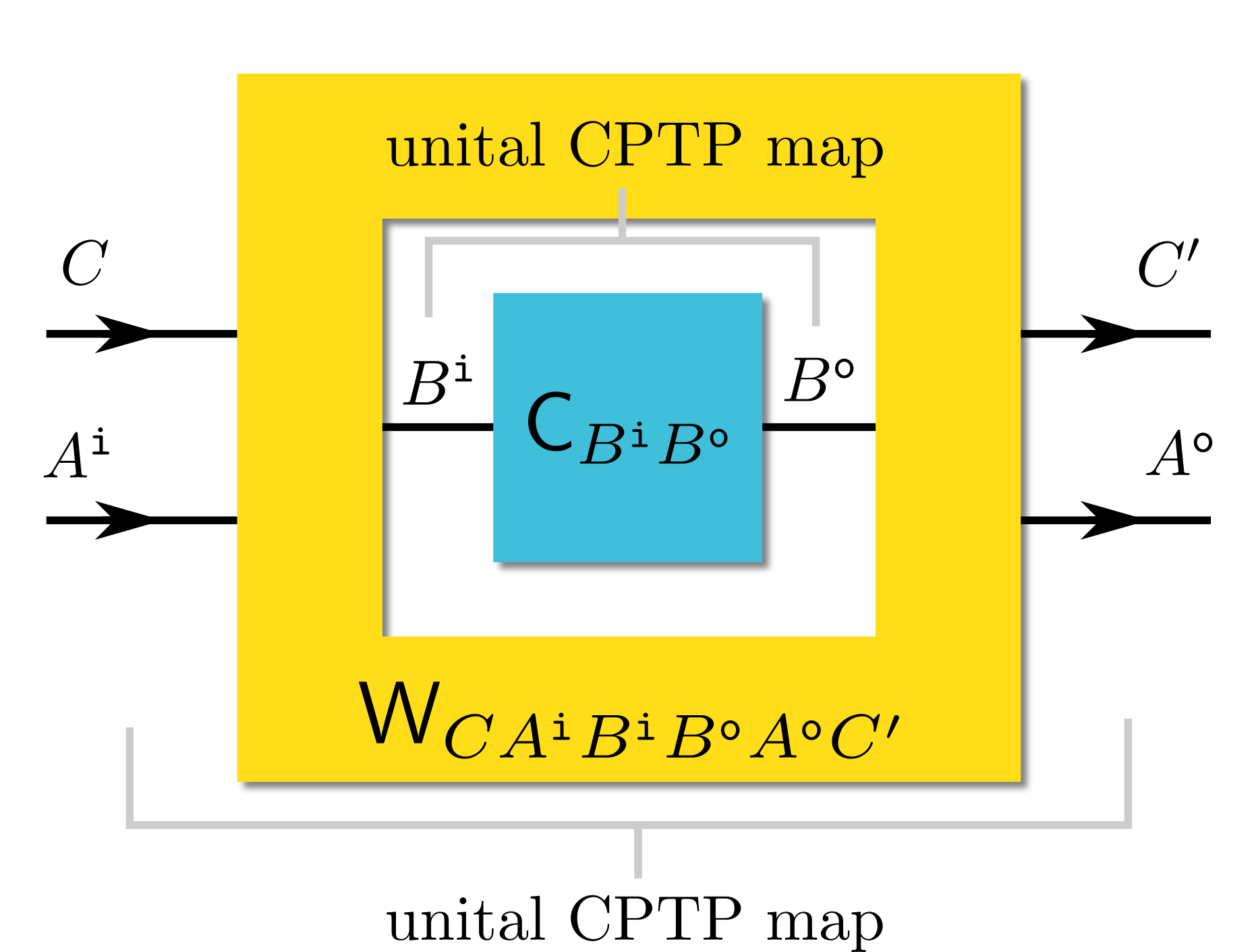}
\label{fig::ico-unital_to_unital}
}\hfill 
\subfigure[\textbf{Quantum Time Flip}]
{
\includegraphics[scale=0.75]{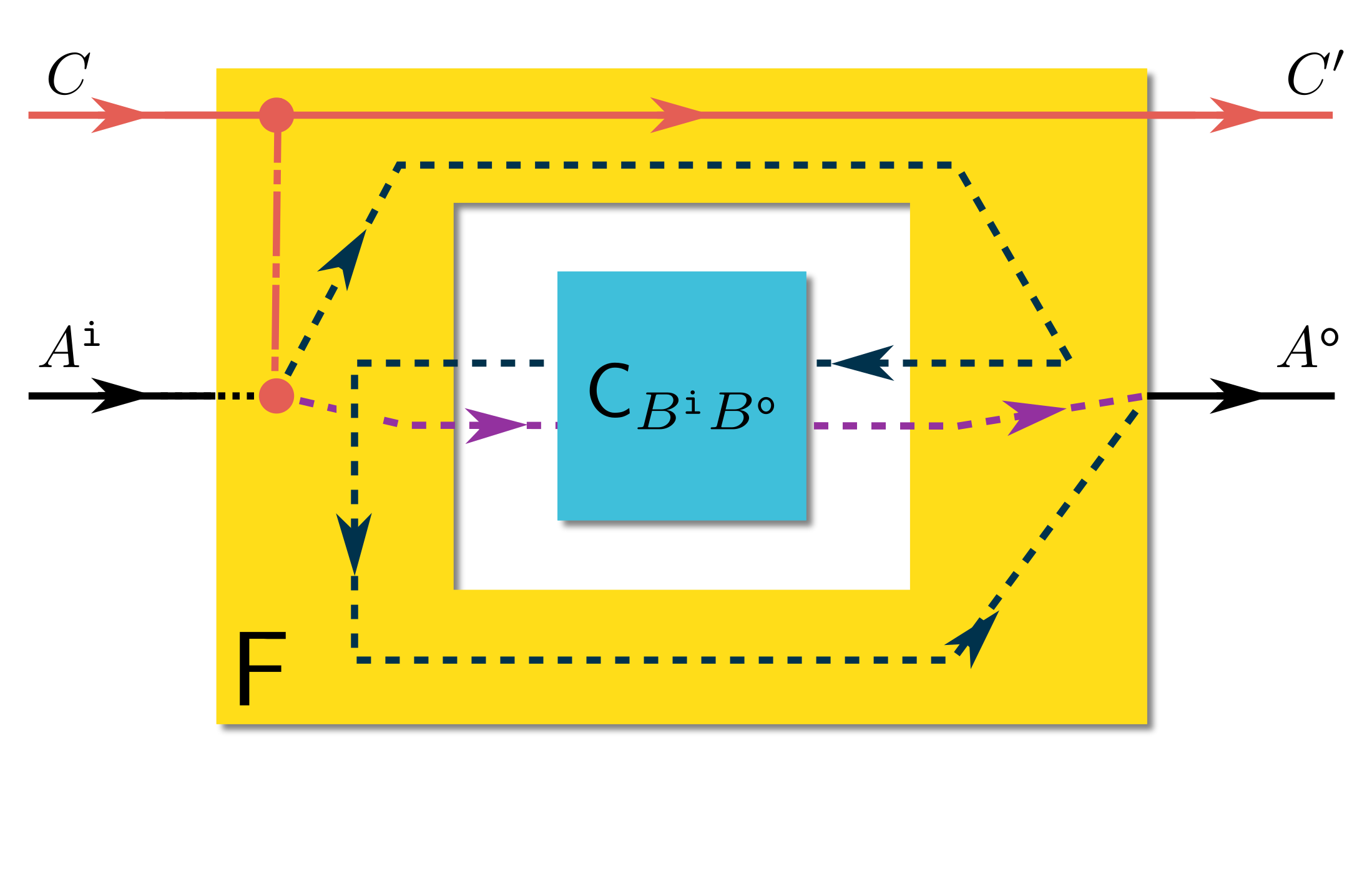}
\label{fig::ico-time_flip}
}\hspace{0.5cm}
\caption{\textbf{$\mathbf{1}$-Slot HOQOs Outside the Comb Formalism.} \textbf{(a)} Limiting the domain and image of $1$-slot HOQOs to, e.g., the set of unital quantum channels, permits HOQOs that lie outside the quantum comb formalism. Note that the wire through the channel $\mathsf{C}_{B^\inp B^\out}$ does not have arrows denoting its direction since it can be traversed in two directions, like in the quantum time flip $\mathsf{F}$. \textbf{(b)} Depending on the state of the control $C$, any state that is inserted on system $A^\inp$ either goes through the channel $\mathsf{C}_{B^\inp B^\out}$ in the forward direction (if the control is in state $\ket{0_C}$; violet dotted path) or in the backward direction (if the control is in state $\ket{1_C}$; navy dotted path). The influence of the control on the time direction is depicted by the red dotted line: If the control is in a superposition of $\ket{0_C}$ and $\ket{1_C}$, then the the map $\mathsf{C}_{B^\inp B^\out}$ is traversed in a superposition of both time directions and the final state of the control is generally entangled with that of the system $A^\out$. Since only unital maps can be understood as channels capable of acting in either time direction, $\mathsf{F}$ only maps unital channels to valid (unital) channels.}
\label{fig::time-flip}
\end{figure}

\vspace{0.25cm}\noindent \textbf{\textul{Quantum Time Flip.}} For example, a HOQO $\mathsf{W}_{CA^\inp B^\inp B^\out A^\out C'}$ (where, again, we split the global past and the global future into two spaces) could be such that it maps all \textit{unital} channels $\mathsf{C}_{B^\inp B^\out}$ onto \textit{unital} channels $\mathsf{C}'_{C A^\inp A^\out  C'}$, but does not necessarily map \textit{all} CPTP maps onto CPTP maps [see Fig.~\ref{fig::ico-unital_to_unital}]. Since set of unital maps is defined by a positivity constraint, a projector, and a trace constraint [see Eq.~\eqref{eq::unital_proj}], the set of all admissible mappings between them can be readily characterised by means of the projector methods presented in Sec.~\ref{subsubsec::axiomatichoqos-qc-transtranstrans}. Here, we focus on a particularly important example of this type of HOQO: the \textit{quantum time flip} $\mathsf{F}_{C A^\inp B^\inp B^\out A^\out C'}$ [see Fig.~\ref{fig::ico-time_flip}], which has been used to study time reversal in quantum mechanics~\cite{chiribella_quantum_2022}.\footnote{In what follows, we assume that $d_{{A^\inp}} = d_{A^\out} = d_{B^\inp} = d_{B^\out}$ and $d_C = d_{C^\prime} = 2$.} This object allows one to superpose different \textit{time directions} (as opposed to different causal orderings of operations, like in the case of the quantum switch). Here, we first introduce the quantum time flip via its action on unital maps and later show that this action can indeed be uniquely represented by a HOQO, i.e., there is a Choi state representation for the time flip.

In particular, if $\{K^{(\alpha)}_B: \Hscr_{B^\inp} \rightarrow \Hscr_{B^\out} \}$ are the Kraus operators corresponding to the channel $\mathsf{C}_{B^\inp B^\out}$, then the Kraus operators $F^{(\alpha)}: \Hscr_{C} \otimes \Hscr_{A^\out} \rightarrow \Hscr_{C'} \otimes \Hscr_{A^\inp}$ of the resulting map $\mathsf{F}_{C A^\inp B^\inp B^\out A^\out C'} \star \mathsf{C}_{B^\inp B^\out} = \sum_\alpha \kketbra{F^{(\alpha)}}{F^{(\alpha)}}$ are given by\footnote{An alternative option would be to take the adjoint instead of the transpose in this equation~\cite{chiribella_quantum_2022}. In this case, the resulting channel and the corresponding time flip would depend on the explicit choice of Kraus operators $\{K^{(\alpha)}_B\}$ and we thus do not consider it here.}
\begin{gather}
\label{eq::Kraus_flip}
F^{(\alpha)} = \ketbra{0_{C'}}{0_C} \otimes K^{(\alpha)} + \ketbra{1_{C'}}{1_C} \otimes K^{(\alpha)\mathrm{T}}, 
\end{gather}
with the understanding that both $K^{(\alpha)}$ and $K^{(\alpha)\mathrm{T}}$ map from $\Hscr_{A^\inp}$ to $\Hscr_{A^\out}$. Similar to the case of the quantum switch, the choice of input state on $\Hscr_{C}$ defines the behaviour of the time flip. In particular, we have 
\begin{align}
    &\mathsf{F}_{C A^\inp B^\inp B^\out A^\out C'} \star \mathsf{C}_{B^\inp B^\out} \star \ketbra{0_C}{0_C} = \mathsf{C}_{A^\inp A^\out} \otimes  \ketbra{0_{C'}}{0_{C'}}, \\
    \text{and} \quad & \mathsf{F}_{C A^\inp B^\inp B^\out A^\out C'} \star \mathsf{C}_{B^\inp B^\out} \star \ketbra{1_C}{1_C} = \$_{A^\inp A^\out}\mathsf{C}_{A^\inp A^\out} \$_{A^\inp A^\out} \otimes  \ketbra{1_{C'}}{1_{C'}}, 
\end{align}
where $\$_{A^\inp A^\out}\mathsf{C}_{A^\inp A^\out} \$_{A^\inp A^\out}$ swaps the input and output spaces of the input channel, therefore corresponding to a time-reversed version of $\mathsf{C}_{B^\inp B^\out}$. Inserting a superposition of $\ket{0_C}$ and $\ket{1_C}$ into the quantum time flip then yields a superposition of the map $\mathsf{C}_{A^\inp A^\out}$ and its time reversed version, i.e., a superposition of time directions. Similarly to the case of the quantum switch, it can be shown that the time flip lies outside of the set of HOQOs with a fixed time direction, and that it provides an advantage in information theoretic tasks over all processes with a fixed time direction; we refer the Reader to Ref.~\cite{chiribella_quantum_2022} for details. Notably, despite the superposition of causal orderings that the quantum switch allows for, all maps in this superposition are individually traversed in a \textit{fixed} time direction, rendering the `exotic' temporal effects that the quantum switch and the quantum time flip display fundamentally different in nature.

It remains to show that, as claimed above, the quantum time flip indeed maps unital CPTP maps to unital CPTP maps and that it can actually be represented by a HOQO $\mathsf{F}_{C A^\inp B^\inp B^\out A^\out C'}$. The former is easily shown by direct insertion. If $\mathsf{C}_{B^\inp B^\out}$ is unital and CPTP, then $\sum_\alpha K^{(\alpha)\dagger}K^{(\alpha)} = \sum_\alpha K^{(\alpha)}K^{(\alpha)\dagger} = \ident$. With this, it follows that the Kraus operators of Eq.~\eqref{eq::Kraus_flip} satisfy
\begin{align}
    &\sum_\alpha F^{(\alpha)\dagger}F^{(\alpha)} = \ketbra{0_{C}}{0_C} \otimes \sum_\alpha K^{(\alpha)\dagger}K^{(\alpha)} + \ketbra{1_{C}}{1_C} \otimes \sum_\alpha K^{(\alpha)*} K^{(\alpha)\mathrm{T}} = \mathds{1} \\
    \text{and} \quad &\sum_\alpha F^{(\alpha)}F^{(\alpha)\dagger} = \ketbra{0_{C'}}{0_{C'}} \otimes \sum_\alpha K^{(\alpha)}K^{(\alpha)\dagger} + \ketbra{1_{C'}}{1_{C'}} \otimes \sum_\alpha K^{(\alpha)\mathrm{T}} K^{(\alpha)*} = \mathds{1}. 
\end{align}
Thus, if $\mathsf{C}_{B^\inp B^\out}$ is unital and CPTP, then so is the resulting map $\mathsf{F}_{C A^\inp B^\inp B^\out A^\out C'} \star \mathsf{C}_{B^\inp B^\out}$. 

To explicitly express the time flip as a HOQO, we note that Eq.~\eqref{eq::Kraus_flip} implies that the action of the time flip 
$\mathsf{F}_{C A^\inp B^\inp B^\out A^\out C'}$ on $\mathsf{C}_{B^\inp B^\out} := \sum_\alpha \kketbra{K^{(\alpha)}_B}{K^{(\alpha)}_B}$ maps every vector $\kket{K^{(\alpha)}_B}$ in the decomposition of $\mathsf{C}_{B^{\inp}B^\out}$ to $\kket{K^{(\alpha)}_A}\ket{0_C 0_{C'}} + \kket{K^{(\alpha)\mathrm{T}}_A}\ket{1_C 1_{C'}}$. Since $\kket{K^{(\alpha)\mathrm{T}}_A} = \$_{B^\inp B^\out \rightarrow A^\out A^\inp} \kket{K^{(\alpha)\mathrm{T}}_B}$, where $\$_{B^\inp B^\out \rightarrow A^\out A^\inp}$ swaps the spaces $B^\inp \rightarrow A^\out$ and $B^\out \rightarrow A^\inp$, we see that~\cite{chiribella_quantum_2022} 
\begin{align}
 &\mathsf{F}_{C A^\inp B^\inp B^\out A^\out C'} \star \mathsf{C}_{B^\inp B^\out} = V \mathsf{C}_{B^\inp B^\out} V^\dagger, \\ \quad \text{with the isometry} \  &V := \ident_{B^\inp B^\out \rightarrow A^\inp A^\out} \otimes \ket{0_C 0_{C'}} + \$_{B^\inp B^\out \rightarrow A^\out A^\inp} \otimes \ket{1_C 1_{C'}}.
\end{align}
Finally, this implies that $\mathsf{F}_{C A^\inp B^\inp B^\out A^\out C'} \star \mathsf{C}_{B^\inp B^\out} = \kketbra{V}{V} \star \mathsf{C}_{B^\inp B^\out}$ with 
\begin{gather}
 \kket{V} := \ket{0_C}\ket{\Phi^+_{B^\inp A^\inp}}\ket{\Phi^+_{B^\out A^\out}} \ket{0_{C'}} + \ket{1_C}\ket{\Phi^+_{B^\inp A^\out}}\ket{\Phi^+_{B^\out A^\inp}} \ket{1_{C'}}. 
\end{gather}
Hence, we have a representation of the quantum time flip as a HOQO. 

Lastly, note that similarly to the quantum switch, discarding the final control system $C'$ destroys the `exotic' properties of the time flip, since this leads to a classical control over forwards and a backwards map (instead of a quantum control). Moreover, due to the fact that the quantum time flip cannot be meaningfully applied to \textit{all} CPTP maps, but only to unital ones, implementations/simulations of the time flip must be either probabilistic or require additional information about the input channel (i.e., they cannot be implemented universally in a fully black-box setting)~\cite{chiribella_quantum_2022}---see Ref.~\cite{stromberg_experimental_2024} for an experimental implementation of the quantum time flip on a photonic setup.

\FloatBarrier


\subsection{Summary}

Throughout this Tutorial, we have strived to convey the usefulness and versatility of HOQOs. Indeed, they allow one to transform between different quantum objects (e.g., states, measurements, channels, instruments, probability distributions, etc.); can be easily adapted to suit almost any conceivable experimental or theoretical situation; and permit an axiomatic derivation and characterisation. Moreover, by making use of the Choi-Jamio{\l}kowski isomorphism, one can represent all types of quantum operations on the same footing as positive operators (see Tab.~\ref{tab::hoqosasmaps}), which subsequently permits direct application of many tools from quantum information theory such as semidefinite programming~\cite{Skrzypczyk_2023} and tensor network methods~\cite{Orus_2014,Cirac_2021}. The general importance of HOQOs is, to a large extent, corroborated by the fact that they keep being rediscovered in a variety of seemingly disconnected fields as the natural descriptor concerning quantum information processing in space and time, as we will now review. 


\begin{table}[H]
\caption{\label{tab::hoqosasmaps}\textbf{Higher-Order Quantum Operations.} Just as the standard ingredients of quantum theory can be seen as linear maps (see Tab.~\ref{tab::tf-cji-basicquantumobjects}), so too can HOQOs, including quantum combs (of fixed causal order), superinstruments (i.e., probabilistically-occurring quantum combs), and process matrices (with no \textit{a priori} causal ordering). All such objects can be represented as supernormalised quantum states via the CJI and characterised accordingly.}
\footnotesize\rm\vspace{-1em}
\begin{tabular*}{\textwidth}{llllll}
\br
Object & Map & Action & Choi & CP & TP \\
\mr
Quantum Comb & $\Tcal_{n:1}$ & $\Tcal_{n:1}[\mathcal{G}^{(x)}_{n:1}] = \mathsf{T}_{n:1} \star \mathsf{G}^{(x)}_{n:1}$ & $\mathsf{T}_{n:1}$ & $\mathsf{T}_{n:1} \geq 0$ & $\mathsf{T}_{n:1}$ satisfies Eq.~\eqref{eq::toqp-dethoqo-def-quantumcombtp} \\
Superinstrument & $\{ \mathcal{G}^{(x)}_{n:1}\}$ & $\mathcal{G}^{(x)}_{n:1}[\Tcal_{n:1}] = \mathsf{T}_{n:1} \star \mathsf{G}^{(x)}_{n:1}$ & $\mathsf{G}^{(x)}_{n:1}$ & $\mathsf{G}^{(x)}_{n:1} \geq 0$ & $\sum_x \mathsf{G}^{(x)}_{n:1}$ satisfies Eq.~\eqref{eq::toqp-dethoqo-def-quantumcombtp} \\
Process Matrix & $\mathcal{W}_k$ & $\mathcal{W}_k[\otimes_{i=1}^k \mathcal{M}_i^{(x_i)}] = \mathsf{W}_k \star_{i=1}^k \mathsf{M}_i^{(x_i)}$ & $\mathsf{W}_k$ & $\mathsf{W}_k \geq 0$ & $W_k \star \otimes_{i=1}^k \mathsf{M}_i = 1 \, \forall \, \mathsf{M}_i \in \mathsf{CPTP}$ \\
\br
\end{tabular*}
\end{table}


\newpage
\section{Review: Applications of Higher-Order Quantum Operations}
\label{pt::review}

The Tutorial part has hopefully provided the Reader with a solid understanding regarding some of the motivations and uses of HOQOs. We will now move to present a Review of the current state-of-the-art surrounding their development and application. As mentioned previously, due to the versatility of the framework, it has been somewhat rediscovered in various sub-fields and with slightly different nomenclature, assumptions, and conventions. Here, we cover five main focus areas, namely higher-order quantum information tasks (Sec.~\ref{subsec::higherorderquantumsubroutines}), open system dynamics \& memory effects (Sec.~\ref{subsec::opensystemdynamicsquantummemory}), many-time quantum physics (Sec.~\ref{subsec::manytimequantumphysics}), causality \& quantum foundations (Sec.~\ref{subsec::causalityquantumfoundations}), and characterisation \& experimental demonstrations (Sec.~\ref{subsec::characterisationexperimentaldemonstrations}). Although by no means exhaustive, we deem these areas to be particularly representative and covering a broad array of key results. We emphasise that, naturally, we make an explicit choice with respect to the level of detail in which we cover the discussed topics, striving to appropriately reflect the breadth of the relevant literature. Our aim for this part is to highlight the wide applicability of HOQOs and demonstrate the connectedness of their use in different sub-fields; accordingly, we will often jump between various perspectives in order to present a holistic picture of many core concepts and results. 

\FloatBarrier


\subsection{Higher-Order Quantum Information Tasks}
\label{subsec::higherorderquantumsubroutines}

\noindent Higher-order quantum operations (HOQOs) play a fundamental role in quantum information processing, particularly in scenarios involving the transformation of quantum states and operations. Traditional quantum information tasks often focus on state transformations---such as the creation of entangled states from uncorrelated ones. Of course, the ability to achieve a certain task depends upon the resources at hand: Given an uncorrelated state, if one can apply arbitrary global unitary operations, then one can generate entanglement; on the other hand, if restricted to \textit{local operations and classical communication (LOCC)}, then no entanglement can be created. HOQOs extend this paradigm of quantum information protocols to more complex scenarios involving the manipulation of quantum operations themselves. For instance, the well-known quantum teleportation protocol can be viewed through this lens as the construction of an identity channel between sender and receiver using shared entanglement, local measurements, classical communication, and local unitary operations.

The higher-order framework provides a natural setting for analysing transformations of quantum channels through pre- and post-processing, potentially involving auxiliary memory systems. Such quantum operations are described by superchannels, which represent the most general objects in quantum theory capable of such channel-to-channel mappings. Key questions in this field are often concerned with the possibility of transforming one channel into another under various physically motivated constraints on the superchannel, such as restrictions on encoder/decoder channels (e.g., requiring separable or LOCC structure), limitations on auxiliary system size of type, or constraints on the number of times the input operation can be queried.

The practical applications of using HOQOs for describing quantum strategies for games and information-processing tasks have been far-reaching. These concepts have enabled optimisation of protocols in quantum networks and improved our understanding of metrology / parameter estimation and discrimination tasks. For quantum computing, the framework has proven valuable in circuit optimisation. 

For instance, the derivation of complexity bounds on quantum algorithms, e.g., Grover's algorithm, can be seen as an early example of the usage of HOQOs in the field of quantum computation~\cite{boyer_tight_1998, ambainis_quantum_2000}. An early explicit application of quantum combs and HOQOs appeared in the seminal work by Gutoski \& Watrous, which established a rigorous mathematical framework for analysing quantum strategies in multi-round quantum games~\cite{Gutoski_2007}. In this work, the authors introduced quantum strategies as collections of quantum operations performed sequentially, with memory spaces maintained between rounds of interaction, i.e., quantum interactive proof systems described as quantum combs. Perhaps most importantly, they demonstrated that the optimal strategy in a variety of quantum games can be found via semidefinite programming and proved strong duality for many cases. Research along similar lines continues, e.g., Ref.~\cite{Thompson_2017, Elliott_2021} demonstrates the importance of memory for quantum adaptive agents and Ref.~\cite{Xing_2023} highlights similar advantages to be gained concerning communication protocols distributed over quantum networks/repeaters.

In terms of quantum computing, the HOQO framework has found particularly important applications in circuit optimisation and verification. One key development came from Chiribella, D'Ariano \& Perinotti, who showed how quantum strategies could be used to optimise quantum circuits with memory constraints~\cite{Chiribella_2008_PRL}—a critical consideration for \textbf{noisy intermediate-scale quantum (NISQ)} devices. The computational complexity of HOQOs as well as their computational power and limitations have been discussed in~\cite{Baumeler_2016, araujo_quantum_2017}. The framework of HOQOs has proven especially valuable in analysing `plug-and-play' elements~\cite{Thompson_2018}, i.e., adaptive quantum circuits where measurements during computation determine subsequent operations. More recent work has focused on using quantum strategies to develop robust verification protocols for quantum computations. For instance, in the context of blind quantum computation~\cite{Fitzsimons_2017_PRA, Fitzsimons_2017}, where a quantum server performs calculations for a client with limited quantum capabilities, the quantum comb formalism has been applied to efficiently verify the computation of partially unknown devices~\cite{Smith_2023}. These protocols exploit the game-theoretic nature of quantum strategies to ensure that any deviation from the intended computation can be detected with high probability. 

In this section, we will focus on a few particular cases of quantum information processing tasks for which key insights have been developed through the use of HOQOs. We will begin by analysing the possibility of transforming unitary operations before moving on to tasks broadly known as `unitary storage-and-retrieval', `unitary learning', and `unitary estimation'. Following this, we analyse the question of transforming Hamiltonian dynamics before finally discussing higher-order quantum resource theories and applications to quantum metrology strategies, parameter estimation, and channel discrimination. 

\FloatBarrier


\subsubsection{Physically Implementing Functions of Unitary Operations}\label{subsubsec::hoqs-functionsunitaries}\hfill\\

\noindent A natural application of HOQOs is to design quantum circuits that transform unknown quantum gates, which can be seen as an instance of quantum functional programming~\cite{Selinger_2004,Valiron_2005}. In this context, one typically has access to $k\in\mathds{N}$ calls of an unknown unitary channel $\mathcal{U}[\,\bullet\,]:= U \bullet U^\dagger$ and aims to transform it in a desired way. This task is usually considered from a `black-box' perspective, assuming no specific knowledge of the input unitary (beyond its dimension $d$), and must be \textit{universal}, working for all possible input unitaries. Accordingly, the quantum operation that is to be implemented can mathematically be represented by a \textit{function} $f:SU(d_\inp)\to SU(d_\out)$ which maps unitary operators to unitary operators, i.e., $U\mapsto f(U)$. Since unitary operators are physically equivalent up to a global phase, there is no ambiguity concerning the distinction between such a function applied to the unitary map $\mathcal{U}$ or its unitary operators $U$, since we can define $f(\mathcal{U})[\,\bullet\,] := f(U) \bullet f(U)^\dagger$; we will use this notation interchangeably.

Several important functions of unitaries have been studied extensively, including unitary inversion $f(U)=U^{-1}$~\cite{Chiribella_2016,Quintino_2019_PRL,Yoshida_2023,Quintino_2022_Quantum,Chen_2024,Mo_2024,Zhu_2024_Reversing}, complex conjugation $f(U)=U^*$~\cite{Chiribella_2016,Miyazaki_2019,Ebler_2022}, transposition $f(U)=U^{\textup{T}}$~\cite{Quintino_2019_PRA,Quintino_2022_Quantum}, cloning $f(U)=U^{\otimes n}$ (for $n>k$)~\cite{Chiribella2008cloning,Chiribella2012replication,Bisio_2014,Dur2015Replication}, controlisation $f(U)=\texttt{ctrl}(U):=\ketbra{0}{0}\otimes \ident +\ketbra{1}{1}\otimes U$~\cite{Chiribella_2016,Dong_2019}, and iteration $f(U)=U^n$ for $n\in \mathds{N}$~\cite{Soleimanifar_2016}. Moreover, functions of isometries have also been considered~\cite{Yoshida_2023,Yoshida_2024}. Apart from unitary cloning and functions of isometries, all cases we analyse here satisfy $d_\inp=d_\out=:d$.


\begin{figure}[t!]
    \centering
    \includegraphics[scale=0.45]{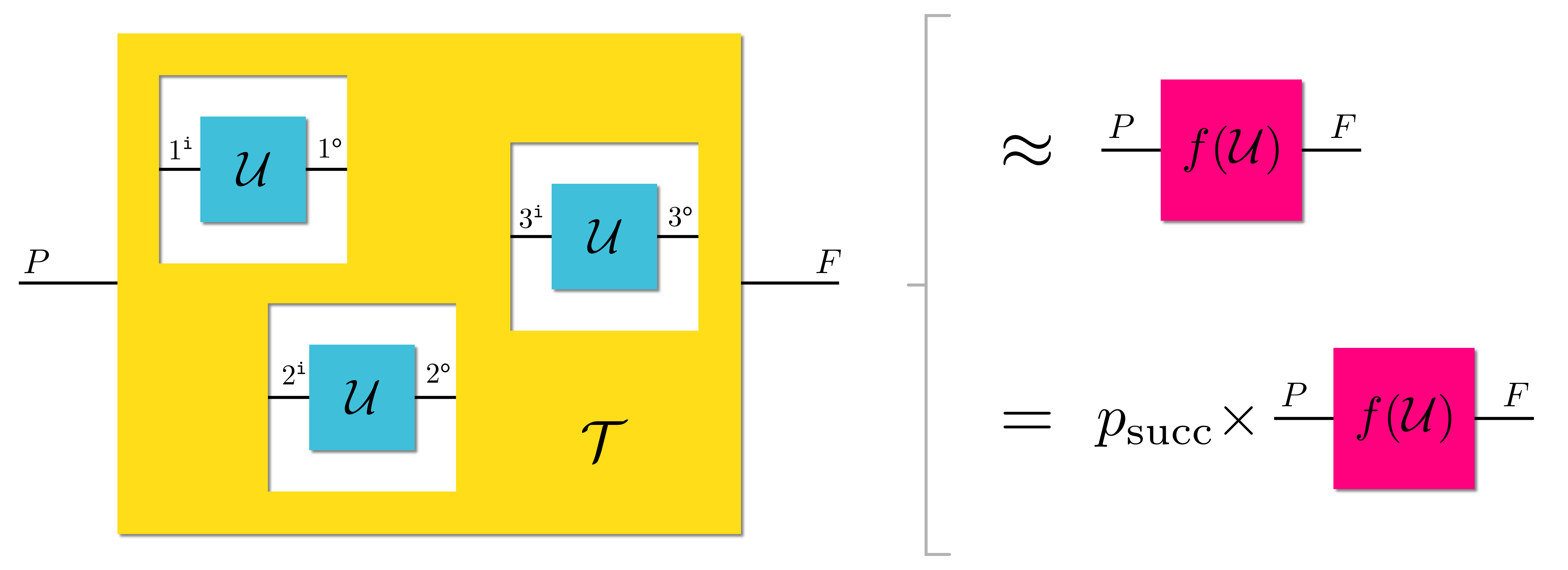}
    \caption{\textbf{Physically Implementing Functions of Unitary Operations.} Pictorial illustration of the general problem described in Eq.~\eqref{eq::hoqs-fu-unitarytransformation} of transforming $k$ calls of a unitary operation  $\mathcal{U}$ into another unitary operation $f(\mathcal{U})$ for some arbitrary function $f:SU(d_\inp)\to SU(d_\out)$. For the moment, we do not assume any particular structure on the HOQO $\mathcal{T}$ (l.h.s., yellow), which may, in principle, not even respect a definite causal order. As we discuss later, it is often not possible to have a deterministic HOQO $\mathcal{T}$ such that $\mathcal{T}[\mathcal{U}^{\otimes k}]=f(\mathcal{U})$ exactly; hence one seeks optimal deterministic \textit{approximations} (upper r.h.s.) or \textit{probabilistic} exact ones (lower r.h.s.). }
    \label{fig::hoqs-functionsunitaries}
\end{figure}


In general, one aims to construct \textit{physical implementations} of such functions: Given a function $f$ and a fixed number of calls $k$ to a $d$-dimensional unitary operation, one seeks a $k$-slot HOQO $\Tcal$ such that (see Fig.~\ref{fig::hoqs-functionsunitaries})
\begin{align}\label{eq::hoqs-fu-unitarytransformation}
    \mathcal{T}[\mathcal{U}^{\otimes k}]=f(\mathcal{U}),
\end{align}
for all unitary operations $\mathcal{U}$. The feasibility of implementing such transformations depends on various factors: the number of available calls $k$ to the unknown unitary, the dimension $d$ of the Hilbert space, the permitted structure of the HOQO (parallel, sequential, or causally indefinite), and the specific function $f$ to be implemented. Note that if $k\to\infty$, one can perform full quantum gate tomography to first characterise any input $\mathcal{U}$ and then simply evaluate $f(\mathcal{U})$ from its classical description. To avoid trivialising the problem in this way, one typically restricts considerations to finite $k$. These problems can often be formulated and solved (in cases of small $d$ and $k$) using semidefinite programming (SDP)~\cite{Skrzypczyk_2023}, thanks to the characterisation of HOQOs (in the Choi representation) through positive semidefinite and affine linear constraints (see Tab.~\ref{tab::hoqosasmaps}).

In few cases, the desired operation can be obtained both \textit{deterministically} and \textit{exactly}. However, in many situations, one must often choose between \textit{deterministic \& approximate} implementations and \textit{probabilistic \& exact} ones. In the deterministic \& approximate case, the standard figure of merit is the average fidelity between the desired and achieved operation (calculated with respect to the Haar measure). For probabilistic \& exact implementations, the performance is judged by the success probability, with the protocol providing a flag (i.e., a heralding measurement outcome) indicating whether the implementation of $f$ succeeded or failed. 

\vspace{0.25cm}\noindent
\textbf{\textul{Figures of Merit: Deterministic \& Approximate.}} For deterministic \& approximate implementations, the standard figure of merit considered is the average fidelity between the desired and achieved transformation. The fidelity $\Xi$ between a unitary channel $\mathcal{U}$ and an arbitrary quantum channel $\mathcal{C}$ in the Choi representation is
\begin{align}\label{eq::hoqs-fu-unitaryfidelity}
    \Xi(\mathcal{U},\mathcal{C}):=&\frac{1}{d^2}\tr{\mathsf{U}\mathsf{C}}
    =\frac{1}{d^2} \mathsf{U}^{\textup T} \star \mathsf{C}.
\end{align}
Since the desired output $f(\Ucal)$ is unitary, it follows that given a $k$-slot HOQO $\mathcal{T}_k$, the average fidelity (with respect to the Haar measure) for implementing the function $f$ is
\begin{align}\label{eq::hoqs-fu-averagefidelity}
   \langle \Xi_{f}(d,k) \rangle := \int_\text{Haar} \Xi \Big(f(\mathcal{U}),\mathcal{T}_k[\mathcal{U}^{\otimes k}]\Big) \,\text{d}\,\mathcal{U} 
   &= \frac{1}{d^2} \int_\text{Haar} f(\mathsf{U})^{\textup T} \star \left(\mathsf{T}_k \star \mathsf{U}^{\otimes k}\right)\, \text{d}\mathsf{U} \notag \\
   &=\mathsf{T}_k \star \frac{1}{d^2} \int_\text{Haar} f(\mathsf{U})^{\textup T}  \star  \mathsf{U}^{\otimes k}\, \text{d}\mathsf{U} \notag \\
   &=:\mathsf{T}_k \star \Omega_{f}(d,k), 
\end{align}
where
\begin{align}\label{eq::hoqs-fu-performanceoperator}
    \Omega_{f}(d,k):=\frac{1}{d^2} \int_\text{Haar} f(\mathsf{U})_{PF}^{\textup T}\otimes \mathsf{U}^{\otimes k}_{IO} \; \text{d}\mathsf{U}
\end{align}
is called the \textit{performance operator}~\cite{Chiribella_2016,Quintino_2022_Quantum}. Here, for the sake of clarity, we explicitly label the past $P$ and future $F$ spaces on which the function to be implemented acts and the collective input $I := \{ 1^{\inp}, \hdots, k^{\inp}\}$ and output $O := \{ 1^{\out}, \hdots, k^{\out}\}$ spaces of the $k$ copies of input unitary (see Fig.~\ref{fig::hoqs-functionsunitaries}). The problem of finding the optimal performance can be phrased as the following SDP\footnote{Equations~\eqref{eq::hoqs-fu-averagefidelity} and~\eqref{eq::hoqs-fu-performancesdp} differ up to a transpose on $\mathsf{T}_k$; however, since the optimisation is taken over a class of HOQOs (e.g., parallel, sequential, or causally indefinite), and transposition leaves any such class of HOQOs invariant, it follows that $\max_{\mathsf{T}_k}\; \tr{\Omega_{f}(d,k)\mathsf{T}_k^{\textup T}}=\max_{\mathsf{T}_k}\; \tr{\Omega_{f}(d,k)\mathsf{T}_k}$.}
\begin{align}\label{eq::hoqs-fu-performancesdp}
    \max_{\mathsf{T}_k}\; &\tr{\Omega_{f}(d,k)\mathsf{T}_k} \\
    \text{subject to: } &\mathsf{T}_k \text{ being a $k$-slot HOQO}.
\end{align}


\begin{figure}[t]
    \centering
    \includegraphics[width = 0.95\linewidth]{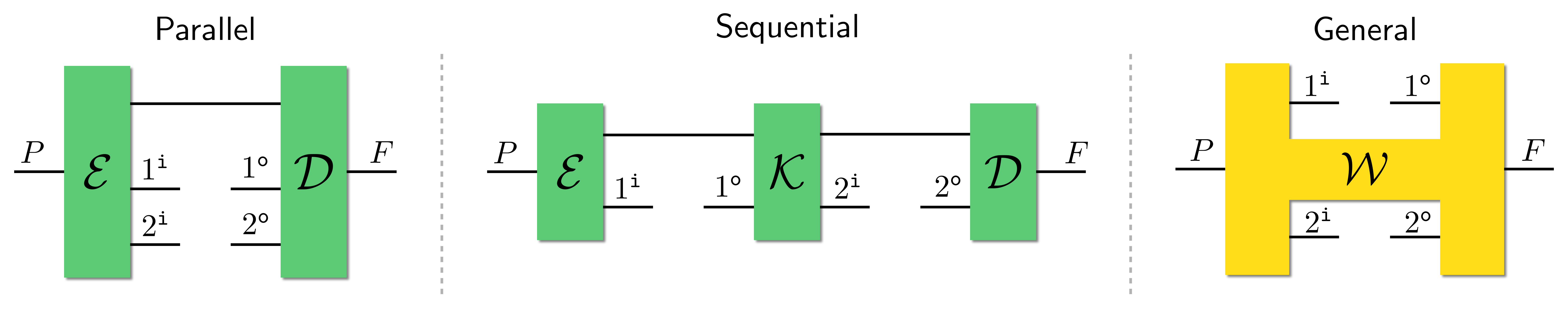}
    \caption{\textbf{Structure of Higher-Order Quantum Strategies.} Three different types of quantum strategies can be used to transform $k$ calls of an arbitrary operations $\mathcal{U}$ into $f(\mathcal{U})$. In particular, the input operations can be processed in parallel (left), sequence (middle), or more generally without definite causal order (right). Without any additional restrictions (say, on the size of the available auxiliary system), the parallel case is included in the sequential one, which, in turn, is included in the general case.}
    \label{fig::hoqs-fu-structurestrategies}
\end{figure}


\noindent In many situations, it is meaningful to further restrict the set of HOQOs maximised over to reflect different implementation \textit{strategies}. For instance, one may consider parallel strategies in which all input unitaries are processed simultaneously; or more general sequential strategies in which they are called in succession; or more generally still using causally indefinite structures (see Fig.~\ref{fig::hoqs-fu-structurestrategies}). We will denote such strategies by labelling objects with superscripts `par', `seq', and `gen', respectively. Since restricting the implementation strategy to such classes amounts to adding extra linear constraints on the possible HOQOs, finding the optimal protocol with a given implementation strategy remains an SDP.

\vspace{0.25cm}\noindent
\textbf{\textul{Figures of Merit: Probabilistic \& Exact.}} Another relevant situation concerns the probabilistic exact setting (also sometimes referred to as the `probabilistic heralded' case). Here, with some probability of success $p$, the $k$ calls of the input operation are transformed to $f(U)$ exactly, and the protocol outputs a `flag' indicating that it worked correctly; with probability $1-p$, the protocol failed. The heralding of success or failure can be realised by performing a dichotomic measurement on the flag state. The figure of merit here is the success probability $p$, which depends upon the desired function and the number of calls to the unitary (of a fixed dimension). Probabilistic exact protocols for unitary transformations have been analysed extensively in Refs.~\cite{Sedlak_2019,Quintino_2019_PRA,Quintino_2019_PRL,Dong21Draw}. 

\vspace{0.25cm}\noindent
\textbf{\textul{General Structural Results.}} Certain structural results emerge when considering specific classes of transformations, particularly for functions that are either \textit{homomorphisms} $f(UV)=f(U)f(V)$ (such as complex conjugation and cloning) or \textit{anti-homomorphisms} $f(UV)=f(V)f(U)$ (such as inversion and transposition). Such structural results have been developed predominately in Ref.~\cite{Bisio_2014,Quintino_2022_Quantum}. 

For both of these cases, group representation theory provides powerful tools for simplifying the mathematical analysis. For instance, the Haar integral appearing in the performance operator $\Omega_f(d,k)$ [see Eq.~\eqref{eq::hoqs-fu-performanceoperator}] can be evaluated analytically find optimal deterministic approximate solutions using standard group representation methods~\cite{Chiribella_2016,Quintino_2022_Quantum}. Similarly, one can assume that the HOQO $\mathsf{T}_k$ implementing $f$ takes a particular symmetric form, namely it is invariant under group twirling. Moreover, the average fidelity coincides with the worst case fidelity, implying that optimising the average fidelity is equivalent to optimising over robustness to white noise~\cite{Quintino_2022_Quantum}; other potential figures of merit (such as the diamond norm distance) also lead to the same optimal solution~\cite{Yang2020Optimal}. Such symmetries and reductions of the problem often help make it more manageable to be solved via SDP methods.

Notably, when the desired function $f$ is a homomorphism, the optimal implementation always takes the form of a parallel strategy, implying that sequential or causally indefinite processes offer no advantage~\cite{Bisio_2014,Quintino_2022_Quantum}. On the other hand, when $f$ is an anti-homomorphism, sequential protocols can exponentially outperform parallel protocols with the same number of calls in both the probabilistic exact~\cite{Quintino_2019_PRL} and deterministic approximate settings~\cite{Quintino_2022_Quantum}.

The implementation of functions of unknown unitary operations represents a fundamental challenge in quantum information processing. These quantum operations, which include functions like complex conjugation, cloning, transposition, and inversion, exhibit distinct characteristics in terms of their feasibility and optimal implementation strategies, as we now review in detail.

\vspace{0.25cm}\noindent
\textbf{\textul{Unitary Complex Conjugation.}} The problem of unitary complex conjugation concerns transforming $k$ calls of an arbitrary $d$-dimensional unitary $U$ into $f(U)=U^*$. Since complex conjugation is a homomorphism, all implementations can be assumed to be of parallel structure without harming performance~\cite{Bisio_2014,Quintino_2022_Quantum}. Complex conjugation displays a sharp threshold behaviour: The transformation $U^{\otimes k} \mapsto U^*$ can be implemented deterministically and exactly given access to $k=d-1$ calls of the unitary by exploiting properties of the totally anti-symmetric state~\cite{Miyazaki_2019}. However, the task becomes impossible---even probabilistically---when fewer calls are available (i.e., for $k<d-1$)~\cite{Quintino_2019_PRA}. Concretely, the optimal success probability $p_\text{conj}(d,k)$ for transforming $k$ calls of a $d$-dimensional unitary $U$ into its complex conjugate $U^*$ behaves as
\begin{align}
    p_\text{conj}(d,k)=
    \begin{cases}
        1 & \quad\text{if } k\geq d-1\\
        0 & \quad\text{if } k< d-1 .
    \end{cases}
\end{align}

\noindent Concerning deterministic and approximate implementations of unitary conjugation, the special case of $k=1$ was first analysed in Ref.~\cite{Chiribella_2016}, where the authors leveraged the performance operator to derive the expression $\langle \Xi_\text{conj}(d,k=1)\rangle=\frac{2}{d(d-1)}$. This result was then generalised to an arbitrary number of calls $k$ by combining techniques from SDP and group theory to prove that the optimal average fidelity in general is given by
\begin{align}
     \langle \Xi_\text{conj}(d,k)\rangle = \frac{k+1}{d(d-k)}.
\end{align}

\begin{figure}[t!]
    \centering
    \includegraphics[scale=.75]{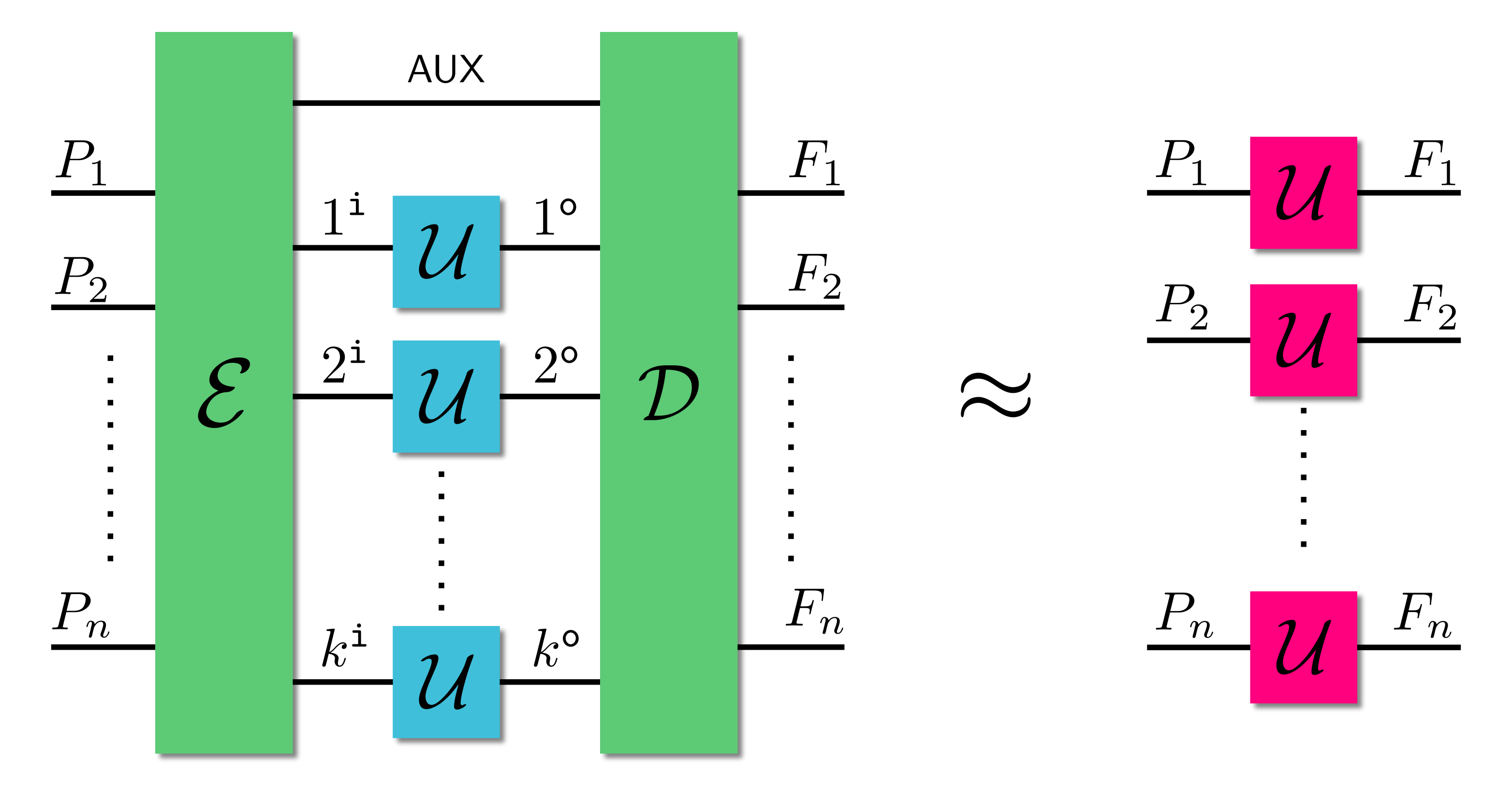}
    \caption{\textbf{Unitary Cloning.} The goal of unitary cloning is to transform $k$ calls of an arbitrary input unitary $U$ (blue) into $n>k$ uses of $U$ (pink), i.e., to approximate the transformation $U^{\otimes k}\mapsto U^{\otimes n}$. Since $f(U)=U^{\otimes n}$ is a homomorphism, optimal cloning can always be achieved with a parallel strategy.}
    \label{fig::hoqs-fu-unitarycloning}
\end{figure}

\vspace{0.25cm}\noindent
\textbf{\textul{Unitary Cloning.}} The problem of cloning a unitary operation concerns transforming $k$ calls of an arbitrary $d$-dimensional unitary $U$ into $f(U)=U^{\otimes n}$, i.e., $U^{\otimes k}\mapsto U^{\otimes n}$ with $n>k$ (see Fig.~\ref{fig::hoqs-fu-unitarycloning}). Since the function $f(U)=U^{\otimes  n}$ is a homomorphism, parallel implementation strategies attain the optimal performance~\cite{Bisio_2014,Quintino_2022_Quantum}. 

While exact unitary cloning is impossible (even probabilistically) due to its inherent non-linearity, deterministic and approximate implementations have been extensively studied. For the simplest case of copying a $d$-dimensional unitary, i.e., $f(U) =U \otimes U$ ($k=1$ and $n=2$), the optimal average fidelity is $\langle \Xi \rangle=\frac{d+\sqrt{d^2-1}}{d^3}$~\cite{Chiribella2008cloning}. When transforming $k=1$ calls into $n$ copies of a \textit{qubit} unitary ($d=2$), the optimal fidelity takes the form of a simple optimisation problem involving dimensions and multiplicities of representations of $SU(2)$~\cite{Bisio_2014}. Finally, the general problem of transforming $k$ calls of an arbitrary $d$-dimensional unitary to $n>k$ copies has been analysed in Refs.~\cite{Chiribella2012replication,Dur2015Replication}, where the remarkable `gate super-replication' phenomenon was derived as
\begin{align}
    \langle \Xi_{\text{clone}} (d,k\to n)\rangle \geq  1-  \frac{2(n+1)^{\frac{d(d-1)}{2}}}{e^{\frac{2k^2}{n d^2}}}.
\end{align}
If $n$ is chosen to be less than $\frac{2k^2}{d^2}$, the fidelity approaches unity exponentially; thus, if $k$ is large and $n$ is of the order of $k^2$, one can clone arbitrary unitaries with very high average fidelity.

\vspace{0.25cm}\noindent\textbf{\textul{Unitary Transposition.}} The goal of unitary transposition is to transform $k$ calls of an arbitrary $d$-dimensional unitary $U$ into $f(U)=U^{\textup{T}}$. In contrast to the previous two tasks, $f(U)=U^\textup{T}$ is an anti-homomorphism, which allows for potential advantages in sequential and general implementation strategies over parallel ones.

Probabilistic and exact parallel implementations of unitary transposition achieve an optimal success probability~\cite{Quintino_2019_PRA}
\begin{align} \label{eq::hoqs-fu-transpositionprobpar}
p_\text{trans}^\text{par}(d,k)=1-\frac{d^2-1}{k+d^2-1}.
\end{align}
This optimal value can be achieved by a unitary store-and-retrieve protocol~\cite{Sedlak_2019} (see Sec.~\ref{subsubsec::hoqs-unitarystoreandretrieve}) or by probabilistic port-based gate-teleportation~\cite{Ishizaka2008PBT,Ishizaka2009PBT,Mozrzymas2018PBT}. When sequential protocols are permitted, a `repeat-until-success' strategy~\cite{Dong21Draw} can be harnessed to construct a protocol that attains~\cite{Quintino_2019_PRA}
\begin{align} \label{eq::hoqs-fu-transpositionprobseq}
p_\text{trans}^\text{seq}(d,k)\geq1-\left(1-\frac{1}{d^2}\right)^{\left\lceil\frac{k}{d} \right\rceil},
\end{align}
where $\left\lceil\bullet \right\rceil$ denotes the ceiling function. The dramatic difference between parallel and sequential protocols for unitary transposition is exemplified through the special case of $d=2$: If $k=4$, a deterministic \textit{and} exact sequential protocol can be constructed~\cite{Yoshida_2023}; on the other hand, Eq.~\eqref{eq::hoqs-fu-transpositionprobpar} implies that parallel strategies with the same number of calls are limited to $p_\text{trans}^\text{par}(2,4)=\tfrac{4}{7}$. 
Moreover, Refs.~\cite{Chen_2024,Mo_2024} show that if one is given access to $\mathcal{O}(k=d^2)$ calls of an arbitrary $d$-dimension unitary $U$, there exists a quantum circuit that deterministically and exactly implements unitary transposition. 
Lastly, for the case of $d=2$, computational methods have shown that for $k=2$ and $k=3$, general strategies with indefinite causal order can outperform sequential ones~\cite{Quintino_2019_PRA}. 

We now move to analyse the deterministic and approximate setting. Here, when parallel processes are considered, the optimal average fidelity $\langle \Xi_\text{trans}^\text{par}(d,k)\rangle$ of transforming $k$ calls of a $d$-dimensional unitary $U$ into its transposition $U^{\textup{T}}$ is always obtained by via \textit{unitary estimation} strategies~\cite{Chiribella2005Estimation,Bisio_2009,Quintino_2022_Quantum} (see Sec.~\ref{subsubsec::hoqs-unitarystoreandretrieve}). In general, there is no simple closed-formula for the optimal performance of deterministic parallel unitary transposition for arbitrary $d$ and $k$, but for the special case of qubits ($d=2$), it is given by~\cite{Quintino_2022_Quantum}
\begin{align}
\langle \Xi_\text{trans}^\text{par}(d=2,k)\rangle =1-\sin^2\left(\frac{\pi}{k+3}\right).
\end{align}
With $k=1$ call, we have $\langle \Xi_\text{trans}(d=2,k=1)\rangle=\frac{2}{d^2}$. Since probabilistic exact protocols always provide a lower bound to the average fidelity of deterministic ones~\cite{Quintino_2022_Quantum}, it follows from Eq.~\eqref{eq::hoqs-fu-transpositionprobseq} that
\begin{align}
\langle \Xi_\text{trans}^\text{seq}(d,k)\rangle\geq 1-\left(1-\frac{1}{d^2}\right)^{\left\lceil\frac{k}{d} \right\rceil}.
\end{align}
Similarly to the probabilistic exact case, numerical methods have demonstrated that when $k=2$ and $k=3$, general strategies with indefinite causal order can outperform sequential ones~\cite{Quintino_2022_Quantum}.

\vspace{0.25cm}\noindent
\textbf{\textul{Unitary Inversion.}} The goal of unitary inversion is to transform $k$ calls of an arbitrary $d$-dimensional unitary $U$ into $f(U)=U^{-1}=U^\dagger$. Similarly to transposition, $f(U)=U^{-1}$ is an anti-homomorphism and parallel protocols are not necessarily optimal; hence, sequential and general strategies may be advantageous. 

Given that probabilistic \& exact unitary complex conjugation is impossible whenever the number of calls is restricted to $k< d-1$, but transposition is not, it follows that unitary inversion (with any implementation strategy) also has a success probability of zero, since it is the composition of complex conjugation and transposition:
\begin{align}
 \text{if } k< d-1: \quad\quad\quad    p_\text{inv}^\text{gen}(d,k)= p_\text{inv}^\text{seq}(d,k)=  p_\text{inv}^\text{par}(d,k)=0.
\end{align}
As mentioned above, in the case where $k\geq d-1$, deterministic and exact complex conjugation is possible; thus, a guaranteed implementation of unitary inversion can be constructed by grouping the number of available calls into blocks of size $k$, first performing unitary conjugation (deterministically and exactly) and then performing unitary transposition. This concatenation of protocols is not guaranteed to be optimal, but nonetheless provides a lower bound for the success probability. In the case of $d=2$, we have minimal groups of size $k=1$; hence, for qubits, unitary inversion and transposition are equivalent problems; this can also be seen by noting that for qubits, complex conjugation can be achieved essentially for free by performing a Pauli-$y$ gate before and after the input unitary, and the assertion follows directly since $U^{-1}={(U^*)}^T$. 

For cases where $d>2$, this aforementioned concatenation strategy leads to the lower bound
\begin{align}
    p_\text{inv}^\text{par}(d,k)\geq 1-\frac{d^2-1}{\left\lfloor \frac{k}{d-1} \right\rfloor+d^2-1}.
\end{align}
For sequential probabilistic scenarios, Refs.~\cite{Quintino_2019_PRA,Quintino_2019_PRL} present a protocol that approaches unity exponentially quickly, achieving $p_\text{inv}^\text{seq}(d,k)\geq 1-\left(1-\frac{1}{d^2}\right)^{\left\lfloor \frac{k}{d-1} \right\rfloor}$. In fact, in the qubit case, unitary inversion can be implemented in a deterministic and exact manner with $k=4$ calls applied in sequence~\cite{Yoshida_2023}. Finally,
given access to $\mathcal{O}(k=d^2)$ calls of an arbitrary $d$-dimension unitary $U$, an explicit quantum circuit implementing deterministic and exact unitary inversion has been derived~\cite{Chen_2024,Mo_2024}.

In the deterministic and approximate setting, when parallel processes are considered, the optimal average fidelity $\langle \Xi_\text{inv}^\text{par}(d,k)\rangle$ for transforming $k$ calls of a $d$-dimensional unitary $U$ into its inverse $U^{-1}$ behaves similarly to that of unitary transposition: it is always obtained by via unitary estimation~\cite{Chiribella2005Estimation,Bisio_2009,Quintino_2022_Quantum} and we have $\langle \Xi_\text{inv}^\text{par}(d=2,k)\rangle=\langle \Xi_\text{trans}^\text{par}(d=2,k)\rangle$ (see Sec.~\ref{subsubsec::hoqs-unitarystoreandretrieve}). Again, numerical methods show that when $k=2, d=2$ and $k=3, d=3$, strategies without definite causal order strictly outperform sequential ones~\cite{Quintino_2022_Quantum}.

\begin{figure}[t!]
    \centering
    \includegraphics[scale = 0.65]{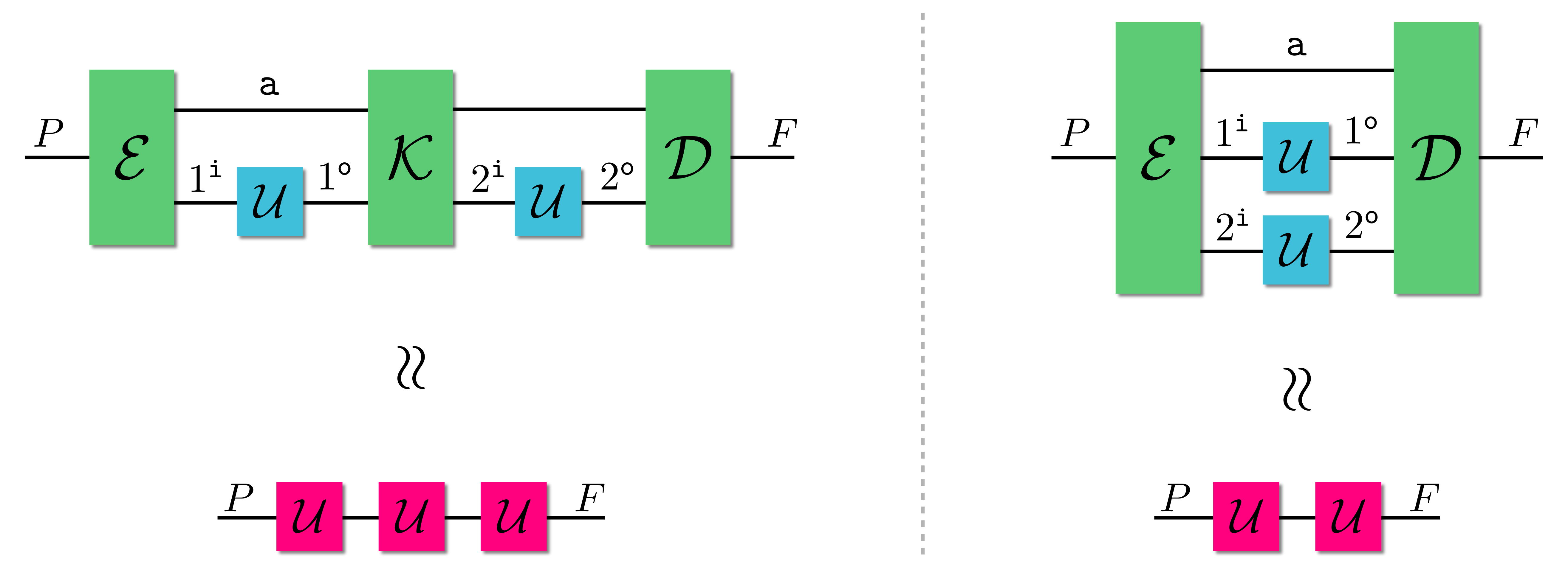}
    \caption{\textbf{Unitary Iteration.}  The goal of unitary iteration is to take $n$ input unitaries $U$ (blue) and apply them sequentially (pink), i.e., approximate the transformation $U^{\otimes k}\mapsto U^n$. For sequential protocols (depicted on the left), this task is non-trivial whenever $n>k$; for parallel protocols (right), the task is non-trivial for every $n>1$.   }
    \label{fig::hoqs-fu-unitaryiteration}
\end{figure}

\vspace{0.25cm}\noindent
\textbf{\textul{Unitary Iteration.}} Unitary iteration aims to transform $k$ calls of an arbitrary unitary operation $U$ into $f(U)=U^n$ for $n\in\mathds{N}$, e.g., $U \mapsto U^2$ (see Fig.~\ref{fig::hoqs-fu-unitaryiteration}). This presents unique challenges as the function is neither homomorphic nor anti-homomorphic. While sequential protocols can trivially achieve the task $U^{\otimes k}\mapsto U^n$ whenever $k\geq n$ (by discarding $k-n$ uses of $U$ and performing $n$ uses in sequence), the problem becomes non-trivial whenever $k<n$. In this case, unitary iteration cannot be achieved exactly (even probabilistically) since the transformation is non-linear in its inputs. 


The special case of deterministic approximate iteration of a single unitary ($k=1$) has been studied in Ref.~\cite{Soleimanifar_2016}. By approximating the performance operator $\Omega_{\textup{iter}}(d,k=1)$ via Monte-Carlo methods, the authors estimated the optimal performance for $d\in\{2,3\}$ and $n\in\{2,\ldots,7\}$. Interestingly, for $n>d$, the numerical simulations show that on average, the optimal strategy is to discard the input operation $\mathcal{U}$ and output the identity channel $\mathcal{I}$.

\begin{figure}[t!]
    \centering
    \includegraphics[scale=.7]{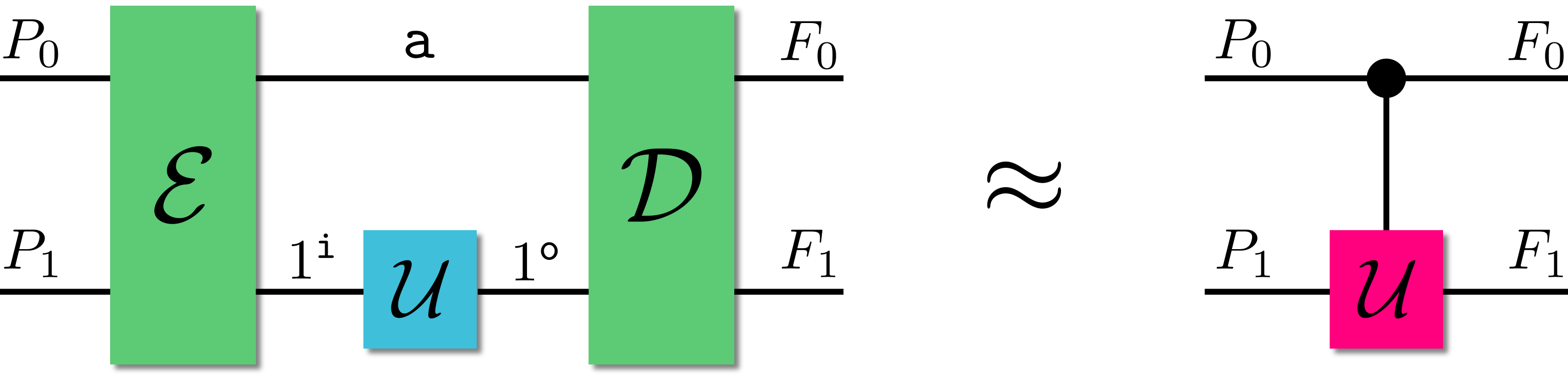} 
    \caption{\textbf{Unitary Controlisation.} The goal of unitary controlisation is to take some number of calls of a unitary $U$ (blue) to the operation $\texttt{ctrl}(U)$ (pink). }
    \label{fig::hoqs-fu-unitarycontrolisation}
\end{figure}

\vspace{0.25cm}\noindent
\textbf{\textul{Unitary Controlisation.}} The \textit{controlisation} of an unknown unitary---universally transforming $U$ into $f(U)=\texttt{ctrl}(U):=\ketbra{0}{0}\otimes \ident + e^{i \theta_U} \ketbra{1}{1} \otimes U$, where $\theta_U\in\mathbb{R}$ (see Fig.~\ref{fig::hoqs-fu-unitarycontrolisation})---is of utmost importance, since many quantum algorithms draw power from such controlled unitary gates (e.g., to generate entanglement). Although any unitary operator has freedom up to a global phase $\theta_U$ of describing the same unitary channel (i.e., $U$ and $e^{i \theta_U}U$ are physically indistinguishable), upon controlisation, such a global phase becomes a relative phase (which could be detected). For this reason, any operator $f(U)=\ketbra{0}{0}\otimes \ident + e^{i \theta_U} \ketbra{1}{1} \otimes U$ is considered to be a faithful controlisation of $U$, regardless the value of $\theta_U$. References~\cite{Soeda2013controlisation, araujo14control, Thompson_2018} have shown that it is impossible to design a universal quantum circuit that exactly controlises an arbitrary unitary operation (even probabilistically); see also Refs.~\cite{arXiv:1006.2670, friis2014control,bisio2016conditional}. Moreover, a considerably stronger no-go result is known: Even when $k$ calls of the unitary are available, any probabilistic universal protocol for controlisation necessarily has a success probability of zero; this result even extends to an approximate non-exact regime~\cite{Gavorova2024topological}. 

Nonetheless, there exist strategies to controlise a unitary operation in a \textit{non-universal} manner, i.e., given partial knowledge of the unitary. For instance, given knowledge of an eigenstate of the unitary, it can be controlised deterministically and exactly with just a single call~\cite{kitaev1995Control,Dong_2019,Gavorova2024topological}. Alternative approaches include implementing related functions: Ref.~\cite{Dong_2019} demonstrated a sequential strategy that transforms $k=d$ calls of an arbitrary $d$-dimensional unitary $U$ into $f(U)=\ketbra{0}{0}\otimes \ident + e^{i \theta_U} \ketbra{1}{1} \otimes \sqrt[d]{U}$, where $\sqrt[d]{U}$ is a unitary operator such that $(\sqrt[d]{U})^d=U$. This construction makes use of the relation $U^{\otimes d}\ket{A_d}=e^{\theta_U} \ket{A_d}$ that holds for the $d$-dimensional totally anti-symmetric state $\ket{A_d}\in {\mathbb{C}_d}^{\otimes d}$ and for any $d$-dimensional unitary $d$. In other words $\ket{A_d}$ is always an eigenstate of $U^{\otimes d}$.


\vspace{0.25cm}\noindent
\textbf{\textul{Functions of Isometries.}} More broadly, the above concepts extend to isometries: operators $V \in \mathscr{L}(\mathscr{H}_d, \mathscr{H}_D)$ such that $V^\dagger V = \mathds{1}_d$, preserving inner products whilst mapping between spaces of different dimensions. Such operations transform pure input states in a $d$-dimensional system to pure output states in a $D$-dimensional one; the special case $d=D$ corresponds to unitary operations. The ability to invert an unknown isometry $\mathcal{V}$, i.e., construct $\mathcal{V}^{-1}$ such that $\mathcal{V}^{-1} \mathcal{V} = \mathcal{I}_d$ has been analysed in Ref.~\cite{Yoshida_2023}, where it was shown that the success probability of any parallel protocol is independent of the output dimension $D$ and satisfies
\begin{align}
    p_{\textup{iso inv}}^{\textup{par}}(d,k) \geq \frac{\lfloor\frac{k}{d-1}\rfloor}{d^2 + \lfloor\frac{k}{d-1}\rfloor-1},
\end{align}
which generalises the unitary case and is optimal for $d=2$. Whether or not the optimal success probability for isometry inversion depends upon $D$ for sequential protocols remains an open question, as does the potential enhancement from causally indefinite strategies. Lastly, the task of isometry adjointation---which includes isometry inversion as a special case---has been analysed in Ref.~\cite{Yoshida_2024}.

\FloatBarrier


\subsubsection{Unitary Storage-and-Retrieval}\label{subsubsec::hoqs-unitarystoreandretrieve}\hfill\\

\begin{figure}[t!]
    \centering
    \includegraphics[width=0.98\linewidth]{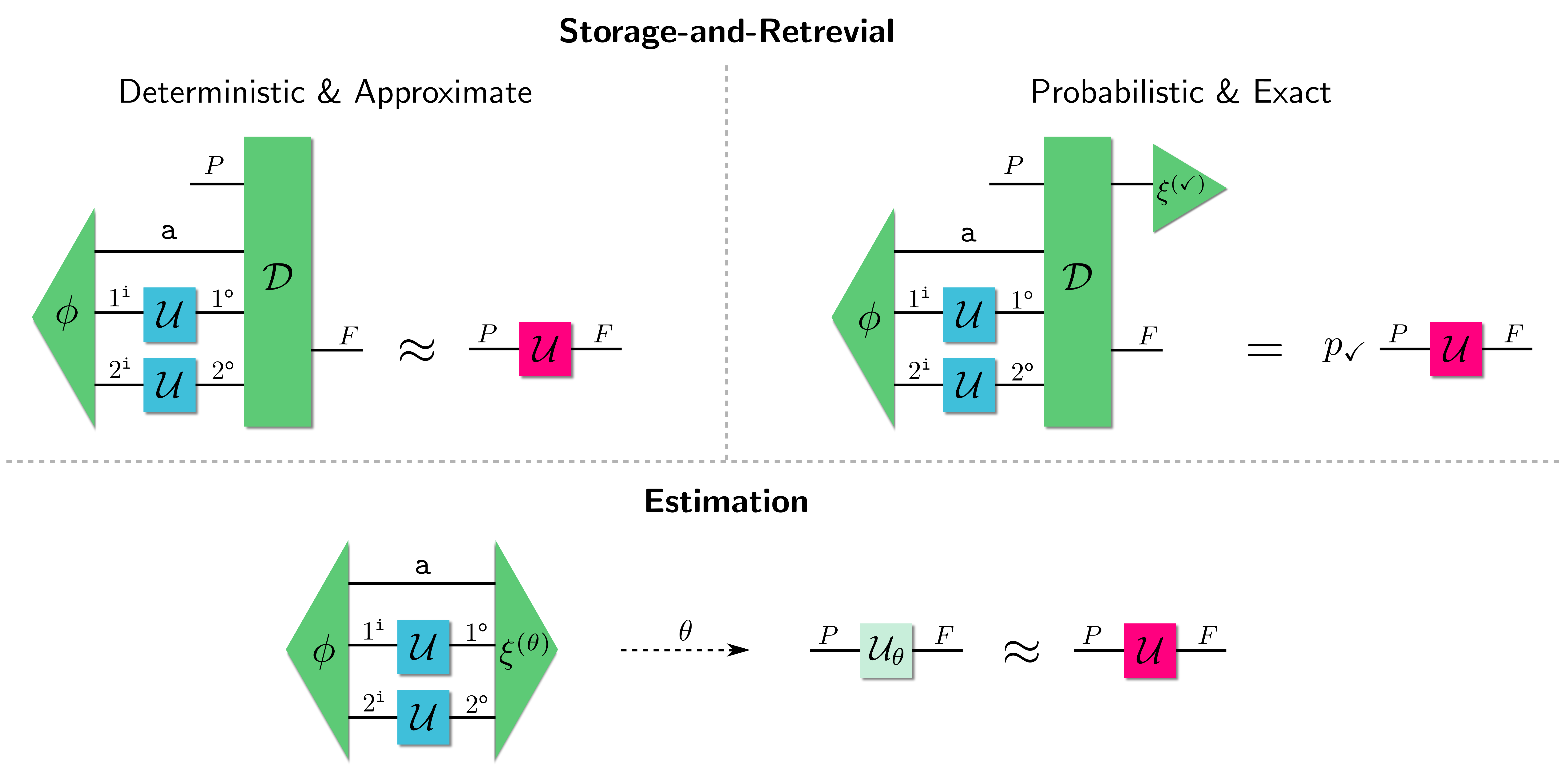}
    \caption{\textbf{Unitary Storage-and-Retrieval and Unitary Estimation.} For unitary storage-and-retrieval (upper panel), one first prepares a state $\phi$ that is subject to $k$ calls of the unitary $U$. The memory state $(U^{\otimes k}\otimes \ident_\aux)\, \ketbra{\phi}{\phi} \left(U^{\otimes k}\otimes \ident_\aux\right)^\dagger$ is \textit{stored} until a later time when one desires to \textit{retrieve} the action of $U$ on a arbitrary state corresponding to the space $\mathscr{H}_P$; this retrieval process is achieved via a quantum channel (referred to as a decoder $\mathcal{D}$). This task can be achieved in two ways: in the deterministic \& approximate (upper left), the action of $U$ is only required to be approximately recovered; in the probabilistic \& exact (upper right), an additional measurement is performed for which a desired outcome $\checkmark$ occurs with probability $p_\checkmark < 1$, heralding that $U$ has been enacted exactly. In the unitary estimation problem (lower panel), one is required to perform a quantum measurement on the memory state $(U^{\otimes k}\otimes \ident_\aux) \, \ketbra{\phi}{\phi} \left(U^{\otimes k}\otimes \ident_\aux\right)^\dagger$ to obtain a classical outcome $\theta$. From this estimator, one aims to construct an associated unitary operation $U_\theta$ that approximates $U$. In Ref~\cite{Bisio_2010}, it is proven that optimal deterministic \& approximate unitary SAR is attainable by an estimation strategy. We stress that here, we follow the same colour coding as previously, with unknown inputs in blue and outputs in pink.}
    \label{fig::hoqs-unitarysar}
\end{figure}


\noindent A particularly interesting and well-studied task is that of unitary \textbf{storage-and-retrieval (SAR)}. It is illustrated in Fig.~\ref{fig::hoqs-unitarysar} and formalised as follows: Alice is given access to $k$ calls of an unknown $d$-dimensional unitary operation $U:\Hscr_I\to\Hscr_O$. Alice seeks to store the action of this unitary---which may correspond to some quantum oracle that solves a particular problem of interest---in some quantum state, so that she may retrieve the use of this unitary at a later time. 

\begin{enumerate}
    \item \textbf{Storage:} First, Alice prepares a multi-partite resource state $\ket{\phi}\in\Hscr_I^{\otimes k}\otimes \Hscr_\aux$, and applies $k$ calls of the unitary $U$ part of this state to obtain the so-called \textit{memory state} $(U^{\otimes k}\otimes \ident_\aux)\ket{\phi}$. 
    \item \textbf{Retrieval:} At a later time, when Alice no longer has access to the unitary $U$, but rather only the memory state $(U^{\otimes k}\otimes \ident_\aux)\ket{\phi}$, she desires to retrieve the action of $U$ on an arbitrary \textit{target} state $\ket{\psi}\in\Hscr_P$ to obtain $U\ket{\psi}$. For this goal, Alice performs a global operation $\mathcal{D}:\mathscr{L}(\Hscr_O^{\otimes k} \otimes\Hscr_\aux\otimes\Hscr_P)\to\mathscr{L}(\Hscr_F)$ on $\left(U^{\otimes k}\otimes \ident_\aux\ket{\phi}\right)\otimes\ket{\psi}$ dubbed a \textit{decoder}. This operation may be a deterministic one, hence achieved by a quantum channel and leading to a deterministic \& approximate protocol; or a probabilistic \& exact one, hence achieved by performing a measurement on an additional auxiliary output of the decoder with a nominal outcome $\checkmark$ heralding exact retrieval of $U$ with some success probability $p_\checkmark <1$. 
\end{enumerate}

\noindent The no-programming theorem~\cite{Nielsen_1997} implies that there is no deterministic operation such that the aforementioned task can be achieved perfectly, i.e., there does not exist a quantum channel \\ $\mathcal{D}:\mathscr{L}(\Hscr_\out^{\otimes k} \otimes\Hscr_\aux\otimes\Hscr_P)\to\mathscr{L}(\Hscr_{F})$ such that
\begin{align}
    \mathcal{D} \left( \left[(U^{\otimes k}\otimes \ident_\aux)\ketbra{\phi}{\phi}({U^\dagger}^{\otimes k}\otimes \ident_\aux)\right] \otimes \ketbra{\psi}{\psi}\right)=U\ketbra{\psi}{\psi}U^\dagger
\end{align} 
for all quantum states $\ket{\psi}\in\Hscr_P$ and all unitaries $U$. Hence, just like for many functions of unitaries considered in the previous section, the task of unitary storage-and-retrieval must be considered from two angles (see Fig.~\ref{fig::hoqs-unitarysar}): For \textit{deterministic \& approximate} strategies, one aims to construct a quantum channel $\mathcal{D}$ that leads to the best approximation  $\mathcal{D}\left( \left[(U^{\otimes k}\otimes \ident_\aux)\ketbra{\phi}{\phi}({U^\dagger}^{\otimes k}\otimes \ident_\aux)\right] \otimes \ketbra{\psi}{\psi}\right)\approx U\ketbra{\psi}{\psi}U^\dagger$. For \textit{probabilistic \& exact} implementations, one must find an additional measurement to be performed on an auxiliary system output by the decoder $\mathcal{D}$ such that a nominal outcome $\checkmark$---which occurs with probability $p_{\checkmark}<1$---heralds that the recovery step perfectly reproduces the action of any $U$ on an arbitrary state $\ket{\psi}$.

\vspace{0.25cm}\noindent\underline{\textbf{Probabilistic \& Exact.}} In Ref.~\cite{Sedlak_2019} it was proven that the optimal success probability of probabilistic and exact $d$-dimensional unitary SAR when $k$ calls are available is given by
\begin{align} \label{eq::hoqs-sar-probabilistic}
    p^{\text{SAR}}_{\checkmark}(d,k) = 1 - \frac{d^2}{k+d^2-1},
\end{align}
and similarly to its deterministic non-exact counterpart~\cite{Bisio_2009}, optimal storage is achieved via $k$ parallel calls of the input operation. Interestingly, this maximum success probability can also be attained by a probabilistic \textbf{port-based teleportation (PBT)}~\cite{Mozrzymas2018PBT}. However, implementing a probabilistic unitary SAR procedure via a probabilistic PBT protocol requires exponentially larger memory than the construction presented in Ref.~\cite{Sedlak_2019}.

This shows that the task of unitary SAR is related to that of PBT. Another interesting connection is to the task of unitary transposition: Ref.~\cite{Quintino_2019_PRA} showed that the problem of probabilistic exact unitary SAR protocol is equivalent to parallel probabilistic unitary transposition. More precisely, any circuit that attains probabilistic exact unitary storage-and-retrieval with probability $p$ can be converted into a parallel quantum circuit for probabilistic exact unitary transposition with the same probability $p$ and \textit{vice versa}. Lastly, variations of probabilistic unitary SAR have also recently been considered, including cases where the unitaries are assumed to be qubit phase gates~\cite{sedlak2020phase}, pairs of unitary channels~\cite{Sedlak_2024}, or cases where access to multiple copies of the input state is assumed~\cite{Grosshans_2024}.

\vspace{0.25cm}\noindent\underline{\textbf{Deterministic \& Approximate.}} When considering deterministic and approximate protocols, the problem of unitary SAR is often referred to as `unitary learning'~\cite{Bisio_2010}. Reference~\cite{Bisio_2009} proved that the optimal average fidelity for $d$-dimensional unitary SAR when $k$ calls are available is given by the maximal eigenvalue of a matrix whose coefficients depend on group representation theory quantities. For qubits, these eigenvalues can be computed analytically, and the optimal average fidelity admits the closed expression
\begin{align}\label{eq::hoqs-sar-deterministic}
\langle \Xi^\text{SAR}(d=2,k)\rangle =1-\sin^2\left(\frac{\pi}{k+3}\right).
\end{align}
Reference~\cite{Bisio_2009} also proved that optimal deterministic unitary learning strategy is obtained by a \textit{unitary estimation} protocol (see Fig.~\ref{fig::hoqs-unitarysar}). The unitary estimation procedure consists of performing a measurement on the state $(U^{\otimes k}\otimes \ident_\aux)\ket{\phi}$ and subsequently using the measurement outcome to guess the unitary $U$. Denoting the guessed operation by ${U}_\theta$, the performance of unitary estimation is evaluated by the average fidelity between the guess ${U}_\theta$ and the given $U$.

As is the case with probabilistic and exact unitary SAR, the task of deterministic and approximate unitary SAR exhibits interesting connections to other tasks concerning unitary transformations. In Ref.~\cite{Quintino_2022_Quantum}, deterministic SAR was shown to be equivalent to parallel unitary inversion and parallel unitary transposition: More precisely, any quantum comb that achieves deterministic SAR with an average fidelity of $\langle \Xi(d,k) \rangle $ can be converted to a circuit that attains deterministic parallel unitary inversion/transposition with the same average fidelity $\langle \Xi(d,k) \rangle $ and \textit{vice versa}. 

Moreover, in Ref.~\cite{Yoshida_2024_Estimation}, it was shown that the problem of deterministic approximate SAR with $k$ calls of the input operation has a one-to-one correspondence to the problem of deterministic PBT with $N=k+1$ ports. That is, any process that performs deterministic SAR with $k$ calls of $U$ with an average fidelity of $\langle \Xi_\text{SAR}(d,k) \rangle $ can be converted to a deterministic PBT protocol with $N=k+1$ ports and the average fidelity $\langle \Xi_\text{PBT}(d,N=k+1) \rangle = \langle \Xi_\text{SAR}(d,k) \rangle $ and \textit{vice versa}. Reference~\cite{Yoshida_2024_Estimation} then built on the results of Refs.~\cite{Yang2020Optimal,christandl2021asymptotic,haah2023query} to show that the asymptotic behaviour of optimal average fidelity for deterministic SAR is  
\begin{align}
\langle \Xi^\text{SAR}(d,k)\rangle = 1-\Theta\left(\frac{d^4}{k^2}\right),    
\end{align}
where $f(x) = \Theta(g(x))$ indicates that $f$ is asymptotically bounded by $g$ both from above and below, i.e., they are asymptotically of the same order.

Lastly, the task of deterministic SAR of \textit{projective measurements} has been analysed in Refs.~\cite{bisio2011learningMeasurements,lewandowska2022measurementSAR,lewandowska2024SARmeasurement2}. In such cases, adaptive strategies can outperform parallel ones~\cite{lewandowska2022measurementSAR}, and strategies with indefinite causal order can in turn outperform adaptive ones~\cite{lewandowska2024SARmeasurement2}. 

\FloatBarrier

\subsubsection{Unitary Resetting, Rewinding / Fast-Forwarding, and Refocusing}\label{subsubsec::unitaryresetting}\hfill\\

\noindent In Ref.~\cite{Navascues_2018} the author considered the task of \textit{resetting} an unknown unitary evolution, i.e., transforming $k$ calls of an arbitrary unitary operation $U$ into the identity channel. Unlike the cases discussed in the previous sections, this work considers a more restricted scenario where the HOQO acts only on an auxiliary system and not on the target itself. Before the protocol begins, the target system interacts with an auxiliary system via the unknown unitary $U$, which entangles the target and the auxiliary system. Due to the built up correlations, probabilistic and exact unitary resetting can be achieved with non-zero probability by manipulating only the auxiliary degrees of freedom, with no access to the target system itself. Considering the same setting, Ref.~\cite{Trillo_2020} developed a universal protocol that enables probabilistic \textit{rewinding} and \textit{fast-forwarding} of an unknown unitary quantum evolution, i.e., transforming $k$ calls of an input unitary operation $U$ into $U^{x}$ where $x\in\mathbb{R}$; this problem includes unitary inversion as a special case but does not coincide with the protocols discussed above, since the considered setup differs in terms of the degrees of freedom the protocol can act upon. Subsequently, Ref.~\cite{Trillo_2023} presented a qubit unitary rewinding protocol whose success probability approaches one as the number of calls to the input unitary operation increases. This probabilistic qubit rewinding protocol was later experimentally realised in an optical setup using the quantum switch~\cite{Schiansky_2023}.

In contrast to the setting described in the previous paragraph, Ref.~\cite{sardharwalla2016refocusing} considered a scenario without auxiliary systems, i.e., where all interventions are performed directly on the target system and no access to an additional auxiliary system is possible. In this case, the authors presented a protocol called quantum \textit{refocusing}, which transforms $n=\log^2(1/\epsilon)$ calls of an arbitrary unitary operation into an operation that is $\epsilon$-close to the identity operator.

\FloatBarrier


\subsubsection{Transforming Hamiltonian Dynamics}\label{subsubsec::hoqs-hamiltoniandynamics}\hfill\\

\noindent So far, we have considered information processing tasks that are implementable via a HOQO that takes quantum maps as its inputs. However, if additional assumptions on the input maps are imposed, then more power is granted. One natural such assumption is that the input channel is generated by some continuous-time dynamics, e.g., the promise that an input unitary is generated by a Hamiltonian dynamics $U = \textup{exp}(-iHt)$ for some Hamiltonian $H$. The additional power gained from such a promise can be seen by the following simple example. Consider the task of implementing the transformation $U \mapsto f(U) = \sqrt{U}$. Clearly, with a single call of the unknown input unitary, this transformation is impossible to realise as a quantum operation, since it is non-linear. On the other hand, if one has the additional information that the input unitary is generated by Hamiltonian dynamics, then the square root function can be universally implemented without knowledge of the input $U = \textup{exp}(-iHt)$ by simply running the dynamics for half of the original time. 

The fact that Hamiltonian dynamics is (infinitely) divisible allows one to consider `breaking up' the dynamics into smaller pieces that can then be processed in a parallel, sequential, or causally indefinite manner. This concept of fractional query access permits access to $U^{\frac{1}{n}}$ as an input, and has subsequently been leveraged to demonstrate HOQOs that universally, deterministically and exactly achieve controlisation of an arbitrary $d$-dimension unitary given $d$ calls of the dynamics~\cite{Dong_2019}. As discussed in Sec.~\ref{subsubsec::hoqs-functionsunitaries}, such universal unitary controlisation is impossible without the promise of a Hamiltonian generator for the dynamics, even if multiple calls are available. This power of fractional query access also permits the projective measurement of the energy of a physical system up to arbitrary accuracy without any dimension dependence on its time cost~\cite{Nakayama_2015}. Additionally, the ability to implement general \textit{functions} of Hamiltonians $H \mapsto f(H)$ given access to only the dynamics $\textup{exp}(-iHt)$ has been demonstrated in Refs.~\cite{Odake_2023_Linear,Odake_2023}. Such algorithms can universally perform, for instance, negative time evolution $H \mapsto -H$, time reversal $H \mapsto H^{\textup{T}}$, and single parameter estimation, and learning~\cite{Zhao_2024_Learning}, amongst other tasks; all without any knowledge of the Hamiltonian $H$.


\FloatBarrier


\subsubsection{Metrology Strategies, Discrimination Tasks \& Higher-Order Resources}\hfill\\
\label{subsubsec::hoqs-metrologydiscrimination}

\vspace{0.25cm}\noindent \textbf{\textul{Metrology \& Parameter Estimation.}} Beyond these above tasks, inspired by the construction of quantum computation subroutines, HOQOs---be they time-ordered quantum combs or processes in superpositions of causal orders---have shown significant promise in enhancing quantum metrology and parameter estimation. Such advantages arise by allowing the construction of \textit{adaptive} metrology strategies that mitigate the impact of environmental noise, improving precision limits beyond those achievable by conventional sequential or parallel protocols~\cite{Zhao_2020_Metrology,Altherr_2021,liu_optimal_2023,Kurdzialek_2023,Mothe_2024}. Notably, causal superpositions enable flexibility in measurement order, further enhancing robustness against noise. In terms of estimating unknown parameters of a quantum comb, Ref.~\cite{Chiribella_2012_Optimal} put forth a framework for deriving the optimal metrology strategy; said framework was subsequently expanded in Refs.~\cite{Yang_2019_Memory,Altherr_2021}, where the authors derived an SDP to bound the quantum Fisher information of a non-Markovian quantum process, which has an operational interpretation as the maximal amount of information that can be extracted from it by an optimally controlled probe state. In cases where the noise is structured or correlated, Ref.~\cite{Kurdzialek_2024} applied tensor network techniques to efficiently derive optimal metrological strategies; see also Ref.~\cite{Liu_2024_Metrology} for an in-depth review of such techniques. HOQOs have also been applied to the setting of Bayesian parameter estimation~\cite{Meyer_2023,Bavaresco_2024}; by explicitly modelling the Bayesian cost function, the framework provides a means to optimise both the prior information and the resource allocation, achieving superior precision compared to traditional methods. Lastly, HOQOs provide the basis for the framework of `Hamiltonian recognition', which aims to identify the Hamiltonian governing a quantum dynamics from a known set of Hamiltonians~\cite{Zhu_2024_Optimal}.\footnote{See also Refs.~\cite{Wiebe_2014,Huang_2023,Bakshi_2024} for results on the task of `Hamiltonian learning', which aims to learn an unknown Hamiltonian given only access to its dynamics. However, these works do not explicitly employ the framework of HOQOs.} By leveraging HOQOs, these works develop systematic methods to design protocols that outperform standard metrological techniques, especially in multi-parameter estimation tasks, and emphasise practical implementations where experimental constraints or limited resources necessitate efficient and adaptive strategies. 

\vspace{0.25cm} \noindent \textbf{\textul{Quantum Channels with Memory.}} Early works on quantum combs focused on studying the channel capacity of quantum channels with memory~\cite{PhysRevA.65.050301, Bowen_2004, Kretschmann_2005, PhysRevA.71.062304}. In this setting, Alice aims to transmit a message to Bob by sending quantum states via a sequential use of a communication channel. Unlike standard scenarios, where these channels are assumed to be identical and independent, here, the channels in each run are not independent of each other, but the channel that Alice uses at time $t_n$ can depend (both in a classical and quantum manner) on the states she sent to Bob at times $t_{n-1}, t_{n-2}, \dots$ Put differently, the channel from Alice to Bob is subject to noise that has memory, which can be modelled, for instance, via a quantum many-body system that serves as an environment~\cite{PhysRevLett.99.120504, Plenio_2008}. This construction is remarkably close to the process tensor formalism used to study quantum non-Markovianity [see Sec.~\ref{subsec::opensystemdynamicsquantummemory}]. The topic of quantum channels with memory is far too rich to properly cover here; fortunately, there exists an extensive review for interested Readers to turn towards~\cite{Caruso_2014}.

\begin{figure}[t!]
    \centering
    \includegraphics[scale=.6]{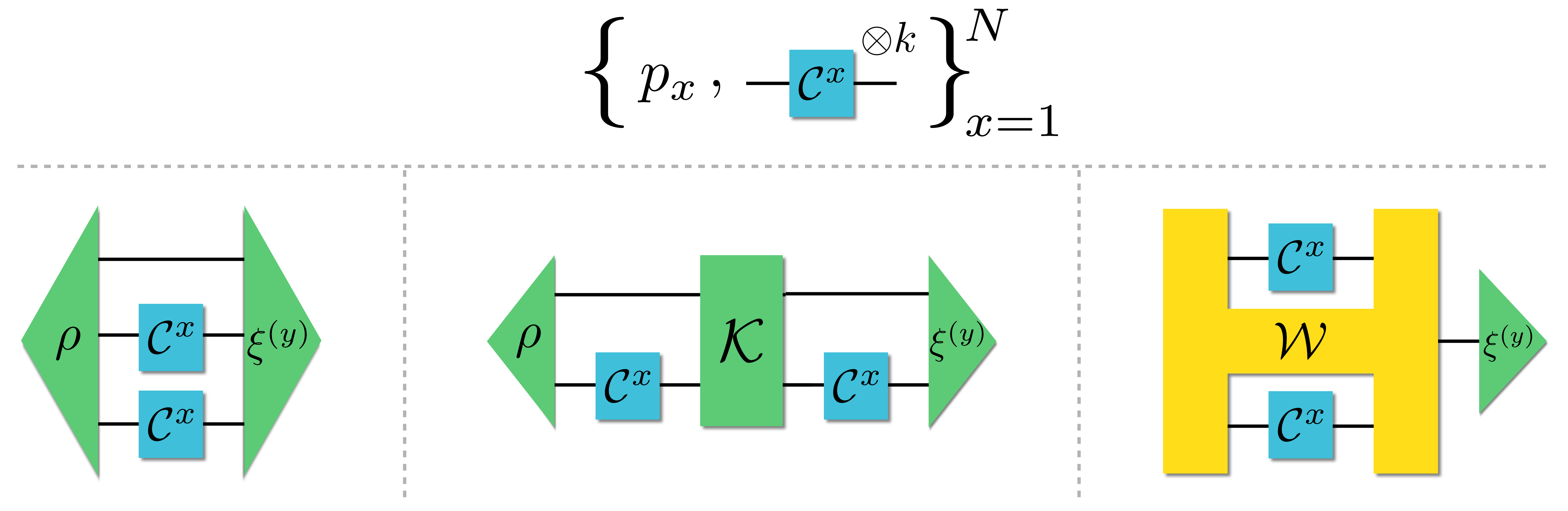}
    \caption{\textbf{Quantum Channel Discrimination.} Alice is provided with one of a set of candidate channels $\{ \Ccal^x\}$ chosen with probability $p_x$ (upper panel). Her task is to process these channels either in parallel (lower panel, left), sequentially (lower middle), or with a general tester (lower right) to determine which channel she was given. The task is successful for a given run whenever the outcome $y$ corresponding to the concluding POVM element $\xi^{(y)}$ coincides with the label $x$ of the channel $\mathcal{C}^x$ of said run.}
    \label{fig::hoqs-md-channeldiscrimination}
\end{figure}

\vspace{0.25cm} \noindent \textbf{\textul{Discrimination Tasks.}} We now move to discuss the task of channel discrimination. In Sec.~\ref{subsubsec::hoqs-functionsunitaries}, we considered quantum information processing protocols in which the respective HOQOs acted on \textit{unitary} input operations. Here, we will consider the setting where the input objects are arbitrary quantum channels: the task is to discriminate a set of quantum channels. Intuitively, this task can be seen as a higher-order extension of the well-known task of quantum state discrimination~\cite{Helstrom_1969}. There, one aims to find an optimal measurement to distinguish between candidates of a set of states; in quantum channel discrimination, one aims to find an optimal tester to distinguish between candidates of a set of channels (see Fig.~\ref{fig::hoqs-md-channeldiscrimination}). Again, there are various possible figures of merit, including maximising the success probability with respect to minimum error discrimination (in which a particular candidate is always guessed) or with respect to unambiguous discrimination (in which the agent is permitted an uninformative outcome); here, we mostly focus on the former.


More formally, the task is the following: With some probability $p_x$, Alice is provided some channel $\mathcal{C}^{x}: \Lcal(\Hscr_{\inp}) \mapsto \Lcal(\Hscr_{\out})$ drawn from a set of candidates $\{ \mathcal{C}^{x} \}_{x=1}^N$ which is \textit{a priori} known to her. She can then use a finite number $k$ of copies to try to learn which particular channel she received, i.e., the classical label $x \in \{ 1, \hdots, N\}$. 

Suppose, e.g., that Alice is allowed only a single use of the channel. Then, the most general quantum strategy she could implement involves feeding part of a joint state $\rho_{\inp \aux}$ into the channel and then performing some joint measurement $\mathcal{J} = \{ \xi^{(y)} \}$ on the joint output and auxiliary space. Her success probability is determined by how often she guesses correctly, i.e., when outcome $y$ coincides with the label $x$ of the channel, which is given by 
\begin{align}
    p_{\checkmark}(\rho, \mathcal{J}) := \sum_x p_x \sum_y \delta_{xy}\tr{(\mathcal{C}^x \otimes \mathcal{I}_{\aux})[\rho] \, \xi^{(y)}},
\end{align}
where $\delta_{xy}$ denotes the Kronecker delta distribution. The optimal strategy involves finding a pair of input states $\rho$ and measurements $\mathcal{J} = \{ \xi^{(y)} \}$ such that the above quantity is maximised, i.e.,
\begin{align}
    p_{\checkmark}^* := \max_{\rho, \mathcal{J}} \; p_{\checkmark}(\rho, \mathcal{J}).
\end{align}
Since every such initial state and global measurement pair corresponds to a tester, the 
above optimisation is an SDP optimisation over a one-slot testers $\mathcal{T} := \{ \mathsf{T}^{(y)} \}$
\begin{align}
    p_{\checkmark}^* := \max_{\mathcal{T}} \sum_x p_x \sum_{y} \delta_{xy} \tr{\mathsf{C}^x \mathsf{T}^{(y) \textup{T}}}.
\end{align}
More generally, however, Alice can be permitted multiple calls of the channel. In such cases, the strategy with which she processes them becomes of pivotal importance. For instance, she could process them all in parallel or sequentially (potentially with some auxiliary system in both cases), or even with a more general causally indefinite strategy (see Fig.~\ref{fig::hoqs-md-channeldiscrimination}). For the fully parallel case, the most general strategy corresponds to preparing a global state, feeding parts into the different copies of the channels, and then performing a global measurement; thus, the optimisation becomes one over single-slot testers (albeit with many input and output wires) as above. For sequential strategies, Alice can plug the channels into a more general comb and in the causally indefinite case, she could use a general process matrix (see Fig.~\ref{fig::hoqs-fu-structurestrategies}).

When considering discrimination between pairs of unitary channels, it has been shown that parallel strategies cannot be outperformed by sequential ones~\cite{Chiribella_2008_ChannelDisc}, and it is unknown whether indefinite causal order may provide an advantage. However, in the special case of discriminating a set of uniformly distributed unitary channels that obey a group structure, parallel strategies are indeed optimal and neither sequential~\cite{Chiribella_2008_ChannelDisc} nor causally indefinite strategies can provide any advantage~\cite{Bavaresco_2022}. Indeed, whenever discriminating a set of $N$ uniformly distributed $d$-dimensional unitary operations with $k$ calls, the optimal success probability (for minimum error state discrimination) for any class of strategies respects~\cite{Bavaresco_2022}
\begin{align}
    p^*_{\checkmark}(d,k,N)\leq \frac{1}{N} \dbinom{k+d^2-1}{k}.
\end{align}
Moreover, whenever the set of unitaries to discriminate form a group $k$-design, the bound is attainable with a parallel strategy~\cite{hayashi05color,chiribella04covariant,chiribella06likelihood}. 

Beyond the special case of discriminating uniformly distributed unitary channels, sequential strategies have been shown to provide an advantage. In particular, enhancements are known for the cases of distinguishing between entanglement breaking channels~\cite{Harrow_2010}, pairs of qubit amplitude damping ones~\cite{Bavaresco_2021,Katariya2021}, sets of unitary operations without a uniform distribution or without a group structure~\cite{Bavaresco_2022}, amongst others~\cite{Pirandola_2019,Zhuang_2020}. Moreover, the advantage of indefinite causal order over sequential strategies (of a fixed causal order) has been demonstrated for certain tasks, such as distinguishing amongst pairs of non-signalling channels~\cite{Chiribella_2012,Bavaresco_2021}. In summary, there is a strict hierarchy between the power of the three distinct types of strategies for channel discrimination.

A related task to channel discrimination is that of \textit{channel comparison}, where one aims to compare whether a number of channels drawn from some candidate set are the same or different. This task has been analysed in the context of unitary channels in Refs.~\cite{Sedlak_2009, Shimbo_2018, Soeda_2021,Hashimoto_2022} (see also Sec.~VII of Ref.~\cite{Wechs_2021}). Moreover, HOQOs have been used to develop a framework for hypothesis testing of symmetries in quantum dynamics~\cite{Chen_2024_Hypothesis} as well as for the identification of cause-effect relations~\cite{Chiribella_2019speedup}.

Lastly, Ref.~\cite{Lewandowska_2023Strategies} has considered the problem of discriminating process matrices in a single shot regime, where the measurements on process matrices are given by a set of positive semidefinite operators that add up to a non-signalling quantum channel. Here the authors showed that---somewhat surprisingly---for the case of discriminating amongst pairs of \textit{causally ordered} processes, such `non-signalling' testers are optimal and cannot be outperformed by more general testers.

\vspace{0.25cm}\noindent\underline{\textbf{Resourcefulness of Higher-Order Quantum Operations.}} 
The capabilities of HOQOs can be systematically investigated by means of resource theories. These theories attempt to formalise the notion that certain objects---typically states or channels, but  these can also be higher-order objects---are cheap, easy, or useless; examples include thermal states in the resource theory of athermality and local operations and classical communication (LOCC) channels in the resource theory of entanglement, to name but a few. A resource theory typically formally comprises a set of free objects, a set of transformations that preserves said set, and a set of monotones that characterise the resourcefulness of any non-trivial process~\cite{Chitambar_2019}. At this level of abstraction, there is no problem with considering scenarios where the free objects are themselves CPTP maps ~\cite{theurer_quantifying_2019, liu_resource_2019, gour_dynamical_2020, gour_dynamical_2020a, gour_entanglement_2021, gour_inevitable_2024} or HOQOs~\cite{Araujo_2015, Berk_2021, Milz_2022}, leading to dynamic, rather than static resource theories. Here, the possible transformations are those that map certain classes of HOQOs to other HOQOs, and the monotones are defined upon such objects accordingly. For instance, these approaches have been employed to 
investigate the resourcefulness of non entanglement-breaking channels~\cite{gour_dynamical_2020, gour_entanglement_2021} and coherence preserving channels~\cite{theurer_quantifying_2019}, where channels that are \textit{not} entanglement breaking and those that do not destroy quantum coherence are considered resources, respectively. Further up the hierarchy of HOQOs, resource theories where process matrices and transformations thereof have been considered, leading to resource theories of causal non-separability~\cite{Araujo_2015, Milz_2022} and causal connection~\cite{Milz_2022}. In the former, causally non-separable process matrices play the role of resourceful objects, while in the latter, this role is played by process matrices that permit signalling between the different parties.

In a similar vein, in Ref.~\cite{Berk_2021} the authors investigate the general conditions under which some uncontrolled background process---corresponding to a quantum comb / process tensor---could be harnessed and exploited by an agent to perform a task that would otherwise be impossible; for instance, they consider quantum operations that might take memoryless processes to ones that have memory. To this end, they develop nine distinct resource theories with a hierarchical relationship implied by the restriction of the quantum of classical communication scenarios in the allowed transformations, one of which is a truly quantum resource theory of non-Markovianity.

Beyond the identification of higher order resources, such approaches have been employed to generalise concepts from the static case to the dynamical one. For example, considering superchannels that preserve the set of marginally uniform channels invariant leads to a natural definition of entropy for quantum channels (instead of quantum states). Similarly, considering transformations of process matrices has led to a notion of dynamics of causal structures~\cite{Castro-Ruiz_2018, Selby_2020}.

The framework of HOQOs thus provides a unified approach to understanding and implementing quantum channel and HOQO transformations, with applications ranging from basic quantum control to complex information processing tasks. The interplay between different implementation strategies, resource constraints, and achievable performance continues to reveal fundamental aspects of quantum information processing while also suggesting practical protocols for quantum computation and communication.


\subsection{Open System Dynamics \& Memory Effects}
\label{subsec::opensystemdynamicsquantummemory}

In examining open system dynamics and memory effects, the HOQO framework offers a distinct perspective from traditional approaches. The fundamental premise is that the universe can be effectively divided into two parts: an experimentally accessible \textit{system} of interest (say, a qubit in a quantum computer) and an inaccessible \textit{environment}. The goal is to characterise system behaviour using only system-level information without explicitly referencing environmental degrees of freedom.

We emphasise that the HOQO approach to open quantum systems is not the only possible framework for this task and differs markedly from master equations descriptions that model the evolution of a quantum state $\rho_t$, most often assumed to be weakly coupled to an external environment. Master equations, say, of the form $\dot{\rho}_t = \Lcal_t[\rho_0]$ (for some superoperator $\Lcal_t$ that depends on the coupling between system and environment) describe \textit{continuous} evolution of the system of interest. At the same time, the HOQO approach that we present here only applies to discrete points in time. On the other hand, HOQOs capture genuine multi-time correlations that cannot be derived from master equations, even in cases where the master equation is of \textbf{Gorini-Kossakowski-Sudarshan-Lindblad (GKSL)} form~\cite{arenz_distinguishing_2015, Milz_2019_CPDiv}. Master equation descriptions (and related approaches) have been thoroughly discussed in Refs.~\cite{Carmichael, BreuerPetruccione, gardiner_quantum_2010, rivas_quantum_2014, Breuer2016, Devega_2017, PRXQuantum.5.020202}; for a comparison of different types of memory effects therein, see also Refs.~\cite{Pollock_2018_PRL, Li_2018, quantumstochasticprocessesandquantumnon}. Here, we focus exclusively on the HOQO description of open system dynamics. 

\FloatBarrier


\subsubsection{Memory Effects in Quantum Processes}\label{subsubsec::osdme-memoryeffects}\hfill\\

\noindent \textbf{\textul{Open System Dynamics with Initial Correlations.}} The aim of open quantum system dynamics is to model the behaviour of a system of interest that is coupled to an environment. Under the assumption of an initially uncorrelated system-environment state, any global unitary dynamics where the environment is eventually discarded produces a linear CPTP map on the system alone[see Eq.~\eqref{eq::me-osdic-uncorrelatedcase}]. Crucially, all information needed to reconstruct such a quantum channel---and thus gather a complete description of the open system evolution---resides on the level of the system, despite the environment's critical role in the dynamics. Conversely, the Stinespring dilation theorem~\cite{Stinespring_1955} ensures that for any CPTP map, there exists a corresponding global unitary and environment state that generates it [see Eq.~\eqref{eq::me-osdic-uncorrelatedcase}], making CPTP maps natural descriptors of open quantum system dynamics whenever the system can be prepared independently of the environment.

The situation becomes more layered when considering initially correlated states $\eta_{SE}$. When $\eta_{SE} \neq \rho \otimes \tau$, the na{\" i}ve extension of Eq.~\eqref{eq::me-osdic-uncorrelatedcase} would read
\begin{align}\label{eq::osdme-me-initialuncorrelated}
    \rho_t = \ptr{E}{\mathcal{U}_t[\eta_{SE}]} \stackrel{?}{=} \widetilde{\mathcal{C}}_t[\rho]\,.
\end{align}
Here, physical arguments dictate that $\widetilde{\Ccal}_t$ should be (at least) linear in its input and CPTP; however, demanding this leads to inconsistencies, as we now detail. Writing the initial state as $\eta_{SE} = \rho \otimes \tau + \chi$, where $\tau := \ptr{S}{\eta_{SE}}$ yields 
\begin{align}
\rho_t = \ptr{E}{\mathcal{U}_t[\rho \otimes \tau]} + \ptr{E}{\mathcal{U}_t[\chi]} =: \Ccal_t[\rho] + \xi_t^{\chi}\, ,
\end{align}
where $\Ccal_t$ is CPTP by construction and $\xi_t^{\chi}$ is traceless. The above form seems to suggest that system dynamics could be described by a trace-preserving \textit{affine} map that depends on the correlation matrix $\chi$. Besides losing it's CPTP property, as alluded to in Sec.~\ref{subsubsec::me-opensystemdynamicswithinitialcorrelations}, the matrix $\xi_t^{\chi}$ depends on \textit{how} the system state $\rho$ was prepared starting from $\eta_{SE}$~\cite{modi_role_2010, modi_preparation_2011}, seemingly making the above mapping $\rho \mapsto \rho_t$ non-linear. 

More generally, as Pechukas and Alicki notably demonstrated, three crucial properties of quantum dynamics---linearity, complete positivity, and consistency of assignment maps---cannot hold simultaneously when initial correlations are present~\cite{Pechukas_1994, Alicki_1995, Pechukas_1995}. In such cases, it impossible to define a linear (let alone completely positive) map that faithfully takes arbitrary input states of the system to their corresponding output states; an issue that has been observed in experiments~\cite{obrien_quantum_2004, Weinstein2004, myrskog_quantum_2005, kavanthesis}. This assigns a peculiar role to the initial time in two-time quantum experiments, as initial correlations evidence memory of previous system-environment interactions. Amongst others, this breakdown in formalism extends to multi-time quantum experiments, where even initially uncorrelated states typically become correlated at later times.

Consequently, there have been a plethora of mathematical approaches to try to describe open quantum dynamics with initial correlations, most prominently among them: not completely positive maps~\cite{stelmachovic_dynamics_2001, Jordan_2004, shaji_whos_2005, Jordan_2006, carteret_dynamics_2008, vacchini_reduced_2016} and maps that are only CP on restricted domains~\cite{shajithesis, RodriguezRosario_2008, RodriguezRosario_2010, brodutch_vanishing_2013}. However, most of these suffer from an unclear operational interpretation~\cite{modi_role_2010, modi_preparation_2011} and are mutually incompatible (see, e.g., Refs.~\cite{Vacchini_2011,Chrusinski_2011,Li_2018}). 

The solution lies in reconceptualising the problem, as discussed in Sec.~\ref{subsubsec::me-opensystemdynamicswithinitialcorrelations}: Instead of mapping initial system states to output states, initial \textit{preparation procedures} should be mapped to final output states via a superchannel~\cite{Modi_2012_SciRep,Modi_2012_PRA}. This shift to HOQOs not only provides mathematical remedy but establishes the correct understanding of mappings in open quantum systems: they should take experimenter-controllable inputs to measurable outputs [see Fig.~\ref{fig::me-osdic-superchannel}]. In the uncorrelated case, input (output) states of the system can be freely chosen (measured); in the correlated case, the initial preparation procedure (output state) can be freely chosen (measured).

Due to linearity, the resulting superchannel $\mathcal{T}_t$ can be tomographically reconstructed (see Sec.~\ref{subsubsec::ced-characterisingtamingquantumprocesses} for a detailed discussion of experimental reconstructions) through a finite number of experiments using $d^4$ linearly independent input CP maps $\Mcal^{(x)}$ and determining the corresponding output states $\rho_t^{(x)}$ (which in turn takes another $d^2$ measurements)~\cite{Modi_2012_SciRep, Milz_2017}.\footnote{The number $d^4$ is the dimension of the space spanned by the set of CP maps acting on $\Lcal(\Hscr_d)$, which can be seen by noting that the Choi states of CP maps are $d^2\times d^2$ matrices that span a $d^4$-dimensional vector space.} Thus, the superchannel provides an operational description of the situation at hand (i.e., open quantum dynamics in the presence of initial correlations) that preserves the desired physical properties of linearity, complete positivity, and trace preservation while maintaining an operational description that can be uniquely characterised on the level of the system alone. \\

\begin{figure}[t!]
    \centering
    \includegraphics[scale=.65]{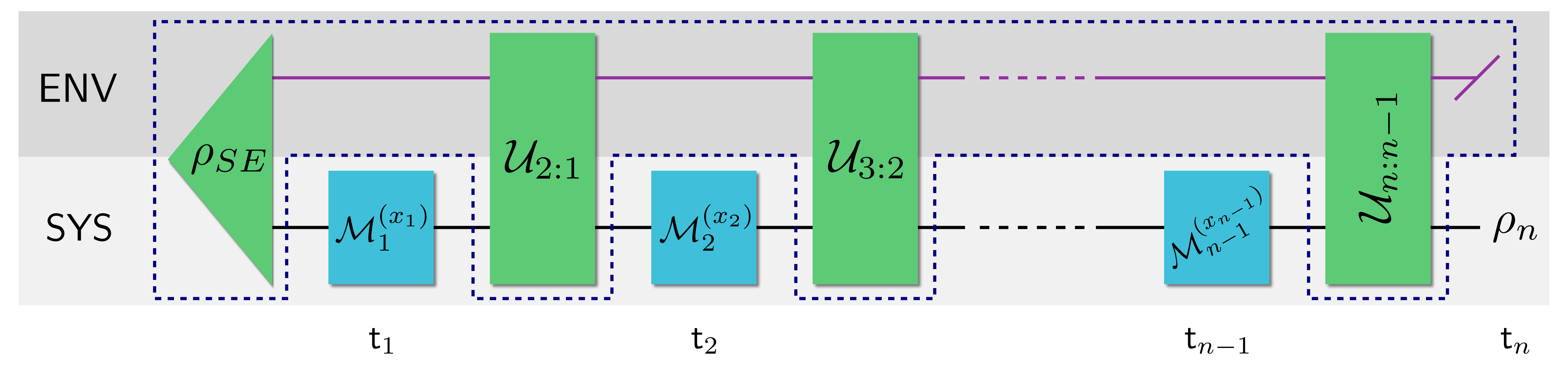}
    \caption{\textbf{Process Tensor.} The process tensor (dashed navy outline) provides an operational description of the multi-time open quantum dynamics of a system (SYS, shaded below in light grey) coupled to an environment (ENV, shaded above in dark grey). It maps arbitrary sequences of operations $\{ \mathcal{M}_i^{(x_i)} \}_{i=1}^{n-1}$ (blue) applied to the system at times $t_1, t_2, \dots, t_{n-1}$ to the corresponding output state $\rho_n$ at time $t_n$.}
    \label{fig::osdme-me-processtensor}
\end{figure}

\noindent \textbf{\textul{Multi-Time Quantum Processes: Process Tensor.}} The framework developed for two-time processes naturally extends to multi-time scenarios where a system is probed sequentially at multiple times. When a system is interrogated at times $\{t_1, \dots, t_{n-1}\}$ using instruments $\Jcal_1, \dots, \Jcal_{n-1}$, the resulting dynamics can be described uniquely by a unique multi-linear map $\mathcal{T}_{n:1}$ (see Fig.~\ref{fig::osdme-me-processtensor})~\cite{Pollock_2018_PRA, quantumstochasticprocessesandquantumnon}. This map transforms any sequence of operations applied to the system into the corresponding final output state at time $t_n$. Consider an experimenter probing a system at successive times, obtaining a sequence of outcomes $x_{1}, \dots, x_{n-1}$ corresponding to CP maps $\Mcal^{(x_{1})}, \dots, \Mcal^{(x_{n-1})}$. The resulting output state can be expressed as a composition of these operations and the intervening global unitary evolutions via
\begin{align}\label{eq::osdme-me-processtensor}
\rho_n(x_{n-1}, \hdots, x_1|\mathcal{J}_{n-1},\hdots,\mathcal{J}_1) &= \ptr{E}{\mathcal{U}_{n:n-1} \mathcal{M}^{(x_{n-1})}_{n-1} \mathcal{U}_{n-1:n-2}\hdots \mathcal{M}^{(x_2)}_2 \mathcal{U}_{2:1} \mathcal{M}^{(x_1)}_1 \rho_{SE}} \notag \\ 
&=: \mathcal{T}_{n:1}[\mathcal{M}^{(x_{n-1})}_{n-1},\hdots,\mathcal{M}^{(x_1)}_1] = \mathsf{T}_{n:1} \bigstar_{i=1}^{n-1} \mathsf{M}_{i}^{(x_i)}. 
\end{align}
Here, identity maps on the environment space are implied and the CP maps $\{ \mathcal{M}^{(x_i)}_i \}$ act only on the system (whereas the unitary evolutions $\mathcal{U}_{i+1:i}$ between interrogations are global). 


From the system-environment dynamics above (see also Sec.~\ref{subsec::timeorderedquantumprocesses}), it is clear that the HOQO $\mathsf{T}_{n:1}$ is a quantum comb (defined in Def.~\ref{def::toqp-def-quantumcombs}). Its Choi matrix is thus positive semidefinite and satisfies the the hierarchy of trace conditions outlined in  Eq.~\eqref{eq::toqp-dethoqo-def-quantumcombtp}.

In the context of open quantum systems, this object has been dubbed the \textit{process tensor}. It provides a consistent framework for properly analysing multi-time quantum processes that develop correlations with the environment and the resulting memory effects~\cite{Pollock_2018_PRA,Pollock_2018_PRL, Taranto_2019_PRL, Taranto_2019_PRA, Taranto_2021_npj}. Multi-time process tensors possess the full richness of many-body quantum states~\cite{Luchnikov_2019, White_2021_Many}; this includes genuine multipartite entanglement~\cite{Giarmatzi_2021, milz-spee-ent_2021} for processes with genuine quantum memory, quantum complexity~\cite{Aloisio_2023, Dowling_2024}, and complex memory structures~\cite{Burgarth_2021}.

A key feature of the process tensor is that it permits the calculation of \textit{joint} probability distributions over outcomes at multiple times. These distributions are given by the trace of the final state (or equivalently by the link product), i.e., 
\begin{align}\label{eq::osdme-me-processtensorprobability}
    \Pprob(x_{n-1},\dots,x_1|\Jcal_{n-1},\dots,\Jcal_1) &= \tr{\rho_n(x_{n-1}, \hdots, x_1|\mathcal{J}_{n-1},\hdots,\mathcal{J}_1)} = \mathsf{T}_{n:1} \star (\mathds{1}_n \bigstar_{i=1}^{n-1} \mathsf{M}_{i}^{(x_i)}) .
\end{align}
In contrast to joint probability distributions describing classical processes, these distributions depend explicitly upon the instruments employed---a reflection of the multiple inequivalent ways a quantum process can be probed. These instrument-dependent probability distributions encapsulate all possible accessible memory effects of a process, making process tensors particularly suitable for extending concepts like Markovianity and Markov order to the quantum realm~\cite{Pollock_2018_PRA, Pollock_2018_PRL,Taranto_2019_PRL, Taranto_2019_PRA, Taranto_2021_npj}, as we review in Sec.~\ref{subsubsec::osdme-quantummarkovorder}.

\begin{figure}[t!]
    \centering
    \includegraphics[scale=.65]{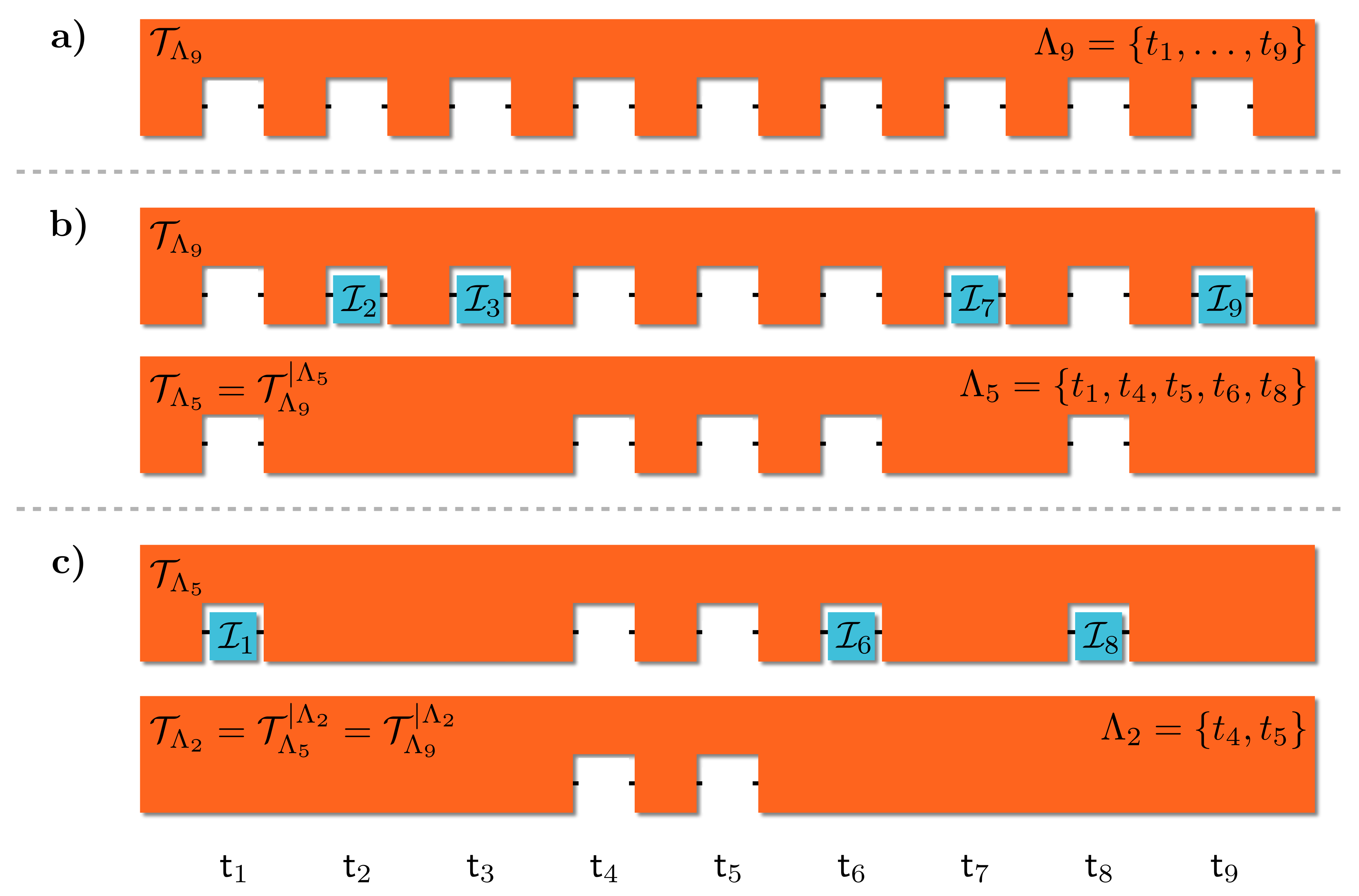}
    \caption{\textbf{Containment Property for the Process Tensor.}  The process tensor satisfies a natural containment property. For concreteness, in panel a) we depict a process tensor over nine timesteps $\mathcal{T}_{\Lambda_{9}}$. From this, the correct descriptor on any subset of timesteps can be derived by letting the it act on identity maps at the appropriate times. Such a restriction of a process tensor on times $\Lambda_\ell$ to times $\Lambda_{k}$ is denoted by $\mathcal{T}^{\Lambda_k}_{\Lambda_\ell}$ in the figure. In panel b) we show how the correct description $\mathcal{T}_{\Lambda_{5}}$ over times $\Lambda_5 = \{t_1, t_4, t_5, t_6, t_8\}$ can be obtained in this way from that defined on $\Lambda_{9} = \{t_1, \hdots , t_{9}\}$. Moreover, in panel c), we show the containment of $\mathcal{T}_{\Lambda_2}$ in both descriptors
$\mathcal{T}_{\Lambda_5}$ and $\mathcal{T}_{\Lambda_{9}}$, where $\Lambda_2 = \{t_4, t_5\}$. The crucial point is that the unique maximal description contains within it the proper description of the process over any subset of timesteps.}
    \label{fig::osdme-me-qolmogorov}
\end{figure}

The framework also provides a natural notion of \textit{marginalisation} in quantum settings via a \textit{containment} property: From a description involving a larger number of time points, one can derive all possible behaviours on any subset of times by inserting identity maps at the superfluous times~\cite{Milz_2020_Quantum} (see Fig.~\ref{fig::osdme-me-qolmogorov}). This ability to obtain reduced descriptions in quantum mechanics is crucial, as such descriptions generally break down unless all multi-time behaviour is captured; see Sec.~\ref{subsubsec::osdme-classicalquantumprocesses}. While here, we describe multi-time quantum processes in terms of process tensors, it is worth noting that this natural extension of classical joint probabilities to \textit{families} of joint probabilities in the quantum case was first expressed using so-called \textit{correlation kernels}~\cite{Lindblad_1979, accardi_quantum_1982}, which provide a description of multi-time quantum processes that is equivalent to that in terms of HOQOs~\cite{9683765}. 


\subsubsection{Markovian Quantum Processes \& Quantum Markov Order}\label{subsubsec::osdme-quantummarkovorder}\hfill\\

\noindent A process tensor $\mathsf{T}_{n:1}$ enables the computation of all conceivable joint probability distributions and well-defined conditional probabilities according to Eq.~\eqref{eq::osdme-me-processtensorprobability}. As a consequence, it allows for an unambiguous description of memory effects, providing clarity in a field where the definition of quantum memory has seen numerous non-equivalent interpretations (see, e.g., Ref.~\cite{Li_2018}).

\vspace{0.25cm}\noindent
\textbf{\textul{Memory Length.}} Memory corresponds to the length of history required to correctly predict future outcomes, formalised through the concept of \textit{Markov order}. A classical process has Markov order (and thus memory length) $\ell$ if its probability distribution satisfies 
\begin{align}
\Pprob(x_{n}|x_{n-1},\dots, x_1) = \Pprob(x_{n}|x_{n-1},\dots, x_{n-\ell}) \quad \forall \, x_{n}, \dots, x_1.
\end{align}
Markovianity represents the special case $\ell \leq 1$. The above expression amounts to conditional independence between outcomes observed in the future (here, $x_{n}$) and history (here, $x_1, \dots, x_{n-1-\ell}$) given those on the memory (here, $x_{n-\ell}, \dots, x_{n-1}$). 

Na\"ively, this notion directly extends to the quantum setting by incorporating the respective instruments with which the system is probed, such that a quantum process with Markov order $\ell$ should satisfy
\begin{align}
    \Pprob(x_{n};\Jcal_{n}|x_{n-1};\Jcal_{n-1},\dots, x_1;\Jcal_1) = \Pprob(x_{n};\Jcal_{n}|x_{n-1};\Jcal_{n-1},\dots, x_{n-\ell};\Jcal_{n-\ell})\, ,
\end{align}
for all outcomes $x_{n}, \dots, x_1$ \textit{and} all instruments $\Jcal_{n},\dots, \Jcal_1$. The object on the r.h.s.\ contains all of the relevant information to determine the behaviour of the process and requires exponentially less (in the number of truncated timesteps) resources to model as compared to the full distribution~\cite{Taranto_2021_npj}. 

However, such a definition in the quantum case is too restrictive, both for Markovianity~\cite{capela_quantum_2022} (i.e., Markov order $\ell \leq 1$) as well as general Markov order~\cite{Taranto_2019_PRA, Taranto_2019_PRL}, leaving only trivial processes and a notion of quantum Markov order that does not coincide with the classical one in the correct limit~\cite{Taranto_Thesis}. This discrepancy arises because although (sharp) classical measurements leave a system in a known state (e.g., finding a classical particle at position $x$ implies that, immediately afterwards, it is indeed at $x$), quantum CP maps do not necessarily decouple the system from its environment, making it impossible to deduce the post-measurement system state with certainty (e.g., a weak measurement will not reset the system to a known state). Consequently, measurements in the classical setting ensure that any conditional dependence of outcomes can be attributed to memory effects in the process \textit{per se}; in contradistinction, such conditional dependence can potentially be transmitted in the quantum case through the measurement instruments themselves\cite{Pollock_2018_PRA, Pollock_2018_PRL, PhysRevLett.121.240401,PhysRevA.99.052125,Taranto_Thesis}.\footnote{Interestingly, similar phenomena can arise in classical stochastic processes where measurements are noisy and cannot be resolved to a sufficient level of granularity~\cite{Siefert2003, Bottcher2006, Lehle2011}; although these can be `argued away' on grounds of fundamentality in the classical realm, it remains an important issue from a practical standpoint.} On the other hand, limiting the employed interrogations of a quantum process to sharp measurements in a fixed basis overlooks experimentally accessible memory effects~\cite{Taranto_2019_PRA}, potentially masking genuinely non-Markovian behaviour by leading to seemingly Markovian statistics that nonetheless have hidden memory effects~\cite{Taranto_2022} (for a discussion of genuine multitime memory effects in concrete dynamical models of open quantum systems, see, e.g., Refs.~\cite{PhysRevLett.121.240401, PhysRevA.99.052125, PhysRevA.102.022216, PhysRevA.103.012221,budini_quantum_2022}).

The notions of Markovianity and Markov order in quantum mechanics thus require particular care with respect to the choice of instruments (in the case of Markovianity~\cite{Pollock_2018_PRL}) or become fundamentally \textit{instrument dependent} (in the case of Markov order~\cite{Taranto_2019_PRL, Taranto_2019_PRA}). Both of these issues can readily be addressed by means of HOQOs.

\begin{figure}[t!]
    \centering
    \includegraphics[scale=.65]{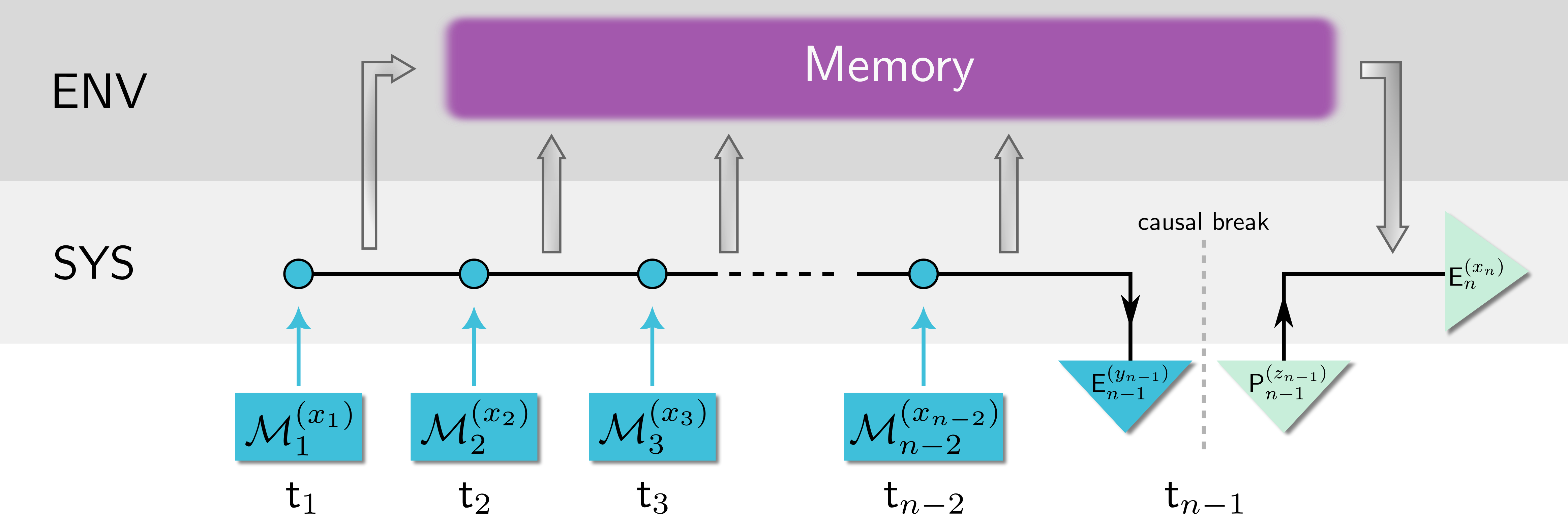}
    \caption{\textbf{Quantum Markovianity.} General quantum operations $\mathcal{M}_{n-2:0}$ are applied to the system during an open quantum process. At the penultimate time $t_{n-1}$, we make a causal break by measuring the system (with corresponding POVM element $\mathsf{E}^{(y_{n-1})}_{n-1}$ and re-preparing it in a fresh state $\mathsf{P}^{(z_{n-1})}_{n-1}$. The process is said to be Markovian iff the final output state $\rho_{n}(\mathsf{P}^{(z_{n-1})}_{n-1},\mathsf{E}^{(y_{n-1})}_{n-1},\mathcal{M}_{n-2:0}) = \rho_n(\mathsf{P}^{(z_{n-1})}_{n-1})$ at all times $t_n$, for all inputs $\mathsf{P}^{(z_{n-1})}_{n-1}$, measurements $\{ \mathsf{E}^{(y_{n-1})}_{n-1}\}$, and control operations $\mathcal{M}_{n-2:0}$. If this conditional independence does not hold, then information about past operations has been transported to the state at time $t_n$ through the memory in the environment.}
    \label{fig::osdme-qmo-quantummarkovianity}
\end{figure}

\vspace{0.25cm}\noindent
\textbf{\textul{Quantum Markovianity.}} The solution to developing a quantum generalisation of \textit{Markovianity} lies in introducing `causal breaks': CP maps that actively reset the system state, allowing for the attribution of temporal correlations exclusively to environmental effects~\cite{Modi_2012_SciRep, Pollock_2018_PRL,Costa_2016, PhysRevA.99.052125, PhysRevLett.121.240401}.\footnote{Such causal breaks are a generalisation of `do-operations' that are used in the field of classical causal modelling~\cite{Pearl} to discover causal relations by actively resetting the state of the system to a known one.} A causal break consists of measurement followed by state preparation, independent of the measurement outcome, which is described by
\begin{align}\label{eq::osdme-qmo-measureprepare}
\Mcal^{(x)}[\rho] := \mathcal{P}^{(z)}\circ\mathcal{E}^{(y)}[\rho] = \tr{\rho \,\mathsf{E}_{{\inp}}^{(y)\mathrm{T}}} \mathsf{P}_{{\out}}^{(z)} = \rho \star \mathsf{E}_{\inp}^{(y)} \star \mathsf{P}^{(z)}_{{\out}} \, ,
\end{align}
where $\rho$ denotes an arbitrary state, $\mathsf{E}^{(y)\mathrm{T}}$ is the POVM element corresponding to the measurement $\mathcal{E}^{(y)}$, $\mathsf{P}^{(z)}$ is the state corresponding to the subsequent preparation $\mathcal{P}^{(z)}$, and the `outcome' label $x$ of the causal break is split into a pair $(z, y)$ to distinguish measurement outcomes ($y$) and re-preparation labels ($z$) (see Fig.~\ref{fig::osdme-qmo-quantummarkovianity}). This operation ensures the system enters a known state and renders the system-environment state into product form. Consequently, any influence of (arbitrary) past measurements detected after a causal break \textit{must} stem from memory effects transmitted through the environment~\cite{Pollock_2018_PRL, Costa_2016, PhysRevA.99.052125, PhysRevLett.121.240401}. 

To test for the presence of non-Markovianity, an experimenter would first choose a sequence of instruments $\mathcal{J}_k = \{ \mathcal{M}^{(x_k)}_k \}$ to probe the system at times $t_k \in \{ t_1, \hdots, t_{n-2} \}$. Then, at the penultimate time $t_{n-1}$, perform an instrument that only contains causal breaks $\Mcal^{(z_{n-1},y_{n-1})}_{n-1}= \mathcal{P}_{n-1^\out}^{(z_{n-1})} \circ \mathcal{E}_{n-1^\inp}^{(y_{n-1})}$, before concluding with a POVM $\Jcal_n = \{E_{n}^{(x_{n}) \textup{T}}\}$ at time $t_n$ (see Fig.~\ref{fig::osdme-qmo-quantummarkovianity}). The resulting conditional probability to obtain outcome $x_n$ at $t_n$ given the history is 
\begin{align}\label{eq::osdme-qmo-nonmarkovianconditionalstate}
    \Pprob(x_{n};\Jcal_{n}|(z_{n-1};y_{n-1});&\Jcal_{n-1},x_{n-2};\Jcal_{n-2}, \dots, x_{1};\Jcal_1) \notag \\
    &=: \Pprob(x_{n};\Jcal_{n}|\mathcal{P}_{n-1}^{(z_{n-1})} \circ \mathcal{E}_{n-1}^{(y_{n-1})}, \Mcal_{n-2}^{(x_{n-2})},\dots, \Mcal_{1}^{(x_{1})}) \notag\\
    &= \frac{\tr{E_{n}^{(x_{n})\textup{T}} \Tcal_{n:1}[\mathcal{P}_{n-1}^{(z_{n-1})} \circ \mathcal{E}_{n-1}^{(y_{n-1})}, \Mcal_{n-2}^{(x_{n-2})},\dots, \Mcal_{1}^{(x_{1})}]}}{\tr{E_{n-1}^{(y_{n-1})\mathrm{T}} \Tcal_{n-1:1}[\Mcal_{n-2}^{(x_{n-2})},\dots, \Mcal_{1}^{(x_{1})}]}} \notag \\
    &= \frac{\mathsf{T}_{n:1} \star (\mathsf{E}_n^{(x_n)} \otimes \mathsf{P}_{n-1^{\out}}^{(z_{n-1})} \otimes \mathsf{E}_{n-1^{\inp}}^{(y_{n-1})} \bigotimes_{\alpha=1}^{n-2} \mathsf{M}_\alpha^{(x_{\alpha})})}{\mathsf{T}_{n-1:1} \star (\mathsf{E}_{n-1^{\inp}}^{(y_{n-1})} \bigotimes_{\alpha=1}^{n-2} \mathsf{M}_\alpha^{(x_{\alpha})})}
\end{align}
where, due to causality, the `reduced' quantum comb $\Tcal_{n-1:1}$ can be obtained uniquely from $\Tcal_{n:1}$. Here, we use both the map description as well as that in the Choi representation to highlight their interchangeability and cater to Readers who may prefer either formalism.


If the above conditional probabilities only depend on $\mathsf{P}_{n-1^\out}$ for all possible instruments, then no information was transported through the environment and to $t_n$ and the process is considered Markovian~\cite{Costa_2016,Pollock_2018_PRL}.\footnote{This definition closely resembles the causal Markov condition employed for classical stochastic processes where interventions are allowed, such as in causal modelling~\cite{Pearl}. There, too, one requires a notion such as a causal break, since active interventions can be unsharp, as opposed to the assumption of standard classical probability theory where the measurements are assumed to be sharp. See Refs.~\cite{Costa_2016, Allen_2017,Cotler_2019,Barrett_2019} for related discussion regarding quantum causal modelling.}Demanding such conditional independence to hold for all times $t_1, \dots, t_n$ forces the process tensor to take the special form of a concatenation of \textit{independent} quantum channels between neighbouring times 
\begin{align}\label{eq::osdme-qmo-markovcombmaps}
    \Tcal_{n:1}^{\textup{Markov}} = \Ccal_{n:n-1} \circ \Ccal_{n-1:n-2} \circ \cdots \circ \Ccal_{2:1} \circ \Rcal_1\, ,
\end{align}
where each $\mathcal{C}_{k+1:k}$ is a CPTP map from $t_k$ to $t_{k-1}$ and $\Rcal_1$ is an initial state preparation at time $t_1$. This structure implies a tensor product form for the Choi matrix of a Markovian process~\cite{Costa_2016,Pollock_2018_PRL}
\begin{align}\label{eq::osdme-qmo-markovcombchoi}
    \mathsf{T}_{n:1}^{\textup{Markov}} = \mathsf{C}_{n:n-1} \otimes \mathsf{C}_{n-1:n-2} \otimes \hdots \otimes \mathsf{C}_{2:1} \otimes \rho_1.
\end{align}
This represents a strong \textit{multi-time} condition that subsumes other definitions of quantum Markovianity in the literature~\cite{Pollock_2018_PRL, Li_2018, Milz_2019_CPDiv, budini_quantum_2022}. For sequences of sharp (rank-1 projective) measurements, processes described by Markovian process tensors (as above) guarantee classical Markovian probability distributions, ensuring the definition appropriately reduces to the classical case~\cite{Pollock_2018_PRL}. On the other hand, probing a Markovian quantum process with instruments that do not only comprise causal breaks can lead to non-Markovian conditional probability distributions~\cite{Taranto_2019_PRA, Taranto_Thesis}.

While detecting non-Markovianity requires finding only a single pair of instrument sequences that produce distinct conditional probabilities, confirming Markovianity demands exponentially many experiments. Lastly, unambiguous measures $\mathcal{N}$ of non-Markovianity can be defined in terms of the minimum distance $\mathcal{D}$ to the set of Markovian processes~\cite{Pollock_2018_PRL}
\begin{align}\label{eq::osdme-qmo-nonmarkovianitymeasure}
    \mathcal{N}(\mathsf{T}_{n:1}) := \min_{\mathsf{T}_{n:1}^{\textup{Markov}} }\mathcal{D}(\mathsf{T}_{n:1}\|\mathsf{T}_{n:1}^{\textup{Markov}}).
\end{align}
Such measures have been given operational meaning as monotones in the resource theory of quantum non-Markovianity~\cite{Berk_2021} and quantifiers for the resourcefulness of processes with memory for quantum information tasks such as dynamical decoupling~\cite{berk_extracting_2021}. However, care must be taken when working with \textit{processes} (as opposed to states), which requires optimisation accounting for ancillary systems (e.g., the gap between the diamond norm and the trace norm~\cite{Zambon_2024, Berk_2024}).

\begin{figure}[t!]
    \centering
    \includegraphics[scale=.6]{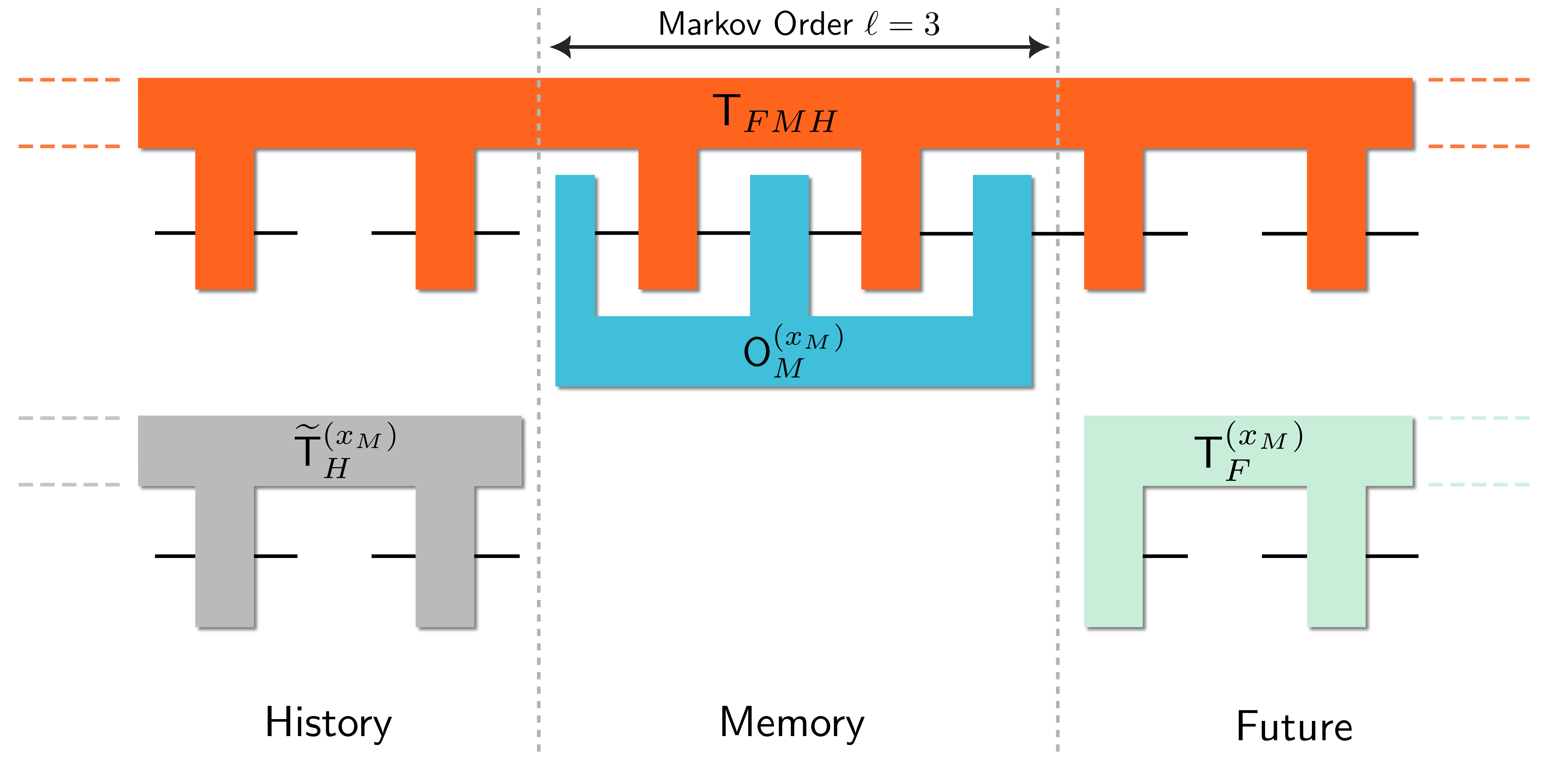}
    \caption{\textbf{Quantum Markov Order.} A tester $\Jcal_M = \{ \mathsf{O}_M^{(x_M)}\}$ (blue) of length $\ell=3$ is applied across a sequence of timesteps to a process tensor $\mathsf{T}_{FMH}$ (orange) spanning the history, memory, and future times. The process is said to have Markov order $\ell$ with respect to this tester if, for each possible realisation $x_M$, the history (grey, $\widetilde{\mathsf{T}}_H^{(x_M)}$) and the future (mint green, $\mathsf{T}_F^{(x_M)}$) are rendered conditionally independent [see Eq.~\eqref{eq::osdme-qmo-quantummarkovorderstructure}].}
    \label{fig::osdme-qmo-quantummarkovorder}
\end{figure}

\vspace{0.25cm}\noindent
\textbf{\textul{Quantum Markov Order.}} Beyond Markovianity, the description of quantum processes in terms of HOQOs also allows for an unambiguous definition of \textit{Markov order}, which, like Markovianity itself, presents itself as an instrument-dependent property in the quantum setting~\cite{Taranto_2019_PRL, Taranto_2019_PRA, Taranto_2021_npj, Taranto_Thesis}. When an operation sequence is applied to an open quantum process, a valid probability distribution is observed; a meaningful definition of quantum Markov order then would be to stipulate that said probability distribution has Markov order $\ell$. By partitioning timesteps into future $F := \{ t_n, \hdots, t_{k}\}$, memory $M = \{ t_{k-1},\hdots,t_{k-\ell}\}$, and history blocks $H = \{ t_{k-\ell-1}, \hdots, t_1\}$, a quantum process exhibits Markov order $\ell$ when the observed statistics satisfy classical Markov order condition~\cite{Taranto_2019_PRL} 
\begin{align}
    \mathds{P}(x_F|\mathcal{J}_F,x_M,\mathcal{J}_M; x_H,\mathcal{J}_H) = \mathds{P}(x_F|\mathcal{J}_F, x_M,\mathcal{J}_M).
\end{align}
This definition translates to a structural constraint on the process tensor $\mathsf{T}_{FMH}$, where observing any length-$\ell$ sequence of outcomes renders the history and future conditionally independent~\cite{Taranto_2019_PRA}
\begin{align}\label{eq::osdme-qmo-quantummarkovorderstructure}
    \mathsf{T}_{FH}^{(x_M)} := \ptr{M}{\mathsf{O}^{(x_M)}_M \mathsf{T}_{FMH}} = \mathsf{T}_F^{(x_M)} \otimes \mathsf{T}_H^{(x_M)}
\end{align}
for all $\mathsf{O}^{(x_M)}_M \in \mathcal{J}_M$ (see Fig.~\ref{fig::osdme-qmo-quantummarkovorder}). 


While this definition of quantum Markov order reduces to its classical counterpart for classical stochastic processes probed with sharp measurements~\cite{Taranto_2019_PRL}, the quantum realm exhibits richer finite memory effects. Classical processes have a fixed Markov order \textit{independent} of measurement choice, owing to the unique deterministic effect of classical instruments. In contrast, quantum theory allows for infinitely many distinct instrument choices. This freedom subsequently implies an important consequence: Quantum processes with finite Markov order for all possible instruments must be completely memoryless~\cite{Taranto_2019_PRL}. In other words, an instrument-\textit{in}dependent notion of quantum Markov order only admits processes with either infinite or zero memory length.

Thus, in the quantum setting, one has no choice but to define memory length in an instrument-dependent manner. The instrument-dependent nature of quantum memory has important practical implications. A process may exhibit different Markov orders for different measurement sequences, as exemplified by memoryless quantum processes showing Markovian behaviour under sharp measurements but non-Markovian behaviour under unsharp ones. While specific instrument sequences may not reveal the complete memory structure, they can nonetheless uncover important structural properties: Ref.~\cite{Taranto_2019_PRA} has characterised processes with finite Markov order for various classes of instruments, including projective measurements, unitary sequences, and informationally complete instruments. Additionally, methods have been developed to `stitch together' finite-length portions of a process tensor with finite memory, demonstrating an efficient way to model the process exactly by only keeping track of descriptions bound by the Markov order~\cite{White_2022}.

In the classical setting, a process that \textit{approximately} satisfies the Markov order $\ell$ condition can always be accurately simulated by a model of Markov order $\ell$, thereby providing an efficient simulation. Recent work considering quantum processes has extended beyond memory length to quantify memory strength across finite timesteps, again relative to specific instrument sequences~\cite{Taranto_2021_npj}. As per the classical case, this quantification has operational significance, as it bounds the error when using approximate processes to simulate expectation values of observables within the span of said instrument. Experimental demonstrations~\cite{Guo_2021,White_2022} and numerical studies~\cite{Taranto_Thesis} have further illuminated the length, strength, and structure of quantum memory effects, advancing our understanding of temporal correlations in quantum systems.

\FloatBarrier


\subsubsection{Classicality \& Memory}\label{subsubsec::osdme-classicalquantumprocesses}\hfill\\

\noindent
\textbf{\textul{Classical Quantum Processes.}} The relationship between quantum processes and measurement sequences extends naturally to defining an operational notion\footnote{We emphasise that the notion of classicality discussed in this section is not the only way of conceptualising \textit{classical} processes (see, e.g., Ref.~\cite{10.21468/SciPostPhys.15.1.024} for an overview). It is, however, the one that is most amenable to being phrased in terms of HOQOs, which is why we opt to present it here.} of \textit{classicality} in quantum processes~\cite{Smirne_2018,Strasberg_2019_Classical, Milz_2020_Classical, PhysRevA.108.012225, 202300304, budini_violation_2023}. The setting is as follows: One chooses a basis in which the target system is to be measured projectively at each time via the instruments $\Jcal_i = \{ \mathcal{P}_i^{(x_i)}\}$, recording the probability distribution (phrased in terms of the Choi matrices of the process $\mathsf{T}_{n:1}$ and the projective measurements $\{ \mathsf{P}_{i}^{(x_i)} \}$)
\begin{align}
    \mathds{P}(x_n,\hdots,x_1|\Jcal_n,\hdots,\Jcal_1) = \tr{\mathsf{T}_{n:1} \bigotimes_{i=1}^{n} \mathsf{P}_{i}^{(x_i)} }.
\end{align} 
This approach distinguishes between classical and quantum processes through the notion of \textit{Kolmogorov consistency}, as we now discuss.

In the classical realm, multi-time probability distributions contain complete information about all possible subsets of measurements through marginalisation. Specifically, consider a classical stochastic process with joint probability distribution $\mathds{P}_\Gamma(x_\Gamma)$ over times $\Gamma:=\{t_1,\hdots,t_n\}$, where we write $x_\Gamma := \{ x_i \}_{i\in\Gamma}$. The `reduced' distribution over any subset of times $\Lambda \subseteq \Gamma$ can be obtained by summing over outcomes at excluded times
\begin{align}\label{eq::osdme-cqp-marginalisation}
    \mathds{P}_\Lambda(x_\Lambda) = \sum_{x \in \Gamma \setminus \Lambda} \mathds{P}_\Gamma(x_\Gamma).
\end{align}
The object on the l.h.s.\ describes the behaviour observed in an experiment in which the agent chooses \textit{not} to measure at times $t_i \in \Gamma \setminus \Lambda$. For classical stochastic processes, this coincides with the marginalisation on the r.h.s.; mathematically, this holds because the classical `do nothing' operation $\Ical$ (implied on the l.h.s.) is equal to the sum over projective measurements $\sum_x \mathcal{P}^{(x)} \bullet \mathcal{P}^{(x)}$ (implied on the r.h.s.). This property, known as \textit{Kolmogorov consistency}, characterises classical stochastic processes: All classical stochastic processes (probed with sharp, projective measurements) yield Kolmogorovian statistics, and conversely, any Kolmogorov-consistent statistics can be reproduced by some classical stochastic process~\cite{Kolmogorov_1956}.

\begin{figure}[t!]
    \centering
    \includegraphics[scale=.55]{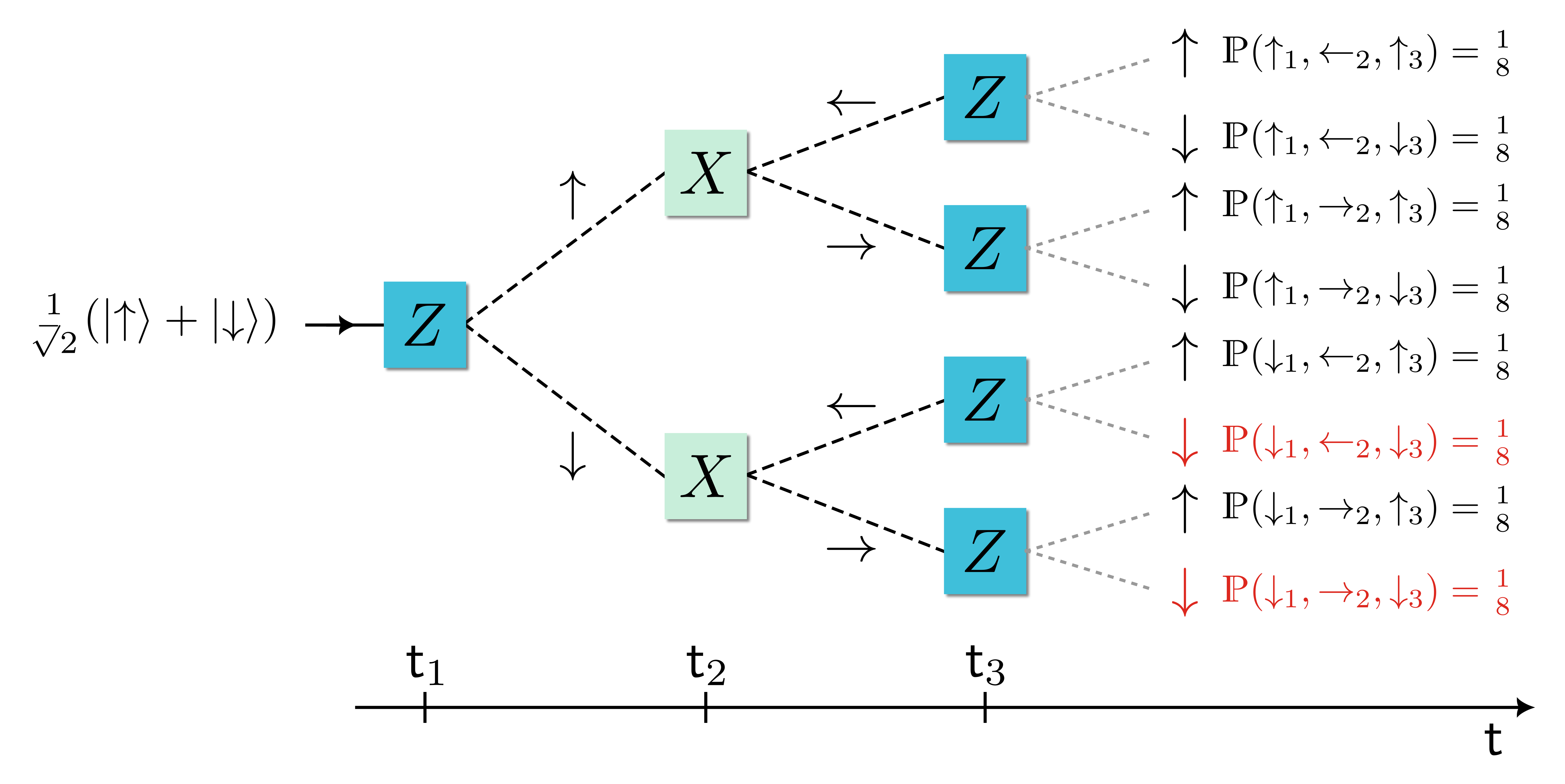}
    \caption{\textbf{Breakdown of Kolmogorov Consistency in a Stern-Gerlach Experiment.} An initial qubit state prepared in an even superposition is subject to three sequential measurements, in the $Z$, $X$, and $Z$ directions respectively (at times $t_1, t_2$, and $t_3$). The joint statistics for each possible sequence of outcomes are equal to $\tfrac{1}{8}$ (shown on the right). However, marginalising over the outcomes observed at the second timestep does not provide the correct probabilities that are predicted by theory in the case where \textit{no} measurement is performed there, highlighting the breakdown of Kolmogorov consistency. If the first outcome is $\downarrow$ and no intervention is made at $t_2$, the measurement at $t_3$ yields $\downarrow$ with certainty. This gives $\mathds{P}(\downarrow_3, \downarrow_1) = \tfrac{1}{2}$, which is in contradiction with the marginalised statistics computed as the sum of probabilities displayed in red (which would give a value of $\tfrac{1}{4}$ for the same sequence).}
    \label{fig::osdme-cqp-sterngerlach}
\end{figure}

However, this consistency typically breaks down in processes involving \textit{invasive} measurements, including both classical causal models~\cite{Pearl} and quantum processes~\cite{Smirne_2018, Strasberg_2019_Classical, Milz_2020_Classical, PhysRevA.108.012225, 202300304, budini_violation_2023}. A canonical example involves sequential Stern-Gerlach measurements of a spin-$\tfrac{1}{2}$ particle---initially in the state $\ket{+} = \tfrac{1}{\sqrt{2}}(\ket{0} + \ket{1}) $---in the Pauli-$z$, Pauli-$x$ and Pauli-$z$ basis (see Fig.~\ref{fig::osdme-cqp-sterngerlach}). Due to the active disturbance of the state by the respective measurements~\cite{Piron1981}, the resulting statistics do \textit{not} satisfy the classical Kolmogorov conditions ~\cite{Smirne_2018, Strasberg_2019_Classical,Milz_2020_Classical, PhysRevA.108.012225, 202300304, budini_violation_2023}. This breakdown of Kolmogorovianity forms the basis for the violation of Leggett-Garg inequalities~\cite{leggett_quantum_1985, leggett_realism_2008}---a temporal analogue of Bell inequalities---observed in quantum mechanics (see~\cite{emary_leggettgarg_2013} for a detailed review).


The Kolmogorov consistency conditions serve a deeper purpose in classical probability theory. Through the Kolmogorov extension theorem~\cite{Kolmogorov_1956}, they establish that any set of consistent statistics over finite time subsets can be extended to a well-defined classical stochastic process over the full time domain. While not satisfying the Kolmogorov consistency conditions in general, the HOQO description of quantum processes displays a \textit{containment} property~\cite{Pollock_2018_PRA, Milz_2020_Quantum} (see Fig.~\ref{fig::osdme-me-qolmogorov}), allowing one to correctly `marginalise' quantum processes. Leveraging this containment property, a generalised Kolmogorov extension theorem for quantum processes has been derived in Refs.~\cite{accardi_quantum_1982, Milz_2020_Quantum, double_nothing}, providing a rigorous mathematical underpinning for the description of quantum stochastic processes.

The breakdown of the Kolmogorov consistency conditions provides a natural delineation between classical and quantum processes: Although generically, quantum processes violate Kolmogorov consistency due to measurement invasiveness, a subset of them can be measured non-invasively for specific instruments; in such cases, the statistics observed are Kolmogorov consistent and the process can meaningfully be considered classical.

\begin{figure}[t!]
    \centering
    \includegraphics[scale=0.55]{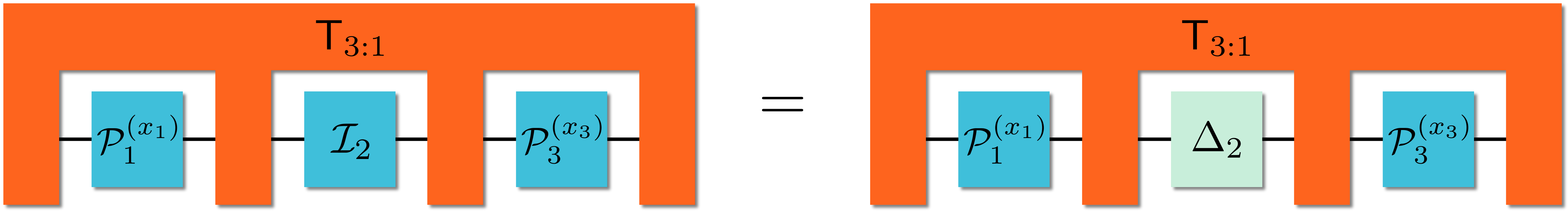}
    \caption{\textbf{Non-Markovian Classical Quantum Comb.} A quantum comb is said to be classical with respect to a fixed (but otherwise arbitrary) basis iff projective measurements in said basis---denoted by $\mathcal{P}_1^{(x_1)}$ and $\mathcal{P}_3^{(x_3)}$ here---cannot distinguish a map $\Delta$ that dephases the system in said basis from `doing nothing' at the same time (denoted by $\mathcal{I}$).}
    \label{fig::osdme-cqp-nonmarkovclassicality}
\end{figure}

Using this operational definition, Refs.~\cite{Strasberg_2019_Classical, Milz_2020_Classical} have provided a structural characterisation of the set of quantum processes (with memory) that produce Kolmogorovian statistics; more physically-motivated underlying dynamics that lead to Kolmogorovian statistics have also been analysed in Refs.~\cite{PhysRevA.108.012225, 10.21468/SciPostPhys.15.1.024, PhysRevX.14.041027, 2406.15577}. These latter works also highlight the close connection between between classically consistent statistics (in the Kolmogorov sense) and the notion of \textit{decoherent histories}, a well-studied approach\footnote{Since this approach does not explicitly rely on the formalism of HOQOs, we will not discuss it in detail here. For an overview, see, e.g., Refs.~\cite{omnes_consistent_1992, paz_environment-induced_1993}.} to the emergence of classical phenomena in quantum processes that provides a strictly stronger notion of classicality than Kolmogorov consistency (see Refs.~\cite{griffiths_consistent_1984, omnes_logical_1988, omnes_consistent_1992, paz_environment-induced_1993, gell-mann_quantum_1997}). Such \textit{classical quantum processes} are precisely those where the `do nothing' operation yields equivalent results to measuring and averaging over outcomes $\mathcal{D}[\bullet] := \sum_x \mathcal{P}^{(x)} \bullet \mathcal{P}^{(x)}$, i.e., processes $\mathsf{T}_{n:1}$ satisfying (see Fig.~\ref{fig::osdme-cqp-nonmarkovclassicality})
\begin{align}
    \tr{\mathsf{T}_{n:1} \left( \bigotimes_{t_i \in \Lambda} \Phi^+_i \bigotimes_{t_j \in \Gamma \setminus \Lambda} \mathsf{P}_{j}^{(x_j)}\right) } = \tr{\mathsf{T}_{n:1} \left( \bigotimes_{t_i \in \Lambda} \mathsf{D}_i \bigotimes_{t_j \in \Gamma \setminus \Lambda} \mathsf{P}_{j}^{(x_j)}\right) },
\end{align}
where $\Phi^+_i$ on the l.h.s.\ represents the identity map whereas $\mathsf{D}_i$ means measure and sum over outcomes.


This condition imposes structural constraints for the quantum comb $\mathsf{T}_{n:1}$~\cite{Milz_2020_Classical}. For memoryless processes $\mathsf{T}_{n:1} = \bigotimes_{i=1}^{n-1} \mathsf{C}_{i+1:i}$, these constraints relate to the ability of the channels between timesteps $\{ \mathsf{C}_{i+1:i} \}$ to generate and detect coherences, establishing a one-to-one correspondence between Kolmogorovian statistics and dynamical properties~\cite{Smirne_2018,Strasberg_2019_Classical, Milz_2020_Classical}. Notably, classical statistics can arise from dynamics that are neither exclusively non-coherence-generating nor non-coherence-detecting, demonstrating that classical behaviour does not require completely incoherent dynamics~\cite{Smirne_2018}. More generally, employing the tools of random matrix theory, it has been shown that for large sets of \textit{closed} dynamics, Kolmogorovianity holds to large precision~\cite{10.21468/SciPostPhys.15.1.024}: For a randomly chosen Hamiltonian as well as a sufficiently degenerate measurement observable, the resulting multi-time statistics observed by probing the system are almost indistinguishable from classical ones, despite the measurements themselves being invasive, i.e., changing the state of the system.

The relationship between dynamics and classicality becomes more subtle when memory is present. While system-environment dynamics that neither generate nor detect quantum \textit{discord} (with respect to a chosen measurement basis) always produce classical statistics, the converse does not hold~\cite{Milz_2020_Classical}. Nonetheless, any classical statistics can always be realised by a dilation involving non-discord-generating-and-detecting dynamics. Further developments include an efficiently computable measure of non-classicality and the identification of `genuinely quantum processes' that generate non-Kolmogorovian statistics for all possible measurement sequences~\cite{Milz_2020_Classical}. Such phenomena necessarily involve memory effects, as memoryless processes always admit at least one non-invasive measurement sequence (namely, in the instantaneous eigenbasis).

\vspace{0.25cm}\noindent
\textbf{\textul{Quantum Regression Formula and Hidden Quantum Memory.}} The above results concern the interplay between the process \textit{per se} and the instruments used to probe it. This relationship between processes and their observable statistics reveals fundamental differences between classical and quantum processes. While \textit{structural} properties of processes (e.g., memorylessness, non-coherence/discord-generating-and-detecting, etc.) often imply specific \textit{operational} expressions for the observed statistics (e.g., Markovianity, classicality, etc.), the reverse implications are more nuanced. Unlike classical systems, where both the operational and structural descriptions coincide in a single probability distribution, quantum processes maintain a rich structure beyond their observed statistics for any given measurement sequence; in other words, a quantum process is not fully characterised by the joint probability distribution over measurement outcomes (for any given instrument sequence).

This quantum-classical divergence manifests in several ways, particularly in how structural and operational concepts relate. A prime example is the relationship between memoryless dynamics and Markovian statistics. While classically these concepts are equivalent, they differ in the quantum realm~\cite{Taranto_2022}. In classical physics, all Markovian statistics can be described using a memoryless dynamical model (that is, as arising from a sequence of independent stochastic matrices). The quantum case is more nuanced: measuring a fixed observable does not provide tomographically complete information. As such, it is unsurprising that there exist processes with memory that nonetheless appear Markovian when probed by that observable~\cite{Taranto_2019_PRL,Taranto_2019_PRA,Taranto_2021_npj}. Despite this complexity, it has been widely assumed that for any quantum experiment yielding Markovian statistics, there must exist some memoryless quantum dynamics that can reproduce the observed data. This assumption, known as the \textbf{quantum regression formula (QRF)}~\cite{Lax_1963,Carmichael,BreuerPetruccione}, provides a crucial link between operational quantities (recorded statistics) and dynamical ones (models of the underlying system).

In Ref.~\cite{Taranto_2022}, the authors investigate whether Markovian statistics can always be faithfully reproduced by a memoryless dynamical model and show that this is generally not the case: Some processes exhibiting Markovian statistics fundamentally \textit{require} memory in their underlying system-environment description. In other words, although the statistics observed are Markovian, the underlying process \textit{must} have memory to faithfully reproduce them, revealing a new quantum phenomenon dubbed \textit{hidden quantum memory}. Such nuances naturally come to light when viewed through a higher-order lens, in which all multi-time phenomena are captured.

\begin{figure}
    \centering
    \includegraphics[width=0.7\linewidth]{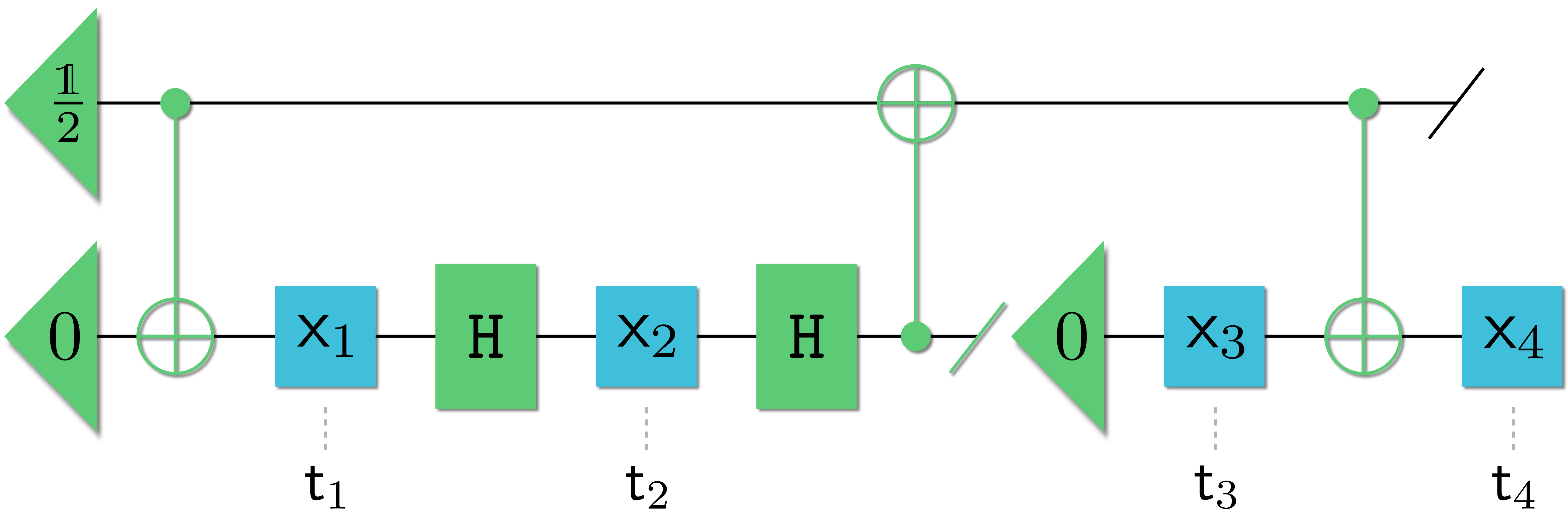}
    \caption{\textbf{Incompatible Markovian Statistics.} Here, the blue boxes denote times at which an agent can interrogate the system; $\mathtt{H}$ represents a Hadamard gate. When the $\sigma_z$ observable with possible outcomes $\textsf{x}_i \in \{0,1\}$ is measured on the system at all subsets of times, the circuit yields Markovian statistics and sub-statistics. Despite this Markovianity, the respective conditional probabilities are incompatible, i.e., they depend on whether or not previous measurements were performed. Such behaviour is \emph{only} possible in the presence of underlying, hidden memory.}
    \label{fig::osdme-cqp-hiddenquantummemory}
\end{figure}

This discovery suggests that quantum memory is an emergent phenomenon: Observing Markovian behaviour with respect to a fixed measurement basis does not guarantee the existence of a memoryless dynamical description. This hidden quantum memory joins other quantum phenomena that require precisely the resource in their implementation that they ultimately conceal, such as quantum channels that preserve separable states but require entanglement to implement~\cite{Plenio_2007,Horodecki_2009,akibue_thesis,Chitambar_2020}, non-signalling maps that require signalling to realise~\cite{beckman_causal_2001}, and maximally incoherent operations that need coherent resources~\cite{Chitambar_2016_Critical,Chitambar_2016_Comparison,Marvian_2016}. 

\begin{figure}[t!]
    \centering
    \includegraphics[width = 0.95\linewidth]{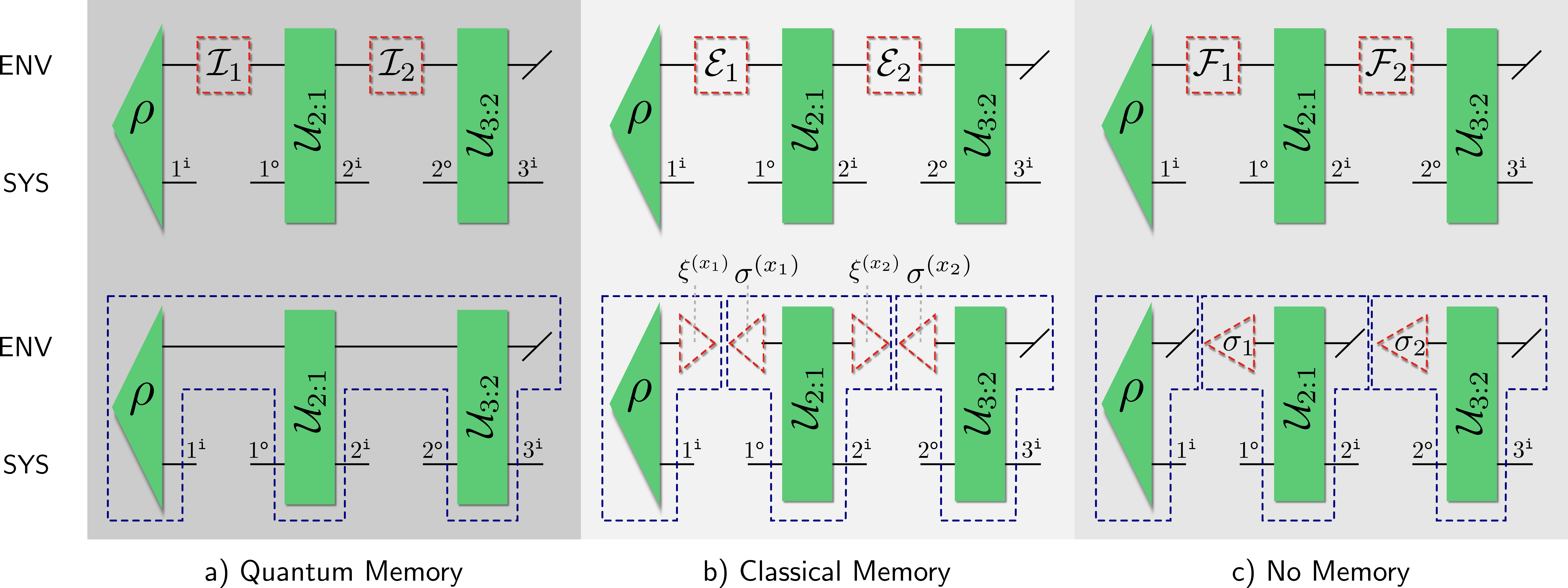}
    \caption{\textbf{Classical Memory Quantum Processes.} Quantum processes can have: a) quantum memory; b) classical memory; or c) no memory (depicted for $n=3$ times). In the upper panels, we show the system-environment representation of the dynamics: in all scenarios, the system-environment evolves unitarily according to $\mathcal{U}_{k:k-1}$ between times. The distinct memory effects depend upon what happens to the environment between times: in the case of quantum memory, the environment evolves coherently, represented by the identity channel $\mathcal{I}$; in the case of classical memory, it is subject to an entanglement-breaking channel $\mathcal{E}$ given by a POVM $\{\xi^{(x_k)}\}$ and a set of states $\{\sigma^{(x_k)}\}$; in the case of no memory, it is discarded and freshly re-prepared, represented by $\mathcal{F}$. In the lower panels, we invoke the structure of the relevant environment channels to deduce the process tensor form (blue dashed outlines). The general case cannot be broken up; the classical memory case leads to a sequence of conditional instruments as each future entanglement-breaking channel can depend upon previous outcomes, which is both more general than convex mixtures of memoryless processes and a special case of separable process tensors; the memoryless case leads to a sequence of independent CPTP channels.}
    \label{fig::osdme-cqp-classicalmemoryquantumprocess}
\end{figure}

\vspace{0.25cm}\noindent
\textbf{\textul{Classical Memory Quantum Processes.}} The characterisation of quantum processes with classical memory mechanisms provides another perspective on this quantum-classical boundary.\footnote{In Refs.~\cite{szankowski_noise_2020, szankowski_objectivity_2024}, the possibility to model external quantum noise a classical noise field was employed as an explicit notion of classicality based on quantum processes. Here, concerning classical memory effects, we instead focus on structural properties of HOQOs.} Such processes can be meaningfully defined as those where the environment undergoes entanglement-breaking channels between timesteps, permitting only classical information to feed forward~\cite{Giarmatzi_2021,Nery_2021,Taranto_2024} (see Fig.~\ref{fig::osdme-cqp-classicalmemoryquantumprocess}). However, this seemingly straightforward characterisation reveals unexpected complexity: These processes form a distinct set from separable process tensors~\cite{Nery_2021} and cannot be represented simply as probabilistic mixtures of Markovian processes~\cite{Taranto_2024}. Recent work has mapped out the relationships between different classes of memory mechanisms in quantum processes and developed computational methods for approximating these sets~\cite{Taranto_2024} (see also Ref.~\cite{Ohst_2024} for a discussion and computational characterisation of different types of memory applied to channel discrimination tasks). Furthermore, the types of underlying continuous-time dynamics that give rise to such classical-memory quantum processes have recently been analysed~\cite{goswami_2024}.


The persistence of quantum effects even under classical memory constraints is particularly noteworthy. Studies have shown that even when memory undergoes entanglement-breaking channels between timesteps, the resulting statistics can still exhibit features impossible to replicate with classical systems of equivalent dimension~\cite{Vieira_2024}. This demonstrates that even when memory appears classical in nature, quantum resources can enable temporal correlations beyond classical bounds, highlighting the subtle interplay between classical and quantum information in multi-time processes.

Lastly, there is a related effort to simulate non-Markovian classical stochastic processes using quantum resources. In Ref.~\cite{Gu_2012}, the authors showed that a certain class of classical hidden Markov models can be simulated using less quantum memory. Crucially, this result extends to multi-time correlations, which are fundamental to classical stochastic processes~\cite{Binder_2018_Practical}. Unsurprisingly, structures like matrix product states arise naturally~\cite{Yang_2018_Matrix}, which have applications to many-body physics~\cite{Suen_2018} and agent-based modelling~\cite{Thompson_2017,Elliott_2021}. This turns out to be a rich research area with many facets (see also Refs.~\cite{Thompson_2018_Causal, Korzekwa_2021} for related results); the role of HOQOs here has not been fully explored and has the potential to be rather fruitful.

\FloatBarrier


\subsection{Many-Time Quantum Physics} \label{subsec::manytimequantumphysics}

\noindent A key desire in physics is that of \textit{simulation}, where one aims to reproduce some observed behaviour by controlling a (typically less complex than the original) process. There are a number of distinct behaviours one might wish to simulate; here we focus on some particularly relevant ones which can most naturally be phrased in terms of HOQOs. Below, we discuss how process tensors / quantum combs are increasingly used for simulating the dynamics of many-body physics, understanding quantum chaos, and studying strong coupling thermodynamics.


\subsubsection{Simulating Complex Quantum Processes}
\label{subsubsec::mtqp-simulatingcomplexquantumprocesses}\hfill\\

\noindent This story begins with the Feynman-Vernon influence functional, a foundational framework for studying complex non-equilibrium physics that forms the backbone of several powerful numerical methods. Implementations like the \textbf{quasi-adiabatic path integral (QuAPI)}~\cite{Makri_1995a, Makri_1995b} and InchWorm~\cite{Cohen_2015} techniques demonstrated the utility of this approach, particularly in capturing complex non-Markovian effects in open quantum processes. The mathematical structure of these methods bears striking resemblance to the formalism of HOQOs, with this correspondence becoming concrete when tensor network methods are incorporated into influence functional calculations. 


The combination of influence functionals with tensor network approaches results in an object known as the \textit{influence matrix}~\cite{Lerose_2021_Influence}. This is a 2D tensor network where one dimension represents time and the other space. By contracting over spatial degrees of freedom, one can study the dynamics of any subsystem of interest, which yields the process tensor. A pivotal breakthrough in this field has been the \textbf{time-evolving matrix-product operator (TEMPO)} method, which reformulates the path integral description into an efficient tensor network representation~\cite{Strathearn_2018}. This approach maps non-Markovian quantum dynamics onto a one-dimensional tensor network where the bond dimension effectively captures the memory depth of the environment~\cite{Pollock_2018_PRA, Strathearn_2018}. The method's implementation in open-source packages such as OQuPy~\cite{OQuPy_2024, OQuPy_package} and ACE~\cite{Cygorek_2024_ACE} has made these sophisticated techniques accessible to the broader scientific community. Incorporating such powerful tensor network methods has indeed advanced our ability to simulate non-Markovian open quantum dynamics, especially in regimes where strong system-environment coupling or complex environmental structures preclude conventional perturbative treatments. 

The power of combining influence functional methods with tensor networks lies in their ability to efficiently represent temporal correlations while enabling systematic compression of memory effects. These methods introduce controlled approximations through truncation schemes, typically implemented via singular value decomposition where singular values below a specified threshold are discarded~\cite{Strathearn_2017, Ye_2021, Gribben_2022, Cygorek_2022, Fux_2023, Ng_2023,Park_2024, nguyen2024, Cygorek_2024_Understanding,Cygorek_2024, Cygorek_2024_ACE,Cygorek_2024_Sublinear, Chen_2024a,Guo_2024,Guo_2024b,Sun_2024}. This approach has proven particularly effective for processes with long memory times, leading to breakthroughs in learning~\cite{Gribben_2022_Quantum, Wang_2024} complex open quantum dynamics (including steady-state properties~\cite{Chen_2024a,Guo_2024b} of time-dependent impurity Hamiltonians~\cite{Sun_2024}) and designing optimal quantum control procedures~\cite{Fux_2021, Ortega_2024, Butler_2024}.

The framework has also been extended to systematically compute higher-order influence functionals, leveraging the mathematical machinery developed for analysing non-Markovian correlations in matrix-product operators~\cite{Pollock_2018_Quantum, Jorgensen_2019}. This extension allows for the study of complex spatiotemporal correlations naturally encoded in the Choi state of a process tensor, moving beyond traditional Markovian and two-point approximations. The matrix-product operator structure of process tensors has been exploited to efficiently simulate memory kernels for master equations, non-Markovian path integrals, and multi-time correlations~\cite{Jorgensen_2020} (see also Ref.~\cite{Cygorek_2024_Understanding} for a related `time-local' approach). Complementary approaches include the transfer tensor technique~\cite{cerrillo_non-markovian_2014, rosenbach_efficient_2016, Pollock_2018_Quantum, Pollock_2021_FCS}, which builds up multi-time process descriptions by systematically incorporating higher orders of correlations~\cite{Strathearn_2018, Jorgensen_2020, Chen_2020, Gherardini_2022}, as well as highly efficient tensor network contractions for Gaussian environments~\cite{Link_2024,Cygorek_2024_Sublinear}. 

\begin{figure}
    \centering
    \includegraphics[width=0.85\linewidth]{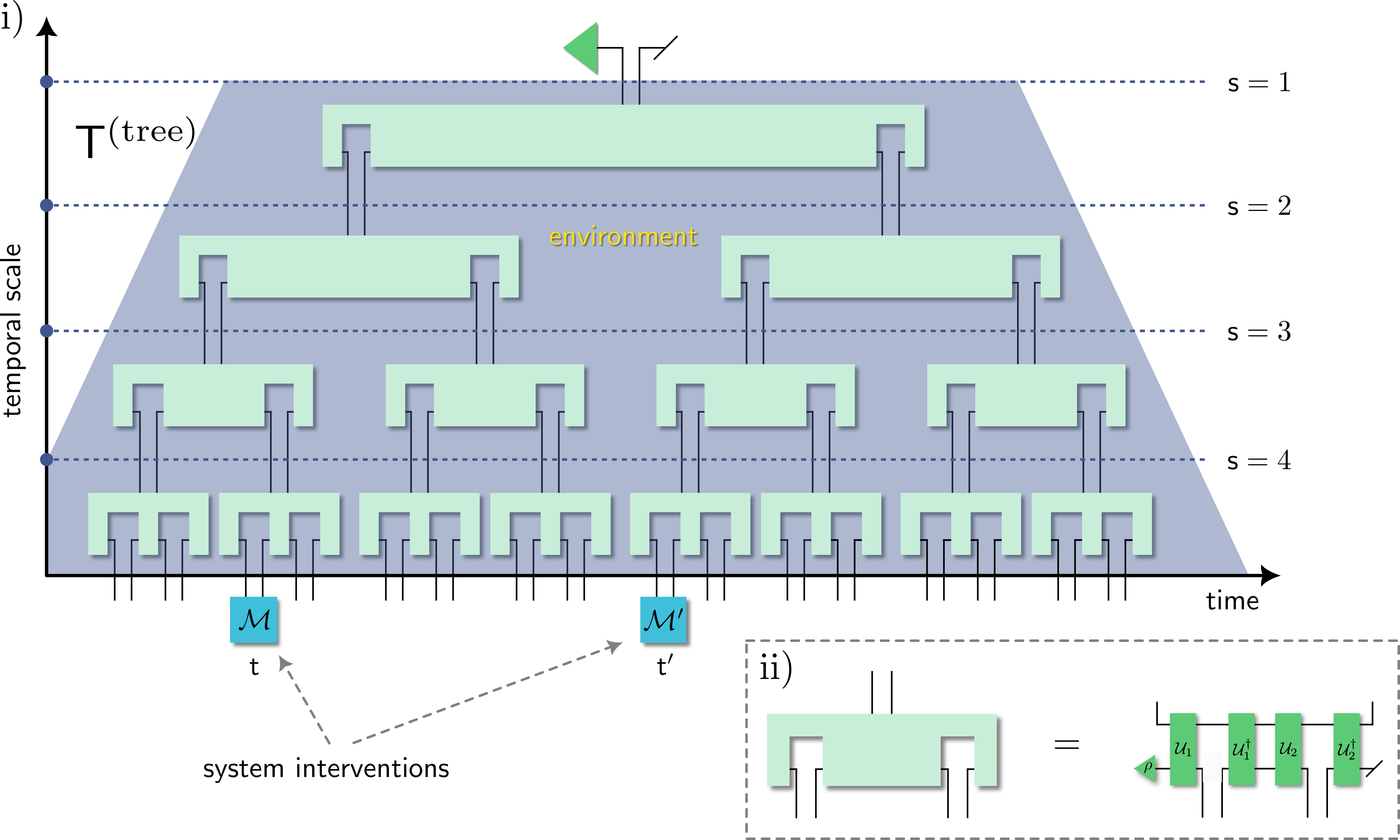}
    \caption{\textbf{Tree Process. i)} A process tree describing an open quantum system $S$ interacting with its environment. The process is constructed by a series of iteratively-connected \textit{causality preserving} maps (tensors). These maps implement a temporally-consistent fine-graining operation where one intervention at a single time is mapped to two interventions at two times while preserving the causal ordering of the process. The vertical extent of the tensor network hence corresponds to a timescale $s$ where interventions may be exponentially more frequent. Each pair of open indices at the bottom of the tree, and pairs of open indices intersected by the dashed lines at any time scale, is an intervention slot where an instrument may be applied. \textbf{ii)} Each causality-preserving map (depicted in mint green) is parameterised as shown by two unitary maps $\mathcal{U}_1$ and $\mathcal{U}_2$ and a density matrix $\rho$. Due to this particular structure, the causality preserving maps do not change the overall causal order of the process. Inserting identity maps into all slots of the causality-preserving map yields an identity channel, thus guaranteeing temporal consistency of the tree process.}
    \label{fig::mtqp-scqp-processtree}
\end{figure}

In most physical processes, such temporal correlations are expected to decay over time; however, capturing slowly decaying correlations has proved problematic, especially in regimes where multiple timescales place a role. To remedy this, Ref.~\cite{Dowling_2024} introduced a tensor network with tree-like geometry, dubbed a \textit{tree process}, to capture slowly decaying temporal correlations (see Fig.~\ref{fig::mtqp-scqp-processtree}). Here, the vertical layers correspond to different timescales yielding a compact description for multi-scale dynamical phenomena. The correlations between observables---depicted at the bottom of the tree---traverse upwards through the tensor; the tree-like geometry implies that the length of the path taken is logarithmic in the separation time between observables, thereby capturing slowly-decaying correlations. When combined with tensor network techniques employed for the simulation of open quantum system dynamics, such as those reported in Refs.~\cite{PhysRevLett.121.227401, Strathearn_2018, chin-natcom, Cygorek_2022, Fux_2023, Cygorek_2024_ACE}, the physically motivated geometric structure of the process tree promises to open up new research frontiers in many-body and many-time physics. See Ref.~\cite{Cerezo-Roquebrun_2025} for a review connecting process tensors, influence functionals, and transfer tensors through the lens of tensor networks and Ref.~\cite{Berezutskii_2005} for a review regarding the application of tensor network techniques for quantum computing.


Recent practical implementations have demonstrated the versatility of these methods in specific physical systems. For instance, these techniques have been successfully applied to model organic polaritons~\cite{Fowler-Wright_2022}, superradiance of quantum emitters~\cite{Wiercinski_2023,Wiercinski_2024}, temperature effects in the biexciton-exciton cascade of a quantum dot embedded in a microcavity~\cite{Bracht_2023}, and complex spectroscopy experiments~\cite{deWit_2024}, where traditional perturbative approaches fail to capture the essential physics. Moreover, Ref.~\cite{Boos_2024} adapted methods from Ref.~\cite{Cygorek_2024_Sublinear} to measure the emission spectra after strong pulsed driving, which present a dynamical analogue of Mollow triplets that require multi-scale resolution, for which matrix-product-operator process tensors are the perfect tool. Despite these advances, significant challenges remain in simulating many complex processes relevant to chemistry, condensed matter physics, and materials science, particularly as quantum technologies continue to evolve and push the boundaries of what constitutes a quantum advantage.

In particular, there exist quantum processes that are highly complex, where classical methods are expected to fundamentally struggle to simulate them~\cite{Aloisio_2023}. The boundary between simulatable and inherently non-simulatable processes is not well-explored and remains a highly active area for research. The ongoing development of these methods serves a dual purpose: they not only expand our ability to simulate complex quantum phenomena using classical resources but also provide benchmarks against which claims of quantum advantage can be evaluated. As these techniques continue to mature, they increasingly challenge the threshold at which quantum devices might demonstrate genuine computational advantages over classical simulation methods. 


\subsubsection{Quantum Chaos} \label{subsubsec::mtqp-quantumchaos}\hfill\\

\noindent The field of quantum chaos investigates the fundamental nature of chaotic behaviour in quantum processes, addressing two primary questions: \textit{What types of underlying dynamics lead to signatures of chaos (such as information scrambling) and what properties can serve as reliable witnesses of chaotic behaviour?} These questions have evolved significantly over time, from early studies of chaotic Hamiltonians to modern investigations of dynamical signatures in terms of HOQOs. The former usually refers to level spacing statistics of the generating Hamiltonian. The latter corresponds to structures emerging from two core ideas: randomness and sensitivity to perturbations, which have related volumetric entanglement scaling of the process. 

\vspace{0.25cm}\noindent
\textbf{\textul{Out-of-Time-Order Tensors.}} Understanding the structure of quantum chaos has been an ongoing challenge since at least the 1980s. Back then, research primarily focused on characterising the properties of chaotic Hamiltonians~\cite{reichl2021transition}. Recent work has shifted toward uncovering operational signatures of chaos from the underlying dynamics~\cite{Sekino_2008, Lashkari_2013, Vikram_2024}, motivated by the fundamental connections between chaos and several physical phenomena, such as sensitivity to perturbation, ergodicity, and thermalisation~\cite{Gogolin_2016, Deutsch_2018, PhysRevResearch.5.033126, Dowling_2022}. In particular, \textbf{out-of-time-order correlators (OTOCs)} have emerged as a prominent tool for analysing quantum chaos via the analysis of two-time correlations~\cite {Hayden_2007, Shenker_2014, Maldacena_2016, Hosur_2016, Roberts_2017, COTLER2018318, Garcia-Mata_2023} (see Ref.~\cite{Hashimoto_2017} for a review). However, OTOCs can sometimes yield false positives, i.e., flagging non-chaotic dynamics as chaotic~\cite{PhysRevLett.118.086801, Balachandran2021-af, e25010059, PhysRevB.107.235421}; thus, the picture remains incomplete. This issue can be somewhat ameliorated by considering \textbf{local operator entanglement (LOE)}, which has been shown to be a stronger, more precise indicator of chaos~\cite{Dowling_2023}. 


\begin{figure}
    \centering
    \includegraphics[width=0.55\linewidth]{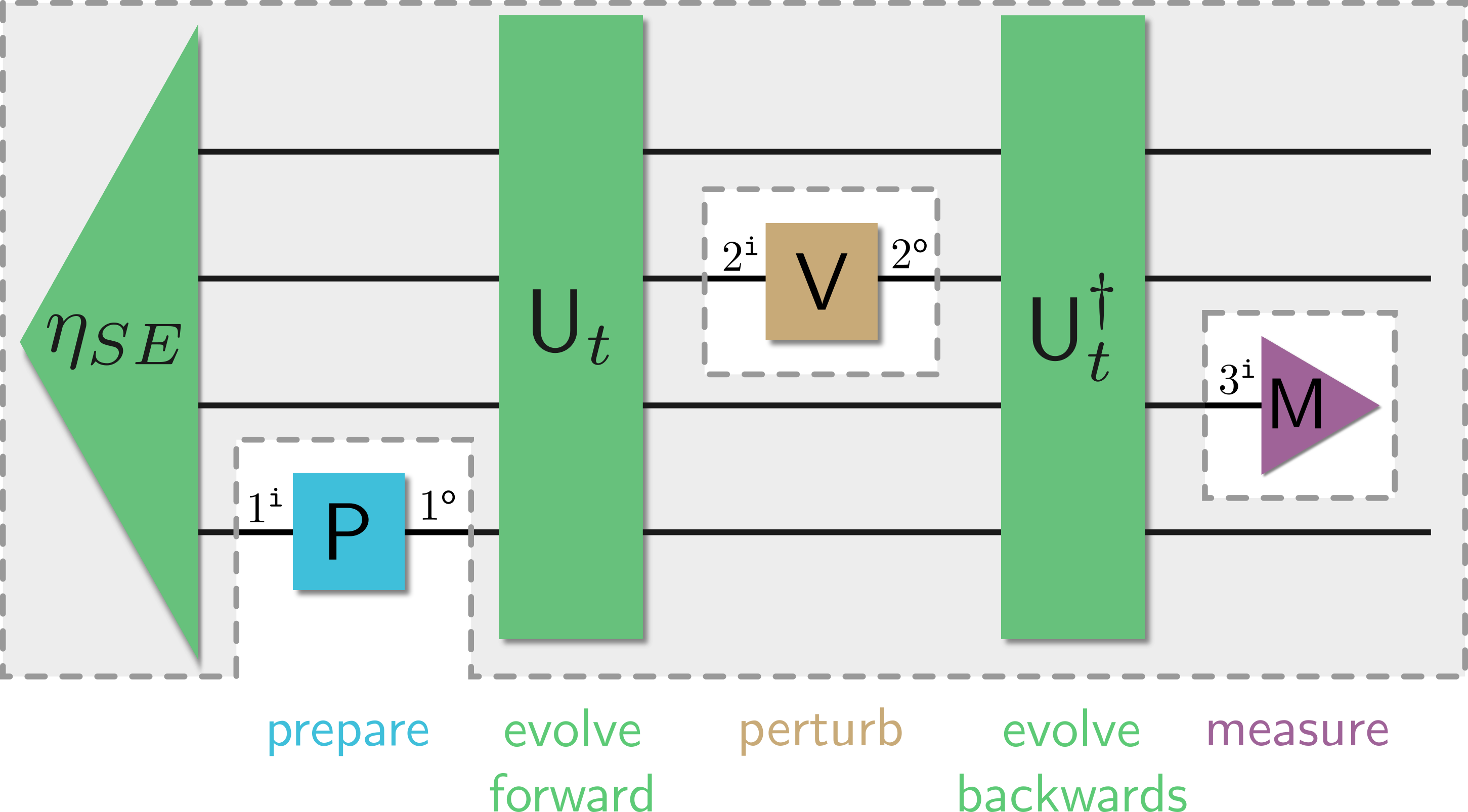}
    \caption{\textbf{Out-of-Time-Order Tensor.} A quantum chaos experiment can be described as a higher order map (the object outlined in grey), which we refer to as an OTOT. The correlations observed from the OTOT, i.e., those between the preparation $\mathsf{P}$, the perturbation $\mathsf{V}$ and the measurement $\mathsf{M}$ of a small probe interacting with a large environment (corresponding to all the wires with no operation on them) encapsulate various signifiers of quantum chaos, such as OTOCs and LOE. This setup is akin to an ink-drop experiment: ink is dropped into a viscous fluid (corresponding to the preparation $\mathsf{P}$), mixed by rotating the fluid (corresponding to the dynamics $\mathsf{U}_t$, locally perturbed (corresponding to $\mathsf{V})$, and subsequently unmixed by rotating in the opposite direction (corresponding to $\mathsf{U}^\dagger_t$). The indistinguishability upon measurement of the initial and final states of the ink droplets, subject to a perturbation prior to the reverse rotation, quantifies the chaoticity of the fluid dynamics.
    }
    \label{fig::mtqp-qc-otot}
\end{figure}


The relationship between these two measures can be elegantly understood through the framework of HOQOs. Ref.~\cite{Zonnios_2022} introduced an \textbf{out-of-time-order tensor (OTOT)}, which describes a procedure in which the system and its environment are evolved forward in time, the system is perturbed, and finally, the global state is evolved back in time (see Fig.~\ref{fig::mtqp-qc-otot}). Let us denote the forward evolution by $\Ucal_t (\, \bullet\, ) := U_t  \, \bullet\,   U^\dag_t$, the backward evolution by $\Ucal^{\dagger}_t ( \, \bullet\, ) := U_t^\dag \, \bullet\,  U_t$, and the initial system-environment state by $\eta_{SE}$. We then define
\begin{align}\label{eq::mtqp-qc-otot}
    \mathsf{T}_{3^\inp 2^\out 2^\inp 1^\out 1^\inp} := \mathds{1}_{E_3}
    \star \mathsf{U}^{\dagger}_{S_{3^\inp} E_{3} S_{2^\out} E_{2}} 
    \star \mathsf{U}_{S_{2^\inp} E_{2} S_{1^\out} E_{1}}
    \star \eta_{S_{1^\inp} E_{1}}.
\end{align}
Here, both $\mathsf{T}$ and $\mathsf{U}$ depend on time; however, to minimise notational clutter, we drop the index $t$ from here on. Moreover, since the OTOT acts only on the system level, we drop the superfluous $S$ label. Both LOE and OTOC stem from this tensor
\begin{align}\label{eq::mtqp-qc-otoc-loe}
    \mathrm{LOE} := -\log\left[({\mathsf{T}_{{3^\inp} {2^\out} {2^\inp} {1^\out} {1^\inp}} 
    \star \mathsf{V}_{{2^\out}  {2^\inp}} \star  \mathds{1}_{1^\inp})}^2\right]
    \quad \mbox{and} \quad
    \mathrm{OTOC} := \mathsf{T}_{{3^\inp} {2^\out} {2^\inp} {1^\out} {1^\inp}} \star \mathsf{M}_{{3^\inp}} \star \mathsf{V}_{{2^\out} {2^\inp}} \star \mathsf{P}_{1^\out{1^\inp}}.
\end{align}
Above, $\mathsf{V}_{{2^\out} {2^\inp}}$ is a perturbation on the system---which is typically taken to be unitary. Moreover, for LOE, the initial state $\eta_{SE}$ is always maximally mixed and the preparation is a non-Hermitian operation. For OTOC this is not required (but nonetheless usually imposed). The LOE quantifies the operator entanglement~\cite{Prosen2007, Prosen2007a, Bertini_2020, Bertini_2020-2} by means  R\'enyi-2 entropy across two equally-sized partitions. On the other hand, OTOC measures the scrambling that occurs in the `back-and-forth' dynamics due to the perturbation.

A key difference between LOE and OTOC is how the final space $E_3$ in Eq.~\eqref{eq::mtqp-qc-otot} is chosen: For LOE, $E_3$ is extensive and includes half of the involved systems that $\eta_{SE}$ is defined on; for OTOC, $E_3$ represents a constant number of subsystems, typically taken to correspond to the same space as the domain of the perturbation. This immediately implies that LOE is more informative than OTOC, which has indeed been shown to be the case in Ref.~\cite{Dowling_2023}. The ability of the OTOT to encompass both OTOC and LOE makes it a powerful theoretical tool for studying quantum chaos, providing a more complete picture than either measure alone. This unified framework helps explain why LOE can serve as a more reliable indicator of chaos than OTOC, while also suggesting new directions for investigating the nature of quantum chaos through the rigorous lens of HOQOs.

Intuitively, LOE growth is expected to be linear for chaotic processes and logarithmic for non-chaotic processes; OTOC is expected to decay quickly for chaotic processes and slowly for non-chaotic ones. However, there are instances where OTOC decay is fast even for non-chaotic processes. While we have used a HOQO description here, one could also think the above quantities stemming from the \textit{family} of quantum channels $\mathsf{C}_{{3^\out}{1^\inp}}(\mathsf{V}):= \mathsf{T}_{{3^\out} {2^\inp} {2^\out} 1^\out {1^\inp}} \star \mathsf{V}_{{2^\out} {2^\inp}} \star \mathds{1}_{1^\out}$, which are a function of the perturbation $\mathsf{B}$. One must be careful in choosing the perturbation in this case: For chaotic processes, the behaviour of both LOE and OTOC should not depend on the perturbation; on the other hand, for non-chaotic processes, this dependence may be non-trivial and may lead to incorrect conclusions by falsely flagging a non-chaotic process as a chaotic one. We now move to study genuinely spatiotemporal HOQOs that are routinely used to understand chaos and go significantly beyond what can be learnt from such a family of perturbation-dependent quantum channels.


\vspace{0.25cm}\noindent
\textbf{\textul{Spatiotemporal Process Tensor.}} Consider a spatiotemporal version of the process tensor $\mathsf{T}^{SE}_{n:1}$ introduced in Ref.~\cite{Dowling_2022}. This generalises the temporal version given in Eq.~\eqref{eq::toqp-dethoqo-quantumcombconstruction}: Here, the final environment space is not traced out but rather represents the `spatial' component of the spatiotemporal process tensor
\begin{align}\label{eq::mtqp-qc-spatiotemporalprocesstensor}
\mathsf{T}^{SE}_{n:1} := 
\mathsf{U}_{S_{n^\inp} E_n S_{n-1^\out} E_{n-1}} \star \mathsf{U}_{S_{n-1^\inp} E_{n-1} S_{n-2^\out} E_{n-2}} \star \hdots \star 
\mathsf{U}_{S_{3^\inp} E_3 S_{2^\out} E_2} \star 
\mathsf{U}_{S_{2^\inp} E_2 S_{1^\out} E_1} \star 
\Psi_{S_{1^\inp} E_1}. 
\end{align}
Above, we have taken the initial state of $SE$ to be pure, i.e., $\Psi_{S_{1^\inp} E_1} := \ket{\psi_{S_{1^\inp}E_1}} \! \bra{\psi_{S_{1^\inp}E_1}}$. In principle, an agent can apply arbitrary interrogation sequences to the system part of such a process. However, for a specific class of instruments, the temporal information that resides on the level of the system gets mapped onto the spatial information encoded in the environment. If the probing instrument sequence $\{ \mathcal{M}_{k}^{(x_k)} \}_{k=1}^{n}$ is chosen such that all its elements are \textit{pure} operations, then the final conditional state will also be pure\footnote{Here, a pure operation is any rank-one operation, which could be proportional to a unitary operation, projective measurement, or projective measurement followed by a pure state preparation.}
\begin{align} \label{eq::mtqp-qc-pureprocessbranch}
    \ket{\psi^{(\vec{x})}_{S_{n^\inp}E}} := U_{n,n-1} M_{n-1}^{(x_{n-1})} U_{n-1,n-2} \hdots U_1 M_{1}^{(x_1)} \ket{\psi_{S_{1^\inp}E}}.
\end{align}
Here, the instrument operators $M_{k}^{(x_k)}$ correspond to rank-one Kraus operators and we collect the sequence of outcomes in the vector $\vec{x} := (x_1,\hdots,x_{n-1})$.

The conditional states above are known as `branches' or `trajectories' of the process~\cite{sakuldee_non-markovian_2018}. An informationally complete set of quantum trajectories provides a full characterisation the spatiotemporal process tensor. To construct such a representation, it suffices to choose Kraus operators of the form $M_{k}^{(x_k)}:= \ket{x_{k^\out}} \bra{x_{(k-1)^{\inp}}}$, where $\{\ket{x_{k^\out}} \}$ and $\{\ket{x_{(k-1)^\inp}}\}$ form an orthonormal basis on the relevant Hilbert spaces; this is similar in spirit to the `measure and prepare' channels (causal breaks) as defined in Eq.~\eqref{eq::osdme-qmo-measureprepare} (albeit with the restriction of being pure). Inserting such operators at each timestep in Eq.~\eqref{eq::mtqp-qc-pureprocessbranch} leads to all possible combinations of trajectories. After some elementary manipulation, we yield
\begin{align} \label{eq::mtqp-qc-stpttrajectories}
    \mathsf{T}^{SE}_{n:1} = \sum_{\vec{x}\vec{y}} 
     \kket{{M}^{(\vec{x})}} \bbra{{M}^{(\vec{y})}} \otimes \ket{\psi^{(\vec{x})}_{S_{n^\inp}E}}
     \bra{\psi^{(\vec{y})}_{S_{n^\inp}E}},
\end{align}
with $\kket{M^{(x_k)}}:= \ket{x_{k^\out}} \otimes \ket{x_{(k-1)^\inp}}$. Above, the second space corresponds to the `spatial' part of the process and the first space the `temporal' part. The above process tensor has a pure Choi state, unlike the temporal process tensor, which is usually mixed. One can derive the temporal process tensor [see, e.g., Eq.~\eqref{eq::osdme-me-processtensor}] by tracing over the spatial part $\mathsf{T}^S_{n:1} = \sum_{\vec{x}\vec{y}} 
    \braket{\psi^{(\vec{y})}_{S_{n^\inp}E}|
    \psi^{(\vec{x})}_{S_{n^\inp}E}} \kket{{M}^{(\vec{x})}} \bbra{{M}^{(\vec{y})}}$.
The spatiotemporal process tensor naturally arises in the context of chaos in several independent studies; we briefly outline these below.

We first focus on applications of the temporal process tensor and will return to the spatiotemporal process tensor later. A first application of the process tensor concerns the notion of \textbf{quantum dynamical entropy (QDE)}~\cite{Lindblad_1979, Alicki1994-dc,Slomczynski1994-ns, Benatti_2004, cotler_superdensity_2017}. Although a rich topic in the classical realm, this concept has been difficult to generalise to the quantum setting. To do so required uncovering the full multi-time structure of quantum stochastic processes, e.g., the process tensor. Concretely, QDE is defined as the entropy of the temporal process tensor.\footnote{Throughout this section, we take the process tensor to be normalised such that entropic quantities are well-behaved.} Choosing the R\'enyi-2 entropy, we have
\begin{align}\label{eq::mtqp-qc-dynamicalentropy}
    \text{QDE} := \lim_{n \to \infty} \frac{-\log(\tr{\mathsf{T}_{n:1}^2})}{n}.
\end{align}
Firstly, this entropy quantifies the entanglement across the spatial and temporal components of the pure spatiotemporal density matrix. For a non-chaotic system, the numerator grows slowly and QDE vanishes. When the numerator grows extensively, the QDE approaches a constant, indicating that the process may be chaotic. However, exceptions to this case exist; see Refs.~\cite{Lindblad1986, Dowling_2022} for examples and relevant discussion.

\textbf{Temporal entanglement (TE)} has emerged as another classifier for quantum chaos. Temporal entanglement was introduced in Ref.~\cite{Banuls_2009}, with significant developments in subsequent works~\cite{Muller-Hermes_2012, Hastings_2015}, for studying many-body dynamics using tensor network tools. This approach uses matrix product operators to estimate dynamical observables, treating dynamics discretely in order to construct a two-dimensional tensor network in spacetime. This construction, known as an influence matrix, amounts to a discrete version of the influence functional~\cite{Strathearn_2018}, highlighting connections with the HOQO formalism. This relationship was further elaborated in Refs.~\cite{Lerose_2021_Influence, PhysRevResearch.6.033021, Chen_2024a}, which established crucial links between quantum chaos and dissipative dynamics through the lens of temporal entanglement.

Temporal entanglement is defined in terms of correlations between the past and future of a quantum process. To compute it, first, the process tensor is vectorised and normalised: $\mathsf{T}_{n:1} \to \kket{\mathsf{T}_{n:1}}/\sqrt{N}$, where $N:=|\bbrakket{\mathsf{T}_{n:1}| \mathsf{T}_{n:1}}|^2$ provides the normalisation. The normalisation is needed because the temporal process tensor is not a pure density matrix. Thus, TE is defined as
\begin{align}\label{eq::mtqp-qc-temporalentanglement}
\text{TE} := -\log \ \tr{\left(\frac{1}{N} \,\ptr{1:\frac{n}{2}}{\kket{\mathsf{T}_{n:1}} \! \bbra{\mathsf{T}_{n:1}}}\right)^2}.
\end{align}
This measure quantifies correlations between past (timesteps $1:\frac{n}{2}$) and future (timesteps $\frac{n}{2}+1:n$) and is reminiscent of measures of non-Markovianity. For chaotic systems, TE grows linearly whereas for non-chaotic ones, it grows logarithmically. Thus, it has proven effective in distinguishing between chaotic and integrable models~\cite{Lerose_2021, Sonner_2021, Thoenniss_2023, Foligno_2023}, a capability that naturally emerges from the inherent capacity of HOQOs to faithfully capture signatures of chaos. As it turns out, TE is a stronger measure of dynamical complexity than QDE: This is because it captures correlations through a quartic function of $\mathsf{T}_{n:1}$, whereas QDE only essentially measures how noisy a process is.

A related concept, known as the \textbf{butterfly flutter fidelity (BFF)}, was introduced in Ref.~\cite{Dowling_2022}. The protocol to assess quantum chaos makes use of the branches of the spatiotemporal process tensor in Eq.~\eqref{eq::mtqp-qc-pureprocessbranch}. This approach aims to understand quantum chaos in terms of the sensitivity of a process to perturbations on a small part of the system. Namely, for two orthogonal perturbations, i.e., $\bbrakket{M^{(\vec{x})}|M^{(\vec{y})}} = 0$, chaotic processes should satisfy
\begin{align}
    \braket{\psi^{(\vec{x})}_{S_{n^\out}E}|
    \psi^{(\vec{y})}_{S_{n^\out}E}} \approx 0.
\end{align}
It is interesting to note that the above criterion naturally arises in the context of decoherent histories~\cite{PhysRevX.14.041027, 2406.15577}, which we alluded to in Sec.~\ref{subsubsec::osdme-classicalquantumprocesses} above. Demanding this behaviour of chaotic processes implies that the spatiotemporal process tensor for a chaotic process displays a volumetrically scaling spatiotemporal entanglement~\cite{Dowling_2022}. Closely related to this study is the multi-time generalisation of OTOC, known as $2k-$OTOC~\cite{Roberts_2017, Leone_2021}. Here, the dynamics go back and forth in time, with perturbations applied in between. This analysis is useful for classifying the chaotic nature of an ensemble of unitaries. 

Subsequently, Ref.~\cite{Dowling_2022} proposed a stronger condition for quantum chaos in terms of the complexity of a `correction' unitary. The idea is that in chaotic systems, small perturbations lead to vastly different branches. Suppose a unitary $V_{\vec{x}\vec{y}}$ is a correction operator that brings two divergent states together:
\begin{align}
    \braket{\psi^{(\vec{x})}_{S_{n^\inp}E}| V_{\vec{x}\vec{y}}|
    \psi^{(\vec{y})}_{S_{n^\inp}E}} \approx 1.
\end{align}
Then, processes that only require corrections $V_{\vec{x}\vec{y}}$ with low complexity are deemed not chaotic. Notably, some of these processes themselves could still have high dynamical entropy. A similar expression arises in Ref.~\cite{1910.14646}, where the authors were concerned with proving the conjecture `complexity = volume', i.e., correlations in a chaotic process grow according to a volume law instead of an area law. They also concluded that the circuit complexity of such processes has to be exponential in the spatiotemporal size of the system under minimal assumptions. Another topic that aims to draw a similar connection between chaos and complexity is \textit{quantum branching}~\cite{2308.04494, 2310.06755}, which provides a mechanism for classical mechanics to arise from quantum mechanics. Essentially, when two branches are sufficiently different that a high-complexity correction is required to bring them together, they cannot interfere and behave as classical probabilistic states. The distinct notions outlined above all highlight that quantifying the complexity of the correction unitary may be an appropriate way to enable meaningful statements about chaos. This is, however, an arduous task that remains unresolved.

Finally, an even stronger class of chaos quantifiers can be built from the process branches: This is known as \textit{deep thermalisation} and is related to the highly active research area of measurement-induced phase transitions. Research into measurement-induced phase transitions~\cite{PhysRevX.9.031009} has shown that measurements in a quantum circuit could potentially steer the final state into two vastly different phases: highly entangled and low entanglement states. Naturally, one can ask about the entanglement properties of the ensemble made up of branches in Eq.~\eqref{eq::mtqp-qc-pureprocessbranch}: $\{p_{\vec{x}} \, , \ket{\psi^{(\vec{x})}_{S_{n^\inp}E}} \}$. Deep thermalisation~\cite{PRXQuantum.4.030322} studies higher moments of this distribution, i.e., the properties of $\{p_{\vec{x}} \, , \ket{\psi^{(\vec{x})}_{S_{n^\inp}E}}^{\otimes k} \}$. These concepts, along with dynamical entropy, are numerically investigated for various models in Ref.~\cite{odonovan2025_diagnosing}.

Our review of the topic of quantum chaos does not do justice to its depth; however, it emphasises how the process tensor naturally arises in this context. This should not be surprising as the process tensor is endowed with the full details of the dynamics. While each of these approaches may have been originally developed using slightly different formalisms, they all find natural expressions within the framework of HOQOs. This convergence highlights the fundamental role of HOQOs in quantum dynamics, providing a unified language for describing diverse phenomena from temporal correlations to measurement-induced transitions. The framework's ability to naturally accommodate these various measures demonstrates its power as a theoretical tool for understanding complex quantum processes. 

\FloatBarrier


\subsubsection{Operational \& Strong-Coupling Quantum Thermodynamics}\label{subsubsec::mtqp-operationalquantumthermodynamics}\hfill\\

\noindent Typical, equilibrium thermodynamics assumes that some system of interest interacts weakly with an environment in thermal equilibrium. The weakness of the interaction at all times formally leads the description of the system-environment state to remain independent (i.e., of tensor product form) and therefore any potential memory effects are neglected. This is not the case, however, for systems that strongly couple to the environment, even if the latter can locally be described in terms of an equilibrium state. In such situations, the system and environment build up non-negligible correlations and therefore the description of the open dynamics in terms of HOQOs is required.

First steps towards a thermodynamically consistent description of weakly coupled systems\footnote{See also the early work Ref.~\cite{van_kampen_quantum_1954} for discussion on the thermodynamics of irreversible quantum processes.} in terms of process tensors were provided in Refs.~\cite{Strasberg_2019_PRE_Operational, Strasberg_2019_PRE_Stochastic}. A fully general framework based on the description of HOQOs to describe strong-coupling thermodynamics has been developed in Refs.~\cite{Strasberg_2019_PRL, strasberg_thermodynamics_2020} (see, in particular, Ref.~\cite{strasberg_quantum_2022} for a comprehensive introduction to strong coupling quantum thermodynamics and process tensors). By construction, this framework provides a consistent interpretation of many quantities relevant to thermodynamics, such as the average work and equilibrium free energy difference. In addition, it also provides sensible non-equilibrium extensions of such quantities, which are justified by their reduction to their equilibrium counterparts in the appropriate limit~\cite{Strasberg_2019_PRL}. Moreover, the framework readily leads to the notion of \textit{quantum trajectories} over sequences of measurement outcomes, which can be used to analyse multi-time behaviour such as energetic fluctuations both at the stochastic and ensemble level~\cite{Strasberg_2019_PRE_Stochastic}. By its very nature, the HOQO description of an open quantum process as a comb allows one to treat any kind of incomplete information such as noisy measurement devices, which has particular relevance in the thermodynamic setting where one may typically only be interested in some coarse-grained quantities~\cite{Strasberg_2019_PRE_Operational}. Lastly, one may also consider non-Markovianity as a resource that can enhance the performance of thermodynamic tasks~\cite{Zambon_2024_Thermodynamic}.

To this end, note that thermodynamics more broadly is concerned with the fact that microscopically relevant details, such as memory effects or specific interaction parameters, tend to wash out at larger scales. Many dynamical thermodynamic phenomena such as equilibration, thermalisation, and ergodicity (to name but a few) have been analysed from various perspectives. For instance, amongst the variety of notions of thermalisation that abound, some make explicit mention of the types of observables being measured (typically highly degenerate macroscopic ones), whereas others do not; the former type falls closely in line with the operational perspective suited to the HOQO paradigm. In Refs.~\cite{Figueroa-Romero_2020, Dowling_2021_Equilibration, Dowling_2021_Relaxation}, the authors take such a viewpoint and characterise the conditions under which a multi-time process with finite temporal resolution can be approximated by an equilibrium one, in particular presenting sufficient conditions for multi-time expectation values to relax close to their equilibrium values whenever the operations are implemented with a noisy clock. 

In a slightly different vein, we also note that although almost all processes have memory, such effects also tend to wash out on average; in other words, Markovianity is an emergent phenomenon. Such behaviour is formalised in Refs.~\cite{Figueroa-Romero_2019, Figueroa-Romero_2021}, where the authors demonstrate that whenever the system of interest is small compared to its environment, almost all processes are typically very close to Markovian ones, independently of any further assumptions on the type of dynamics or coupling strength. Recent investigations on the validity of master equations have shed light on the role of non-integrable environment~\cite{Xu_2018, Xu_2023, Zhang_2024, odonovan2024_qme}. Here, HOQOs can serve as a new numerical tool to explore open dynamics in the realms that were not possible previously.


\subsection{Causality \& Quantum Foundations}
\label{subsec::causalityquantumfoundations}

Up to this point, most of the HOQOs that we have considered in the Review part of this work abode---either because of the physical situation they described or because we inserted it `by hand' as an axiomatic demand---by a fixed causal order. However, the requirement of a fixed \textit{global} causal order can be dropped without introducing logical paradoxes or violating the rules of quantum mechanics. Here, we will not discuss the ontological status of such `exotic' spatiotemporal situations (see, e.g, Refs.~\cite{Araujo_2017, Dariano_2018, Oreshkov_2019} for a discussion thereof), but rather lay out how HOQOs are the natural tool to describe such situations, since they provide a natural framework for the characterisation of spacetime correlations---causally ordered or not---between disjoint laboratories/observers. While the study of causally indefinite processes in terms of HOQOs was first carried out in the context of computational advantages in the absence of definite causal order~\cite{Hardy_2009, Chiribella_2012, Chiribella_2013}, here, we first discuss the later introduced notion of process matrices~\cite{Oreshkov_2012} and subsequently the more experimentally amenable case of the quantum switch---both of which we have already encountered in the Tutorial part of this Review Article (see Sec.~\ref{subsec::indefinitecausalorder}).


\subsubsection{Process Matrices}\label{subsubsec::cqf-processmatrices}\hfill\\

\noindent In the study of indefinite causal order in quantum theory, the notion that events must occur in a definite temporal sequence is challenged and replaced by the weaker notion that both quantum mechanics\footnote{The consequences of quantum mechanics holding locally, i.e., in Alice's and Bob's laboratory, combined with an additional non-signalling constraint between them (which we do \textit{not} impose here) has been investigated in Ref.~\cite{barnum_local_2010}, which demonstrated that correlations obtained in such non-signalling scenarios can always be explained by means of a global quantum state.} and causality need only hold \textit{locally}~\cite{Oreshkov_2012}. That is, while one no longer requires that there is some fixed causal ordering of events between different laboratories, one at least demands that if parties in different laboratories act freely and independently, the overall result is well-behaved and does not lead to logical inconsistencies. This requirement can come in different `flavours', depending on the situation that is to be investigated, but generally boils down to demanding that independent deterministic operations (e.g., CPTP maps) are mapped onto a deterministic object.\footnote{A notable exception is the quantum time flip~\cite{chiribella_quantum_2022}, whose action is only well-defined on a subset of all CPTP maps (see Sec.~\ref{subsubsec::ico-quantumtimeflip}).}

As an example, note that this requirement is imposed (and holds true) for the quantum switch---one of the earliest applications of HOQOs in the study of causal order---which maps pairs~\cite{Chiribella_2013, Chiribella_2012} or $k$-tuples~\cite{colnaghi_quantum_2012, araujo_computational_2014} of quantum channels (i.e., CPTP maps) to a valid quantum channel; similarly, process matrices map pairs~\cite{Oreshkov_2012} or $k$-tuples~\cite{Araujo_2015, Oreshkov_2016} of quantum channels to unit probability. For process matrices $\mathsf{W}\geq 0$ in particular, this implies\footnote{Due to differences in the definition of the CJI, our definition here may differ by a transposition from that found in the literature (see, e.g., Refs.~\cite{Oreshkov_2012, Araujo_2015}); this difference is merely notational.}
\begin{gather}
\label{eqn::def_proc_mat_rev}
\begin{split}
    &    \mathsf{W} \star (\mathsf{M}_A \otimes \mathsf{N}_B) = 1 \quad \text{(two-parties)} \\ \text{or} \qquad
    &\mathsf{W} \star (\mathsf{M}_A \otimes \mathsf{N}_B \otimes \mathsf{O}_C \otimes \cdots ) = 1 \qquad \text{(multiple parties)}
\end{split}
\end{gather}
for all CPTP maps $\mathsf{M}_A, \mathsf{N}_B, \mathsf{O}_C, \dots$ performed in the laboratories of Alice, Bob, Charlie, etc. Linearity, positivity, and the above normalisation condition on $\mathsf{W}$ embody the fundamental requirements that: i) quantum mechanics holds locally~\cite{barnum_local_2010, Oreshkov_2012}, such that probabilities are computed via a `Born rule' like in Eq.~\eqref{eqn::def_proc_mat_rev}; ii) the process matrix yields positive outputs even when only acting on a parts of the respective maps~\cite{Oreshkov_2012, Castro-Ruiz_2018}; and iii) probabilities are normalised---a requirement equivalent to \textit{local causality} holding (i.e., none of the parties can obtain super- or subnormalised `probability distributions' in their individual laboratories). 

As such, process matrices can describe the most general spatiotemporal correlations that can be established between disjoint laboratories under the assumption that in each laboratory, both quantum mechanics and causality hold. Notably, the requirement of a \textit{global} causal order amongst laboratories is absent in the requirements of process matrices, allowing them to describe spatiotemporal scenarios without logical paradoxes that nonetheless lie outside the set of causally ordered processes (i.e., those described by quantum combs / process tensors). As a result, such processes have been conjectured to be relevant for an understanding of quantum gravity~\cite{Hardy_2001, hardy_probability_2005, Hardy_2009, Hardy_2012, Chiribella_2013, Oreshkov_2012, Hardy_2015, baumeler_reversible_2019, zych_bells_2019}, where causal orders are likely to be superposed or potentially even more `exotic'.

A central result in the study of process matrices was the discovery of \textit{causally non-separable} processes. As discussed previously, in the bipartite setting, a process matrix $\mathsf{W}_{AB} \in \Lscr(\Hscr_{A^\inp} \otimes \Hscr_{A^\out} \otimes \Hscr_{B^\inp} \otimes \Hscr_{B^\out})$ is causally separable iff it cannot be decomposed as a probabilistic mixture of quantum processes with fixed causal order, i.e., 
\begin{gather}
\mathsf{W}_{AB} \neq p\mathsf{T}^{A\prec B} + (1-p)\mathsf{T}^{B\prec A}, 
\end{gather}
where $\mathsf{T}^{A\prec B}$ ($\mathsf{T}^{B\prec A}$) is a process with fixed causal order $A\prec B$ ($B\prec A$).

In the multipartite case, the situation becomes more involved since concepts such as dynamical ordering---i.e., the causal order is determined on-the-fly as the parties act, and their respective actions can potentially influence the causal ordering of the other---imply that the definition of causal separability is \textit{inequivalent} to convex combinations of fixed causal order. As a consequence, different (and possibly non-equivalent) definitions of multipartite causal non-separability have been proposed, each one focusing on different aspects of multipartite causality~\cite{Baumeler_2014_PefectSignalling, Oreshkov_2016, Wechs2019MultipartiteCausality,Wechs_2021}.

The deviation of process matrices from the set of causally ordered processes can not only be characterised in terms of causal non-separability---a manifestly device-dependent marker of causal indefiniteness---but also in a stronger sense: via the violation of \textit{causal inequalities}~\cite{Oreshkov_2012,  baumeler_maximal_2014, Baumeler_2014_PefectSignalling, branciard_simplest_2015, abbott_multipartite_2016, Oreshkov_2016, Abbott_2017, wechs_existence_2023}. These are inequalities that must be satisfied by \textit{any} process with a fixed underlying causal order. As a concrete example, consider the simplest bipartite \textbf{guess your neighbours input (GYNI)} game~\cite{almeida_guess_2010, branciard_simplest_2015}, where two parties---Alice and Bob---are sent uniformly sampled input bits $x$ and $y$ respectively and their task is to guess each other's input, i.e., to respectively produce output bits $a$ and $b$ such that $a=y$ and $b=x$. If Alice and Bob's laboratories are connected in a causally ordered way, say $A\prec B$, then $\mathds{P}(a=y) = \frac{1}{2}$ since her best strategy is to make a random guess about Bob's bit; analogously, $\mathds{P}(b=x) = \frac{1}{2}$ for the order $B\prec A$. In this bipartite scenario, causal distributions are given by convex combinations of fixed ordered processes. Since taking convex combinations of causal orders cannot increase these guessing probabilities, it follows that in the \textbf{causally ordered (co)} case  
\begin{gather}
    \mathds{P}_{\text{GYNI}} := \mathds{P}(a=y, b=x) \overset{\text{co}}{\leq} \frac{1}{2}
\end{gather}
holds for all (convex combinations of) causally ordered processes. However, allowing for general process matrices, one can show that a value of at least $\mathds{P}_{\text{GYNI}} \approx 0.6218 > \tfrac{1}{2}$ can be achieved~\cite{branciard_simplest_2015}.\footnote{In addition to advantages for such information theoretic games, causally non-separable process matrices have also been shown to be resourceful for implementing quantum functions, as well as for metrological and channel discrimination tasks; see Sec.~\ref{subsec::higherorderquantumsubroutines}.} More generally, the set of causal correlations for a given number of inputs and outputs forms a polytope (the so-called \textit{causal polytope}). The analysis of causal polytopes allows for a systematic construction of such causal inequalities~\cite{branciard_simplest_2015, baumeler_space_2016, abbott_multipartite_2016} and their possible violation by general process matrices has been demonstrated both in the bi- and multipartite case~\cite{Oreshkov_2012,  baumeler_maximal_2014, branciard_simplest_2015, abbott_multipartite_2016, Oreshkov_2016, Abbott_2017, wechs_existence_2023, kunjwal_nonclassicality_2024,kunjwal_generalizing_2024}. The ultimate boundaries of these violations have been explored for different specific scenarios in Refs.~\cite{brukner_bounding_2015, bhattacharya_biased_2015, kunjwal_nonclassicality_2024,kunjwal_generalizing_2024}. Moreover, a general method to upper bound the achievable violations of causal inequalities in quantum mechanics---together with an achievable Tsirelson-like bound for the OCB inequality and so-called \textit{single-trigger causal inequalities}---has been provided in Ref.~\cite{liu_tsirelson_2024} (see also Refs.~\cite{bavaresco_indefinite_2024} and~\cite{Vilasini2019} for a respective discussion of violations of causal inequalities and multi-agent paradoxes in so-called `box-world theories', i.e., higher order generalised probabilistic theories that go beyond quantum mechanics). These methods can be used to show that quantum strategies based on process matrices necessarily respect the upper bound $\mathds{P}_{\text{GYNI}}\leq 0.7592$ (which may not be tight).

\textit{A priori}, the violation of a causal inequality is a stronger condition than causal non-separability, since the latter implies that \textit{no} causally ordered process---independent of its dimension or the physical theory it abides by---can recreate the correlations exhibited by the process that violates the causal inequality. In contrast, causal non-separability merely implies that a process matrix cannot be expressed as a convex combination of causally ordered process matrices of the same dimension.\footnote{We emphasise again that the definition of causal non-separability is more involved in the multipartite case; nonetheless, the distinction between causally non-separable processes and those that violate a causal inequality remains the same conceptually.} Indeed---similar to the existence of entangled quantum states that admit a hidden variable model---there exist causally non-separable processes that nonetheless do not violate any causal inequalities~\cite{Araujo_2015, feix_causally_2016, purves_quantum_2021,Wechs_2021}. Consequently, in order to distinguish causally non-separable processes from those that indeed violate causal inequalities, the latter are sometimes called `acausal'~\cite{Oreshkov_2016}. We finally emphasise that quantum correlations are \textit{not} a prerequisite for the violation of causal inequalities. While this is the case in the bipartite scenario~\cite{Oreshkov_2012}, in the tripartite setting there exist valid process matrices with entirely classical correlations that violate causal inequalities~\cite{baumeler_maximal_2014, baumeler_space_2016,Araujo_2017}. 

General process matrices with multiple slots have also been shown to outperform causal processes in tasks where the quantum switch, switch-like processes, and even quantum circuits with quantum control of causal order~\cite{Wechs_2021} do not offer any advantage over fixed order processes. The causal inequalities for instance, cannot be violated by performing local instruments on switch-like processes~\cite{Araujo_2015,Wechs_2021,purves_quantum_2021}. Other tasks where general process matrices offer an advantage over causal processes, but the switch-like processes do not, are the discrimination of unitary channels~\cite{Bavaresco_2022}, quantum query complexity of Boolean Functions~\cite{Abbott_2024}, as well as inverting and transposing unitary operations~\cite{Quintino_2019_PRA,Quintino_2019_PRL,Quintino_2022_Quantum}.

While not forbidden \textit{a priori} by the laws of quantum mechanics, causally non-separable---and in particular acausal processes---are starkly at odds with the fundamental physical notion of causal ordering. Consequently, further axiomatic requirements for valid process matrices, such as an additional purification postulate---corresponding to a requirement of preserving reversibility---have been proposed~\cite{Araujo_2017}.
A process is \textit{reversibility preserving} (or \textit{pure}, using the nomenclature of~\cite{Araujo_2017}) if it transforms reversible operations to reversible operations~\cite{baumeler_device-independent_2016, baumeler_reversible_2019} (i.e., unitary operations into a unitary operations in the quantum case), even when acting only on part of the input operations. This property is motivated by the desire for uncovering a physical principle that renders HOQOs physically implementable. Reference~\cite{Araujo_2017} also shows that a process matrix $W$ is reversibility preserving if it can be written as $W=\ketbra{\mathsf{U}}{\mathsf{U}}$, where $\ket{\mathsf{U}} = \kket{U}$ and $U$ is a unitary operator.

Reference~\cite{Araujo_2017} then argued that `physical' process matrices should arise from reversibility preserving process by inputting a quantum state in some auxiliary past space $P'$ and then discarding some auxiliary future space $F'$. That is, it claims that all `physical' process matrices respect $\mathsf{W} = \ptr{F'}{\ketbra{0_{P'}}{0_{P'}} \star \ketbra{\mathsf{U}}{\mathsf{U}}}$ 
for some reversibility preserving process $\ketbra{\mathsf{U}}{\mathsf{U}}$, with the requirement that this property is not fine tuned; that is, $\ketbra{\mathsf{U}}{\mathsf{U}}$ has to be such that $\mathsf{W}(\Psi) = \ptr{F'}{\ketbra{\Psi_{P'}}{\Psi_{P'}} \star \ketbra{\mathsf{U}}{\mathsf{U}}}$ is a valid process matrix for \textit{all} quantum states $\ket{\Psi}$. 

While the purification requirement excludes certain acausal process matrices such as $\mathsf{W}^{\text{OCB}}$ [see Eq.~\eqref{eq::WOCB}], it is insufficient to exclude \textit{all} acausal processes~\cite{baumeler_device-independent_2016, baumeler_reversible_2019, Araujo_2017}. In particular, it does not exclude the tripartite classical process known as the \textit{Lugano process}~\cite{baumeler_maximal_2014,baumeler_space_2016,Araujo_2017}, also called the \textit{Baumeler-Wolf} or \textit{Ara{\' u}jo-Feix process}. 

Following the line of research to retain certain `physical' aspects, particular extensions to the quantum comb formalism that can model indefinite causality while displaying desired properties have been proposed. For instance, Refs.~\cite{Wechs_2021,purves_quantum_2021} put forth a generalisation of quantum circuits that allows for quantum control of causal orders, which in turn admits the quantum switch and generalises the concept of quantum control to multipartite scenarios. In such scenarios, the possibility of quantum control leads to processes beyond those that are unitarily equivalent to the $N$-partite quantum switch~\cite{araujo_computational_2014}, thereby admitting an even larger class of causally non-separable processes. Similarly, notions such as \textit{routed circuits}~\cite{Vanrietvelde_2021,Vanrietvelde_2022} and \textit{addressable gates}~\cite{Arrighi_2021} have also been introduced to capture the behaviour of the quantum switch (as well as more `exotic' processes).

While it is unclear how to physically implement most causally non-separable processes in a deterministic manner, there exist schemes to simulate their action probabilistically~\cite{Chiribella_2008, Chiribella_2009, oreshkov_operational_2016, araujo_quantum_2017, silva_connecting_2017, Milz_2017_NJP,Bavaresco_2024_Switch}; such simulations fundamentally require the resource of multipartite entanglement and non-local unitary gates~\cite{Milz_2017_NJP}. As a consequence of this probabilistic simulability, the formalism of process matrices is tightly connected to the concept of multi-time states~\cite{aharonov2005, silva_connecting_2017}, with process matrices being equivalent to those multi-time states for which observed probabilities are linear functions of the states and of the measurements~\cite{silva_connecting_2017}.

Process matrices---both in the bi- and the multipartite case---provide the natural framework for the investigation of spatiotemporal correlations between distributed observers. As such, they have not only been employed in the explicit study of causal indefiniteness, but also in the field of (quantum) causal modelling~\cite{Costa_2016, Allen_2017,Cotler_2019,Barrett_2019} and to derive uncertainty relations for quantum processes~\cite{Xiao_2021,Xiao_2023} (in both of these cases, process matrices become equivalent to the quantum comb / process tensor formalism without a final output wire, since a fixed causal ordering is assumed). Here, the aim is to uncover causal relations between different events in space and time, i.e., to uncover the manner in which different events influence each other~\cite{Cotler_2019}. Algorithms for the discovery of quantum causal models / causal relations have been proposed in Refs.~\cite{giarmatzi_quantum_2018, bai_quantum_2022}. Analysis of causal relations by means of process matrices / quantum combs reveals that quantum causal models display more general causal relations---even when they are in a fixed order---than classical ones. While two classical events are \textit{either} correlated via a common or a direct cause (or a convex mixture thereof), quantum processes can be in a \textit{superposition} of common and direct causes~\cite{maclean_quantum-coherent_2017, Feix_2017, Nery_2021}.\footnote{While the possibility of superposition distinguishes causal relations in quantum mechanics from those in classical physics, the admissible signalling \textit{structures}, i.e., the sets of directed graphs depicting how events can influence each other, are likely the same in both theories~\cite{Tselentis_2023}.} The process matrix formalism has been further applied to understand causally ordered processes in Refs.~\cite{Allen_2017, Barrett_2019} and quantum measurements in space and time~\cite{Shrapnel_2017}. 

Finally, process matrices are not the end of the hierarchy of causally indefinite processes. Firstly, more general `users' of quantum processes can be conceived: For instance, situations where the individual parties interact in a sequential, multi-round way with the corresponding \textit{multi-round process matrix} have been considered~\cite{Hoffreumon_2021}. This paradigm, where the operations in individual laboratories are themselves described by quantum combs (instead of quantum channels) naturally leads to a refined notion of causal non-separability. In addition, considering \textit{transformations} of process matrices allows for the analysis of the dynamics of causal order, uncovering symmetries in `time' evolution on a higher level~\cite{Castro-Ruiz_2018,Selby_2020}. Similarly, such investigation of transformations of process matrices enables one to consider both their signalling content as well as their causal non-separability as a resource, which has led to the formulation of a resource theory of signalling / causal connection~\cite{Milz_2022} and causal non-separability~\cite{Araujo_2015, Milz_2022}.

As mentioned previously, questions concerning causal relations in the multipartite setting are significantly more nuanced than in the bipartite case. For instance, multipartite causally non-separable processes may not be fully ascribed to quantum phenomena, since entirely classical processes with \textit{dynamical} control of causal orders can lead to such behaviour~\cite{baumeler_maximal_2014,baumeler_space_2016}. Such situations have been analysed within the framework of \textit{process functions}, which have found utility in delineating the quantum/classical boundary in causal modelling~\cite{Baumeler_2020,Tselentis_2023} and can be seen as the classical limit of process matrices~\cite{kunjwal_nonclassicality_2024,kunjwal_generalizing_2024}. The ability for seemingly benign classical processes to exhibit causally exotic behaviours here is reminiscent of other `counter-intuitive' results such as the existence of bipartite channels with separable Kraus operators that cannot be implemented with LOCC  \cite{Bennett_1999,Plenio_2007,Horodecki_2009,akibue_thesis,Chitambar_2020} (a phenomenon often phrased as `quantum non-locality without entanglement'); indeed, concrete connections have been made amongst these areas by way of process functions~\cite{akibue_thesis,Akibue_2017,Kunjwal_2023,Steffinlongo_2025}.

Foundational work on such higher order extensions and the ensuing analysis of causality continues to date. A theoretical framework for higher-order quantum theory has been constructed on axiomatic grounds in Ref.~\cite{Bisio_2019}, and Refs.~\cite{lmcs:4426,simmons_2024} have developed a categorical semantics for causal order. Many of these ideas connect to research on quantum gravity, with the core idea being that quantum superpositions of spacetime itself might necessarily lead to indefinite causal order. This has stimulated the exploration of causally non-separable processes as potential signatures of quantum gravitational effects~\cite{Hardy_2001, Hardy_2009, Hardy_2012, Hardy_2015}.

Unlike general process matrices---for which it is not clear how to implement them deterministically---the quantum switch permits deterministic experimental implementations/simulations
\footnote{A discussion of whether experiments \textit{implement} or rather \textit{simulate} a causally non-separable HOQOs, and what \textit{implementation} means in the first place, can be can be found, e.g., in Refs.~\cite{chiribella_quantum_2019, Oreshkov_2019,Ormrod2023CausalStructure,Rozema_2024,Kabel2024_Gravity,Vilasini_2024Realizing}. Here, we shall stay agnostic with respect to this question.}
and has thus seen much theoretical and experimental interest---not least because it provides a powerful resource for many information theoretic tasks. We thus finish our review of HOQOs regarding the investigation of causal order with a discussion of the quantum switch, before turning to consider the experimental characterisation and implementation of HOQOs.

\FloatBarrier


\subsubsection{Quantum Switch}\label{subsubsec::cqf-quantumswitch}\hfill\\

\noindent The practical significance of indefinite causal order is that it has been shown to provide advantages in various quantum information tasks. The most famous example of a causally indefinite process is the `quantum switch' (see Fig.~\ref{fig::ico-quantumswitch}), where the order of two operations is controlled by an auxiliary quantum system. The quantum switch was introduced in Ref.~\cite{Chiribella_2013} as a HOQO that transforms a pair of independent quantum channels (or equivalently bipartite non-signalling channels) into a valid quantum channel, but nonetheless does not admit a realisation in terms of a standard quantum circuit; that is, it cannot be written as a quantum comb (or convex combinations thereof). Subsequently, the quantum switch has been demonstrated to be a useful resource for discriminating amongst non-signalling channels that have commuting and anti-commuting structure~\cite{Chiribella_2012}. This task was then revisited in Ref.~\cite{Araujo_2015}, which considered the discrimination of pairs of independent unitaries that are promised to either commute or anti-commute. This commutation task then motivated several computational problems where the multipartite versions of the quantum switch were shown to outperform standard quantum circuits~\cite{araujo_computational_2014,taddei_computational_2021,Renner_2021,Renner_2022}. Returning to channel discrimination, two-slot `switch-like' processes\footnote{`Switch-like' processes refer to the quantum switch a simple generalisation thereof, where one can perform arbitrary unitary operations both before and after the input operations, and before and after the global past and global future.} were shown to be useful for discriminating pairs of non-unitary channels in Ref.~\cite{Bavaresco_2021}, even if such processes cannot outperform sequential strategies in \textit{unitary} channel discrimination tasks~\cite{Bavaresco_2022,Abbott_2024}. 

The quantum switch has also been shown to offer advantages in metrological tasks~\cite{Goldberg_2023,Mothe_2024}; in a continuous variable scenario, it can be used to overcome the Heisenberg scaling limit~\cite{Zhao_2020_Metrology}. Further advantages have been reported in complex communication tasks~\cite{guerin_exponential_2016}, as well as activating and enhancing both quantum and classical communication capacities over noisy channels~\cite{ebler_enhanced_2018,Salek_2018,chiribella_quantum_2019,kristjansson_resource_2020,Chiribella_2021}, which has also been revisited in the context of coherent interferometers~\cite{abbott_communication_2020,branciard_coherent_2021}. Moreover, the quantum switch has been shown to be able to activate `non-stabiliserness'; in other words, completely stabiliser-preserving operations---which cannot generate magic states under standard conditions---can be transformed to do so when processed by the quantum switch~\cite{Yin_2024_Enhancement}. Lastly, in the context of thermodynamics, such results lead to enhanced refrigeration rates~\cite{Felce_2020}.

Regarding correlations obtained by applying quantum instruments to the quantum switch, as mentioned above, the quantum switch does not lead to device-independent non-causal correlations, i.e, it cannot be used violate causal inequalities~\cite{Araujo_2015}. Nonetheless, the indefinite causal behaviour of the switch can be certified in a \textit{semi}-device-independent setting, where one of the parties has fully characterised instruments~\cite{Bavaresco_2019}, as well as in a semi-quantum scenario, where the inputs are trusted quantum states~\cite{Dourdent_2022}. Also, by invoking extra non-signalling assumptions, a weaker notion of indefinite causality of the quantum switch can be certified in a device-independent manner~\cite{Gogioso_2023,Lugt_2023,vanderLugt2024possibilistic} (see also Ref.~\cite{Dourdent_2024} for a distinct device-independent certification, and Ref.~\cite{baumann_no_2025} for a discussion of other possible avenues of certifying the quantum switch device-independently).

Naturally, the question of simulating the quantum switch in a time-ordered manner by making use of extra resources (e.g., more calls to the input instruments) has garnered significant attention. The quantum switch can always be simulated exactly by causally ordered circuits (i.e., quantum combs) in a probabilistic manner~\cite{Chiribella_2013,Milz_2017_NJP}. When the input channels are restricted to be unitary, the quantum switch can be simulated by a quantum comb whenever just one extra call of one of the input channels is permitted~\cite{Chiribella_2013}. However, when considering general quantum channels as inputs, even given access to an extra call of each of the input channels, the quantum switch cannot be deterministically simulated~\cite{Bavaresco_2024_Switch}. Lastly, Refs.~\cite{Kristjansson_2024_Simulating,Bavaresco_2024_Switch} have proved that the action of the quantum switch on a pair of $n$-qubit systems (i.e., each of dimension $d=2^n$) \textit{cannot} be deterministically simulated by causally ordered circuits / quantum combs given access to a single call of one input channel and $k<2^n$ calls of the other.

In the two-slot scenario, up to unitary equivalencies, the quantum switch has been shown to be the only reversibility preserving (see previous section) HOQO that is not causally ordered~\cite{Barrett_2021,Yokojima_2021}; see also Ref.~\cite{Costa_2022} (which refers to reversibility preserving process as \textit{unitary processes}). Lastly, as we will detail in Sec.~\ref{subsubsec::ced-experimentaldemonstrations}, various experimental realisations/simulations of the quantum switch---mostly using optical setups---have been demonstrated. 

\FloatBarrier


\subsection{Characterisation \& Experimental Demonstrations}\label{subsec::characterisationexperimentaldemonstrations}

Processes---be they quantum or classical---are notoriously difficult to characterise. Fundamentally, this is due to the exponentially growing number of parameters required to fully describe a quantum object, be it a state, a channel, or even a higher-order quantum operation (or, in the classical case, a joint probability distribution). Given access to some such black-box that describes any such object, one can tomographically reconstruct the unknown object by probing it; however, this is quite costly. Thus, naturally, methods to perform efficient tomographies and/or characterise key properties of quantum objects have been developed. For quantum states, examples of such techniques include shadow tomography, entanglement witnesses, etc. As we will discuss, certain aspects of these methods have been generalised to quantum channels and HOQOs, although ensuring properties such as (generalised) complete positivity, trace preservation, and causality make these extensions non-trivial in many cases. Moreover, in practice, one is often concerned with understanding the noise profile of a process and its subsequent impact, which can be achieved via randomised benchmarking, as well as possible remedies for complex, correlated errors in practical quantum computers. Recent developments in these areas are the focus of this section. 


\subsubsection{Process Tomography}\label{subsubsec::ced-processtomography}\hfill\\

\noindent
\textbf{\textul{General Theory.}} As we have seen, the operational consequences of general quantum processes with memory come down to how the comb acts upon the sequence of probing instruments employed. It is precisely from such observed data that one typically wishes to infer definitive structural properties of the underlying process. In the most informative case, one could perform an informationally complete sequence of instruments and therefore perform a full multi-time quantum process tomography to reconstruct the comb~\cite{White_2022}. However, one problem with such tomography is that it requires a lot of resources: In typical channel tomography (of a $d$-dimensional system), one must prepare $d^2$ input states and then perform a measurement containing $d^2$ outcomes, leading to a resource cost that scales like $d^4$. Tomographically reconstructing an $n$-step comb (where, for simplicity, the system dimension throughout is assumed to be $d$) requires implementing informationally complete set of CP maps at each step; each one of these has $d^4$ elements and so the resources scale like $d^{4n}$~\cite{Pollock_2018_PRA,Milz_2018,White_2022}.

Overall, practically feasible complete process tomography is a special case of `partial reconstruction' of a quantum comb, namely when one can implement an informationally complete set of tester elements~\cite{Milz_2018}, which can be seen as follows. Consider the standard notion of quantum state tomography: Here, one aims to reconstruct the (unknown) density operator $\rho$ from the observed probability distribution over a set of measurement outcomes corresponding to the POVM $\mathcal{J} = \{ \xi^{(x)} \}$; the probability distribution is calculated according to the Born rule
\begin{align}
    \mathds{P}(x|\mathcal{J}) = \tr{\rho \, \xi^{(x)}}.
\end{align}
Now, for any set of linearly independent objects, such as can be assumed without loss of generality for $\{ \xi^{(x)} \}$, there exists a dual set $\{ \Delta^{(y)}\}$ such that $\tr{\xi^{(x)} \Delta^{(y) \textup{T}}} = \delta_{xy}$~\cite{Modi_2012_PRA,Milz_2017}. One then performs the following linear inversion to construct the object\footnote{Note that a similar construction works in the continuous variable case, with the sum replaces by an integral, probability distribution replaced by a probability density function, etc.\ (see, e.g., Ref.~\cite{Bisio_2009}).}
\begin{align}
    \hat{\rho} = \sum_x \mathds{P}(x|\mathcal{J}) \, \Delta^{(x)}.
\end{align}
By construction (and as a consequence of linearity), $\hat{\rho}$ contains all the necessary information to compute the correct probability distribution over outcomes for any measurement whose elements lie entirely within the span of the original one; thus, whenever $\mathcal{J} = \{ \xi^{(x)} \}$ is informationally complete (i.e., spans the entire space), then the reconstruction $\hat{\rho}$ faithfully describes the unknown quantum state $\rho$. However, if one is only interested in the operational behaviour expected on a restricted set of measurements, e.g., those in some fixed basis, then one need not perform a complete tomographic reconstruction, but rather use any partial tomography built from measurements that span (at least) the subspace of interest. Such a partial reconstruction does not yield a positive semidefinite operator in general, although it acts like one on all objects that lie in its domain, constituting a good approximation for many practical purposes. 

The story is similar for channel tomography~\cite{Chuang_1997,Mohseni_2008} and indeed multi-time quantum process tomography~\cite{Pollock_2018_PRA,Milz_2017,White_2022}; this is because all of the logic above relies only upon the linearity of the objects involved. Thus, for any linearly independent tester $\mathcal{J} = \{ \mathsf{O}^{(x)} \}$ (which can, in general, be temporally correlated) applied to a process tensor / quantum comb $\mathsf{T}$, one observes the probability distribution
\begin{align}
    \mathds{P}(x|\mathcal{J}) = \tr{\mathsf{T} \, \mathsf{O}^{(x)}},
\end{align}
from which one can reconstruct the object
\begin{align}
    \widehat{\mathsf{T}} = \sum_x \mathds{P}(x|\mathcal{J}) \, \Delta^{(x)},
\end{align}
where $\{ \Delta^{(x)} \}$ is the dual set to the tester $\{ \mathsf{O}^{(x)} \}$. Indeed, for the reason above, the term `process POVM' was used in place of `tester' in Ref.~\cite{Ziman_2008}. Again, whenever the tester forms an informationally complete set, then the reconstructed $\widehat{\mathsf{T}}$ is positive semidefinite (representing a completely positive multi-time system-environment dynamics). A convenient informationally complete tester comprises a sequence of informationally complete causal breaks instruments comprising an informationally complete set of measurements followed by the preparation of a set of spanning states at each timestep~\cite{Pollock_2018_PRA}. Moreover, whenever the tester is not informationally complete, one can still use the above reconstructed object to meaningfully compute behaviours for superinstruments whose elements lie within the span of the original tester~\cite{Kuah_2007,Milz_2018,Taranto_2021_npj}. 


\subsubsection{Characterising \& Taming Quantum Processes}\label{subsubsec::ced-characterisingtamingquantumprocesses}\hfill\\

\noindent The ascension of sophisticated quantum computing hardware means that it is easy to access complex quantum processes over the cloud. There are significant needs and interests to characterise such processes from both engineering and foundational perspectives. However, characterisation of complex quantum processes presents several fundamental challenges, from practical experimental constraints to computational feasibility~\cite{white-rev}. These challenges have spurred the development of various approaches to process characterisation, each suited to different experimental scenarios and requirements. 

\vspace{0.25cm}\noindent
\textbf{\textul{Efficient Tomography.}} Developing efficient tomographic methods for characterising multi-time quantum processes under reasonable assumptions while managing resource constraints provides a pressing challenge~\cite{White_2022}. The practical limitations of experimental implementations mean that any reconstructed probability distributions are actually derived from finite-sample frequencies: Such finite sampling, combined with potential deviations from \textbf{independent and identically distributed (i.i.d.)} conditions, can lead to reconstructed processes that violate complete positivity and therefore cannot be ascribed to a proper quantum circuit. This issue can be addressed by incorporating appropriate constraints in the maximum likelihood estimators used to reconstruct quantum combs from experimental data~\cite{chiribella06likelihood, White_2022}.

Beyond such issues, the computational complexity of full process tomography often proves prohibitive, either due to the complexity of the process or engineering limitations. For instance, mid-circuit measurements are often not available, posing a serious limitation for characterisation. Nonetheless, it is still possible to partially characterise multi-time quantum process~\cite{Kuah_2007,Milz_2018}. Below we discuss some of the methods that are employed in taming the complexities in characterisation. In many practical scenarios, a complete process description may be unnecessary; prominent examples include situations where one only wishes to predict expectation values of coarse-grained, local, or otherwise restricted observables. Shadow tomography, originally developed for quantum states~\cite{Cramer_2010, Aaronson_2017, huang-shadow}, has been extended to quantum channels and process tensors/quantum combs~\cite{Ringbauer_2015, White_2021_Many, PhysRevLett.130.160401, Zhao_2024}, providing efficient methods for extracting relevant information about restricted or local observables. Another technique for taming process complexity employs the concept of quantum Markov order~\cite{Taranto_2019_PRA, Taranto_2019_PRL, White_2022} to truncate the size of non-Markovian memory to a finite number of timesteps~\cite{Taranto_2021_npj, whitethesis}. Finally, as the process tensor has a natural tensor network representation, one can always control its complexity by controlling the bond dimension of the model~\cite{Luchnikov_2019, Luchnikov_2020, 10.21468/SciPostPhys.13.2.028, white2023unify}, which builds on the notion of matrix product operator tomography put forth in Refs.~\cite{Baumgratz_2013,Holzapfel_2015}. 

\vspace{0.25cm}\noindent
\textbf{\textul{Machine Learning Approaches.}} Machine learning approaches offer an alternative route to process characterisation, focusing on constructing approximate processes that accurately reproduce observed behaviour while minimising complexity---typically considered in terms of the size of the environment or memory depth of the model. Ensemble learning methods have been successfully employed to estimate environmental size~\cite{Shrapnel_2018}, while maximum likelihood estimation techniques enable the embedding of non-Markovian processes into Markovian ones with minimal auxiliary systems~\cite{Luchnikov_2019, Luchnikov_2020}. These approaches have been complemented by algorithms specifically designed to learn memory structures in open quantum processes~\cite{PhysRevE.98.042114, Guo_2020, torlai2020quantum, white2023unify, PhysRevA.106.022411, guo2023} based on techniques developed to learn matrix product states~\cite{PhysRevE.98.042114}.

\vspace{0.25cm}\noindent
\textbf{\textul{Randomised Benchmarking and Gate Set Tomography.}} The characterisation of quantum noise presents a fundamental challenge in the development of fault-tolerant quantum computation. To this end, \textit{randomised benchmarking} has emerged as a robust protocol for estimating average noise strength in quantum devices~\cite{Eisert_2020, Kliesch_2021}, with recent theoretical advances extending their applicability to more realistic noise models including spatiotemporally correlated errors~\cite{Figueroa-Romero_2021_PRXQ, Figueroa-Romero_2022, guo2023, Figueroa-Romero_2024, liu2024}. These developments are particularly crucial as quantum control achieves ever-shorter timescales, where the assumption of Markovian noise becomes increasingly inadequate. The integration of non-Markovian noise models with Clifford group operations has provided new diagnostic tools for identifying non-Markovian characteristics and estimating memory timescales (such as decoherence times), which can be further enhanced by machine learning techniques to improve efficiency~\cite{Yang_2022}.

These diverse approaches to characterising complex quantum processes reflect the multifaceted nature of the challenge, from practical experimental constraints to fundamental theoretical considerations. As quantum devices continue to advance in sophistication and scale, the ability to accurately characterise and manage complex quantum processes becomes increasingly critical to their practical implementation. In particular, recent proposals have extended the tools of self-consistent characterisation---such as gate set tomography~\cite{GST-quantum}---to the non-Markovian setting~\cite{white2023unify, Li_2024}.

\vspace{0.25cm}\noindent
\textbf{\textul{Quantum Error Correction.}} The challenge of correlated errors extends to quantum error correction, where standard approaches typically assume uncorrelated errors throughout a computation. This assumption is becoming increasingly far from the truth as we develop control over more qubits, packed closer together, and at shorter timescales, which leads to complex errors that are correlated in both space and time. Recent work has begun addressing this challenge by reformulating error correction in the language of higher-order quantum operations~\cite{Tanggara_2024}, providing a natural framework for describing and managing such spatiotemporally correlated noise. Investigations into the effects of correlated noise models on specific instances of quantum codes have shown their vulnerability~\cite{kam_2024, Kobayashi_2024}. This likely means that this will be an active area of research that will employ methods of HOQOs.


\subsubsection{Experimental Demonstrations}\label{subsubsec::ced-experimentaldemonstrations}\hfill\\

\noindent We now move to discuss a number of experimental demonstrations related to HOQOs; that such experiments have been conducted within a relatively short timeframe as the development of the theory itself is a testament to its immediate relevance.

\vspace{0.25cm}\noindent
\textbf{\textul{Multi-Time Quantum Processes.}} A superchannel was first tomographically reconstructed in Ref.~\cite{Ringbauer_2015}, where the authors fully characterised an initially correlated open system dynamics using a photonic qubit coupled to a simulated environment. Subsequently, similar photonic setups have been exploited to characterise certain types of multi-time processes such as `common-cause' processes~\cite{Guo_2021}. Ref.~\cite{White_2020} reconstructed a restricted process tensor on a multi-qubit superconducting quantum device, reporting average infidelity as low as $10^{-3}$. Next, the development of \textit{process tensor tomography} using maximum likelihood estimators in Ref.~\cite{White_2022} led to a significant reduction in resource requirements. Namely, whereas Ref.~\cite{White_2020} required $24^k$ circuits for optimal tomography of a $k$-step process, Ref.~\cite{White_2022} reduced that to $10^k$ circuits. The same reference employed quantum Markov order to truncate the exponential complexity of a generic process to be linear in the number of timesteps and constant in the size of the memory, i.e., further reducing the number of circuits required to $k \times 10^\ell$, where $\ell$ is the Markov order. Reference~\cite{White_2021_Many} developed methods to learn important properties of the process, including estimating non-Markovian memory and temporal entanglement, from partial reconstructions. 

Methods to extend classical shadow tomography techniques to estimate multi-time correlation functions in situations allowing for mid-circuit measurements have been developed in Ref.~\cite{White_2024_Many}, where a $20$-step process tensor was reconstructed so that low Pauli weight marginals are readily accessible. 
References~\cite{2105.03333, Giarmatzi_2023} further devised mid-circuit measurements to enable full characterisation of multi-time quantum processes. Lastly, Ref.~\cite{white2023unify} has developed crucial tools for \textit{self-consistent} process tensor tomography: Here, no assumptions are made about the control operations---they are assumed to be noisy and flawed. This tensor network-based method is designed to characterise the non-Markovian quantum process and the control operations simultaneously; very much in the spirit of gate set tomography~\cite{GST-quantum}. Finally, the efficient simulation of classical hidden Markov models using quantum memory has been demonstrated in Ref.~\cite{Palsson_2017}.

\vspace{0.25cm}\noindent
\textbf{\textul{Causality Experiments.}} The derivation of causally indefinite quantum processes has led to significant interest in performing causality-related experiments. As this research area has developed so rapidly and several review articles have already been written, we refer the Reader to Refs.~\cite{Goswami_2020, Rozema_2024} for a broad overview; here, we will simply highlight some of the main results to this end. 

Recent experimental work has demonstrated fundamental capabilities of information processing with the quantum switch / coherent control of causal orders~\cite{goswami_2020_increasing, Guo_2020a}. For example, Ref.~\cite{goswami_2020_increasing} showed that the quantum switch could `activate' channel capacity to enable information transmission through completely noisy (depolarising) channels---achieving $3.4\times 10^{-2}$ bits through two fully depolarising channels. Later experiments focused on more sophisticated implementations, demonstrating high-fidelity quantum information transmission through noisy channels arranged across superposed trajectories, comparing quantum-controlled parallel channels, series channels with quantum-controlled operations, and quantum-controlled channel ordering~\cite{rubino2021experimental}. In Ref.~\cite{procopio_experimental_2015}, the authors experimentally demonstrated an advantage for discriminating quantum channels that are promised to commute or anti-commute; a task also considered in Ref.~\cite{Stromberg_2023} that implemented the quantum switch based on a Sagnac interferometer and Ref.~\cite{taddei_computational_2021} which used multi-mode fibres for a $4$-party generalisation of the quantum switch. In the context of quantum metrology, Ref.~\cite{Yin_2023} employed the quantum switch to bypass the Heisenbarg scaling limit. Lastly, Refs.~\cite{Nie_2022,xi_experimental_2024} experimentally verified advantages of the quantum switch in thermodynamic scenarios, with Ref.~\cite{Zhu_2023} demonstrating enhanced quantum battery charging with the quantum switch.

At the same time, research has pushed towards more rigorous experimental verification and characterisation of quantum processes with indefinite causal order~\cite{Ringbauer_2016, rubino2017experimental, cao2023semideviceindependent, antesberger2024higherorder}. Early witnesses of causal indefiniteness were reported using photonic setups in Refs.~\cite{Ringbauer_2016, rubino2017experimental,goswami_indefinite_2018}; this led to more sophisticated experiments that attempted to close certain `loopholes', such as Ref.~\cite{cao2023semideviceindependent}, where the authors achieved a semi-device-independent certification of indefinite causal order (requiring less assumptions about device characterisation than previously). A significant breakthrough came with the development of a passively stable fibre-based quantum switch using time-bin encoded qubits, which enabled the first complete experimental characterisation of an optical passive-stable quantum switch via process tomography~\cite{antesberger2024higherorder}. In addition, a deterministic experimental implementation of the quantum time flip together with an explicit certification of its indefinite time direction was carried out in~\cite{stromberg_experimental_2024}, exploiting device dependent symmetries of the employed experimental setup. Notably, all of these experiments are based on photonic/optical platforms.

\FloatBarrier


\section{Summary \& Outlook}
\label{part::summaryoutlook}

\vspace{0.25cm}\noindent
\textbf{\textul{Higher-Order Quantum Information Tasks.}} Higher-order quantum operations (HOQOs) represent powerful tools for processing quantum information, yet their potential remains to be fully explored in several key areas. A major open problem is determining \textit{tight} upper and lower bounds on the number of calls required to implement specific functions of quantum operations. While progress has been made in understanding the efficiency of certain tasks, many questions remain about the minimal resources needed for general classes of functions. For example, in classical computation, comparator functions are ubiquitous and essential; defining quantum analogues of such \textit{multi-input} functions and determining their optimal implementation costs is a pressing challenge. Addressing these questions could provide a clearer roadmap for realising a quantum functional programming paradigm that uses higher-order functions as primitives in quantum computation.

Another key avenue is exploring the robustness of such higher-order functions. Many existing results focus on idealised scenarios, but real-world implementations require an understanding of how noise and imperfections impact performance. Developing robust `go’ results---i.e., proof-of-concept demonstrations for specific, practically relevant functions---could bridge the gap between theoretical constructs and experimental applications. Furthermore, an open question remains about the interplay between known and unknown quantum operations: How can HOQOs be effectively applied in scenarios where the properties of input operations are partially known? This challenge ties into the broader goal of learning the properties of unknown objects through higher-order processes under certain assumptions. Investigating these questions not only deepens the theoretical foundations of HOQOs but also has implications for practical domains, such as quantum cryptography, where composing complex protocols while maintaining security remains a critical issue.

Finally, extending the scope of HOQOs to continuously controlled quantum systems presents a rich avenue for exploration. Hamiltonian supermaps---i.e., higher-order maps that act on time-continuous evolutions---could unlock new paradigms for dynamically controlled quantum processes. This area holds promise for applications in quantum simulation and continuous-variable quantum computation, but a formal understanding of the operations enabled by such supermaps remains in its infancy. 

\vspace{0.25cm}\noindent
\textbf{\textul{Open System Dynamics \& Memory Effects.}} In the context of open quantum systems, HOQOs offer a compelling lens through which to understand and leverage memory effects. One fundamental open question concerns the origin of Markovianity in quantum processes: How and under what conditions can memoryless dynamics emerge from more general non-Markovian behaviour? While Markovianity is often an idealised assumption, real-world systems exhibit varying degrees of memory, shaped by interactions with structured environments. As we have discussed, process tensors---which naturally encode temporal correlations---provide a rigorous framework for probing the conditions under which memory effects can be quantified, suppressed, or harnessed. Extending these insights to different noise structures---such as coloured or otherwise correlated noise---remains a critical challenge, with implications for designing more robust quantum devices.

This issue has implications for understanding how temporal correlations spread throughout a quantum process. For instance, in analogy to the spatial case where correlations are restricted by monogamy constraints, one would expect similar restrictions of temporal correlations. HOQOs could provide a means to quantify and exploit these restrictions, with potential applications in resource allocation for optimising quantum circuit architectures.

Finally, the interplay between memory effects and optimal control provides a rich area for exploration. By integrating process tensors with quantum control theory, it may be possible to design protocols that explicitly account for memory effects to achieve enhanced control over quantum systems: Instead of mitigating non-Markovianity as a source of noise, one could harness the memory as a resource for tasks such as information storage or coherent quantum feedback. A particularly intriguing question is whether certain classes of non-Markovian dynamics are inherently more advantageous for specific quantum information tasks and how HOQOs can uncover such advantages. Addressing these problems not only advances our understanding of open system dynamics but also has practical implications for developing fault-tolerant quantum technologies and memory-enhanced quantum computation.

\vspace{0.25cm}\noindent
\textbf{\textul{Many-Time Quantum Physics.}} Higher-order quantum operations also provide a powerful framework for exploring many-time physics and the intricate dynamics of complex quantum processes. A fundamental question concerns the relationship between these dynamics and phenomena such as thermalisation, equilibration, and ergodicity. While traditional approaches often focus on the long-time behaviour of systems under specific assumptions (e.g., weak interactions or large environments), HOQOs offer a way to analyse and quantify how temporal correlations behave across multiple timesteps. For example, it would be interesting to see how process tensors can encode and distinguish between ergodic and non-ergodic behaviours, shed light on the conditions under which a system reaches equilibrium, and witness deviations from thermalisation.

Another challenge is efficiently simulating complex dynamics, particularly in regimes where the dynamics are highly correlated or driven by intricate interactions. While many simulation techniques rely on simplifying assumptions to make such problems tractable, HOQOs enable a way to incorporate these assumptions directly into the model. This approach could make simulations more efficient by leveraging the natural temporal structures present in the dynamics, such as periodicity or symmetry. Exploring how assumptions like approximate Markovianity or specific interaction patterns affect the computational complexity of simulating many-time processes is a promising direction for future research.

The ability of HOQOs to learn properties of complex dynamics also opens up exciting avenues. By employing quantum combs to analyse experimental data, one could infer hidden characteristics of an open quantum process, such as the presence of conserved quantities, emergent symmetries, or dynamical phases. A significant open question is how to optimise this learning process: What are the most efficient measurement strategies for extracting information about complex dynamics using HOQOs? Similarly, can process tensors help identify universal features of dynamics, such as signatures of chaos or localisation, in a way that generalises across different systems? Addressing these problems would deepen our understanding of multi-time phenomena in the quantum realm and provide new tools for investigating their behaviour.

These directions underscore the potential of HOQOs to act as a bridge between the theoretical and experimental realms, offering practical insights into the behaviour of systems that are otherwise challenging to describe with traditional methods. Their application to many-time quantum physics has the potential to reshape our understanding of how complex dynamics emerge and evolve.

\vspace{0.25cm}\noindent
\textbf{\textul{Causality \& Quantum Foundations.}} As we have discussed, HOQOs provide a versatile framework for investigating the fundamental principles of causality in quantum mechanics and their implications for quantum foundations. One intriguing direction is the study of causal structures beyond traditional circuit diagrams, such as routed circuits or general acyclic graphs. Conventional circuit models presuppose fixed causal orders, but HOQOs—especially those involving quantum processes in superpositions of causal orders (e.g., the quantum switch)—challenge these assumptions. A key open question is whether routed circuits or other related frameworks can offer a more natural or complete description of causally indefinite processes. Addressing this could help elucidate how HOQOs generalise classical notions of causality and provide insights into their potential physical realisation.

Another exciting avenue is the simulation and characterisation of causally indefinite quantum processes. While the quantum switch is a well-studied example of indefinite causal order, its full range of applications and generalisations remains under-explored. For instance, can HOQOs simulate more complex causally indefinite scenarios that go beyond two input operations, and what computational or physical resources are required for such simulations? Furthermore, determining the fundamental limits of information processing and control within these settings remains an open challenge. Exploring these questions could have far-reaching implications for quantum communication and computation in scenarios where fixed causal structures are no longer appropriate.

Lastly, HOQOs provide a rigorous starting point for the exploration of acausality as a resource in quantum protocols. For example, under what conditions does causal indefiniteness provide a genuine advantage, and how can it be quantified or harnessed in practical settings? This relates to broader foundational questions, such as whether causality is an emergent property of quantum systems or a fundamental aspect of their description. 

By pushing the boundaries of traditional causal models and exploring their quantum generalisations, HOQOs offer a unique perspective on the foundations of quantum mechanics. These tools not only challenge classical intuitions about cause and effect but also provide a fertile ground for developing new theoretical and experimental approaches to quantum information processing.

\vspace{0.25cm}\noindent
\textbf{\textul{Characterisation \& Experimental Demonstrations.}} Regarding the characterisation of complex quantum processes, several challenges remain in developing efficient and experimentally feasible tools for this purpose. A critical open problem is advancing multi-time quantum process tensor tomography, the task of reconstructing the full process tensor that captures the dynamics of a quantum system over multiple timesteps. While existing methods provide a theoretical blueprint, they often suffer from high resource costs due to the exponential scaling of required measurements with the number of time steps and the Hilbert space dimension. Developing more efficient reconstruction techniques---such as those leveraging sparsity, symmetries, or prior knowledge about the dynamics---remains an urgent goal. The integration of adaptive protocols, where intermediate measurement outcomes guide subsequent choices, could significantly enhance efficiency and scalability.

Another open challenge is extending these characterisation tools to noisy or incomplete data regimes. Practical experiments are often hampered by noise, finite sampling, and other imperfections, which can obscure the reconstruction of processes at hand. HOQOs could be harnessed to design robust tomography methods that are resilient to such imperfections, for instance, through noise-aware algorithms or Bayesian approaches that incorporate error models. Additionally, defining appropriate error metrics for reconstructed multi-time processes and understanding their implications for process validation and benchmarking are important tasks for bridging the gap between theory and experiment.

A related open avenue is the use of HOQOs for on-the-fly tomography of dynamic processes. Traditional approaches to characterising quantum systems typically assume static conditions (e.g., identical and independently distributed samples), but many scenarios require adaptive strategies to deal with more complex situations. In principle, process tensors can facilitate such real-time reconstruction of time-dependent dynamics, enabling efficient characterisation of complex systems under minimal assumptions, although the precise details are yet to be demonstrated. 

Further directions involve experimental demonstrations of complex, temporally correlated quantum processes. While some experiments have successfully demonstrated specific higher-order operations, realising more intricate process tensors in the laboratory is a largely untapped frontier. This requires addressing practical challenges such as maintaining coherence across multiple timesteps, controlling environmental interactions, and scaling up the number of accessible quantum states and operations. Exploring how HOQOs can be realised in cutting-edge platforms---such as superconducting qubits, trapped ions, or photonic systems---while optimising experimental overheads is a vital step forward.

Lastly, the validation and certification of experimentally reconstructed process tensors represent another critical open question. How can we ensure that a reconstructed process tensor accurately reflects the underlying physical dynamics, especially when it exhibits complex temporal correlations? Addressing this requires developing rigorous, efficient tools for comparing reconstructed processes with theoretical predictions or other experimentally reconstructed tensors. By tackling these challenges, the application of HOQOs to characterising complex quantum processes can significantly advance both the foundational understanding and practical capabilities of quantum technologies.

\FloatBarrier

\section*{Acknowledgments}
We thank Alastair Abbott, Matheus Ara{\' u}jo, {\" A}min Baumeler, {\v C}aslav Brukner, Moritz Cygorek, Jordan Cotler, Guo Chu, Otfried G{\" u}hne, Jonathan Keeling, Sumeet Khatri, Ravi Kunjwal, Rapha{\" e}l Mothe, Martin Plenio, Philipp Strasberg, Xin Wang, Yunlong Xiao, and Yuxiang Yang for valuable feedback on an early draft of this Review Article.

PT acknowledges support from the Japan Society for the Promotion of Science (JSPS) through the Postdoctoral Fellowship for Research in Japan and KAKENHI grant No.\ 21H03394, and the IBM-UTokyo Laboratory. SM acknowledges funding from the European Union's Horizon Europe research and innovation programme under the Marie Sk{\l}odowska-Curie grant agreement No.\ 101068332. MM acknowledges support from MEXT Quantum Leap Flagship Program (MEXT QLEAP) JPMXS0118069605, JPMXS0120351339, JSPS KAKENHI grant No.\ 23K21643, and IBM Quantum. KM acknowledges support from the Australian Research Council Future Fellowship FT160100073 and Discovery Projects DP210100597 and DP220101793. 

\cleardoublepage
\phantomsection
\addcontentsline{toc}{section}{References}


\begin{thebibliography}{100}
\expandafter\ifx\csname url\endcsname\relax
  \def\url#1{{\tt #1}}\fi
\expandafter\ifx\csname urlprefix\endcsname\relax\def\urlprefix{URL }\fi
\providecommand{\eprint}[2][]{\url{#2}}

\bibitem{Landauer_1961}
R~Landauer,  \emph{{Irreversibility and Heat Generation in the Computing
  Process}},  \href{https://doi.org/10.1147/rd.53.0183}{IBM J. Res. Dev. {\bf
  5}, 183 (1961)}

\bibitem{Hardy_2001}
L~Hardy,  \emph{{Quantum Theory From Five Reasonable Axioms}} (2001) (Preprint
  {\href{https://arxiv.org/abs/quant-ph/0101012}{arXiv:quant-ph/0101012}})

\bibitem{Clifton_2003}
R~Clifton, J~Bub and H~Halvorson,  \emph{{Characterizing Quantum Theory in
  Terms of Information-Theoretic Constraints}},
  \href{https://doi.org/10.1023/A:1026056716397}{Found. Phys. {\bf 33}, 1561
  (2003)} (Preprint
  {\href{https://arxiv.org/abs/quant-ph/0211089}{arXiv:quant-ph/0211089}})

\bibitem{Chiribella_2011}
G~Chiribella, G~M D'Ariano and P~Perinotti,  \emph{{Informational Derivation of
  Quantum Theory}},  \href{https://doi.org/10.1103/PhysRevA.84.012311}{Phys.
  Rev. A {\bf 84}, 012311 (2011)} (Preprint
  {\href{https://arxiv.org/abs/1011.6451}{arXiv:1011.6451}})

\bibitem{D'ariano_Chiribella_Perinotti_2017}
G~M D'Ariano, G~Chiribella and P~Perinotti,
  {\href{https://doi.org/10.1017/9781107338340}{\textit{{Quantum Theory from
  First Principles: An Informational Approach}}}} (Cambridge, UK: Cambridge
  University Press) (2017)

\bibitem{Preskill_1997}
J~Preskill,
  {\href{https://doi.org/10.1142/9789812385253_0008}{\textit{{Fault-tolerant
  quantum computation}}}}, {\em Introduction to Quantum Computation\/}, ed H-K
  Lo, S~Popescu and T~P Spiller (Singapore: World Scientific), p 213 (1997)
  (Preprint
  {\href{https://arxiv.org/abs/quant-ph/9712048}{arXiv:quant-ph/9712048}})

\bibitem{Preskill_2018}
J~Preskill,  \emph{{Quantum Computing in the NISQ era and beyond}},
  \href{https://doi.org/10.22331/q-2018-08-06-79}{Quantum {\bf 2}, 79 (2018)}
  (Preprint {\href{https://arxiv.org/abs/1801.00862}{arXiv:1801.00862}})

\bibitem{Preskill_2025}
J~Preskill,  \emph{{Beyond NISQ: The Megaquop Machine}} (2025) (Preprint
  {\href{https://arxiv.org/abs/2502.17368}{arXiv:2502.17368}})

\bibitem{Gisin_2002}
N~Gisin, G~Ribordy, W~Tittel and H~Zbinden,  \emph{{Quantum Cryptography}},
  \href{https://doi.org/10.1103/RevModPhys.74.145}{Rev. Mod. Phys. {\bf 74},
  145 (2002)} (Preprint
  {\href{https://arxiv.org/abs/quant-ph/0101098}{arXiv:quant-ph/0101098}})

\bibitem{Pirandola_2020}
S~Pirandola, U~L Andersen, L~Banchi, M~Berta, D~Bunandar, R~Colbeck, D~Englund,
  T~Gehring, C~Lupo, C~Ottaviani, J~L Pereira, M~Razavi, J~Shamsul~Shaari,
  M~Tomamichel, V~C Usenko, G~Vallone, P~Villoresi and P~Wallden,
  \emph{{Advances in Quantum Cryptography}},
  \href{https://doi.org/10.1364/AOP.361502}{Adv. Opt. Photonics {\bf 12}, 1012
  (2020)} (Preprint
  {\href{https://arxiv.org/abs/1906.01645}{arXiv:1906.01645}})

\bibitem{Giovannetti_2011}
V~Giovannetti, S~Lloyd and L~Maccone,  \emph{{Advances in Quantum Metrology}},
  \href{https://doi.org/10.1038/nphoton.2011.35}{Nat. Photonics {\bf 5}, 222
  (2011)} (Preprint {\href{https://arxiv.org/abs/1102.2318}{arXiv:1102.2318}})

\bibitem{Toth_2014}
G~T{\'o}th and I~Apellaniz,  \emph{{Quantum Metrology from a Quantum
  Information Science Perspective}},
  \href{https://doi.org/10.1088/1751-8113/47/42/424006}{J. Phys. A: Math.
  Theor. {\bf 47}, 424006 (2014)} (Preprint
  {\href{https://arxiv.org/abs/1405.4878}{arXiv:1405.4878}})

\bibitem{Goold_2016}
J~Goold, M~Huber, A~Riera, L~{del Rio} and P~Skrzypczyk,  \emph{{The Role of
  Quantum Information in Thermodynamics---A Topical Review}},
  \href{https://doi.org/10.1088/1751-8113/49/14/143001}{J. Phys. A: Math.
  Theor. {\bf 49}, 143001 (2016)} (Preprint
  {\href{https://arxiv.org/abs/1505.07835}{arXiv:1505.07835}})

\bibitem{Binder_2018}
F~Binder, L~A Correa, C~Gogolin, J~Anders and G~Adesso (eds)
  {\href{https://doi.org/10.1007/978-3-319-99046-0}{\textit{{Thermodynamics in
  the Quantum Regime: Fundamental Aspects and New Directions}}}} (Cham,
  Switzerland: Springer International Publishing) (2018)

\bibitem{Devega_2017}
I~{de Vega} and D~Alonso,  \emph{{Dynamics of Non-Markovian Open Quantum
  Systems}},  \href{https://doi.org/10.1103/RevModPhys.89.015001}{Rev. Mod.
  Phys. {\bf 89}, 015001 (2017)} (Preprint
  {\href{https://arxiv.org/abs/1511.06994}{arXiv:1511.06994}})

\bibitem{Weimer_2021}
H~Weimer, A~Kshetrimayum and R~Or{\'u}s,  \emph{{Simulation Methods for Open
  Quantum Many-Body Systems}},
  \href{https://doi.org/10.1103/RevModPhys.93.015008}{Rev. Mod. Phys. {\bf 93},
  015008 (2021)} (Preprint
  {\href{https://arxiv.org/abs/1907.07079}{arXiv:1907.07079}})

\bibitem{Bruzewicz_2019}
C~D Bruzewicz, J~Chiaverini, R~McConnell and J~M Sage,  \emph{{Trapped-Ion
  Quantum Computing: Progress and Challenges}},
  \href{https://doi.org/10.1063/1.5088164}{Appl. Phys. Rev. {\bf 6}, 021314
  (2019)} (Preprint
  {\href{https://arxiv.org/abs/1904.04178}{arXiv:1904.04178}})

\bibitem{Siddiqi_2021}
I~Siddiqi,  \emph{{Engineering High-Coherence Superconducting Qubits}},
  \href{https://doi.org/10.1038/s41578-021-00370-4}{Nat. Rev. Mater. {\bf 6},
  875 (2021)}

\bibitem{Chatterjee_2021}
A~Chatterjee, P~Stevenson, S~{De Franceschi}, A~Morello, N~P {de Leon} and
  F~Kuemmeth,  \emph{{Semiconductor Qubits in Practice}},
  \href{https://doi.org/10.1038/s42254-021-00283-9}{Nat. Rev. Phys. {\bf 3},
  157 (2021)} (Preprint
  {\href{https://arxiv.org/abs/2005.06564}{arXiv:2005.06564}})

\bibitem{Kok_2007}
P~Kok, W~J Munro, K~Nemoto, T~C Ralph, J~P Dowling and G~J Milburn,
  \emph{{Linear Optical Quantum Computing with Photonic Qubits}},
  \href{https://doi.org/10.1103/RevModPhys.79.135}{Rev. Mod. Phys. {\bf 79},
  135 (2007)} (Preprint
  {\href{https://arxiv.org/abs/quant-ph/0512071}{arXiv:quant-ph/0512071}})

\bibitem{OBrien_2007}
J~L O'Brien,  \emph{{Optical Quantum Computing}},
  \href{https://doi.org/10.1126/science.1142892}{Science {\bf 318}, 1567
  (2007)} (Preprint {\href{https://arxiv.org/abs/0803.1554}{arXiv:0803.1554}})

\bibitem{Acin_2018}
A~Ac{\' i}n, I~Bloch, H~Buhrman, T~Calarco, C~Eichler, J~Eisert, D~Esteve,
  N~Gisin, S~J Glaser, F~Jelezko, S~Kuhr, M~Lewenstein, M~F Riedel, P~O
  Schmidt, R~Thew, A~Wallraff, I~Walmsley and F~K Wilhelm,  \emph{{The Quantum
  Technologies Roadmap: A European Community View}},
  \href{https://doi.org/10.1088/1367-2630/aad1ea}{New J. Phys. {\bf 20}, 080201
  (2018)} (Preprint
  {\href{https://arxiv.org/abs/1712.03773}{arXiv:1712.03773}})

\bibitem{awschalom2022roadmap}
D~D Awschalom, H~Bernien, R~Brown, A~Clerk, E~Chitambar, A~Dibos, J~Dionne,
  M~Eriksson, B~Fefferman, G~D Fuchs, J~Gambetta, E~Goldschmidt, S~Guha, F~J
  Heremans, K~D Irwin, A~B Jayich, L~Jiang, J~Karsch, M~Kasevich, S~Kolkowitz,
  P~G Kwiat, T~Ladd, J~Lowell, D~Maslov, N~Mason, A~Y Matsuura, R~McDermott,
  R~Meter, A~Miller, J~Orcutt, M~Saffman, M~Schleier-Smith, M~K Singh, P~Smith,
  M~Suchara, F~Toudeh-Fallah, M~Turlington, B~Woods and T~Zhong (2022)
  \emph{{\href{https://doi.org/10.2172/1900586}{A Roadmap for Quantum
  Interconnects}}} Tech. rep. Argonne National Laboratory (ANL), Argonne, IL,
  USA

\bibitem{Milz_2020_Quantum}
S~Milz, F~Sakuldee, F~A Pollock and K~Modi,  \emph{{Kolmogorov Extension
  Theorem for (Quantum) Causal Modelling and General Probabilistic Theories}},
  \href{https://doi.org/10.22331/q-2020-04-20-255}{Quantum {\bf 4}, 255 (2020)}
  (Preprint {\href{https://arxiv.org/abs/1712.02589}{arXiv:1712.02589}})

\bibitem{Chiribella_2008_PRL}
G~Chiribella, G~M D'Ariano and P~Perinotti,  \emph{{Quantum Circuit
  Architecture}},  \href{https://doi.org/10.1103/PhysRevLett.101.060401}{Phys.
  Rev. Lett. {\bf 101}, 060401 (2008)} (Preprint
  {\href{https://arxiv.org/abs/0712.1325}{arXiv:0712.1325}})

\bibitem{Chiribella_2009}
G~Chiribella, G~M D'Ariano and P~Perinotti,  \emph{{Theoretical Framework for
  Quantum Networks}},  \href{https://doi.org/10.1103/PhysRevA.80.022339}{Phys.
  Rev. A {\bf 80}, 022339 (2009)} (Preprint
  {\href{https://arxiv.org/abs/0904.4483}{arXiv:0904.4483}})

\bibitem{Pollock_2018_PRA}
F~A Pollock, C~Rodr{\' i}guez-Rosario, T~Frauenheim, M~Paternostro and K~Modi,
  \emph{{Non-Markovian Quantum Processes: Complete Framework and Efficient
  Characterization}},  \href{https://doi.org/10.1103/PhysRevA.97.012127}{Phys.
  Rev. A {\bf 97}, 012127 (2018)} (Preprint
  {\href{https://arxiv.org/abs/1512.00589}{arXiv:1512.00589}})

\bibitem{Pollock_2018_PRL}
F~A Pollock, C~Rodr{\' i}guez-Rosario, T~Frauenheim, M~Paternostro and K~Modi,
  \emph{{Operational Markov Condition for Quantum Processes}},
  \href{https://doi.org/10.1103/PhysRevLett.120.040405}{Phys. Rev. Lett. {\bf
  120}, 040405 (2018)} (Preprint
  {\href{https://arxiv.org/abs/1801.09811}{arXiv:1801.09811}})

\bibitem{Oreshkov_2012}
O~Oreshkov, F~Costa and {\v C}~Brukner,  \emph{{Quantum Correlations with No
  Causal Order}},  \href{https://doi.org/10.1038/ncomms2076}{Nat. Commun. {\bf
  3}, 1092 (2012)} (Preprint
  {\href{https://arxiv.org/abs/1105.4464}{arXiv:1105.4464}})

\bibitem{Chiribella_2013}
G~Chiribella, G~M D'Ariano, P~Perinotti and B~Valiron,  \emph{{Quantum
  Computations without Definite Causal Structure}},
  \href{https://doi.org/10.1103/PhysRevA.88.022318}{Phys. Rev. A {\bf 88},
  022318 (2013)} (Preprint
  {\href{https://arxiv.org/abs/0912.0195}{arXiv:0912.0195}})

\bibitem{Shrapnel_2017}
S~Shrapnel, F~Costa and G~Milburn,  \emph{{Updating the Born Rule}},
  \href{https://doi.org/10.1088/1367-2630/aabe12}{New J. Phys. {\bf 20}, 053010
  (2018)} (Preprint
  {\href{https://arxiv.org/abs/1702.01845}{arXiv:1702.01845}})

\bibitem{White_2020}
G~A~L White, C~D Hill, F~A Pollock, L~C~L Hollenberg and K~Modi,
  \emph{{Demonstration of Non-Markovian Process Characterisation and Control on
  a Quantum Processor}},
  \href{https://doi.org/10.1038/s41467-020-20113-3}{Nat. Commun. {\bf 11}, 6301
  (2020)} (Preprint
  {\href{https://arxiv.org/abs/2004.14018}{arXiv:2004.14018}})

\bibitem{White_2021_Many}
G~A~L White, F~A Pollock, L~C~L Hollenberg, C~D Hill and K~Modi,  \emph{{What
  can unitary sequences tell us about multi-time physics?}} (2021) (Preprint
  {\href{https://arxiv.org/abs/2107.13934}{arXiv:2107.13934}})

\bibitem{Figueroa-Romero_2021_PRXQ}
P~Figueroa-Romero, K~Modi, R~J Harris, T~M Stace and M-H Hsieh,
  \emph{{Randomized Benchmarking for Non-Markovian Noise}},
  \href{https://doi.org/10.1103/PRXQuantum.2.040351}{PRX Quantum {\bf 2},
  040351 (2021)} (Preprint
  {\href{https://arxiv.org/abs/2107.05403}{arXiv:2107.05403}})

\bibitem{Figueroa-Romero_2022}
P~Figueroa-Romero, K~Modi and M-H Hsieh,  \emph{{Towards a General Framework of
  Randomized Benchmarking Incorporating non-Markovian Noise}},
  \href{https://doi.org/10.22331/q-2022-12-01-868}{Quantum {\bf 6}, 868 (2022)}
  (Preprint {\href{https://arxiv.org/abs/2202.11338}{arXiv:2202.11338}})

\bibitem{Morris_2022}
J~Morris, F~A Pollock and K~Modi,  \emph{{Quantifying non-Markovian Memory in a
  Superconducting Quantum Computer}},
  \href{https://doi.org/10.1142/S123016122250007X}{Open Sys. Inf. Dyn. {\bf
  29}, 2250007 (2022)} (Preprint
  {\href{https://arxiv.org/abs/1902.07980}{arXiv:1902.07980}})

\bibitem{Chiribella_2008}
G~Chiribella, G~M D'Ariano and P~Perinotti,  \emph{{Transforming Quantum
  Operations: Quantum Supermaps}},
  \href{https://doi.org/10.1209/0295-5075/83/30004}{EPL {\bf 83}, 30004 (2008)}
  (Preprint {\href{https://arxiv.org/abs/0804.0180}{arXiv:0804.0180}})

\bibitem{Quintino_2019_PRL}
M~T Quintino, Q~Dong, A~Shimbo, A~Soeda and M~Murao,  \emph{{Reversing Unknown
  Quantum Transformations: Universal Quantum Circuit for Inverting General
  Unitary Operations}},
  \href{https://doi.org/10.1103/PhysRevLett.123.210502}{Phys. Rev. Lett. {\bf
  123}, 210502 (2019)} (Preprint
  {\href{https://arxiv.org/abs/1810.06944}{arXiv:1810.06944}})

\bibitem{Quintino_2019_PRA}
M~T Quintino, Q~Dong, A~Shimbo, A~Soeda and M~Murao,  \emph{{Probabilistic
  Exact Universal Quantum Circuits for Transforming Unitary Operations}},
  \href{https://doi.org/10.1103/PhysRevA.100.062339}{Phys. Rev. A {\bf 100},
  062339 (2019)} (Preprint
  {\href{https://arxiv.org/abs/1909.01366}{arXiv:1909.01366}})

\bibitem{Quintino_2022_Quantum}
M~T Quintino and D~Ebler,  \emph{{Deterministic Transformations between Unitary
  Operations: Exponential Advantage with Adaptive Quantum Circuits and the
  Power of Indefinite Causality}},
  \href{https://doi.org/10.22331/q-2022-03-31-679}{Quantum {\bf 6}, 679 (2022)}
  (Preprint {\href{https://arxiv.org/abs/2109.08202}{arXiv:2109.08202}})

\bibitem{Kitaev2000Parallelization}
A~Kitaev and J~Watrous,
  {\href{https://doi.org/10.1145/335305.335387}{\textit{{Parallelization,
  amplification, and exponential time simulation of quantum interactive proof
  systems}}}}, {\em Proceedings of the 32nd Annual ACM Symposium on Theory of
  Computing (STOC)\/}, (New York, NY, USA: Association for Computing Machinery)
  p 608 (2000)

\bibitem{ambainis_quantum_2000}
A~Ambainis, {\href{https://doi.org/10.1145/335305.335394}{\textit{{Quantum
  lower bounds by quantum arguments}}}}, {\em Proceedings of the 32nd Annual
  ACM Symposium on Theory of Computing (STOC)\/}, (New York, NY, USA:
  Association for Computing Machinery) p 636 (2000) (Preprint
  {\href{https://arxiv.org/abs/quant-ph/0002066}{arXiv:quant-ph/0002066}})

\bibitem{Gutoski_2007}
G~Gutoski and J~Watrous,
  {\href{https://doi.org/10.1145/1250790.1250873}{\textit{{Toward a General
  Theory of Quantum Games}}}}, {\em Proceedings of the 39th Annual ACM
  Symposium on Theory of Computing (STOC)\/}, (New York, NY, USA: Association
  for Computing Machinery) p 565 (2007) (Preprint
  {\href{https://arxiv.org/abs/quant-ph/0611234}{arXiv:quant-ph/0611234}})

\bibitem{Portmann_2015}
C~Portmann, C~Matt, U~Maurer, R~Renner and B~Tackmann,  \emph{{Causal Boxes:
  Quantum Information-Processing Systems Closed Under Composition}},
  \href{https://doi.org/10.1109/TIT.2017.2676805}{IEEE Trans. Inf. Theory {\bf
  63}, 3277 (2017)} (Preprint
  {\href{https://arxiv.org/abs/1512.02240}{arXiv:1512.02240}})

\bibitem{Baumeler2017noncausal_computation}
{\"A}~{Baumeler} and S~{Wolf},  \emph{{Non-Causal Computation}},
  \href{https://doi.org/10.3390/e19070326}{Entropy {\bf 19}, 326 (2017)}
  (Preprint {\href{https://arxiv.org/abs/1601.06522}{arXiv:1601.06522}})

\bibitem{araujo_quantum_2017}
M~Ara{\'u}jo, P~A Gu{\'e}rin and {\"A}~Baumeler,  \emph{{Quantum computation
  with indefinite causal structures}},
  \href{https://doi.org/10.1103/PhysRevA.96.052315}{Phys. Rev. A {\bf 96},
  052315 (2017)} (Preprint
  {\href{https://arxiv.org/abs/1706.09854}{arXiv:1706.09854}})

\bibitem{Baumeler_2016}
{\" A}~Baumeler and S~Wolf,  \emph{{Computational tameness of classical
  non-causal models}},  \href{https://doi.org/10.1098/rspa.2017.0698}{Proc. R.
  Soc. Lond. {\bf 474}, 20170698 (2018)} (Preprint
  {\href{https://arxiv.org/abs/1611.05641}{arXiv:1611.05641}})

\bibitem{Araujo_2015}
M~Ara{\'u}jo, C~Branciard, F~Costa, A~Feix, C~Giarmatzi and {\v C}~Brukner,
  \emph{{Witnessing causal nonseparability}},
  \href{https://doi.org/10.1088/1367-2630/17/10/102001}{New J. Phys. {\bf 17},
  102001 (2015)} (Preprint
  {\href{https://arxiv.org/abs/1506.03776}{arXiv:1506.03776}})

\bibitem{Bavaresco_2019}
J~Bavaresco, M~Ara{\' u}jo, {\v{C}}~Brukner and M~T Quintino,
  \emph{{Semi-Device-Independent Certification of Indefinite Causal Order}},
  \href{https://doi.org/10.22331/q-2019-08-19-176}{Quantum {\bf 3}, 176 (2019)}
  (Preprint {\href{https://arxiv.org/abs/1903.10526}{arXiv:1903.10526}})

\bibitem{Milz_2022}
S~Milz, J~Bavaresco and G~Chiribella,  \emph{{Resource Theory of Causal
  Connection}},  \href{https://doi.org/10.22331/q-2022-08-25-788}{Quantum {\bf
  6}, 788 (2022)} (Preprint
  {\href{https://arxiv.org/abs/2110.03233}{arXiv:2110.03233}})

\bibitem{Costa_2016}
F~Costa and S~Shrapnel,  \emph{{Quantum causal modelling}},
  \href{https://doi.org/10.1088/1367-2630/18/6/063032}{New J. Phys. {\bf 18},
  063032 (2016)} (Preprint
  {\href{https://arxiv.org/abs/1512.07106}{arXiv:1512.07106}})

\bibitem{Taranto_2019_PRL}
P~Taranto, F~A Pollock, S~Milz, M~Tomamichel and K~Modi,  \emph{{Quantum Markov
  Order}},  \href{https://doi.org/10.1103/PhysRevLett.122.140401}{Phys. Rev.
  Lett. {\bf 122}, 140401 (2019)} (Preprint
  {\href{https://arxiv.org/abs/1805.11341}{arXiv:1805.11341}})

\bibitem{Taranto_2019_PRA}
P~Taranto, S~Milz, F~A Pollock and K~Modi,  \emph{{Structure of quantum
  stochastic processes with finite Markov order}},
  \href{https://doi.org/10.1103/PhysRevA.99.042108}{Phys. Rev. A {\bf 99},
  042108 (2019)} (Preprint
  {\href{https://arxiv.org/abs/1810.10809}{arXiv:1810.10809}})

\bibitem{Taranto_2021_npj}
P~Taranto, F~A Pollock and K~Modi,  \emph{{Non-Markovian memory strength bounds
  quantum process recoverability}},
  \href{https://doi.org/10.1038/s41534-021-00481-4}{npj Quantum Inf. {\bf 7},
  149 (2021)} (Preprint
  {\href{https://arxiv.org/abs/1907.12583}{arXiv:1907.12583}})

\bibitem{Kretschmann_2005}
D~Kretschmann and R~F Werner,  \emph{{Quantum channels with memory}},
  \href{https://doi.org/10.1103/PhysRevA.72.062323}{Phys. Rev. A {\bf 72},
  062323 (2005)} (Preprint
  {\href{https://arxiv.org/abs/quant-ph/0502106}{arXiv:quant-ph/0502106}})

\bibitem{Modi_2012_SciRep}
K~Modi,  \emph{{Operational approach to open dynamics and quantifying initial
  correlations}},  \href{https://doi.org/10.1038/srep00581}{Sci. Rep. {\bf 2},
  581 (2012)} (Preprint
  {\href{https://arxiv.org/abs/1011.6138}{arXiv:1011.6138}})

\bibitem{Modi_2012_PRA}
K~Modi, C~A Rodr{\' i}guez-Rosario and A~Aspuru-Guzik,  \emph{{Positivity in
  the presence of initial system-environment correlation}},
  \href{https://doi.org/10.1103/PhysRevA.86.064102}{Phys. Rev. A {\bf 86},
  064102 (2012)} (Preprint
  {\href{https://arxiv.org/abs/1203.5209}{arXiv:1203.5209}})

\bibitem{Ringbauer_2015}
M~Ringbauer, C~J Wood, K~Modi, A~Gilchrist, A~G White and A~Fedrizzi,
  \emph{{Characterizing Quantum Dynamics with Initial System-Environment
  Correlations}},  \href{https://doi.org/10.1103/PhysRevLett.114.090402}{Phys.
  Rev. Lett. {\bf 114}, 090402 (2015)} (Preprint
  {\href{https://arxiv.org/abs/1410.5826}{arXiv:1410.5826}})

\bibitem{Oreshkov_2016}
O~Oreshkov and C~Giarmatzi,  \emph{{Causal and causally separable processes}},
  \href{https://doi.org/10.1088/1367-2630/18/9/093020}{New J. Phys. {\bf 18},
  093020 (2016)} (Preprint
  {\href{https://arxiv.org/abs/1506.05449}{arXiv:1506.05449}})

\bibitem{Nielsen_Chuang_2010}
M~A Nielsen and I~L Chuang,
  {\href{https://doi.org/10.1017/CBO9780511976667}{\textit{{Quantum Computation
  and Quantum Information}}}} (Cambridge, UK: Cambridge University Press)
  (2000)

\bibitem{Stinespring_1955}
W~F Stinespring,  \emph{{Positive functions on $C^{\ast}$-algebras}},
  \href{https://doi.org/10.1090/S0002-9939-1955-0069403-4}{Proc. Am. Math. Soc.
  {\bf 6}, 211 (1955)}

\bibitem{Chiribella_2016}
G~Chiribella and D~Ebler,  \emph{{Optimal quantum networks and one-shot
  entropies}},  \href{https://doi.org/10.1088/1367-2630/18/9/093053}{New J.
  Phys. {\bf 18}, 093053 (2016)} (Preprint
  {\href{https://arxiv.org/abs/1606.02394}{arXiv:1606.02394}})

\bibitem{Miyazaki_2019}
J~Miyazaki, A~Soeda and M~Murao,  \emph{{Complex conjugation supermap of
  unitary quantum maps and its universal implementation protocol}},
  \href{https://doi.org/10.1103/PhysRevResearch.1.013007}{Phys. Rev. Res. {\bf
  1}, 013007 (2019)} (Preprint
  {\href{https://arxiv.org/abs/1706.03481}{arXiv:1706.03481}})

\bibitem{Ebler_2022}
D~Ebler, M~Horodecki, M~Marciniak, T~M{\l}ynik, M~T Quintino and M~Studzi{\'
  n}ski,  \emph{{Optimal Universal Quantum Circuits for Unitary Complex
  Conjugation}},  \href{https://doi.org/10.1109/TIT.2023.3263771}{IEEE Trans.
  Inf. Theory {\bf 69}, 5069 (2023)} (Preprint
  {\href{https://arxiv.org/abs/2206.00107}{arXiv:2206.00107}})

\bibitem{Thompson_2018}
J~Thompson, K~Modi, V~Vedral and M~Gu,  \emph{{Quantum plug n' play: modular
  computation in the quantum regime}},
  \href{https://doi.org/10.1088/1367-2630/aa99b3}{New J. Phys. {\bf 20}, 013004
  (2018)} (Preprint {\href{https://arxiv.org/abs/1310.2927}{arXiv:1310.2927}})

\bibitem{Yoshida_2022}
S~Yoshida, A~Soeda and M~Murao,  \emph{{Reversing Unknown Qubit-Unitary
  Operation, Deterministically and Exactly}},
  \href{https://doi.org/10.1103/PhysRevLett.131.120602}{Phys. Rev. Lett. {\bf
  131}, 120602 (2023)} (Preprint
  {\href{https://arxiv.org/abs/2209.02907}{arXiv:2209.02907}})

\bibitem{Yoshida_2023}
S~Yoshida, A~Soeda and M~Murao,  \emph{{Universal construction of decoders from
  encoding black boxes}},
  \href{https://doi.org/10.22331/q-2023-03-20-957}{Quantum {\bf 7}, 957 (2023)}
  (Preprint {\href{https://arxiv.org/abs/2110.00258}{arXiv:2110.00258}})

\bibitem{Yoshida_2024}
S~Yoshida, A~Soeda and M~Murao,  \emph{{Universal adjointation of isometry
  operations using transformation of quantum supermaps}} (2024) (Preprint
  {\href{https://arxiv.org/abs/2401.10137}{arXiv:2401.10137}})

\bibitem{Chen_2024}
Y-A Chen, Y~Mo, Y~Liu, L~Zhang and X~Wang,  \emph{{Quantum Advantage in
  Reversing Unknown Unitary Evolutions}} (2024) (Preprint
  {\href{https://arxiv.org/abs/2403.04704}{arXiv:2403.04704}})

\bibitem{Mo_2024}
Y~Mo, L~Zhang, Y-A Chen, Y~Liu, T~Lin and X~Wang,  \emph{{Parameterized quantum
  comb and simpler circuits for reversing unknown qubit-unitary operations}},
  \href{https://doi.org/10.1038/s41534-025-00979-1}{npj Quantum Inf. {\bf 11},
  32 (2025)} (Preprint
  {\href{https://arxiv.org/abs/2403.03761}{arXiv:2403.03761}})

\bibitem{Zhu_2024_Reversing}
C~Zhu, Y~Mo, Y-A Chen and X~Wang,  \emph{{Reversing Unknown Quantum Processes
  via Virtual Combs for Channels with Limited Information}},
  \href{https://doi.org/10.1103/PhysRevLett.133.030801}{Phys. Rev. Lett. {\bf
  133}, 030801 (2024)} (Preprint
  {\href{https://arxiv.org/abs/2401.04672}{arXiv:2401.04672}})

\bibitem{Bavaresco_2021}
J~Bavaresco, M~Murao and M~T Quintino,  \emph{{Strict Hierarchy between
  Parallel, Sequential, and Indefinite-Causal-Order Strategies for Channel
  Discrimination}},
  \href{https://doi.org/10.1103/PhysRevLett.127.200504}{Phys. Rev. Lett. {\bf
  127}, 200504 (2021)} (Preprint
  {\href{https://arxiv.org/abs/2011.08300}{arXiv:2011.08300}})

\bibitem{Bavaresco_2022}
J~Bavaresco, M~Murao and M~T Quintino,  \emph{{Unitary channel discrimination
  beyond group structures: Advantages of sequential and indefinite-causal-order
  strategies}},  \href{https://doi.org/10.1063/5.0075919}{J. Math. Phys. {\bf
  63}, 042203 (2022)} (Preprint
  {\href{https://arxiv.org/abs/2105.13369}{arXiv:2105.13369}})

\bibitem{Bisio_2009}
A~Bisio, G~Chiribella, G~M D'Ariano, S~Facchini and P~Perinotti,
  \emph{{Optimal Quantum Tomography}},
  \href{https://doi.org/10.1109/JSTQE.2009.2029243}{IEEE J. Sel. Top. Quantum
  Electron. {\bf 15}, 1646 (2009)} (Preprint
  {\href{https://arxiv.org/abs/1702.08751}{arXiv:1702.08751}})

\bibitem{Bisio_2010}
A~Bisio, G~Chiribella, G~M D'Ariano, S~Facchini and P~Perinotti,
  \emph{{Optimal quantum learning of a unitary transformation}},
  \href{https://doi.org/10.1103/PhysRevA.81.032324}{Phys. Rev. A {\bf 81},
  032324 (2010)} (Preprint
  {\href{https://arxiv.org/abs/0903.0543}{arXiv:0903.0543}})

\bibitem{Zhao_2024}
X~Zhao, X~Wang and G~Chiribella,  \emph{{Shadow Simulation of Quantum
  Processes}},  \href{https://doi.org/10.1103/PhysRevLett.133.120804}{Phys.
  Rev. Lett. {\bf 133}, 120804 (2024)} (Preprint
  {\href{https://arxiv.org/abs/2401.14934}{arXiv:2401.14934}})

\bibitem{Raza_2024}
A~Raza, M~C Caro, J~Eisert and S~Khatri,  \emph{{Online learning of quantum
  processes}} (2024) (Preprint
  {\href{https://arxiv.org/abs/2406.04250}{arXiv:2406.04250}})

\bibitem{Sedlak_2019}
M~Sedl{\'a}k, A~Bisio and M~Ziman,  \emph{{Optimal Probabilistic Storage and
  Retrieval of Unitary Channels}},
  \href{https://doi.org/10.1103/PhysRevLett.122.170502}{Phys. Rev. Lett. {\bf
  122}, 170502 (2019)} (Preprint
  {\href{https://arxiv.org/abs/1809.04552}{arXiv:1809.04552}})

\bibitem{Chen_2024_Hypothesis}
Y-A Chen, C~Zhu, K~He, Y~Liu and X~Wang,  \emph{{Hypothesis testing of symmetry
  in quantum dynamics}} (2024) (Preprint
  {\href{https://arxiv.org/abs/2411.14292}{arXiv:2411.14292}})

\bibitem{Zhu_2024_Optimal}
C~Zhu, S~He, Y-A Chen, L~Zhang and X~Wang,  \emph{{Optimal Hamiltonian
  recognition of unknown quantum dynamics}} (2024) (Preprint
  {\href{https://arxiv.org/abs/2412.13067}{arXiv:2412.13067}})

\bibitem{Feix_2017}
A~Feix and {\v C}~Brukner,  \emph{{Quantum superpositions of `common-cause' and
  `direct-cause' causal structures}},
  \href{https://doi.org/10.1088/1367-2630/aa9b1a}{New J. Phys. {\bf 19}, 123028
  (2017)} (Preprint
  {\href{https://arxiv.org/abs/1606.09241}{arXiv:1606.09241}})

\bibitem{Allen_2017}
J-M~A Allen, J~Barrett, D~C Horsman, C~M Lee and R~W Spekkens,  \emph{{Quantum
  Common Causes and Quantum Causal Models}},
  \href{https://doi.org/10.1103/PhysRevX.7.031021}{Phys. Rev. X {\bf 7}, 031021
  (2017)} (Preprint
  {\href{https://arxiv.org/abs/1609.09487}{arXiv:1609.09487}})

\bibitem{Guo_2021}
Y~Guo, P~Taranto, B-H Liu, X-M Hu, Y-F Huang, C-F Li and G-C Guo,
  \emph{{Experimental Demonstration of Instrument-Specific Quantum Memory
  Effects and Non-Markovian Process Recovery for Common-Cause Processes}},
  \href{https://doi.org/10.1103/PhysRevLett.126.230401}{Phys. Rev. Lett. {\bf
  126}, 230401 (2021)} (Preprint
  {\href{https://arxiv.org/abs/2003.14045}{arXiv:2003.14045}})

\bibitem{Nery_2021}
M~Nery, M~T Quintino, P~A Gu{\'{e}}rin, T~O Maciel and R~O Vianna,
  \emph{{Simple and maximally robust processes with no classical common-cause
  or direct-cause explanation}},
  \href{https://doi.org/10.22331/q-2021-09-09-538}{Quantum {\bf 5}, 538 (2021)}
  (Preprint {\href{https://arxiv.org/abs/2101.11630}{arXiv:2101.11630}})

\bibitem{Araujo_2017}
M~Ara{\'{u}}jo, A~Feix, M~Navascu{\'{e}}s and {\v{C}}~Brukner,  \emph{{A
  purification postulate for quantum mechanics with indefinite causal order}},
  \href{https://doi.org/10.22331/q-2017-04-26-10}{Quantum {\bf 1}, 10 (2017)}
  (Preprint {\href{https://arxiv.org/abs/1611.08535}{arXiv:1611.08535}})

\bibitem{Oreshkov_2019}
O~Oreshkov,  \emph{{Time-delocalized quantum subsystems and operations: on the
  existence of processes with indefinite causal structure in quantum
  mechanics}},  \href{https://doi.org/10.22331/q-2019-12-02-206}{Quantum {\bf
  3}, 206 (2019)} (Preprint
  {\href{https://arxiv.org/abs/1801.07594}{arXiv:1801.07594}})

\bibitem{Wechs_2021}
J~Wechs, H~Dourdent, A~A Abbott and C~Branciard,  \emph{{Quantum Circuits with
  Classical Versus Quantum Control of Causal Order}},
  \href{https://doi.org/10.1103/PRXQuantum.2.030335}{PRX Quantum {\bf 2},
  030335 (2021)} (Preprint
  {\href{https://arxiv.org/abs/2101.08796}{arXiv:2101.08796}})

\bibitem{wechs_existence_2023}
J~Wechs, C~Branciard and O~Oreshkov,  \emph{{Existence of processes violating
  causal inequalities on time-delocalised subsystems}},
  \href{https://doi.org/10.1038/s41467-023-36893-3}{Nat. Commun. {\bf 14}, 1471
  (2023)} (Preprint
  {\href{https://arxiv.org/abs/2201.11832}{arXiv:2201.11832}})

\bibitem{Dupuis_2020}
F~Dupuis, O~Fawzi and R~Renner,  \emph{{Entropy Accumulation}},
  \href{https://doi.org/10.1007/s00220-020-03839-5}{Commun. Math. Phys. {\bf
  379}, 867 (2020)} (Preprint
  {\href{https://arxiv.org/abs/1607.01796}{arXiv:1607.01796}})

\bibitem{Metger_2022}
T~Metger, O~Fawzi, D~Sutter and R~Renner,
  {\href{https://doi.org/10.1109/FOCS54457.2022.00085}{\textit{{Generalised
  entropy accumulation}}}}, {\em IEEE 63rd Annual Symposium on Foundations of
  Computer Science (FOCS)\/}, p 844 (2022) (Preprint
  {\href{https://arxiv.org/abs/2203.04989}{arXiv:2203.04989}})

\bibitem{Cotler_2019}
J~Cotler, X~Han, X-L Qi and Z~Yang,  \emph{{Quantum causal influence}},
  \href{https://doi.org/10.1007/JHEP07(2019)042}{J. High Energy Phys. {\bf
  2019}, 42 (2019)} (Preprint
  {\href{https://arxiv.org/abs/1811.05485}{arXiv:1811.05485}})

\bibitem{Barrett_2019}
J~Barrett, R~Lorenz and O~Oreshkov,  \emph{{Quantum Causal Models}} (2019)
  (Preprint {\href{https://arxiv.org/abs/1906.10726}{arXiv:1906.10726}})

\bibitem{Bisio_2011}
A~Bisio, G~Chiribella, G~M D'Ariano and P~Perinotti,  \emph{{Quantum Networks:
  general theory and applications}},
  \href{https://www.physics.sk/aps/pub.php?y=2011&pub=aps-11-03}{Acta Phys.
  Slovaca {\bf 61}, 273 (2011)} (Preprint
  {\href{https://arxiv.org/abs/1601.04864}{arXiv:1601.04864}})

\bibitem{Caruso_2014}
F~Caruso, V~Giovannetti, C~Lupo and S~Mancini,  \emph{{Quantum channels and
  memory effects}},  \href{https://doi.org/10.1103/RevModPhys.86.1203}{Rev.
  Mod. Phys. {\bf 86}, 1203 (2014)} (Preprint
  {\href{https://arxiv.org/abs/1207.5435}{arXiv:1207.5435}})

\bibitem{Hardy_2009}
L~Hardy, {\href{https://doi.org/10.1007/978-1-4020-9107-0_21}{\textit{{Quantum
  Gravity Computers: On the Theory of Computation with Indefinite Causal
  Structure}}}}, {\em {Quantum Reality, Relativistic Causality, and Closing the
  Epistemic Circle: Essays in Honour of Abner Shimony}\/}, ed W~C Myrvold and
  J~Christian (Dordrecht, Netherlands: Springer), p 379 (2009) (Preprint
  {\href{https://arxiv.org/abs/quant-ph/0701019}{arXiv:quant-ph/0701019}})

\bibitem{Hardy_2012}
L~Hardy,  \emph{{The operator tensor formulation of quantum theory}},
  \href{https://doi.org/10.1098/rsta.2011.0326}{Phil. Trans. R. Soc. A {\bf
  370}, 3385 (2012)} (Preprint
  {\href{https://arxiv.org/abs/1201.4390}{arXiv:1201.4390}})

\bibitem{Hardy_2015}
L~Hardy,  \emph{{Quantum theory with bold operator tensors}},
  \href{https://doi.org/10.1098/rsta.2014.0239}{Phil. Trans. R. Soc. A {\bf
  373}, 20140239 (2015)}

\bibitem{Perinotti_2017}
P~Perinotti,
  {\href{https://doi.org/10.1007/978-3-319-68655-4_7}{\textit{{Causal
  Structures and the Classification of Higher Order Quantum Computations}}}},
  {\em {Time in Physics}\/}, ed R~Renner and S~Stupar (Cham, Switzerland:
  Springer International Publishing), p 103 (2017) (Preprint
  {\href{https://arxiv.org/abs/1612.05099}{arXiv:1612.05099}})

\bibitem{Bisio_2019}
A~Bisio and P~Perinotti,  \emph{{Theoretical framework for higher-order quantum
  theory}},  \href{https://doi.org/10.1098/rspa.2018.0706}{Proc. R. Soc. A {\bf
  475}, 20180706 (2019)} (Preprint
  {\href{https://arxiv.org/abs/1806.09554}{arXiv:1806.09554}})

\bibitem{Barrett_2005}
J~Barrett,  \emph{{Information processing in generalized probabilistic
  theories}} (2005) (Preprint
  {\href{https://arxiv.org/abs/quant-ph/0508211}{arXiv:quant-ph/0508211}})

\bibitem{Masanes_2011}
L~Masanes and M~P M{\"u}ller,  \emph{{A derivation of quantum theory from
  physical requirements}},
  \href{https://doi.org/10.1088/1367-2630/13/6/063001}{New J. Phys. {\bf 13},
  063001 (2011)} (Preprint
  {\href{https://arxiv.org/abs/1004.1483}{arXiv:1004.1483}})

\bibitem{Deutsch_1991}
D~Deutsch,  \emph{{Quantum mechanics near closed timelike lines}},
  \href{https://doi.org/10.1103/PhysRevD.44.3197}{Phys. Rev. D {\bf 44}, 3197
  (1991)}

\bibitem{Bennett_2005}
C~H Bennett,
  \emph{\href{https://web.archive.org/web/20070206131550/https://www.research.ibm.com/people/b/bennetc/QUPONBshort.pdf}{Simulated
  Time Travel, Teleportation Without Communication, and How to Conduct a
  Romance with Someone who has Fallen into a Black Hole}} (2005)

\bibitem{Lloyd_2011_PRL}
S~Lloyd, L~Maccone, R~Garcia-Patron, V~Giovannetti, Y~Shikano, S~Pirandola, L~A
  Rozema, A~Darabi, Y~Soudagar, L~K Shalm and A~M Steinberg,  \emph{{Closed
  Timelike Curves via Postselection: Theory and Experimental Test of
  Consistency}},  \href{https://doi.org/10.1103/PhysRevLett.106.040403}{Phys.
  Rev. Lett. {\bf 106}, 040403 (2011)} (Preprint
  {\href{https://arxiv.org/abs/1005.2219}{arXiv:1005.2219}})

\bibitem{Lloyd_2011}
S~Lloyd, L~Maccone, R~Garcia-Patron, V~Giovannetti and Y~Shikano,
  \emph{{Quantum mechanics of time travel through post-selected
  teleportation}},  \href{https://doi.org/10.1103/PhysRevD.84.025007}{Phys.
  Rev. D {\bf 84}, 025007 (2011)} (Preprint
  {\href{https://arxiv.org/abs/1007.2615}{arXiv:1007.2615}})

\bibitem{Aharonov_1964}
Y~Aharonov, P~G Bergmann and J~L Lebowitz,  \emph{{Time Symmetry in the Quantum
  Process of Measurement}},
  \href{https://doi.org/10.1103/PhysRev.134.B1410}{Phys. Rev. {\bf 134}, 1410
  (1964)}

\bibitem{Aharonov_1990}
Y~Aharonov, J~Anandan, S~Popescu and L~Vaidman,  \emph{{Superpositions of time
  evolutions of a quantum system and a quantum time-translation machine}},
  \href{https://doi.org/10.1103/PhysRevLett.64.2965}{Phys. Rev. Lett. {\bf 64},
  2965 (1990)}

\bibitem{silva_connecting_2017}
R~Silva, Y~Guryanova, A~J Short, P~Skrzypczyk, N~Brunner and S~Popescu,
  \emph{Connecting processes with indefinite causal order and multi-time
  quantum states},  \href{https://doi.org/10.1088/1367-2630/aa84fe}{New J.
  Phys. {\bf 19}, 103022 (2017)} (Preprint
  {\href{https://arxiv.org/abs/1701.08638}{arXiv:1701.08638}})

\bibitem{Kraus_1983}
K~Kraus, {\textit{{States, Effects, and Operations: Fundamental Notions of
  Quantum Theory}}} (Berlin, Germany: Springer Berlin Heidelberg) (1983)

\bibitem{Choi_1975}
M-D Choi,  \emph{{Completely positive linear maps on complex matrices}},
  \href{https://doi.org/10.1016/0024-3795(75)90075-0}{Linear Algebra Its Appl.
  {\bf 10}, 285 (1975)}

\bibitem{PhysRev.121.920}
E~C~G Sudarshan, P~M Mathews and J~Rau,  \emph{{Stochastic Dynamics of
  Quantum-Mechanical Systems}},
  \href{https://doi.org/10.1103/PhysRev.121.920}{Phys. Rev. {\bf 121}, 920
  (1961)}

\bibitem{Lueders_2006}
G~L{\"u}ders,  \emph{{Concerning the state-change due to the measurement
  process}},  \href{https://doi.org/10.1002/andp.20065180904}{Ann. Phys. {\bf
  15}, 663 (2006)} [Original Text:
  \href{https://doi.org/10.1002/andp.19504430510}{Ann. Phys. \textbf{443}, 322
  (1950)}] (Preprint
  {\href{https://arxiv.org/abs/quant-ph/0403007}{arXiv:quant-ph/0403007}})

\bibitem{Jamiolkowski_1972}
A~Jamio{\l}kowski,  \emph{{Linear transformations which preserve trace and
  positive semidefiniteness of operators}},
  \href{https://doi.org/10.1016/0034-4877(72)90011-0}{Rep. Math. Phys. {\bf 3},
  275 (1972)}

\bibitem{Holevo_2010}
A~S Holevo,  \emph{{The Choi--Jamiolkowski forms of quantum Gaussian
  channels}},  \href{https://doi.org/10.1063/1.3581879}{J. Math. Phys. {\bf
  52}, 042202 (2011)} (Preprint
  {\href{https://arxiv.org/abs/1004.0196}{arXiv:1004.0196}})

\bibitem{Dariano_2018}
G~M D'Ariano,  \emph{{Causality re-established}},
  \href{https://doi.org/10.1098/rsta.2017.0313}{Phil. Trans. R. Soc. A {\bf
  376}, 20170313 (2018)} (Preprint
  {\href{https://arxiv.org/abs/1804.10810}{arXiv:1804.10810}})

\bibitem{chiribella_probabilistic_2010}
G~Chiribella, G~M D'Ariano and P~Perinotti,  \emph{{Probabilistic theories with
  purification}},  \href{https://doi.org/10.1103/PhysRevA.81.062348}{Phys. Rev.
  A {\bf 81}, 062348 (2010)} (Preprint
  {\href{https://arxiv.org/abs/0908.1583}{arXiv:0908.1583}})

\bibitem{Brukner_2014_NBTS}
{\v{C}}~Brukner,  \emph{{Talk given at 554. WE-Heraeus-Seminar conference
  Quantum Contextuality, Non-Locality, and the Foundations of Quantum
  Mechanics.}} (2014)

\bibitem{Guryanova_2019}
Y~Guryanova, R~Silva, A~J Short, P~Skrzypczyk, N~Brunner and S~Popescu,
  \emph{{Exploring the limits of no backwards in time signalling}},
  \href{https://doi.org/10.22331/q-2019-12-09-211}{Quantum {\bf 3}, 211 (2019)}
  (Preprint {\href{https://arxiv.org/abs/1708.00669}{arXiv:1708.00669}})

\bibitem{Milz_2017}
S~Milz, F~A Pollock and K~Modi,  \emph{{An introduction to operational quantum
  dynamics}},  \href{https://doi.org/10.1142/S1230161217400169}{Open Syst. Inf.
  Dyn. {\bf 24}, 1740016 (2017)} (Preprint
  {\href{https://arxiv.org/abs/1708.00769}{arXiv:1708.00769}})

\bibitem{Ziman_2008}
M~Ziman,  \emph{{Process positive-operator-valued measure: A mathematical
  framework for the description of process tomography experiments}},
  \href{https://doi.org/10.1103/PhysRevA.77.062112}{Phys. Rev. A {\bf 77},
  062112 (2008)} (Preprint
  {\href{https://arxiv.org/abs/0802.3862}{arXiv:0802.3862}})

\bibitem{milz_characterising_2024}
S~Milz and M~T Quintino,  \emph{{Characterising transformations between quantum
  objects, `completeness' of quantum properties, and transformations without a
  fixed causal order}},
  \href{https://doi.org/10.22331/q-2024-07-17-1415}{Quantum {\bf 8}, 1415
  (2024)} (Preprint
  {\href{https://arxiv.org/abs/2305.01247}{arXiv:2305.01247}})

\bibitem{Chiribella_2012}
G~Chiribella,  \emph{{Perfect discrimination of no-signalling channels via
  quantum superposition of causal structures}},
  \href{https://doi.org/10.1103/PhysRevA.86.040301}{Phys. Rev. A {\bf 86},
  040301 (2012)} (Preprint
  {\href{https://arxiv.org/abs/1109.5154}{arXiv:1109.5154}})

\bibitem{Hoffreumon_2021}
T~Hoffreumon and O~Oreshkov,  \emph{{The Multi-round Process Matrix}},
  \href{https://doi.org/10.22331/q-2021-01-20-384}{Quantum {\bf 5}, 384 (2021)}
  (Preprint {\href{https://arxiv.org/abs/2005.04204}{arXiv:2005.04204}})

\bibitem{simmons_higher-order_2022}
W~Simmons and A~Kissinger,  \emph{{Higher-order causal theories are models of
  BV-logic}} (2022) (Preprint
  {\href{https://arxiv.org/abs/2205.11219}{arXiv:2205.11219}})

\bibitem{ChiribellaReal2009}
G~Chiribella, G~M D'Ariano and P~Perinotti,  \emph{{Realization schemes for
  quantum instruments in finite dimensions}},
  \href{https://doi.org/10.1063/1.3105923}{J. Math. Phys. {\bf 50}, 042101
  (2009)} (Preprint {\href{https://arxiv.org/abs/0810.3211}{arXiv:0810.3211}})

\bibitem{Burniston_2020}
J~Burniston, M~Grabowecky, C~M Scandolo, G~Chiribella and G~Gour,
  \emph{{Necessary and Sufficient Conditions on Measurements of Quantum
  Channels}},  \href{https://doi.org/10.1098/rspa.2019.0832}{Proc. R. Soc. A
  {\bf 476}, 20190832 (2020)} (Preprint
  {\href{https://arxiv.org/abs/1904.09161}{arXiv:1904.09161}})

\bibitem{lmcs:4426}
A~Kissinger and S~Uijlen,  \emph{{A categorical semantics for causal
  structure}},  \href{https://doi.org/10.23638/LMCS-15(3:15)2019}{Log. Methods
  Comput. Sci. {\bf 15}, 15 (2019)} (Preprint
  {\href{https://arxiv.org/abs/1701.04732}{arXiv:1701.04732}})

\bibitem{hoffreumon_projective_2022}
T~Hoffreumon and O~Oreshkov,  \emph{{Projective characterization of
  higher-order quantum transformations}} (2022) (Preprint
  {\href{https://arxiv.org/abs/2206.06206}{arXiv:2206.06206}})

\bibitem{apadula2024}
L~Apadula, A~Bisio and P~Perinotti,  \emph{{No-signalling constrains quantum
  computation with indefinite causal structure}},
  \href{https://doi.org/10.22331/q-2024-02-05-1241}{Quantum {\bf 8}, 1241
  (2024)} (Preprint
  {\href{https://arxiv.org/abs/2202.10214}{arXiv:2202.10214}})

\bibitem{chiribella_quantum_2022}
G~Chiribella and Z~Liu,  \emph{{Quantum operations with indefinite time
  direction}},  \href{https://doi.org/10.1038/s42005-022-00967-3}{Commun. Phys.
  {\bf 5}, 1 (2022)} (Preprint
  {\href{https://arxiv.org/abs/2012.03859}{arXiv:2012.03859}})

\bibitem{theurer_quantifying_2019}
T~Theurer, D~Egloff, L~Zhang and M~B Plenio,  \emph{Quantifying {Operations}
  with an {Application} to {Coherence}},
  \href{https://doi.org/10.1103/PhysRevLett.122.190405}{Phys. Rev. Lett. {\bf
  122}, 190405 (2019)} (Preprint
  {\href{https://arxiv.org/abs/1806.07332}{arXiv:1806.07332}})

\bibitem{liu_resource_2019}
Z-W Liu and A~Winter,  \emph{Resource theories of quantum channels and the
  universal role of resource erasure} (2019) (Preprint
  {\href{https://arxiv.org/abs/1904.04201}{arXiv:1904.04201}})

\bibitem{gour_dynamical_2020}
G~Gour and C~M Scandolo,  \emph{{Dynamical Resources}} (2020) (Preprint
  {\href{https://arxiv.org/abs/2101.01552}{arXiv:2101.01552}})

\bibitem{gour_dynamical_2020a}
G~Gour and C~M Scandolo,  \emph{{Dynamical Entanglement}},
  \href{https://doi.org/10.1103/PhysRevLett.125.180505}{Phys. Rev. Lett. {\bf
  125}, 180505 (2020)} (Preprint
  {\href{https://arxiv.org/abs/2009.12304}{arXiv:2009.12304}})

\bibitem{gour_entanglement_2021}
G~Gour and C~M Scandolo,  \emph{{Entanglement of a bipartite channel}},
  \href{https://doi.org/10.1103/PhysRevA.103.062422}{Phys. Rev. A {\bf 103},
  062422 (2021)} (Preprint
  {\href{https://arxiv.org/abs/1907.02552}{arXiv:1907.02552}})

\bibitem{gour_inevitable_2024}
G~Gour, D~Kim, T~Nateeboon, G~Shemesh and G~Yoeli,  \emph{{Inevitable
  Negativity: Additivity Commands Negative Quantum Channel Entropy}} (2024)
  (Preprint {\href{https://arxiv.org/abs/2406.13823}{arXiv:2406.13823}})

\bibitem{Berk_2021}
G~D Berk, A~J~P Garner, B~Yadin, K~Modi and F~A Pollock,  \emph{{Resource
  theories of multi-time processes: A window into quantum non-Markovianity}},
  \href{https://doi.org/10.22331/q-2021-04-20-435}{Quantum {\bf 5}, 435 (2021)}
  (Preprint {\href{https://arxiv.org/abs/1907.07003}{arXiv:1907.07003}})

\bibitem{Wilson_2022_Mathematical}
M~Wilson and G~Chiribella,  \emph{{A Mathematical Framework for Transformations
  of Physical Processes}} (2022) (Preprint
  {\href{https://arxiv.org/abs/2204.04319}{arXiv:2204.04319}})

\bibitem{Wilson_2022_Locality}
M~Wilson, G~Chiribella and A~Kissinger,  \emph{{Quantum Supermaps are
  Characterized by Locality}} (2022) (Preprint
  {\href{https://arxiv.org/abs/2205.09844}{arXiv:2205.09844}})

\bibitem{Wilson_2022_Unitary}
M~Wilson and G~Chiribella,  \emph{{Free Polycategories for Unitary Supermaps of
  Arbitrary Dimension}} (2022) (Preprint
  {\href{https://arxiv.org/abs/2207.09180}{arXiv:2207.09180}})

\bibitem{Hefford_2023}
J~Hefford and C~Comfort,
  {\href{https://doi.org/10.4204/EPTCS.380.4}{\textit{{Coend Optics for Quantum
  Combs}}}}, {\em Proceedings 5th International Conference on Applied Category
  Theory (ACT)\/}, Electronic Proceedings in Theoretical Computer Science
  (EPTCS) (Waterloo NSW, Australia: Open Publishing Association) p~63 (2023)
  (Preprint {\href{https://arxiv.org/abs/2205.09027}{arXiv:2205.09027}})

\bibitem{Hefford_2024}
J~Hefford and M~Wilson,
  {\href{https://doi.org/10.1145/3661814.3662123}{\textit{{A Profunctorial
  Semantics for Quantum Supermaps}}}}, {\em Proceedings of the 39th Annual
  ACM/IEEE Symposium on Logic in Computer Science (LICS)\/}, (New York, NY,
  USA: Association for Computing Machinery) p~1 (2024) (Preprint
  {\href{https://arxiv.org/abs/2402.02997}{arXiv:2402.02997}})

\bibitem{jencova_structure_2024}
A~Jen{\v c}ov{\'a},  \emph{{On the structure of higher order quantum maps}}
  (2024) (Preprint {\href{https://arxiv.org/abs/2411.09256}{arXiv:2411.09256}})

\bibitem{Castro-Ruiz_2018}
E~Castro-Ruiz, F~Giacomini and {\v{C}}~Brukner,  \emph{{Dynamics of Quantum
  Causal Structures}},  \href{https://doi.org/10.1103/PhysRevX.8.011047}{Phys.
  Rev. X {\bf 8}, 011047 (2018)} (Preprint
  {\href{https://arxiv.org/abs/1710.03139}{arXiv:1710.03139}})

\bibitem{genkina_optimal_2012}
D~Genkina, G~Chiribella and L~Hardy,  \emph{{Optimal probabilistic simulation
  of quantum channels from the future to the past}},
  \href{https://doi.org/10.1103/PhysRevA.85.022330}{Phys. Rev. A {\bf 85},
  022330 (2012)} (Preprint
  {\href{https://arxiv.org/abs/1112.1469}{arXiv:1112.1469}})

\bibitem{oreshkov_operational_2016}
O~Oreshkov and N~J Cerf,  \emph{{Operational quantum theory without predefined
  time}},  \href{https://doi.org/10.1088/1367-2630/18/7/073037}{New J. Phys.
  {\bf 18}, 073037 (2016)} (Preprint
  {\href{https://arxiv.org/abs/1406.3829}{arXiv:1406.3829}})

\bibitem{Milz_2017_NJP}
S~Milz, F~A Pollock, T~P Le, G~Chiribella and K~Modi,  \emph{{Entanglement,
  non-Markovianity, and causal non-separability}},
  \href{https://doi.org/10.1088/1367-2630/aaafee}{New J. Phys. {\bf 20}, 033033
  (2018)} (Preprint
  {\href{https://arxiv.org/abs/1711.04065}{arXiv:1711.04065}})

\bibitem{Baumeler_2014_PefectSignalling}
{\" A}~Baumeler and S~Wolf,
  {\href{https://doi.org/10.1109/ISIT.2014.6874888}{\textit{{Perfect signaling
  among three parties violating predefined causal order}}}}, {\em 2014 IEEE
  International Symposium on Information Theory (ISITs)\/}, p 526 (2014)
  (Preprint {\href{https://arxiv.org/abs/1312.5916}{arXiv:1312.5916}})

\bibitem{abbott_multipartite_2016}
A~A Abbott, C~Giarmatzi, F~Costa and C~Branciard,  \emph{Multipartite causal
  correlations: {Polytopes} and inequalities},
  \href{https://doi.org/10.1103/PhysRevA.94.032131}{Phys. Rev. A {\bf 94},
  032131 (2016)} (Preprint
  {\href{https://arxiv.org/abs/1608.01528}{arXiv:1608.01528}})

\bibitem{Wechs2019MultipartiteCausality}
J~{Wechs}, A~A {Abbott} and C~{Branciard},  \emph{{On the definition and
  characterisation of multipartite causal (non)separability}},
  \href{https://doi.org/10.1088/1367-2630/aaf352}{New J. Phys. {\bf 21}, 013027
  (2019)} (Preprint
  {\href{https://arxiv.org/abs/1807.10557}{arXiv:1807.10557}})

\bibitem{klay_tensor_1987}
M~Kl{\"a}y, C~Randall and D~Foulis,  \emph{{Tensor products and probability
  weights}},  \href{https://doi.org/10.1007/BF00668911}{Int. J. Theor. Phys.
  {\bf 26}, 199 (1987)}

\bibitem{Barnum_2005}
H~Barnum, C~A Fuchs, J~M Renes and A~Wilce,  \emph{{Influence-free states on
  compound quantum systems}} (2005) (Preprint
  {\href{https://arxiv.org/abs/quant-ph/0507108}{arXiv:quant-ph/0507108}})

\bibitem{Gutoski09}
G~Gutoski,  \emph{{Properties of local quantum operations with shared
  entanglement}},
  \href{https://dl.acm.org/doi/abs/10.5555/2011804.2011806}{Quantum Inf.
  Comput. {\bf 9}, 739 (2009)} (Preprint
  {\href{https://arxiv.org/abs/0805.2209}{arXiv:0805.2209}})

\bibitem{Skrzypczyk_2023}
P~Skrzypczyk and D~Cavalcanti,
  {\href{https://doi.org/10.1088/978-0-7503-3343-6}{\textit{{Semidefinite
  Programming in Quantum Information Science}}}} (Bristol, UK: IOP Publishing)
  (2023) (Preprint \href{https://arxiv.org/abs/2306.11637}{arXiv:2306.11637})

\bibitem{Branciard_2016}
C~Branciard,  \emph{{Witnesses of causal nonseparability: an introduction and a
  few case studies}},  \href{https://doi.org/10.1038/srep26018}{Sci. Rep. {\bf
  6}, 26018 (2016)} (Preprint
  {\href{https://arxiv.org/abs/1603.00043}{arXiv:1603.00043}})

\bibitem{werner_quantum_1989}
R~F Werner,  \emph{Quantum states with {Einstein}-{Podolsky}-{Rosen}
  correlations admitting a hidden-variable model},
  \href{https://doi.org/10.1103/PhysRevA.40.4277}{Phys. Rev. A {\bf 40}, 4277
  (1989)}

\bibitem{baumeler_maximal_2014}
{\" A}~Baumeler, A~Feix and S~Wolf,  \emph{{Maximal incompatibility of locally
  classical behavior and global causal order in multiparty scenarios}},
  \href{https://doi.org/10.1103/PhysRevA.90.042106}{Phys. Rev. A {\bf 90},
  042106 (2014)} (Preprint
  {\href{https://arxiv.org/abs/1403.7333}{arXiv:1403.7333}})

\bibitem{branciard_simplest_2015}
C~Branciard, M~Ara{\'u}jo, A~Feix, F~Costa and {\v C}~Brukner,  \emph{{The
  simplest causal inequalities and their violation}},
  \href{https://doi.org/10.1088/1367-2630/18/1/013008}{New J. Phys. {\bf 18},
  013008 (2015)} (Preprint
  {\href{https://arxiv.org/abs/1508.01704}{arXiv:1508.01704}})

\bibitem{feix_causally_2016}
A~Feix, M~Ara{\'u}jo and {\v C}~Brukner,  \emph{{Causally nonseparable
  processes admitting a causal model}},
  \href{https://doi.org/10.1088/1367-2630/18/8/083040}{New J Phys. {\bf 18},
  083040 (2016)} (Preprint
  {\href{https://arxiv.org/abs/1604.03391}{arXiv:1604.03391}})

\bibitem{purves_quantum_2021}
T~Purves and A~J Short,  \emph{Quantum {Theory} {Cannot} {Violate} a {Causal}
  {Inequality}},  \href{https://doi.org/10.1103/PhysRevLett.127.110402}{Phys.
  Rev. Lett. {\bf 127}, 110402 (2021)} (Preprint
  {\href{https://arxiv.org/abs/2101.09107}{arXiv:2101.09107}})

\bibitem{Dourdent_2022}
H~Dourdent, A~A Abbott, N~Brunner, I~{{\v{S}}upi{\'c}} and C~Branciard,
  \emph{{Semi-Device-Independent Certification of Causal Nonseparability with
  Trusted Quantum Inputs}},
  \href{https://doi.org/10.1103/PhysRevLett.129.090402}{Phys. Rev. Lett. {\bf
  129}, 090402 (2022)} (Preprint
  {\href{https://arxiv.org/abs/2107.10877}{arXiv:2107.10877}})

\bibitem{Dourdent_2024}
H~Dourdent, A~A Abbott, I~{\v{S}}upi{\'{c}} and C~Branciard,
  \emph{Network-{D}evice-{I}ndependent {C}ertification of {C}ausal
  {N}onseparability},
  \href{https://doi.org/10.22331/q-2024-10-30-1514}{{Quantum} {\bf 8}, 1514
  (2024)} (Preprint
  {\href{https://arxiv.org/abs/2308.12760}{arXiv:2308.12760}})

\bibitem{Lugt_2023}
T~{van der Lugt}, J~Barrett and G~Chiribella,  \emph{{Device-independent
  certification of indefinite causal order in the quantum switch}},
  \href{https://doi.org/10.1038/s41467-023-40162-8}{Nat. Commun. {\bf 14}, 5811
  (2023)} (Preprint
  {\href{https://arxiv.org/abs/2208.00719}{arXiv:2208.00719}})

\bibitem{vanderLugt2024possibilistic}
T~{van der Lugt} and N~Ormrod,  \emph{{Possibilistic and maximal indefinite
  causal order in the quantum switch}},
  \href{https://doi.org/10.22331/q-2024-12-03-1543}{{Quantum} {\bf 8}, 1543
  (2024)} (Preprint
  {\href{https://arxiv.org/abs/2311.00557}{arXiv:2311.00557}})

\bibitem{araujo_computational_2014}
M~Ara{\'u}jo, F~Costa and {\v C}~Brukner,  \emph{Computational {Advantage} from
  {Quantum}-{Controlled} {Ordering} of {Gates}},
  \href{https://doi.org/10.1103/PhysRevLett.113.250402}{Phys. Rev. Lett. {\bf
  113}, 250402 (2014)} (Preprint
  {\href{https://arxiv.org/abs/1401.8127}{arXiv:1401.8127}})

\bibitem{procopio_experimental_2015}
L~M Procopio, A~Moqanaki, M~Ara{\'u}jo, F~Costa, I~Alonso~Calafell, E~G Dowd,
  D~R Hamel, L~A Rozema, {\v C}~Brukner and P~Walther,  \emph{{Experimental
  superposition of orders of quantum gates}},
  \href{https://doi.org/10.1038/ncomms8913}{Nat. Commun. {\bf 6}, 7913 (2015)}
  (Preprint {\href{https://arxiv.org/abs/1412.4006}{arXiv:1412.4006}})

\bibitem{goswami_indefinite_2018}
K~Goswami, C~Giarmatzi, M~Kewming, F~Costa, C~Branciard, J~Romero and A~G
  White,  \emph{Indefinite {Causal} {Order} in a {Quantum} {Switch}},
  \href{https://doi.org/10.1103/PhysRevLett.121.090503}{Phys. Rev. Lett. {\bf
  121}, 090503 (2018)} (Preprint
  {\href{https://arxiv.org/abs/1803.04302}{arXiv:1803.04302}})

\bibitem{rubino2017experimental}
G~Rubino, L~A Rozema, A~Feix, M~Ara{\'u}jo, J~M Zeuner, L~M Procopio,
  {\v{C}}~Brukner and P~Walther,  \emph{{Experimental verification of an
  indefinite causal order}},
  \href{https://doi.org/10.1126/sciadv.1602589}{Sci. Adv. {\bf 3}, 1602589
  (2017)} (Preprint
  {\href{https://arxiv.org/abs/1608.01683}{arXiv:1608.01683}})

\bibitem{Goswami_2020}
K~Goswami and J~Romero,  \emph{{Experiments on quantum causality}},
  \href{https://doi.org/10.1116/5.0010747}{AVS Quantum Sci. {\bf 2}, 037101
  (2020)} (Preprint
  {\href{https://arxiv.org/abs/2009.00515}{arXiv:2009.00515}})

\bibitem{Rozema_2024}
L~A Rozema, T~Str{\"o}mberg, H~Cao, Y~Guo, B-H Liu and P~Walther,
  \emph{{Experimental aspects of indefinite causal order in quantum
  mechanics}},  \href{https://doi.org/10.1038/s42254-024-00739-8}{Nat. Rev.
  Phys. {\bf 6}, 483 (2024)} (Preprint
  {\href{https://arxiv.org/abs/2405.00767}{arXiv:2405.00767}})

\bibitem{stromberg_experimental_2024}
T~Str{\" o}mberg, P~Schiansky, M~T Quintino, M~Antesberger, L~A Rozema,
  I~Agresti, {\v C}~Brukner and P~Walther,  \emph{{Experimental superposition
  of a quantum evolution with its time reverse}},
  \href{https://doi.org/10.1103/PhysRevResearch.6.023071}{Phys. Rev. Res. {\bf
  6}, 023071 (2024)} (Preprint
  {\href{https://arxiv.org/abs/2211.01283}{arXiv:2211.01283}})

\bibitem{Orus_2014}
R~Or{\' u}s,  \emph{{A Practical Introduction to Tensor Networks: Matrix
  Product States and Projected Entangled Pair States}},
  \href{https://doi.org/10.1016/j.aop.2014.06.013}{Ann. Phys. {\bf 349}, 117
  (2014)} (Preprint {\href{https://arxiv.org/abs/1306.2164}{arXiv:1306.2164}})

\bibitem{Cirac_2021}
J~I Cirac, D~P{\' e}rez-Garc{\' i}a, N~Schuch and F~Verstraete,  \emph{{Matrix
  Product States and Projected Entangled Pair States: Concepts, Symmetries,
  Theorems}},  \href{https://doi.org/10.1103/RevModPhys.93.045003}{Rev. Mod.
  Phys. {\bf 93}, 045003 (2021)} (Preprint
  {\href{https://arxiv.org/abs/2011.12127}{arXiv:2011.12127}})

\bibitem{boyer_tight_1998}
M~Boyer, G~Brassard, P~H{\o}yer and A~Tapp,  \emph{Tight {Bounds} on {Quantum}
  {Searching}},
  \href{https://doi.org/10.1002/(SICI)1521-3978(199806)46:4/5<493::AID-PROP493>3.0.CO;2-P}{Fortschr.
  Phys. {\bf 46}, 493 (1998)} (Preprint
  {\href{https://arxiv.org/abs/quant-ph/9605034}{arXiv:quant-ph/9605034}})

\bibitem{Thompson_2017}
J~Thompson, A~J~P Garner, V~Vedral and M~Gu,  \emph{{Using quantum theory to
  simplify input--output processes}},
  \href{https://doi.org/10.1038/s41534-016-0001-3}{npj Quantum Inf. {\bf 3}, 6
  (2017)} (Preprint
  {\href{https://arxiv.org/abs/1601.05420}{arXiv:1601.05420}})

\bibitem{Elliott_2021}
T~J Elliott, M~Gu, A~J~P Garner and J~Thompson,  \emph{{Quantum Adaptive Agents
  with Efficient Long-Term Memories}},
  \href{https://doi.org/10.1103/PhysRevX.12.011007}{Phys. Rev. X {\bf 12},
  011007 (2022)} (Preprint
  {\href{https://arxiv.org/abs/2108.10876}{arXiv:2108.10876}})

\bibitem{Xing_2023}
J~Xing, T~Feng, Z~Fan, H~Ma, K~Bharti, D~E Koh and Y~Xiao,  \emph{{Fundamental
  Limitations on Communication over a Quantum Network}} (2023) (Preprint
  {\href{https://arxiv.org/abs/2306.04983}{arXiv:2306.04983}})

\bibitem{Fitzsimons_2017_PRA}
J~F Fitzsimons and E~Kashefi,  \emph{{Unconditionally verifiable blind quantum
  computation}},  \href{https://doi.org/10.1103/PhysRevA.96.012303}{Phys. Rev.
  A {\bf 96}, 012303 (2017)} (Preprint
  {\href{https://arxiv.org/abs/1203.5217}{arXiv:1203.5217}})

\bibitem{Fitzsimons_2017}
J~F Fitzsimons,  \emph{{Private quantum computation: an introduction to blind
  quantum computing and related protocols}},
  \href{https://doi.org/10.1038/s41534-017-0025-3}{npj Quantum Inf. {\bf 3}, 23
  (2017)} (Preprint
  {\href{https://arxiv.org/abs/1611.10107}{arXiv:1611.10107}})

\bibitem{Smith_2023}
I~D Smith, M~Krumm, L~J Fiderer, H~P Nautrup and H~J Briegel,  \emph{The
  {M}in-{E}ntropy of {C}lassical-{Q}uantum {C}ombs for {M}easurement-{B}ased
  {A}pplications},  \href{https://doi.org/10.22331/q-2023-12-12-1206}{{Quantum}
  {\bf 7}, 1206 (2023)} (Preprint
  {\href{https://arxiv.org/abs/2212.00553}{arXiv:2212.00553}})

\bibitem{Selinger_2004}
P~Selinger,  \emph{{Towards a Quantum Programming Language}},
  \href{https://doi.org/10.1017/S0960129504004256}{Math. Struct. Comput. Sci.
  {\bf 14}, 527 (2004)}

\bibitem{Valiron_2005}
P~Selinger and B~Valiron,
  {\href{https://doi.org/10.1007/11417170_26}{\textit{{A Lambda Calculus for
  Quantum Computation with Classical Control}}}}, {\em {Typed Lambda Calculi
  and Applications}\/}, (Berlin, Germany: Springer), p 354 (2005) (Preprint
  {\href{https://arxiv.org/abs/cs/0404056}{arXiv:cs/0404056}})

\bibitem{Chiribella2008cloning}
G~Chiribella, G~M D'Ariano and P~Perinotti,  \emph{{Optimal Cloning of Unitary
  Transformation}},
  \href{https://doi.org/10.1103/PhysRevLett.101.180504}{Phys. Rev. Lett. {\bf
  101}, 180504 (2008)} (Preprint
  {\href{https://arxiv.org/abs/0804.0129}{arXiv:0804.0129}})

\bibitem{Chiribella2012replication}
G~Chiribella, Y~Yang and C~Huang,  \emph{{Universal Superreplication of Unitary
  Gates}},  \href{https://doi.org/10.1103/PhysRevLett.114.120504}{Phys. Rev.
  Lett. {\bf 114}, 120504 (2015)} (Preprint
  {\href{https://arxiv.org/abs/1412.1349}{arXiv:1412.1349}})

\bibitem{Bisio_2014}
A~Bisio, G~M D'Ariano, P~Perinotti and M~Sedl{\' a}k,  \emph{{Optimal
  Processing of Reversible Quantum Channels}},
  \href{https://doi.org/10.1016/j.physleta.2014.04.042}{Phys. Lett. A {\bf
  378}, 1797 (2014)} (Preprint
  {\href{https://arxiv.org/abs/1308.3254}{arXiv:1308.3254}})

\bibitem{Dur2015Replication}
W~D{\" u}r, P~Sekatski and M~Skotiniotis,  \emph{{Deterministic
  Superreplication of One-Parameter Unitary Transformations}},
  \href{https://doi.org/10.1103/PhysRevLett.114.120503}{Phys. Rev. Lett. {\bf
  114}, 120503 (2015)} (Preprint
  {\href{https://arxiv.org/abs/1410.6008}{arXiv:1410.6008}})

\bibitem{Dong_2019}
Q~Dong, S~Nakayama, A~Soeda and M~Murao,  \emph{{Controlled Quantum Operations
  and Combs, and Their Applications to Universal Controllization of Divisible
  Unitary Operations}} (2019) (Preprint
  {\href{https://arxiv.org/abs/1911.01645}{arXiv:1911.01645}})

\bibitem{Soleimanifar_2016}
M~Soleimanifar and V~Karimipour,  \emph{{No-Go Theorem for Iterations of
  Unknown Quantum Gates}},
  \href{https://doi.org/10.1103/PhysRevA.93.012344}{Phys. Rev. A {\bf 93},
  012344 (2016)} (Preprint
  {\href{https://arxiv.org/abs/1510.06888}{arXiv:1510.06888}})

\bibitem{Dong21Draw}
Q~Dong, M~T Quintino, A~Soeda and M~Murao,  \emph{{Success-or-Draw: A Strategy
  Allowing Repeat-Until-Success in Quantum Computation}},
  \href{https://doi.org/10.1103/PhysRevLett.126.150504}{Phys. Rev. Lett. {\bf
  126}, 150504 (2021)} (Preprint
  {\href{https://arxiv.org/abs/2011.01055}{arXiv:2011.01055}})

\bibitem{Yang2020Optimal}
Y~Yang, R~Renner and G~Chiribella,  \emph{{Optimal Universal Programming of
  Unitary Gates}},  \href{https://doi.org/10.1103/PhysRevLett.125.210501}{Phys.
  Rev. Lett. {\bf 125}, 210501 (2020)} (Preprint
  {\href{https://arxiv.org/abs/2007.10363}{arXiv:2007.10363}})

\bibitem{Ishizaka2008PBT}
S~Ishizaka and T~Hiroshima,  \emph{{Asymptotic Teleportation Scheme as a
  Universal Programmable Quantum Processor}},
  \href{https://doi.org/10.1103/PhysRevLett.101.240501}{Phys. Rev. Lett. {\bf
  101}, 240501 (2008)} (Preprint
  {\href{https://arxiv.org/abs/0807.4568}{arXiv:0807.4568}})

\bibitem{Ishizaka2009PBT}
S~Ishizaka and T~Hiroshima,  \emph{{Quantum Teleportation Scheme by Selecting
  One of Multiple Output Ports}},
  \href{https://doi.org/10.1103/PhysRevA.79.042306}{Phys. Rev. A {\bf 79},
  042306 (2009)} (Preprint
  {\href{https://arxiv.org/abs/0901.2975}{arXiv:0901.2975}})

\bibitem{Mozrzymas2018PBT}
M~Mozrzymas, M~Studzi{\' n}ski, S~Strelchuk and M~Horodecki,  \emph{{Optimal
  Port-Based Teleportation}},
  \href{https://doi.org/10.1088/1367-2630/aab8e7}{New J. Phys. {\bf 20}, 053006
  (2018)} (Preprint
  {\href{https://arxiv.org/abs/1707.08456}{arXiv:1707.08456}})

\bibitem{Chiribella2005Estimation}
G~Chiribella, G~M D'Ariano and M~F Sacchi,  \emph{{Optimal Estimation of Group
  Transformations Using Entanglement}},
  \href{https://doi.org/10.1103/PhysRevA.72.042338}{Phys. Rev. A {\bf 72},
  042338 (2005)} (Preprint
  {\href{https://arxiv.org/abs/quant-ph/0506267}{arXiv:quant-ph/0506267}})

\bibitem{Soeda2013controlisation}
A~Soeda,  \emph{{Limitations on Quantum Subroutine Designing Due to the Linear
  Structure of Quantum Operators}} (2013) presented at the 3rd International
  Conference on Quantum Information and Technology (ICQIT2013)

\bibitem{araujo14control}
M~Ara{\'u}jo, A~Feix, F~Costa and {\v{C}}~Brukner,  \emph{{Quantum circuits
  cannot control unknown operations}},
  \href{https://doi.org/10.1088/1367-2630/16/9/093026}{New J. Phys. {\bf 16},
  093026 (2014)} (Preprint
  {\href{https://arxiv.org/abs/1309.7976}{arXiv:1309.7976}})

\bibitem{arXiv:1006.2670}
X-Q Zhou, T~C Ralph, P~Kalasuwan, M~Zhang, A~Peruzzo, B~P Lanyon and J~L
  O'Brien,  \emph{{Adding control to arbitrary unknown quantum operations}},
  \href{https://doi.org/10.1038/ncomms1392}{Nat. Commun. {\bf 2}, 413 (2011)}
  (Preprint {\href{https://arxiv.org/abs/1006.2670}{arXiv:1006.2670}})

\bibitem{friis2014control}
N~Friis, V~Dunjko, W~D{\" u}r and H~J Briegel,  \emph{{Implementing Quantum
  Control for Unknown Subroutines}},
  \href{https://doi.org/10.1103/PhysRevA.89.030303}{Phys. Rev. A {\bf 89},
  030303 (2014)} (Preprint
  {\href{https://arxiv.org/abs/1401.8128}{arXiv:1401.8128}})

\bibitem{bisio2016conditional}
A~Bisio, M~Dall'Arno and P~Perinotti,  \emph{{Quantum Conditional Operations}},
   \href{https://doi.org/10.1103/PhysRevA.94.022340}{Phys. Rev. A {\bf 94},
  022340 (2016)} (Preprint
  {\href{https://arxiv.org/abs/1509.01062}{arXiv:1509.01062}})

\bibitem{Gavorova2024topological}
Z~Gavorov{\' a}, M~Seidel and Y~Touati,  \emph{{Topological Obstructions to
  Quantum Computation with Unitary Oracles}},
  \href{https://doi.org/10.1103/PhysRevA.109.032625}{Phys. Rev. A {\bf 109},
  032625 (2024)} (Preprint
  {\href{https://arxiv.org/abs/2011.10031}{arXiv:2011.10031}})

\bibitem{kitaev1995Control}
A~Y {Kitaev},  \emph{{Quantum measurements and the Abelian Stabilizer Problem}}
  (1995) (Preprint
  {\href{https://arxiv.org/abs/quant-ph/9511026}{arXiv:quant-ph/9511026}})

\bibitem{Nielsen_1997}
M~A Nielsen and I~L Chuang,  \emph{{Programmable Quantum Gate Arrays}},
  \href{https://doi.org/10.1103/PhysRevLett.79.321}{Phys. Rev. Lett. {\bf 79},
  321 (1997)} (Preprint
  {\href{https://arxiv.org/abs/quant-ph/9703032}{arXiv:quant-ph/9703032}})

\bibitem{sedlak2020phase}
M~Sedl{\' a}k and M~Ziman,  \emph{{Probabilistic Storage and Retrieval of Qubit
  Phase Gates}},  \href{https://doi.org/10.1103/PhysRevA.102.032618}{Phys. Rev.
  A {\bf 102}, 032618 (2020)} (Preprint
  {\href{https://arxiv.org/abs/2008.09555}{arXiv:2008.09555}})

\bibitem{Sedlak_2024}
M~Sedl{\'a}k, R~St{\'a}rek, N~Horov{\'a}, M~Mi{\v c}uda, J~Fiur{\'a}{\v s}ek
  and A~Bisio,  \emph{{Storage and retrieval of two unknown unitary channels}}
  (2024) (Preprint {\href{https://arxiv.org/abs/2410.23376}{arXiv:2410.23376}})

\bibitem{Grosshans_2024}
F~Grosshans, M~Horodecki, M~Murao, T~M{\l}ynik, M~T Quintino, M~Studzi{\' n}ski
  and S~Yoshida,  \emph{{Multicopy quantum state teleportation with application
  to storage and retrieval of quantum programs}} (2024) (Preprint
  {\href{https://arxiv.org/abs/2409.10393}{arXiv:2409.10393}})

\bibitem{Yoshida_2024_Estimation}
S~Yoshida, Y~Koizumi, M~Studzi{\' n}ski, M~T Quintino and M~Murao,
  \emph{{One-to-one Correspondence between Deterministic Port-Based
  Teleportation and Unitary Estimation}} (2024) (Preprint
  {\href{https://arxiv.org/abs/2408.11902}{arXiv:2408.11902}})

\bibitem{christandl2021asymptotic}
M~Christandl, F~Leditzky, C~Majenz, G~Smith, F~Speelman and M~Walter,
  \emph{{Asymptotic Performance of Port-Based Teleportation}},
  \href{https://doi.org/10.1007/s00220-020-03884-0}{Commun. Math. Phys. {\bf
  381}, 379 (2021)} (Preprint
  {\href{https://arxiv.org/abs/1809.10751}{arXiv:1809.10751}})

\bibitem{haah2023query}
J~Haah, R~Kothari, R~O'Donnell and E~Tang,
  {\href{https://doi.org/10.1109/FOCS57990.2023.00028}{\textit{{Query-Optimal
  Estimation of Unitary Channels in Diamond Distance}}}}, {\em IEEE 64th Annual
  Symposium on Foundations of Computer Science (FOCS)\/}, p 363 (2023)
  (Preprint {\href{https://arxiv.org/abs/2302.14066}{arXiv:2302.14066}})

\bibitem{bisio2011learningMeasurements}
A~Bisio, G~M D'Ariano, P~Perinotti and M~Sedl{\' a}k,  \emph{{Quantum Learning
  Algorithms for Quantum Measurements}},
  \href{https://doi.org/10.1016/j.physleta.2011.08.002}{Phys. Lett. A {\bf
  375}, 3425 (2011)} (Preprint
  {\href{https://arxiv.org/abs/1103.0480}{arXiv:1103.0480}})

\bibitem{lewandowska2022measurementSAR}
P~Lewandowska, R~Kukulski, {\L}~Pawela and Z~Pucha{\l}a,  \emph{{Storage and
  Retrieval of von Neumann Measurements}},
  \href{https://doi.org/10.1103/PhysRevA.106.052423}{Phys. Rev. A {\bf 106},
  052423 (2022)} (Preprint
  {\href{https://arxiv.org/abs/2204.03029}{arXiv:2204.03029}})

\bibitem{lewandowska2024SARmeasurement2}
P~Lewandowska and R~Kukulski,  \emph{{Storage and Retrieval of von Neumann
  Measurements via Indefinite Causal Order Structures}},
  \href{https://doi.org/10.1103/PhysRevA.110.042422}{Phys. Rev. A {\bf 110},
  042422 (2024)} (Preprint
  {\href{https://arxiv.org/abs/2405.11202}{arXiv:2405.11202}})

\bibitem{Navascues_2018}
M~Navascu{\'e}s,  \emph{{Resetting Uncontrolled Quantum Systems}},
  \href{https://doi.org/10.1103/PhysRevX.8.031008}{Phys. Rev. X {\bf 8}, 031008
  (2018)} (Preprint
  {\href{https://arxiv.org/abs/1710.02470}{arXiv:1710.02470}})

\bibitem{Trillo_2020}
D~Trillo, B~Dive and M~Navascu{\'e}s,  \emph{{Translating Uncontrolled Systems
  in Time}},  \href{https://doi.org/10.22331/q-2020-12-15-374}{Quantum {\bf 4},
  374 (2020)} (Preprint
  {\href{https://arxiv.org/abs/1903.10568}{arXiv:1903.10568}})

\bibitem{Trillo_2023}
D~Trillo, B~Dive and M~Navascu{\'e}s,  \emph{{Universal Quantum Rewinding
  Protocol with an Arbitrarily High Probability of Success}},
  \href{https://doi.org/10.1103/PhysRevLett.130.110201}{Phys. Rev. Lett. {\bf
  130}, 110201 (2023)} (Preprint
  {\href{https://arxiv.org/abs/2205.01131}{arXiv:2205.01131}})

\bibitem{Schiansky_2023}
P~Schiansky, T~Str{\"{o}}mberg, D~Trillo, V~Saggio, B~Dive, M~Navascu{\'{e}}s
  and P~Walther,  \emph{{Demonstration of universal time-reversal for qubit
  processes}},  \href{https://doi.org/10.1364/OPTICA.469109}{Optica {\bf 10},
  200 (2023)} (Preprint
  {\href{https://arxiv.org/abs/2205.01122}{arXiv:2205.01122}})

\bibitem{sardharwalla2016refocusing}
I~S~B Sardharwalla, T~S Cubitt, A~W Harrow and N~Linden,  \emph{{Universal
  Refocusing of Systematic Quantum Noise}} (2016) (Preprint
  {\href{https://arxiv.org/abs/1602.07963}{arXiv:1602.07963}})

\bibitem{Nakayama_2015}
S~Nakayama, A~Soeda and M~Murao,  \emph{{Quantum Algorithm for Universal
  Implementation of the Projective Measurement of Energy}},
  \href{https://doi.org/10.1103/PhysRevLett.114.190501}{Phys. Rev. Lett. {\bf
  114}, 190501 (2015)} (Preprint
  {\href{https://arxiv.org/abs/1310.3047}{arXiv:1310.3047}})

\bibitem{Odake_2023_Linear}
T~Odake, H~Kristj{\'a}nsson, A~Soeda and M~Murao,  \emph{{Higher-order quantum
  transformations of Hamiltonian dynamics}},
  \href{https://doi.org/10.1103/PhysRevResearch.6.L012063}{Phys. Rev. Res. {\bf
  6}, L012063 (2024)} (Preprint
  {\href{https://arxiv.org/abs/2303.09788}{arXiv:2303.09788}})

\bibitem{Odake_2023}
T~Odake, H~Kristj{\'a}nsson, P~Taranto and M~Murao,  \emph{{Universal algorithm
  for transforming Hamiltonian eigenvalues}} (2023) (Preprint
  {\href{https://arxiv.org/abs/2312.08848}{arXiv:2312.08848}})

\bibitem{Zhao_2024_Learning}
A~Zhao,  \emph{{Learning the structure of any Hamiltonian from minimal
  assumptions}} (2024) (Preprint
  {\href{https://arxiv.org/abs/2410.21635}{arXiv:2410.21635}})

\bibitem{Zhao_2020_Metrology}
X~Zhao, Y~Yang and G~Chiribella,  \emph{{Quantum Metrology with Indefinite
  Causal Order}},  \href{https://doi.org/10.1103/PhysRevLett.124.190503}{Phys.
  Rev. Lett. {\bf 124}, 190503 (2020)} (Preprint
  {\href{https://arxiv.org/abs/1912.02449}{arXiv:1912.02449}})

\bibitem{Altherr_2021}
A~Altherr and Y~Yang,  \emph{{Quantum Metrology for Non-Markovian Processes}},
  \href{https://doi.org/10.1103/PhysRevLett.127.060501}{Phys. Rev. Lett. {\bf
  127}, 060501 (2021)} (Preprint
  {\href{https://arxiv.org/abs/2103.02619}{arXiv:2103.02619}})

\bibitem{liu_optimal_2023}
Q~Liu, Z~Hu, H~Yuan and Y~Yang,  \emph{Optimal {Strategies} of {Quantum}
  {Metrology} with a {Strict} {Hierarchy}},
  \href{https://doi.org/10.1103/PhysRevLett.130.070803}{Phys. Rev. Lett. {\bf
  130}, 070803 (2023)} (Preprint
  {\href{https://arxiv.org/abs/2203.09758}{arXiv:2203.09758}})

\bibitem{Kurdzialek_2023}
S~Kurdzia{\l}ek, W~G{\'o}recki, F~Albarelli and R~Demkowicz-Dobrza{\'n}ski,
  \emph{{Using Adaptiveness and Causal Superpositions Against Noise in Quantum
  Metrology}},  \href{https://doi.org/10.1103/PhysRevLett.131.090801}{Phys.
  Rev. Lett. {\bf 131}, 090801 (2023)} (Preprint
  {\href{https://arxiv.org/abs/2212.08106}{arXiv:2212.08106}})

\bibitem{Mothe_2024}
R~Mothe, C~Branciard and A~A Abbott,  \emph{{Reassessing the advantage of
  indefinite causal orders for quantum metrology}},
  \href{https://doi.org/10.1103/PhysRevA.109.062435}{Phys. Rev. A {\bf 109},
  062435 (2024)} (Preprint
  {\href{https://arxiv.org/abs/2312.12172}{arXiv:2312.12172}})

\bibitem{Chiribella_2012_Optimal}
G~Chiribella,  \emph{{Optimal networks for quantum metrology: semidefinite
  programs and product rules}},
  \href{https://doi.org/10.1088/1367-2630/14/12/125008}{New J. Phys. {\bf 14},
  125008 (2012)} (Preprint
  {\href{https://arxiv.org/abs/1207.6172}{arXiv:1207.6172}})

\bibitem{Yang_2019_Memory}
Y~Yang,  \emph{{Memory Effects in Quantum Metrology}},
  \href{https://doi.org/10.1103/PhysRevLett.123.110501}{Phys. Rev. Lett. {\bf
  123}, 110501 (2019)} (Preprint
  {\href{https://arxiv.org/abs/1904.07267}{arXiv:1904.07267}})

\bibitem{Kurdzialek_2024}
S~Kurdzialek, P~Dulian, J~Majsak, S~Chakraborty and R~Demkowicz-Dobrzanski,
  \emph{{Quantum metrology using quantum combs and tensor network formalism}}
  (2024) (Preprint {\href{https://arxiv.org/abs/2403.04854}{arXiv:2403.04854}})

\bibitem{Liu_2024_Metrology}
Q~Liu, Z~Hu, H~Yuan and Y~Yang,  \emph{{Fully-Optimized Quantum Metrology:
  Framework, Tools, and Applications}},
  \href{https://doi.org/10.1002/qute.202400094}{Adv. Quantum Technol. {\bf 7},
  2400094 (2024)} (Preprint
  {\href{https://arxiv.org/abs/2409.07068}{arXiv:2409.07068}})

\bibitem{Meyer_2023}
J~J Meyer, S~Khatri, D~S Fran{\c c}a, J~Eisert and P~Faist,  \emph{{Quantum
  metrology in the finite-sample regime}} (2023) (Preprint
  {\href{https://arxiv.org/abs/2307.06370}{arXiv:2307.06370}})

\bibitem{Bavaresco_2024}
J~Bavaresco, P~Lipka-Bartosik, P~Sekatski and M~Mehboudi,  \emph{{Designing
  optimal protocols in Bayesian quantum parameter estimation with higher-order
  operations}},  \href{https://doi.org/10.1103/PhysRevResearch.6.023305}{Phys.
  Rev. Res. {\bf 6}, 023305 (2024)} (Preprint
  {\href{https://arxiv.org/abs/2311.01513}{arXiv:2311.01513}})

\bibitem{Wiebe_2014}
N~Wiebe, C~Granade, C~Ferrie and D~Cory,  \emph{{Quantum Hamiltonian learning
  using imperfect quantum resources}},
  \href{https://doi.org/10.1103/PhysRevA.89.042314}{Phys. Rev. A {\bf 89},
  042314 (2014)} (Preprint
  {\href{https://arxiv.org/abs/1311.5269}{arXiv:1311.5269}})

\bibitem{Huang_2023}
H-Y Huang, Y~Tong, D~Fang and Y~Su,  \emph{{Learning Many-Body Hamiltonians
  with Heisenberg-Limited Scaling}},
  \href{https://doi.org/10.1103/PhysRevLett.130.200403}{Phys. Rev. Lett. {\bf
  130}, 200403 (2023)} (Preprint
  {\href{https://arxiv.org/abs/2210.03030}{arXiv:2210.03030}})

\bibitem{Bakshi_2024}
A~Bakshi, A~Liu, A~Moitra and E~Tang,
  {\href{https://doi.org/10.1109/FOCS61266.2024.00069}{\textit{{Structure
  Learning of Hamiltonians from Real-Time Evolution}}}}, {\em IEEE 65th Annual
  Symposium on Foundations of Computer Science (FOCS)\/}, p 1037 (2024)
  (Preprint {\href{https://arxiv.org/abs/2405.00082}{arXiv:2405.00082}})

\bibitem{PhysRevA.65.050301}
C~Macchiavello and G~M Palma,  \emph{{Entanglement-enhanced information
  transmission over a quantum channel with correlated noise}},
  \href{https://doi.org/10.1103/PhysRevA.65.050301}{Phys. Rev. A {\bf 65},
  050301 (2002)} (Preprint
  {\href{https://arxiv.org/abs/quant-ph/0107052}{arXiv:quant-ph/0107052}})

\bibitem{Bowen_2004}
G~Bowen and S~Mancini,  \emph{{Quantum channels with a finite memory}},
  \href{https://doi.org/10.1103/PhysRevA.69.012306}{Phys. Rev. A {\bf 69},
  012306 (2004)} (Preprint
  {\href{https://arxiv.org/abs/quant-ph/0305010}{arXiv:quant-ph/0305010}})

\bibitem{PhysRevA.71.062304}
V~Giovannetti and S~Mancini,  \emph{{Bosonic memory channels}},
  \href{https://doi.org/10.1103/PhysRevA.71.062304}{Phys. Rev. A {\bf 71},
  062304 (2005)} (Preprint
  {\href{https://arxiv.org/abs/quant-ph/0410176}{arXiv:quant-ph/0410176}})

\bibitem{PhysRevLett.99.120504}
M~B Plenio and S~Virmani,  \emph{{Spin Chains and Channels with Memory}},
  \href{https://doi.org/10.1103/PhysRevLett.99.120504}{Phys. Rev. Lett. {\bf
  99}, 120504 (2007)} (Preprint
  {\href{https://arxiv.org/abs/quant-ph/0702059}{arXiv:quant-ph/0702059}})

\bibitem{Plenio_2008}
M~B Plenio and S~Virmani,  \emph{{Many-Body Physics and the Capacity of Quantum
  Channels with Memory}},
  \href{https://doi.org/10.1088/1367-2630/10/4/043032}{New J. Phys. {\bf 10},
  043032 (2008)} (Preprint
  {\href{https://arxiv.org/abs/0710.3299}{arXiv:0710.3299}})

\bibitem{Helstrom_1969}
C~W Helstrom,  \emph{{Quantum Detection and Estimation Theory}},
  \href{https://doi.org/10.1007/BF01007479}{J. Stat. Phys. {\bf 1}, 231 (1969)}

\bibitem{Chiribella_2008_ChannelDisc}
G~Chiribella, G~M D'Ariano and P~Perinotti,  \emph{{Memory Effects in Quantum
  Channel Discrimination}},
  \href{https://doi.org/10.1103/PhysRevLett.101.180501}{Phys. Rev. Lett. {\bf
  101}, 180501 (2008)} (Preprint
  {\href{https://arxiv.org/abs/0803.3237}{arXiv:0803.3237}})

\bibitem{hayashi05color}
A~Hayashi, T~Hashimoto and M~Horibe,  \emph{{Extended Quantum Color Coding}},
  \href{https://doi.org/10.1103/PhysRevA.71.012326}{Phys. Rev. A {\bf 71},
  012326 (2005)} (Preprint
  {\href{https://arxiv.org/abs/quant-ph/0409173}{arXiv:quant-ph/0409173}})

\bibitem{chiribella04covariant}
G~Chiribella, G~M D'Ariano, P~Perinotti and M~F Sacchi,  \emph{{Covariant
  Quantum Measurements That Maximize the Likelihood}},
  \href{https://doi.org/10.1103/PhysRevA.70.062105}{Phys. Rev. A {\bf 70},
  062105 (2004)} (Preprint
  {\href{https://arxiv.org/abs/quant-ph/0403083}{arXiv:quant-ph/0403083}})

\bibitem{chiribella06likelihood}
G~Chiribella, G~M D'Ariano, P~Perinotti and M~F Sacchi,  \emph{{Maximum
  Likelihood Estimation for a Group of Physical Transformations}},
  \href{https://doi.org/10.1142/S0219749906002018}{Int. J. Quantum Inf. {\bf
  4}, 453 (2006)} (Preprint
  {\href{https://arxiv.org/abs/quant-ph/0507007}{arXiv:quant-ph/0507007}})

\bibitem{Harrow_2010}
A~W Harrow, A~Hassidim, D~W Leung and J~Watrous,  \emph{{Adaptive versus
  Nonadaptive Strategies for Quantum Channel Discrimination}},
  \href{https://doi.org/10.1103/PhysRevA.81.032339}{Phys. Rev. A {\bf 81},
  032339 (2010)} (Preprint
  {\href{https://arxiv.org/abs/0909.0256}{arXiv:0909.0256}})

\bibitem{Katariya2021}
V~Katariya and M~M Wilde,  \emph{{Evaluating the Advantage of Adaptive
  Strategies for Quantum Channel Distinguishability}},
  \href{https://doi.org/10.1103/PhysRevA.104.052406}{Phys. Rev. A {\bf 104},
  052406 (2021)} (Preprint
  {\href{https://arxiv.org/abs/2001.05376}{arXiv:2001.05376}})

\bibitem{Pirandola_2019}
S~Pirandola, R~Laurenza, C~Lupo and J~L Pereira,  \emph{{Fundamental Limits to
  Quantum Channel Discrimination}},
  \href{https://doi.org/10.1038/s41534-019-0162-y}{npj Quantum Inf. {\bf 5}, 50
  (2019)} (Preprint
  {\href{https://arxiv.org/abs/1803.02834}{arXiv:1803.02834}})

\bibitem{Zhuang_2020}
Q~Zhuang and S~Pirandola,  \emph{{Ultimate Limits for Multiple Quantum Channel
  Discrimination}},
  \href{https://doi.org/10.1103/PhysRevLett.125.080505}{Phys. Rev. Lett. {\bf
  125}, 080505 (2020)} (Preprint
  {\href{https://arxiv.org/abs/2007.14566}{arXiv:2007.14566}})

\bibitem{Sedlak_2009}
M~Sedl{\'a}k and M~Ziman,  \emph{{Unambiguous comparison of unitary channels}},
   \href{https://doi.org/10.1103/PhysRevA.79.012303}{Phys. Rev. A {\bf 79},
  012303 (2009)} (Preprint
  {\href{https://arxiv.org/abs/0809.4401}{arXiv:0809.4401}})

\bibitem{Shimbo_2018}
A~Shimbo, A~Soeda and M~Murao,  \emph{{Equivalence determination of unitary
  operations}} (2018) (Preprint
  {\href{https://arxiv.org/abs/1803.11414}{arXiv:1803.11414}})

\bibitem{Soeda_2021}
A~Soeda, A~Shimbo and M~Murao,  \emph{{Optimal quantum discrimination of
  single-qubit unitary gates between two candidates}},
  \href{https://doi.org/10.1103/PhysRevA.104.022422}{Phys. Rev. A {\bf 104},
  022422 (2021)} (Preprint
  {\href{https://arxiv.org/abs/2103.08208}{arXiv:2103.08208}})

\bibitem{Hashimoto_2022}
Y~Hashimoto, A~Soeda and M~Murao,  \emph{{Comparison of unknown unitary
  channels with multiple uses}} (2022) (Preprint
  {\href{https://arxiv.org/abs/2208.12519}{arXiv:2208.12519}})

\bibitem{Chiribella_2019speedup}
G~Chiribella and D~Ebler,  \emph{{Quantum speedup in the identification of
  cause–effect relations}},
  \href{https://doi.org/10.1038/s41467-019-09383-8}{Nat. Commun. {\bf 10}, 1472
  (2019)} (Preprint
  {\href{https://arxiv.org/abs/1806.06459}{arXiv:1806.06459}})

\bibitem{Lewandowska_2023Strategies}
P~Lewandowska, {\L}~Pawela and Z~Pucha{\l}a,  \emph{{Strategies for single-shot
  discrimination of process matrices}},
  \href{https://doi.org/10.1038/s41598-023-30191-0}{Sci. Rep. {\bf 13}, 3046
  (2023)} (Preprint
  {\href{https://arxiv.org/abs/2210.14575}{arXiv:2210.14575}})

\bibitem{Chitambar_2019}
E~Chitambar and G~Gour,  \emph{{Quantum resource theories}},
  \href{https://doi.org/10.1103/RevModPhys.91.025001}{Rev. Mod. Phys. {\bf 91},
  025001 (2019)} (Preprint
  {\href{https://arxiv.org/abs/1806.06107}{arXiv:1806.06107}})

\bibitem{Selby_2020}
J~H Selby, A~B Sainz and P~Horodecki,  \emph{{Revisiting Dynamics of Quantum
  Causal Structures---When Can Causal Order Evolve?}},
  \href{https://doi.org/10.3390/e26080643}{Entropy {\bf 26}, 643 (2024)}
  (Preprint {\href{https://arxiv.org/abs/2008.12757}{arXiv:2008.12757}})

\bibitem{arenz_distinguishing_2015}
C~Arenz, R~Hillier, M~Fraas and D~Burgarth,  \emph{{Distinguishing Decoherence
  from Alternative Quantum Theories by Dynamical Decoupling}},
  \href{https://doi.org/10.1103/PhysRevA.92.022102}{Phys. Rev. A {\bf 92},
  022102 (2015)} (Preprint
  {\href{https://arxiv.org/abs/1405.7644}{arXiv:1405.7644}})

\bibitem{Milz_2019_CPDiv}
S~Milz, M~S Kim, F~A Pollock and K~Modi,  \emph{{Completely Positive
  Divisibility Does Not Mean Markovianity}},
  \href{https://doi.org/10.1103/PhysRevLett.123.040401}{Phys. Rev. Lett. {\bf
  123}, 040401 (2019)} (Preprint
  {\href{https://arxiv.org/abs/1901.05223}{arXiv:1901.05223}})

\bibitem{Carmichael}
H~Carmichael, {\href{https://doi.org/10.1007/978-3-540-47620-7}{\textit{An
  {Open} {Systems} {Approach} to {Quantum} {Optics}}}} (Berlin, Germany:
  Springer-Verlag) (1993)

\bibitem{BreuerPetruccione}
H-P Breuer and F~Petruccione,
  {\href{https://doi.org/10.1093/acprof:oso/9780199213900.001.0001}{\textit{{The
  Theory of Open Quantum Systems}}}} (Oxford, UK: Oxford University Press)
  (2002)

\bibitem{gardiner_quantum_2010}
C~W Gardiner and P~Zoller, {\textit{{Quantum Noise: A Handbook of Markovian and
  Non-Markovian Quantum Stochastic Methods with Applications to Quantum
  Optics}}} (Berlin, Germany: Springer Berlin Heidelberg) (2010)

\bibitem{rivas_quantum_2014}
{\' A}~Rivas, S~F Huelga and M~B Plenio,  \emph{{Quantum Non-Markovianity:
  Characterization, Quantification and Detection}},
  \href{https://doi.org/10.1088/0034-4885/77/9/094001}{Rep. Prog. Phys. {\bf
  77}, 094001 (2014)} (Preprint
  {\href{https://arxiv.org/abs/1405.0303}{arXiv:1405.0303}})

\bibitem{Breuer2016}
H-P Breuer, E-M Laine, J~Piilo and B~Vacchini,  \emph{{Colloquium:
  Non-Markovian Dynamics in Open Quantum Systems}},
  \href{https://doi.org/10.1103/RevModPhys.88.021002}{Rev. Mod. Phys. {\bf 88},
  021002 (2016)} (Preprint
  {\href{https://arxiv.org/abs/1505.01385}{arXiv:1505.01385}})

\bibitem{PRXQuantum.5.020202}
F~Campaioli, J~H Cole and H~Hapuarachchi,  \emph{{Quantum Master Equations:
  Tips and Tricks for Quantum Optics, Quantum Computing, and Beyond}},
  \href{https://doi.org/10.1103/PRXQuantum.5.020202}{PRX Quantum {\bf 5},
  020202 (2024)} (Preprint
  {\href{https://arxiv.org/abs/2303.16449}{arXiv:2303.16449}})

\bibitem{Li_2018}
L~Li, M~J~W Hall and H~M Wiseman,  \emph{{Concepts of Quantum Non-Markovianity:
  A Hierarchy}},  \href{https://doi.org/10.1016/j.physrep.2018.07.001}{Phys.
  Rep. {\bf 759}, 1 (2018)} (Preprint
  {\href{https://arxiv.org/abs/1712.08879}{arXiv:1712.08879}})

\bibitem{quantumstochasticprocessesandquantumnon}
S~Milz and K~Modi,  \emph{{Quantum Stochastic Processes and Quantum
  Non-Markovian Phenomena}},
  \href{https://doi.org/10.1103/PRXQuantum.2.030201}{PRX Quantum {\bf 2},
  030201 (2021)} (Preprint
  {\href{https://arxiv.org/abs/2012.01894}{arXiv:2012.01894}})

\bibitem{modi_role_2010}
K~Modi and E~C~G Sudarshan,  \emph{{Role of Preparation in Quantum Process
  Tomography}},  \href{https://doi.org/10.1103/PhysRevA.81.052119}{Phys. Rev. A
  {\bf 81}, 052119 (2010)} (Preprint
  {\href{https://arxiv.org/abs/0904.4663}{arXiv:0904.4663}})

\bibitem{modi_preparation_2011}
K~Modi,  \emph{{Preparation of States in Open Quantum Mechanics}},
  \href{https://doi.org/10.1142/S1230161211000170}{Open Syst. Inf. Dyn. {\bf
  18}, 253 (2011)} (Preprint
  {\href{https://arxiv.org/abs/0903.2027}{arXiv:0903.2027}})

\bibitem{Pechukas_1994}
P~Pechukas,  \emph{{Reduced Dynamics Need Not Be Completely Positive}},
  \href{https://doi.org/10.1103/PhysRevLett.73.1060}{Phys. Rev. Lett. {\bf 73},
  1060 (1994)}

\bibitem{Alicki_1995}
R~Alicki,  \emph{{Comment on ``Reduced Dynamics Need Not Be Completely
  Positive''}},  \href{https://doi.org/10.1103/PhysRevLett.75.3020}{Phys. Rev.
  Lett. {\bf 75}, 3020 (1995)}

\bibitem{Pechukas_1995}
P~Pechukas,  \emph{{Pechukas Replies:}},
  \href{https://doi.org/10.1103/PhysRevLett.75.3021}{Phys. Rev. Lett. {\bf 75},
  3021 (1995)}

\bibitem{obrien_quantum_2004}
J~L O'Brien, G~J Pryde, A~Gilchrist, D~F~V James, N~K Langford, T~C Ralph and
  A~G White,  \emph{{Quantum Process Tomography of a Controlled-NOT Gate}},
  \href{https://doi.org/10.1103/PhysRevLett.93.080502}{Phys. Rev. Lett. {\bf
  93}, 080502 (2004)} (Preprint
  {\href{https://arxiv.org/abs/quant-ph/0402166}{arXiv:quant-ph/0402166}})

\bibitem{Weinstein2004}
Y~S Weinstein, T~F Havel, J~Emerson, N~Boulant, M~Saraceno, S~Lloyd and D~G
  Cory,  \emph{{Quantum Process Tomography of the Quantum Fourier Transform}},
  \href{https://doi.org/10.1063/1.1785151}{J. Chem. Phys. {\bf 121}, 6117
  (2004)} (Preprint
  {\href{https://arxiv.org/abs/quant-ph/0406239}{arXiv:quant-ph/0406239}})

\bibitem{myrskog_quantum_2005}
S~H Myrskog, J~K Fox, M~W Mitchell and A~M Steinberg,  \emph{{Quantum Process
  Tomography on Vibrational States of Atoms in an Optical Lattice}},
  \href{https://doi.org/10.1103/PhysRevA.72.013615}{Phys. Rev. A {\bf 72},
  013615 (2005)} (Preprint
  {\href{https://arxiv.org/abs/quant-ph/0312210}{arXiv:quant-ph/0312210}})

\bibitem{kavanthesis}
K~Modi, {\href{https://arxiv.org/abs/0903.2724}{\textit{{A Theoretical Analysis
  of Experimental Open Quantum Dynamics}}}}, {PhD Thesis}, The University of
  Texas at Austin (2008)

\bibitem{stelmachovic_dynamics_2001}
P~{\v S}telamchovi{\v c} and V~Bu{\v z}ek,  \emph{{Dynamics of Open Quantum
  Systems Initially Entangled with Environment: Beyond the Kraus
  Representation}},  \href{https://doi.org/10.1103/PhysRevA.64.062106}{Phys.
  Rev. A {\bf 64}, 062106 (2001)} (Preprint
  {\href{https://arxiv.org/abs/quant-ph/0108136}{arXiv:quant-ph/0108136}})

\bibitem{Jordan_2004}
T~F Jordan, A~Shaji and E~C~G Sudarshan,  \emph{{Dynamics of initially
  entangled open quantum systems}},
  \href{https://doi.org/10.1103/PhysRevA.70.052110}{Phys. Rev. A {\bf 70},
  052110 (2004)} (Preprint
  {\href{https://arxiv.org/abs/quant-ph/0407083}{arXiv:quant-ph/0407083}})

\bibitem{shaji_whos_2005}
A~Shaji and E~C~G Sudarshan,  \emph{{Who's afraid of not completely positive
  maps?}},  \href{https://doi.org/10.1016/j.physleta.2005.04.029}{Phys. Lett. A
  {\bf 341}, 48 (2005)}

\bibitem{Jordan_2006}
T~F Jordan, A~Shaji and E~C~G Sudarshan,  \emph{{Mapping the Schr{\"o}dinger
  picture of open quantum dynamics}},
  \href{https://doi.org/10.1103/PhysRevA.73.012106}{Phys. Rev. A {\bf 73},
  012106 (2006)} (Preprint
  {\href{https://arxiv.org/abs/quant-ph/0505123}{arXiv:quant-ph/0505123}})

\bibitem{carteret_dynamics_2008}
H~A Carteret, D~R Terno and K~{\.Z}yczkowski,  \emph{{Dynamics beyond
  completely positive maps: Some properties and applications}},
  \href{https://doi.org/10.1103/PhysRevA.77.042113}{Phys. Rev. A {\bf 77},
  042113 (2008)} (Preprint
  {\href{https://arxiv.org/abs/quant-ph/0512167}{arXiv:quant-ph/0512167}})

\bibitem{vacchini_reduced_2016}
B~Vacchini and G~Amato,  \emph{{Reduced dynamical maps in the presence of
  initial correlations}},  \href{https://doi.org/10.1038/srep37328}{Sci. Rep.
  {\bf 6}, 37328 (2016)} (Preprint
  {\href{https://arxiv.org/abs/1605.04159}{arXiv:1605.04159}})

\bibitem{shajithesis}
A~Shaji,
  {\href{https://repositories.lib.utexas.edu/handle/2152/1715}{\textit{{Dynamics
  of Initially Entangled Open Quantum Systems}}}}, {PhD Thesis}, The University
  of Texas at Austin (2005)

\bibitem{RodriguezRosario_2008}
C~A Rodr{\'{\i}}guez-Rosario, K~Modi, A~Kuah, A~Shaji and E~C~G Sudarshan,
  \emph{{Completely positive maps and classical correlations}},
  \href{https://doi.org/10.1088/1751-8113/41/20/205301}{J. Phys. A {\bf 41},
  205301 (2008)} (Preprint
  {\href{https://arxiv.org/abs/quant-ph/0703022}{arXiv:quant-ph/0703022}})

\bibitem{RodriguezRosario_2010}
C~A Rodr{\' i}guez-Rosario, K~Modi and A~Aspuru-Guzik,  \emph{{Linear
  assignment maps for correlated system-environment states}},
  \href{https://doi.org/10.1103/PhysRevA.81.012313}{Phys. Rev. A {\bf 81},
  012313 (2010)} (Preprint
  {\href{https://arxiv.org/abs/0910.5568}{arXiv:0910.5568}})

\bibitem{brodutch_vanishing_2013}
A~Brodutch, A~Datta, K~Modi, {\'{A}}~Rivas and C~A Rodr{\'{i}}guez-Rosario,
  \emph{{Vanishing quantum discord is not necessary for completely positive
  maps}},  \href{https://doi.org/10.1103/PhysRevA.87.042301}{Phys. Rev. A {\bf
  87}, 042301 (2013)} (Preprint
  {\href{https://arxiv.org/abs/1212.4387}{arXiv:1212.4387}})

\bibitem{Vacchini_2011}
B~Vacchini, A~Smirne, E-M Laine, J~Piilo and H-P Breuer,  \emph{{Markovianity
  and non-Markovianity in quantum and classical systems}},
  \href{https://doi.org/10.1088/1367-2630/13/9/093004}{New J. Phys. {\bf 13},
  093004 (2011)} (Preprint
  {\href{https://arxiv.org/abs/1106.0138}{arXiv:1106.0138}})

\bibitem{Chrusinski_2011}
D~Chru{\'{s}}ci{\'{n}}ski, A~Kossakowski and {\'{A}}~Rivas,  \emph{{Measures of
  non-Markovianity: Divisibility versus backflow of information}},
  \href{https://doi.org/10.1103/PhysRevA.83.052128}{Phys. Rev. A {\bf 83},
  052128 (2011)} (Preprint
  {\href{https://arxiv.org/abs/1102.4318}{arXiv:1102.4318}})

\bibitem{Luchnikov_2019}
I~A Luchnikov, S~V Vintskevich, H~Ouerdane and S~N Filippov,  \emph{{Simulation
  Complexity of Open Quantum Dynamics: Connection with Tensor Networks}},
  \href{https://doi.org/10.1103/PhysRevLett.122.160401}{Phys. Rev. Lett. {\bf
  122}, 160401 (2019)} (Preprint
  {\href{https://arxiv.org/abs/1812.00043}{arXiv:1812.00043}})

\bibitem{Giarmatzi_2021}
C~Giarmatzi and F~Costa,  \emph{{Witnessing quantum memory in non-Markovian
  processes}},  \href{https://doi.org/10.22331/q-2021-04-26-440}{Quantum {\bf
  5}, 440 (2021)} (Preprint
  {\href{https://arxiv.org/abs/1811.03722}{arXiv:1811.03722}})

\bibitem{milz-spee-ent_2021}
S~Milz, C~Spee, Z-P Xu, F~A Pollock, K~Modi and O~G{\" u}hne,  \emph{{Genuine
  multipartite entanglement in time}},
  \href{https://doi.org/10.21468/SciPostPhys.10.6.141}{SciPost Phys. {\bf 10},
  141 (2021)} (Preprint
  {\href{https://arxiv.org/abs/2011.09340}{arXiv:2011.09340}})

\bibitem{Aloisio_2023}
I~Aloisio, G~White, C~Hill and K~Modi,  \emph{{Sampling Complexity of Open
  Quantum Systems}},  \href{https://doi.org/10.1103/PRXQuantum.4.020310}{PRX
  Quantum {\bf 4}, 020310 (2023)} (Preprint
  {\href{https://arxiv.org/abs/2209.10870}{arXiv:2209.10870}})

\bibitem{Dowling_2024}
N~Dowling, K~Modi, R~N Mu{\~ n}oz, S~Singh and G~A~L White,  \emph{{Capturing
  Long-Range Memory Structures with Tree-Geometry Process Tensors}},
  \href{https://doi.org/10.1103/PhysRevX.14.041018}{Phys. Rev. X {\bf 14},
  041018 (2024)} (Preprint
  {\href{https://arxiv.org/abs/2312.04624}{arXiv:2312.04624}})

\bibitem{Burgarth_2021}
D~Burgarth, P~Facchi, D~Lonigro and K~Modi,  \emph{{Quantum non-Markovianity
  elusive to interventions}},
  \href{https://doi.org/10.1103/PhysRevA.104.L050404}{Phys. Rev. A {\bf 104},
  L050404 (2021)} (Preprint
  {\href{https://arxiv.org/abs/2108.05750}{arXiv:2108.05750}})

\bibitem{Lindblad_1979}
G~Lindblad,  \emph{{Non-Markovian quantum stochastic processes and their
  entropy}},
  \href{https://projecteuclid.org/journals/communications-in-mathematical-physics/volume-65/issue-3/Non-Markovian-quantum-stochastic-processes-and-their-entropy/cmp/1103904877.full}{Commun.
  Math. Phys. {\bf 65}, 281 (1979)}

\bibitem{accardi_quantum_1982}
L~Accardi, A~Frigerio and J~T Lewis,  \emph{{Quantum Stochastic Processes}},
  \href{https://doi.org/10.2977/prims/1195184017}{Publ. Res. Inst. Math. Sci.
  {\bf 18}, 97 (1982)}

\bibitem{9683765}
H~I Nurdin and J~Gough,
  {\href{https://doi.org/10.1109/CDC45484.2021.9683765}{\textit{{From the
  Heisenberg to the Schr{\" o}dinger Picture: Quantum Stochastic Processes and
  Process Tensors}}}}, {\em 60th IEEE Conference on Decision and Control
  (CDC)\/}, p 4164 (2021) (Preprint
  {\href{https://arxiv.org/abs/2109.09256}{arXiv:2109.09256}})

\bibitem{capela_quantum_2022}
M~Capela, L~C C{\'{e}}leri, R~Chaves and K~Modi,  \emph{{Quantum Markov
  monogamy inequalities}},
  \href{https://doi.org/10.1103/PhysRevA.106.022218}{Phys. Rev. A {\bf 106},
  022218 (2022)} (Preprint
  {\href{https://arxiv.org/abs/2108.11533}{arXiv:2108.11533}})

\bibitem{Taranto_Thesis}
P~Taranto,  \emph{{Memory effects in quantum processes}},
  \href{https://doi.org/10.1142/S0219749919410028}{Int. J. Quantum Inf. {\bf
  18}, 1941002 (2020)} (Preprint
  {\href{https://arxiv.org/abs/1909.05245}{arXiv:1909.05245}})

\bibitem{PhysRevLett.121.240401}
A~A Budini,  \emph{{Quantum Non-Markovian Processes Break Conditional
  Past-Future Independence}},
  \href{https://doi.org/10.1103/PhysRevLett.121.240401}{Phys. Rev. Lett. {\bf
  121}, 240401 (2018)} (Preprint
  {\href{https://arxiv.org/abs/1811.03448}{arXiv:1811.03448}})

\bibitem{PhysRevA.99.052125}
A~A Budini,  \emph{{Conditional past-future correlation induced by
  non-Markovian dephasing reservoirs}},
  \href{https://doi.org/10.1103/PhysRevA.99.052125}{Phys. Rev. A {\bf 99},
  052125 (2019)} (Preprint
  {\href{https://arxiv.org/abs/1903.05259}{arXiv:1903.05259}})

\bibitem{Siefert2003}
M~Siefert, A~Kittel, R~Friedrich and J~Peinke,  \emph{{On a quantitative method
  to analyze dynamical and measurement noise}},
  \href{https://doi.org/10.1209/epl/i2003-00152-9}{EPL {\bf 61}, 466 (2003)}
  (Preprint
  {\href{https://arxiv.org/abs/physics/0108034}{arXiv:physics/0108034}})

\bibitem{Bottcher2006}
F~B{\"o}ttcher, J~Peinke, D~Kleinhans, R~Friedrich, P~G Lind and M~Haase,
  \emph{{Reconstruction of Complex Dynamical Systems Affected by Strong
  Measurement Noise}},
  \href{https://doi.org/10.1103/PhysRevLett.97.090603}{Phys. Rev. Lett. {\bf
  97}, 090603 (2006)} (Preprint
  {\href{https://arxiv.org/abs/nlin/0607002}{arXiv:nlin/0607002}})

\bibitem{Lehle2011}
B~Lehle,  \emph{{Analysis of stochastic time series in the presence of strong
  measurement noise}},  \href{https://doi.org/10.1103/PhysRevE.83.021113}{Phys.
  Rev. E {\bf 83}, 021113 (2011)} (Preprint
  {\href{https://arxiv.org/abs/1010.5641}{arXiv:1010.5641}})

\bibitem{Taranto_2022}
P~Taranto, T~J Elliott and S~Milz,  \emph{{Hidden Quantum Memory: Is Memory
  There When Somebody Looks?}},
  \href{https://doi.org/10.22331/q-2023-04-27-991}{Quantum {\bf 7}, 991 (2023)}
  (Preprint {\href{https://arxiv.org/abs/2204.08298}{arXiv:2204.08298}})

\bibitem{PhysRevA.102.022216}
M~Bonifacio and A~A Budini,  \emph{{Perturbation theory for operational quantum
  non-Markovianity}},  \href{https://doi.org/10.1103/PhysRevA.102.022216}{Phys.
  Rev. A {\bf 102}, 022216 (2020)} (Preprint
  {\href{https://arxiv.org/abs/2004.11813}{arXiv:2004.11813}})

\bibitem{PhysRevA.103.012221}
A~A Budini,  \emph{{Detection of bidirectional system-environment information
  exchanges}},  \href{https://doi.org/10.1103/PhysRevA.103.012221}{Phys. Rev. A
  {\bf 103}, 012221 (2021)} (Preprint
  {\href{https://arxiv.org/abs/2101.10308}{arXiv:2101.10308}})

\bibitem{budini_quantum_2022}
A~A Budini,  \emph{Quantum {Non}-{Markovian} {Environment}-to-{System}
  {Backflows} of {Information}: {Nonoperational} vs. {Operational}
  {Approaches}},  \href{https://doi.org/10.3390/e24050649}{Entropy {\bf 24},
  649 (2022)} (Preprint
  {\href{https://arxiv.org/abs/2205.03333}{arXiv:2205.03333}})

\bibitem{Pearl}
J~Pearl, {\textit{{Causality: Models, Reasoning and Inference}}} (Cambridge,
  UK; New York, USA: Cambridge University Press) (2009)

\bibitem{berk_extracting_2021}
G~D Berk, S~Milz, F~A Pollock and K~Modi,  \emph{{Extracting quantum dynamical
  resources: consumption of non-Markovianity for noise reduction}},
  \href{https://doi.org/10.1038/s41534-023-00774-w}{npj Quantum Inf. {\bf 9},
  104 (2023)} (Preprint
  {\href{https://arxiv.org/abs/2110.02613}{arXiv:2110.02613}})

\bibitem{Zambon_2024}
G~Zambon,  \emph{{Process tensor distinguishability measures}},
  \href{https://doi.org/10.1103/PhysRevA.110.042210}{Phys. Rev. A {\bf 110},
  042210 (2024)} (Preprint
  {\href{https://arxiv.org/abs/2407.15712}{arXiv:2407.15712}})

\bibitem{Berk_2024}
G~D Berk, S~Milz and K~Modi,  \emph{{Monotones in Resource Theories for
  Dynamical Decoupling}} (2024) (Preprint
  {\href{https://arxiv.org/abs/2412.11595}{arXiv:2412.11595}})

\bibitem{White_2022}
G~A~L White, F~A Pollock, L~C~L Hollenberg, K~Modi and C~D Hill,
  \emph{{Non-Markovian Quantum Process Tomography}},
  \href{https://doi.org/10.1103/PRXQuantum.3.020344}{PRX Quantum {\bf 3},
  020344 (2022)} (Preprint
  {\href{https://arxiv.org/abs/2106.11722}{arXiv:2106.11722}})

\bibitem{10.21468/SciPostPhys.15.1.024}
P~Strasberg,  \emph{{Classicality with(out) decoherence: Concepts, relation to
  Markovianity, and a random matrix theory approach}},
  \href{https://doi.org/10.21468/SciPostPhys.15.1.024}{SciPost Phys. {\bf 15},
  024 (2023)} (Preprint
  {\href{https://arxiv.org/abs/2301.02563}{arXiv:2301.02563}})

\bibitem{Smirne_2018}
A~Smirne, D~Egloff, M~G D{\'\i}az, M~B Plenio and S~F Huelga,  \emph{{Coherence
  and non-classicality of quantum Markov processes}},
  \href{https://doi.org/10.1088/2058-9565/aaebd5}{Quantum Sci. Technol. {\bf
  4}, 01LT01 (2018)} (Preprint
  {\href{https://arxiv.org/abs/1709.05267}{arXiv:1709.05267}})

\bibitem{Strasberg_2019_Classical}
P~Strasberg and M~G D{\'\i}az,  \emph{{Classical quantum stochastic
  processes}},  \href{https://doi.org/10.1103/PhysRevA.100.022120}{Phys. Rev. A
  {\bf 100}, 022120 (2019)} (Preprint
  {\href{https://arxiv.org/abs/1905.03018}{arXiv:1905.03018}})

\bibitem{Milz_2020_Classical}
S~Milz, D~Egloff, P~Taranto, T~Theurer, M~B Plenio, A~Smirne and S~F Huelga,
  \emph{{When Is a Non-Markovian Quantum Process Classical?}},
  \href{https://doi.org/10.1103/PhysRevX.10.041049}{Phys. Rev. X {\bf 10},
  041049 (2020)} (Preprint
  {\href{https://arxiv.org/abs/1907.05807}{arXiv:1907.05807}})

\bibitem{PhysRevA.108.012225}
P~Strasberg, A~Winter, J~Gemmer and J~Wang,  \emph{{Classicality, Markovianity,
  and local detailed balance from pure-state dynamics}},
  \href{https://doi.org/10.1103/PhysRevA.108.012225}{Phys. Rev. A {\bf 108},
  012225 (2023)} (Preprint
  {\href{https://arxiv.org/abs/2209.07977}{arXiv:2209.07977}})

\bibitem{202300304}
M~F Richter, A~Smirne, W~T Strunz and D~Egloff,  \emph{{Classical Invasive
  Description of Informationally-Complete Quantum Processes}},
  \href{https://doi.org/10.1002/andp.202300304}{Ann. Phys. {\bf 536}, 2300304
  (2024)} (Preprint
  {\href{https://arxiv.org/abs/2312.06545}{arXiv:2312.06545}})

\bibitem{budini_violation_2023}
A~A Budini,  \emph{Violation of {Diagonal} {Non}-{Invasiveness}: {A} {Hallmark}
  of {Non}-{Classical} {Memory} {Effects}} (2023) (Preprint
  {\href{https://arxiv.org/abs/2301.02500}{arXiv:2301.02500}})

\bibitem{Kolmogorov_1956}
A~N Kolmogorov, {\textit{{Grundbegriffe der Wahrscheinlichkeitsrechnung}}}
  (Berlin, Germany: Springer) (1933) [\textit{{F}oundations of the {Theory} of
  {Probability}} (Chelsea, New York, USA 1956)]

\bibitem{Piron1981}
C~Piron,  \emph{{Ideal measurement and probability in quantum mechanics}},
  \href{https://doi.org/10.1007/BF00211379}{Erkenntnis {\bf 16}, 397 (1981)}

\bibitem{leggett_quantum_1985}
A~J Leggett and A~Garg,  \emph{{Quantum mechanics versus macroscopic realism:
  {Is} the flux there when nobody looks?}},
  \href{https://doi.org/10.1103/PhysRevLett.54.857}{Phys. Rev. Lett. {\bf 54},
  857 (1985)}

\bibitem{leggett_realism_2008}
A~J Leggett,  \emph{{Realism and the physical world}},
  \href{https://doi.org/10.1088/0034-4885/71/2/022001}{Rep. Prog. Phys. {\bf
  71}, 022001 (2008)}

\bibitem{emary_leggettgarg_2013}
C~Emary, N~Lambert and F~Nori,  \emph{Leggett–{Garg} inequalities},
  \href{https://doi.org/10.1088/0034-4885/77/1/016001}{Rep. Prog. Phys. {\bf
  77}, 016001 (2013)} (Preprint
  {\href{https://arxiv.org/abs/1304.5133}{arXiv:1304.5133}})

\bibitem{double_nothing}
D~Lonigro, F~Sakuldee, {\L}~Cywi{\'n}ski, D~Chru{\'s}ci{\'n}ski and
  P~Sza{\'n}kowski,  \emph{{Double or nothing: a Kolmogorov extension theorem
  for multitime (bi)probabilities in quantum mechanics}},
  \href{https://doi.org/10.22331/q-2024-08-27-1447}{Quantum {\bf 8}, 1447
  (2024)} (Preprint
  {\href{https://arxiv.org/abs/2402.01218}{arXiv:2402.01218}})

\bibitem{PhysRevX.14.041027}
P~Strasberg, T~E Reinhard and J~Schindler,  \emph{{First Principles Numerical
  Demonstration of Emergent Decoherent Histories}},
  \href{https://doi.org/10.1103/PhysRevX.14.041027}{Phys. Rev. X {\bf 14},
  041027 (2024)} (Preprint
  {\href{https://arxiv.org/abs/2304.10258}{arXiv:2304.10258}})

\bibitem{2406.15577}
J~Wang and P~Strasberg,  \emph{{Decoherence of Histories: Chaotic Versus
  Integrable Systems}} (2024) (Preprint
  {\href{https://arxiv.org/abs/2406.15577}{arXiv:2406.15577}})

\bibitem{omnes_consistent_1992}
R~Omn{\`e}s,  \emph{{Consistent interpretations of quantum mechanics}},
  \href{https://doi.org/10.1103/RevModPhys.64.339}{Rev. Mod. Phys. {\bf 64},
  339 (1992)}

\bibitem{paz_environment-induced_1993}
J~P Paz and W~H Zurek,  \emph{{Environment-induced decoherence, classicality,
  and consistency of quantum histories}},
  \href{https://doi.org/10.1103/PhysRevD.48.2728}{Phys. Rev. D {\bf 48}, 2728
  (1993)} (Preprint {\href{https://arxiv.org/abs/9304031}{arXiv:9304031}})

\bibitem{griffiths_consistent_1984}
R~B Griffiths,  \emph{{Consistent histories and the interpretation of quantum
  mechanics}},  \href{https://doi.org/10.1007/BF01015734}{J. Stat. Phys. {\bf
  36}, 219 (1984)}

\bibitem{omnes_logical_1988}
R~Omn{\`e}s,  \emph{{Logical reformulation of quantum mechanics. {I}.
  {Foundations}}},  \href{https://doi.org/10.1007/BF01014230}{J. Stat. Phys.
  {\bf 53}, 893 (1988)}

\bibitem{gell-mann_quantum_1997}
M~Gell-Mann and J~B Hartle,
  {\href{https://doi.org/10.1142/9789812819895_0036}{\textit{Quantum
  {Mechanics} in the {Light} of {Quantum} {Cosmology}}}}, {\em Foundations of
  {Quantum} {Mechanics} in the {Light} of {New} {Technology}\/}, vol~4
  (Singapore: World Scientific), p 347 (1997) (Preprint
  {\href{https://arxiv.org/abs/1803.04605}{arXiv:1803.04605}})

\bibitem{Lax_1963}
M~Lax,  \emph{{Formal Theory of Quantum Fluctuations from a Driven State}},
  \href{https://doi.org/10.1103/PhysRev.129.2342}{Phys. Rev. {\bf 129}, 2342
  (1963)}

\bibitem{Plenio_2007}
M~B Plenio and S~Virmani,  \emph{{An Introduction to Entanglement Measures}},
  \href{https://dl.acm.org/doi/10.5555/2011706.2011707}{Quantum Info. Comput.
  {\bf 7}, 1 (2007)} (Preprint
  {\href{https://arxiv.org/abs/quant-ph/0504163}{arXiv:quant-ph/0504163}})

\bibitem{Horodecki_2009}
R~Horodecki, P~Horodecki, M~Horodecki and K~Horodecki,  \emph{{Quantum
  entanglement}},  \href{https://doi.org/10.1103/RevModPhys.81.865}{Rev. Mod.
  Phys. {\bf 81}, 865 (2009)} (Preprint
  {\href{https://arxiv.org/abs/quant-ph/0702225}{arXiv:quant-ph/0702225}})

\bibitem{akibue_thesis}
S~Akibue, {\href{https://arxiv.org/abs/2202.06518}{\textit{{Entanglement and
  Causal Relation in distributed quantum computation}}}}, {PhD Thesis}, The
  University of Tokyo (2015)

\bibitem{Chitambar_2020}
E~Chitambar, J~I {de Vicente}, M~W Girard and G~Gour,  \emph{{Entanglement
  manipulation beyond local operations and classical communication}},
  \href{https://doi.org/10.1063/1.5124109}{J. Math. Phys. {\bf 61}, 042201
  (2020)} (Preprint
  {\href{https://arxiv.org/abs/1711.03835}{arXiv:1711.03835}})

\bibitem{beckman_causal_2001}
D~Beckman, D~Gottesman, M~A Nielsen and J~Preskill,  \emph{{Causal and
  localizable quantum operations}},
  \href{https://doi.org/10.1103/PhysRevA.64.052309}{Phys. Rev. A {\bf 64},
  052309 (2001)} (Preprint
  {\href{https://arxiv.org/abs/quant-ph/0102043}{arXiv:quant-ph/0102043}})

\bibitem{Chitambar_2016_Critical}
E~Chitambar and G~Gour,  \emph{{Critical Examination of Incoherent Operations
  and a Physically Consistent Resource Theory of Quantum Coherence}},
  \href{https://doi.org/10.1103/PhysRevLett.117.030401}{Phys. Rev. Lett. {\bf
  117}, 030401 (2016)} (Preprint
  {\href{https://arxiv.org/abs/1602.06969}{arXiv:1602.06969}})

\bibitem{Chitambar_2016_Comparison}
E~Chitambar and G~Gour,  \emph{{Comparison of incoherent operations and
  measures of coherence}},
  \href{https://doi.org/10.1103/PhysRevA.94.052336}{Phys. Rev. A {\bf 94},
  052336 (2016)} (Preprint
  {\href{https://arxiv.org/abs/1602.06969}{arXiv:1602.06969}})

\bibitem{Marvian_2016}
I~Marvian and R~W Spekkens,  \emph{How to quantify coherence: {Distinguishing}
  speakable and unspeakable notions},
  \href{https://doi.org/10.1103/PhysRevA.94.052324}{Phys. Rev. A {\bf 94},
  052324 (2016)} (Preprint
  {\href{https://arxiv.org/abs/1602.08049}{arXiv:1602.08049}})

\bibitem{szankowski_noise_2020}
P~Sza{\'n}kowski and {\L}~Cywi{\'n}ski,  \emph{{Noise representations of open
  system dynamics}},  \href{https://doi.org/10.1038/s41598-020-78079-7}{Sci.
  Rep. {\bf 10}, 22189 (2020)} (Preprint
  {\href{https://arxiv.org/abs/2003.09688}{arXiv:2003.09688}})

\bibitem{szankowski_objectivity_2024}
P~Sza{\'n}kowski and {\L}~Cywi{\'n}ski,  \emph{{Objectivity of classical
  quantum stochastic processes}},
  \href{https://doi.org/10.22331/q-2024-06-27-1390}{Quantum {\bf 8}, 1390
  (2024)} (Preprint
  {\href{https://arxiv.org/abs/2304.07110}{arXiv:2304.07110}})

\bibitem{Taranto_2024}
P~Taranto, M~T Quintino, M~Murao and S~Milz,  \emph{{Characterising the
  Hierarchy of Multi-time Quantum Processes with Classical Memory}},
  \href{https://doi.org/10.22331/q-2024-05-02-1328}{Quantum {\bf 8}, 1328
  (2024)} (Preprint
  {\href{https://arxiv.org/abs/2307.11905}{arXiv:2307.11905}})

\bibitem{Ohst_2024}
T-A {Ohst}, S~{Zhang}, H~{Chau Nguyen}, M~{Pl{\'a}vala} and M~{T{\'u}lio
  Quintino},  \emph{{Characterising memory in quantum channel discrimination
  via constrained separability problems}} (2024) (Preprint
  {\href{https://arxiv.org/abs/2411.08110}{arXiv:2411.08110}})

\bibitem{goswami_2024}
K~Goswami, A~K Roy, V~Srivastava, B~Perez, C~Giarmatzi, A~Gilchrist and
  F~Costa,  \emph{{Hamiltonian characterisation of multi-time processes with
  classical memory}} (2024) (Preprint
  {\href{https://arxiv.org/abs/2412.01998}{arXiv:2412.01998}})

\bibitem{Vieira_2024}
L~B Vieira, H-Y Ku and C~Budroni,  \emph{{Entanglement-breaking channels are a
  quantum memory resource}} (2024) (Preprint
  {\href{https://arxiv.org/abs/2402.16789}{arXiv:2402.16789}})

\bibitem{Gu_2012}
M~Gu, K~Wiesner, E~Rieper and V~Vedral,  \emph{{Quantum mechanics can reduce
  the complexity of classical models}},
  \href{https://doi.org/10.1038/ncomms1761}{Nat. Commun. {\bf 3}, 762 (2012)}
  (Preprint {\href{https://arxiv.org/abs/1102.1994}{arXiv:1102.1994}})

\bibitem{Binder_2018_Practical}
F~C Binder, J~Thompson and M~Gu,  \emph{{Practical Unitary Simulator for
  Non-Markovian Complex Processes}},
  \href{https://doi.org/10.1103/PhysRevLett.120.240502}{Phys. Rev. Lett. {\bf
  120}, 240502 (2018)} (Preprint
  {\href{https://arxiv.org/abs/1709.02375}{arXiv:1709.02375}})

\bibitem{Yang_2018_Matrix}
C~Yang, F~C Binder, V~Narasimhachar and M~Gu,  \emph{{Matrix Product States for
  Quantum Stochastic Modeling}},
  \href{https://doi.org/10.1103/PhysRevLett.121.260602}{Phys. Rev. Lett. {\bf
  121}, 260602 (2018)} (Preprint
  {\href{https://arxiv.org/abs/1803.08220}{arXiv:1803.08220}})

\bibitem{Suen_2018}
W~Y Suen, T~J Elliott, J~Thompson, A~J~P Garner, J~R Mahoney, V~Vedral and
  M~Gu,  \emph{{Surveying Structural Complexity in Quantum Many-Body Systems}},
   \href{https://doi.org/10.1007/s10955-022-02895-6}{J. Stat. Phys. {\bf 187},
  4 (2022)} (Preprint
  {\href{https://arxiv.org/abs/1812.09738}{arXiv:1812.09738}})

\bibitem{Thompson_2018_Causal}
J~Thompson, A~J~P Garner, J~R Mahoney, J~P Crutchfield, V~Vedral and M~Gu,
  \emph{{Causal Asymmetry in a Quantum World}},
  \href{https://doi.org/10.1103/PhysRevX.8.031013}{Phys. Rev. X {\bf 8}, 031013
  (2018)} (Preprint
  {\href{https://arxiv.org/abs/1712.02368}{arXiv:1712.02368}})

\bibitem{Korzekwa_2021}
K~Korzekwa and M~Lostaglio,  \emph{{Quantum Advantage in Simulating Stochastic
  Processes}},  \href{https://doi.org/10.1103/PhysRevX.11.021019}{Phys. Rev. X
  {\bf 11}, 021019 (2021)} (Preprint
  {\href{https://arxiv.org/abs/2005.02403}{arXiv:2005.02403}})

\bibitem{Makri_1995a}
N~Makri and D~E Makarov,  \emph{{Tensor Propagator for Iterative Quantum Time
  Evolution of Reduced Density Matrices. I. Theory}},
  \href{https://doi.org/10.1063/1.469508}{J. Chem. Phys. {\bf 102}, 4600
  (1995)}

\bibitem{Makri_1995b}
N~Makri and D~E Makarov,  \emph{{Tensor Propagator for Iterative Quantum Time
  Evolution of Reduced Density Matrices. II. Numerical Methodology}},
  \href{https://doi.org/10.1063/1.469509}{J. Chem. Phys. {\bf 102}, 4611
  (1995)}

\bibitem{Cohen_2015}
G~Cohen, E~Gull, D~R Reichman and A~J Millis,  \emph{{Taming the Dynamical Sign
  Problem in Real-Time Evolution of Quantum Many-Body Problems}},
  \href{https://doi.org/10.1103/PhysRevLett.115.266802}{Phys. Rev. Lett. {\bf
  115}, 266802 (2015)} (Preprint
  {\href{https://arxiv.org/abs/1510.03534}{arXiv:1510.03534}})

\bibitem{Lerose_2021_Influence}
A~Lerose, M~Sonner and D~A Abanin,  \emph{{Influence Matrix Approach to
  Many-Body Floquet Dynamics}},
  \href{https://doi.org/10.1103/PhysRevX.11.021040}{Phys. Rev. X {\bf 11},
  021040 (2021)} (Preprint
  {\href{https://arxiv.org/abs/2009.10105}{arXiv:2009.10105}})

\bibitem{Strathearn_2018}
A~Strathearn, P~Kirton, D~Kilda, J~Keeling and B~W Lovett,  \emph{{Efficient
  non-Markovian quantum dynamics using time-evolving matrix product
  operators}},  \href{https://doi.org/10.1038/s41467-018-05617-3}{Nat. Commun.
  {\bf 9}, 3322 (2018)} (Preprint
  {\href{https://arxiv.org/abs/1711.09641}{arXiv:1711.09641}})

\bibitem{OQuPy_2024}
G~E Fux, P~Fowler-Wright, J~Beckles, E~P Butler, P~R Eastham, D~Gribben,
  J~Keeling, D~Kilda, P~Kirton, E~D~C Lawrence, B~W Lovett, E~O'Neill,
  A~Strathearn and R~{de Wit},  \emph{{OQuPy: A Python package to efficiently
  simulate non-Markovian open quantum systems with process tensors}},
  \href{https://doi.org/10.1063/5.0225367}{J. Chem. Phys. {\bf 161}, 124108
  (2024)} (Preprint
  {\href{https://arxiv.org/abs/2406.16650}{arXiv:2406.16650}})

\bibitem{OQuPy_package}
{The TEMPO Collaboration},
  \emph{\href{https://doi.org/10.5281/zenodo.12517767}{{TimeEvolvingMPO: A
  Python 3 Package to Efficiently Compute Non-Markovian Open Quantum Systems}}}
  (2024)

\bibitem{Cygorek_2024_ACE}
M~Cygorek and E~M Gauger,  \emph{{ACE: A General-Purpose Non-Markovian Open
  Quantum Systems Simulation Toolkit Based on Process Tensors}},
  \href{https://doi.org/10.1063/5.0221182}{J. Chem. Phys. {\bf 161}, 074111
  (2024)} (Preprint
  {\href{https://arxiv.org/abs/2405.19319}{arXiv:2405.19319}})

\bibitem{Strathearn_2017}
A~Strathearn, B~W Lovett and P~Kirton,  \emph{{Efficient Real-Time Path
  Integrals for Non-Markovian Spin-Boson Models}},
  \href{https://doi.org/10.1088/1367-2630/aa8744}{New J. Phys. {\bf 19}, 093009
  (2017)} (Preprint
  {\href{https://arxiv.org/abs/1704.04099}{arXiv:1704.04099}})

\bibitem{Ye_2021}
E~Ye and G~K-L Chan,  \emph{{Constructing tensor network influence functionals
  for general quantum dynamics}},  \href{https://doi.org/10.1063/5.0047260}{J.
  Chem. Phys. {\bf 155}, 044104 (2021)} (Preprint
  {\href{https://arxiv.org/abs/2101.05466}{arXiv:2101.05466}})

\bibitem{Gribben_2022}
D~Gribben, D~M Rouse, J~Iles-Smith, A~Strathearn, H~Maguire, P~Kirton, A~Nazir,
  E~M Gauger and B~W Lovett,  \emph{{Exact Dynamics of Nonadditive Environments
  in Non-Markovian Open Quantum Systems}},
  \href{https://doi.org/10.1103/PRXQuantum.3.010321}{PRX Quantum {\bf 3},
  010321 (2022)} (Preprint
  {\href{https://arxiv.org/abs/2109.08442}{arXiv:2109.08442}})

\bibitem{Cygorek_2022}
M~Cygorek, M~Cosacchi, A~Vagov, V~M Axt, B~W Lovett, J~Keeling and E~M Gauger,
  \emph{{Simulation of Open Quantum Systems by Automated Compression of
  Arbitrary Environments}},
  \href{https://doi.org/10.1038/s41567-022-01544-9}{Nat. Phys. {\bf 18}, 662
  (2022)} (Preprint
  {\href{https://arxiv.org/abs/2101.01653}{arXiv:2101.01653}})

\bibitem{Fux_2023}
G~E Fux, D~Kilda, B~W Lovett and J~Keeling,  \emph{{Tensor Network Simulation
  of Chains of Non-Markovian Open Quantum Systems}},
  \href{https://doi.org/10.1103/PhysRevResearch.5.033078}{Phys. Rev. Res. {\bf
  5}, 033078 (2023)} (Preprint
  {\href{https://arxiv.org/abs/2201.05529}{arXiv:2201.05529}})

\bibitem{Ng_2023}
N~Ng, G~Park, A~J Millis, G~K-L Chan and D~R Reichman,  \emph{{Real-time
  evolution of Anderson impurity models via tensor network influence
  functionals}},  \href{https://doi.org/10.1103/PhysRevB.107.125103}{Phys. Rev.
  B {\bf 107} (2023)} (Preprint
  {\href{https://arxiv.org/abs/2211.10430}{arXiv:2211.10430}})

\bibitem{Park_2024}
G~Park, N~Ng, D~R Reichman and G~K-L Chan,  \emph{{Tensor network influence
  functionals in the continuous-time limit: Connections to quantum embedding,
  bath discretization, and higher-order time propagation}},
  \href{https://doi.org/10.1103/PhysRevB.110.045104}{Phys. Rev. B {\bf 110}
  (2024)} (Preprint
  {\href{https://arxiv.org/abs/2401.12460}{arXiv:2401.12460}})

\bibitem{nguyen2024}
H~Nguyen, N~Ng, L~P Lindoy, G~Park, A~J Millis, G~K-L Chan and D~R Reichman,
  \emph{{Correlation Functions From Tensor Network Influence Functionals: The
  Case of the Spin-Boson Model}} (2024) (Preprint
  {\href{https://arxiv.org/abs/2406.15737}{arXiv:2406.15737}})

\bibitem{Cygorek_2024_Understanding}
M~Cygorek and E~M Gauger,  \emph{{Understanding and utilizing the inner bonds
  of process tensors}},
  \href{https://scipost.org/10.21468/SciPostPhys.18.1.024}{SciPost Phys. {\bf
  18}, 024 (2025)} (Preprint
  {\href{https://arxiv.org/abs/2404.01287}{arXiv:2404.01287}})

\bibitem{Cygorek_2024}
M~Cygorek, B~W Lovett, J~Keeling and E~M Gauger,  \emph{{Treelike process
  tensor contraction for automated compression of environments}},
  \href{https://doi.org/10.1103/PhysRevResearch.6.043203}{Phys. Rev. Res. {\bf
  6}, 043203 (2024)} (Preprint
  {\href{https://arxiv.org/abs/2405.16548}{arXiv:2405.16548}})

\bibitem{Cygorek_2024_Sublinear}
M~Cygorek, J~Keeling, B~W Lovett and E~M Gauger,  \emph{{Sublinear Scaling in
  Non-Markovian Open Quantum Systems Simulations}},
  \href{https://doi.org/10.1103/PhysRevX.14.011010}{Phys. Rev. X {\bf 14},
  011010 (2024)} (Preprint
  {\href{https://arxiv.org/abs/2304.05291}{arXiv:2304.05291}})

\bibitem{Chen_2024a}
R~Chen, X~Xu and C~Guo,  \emph{{Grassmann Time-Evolving Matrix Product
  Operators for Quantum Impurity Models}},
  \href{https://doi.org/10.1103/PhysRevB.109.045140}{Phys. Rev. B {\bf 109},
  045140 (2024)} (Preprint
  {\href{https://arxiv.org/abs/2308.05279}{arXiv:2308.05279}})

\bibitem{Guo_2024}
C~Guo and R~Chen,  \emph{{Efficient Construction of the Feynman-Vernon
  Influence Functional as Matrix Product States}},
  \href{https://doi.org/10.21468/SciPostPhysCore.7.3.063}{SciPost Phys. Core
  {\bf 7}, 063 (2024)} (Preprint
  {\href{https://arxiv.org/abs/2402.14350}{arXiv:2402.14350}})

\bibitem{Guo_2024b}
C~Guo and R~Chen,  \emph{{Infinite Grassmann time-evolving matrix product
  operator method in the steady state}},
  \href{https://doi.org/10.1103/PhysRevB.110.045106}{Phys. Rev. B {\bf 110},
  045106 (2024)} (Preprint
  {\href{https://arxiv.org/abs/2403.16700}{arXiv:2403.16700}})

\bibitem{Sun_2024}
Z~Sun, R~Chen, Z~Li and C~Guo,  \emph{{Infinite Grassmann time-evolving matrix
  product operators for non-equilibrium quantum impurity problems}} (2024)
  (Preprint {\href{https://arxiv.org/abs/2412.04702}{arXiv:2412.04702}})

\bibitem{Gribben_2022_Quantum}
D~Gribben, A~Strathearn, G~E Fux, P~Kirton and B~W Lovett,  \emph{{Using the
  Environment to Understand Non-Markovian Open Quantum Systems}},
  \href{https://doi.org/10.22331/q-2022-10-25-847}{Quantum {\bf 6}, 847 (2022)}
  (Preprint {\href{https://arxiv.org/abs/2106.04212}{arXiv:2106.04212}})

\bibitem{Wang_2024}
H-R Wang, X-Y Yang and Z~Wang,  \emph{{Exact Hidden Markovian Dynamics in
  Quantum Circuits}},
  \href{https://doi.org/10.1103/PhysRevLett.133.170402}{Phys. Rev. Lett. {\bf
  133}, 170402 (2024)} (Preprint
  {\href{https://arxiv.org/abs/2403.14807}{arXiv:2403.14807}})

\bibitem{Fux_2021}
G~E Fux, E~P Butler, P~R Eastham, B~W Lovett and J~Keeling,  \emph{{Efficient
  Exploration of Hamiltonian Parameter Space for Optimal Control of
  Non-Markovian Open Quantum Systems}},
  \href{https://doi.org/10.1103/PhysRevLett.126.200401}{Phys. Rev. Lett. {\bf
  126}, 200401 (2021)} (Preprint
  {\href{https://arxiv.org/abs/2101.03071}{arXiv:2101.03071}})

\bibitem{Ortega_2024}
C~Ortega-Taberner, E~O'Neill, E~Butler, G~E Fux and P~R Eastham,
  \emph{{Unifying Methods for Optimal Control in Non-Markovian Quantum Systems
  via Process Tensors}},  \href{https://doi.org/10.1063/5.0226031}{J. Chem.
  Phys. {\bf 161}, 124119 (2024)} (Preprint
  {\href{https://arxiv.org/abs/2406.17719}{arXiv:2406.17719}})

\bibitem{Butler_2024}
E~P Butler, G~E Fux, C~Ortega-Taberner, B~W Lovett, J~Keeling and P~R Eastham,
  \emph{{Optimizing Performance of Quantum Operations with Non-Markovian
  Decoherence: The Tortoise or the Hare?}},
  \href{https://doi.org/10.1103/PhysRevLett.132.060401}{Phys. Rev. Lett. {\bf
  132}, 060401 (2024)} (Preprint
  {\href{https://arxiv.org/abs/2201.05529}{arXiv:2201.05529}})

\bibitem{Pollock_2018_Quantum}
F~A Pollock and K~Modi,  \emph{{Tomographically Reconstructed Master Equations
  for Any Open Quantum Dynamics}},
  \href{https://doi.org/10.22331/q-2018-07-11-76}{Quantum {\bf 2}, 76 (2018)}
  (Preprint {\href{https://arxiv.org/abs/1704.06204}{arXiv:1704.06204}})

\bibitem{Jorgensen_2019}
M~R J{\o}rgensen and F~A Pollock,  \emph{{Exploiting the Causal Tensor Network
  Structure of Quantum Processes to Efficiently Simulate Non-Markovian Path
  Integrals}},  \href{https://doi.org/10.1103/PhysRevLett.123.240602}{Phys.
  Rev. Lett. {\bf 123}, 240602 (2019)} (Preprint
  {\href{https://arxiv.org/abs/1902.00315}{arXiv:1902.00315}})

\bibitem{Jorgensen_2020}
M~R J{\o}rgensen and F~A Pollock,  \emph{{Discrete Memory Kernel for Multitime
  Correlations in Non-Markovian Quantum Processes}},
  \href{https://doi.org/10.1103/PhysRevA.102.052206}{Phys. Rev. A {\bf 102},
  052206 (2020)} (Preprint
  {\href{https://arxiv.org/abs/2007.03234}{arXiv:2007.03234}})

\bibitem{cerrillo_non-markovian_2014}
J~Cerrillo and J~Cao,  \emph{Non-{Markovian} {Dynamical} {Maps}: {Numerical}
  {Processing} of {Open} {Quantum} {Trajectories}},
  \href{https://doi.org/10.1103/PhysRevLett.112.110401}{Phys. Rev. Lett. {\bf
  112}, 110401 (2014)} (Preprint
  {\href{https://arxiv.org/abs/1307.7743}{arXiv:1307.7743}})

\bibitem{rosenbach_efficient_2016}
R~Rosenbach, J~Cerrillo, S~F Huelga, J~Cao and M~B Plenio,  \emph{{Efficient
  simulation of non-{Markovian} system-environment interaction}},
  \href{https://doi.org/10.1088/1367-2630/18/2/023035}{New J. Phys. {\bf 18},
  023035 (2016)} (Preprint
  {\href{https://arxiv.org/abs/1510.03100}{arXiv:1510.03100}})

\bibitem{Pollock_2021_FCS}
F~A Pollock, E~Gull, K~Modi and G~Cohen,  \emph{{Reduced Dynamics of Full
  Counting Statistics}},
  \href{https://doi.org/10.21468/SciPostPhys.13.2.027}{SciPost Phys. {\bf 13},
  027 (2022)} (Preprint
  {\href{https://arxiv.org/abs/2111.08525}{arXiv:2111.08525}})

\bibitem{Chen_2020}
Y-Q Chen, K-L Ma, Y-C Zheng, J~Allcock, S~Zhang and C-Y Hsieh,
  \emph{{Non-Markovian Noise Characterization with the Transfer Tensor
  Method}},  \href{https://doi.org/10.1103/PhysRevApplied.13.034045}{Phys. Rev.
  Appl. {\bf 13}, 034045 (2020)} (Preprint
  {\href{https://arxiv.org/abs/1905.10941}{arXiv:1905.10941}})

\bibitem{Gherardini_2022}
S~Gherardini, A~Smirne, S~F Huelga and F~Caruso,  \emph{{Transfer-Tensor
  Description of Memory Effects in Open-System Dynamics and Multi-Time
  Statistics}},  \href{https://doi.org/10.1088/2058-9565/ac4422}{Quantum Sci.
  Technol. {\bf 7}, 025005 (2022)} (Preprint
  {\href{https://arxiv.org/abs/2101.11662}{arXiv:2101.11662}})

\bibitem{Link_2024}
V~Link, H-H Tu and W~T Strunz,  \emph{{Open Quantum System Dynamics from
  Infinite Tensor Network Contraction}},
  \href{https://doi.org/10.1103/PhysRevLett.132.200403}{Phys. Rev. Lett. {\bf
  132}, 200403 (2024)} (Preprint
  {\href{https://arxiv.org/abs/2307.01802}{arXiv:2307.01802}})

\bibitem{PhysRevLett.121.227401}
J~{del Pino}, F~A Y~N Schr{\" o}der, A~W Chin, J~Feist and F~J Garcia-Vidal,
  \emph{{Tensor Network Simulation of Non-Markovian Dynamics in Organic
  Polaritons}},  \href{https://doi.org/10.1103/PhysRevLett.121.227401}{Phys.
  Rev. Lett. {\bf 121}, 227401 (2018)} (Preprint
  {\href{https://arxiv.org/abs/1804.04511}{arXiv:1804.04511}})

\bibitem{chin-natcom}
F~A Y~N Schr{\" o}der, D~H~P Turban, A~J Musser, N~D~M Hine and A~W Chin,
  \emph{{Tensor Network Simulation of Multi-Environmental Open Quantum Dynamics
  via Machine Learning and Entanglement Renormalisation}},
  \href{https://doi.org/10.1038/s41467-019-09039-7}{Nat. Commun. {\bf 10}, 1062
  (2019)} (Preprint
  {\href{https://arxiv.org/abs/1710.01362}{arXiv:1710.01362}})

\bibitem{Cerezo-Roquebrun_2025}
S~Cerezo-Roquebr{\' u}n, A~Bou-Comas, J~T Schneider, E~L{\' o}pez,
  L~Tagliacozzo and S~Carignano,  \emph{{Spatio-temporal tensor-network
  approaches to out-of-equilibrium dynamics bridging open and closed systems}}
  (2025) (Preprint {\href{https://arxiv.org/abs/2502.20214}{arXiv:2502.20214}})

\bibitem{Berezutskii_2005}
A~Berezutskii, A~Acharya, R~Ellerbrock, J~Gray, R~Haghshenas, Z~He, A~Khan,
  V~Kuzmin, M~Liu, D~Lyakh, D~Lykov, S~Mandr{\` a}, C~Mansell, A~Melnikov,
  A~Melnikov, V~Mironov, D~Morozov, F~Neukart, A~Nocera, M~A Perlin,
  M~Perelshtein, R~Shaydulin, B~Villalonga, M~Pflitsch, M~Pistoia, V~Vinokur
  and Y~Alexeev,  \emph{{Tensor networks for quantum computing}} (2025)
  (Preprint {\href{https://arxiv.org/abs/2503.08626}{arXiv:2503.08626}})

\bibitem{Fowler-Wright_2022}
P~Fowler-Wright, B~W Lovett and J~Keeling,  \emph{{Efficient Many-Body
  Non-Markovian Dynamics of Organic Polaritons}},
  \href{https://doi.org/10.1103/PhysRevLett.129.173001}{Phys. Rev. Lett. {\bf
  129}, 173001 (2022)} (Preprint
  {\href{https://arxiv.org/abs/2112.09003}{arXiv:2112.09003}})

\bibitem{Wiercinski_2023}
J~Wiercinski, E~M Gauger and M~Cygorek,  \emph{{Phonon coupling versus pure
  dephasing in the photon statistics of cooperative emitters}},
  \href{https://doi.org/10.1103/PhysRevResearch.5.013176}{Phys. Rev. Res. {\bf
  5}, 013176 (2023)} (Preprint
  {\href{https://arxiv.org/abs/2208.14549}{arXiv:2208.14549}})

\bibitem{Wiercinski_2024}
J~Wiercinski, M~Cygorek and E~M Gauger,  \emph{{Role of polaron dressing in
  superradiant emission dynamics}},
  \href{https://doi.org/10.1103/PhysRevResearch.6.033231}{Phys. Rev. Res. {\bf
  6}, 033231 (2024)} (Preprint
  {\href{https://arxiv.org/abs/2403.05533}{arXiv:2403.05533}})

\bibitem{Bracht_2023}
T~K Bracht, M~Cygorek, T~Seidelmann, V~M Axt and D~E Reiter,
  \emph{{Temperature-independent almost perfect photon entanglement from
  quantum dots via the SUPER scheme}},
  \href{https://doi.org/10.1364/OPTICAQ.498559}{Opt. Quantum {\bf 1}, 103
  (2023)} (Preprint
  {\href{https://arxiv.org/abs/2307.00304}{arXiv:2307.00304}})

\bibitem{deWit_2024}
R~{de Wit}, J~Keeling, B~W Lovett and A~W Chin,  \emph{{Process Tensor
  Approaches to Modeling Two-Dimensional Spectroscopy}} (2024) (Preprint
  {\href{https://arxiv.org/abs/2402.15454}{arXiv:2402.15454}})

\bibitem{Boos_2024}
K~Boos, S~K Kim, T~Bracht, F~Sbresny, J~M Kaspari, M~Cygorek, H~Riedl, F~W
  Bopp, W~Rauhaus, C~Calcagno, J~J Finley, D~E Reiter and K~M{\"u}ller,
  \emph{{Signatures of Dynamically Dressed States}},
  \href{https://doi.org/10.1103/PhysRevLett.132.053602}{Phys. Rev. Lett. {\bf
  132}, 053602 (2024)} (Preprint
  {\href{https://arxiv.org/abs/2305.15827}{arXiv:2305.15827}})

\bibitem{reichl2021transition}
L~Reichl, {\href{https://doi.org/10.1007/978-3-030-63534-3}{\textit{{The
  Transition to Chaos: Conservative Classical and Quantum Systems}}}} (Cham,
  Switzerland: Springer) (2021)

\bibitem{Sekino_2008}
Y~Sekino and L~Susskind,  \emph{{Fast Scramblers}},
  \href{https://doi.org/10.1088/1126-6708/2008/10/065}{J. High Energy Phys.
  {\bf 2008}, 065 (2008)} (Preprint
  {\href{https://arxiv.org/abs/0808.2096}{arXiv:0808.2096}})

\bibitem{Lashkari_2013}
N~Lashkari, D~Stanford, M~Hastings, T~Osborne and P~Hayden,  \emph{{Towards the
  Fast Scrambling Conjecture}},
  \href{https://doi.org/10.1007/JHEP04(2013)022}{J. High Energy Phys. {\bf
  2013}, 22 (2013)} (Preprint
  {\href{https://arxiv.org/abs/1111.6580}{arXiv:1111.6580}})

\bibitem{Vikram_2024}
A~Vikram and V~Galitski,  \emph{{Exact Universal Bounds on Quantum Dynamics and
  Fast Scrambling}},
  \href{https://doi.org/10.1103/PhysRevLett.132.040402}{Phys. Rev. Lett. {\bf
  132}, 040402 (2024)} (Preprint
  {\href{https://arxiv.org/abs/2212.14021}{arXiv:2212.14021}})

\bibitem{Gogolin_2016}
C~Gogolin and J~Eisert,  \emph{{Equilibration, Thermalisation, and the
  Emergence of Statistical Mechanics in Closed Quantum Systems}},
  \href{https://doi.org/10.1088/0034-4885/79/5/056001}{Rep. Prog. Phys. {\bf
  79}, 056001 (2016)} (Preprint
  {\href{https://arxiv.org/abs/1503.07538}{arXiv:1503.07538}})

\bibitem{Deutsch_2018}
J~M Deutsch,  \emph{{Eigenstate Thermalization Hypothesis}},
  \href{https://doi.org/10.1088/1361-6633/aac9f1}{Rep. Prog. Phys. {\bf 81},
  082001 (2018)} (Preprint
  {\href{https://arxiv.org/abs/1805.01616}{arXiv:1805.01616}})

\bibitem{PhysRevResearch.5.033126}
A~Vikram and V~Galitski,  \emph{{Dynamical Quantum Ergodicity from Energy Level
  Statistics}},  \href{https://doi.org/10.1103/PhysRevResearch.5.033126}{Phys.
  Rev. Res. {\bf 5}, 033126 (2023)} (Preprint
  {\href{https://arxiv.org/abs/2205.05704}{arXiv:2205.05704}})

\bibitem{Dowling_2022}
N~Dowling and K~Modi,  \emph{{Operational Metric for Quantum Chaos and the
  Corresponding Spatiotemporal-Entanglement Structure}},
  \href{https://doi.org/10.1103/PRXQuantum.5.010314}{PRX Quantum {\bf 5},
  010314 (2024)} (Preprint
  {\href{https://arxiv.org/abs/2210.14926}{arXiv:2210.14926}})

\bibitem{Hayden_2007}
P~Hayden and J~Preskill,  \emph{{Black Holes as Mirrors: Quantum Information in
  Random Subsystems}},  \href{https://doi.org/10.1088/1126-6708/2007/09/120}{J.
  High Energy Phys. {\bf 2007}, 120 (2007)} (Preprint
  {\href{https://arxiv.org/abs/0708.4025}{arXiv:0708.4025}})

\bibitem{Shenker_2014}
S~H Shenker and D~Stanford,  \emph{{Black Holes and the Butterfly Effect}},
  \href{https://doi.org/10.1007/JHEP03(2014)067}{J. High Energy Phys. {\bf
  2014}, 67 (2014)} (Preprint
  {\href{https://arxiv.org/abs/1306.0622}{arXiv:1306.0622}})

\bibitem{Maldacena_2016}
J~Maldacena, S~H Shenker and D~Stanford,  \emph{{A Bound on Chaos}},
  \href{https://doi.org/10.1007/JHEP08(2016)106}{J. High Energy Phys. {\bf
  2016}, 106 (2016)} (Preprint
  {\href{https://arxiv.org/abs/1503.01409}{arXiv:1503.01409}})

\bibitem{Hosur_2016}
P~Hosur, X-L Qi, D~A Roberts and B~Yoshida,  \emph{{Chaos in Quantum
  Channels}},  \href{https://doi.org/10.1007/JHEP02(2016)004}{J. High Energy
  Phys. {\bf 2016}, 4 (2016)} (Preprint
  {\href{https://arxiv.org/abs/1511.04021}{arXiv:1511.04021}})

\bibitem{Roberts_2017}
D~A Roberts and B~Yoshida,  \emph{{Chaos and Complexity by Design}},
  \href{https://doi.org/10.1007/JHEP04(2017)121}{J. High Energy Phys. {\bf
  2017}, 121 (2017)} (Preprint
  {\href{https://arxiv.org/abs/1610.04903}{arXiv:1610.04903}})

\bibitem{COTLER2018318}
J~S Cotler, D~Ding and G~R Penington,  \emph{{Out-of-time-order operators and
  the butterfly effect}},
  \href{https://doi.org/10.1016/j.aop.2018.07.020}{Ann. Phys. {\bf 396}, 318
  (2018)} (Preprint
  {\href{https://arxiv.org/abs/1704.02979}{arXiv:1704.02979}})

\bibitem{Garcia-Mata_2023}
I~Garc\'{\i}a-Mata, R~A Jalabert and D~A Wisniacki,  \emph{{Out-of-Time-Order
  Correlations and Quantum Chaos}} (2023) (Preprint
  {\href{https://arxiv.org/abs/2209.07965}{arXiv:2209.07965}})

\bibitem{Hashimoto_2017}
K~Hashimoto, K~Murata and R~Yoshii,  \emph{{Out-of-Time-Order Correlators in
  Quantum Mechanics}},  \href{https://doi.org/10.1007/JHEP10(2017)138}{J. High
  Energy Phys. {\bf 2017}, 138 (2017)} (Preprint
  {\href{https://arxiv.org/abs/1703.09435}{arXiv:1703.09435}})

\bibitem{PhysRevLett.118.086801}
E~B Rozenbaum, S~Ganeshan and V~Galitski,  \emph{{Lyapunov Exponent and
  Out-of-Time-Ordered Correlator's Growth Rate in a Chaotic System}},
  \href{https://doi.org/10.1103/PhysRevLett.118.086801}{Phys. Rev. Lett. {\bf
  118}, 086801 (2017)} (Preprint
  {\href{https://arxiv.org/abs/1609.01707}{arXiv:1609.01707}})

\bibitem{Balachandran2021-af}
V~Balachandran, G~Benenti, G~Casati and D~Poletti,  \emph{{From the eigenstate
  thermalization hypothesis to algebraic relaxation of OTOCs in systems with
  conserved quantities}},
  \href{https://doi.org/10.1103/PhysRevB.104.104306}{Phys. Rev. B {\bf 104},
  104306 (2021)} (Preprint
  {\href{https://arxiv.org/abs/2103.04595}{arXiv:2103.04595}})

\bibitem{e25010059}
V~Balachandran and D~Poletti,  \emph{{Relaxation Exponents of OTOCs and Overlap
  with Local Hamiltonians}},  \href{https://doi.org/10.3390/e25010059}{Entropy
  {\bf 25}, 59 (2023)} (Preprint
  {\href{https://arxiv.org/abs/2211.09965}{arXiv:2211.09965}})

\bibitem{PhysRevB.107.235421}
V~Balachandran, L~F Santos, M~Rigol and D~Poletti,  \emph{{Slow Relaxation of
  Out-of-Time-Ordered Correlators in Interacting Integrable and Nonintegrable
  Spin-$\frac{1}{2}$ XYZ Chains}},
  \href{https://doi.org/10.1103/PhysRevB.107.235421}{Phys. Rev. B {\bf 107},
  235421 (2023)} (Preprint
  {\href{https://arxiv.org/abs/2211.07073}{arXiv:2211.07073}})

\bibitem{Dowling_2023}
N~Dowling, P~Kos and K~Modi,  \emph{{Scrambling Is Necessary but Not Sufficient
  for Chaos}},  \href{https://doi.org/10.1103/PhysRevLett.131.180403}{Phys.
  Rev. Lett. {\bf 131}, 180403 (2023)} (Preprint
  {\href{https://arxiv.org/abs/2304.07319}{arXiv:2304.07319}})

\bibitem{Zonnios_2022}
M~Zonnios, J~Levinsen, M~M Parish, F~A Pollock and K~Modi,  \emph{{Signatures
  of Quantum Chaos in an Out-of-Time-Order Tensor}},
  \href{https://doi.org/10.1103/PhysRevLett.128.150601}{Phys. Rev. Lett. {\bf
  128}, 150601 (2022)} (Preprint
  {\href{https://arxiv.org/abs/2105.08282}{arXiv:2105.08282}})

\bibitem{Prosen2007}
T~Prosen and M~{\v Z}nidari{\v c},  \emph{{Is the efficiency of classical
  simulations of quantum dynamics related to integrability?}},
  \href{https://doi.org/10.1103/PhysRevE.75.015202}{Phys. Rev. E {\bf 75},
  015202 (2007)} (Preprint
  {\href{https://arxiv.org/abs/quant-ph/0608057}{arXiv:quant-ph/0608057}})

\bibitem{Prosen2007a}
T~Prosen and I~Pi{\v z}orn,  \emph{{Operator space entanglement entropy in a
  transverse Ising chain}},
  \href{https://doi.org/10.1103/PhysRevA.76.032316}{Phys. Rev. A {\bf 76},
  032316 (2007)} (Preprint
  {\href{https://arxiv.org/abs/0706.2480}{arXiv:0706.2480}})

\bibitem{Bertini_2020}
B~Bertini, P~Kos and T~Prosen,  \emph{{Operator Entanglement in Local Quantum
  Circuits I: Chaotic Dual-Unitary Circuits}},
  \href{https://doi.org/10.21468/SciPostPhys.8.4.067}{SciPost Phys. {\bf 8},
  067 (2020)} (Preprint
  {\href{https://arxiv.org/abs/1909.07407}{arXiv:1909.07407}})

\bibitem{Bertini_2020-2}
B~Bertini, P~Kos and T~Prosen,  \emph{{Operator Entanglement in Local Quantum
  Circuits II: Solitons in Chains of Qubits}},
  \href{https://doi.org/10.21468/SciPostPhys.8.4.068}{SciPost Phys. {\bf 8},
  068 (2020)} (Preprint
  {\href{https://arxiv.org/abs/1909.07410}{arXiv:1909.07410}})

\bibitem{sakuldee_non-markovian_2018}
F~Sakuldee, S~Milz, F~A Pollock and K~Modi,  \emph{Non-{Markovian} quantum
  control as coherent stochastic trajectories},
  \href{https://doi.org/10.1088/1751-8121/aabb1e}{J. Phys. A {\bf 51}, 414014
  (2018)} (Preprint
  {\href{https://arxiv.org/abs/1802.03190}{arXiv:1802.03190}})

\bibitem{Alicki1994-dc}
R~Alicki and M~Fannes,  \emph{{Defining quantum dynamical entropy}},
  \href{https://doi.org/10.1007/BF00761125}{Lett. Math. Phys. {\bf 32}, 75
  (1994)}

\bibitem{Slomczynski1994-ns}
W~S{\l}omczy{\'n}ski and K~{\.Z}yczkowski,  \emph{{Quantum chaos: An entropy
  approach}},  \href{https://doi.org/10.1063/1.530704}{J. Math. Phys. {\bf 35},
  5674 (1994)}

\bibitem{Benatti_2004}
F~Benatti, V~Cappellini and F~Zertuche,  \emph{{Quantum dynamical entropies in
  discrete classical chaos}},
  \href{https://doi.org/10.1088/0305-4470/37/1/007}{J. Phys. A: Math. Gen. {\bf
  37}, 105 (2003)} (Preprint
  {\href{https://arxiv.org/abs/math-ph/0308033}{arXiv:math-ph/0308033}})

\bibitem{cotler_superdensity_2017}
J~Cotler, C-M Jian, X-L Qi and F~Wilczek,  \emph{{Superdensity operators for
  spacetime quantum mechanics}},
  \href{https://doi.org/10.1007/JHEP09(2018)093}{J. High Energy Phys. {\bf
  2018}, 93 (2018)} (Preprint
  {\href{https://arxiv.org/abs/1711.03119}{arXiv:1711.03119}})

\bibitem{Lindblad1986}
G~Lindblad,
  {\href{https://doi.org/10.1007/978-1-4684-5221-1_22}{\textit{{Quantum
  Ergodicity and Chaos}}}}, {\em {Fundamental Aspects of Quantum Theory}\/}, ed
  V~Gorini and A~Frigerio (Boston, MA: Springer USA), p 199 (1986)

\bibitem{Banuls_2009}
M~C Ba{\~n}uls, M~B Hastings, F~Verstraete and J~I Cirac,  \emph{{Matrix
  Product States for Dynamical Simulation of Infinite Chains}},
  \href{https://doi.org/10.1103/PhysRevLett.102.240603}{Phys. Rev. Lett. {\bf
  102}, 240603 (2009)} (Preprint
  {\href{https://arxiv.org/abs/0904.1926}{arXiv:0904.1926}})

\bibitem{Muller-Hermes_2012}
A~M{\"u}ller-Hermes, J~I Cirac and M~C Ba{\~n}uls,  \emph{{Tensor network
  techniques for the computation of dynamical observables in one-dimensional
  quantum spin systems}},
  \href{https://doi.org/10.1088/1367-2630/14/7/075003}{New J. Phys. {\bf 14},
  075003 (2012)} (Preprint
  {\href{https://arxiv.org/abs/1204.5080}{arXiv:1204.5080}})

\bibitem{Hastings_2015}
M~B Hastings and R~Mahajan,  \emph{{Connecting entanglement in time and space:
  Improving the folding algorithm}},
  \href{https://doi.org/10.1103/PhysRevA.91.032306}{Phys. Rev. A {\bf 91},
  032306 (2015)} (Preprint
  {\href{https://arxiv.org/abs/1411.7950}{arXiv:1411.7950}})

\bibitem{PhysRevResearch.6.033021}
S~Carignano, C~R Marim{\'o}n and L~Tagliacozzo,  \emph{{Temporal entropy and
  the complexity of computing the expectation value of local operators after a
  quench}},  \href{https://doi.org/10.1103/PhysRevResearch.6.033021}{Phys. Rev.
  Res. {\bf 6}, 033021 (2024)} (Preprint
  {\href{https://arxiv.org/abs/2307.11649}{arXiv:2307.11649}})

\bibitem{Lerose_2021}
A~Lerose, M~Sonner and D~A Abanin,  \emph{{Scaling of temporal entanglement in
  proximity to integrability}},
  \href{https://doi.org/10.1103/PhysRevB.104.035137}{Phys. Rev. B {\bf 104},
  035137 (2021)} (Preprint
  {\href{https://arxiv.org/abs/2104.07607}{arXiv:2104.07607}})

\bibitem{Sonner_2021}
M~Sonner, A~Lerose and D~A Abanin,  \emph{{Influence functional of many-body
  systems: Temporal entanglement and matrix-product state representation}},
  \href{https://doi.org/10.1016/j.aop.2021.168677}{Ann. Phys. {\bf 435}, 168677
  (2021)} (Preprint
  {\href{https://arxiv.org/abs/2103.13741}{arXiv:2103.13741}})

\bibitem{Thoenniss_2023}
J~Thoenniss, M~Sonner, A~Lerose and D~A Abanin,  \emph{{Efficient method for
  quantum impurity problems out of equilibrium}},
  \href{https://doi.org/10.1103/PhysRevB.107.L201115}{Phys. Rev. B {\bf 107},
  L201115 (2023)} (Preprint
  {\href{https://arxiv.org/abs/2211.10272}{arXiv:2211.10272}})

\bibitem{Foligno_2023}
A~Foligno, T~Zhou and B~Bertini,  \emph{{Temporal Entanglement in Chaotic
  Quantum Circuits}},  \href{https://doi.org/10.1103/PhysRevX.13.041008}{Phys.
  Rev. X {\bf 13}, 041008 (2023)} (Preprint
  {\href{https://arxiv.org/abs/2302.08502}{arXiv:2302.08502}})

\bibitem{Leone_2021}
L~Leone, S~F~E Oliviero, Y~Zhou and A~Hamma,  \emph{{Quantum Chaos is
  Quantum}},  \href{https://doi.org/10.22331/q-2021-05-04-453}{Quantum {\bf 5},
  453 (2021)} (Preprint
  {\href{https://arxiv.org/abs/2102.08406}{arXiv:2102.08406}})

\bibitem{1910.14646}
A~Bouland, B~Fefferman and U~Vazirani,  \emph{{Computational pseudorandomness,
  the wormhole growth paradox, and constraints on the AdS/CFT duality}} (2019)
  (Preprint {\href{https://arxiv.org/abs/1910.14646}{arXiv:1910.14646}})

\bibitem{2308.04494}
J~K Taylor and I~P McCulloch,  \emph{{Wavefunction branching: when you can't
  tell pure states from mixed states}} (2023) (Preprint
  {\href{https://arxiv.org/abs/2308.04494}{arXiv:2308.04494}})

\bibitem{2310.06755}
P~Strasberg and J~Schindler,  \emph{{Shearing Off the Tree: Emerging Branch
  Structure and Born's Rule in an Equilibrated Multiverse}} (2023) (Preprint
  {\href{https://arxiv.org/abs/2310.06755}{arXiv:2310.06755}})

\bibitem{PhysRevX.9.031009}
B~Skinner, J~Ruhman and A~Nahum,  \emph{{Measurement-Induced Phase Transitions
  in the Dynamics of Entanglement}},
  \href{https://doi.org/10.1103/PhysRevX.9.031009}{Phys. Rev. X {\bf 9}, 031009
  (2019)} (Preprint
  {\href{https://arxiv.org/abs/1808.05953}{arXiv:1808.05953}})

\bibitem{PRXQuantum.4.030322}
M~Ippoliti and W~W Ho,  \emph{{Dynamical Purification and the Emergence of
  Quantum State Designs from the Projected Ensemble}},
  \href{https://doi.org/10.1103/PRXQuantum.4.030322}{PRX Quantum {\bf 4},
  030322 (2023)} (Preprint
  {\href{https://arxiv.org/abs/2204.13657}{arXiv:2204.13657}})

\bibitem{odonovan2025_diagnosing}
P~O'Donovan, N~Dowling, K~Modi and M~T Mitchison,  \emph{{Diagnosing chaos with
  projected ensembles of process tensors}} (2025) (Preprint
  {\href{https://arxiv.org/abs/2502.13930}{arXiv:2502.13930}})

\bibitem{van_kampen_quantum_1954}
N~G {van Kampen},  \emph{{Quantum statistics of irreversible processes}},
  \href{https://doi.org/10.1016/S0031-8914(54)80074-7}{Physica {\bf 20}, 603
  (1954)}

\bibitem{Strasberg_2019_PRE_Operational}
P~Strasberg,  \emph{{Operational approach to quantum stochastic
  thermodynamics}},  \href{https://doi.org/10.1103/PhysRevE.100.022127}{Phys.
  Rev. E {\bf 100}, 022127 (2019)} (Preprint
  {\href{https://arxiv.org/abs/1810.00698}{arXiv:1810.00698}})

\bibitem{Strasberg_2019_PRE_Stochastic}
P~Strasberg and A~Winter,  \emph{{Stochastic thermodynamics with arbitrary
  interventions}},  \href{https://doi.org/10.1103/PhysRevE.100.022135}{Phys.
  Rev. E {\bf 100}, 022135 (2019)} (Preprint
  {\href{https://arxiv.org/abs/1905.07990}{arXiv:1905.07990}})

\bibitem{Strasberg_2019_PRL}
P~Strasberg,  \emph{{Repeated Interactions and Quantum Stochastic
  Thermodynamics at Strong Coupling}},
  \href{https://doi.org/10.1103/PhysRevLett.123.180604}{Phys. Rev. Lett. {\bf
  123}, 180604 (2019)} (Preprint
  {\href{https://arxiv.org/abs/1907.01804}{arXiv:1907.01804}})

\bibitem{strasberg_thermodynamics_2020}
P~Strasberg,  \emph{Thermodynamics of {Quantum} {Causal} {Models}: {An}
  {Inclusive}, {Hamiltonian} {Approach}},
  \href{https://quantum-journal.org/papers/q-2020-03-02-240/}{Quantum {\bf 4},
  240 (2020)} (Preprint
  {\href{https://arxiv.org/abs/1911.01730}{arXiv:1911.01730}})

\bibitem{strasberg_quantum_2022}
P~Strasberg,
  {\href{https://doi.org/10.1093/oso/9780192895585.001.0001}{\textit{Quantum
  {Stochastic} {Thermodynamics}: {Foundations} and {Selected} {Applications}}}}
  (Oxford, UK: OUP Oxford) (2022)

\bibitem{Zambon_2024_Thermodynamic}
G~Zambon and G~Adesso,  \emph{{Quantum processes as thermodynamic resources:
  the role of non-Markovianity}} (2024) (Preprint
  {\href{https://arxiv.org/abs/2411.05559}{arXiv:2411.05559}})

\bibitem{Figueroa-Romero_2020}
P~Figueroa-Romero, K~Modi and F~A Pollock,  \emph{{Equilibration on average in
  quantum processes with finite temporal resolution}},
  \href{https://doi.org/10.1103/PhysRevE.102.032144}{Phys. Rev. E {\bf 102},
  032144 (2020)} (Preprint
  {\href{https://arxiv.org/abs/1905.08469}{arXiv:1905.08469}})

\bibitem{Dowling_2021_Equilibration}
N~Dowling, P~Figueroa-Romero, F~A Pollock, P~Strasberg and K~Modi,
  \emph{{Equilibration of multitime quantum processes in finite time
  intervals}},  \href{https://doi.org/10.21468/SciPostPhysCore.6.2.043}{SciPost
  Phys. Core {\bf 6}, 043 (2023)} (Preprint
  {\href{https://arxiv.org/abs/2112.01099}{arXiv:2112.01099}})

\bibitem{Dowling_2021_Relaxation}
N~Dowling, P~Figueroa-Romero, F~A Pollock, P~Strasberg and K~Modi,
  \emph{{Relaxation of Multitime Statistics in Quantum Systems}},
  \href{https://doi.org/10.22331/q-2023-06-01-1027}{Quantum {\bf 7}, 1027
  (2023)} (Preprint
  {\href{https://arxiv.org/abs/2108.07420}{arXiv:2108.07420}})

\bibitem{Figueroa-Romero_2019}
P~Figueroa-Romero, K~Modi and F~A Pollock,  \emph{{Almost Markovian processes
  from closed dynamics}},
  \href{https://doi.org/10.22331/q-2019-04-30-136}{Quantum {\bf 3}, 136 (2019)}
  (Preprint {\href{https://arxiv.org/abs/1802.10344}{arXiv:1802.10344}})

\bibitem{Figueroa-Romero_2021}
P~Figueroa-Romero, F~A Pollock and K~Modi,  \emph{{Markovianization with
  approximate unitary designs}},
  \href{https://doi.org/10.1038/s42005-021-00629-w}{Commun. Phys. {\bf 4}, 127
  (2021)} (Preprint
  {\href{https://arxiv.org/abs/2004.07620}{arXiv:2004.07620}})

\bibitem{Xu_2018}
X~Xu, C~Guo and D~Poletti,  \emph{{Interplay of interaction and disorder in the
  steady state of an open quantum system}},
  \href{https://doi.org/10.1103/PhysRevB.97.140201}{Phys. Rev. B {\bf 97},
  140201 (2018)} (Preprint
  {\href{https://arxiv.org/abs/1709.08934}{arXiv:1709.08934}})

\bibitem{Xu_2023}
X~Xu, C~Guo and D~Poletti,  \emph{{Emergence of steady currents due to strong
  prethermalization}},
  \href{https://doi.org/10.1103/PhysRevA.107.022220}{Phys. Rev. A {\bf 107},
  022220 (2023)} (Preprint
  {\href{https://arxiv.org/abs/2206.12665}{arXiv:2206.12665}})

\bibitem{Zhang_2024}
P~Zhang, Y~Gao, X~Xu, N~Wang, H~Dong, C~Guo, J~Deng, X~Zhang, J~Chen, S~Xu,
  K~Wang, Y~Wu, C~Zhang, F~Jin, X~Zhu, A~Zhang, Y~Zou, Z~Tan, Z~Cui, Z~Zhu,
  F~Shen, T~Li, J~Zhong, Z~Bao, L~Zhao, J~Hao, H~Li, Z~Wang, C~Song, Q~Guo,
  H~Wang and D~Poletti,  \emph{{Emergence of steady quantum transport in a
  superconducting processor}},
  \href{https://doi.org/10.1038/s41467-024-54332-9}{Nat. Commun. {\bf 15},
  10115 (2024)} (Preprint
  {\href{https://arxiv.org/abs/2411.06794}{arXiv:2411.06794}})

\bibitem{odonovan2024_qme}
P~O'Donovan, P~Strasberg, K~Modi, J~Goold and M~T Mitchison,  \emph{{Quantum
  master equation from the eigenstate thermalization hypothesis}} (2024)
  (Preprint {\href{https://arxiv.org/abs/2411.07706}{arXiv:2411.07706}})

\bibitem{barnum_local_2010}
H~Barnum, S~Beigi, S~Boixo, M~B Elliott and S~Wehner,  \emph{Local {Quantum}
  {Measurement} and {No}-{Signaling} {Imply} {Quantum} {Correlations}},
  \href{https://doi.org/10.1103/PhysRevLett.104.140401}{Phys. Rev. Lett. {\bf
  104}, 140401 (2010)} (Preprint
  {\href{https://arxiv.org/abs/0910.3952}{arXiv:0910.3952}})

\bibitem{colnaghi_quantum_2012}
T~Colnaghi, G~M D'Ariano, S~Facchini and P~Perinotti,  \emph{{Quantum
  computation with programmable connections between gates}},
  \href{https://doi.org/10.1016/j.physleta.2012.08.028}{Phys. Lett. A {\bf
  376}, 2940 (2012)} (Preprint
  {\href{https://arxiv.org/abs/1109.5987}{arXiv:1109.5987}})

\bibitem{hardy_probability_2005}
L~Hardy,  \emph{Probability {Theories} with {Dynamic} {Causal} {Structure}: {A}
  {New} {Framework} for {Quantum} {Gravity}} (2005) (Preprint
  {\href{https://arxiv.org/abs/gr-qc/0509120}{arXiv:gr-qc/0509120}})

\bibitem{baumeler_reversible_2019}
{\"A}~Baumeler, F~Costa, T~C Ralph, S~Wolf and M~Zych,  \emph{Reversible time
  travel with freedom of choice},
  \href{https://doi.org/10.1088/1361-6382/ab4973}{Class. Quantum Gravity {\bf
  36}, 224002 (2019)} (Preprint
  {\href{https://arxiv.org/abs/1703.00779}{arXiv:1703.00779}})

\bibitem{zych_bells_2019}
M~Zych, F~Costa, I~Pikovski and {\v C}~Brukner,  \emph{{Bell's theorem for
  temporal order}},  \href{https://doi.org/10.1038/s41467-019-11579-x}{Nat.
  Commun. {\bf 10}, 3772 (2019)} (Preprint
  {\href{https://arxiv.org/abs/1708.00248}{arXiv:1708.00248}})

\bibitem{Abbott_2017}
A~A Abbott, J~Wechs, F~Costa and C~Branciard,  \emph{{Genuinely multipartite
  noncausality}},  \href{https://doi.org/10.22331/q-2017-12-14-39}{{Quantum}
  {\bf 1}, 39 (2017)} (Preprint
  {\href{https://arxiv.org/abs/1708.07663}{arXiv:1708.07663}})

\bibitem{almeida_guess_2010}
M~L Almeida, J-D Bancal, N~Brunner, A~Ac{\'i}n, N~Gisin and S~Pironio,
  \emph{Guess {Your} {Neighbor}'s {Input}: {A} {Multipartite} {Nonlocal} {Game}
  with {No} {Quantum} {Advantage}},
  \href{https://doi.org/10.1103/PhysRevLett.104.230404}{Phys. Rev. Lett. {\bf
  104}, 230404 (2010)} (Preprint
  {\href{https://arxiv.org/abs/1003.3844}{arXiv:1003.3844}})

\bibitem{baumeler_space_2016}
{\"A}~Baumeler and S~Wolf,  \emph{{The space of logically consistent classical
  processes without causal order}},
  \href{https://doi.org/10.1088/1367-2630/18/1/013036}{New J. Phys. {\bf 18},
  013036 (2016)} (Preprint
  {\href{https://arxiv.org/abs/1507.01714}{arXiv:1507.01714}})

\bibitem{kunjwal_nonclassicality_2024}
R~Kunjwal and O~Oreshkov,  \emph{{Nonclassicality in correlations without
  causal order}} (2023) (Preprint
  {\href{https://arxiv.org/abs/2307.02565}{arXiv:2307.02565}})

\bibitem{kunjwal_generalizing_2024}
R~Kunjwal and O~Oreshkov,  \emph{{Generalizing Bell nonlocality without global
  causal assumptions}} (2024) (Preprint
  {\href{https://arxiv.org/abs/2411.11397}{arXiv:2411.11397}})

\bibitem{brukner_bounding_2015}
{\v C}~Brukner,  \emph{{Bounding quantum correlations with indefinite causal
  order}},  \href{https://doi.org/10.1088/1367-2630/17/8/083034}{New J. Phys.
  {\bf 17}, 083034 (2015)} (Preprint
  {\href{https://arxiv.org/abs/1404.0721}{arXiv:1404.0721}})

\bibitem{bhattacharya_biased_2015}
S~S Bhattacharya and M~Banik,  \emph{Biased {Non}-{Causal} {Game}} (2015)
  (Preprint {\href{https://arxiv.org/abs/1509.02721}{arXiv:1509.02721}})

\bibitem{liu_tsirelson_2024}
Z~Liu and G~Chiribella,  \emph{{Tsirelson bounds for quantum correlations with
  indefinite causal order}} (2024) (Preprint
  {\href{https://arxiv.org/abs/2403.02749}{arXiv:2403.02749}})

\bibitem{bavaresco_indefinite_2024}
J~Bavaresco, {\"A}~Baumeler, Y~Guryanova and C~Budroni,  \emph{Indefinite
  causal order in boxworld theories} (2024) (Preprint
  {\href{https://arxiv.org/abs/2411.00951}{arXiv:2411.00951}})

\bibitem{Vilasini2019}
V~Vilasini, N~Nurgalieva and L~{del Rio},  \emph{{Multi-agent paradoxes beyond
  quantum theory}},  \href{https://doi.org/10.1088/1367-2630/ab4fc4}{New J.
  Phys. {\bf 21}, 113028 (2019)} (Preprint
  {\href{https://arxiv.org/abs/1904.06247}{arXiv:1904.06247}})

\bibitem{Abbott_2024}
A~A Abbott, M~Mhalla and P~Pocreau,  \emph{{Quantum query complexity of Boolean
  functions under indefinite causal order}},
  \href{https://doi.org/10.1103/PhysRevResearch.6.L032020}{Phys. Rev. Res. {\bf
  6}, L032020 (2024)} (Preprint
  {\href{https://arxiv.org/abs/2307.10285}{arXiv:2307.10285}})

\bibitem{baumeler_device-independent_2016}
{\"A}~Baumeler and S~Wolf,  \emph{Device-independent test of causal order and
  relations to fixed-points},
  \href{https://doi.org/10.1088/1367-2630/18/3/035014}{New J. Phys. {\bf 18},
  035014 (2016)} (Preprint
  {\href{https://arxiv.org/abs/1511.05444}{arXiv:1511.05444}})

\bibitem{Vanrietvelde_2021}
A~Vanrietvelde, H~Kristj{\' a}nsson and J~Barrett,  \emph{{Routed quantum
  circuits}},  \href{https://doi.org/10.22331/q-2021-07-13-503}{Quantum {\bf
  5}, 503 (2021)} (Preprint
  {\href{https://arxiv.org/abs/2011.08120}{arXiv:2011.08120}})

\bibitem{Vanrietvelde_2022}
A~Vanrietvelde, N~Ormrod, H~Kristj{\' a}nsson and J~Barrett,  \emph{{Consistent
  circuits for indefinite causal order}} (2022) (Preprint
  {\href{https://arxiv.org/abs/2206.10042}{arXiv:2206.10042}})

\bibitem{Arrighi_2021}
P~Arrighi, C~Cedzich, M~Costes, U~R\'{e}mond and B~Valiron,  \emph{{Addressable
  Quantum Gates}},  \href{https://doi.org/10.1145/3581760}{ACM Trans. Quantum
  Comput. {\bf 4} (2023)} (Preprint
  {\href{https://arxiv.org/abs/2109.08050}{arXiv:2109.08050}})

\bibitem{Bavaresco_2024_Switch}
J~Bavaresco, S~Yoshida, T~Odake, H~Kristj{\'a}nsson, P~Taranto, M~Murao and M~T
  Quintino,  \emph{{Can the quantum switch be deterministically simulated?}}
  (2024) (Preprint {\href{https://arxiv.org/abs/2409.18202}{arXiv:2409.18202}})

\bibitem{aharonov2005}
Y~Aharonov and E~Y Gruss,  \emph{{Two-time interpretation of quantum
  mechanics}} (2005) (Preprint
  {\href{https://arxiv.org/abs/quant-ph/0507269}{arXiv:quant-ph/0507269}})

\bibitem{Xiao_2021}
Y~Xiao, K~Sengupta, S~Yang and G~Gour,  \emph{{Uncertainty principle of quantum
  processes}},  \href{https://doi.org/10.1103/PhysRevResearch.3.023077}{Phys.
  Rev. Res. {\bf 3}, 023077 (2021)} (Preprint
  {\href{https://arxiv.org/abs/2004.05315}{arXiv:2004.05315}})

\bibitem{Xiao_2023}
Y~Xiao, Y~Yang, X~Wang, Q~Liu and M~Gu,  \emph{{Quantum Uncertainty Principles
  for Measurements with Interventions}},
  \href{https://doi.org/10.1103/PhysRevLett.130.240201}{Phys. Rev. Lett. {\bf
  130}, 240201 (2023)} (Preprint
  {\href{https://arxiv.org/abs/2305.07914}{arXiv:2305.07914}})

\bibitem{giarmatzi_quantum_2018}
C~Giarmatzi and F~Costa,  \emph{{A quantum causal discovery algorithm}},
  \href{https://doi.org/10.1038/s41534-018-0062-6}{npj Quantum Inf. {\bf 4}, 17
  (2018)} (Preprint
  {\href{https://arxiv.org/abs/1704.00800}{arXiv:1704.00800}})

\bibitem{bai_quantum_2022}
G~Bai, Y-D Wu, Y~Zhu, M~Hayashi and G~Chiribella,  \emph{{Quantum causal
  unravelling}},  \href{https://doi.org/10.1038/s41534-022-00578-4}{npj Quantum
  Inf. {\bf 8}, 69 (2022)} (Preprint
  {\href{https://arxiv.org/abs/2109.13166}{arXiv:2109.13166}})

\bibitem{maclean_quantum-coherent_2017}
J-P~W MacLean, K~Ried, R~W Spekkens and K~J Resch,  \emph{{Quantum-coherent
  mixtures of causal relations}},
  \href{https://doi.org/10.1038/ncomms15149}{Nat. Commun. {\bf 8}, 15149
  (2017)} (Preprint
  {\href{https://arxiv.org/abs/1606.04523}{arXiv:1606.04523}})

\bibitem{Tselentis_2023}
E-E Tselentis and {\"A}~Baumeler,  \emph{{Admissible Causal Structures and
  Correlations}},  \href{https://doi.org/10.1103/PRXQuantum.4.040307}{PRX
  Quantum {\bf 4}, 040307 (2023)} (Preprint
  {\href{https://arxiv.org/abs/2210.12796}{arXiv:2210.12796}})

\bibitem{Baumeler_2020}
{\"A}~Baumeler and E-E Tselentis,
  {\href{https://doi.org/10.4204/EPTCS.340.1}{\textit{{Equivalence of
  Grandfather and Information Antinomy Under Intervention}}}}, {\em Proceedings
  17th International Conference on Quantum Physics and Logic (QPL)\/},
  Electronic Proceedings in Theoretical Computer Science (EPTCS) (Waterloo NSW,
  Australia: Open Publishing Association) p~1 (2020) (Preprint
  {\href{https://arxiv.org/abs/2004.12921}{arXiv:2004.12921}})

\bibitem{Bennett_1999}
C~H Bennett, D~P DiVincenzo, C~A Fuchs, T~Mor, E~Rains, P~W Shor, J~A Smolin
  and W~K Wootters,  \emph{{Quantum nonlocality without entanglement}},
  \href{https://doi.org/10.1103/PhysRevA.59.1070}{Phys. Rev. A {\bf 59}, 1070
  (1999)} (Preprint
  {\href{https://arxiv.org/abs/quant-ph/9804053}{arXiv:quant-ph/9804053}})

\bibitem{Akibue_2017}
S~Akibue, M~Owari, G~Kato and M~Murao,  \emph{{Entanglement-assisted classical
  communication can simulate classical communication without causal order}},
  \href{https://doi.org/10.1103/PhysRevA.96.062331}{Phys. Rev. A {\bf 96},
  062331 (2017)} (Preprint
  {\href{https://arxiv.org/abs/1602.08835}{arXiv:1602.08835}})

\bibitem{Kunjwal_2023}
R~Kunjwal and {\"A}~Baumeler,  \emph{{Trading Causal Order for Locality}},
  \href{https://doi.org/10.1103/PhysRevLett.131.120201}{Phys. Rev. Lett. {\bf
  131}, 120201 (2023)} (Preprint
  {\href{https://arxiv.org/abs/2202.00440}{arXiv:2202.00440}})

\bibitem{Steffinlongo_2025}
A~Steffinlongo and H~Dourdent,  \emph{{Simulating Noncausality with Quantum
  Control of Causal Orders}} (2025) (Preprint
  {\href{https://arxiv.org/abs/2502.15579}{arXiv:2502.15579}})

\bibitem{simmons_2024}
W~Simmons and A~Kissinger,  \emph{{A complete logic for causal consistency}}
  (2024) (Preprint {\href{https://arxiv.org/abs/2403.09297}{arXiv:2403.09297}})

\bibitem{chiribella_quantum_2019}
G~Chiribella and H~Kristj{\'a}nsson,  \emph{Quantum {Shannon} theory with
  superpositions of trajectories},
  \href{https://doi.org/10.1098/rspa.2018.0903}{Proc. R. Soc. A {\bf 475},
  20180903 (2019)} (Preprint
  {\href{https://arxiv.org/abs/1812.05292}{arXiv:1812.05292}})

\bibitem{Ormrod2023CausalStructure}
N~{Ormrod}, A~{Vanrietvelde} and J~{Barrett},  \emph{{Causal structure in the
  presence of sectorial constraints, with application to the quantum switch}},
  \href{https://doi.org/10.22331/q-2023-06-01-1028}{Quantum {\bf 7}, 1028
  (2023)} (Preprint
  {\href{https://arxiv.org/abs/2204.10273}{arXiv:2204.10273}})

\bibitem{Kabel2024_Gravity}
V~{Kabel}, A-C {de la Hamette}, L~{Apadula}, C~{Cepollaro}, H~{Gomes},
  J~{Butterfield} and {\v{C}}~{Brukner},  \emph{{Identification is Pointless:
  Quantum Reference Frames, Localisation of Events, and the Quantum Hole
  Argument}} (2024) (Preprint
  {\href{https://arxiv.org/abs/2402.10267}{arXiv:2402.10267}})

\bibitem{Vilasini_2024Realizing}
V~Vilasini and R~Renner,  \emph{{Fundamental Limits for Realizing Quantum
  Processes in Spacetime}},
  \href{https://doi.org/10.1103/PhysRevLett.133.080201}{Phys. Rev. Lett. {\bf
  133}, 080201 (2024)} (Preprint
  {\href{https://arxiv.org/abs/2408.13387}{arXiv:2408.13387}})

\bibitem{taddei_computational_2021}
M~M Taddei, J~Cari{\~n}e, D~Mart{\'i}nez, T~Garc{\'i}a, N~Guerrero, A~A Abbott,
  M~Ara{\'u}jo, C~Branciard, E~S G{\'o}mez, S~P Walborn, L~Aolita and G~Lima,
  \emph{Computational {Advantage} from the {Quantum} {Superposition} of
  {Multiple} {Temporal} {Orders} of {Photonic} {Gates}},
  \href{https://doi.org/10.1103/PRXQuantum.2.010320}{PRX Quantum {\bf 2},
  010320 (2021)} (Preprint
  {\href{https://arxiv.org/abs/2002.07817}{arXiv:2002.07817}})

\bibitem{Renner_2021}
M~J Renner and {\v C}~Brukner,  \emph{{Reassessing the computational advantage
  of quantum-controlled ordering of gates}},
  \href{https://doi.org/10.1103/PhysRevResearch.3.043012}{Phys. Rev. Res. {\bf
  3}, 043012 (2021)} (Preprint
  {\href{https://arxiv.org/abs/2102.11293}{arXiv:2102.11293}})

\bibitem{Renner_2022}
M~J Renner and {\v C}~Brukner,  \emph{{Computational Advantage from a Quantum
  Superposition of Qubit Gate Orders}},
  \href{https://doi.org/10.1103/PhysRevLett.128.230503}{Phys. Rev. Lett. {\bf
  128}, 230503 (2022)} (Preprint
  {\href{https://arxiv.org/abs/2112.14541}{arXiv:2112.14541}})

\bibitem{Goldberg_2023}
A~Z Goldberg, K~Heshami and L~L S{\'a}nchez-Soto,  \emph{{Evading noise in
  multiparameter quantum metrology with indefinite causal order}},
  \href{https://doi.org/10.1103/PhysRevResearch.5.033198}{Phys. Rev. Res. {\bf
  5}, 033198 (2023)} (Preprint
  {\href{https://arxiv.org/abs/2309.07220}{arXiv:2309.07220}})

\bibitem{guerin_exponential_2016}
P~A Gu{\'e}rin, A~Feix, M~Ara{\'u}jo and {\v C}~Brukner,  \emph{Exponential
  {Communication} {Complexity} {Advantage} from {Quantum} {Superposition} of
  the {Direction} of {Communication}},
  \href{https://doi.org/10.1103/PhysRevLett.117.100502}{Phys. Rev. Lett. {\bf
  117}, 100502 (2016)} (Preprint
  {\href{https://arxiv.org/abs/1605.07372}{arXiv:1605.07372}})

\bibitem{ebler_enhanced_2018}
D~Ebler, S~Salek and G~Chiribella,  \emph{Enhanced {Communication} with the
  {Assistance} of {Indefinite} {Causal} {Order}},
  \href{https://doi.org/10.1103/PhysRevLett.120.120502}{Phys. Rev. Lett. {\bf
  120}, 120502 (2018)} (Preprint
  {\href{https://arxiv.org/abs/1711.10165}{arXiv:1711.10165}})

\bibitem{Salek_2018}
S~{Salek}, D~{Ebler} and G~{Chiribella},  \emph{{Quantum communication in a
  superposition of causal orders}} (2018) (Preprint
  {\href{https://arxiv.org/abs/1809.06655}{arXiv:1809.06655}})

\bibitem{kristjansson_resource_2020}
H~Kristj{\'a}nsson, G~Chiribella, S~Salek, D~Ebler and M~Wilson,
  \emph{{Resource theories of communication}},
  \href{https://doi.org/10.1088/1367-2630/ab8ef7}{New J. Phys. {\bf 22}, 073014
  (2020)} (Preprint
  {\href{https://arxiv.org/abs/1910.08197}{arXiv:1910.08197}})

\bibitem{Chiribella_2021}
G~Chiribella, M~Banik, S~S Bhattacharya, T~Guha, M~Alimuddin, A~Roy, S~Saha,
  S~Agrawal and G~Kar,  \emph{{Indefinite causal order enables perfect quantum
  communication with zero capacity channels}},
  \href{https://doi.org/10.1088/1367-2630/abe7a0}{New J. Phys. {\bf 23}, 033039
  (2021)} (Preprint
  {\href{https://arxiv.org/abs/1810.10457}{arXiv:1810.10457}})

\bibitem{abbott_communication_2020}
A~A Abbott, J~Wechs, D~Horsman, M~Mhalla and C~Branciard,  \emph{{Communication
  through coherent control of quantum channels}},
  \href{https://doi.org/10.22331/q-2020-09-24-333}{Quantum {\bf 4}, 333 (2020)}
  (Preprint {\href{https://arxiv.org/abs/1810.09826}{arXiv:1810.09826}})

\bibitem{branciard_coherent_2021}
C~Branciard, A~Cl{\'e}ment, M~Mhalla and S~Perdrix,
  {\href{https://doi.org/10.4230/LIPIcs.MFCS.2021.22}{\textit{Coherent
  {Control} and {Distinguishability} of {Quantum} {Channels} via
  {PBS}-{Diagrams}}}}, {\em 46th {International} {Symposium} on {Mathematical}
  {Foundations} of {Computer} {Science} (MFCS)\/}, Leibniz {International}
  {Proceedings} in {Informatics} ({LIPIcs}) (Dagstuhl, Germany: Schloss
  Dagstuhl – Leibniz-Zentrum f{\"u}r Informatik) p~22 (2021) (Preprint
  {\href{https://arxiv.org/abs/2103.02073}{arXiv:2103.02073}})

\bibitem{Yin_2024_Enhancement}
Y~Mo, C~Zhu, Z~Liu, M~Jing and X~Wang,  \emph{{Enhancement of nonstabilizerness
  within indefinite causal order}},
  \href{https://doi.org/10.1103/PhysRevA.109.062428}{Phys. Rev. A {\bf 109},
  062428 (2024)} (Preprint
  {\href{https://arxiv.org/abs/2311.15494}{arXiv:2311.15494}})

\bibitem{Felce_2020}
D~Felce and V~Vedral,  \emph{{Quantum Refrigeration with Indefinite Causal
  Order}},  \href{https://doi.org/10.1103/PhysRevLett.125.070603}{Phys. Rev.
  Lett. {\bf 125}, 070603 (2020)} (Preprint
  {\href{https://arxiv.org/abs/2003.00794}{arXiv:2003.00794}})

\bibitem{Gogioso_2023}
S~Gogioso and N~Pinzani,  \emph{{The Geometry of Causality}} (2023) (Preprint
  {\href{https://arxiv.org/abs/2303.09017}{arXiv:2303.09017}})

\bibitem{baumann_no_2025}
V~Baumann, {\"A}~Baumeler and E-E Tselentis,  \emph{{No quantum advantage for
  violating fixed-order inequalities?}} (2025) (Preprint
  {\href{https://arxiv.org/abs/2412.17551}{arXiv:2412.17551}})

\bibitem{Kristjansson_2024_Simulating}
H~Kristj{\'a}nsson, T~Odake, S~Yoshida, P~Taranto, J~Bavaresco, M~T Quintino
  and M~Murao,  \emph{{Exponential separation in quantum query complexity of
  the quantum switch with respect to simulations with standard quantum
  circuits}} (2024) (Preprint
  {\href{https://arxiv.org/abs/2409.18420}{arXiv:2409.18420}})

\bibitem{Barrett_2021}
J~Barrett, R~Lorenz and O~Oreshkov,  \emph{{Cyclic quantum causal models}},
  \href{https://doi.org/10.1038/s41467-020-20456-x}{Nat. Commun. {\bf 12}, 885
  (2021)} (Preprint
  {\href{https://arxiv.org/abs/2002.12157}{arXiv:2002.12157}})

\bibitem{Yokojima_2021}
W~Yokojima, M~T Quintino, A~Soeda and M~Murao,  \emph{{Consequences of
  preserving reversibility in quantum superchannels}},
  \href{https://doi.org/10.22331/q-2021-04-26-441}{Quantum {\bf 5}, 441 (2021)}
  (Preprint {\href{https://arxiv.org/abs/2003.05682}{arXiv:2003.05682}})

\bibitem{Costa_2022}
F~Costa,  \emph{{A no-go theorem for superpositions of causal orders}},
  \href{https://doi.org/10.22331/q-2022-03-01-663}{Quantum {\bf 6}, 663 (2022)}
  (Preprint {\href{https://arxiv.org/abs/2008.06205}{arXiv:2008.06205}})

\bibitem{Milz_2018}
S~Milz, F~A Pollock and K~Modi,  \emph{{Reconstructing non-Markovian quantum
  dynamics with limited control}},
  \href{https://doi.org/10.1103/PhysRevA.98.012108}{Phys. Rev. A {\bf 98},
  012108 (2018)} (Preprint
  {\href{https://arxiv.org/abs/1610.02152}{arXiv:1610.02152}})

\bibitem{Chuang_1997}
I~L Chuang and M~A Nielsen,  \emph{{Prescription for experimental determination
  of the dynamics of a quantum black box}},
  \href{https://doi.org/10.1080/09500349708231894}{J. Mod. Opt. {\bf 44}, 2455
  (1997)} (Preprint
  {\href{https://arxiv.org/abs/quant-ph/9610001}{arXiv:quant-ph/9610001}})

\bibitem{Mohseni_2008}
M~Mohseni, A~T Rezakhani and D~A Lidar,  \emph{{Quantum-process tomography:
  Resource analysis of different strategies}},
  \href{https://doi.org/10.1103/PhysRevA.77.032322}{Phys. Rev. A {\bf 77},
  032322 (2008)} (Preprint
  {\href{https://arxiv.org/abs/quant-ph/0702131}{arXiv:quant-ph/0702131}})

\bibitem{Kuah_2007}
A-m Kuah, K~Modi, C~A Rodr{\'{\i}}guez-Rosario and E~C~G Sudarshan,  \emph{{How
  state preparation can affect a quantum experiment: Quantum process tomography
  for open systems}},  \href{https://doi.org/10.1103/PhysRevA.76.042113}{Phys.
  Rev. A {\bf 76}, 02113 (2007)} (Preprint
  {\href{https://arxiv.org/abs/0706.0394}{arXiv:0706.0394}})

\bibitem{white-rev}
G~White,  \emph{{Characterization and control of non-Markovian quantum noise}},
   \href{https://doi.org/10.1038/s42254-022-00446-2}{Nat. Rev. Phys. {\bf 4},
  287 (2022)}

\bibitem{Cramer_2010}
M~Cramer, M~B Plenio, S~T Flammia, R~Somma, D~Gross, S~D Bartlett,
  O~Landon-Cardinal, D~Poulin and Y-K Liu,  \emph{{Efficient quantum state
  tomography}},  \href{https://doi.org/10.1038/ncomms1147}{Nat. Commun. {\bf
  1}, 149 (2010)} (Preprint
  {\href{https://arxiv.org/abs/1101.4366}{arXiv:1101.4366}})

\bibitem{Aaronson_2017}
S~Aaronson,  \emph{{Shadow Tomography of Quantum States}} (2017) (Preprint
  {\href{https://arxiv.org/abs/1711.01053}{arXiv:1711.01053}})

\bibitem{huang-shadow}
H-Y Huang, R~Kueng and J~Preskill,  \emph{{Predicting many properties of a
  quantum system from very few measurements}},
  \href{https://doi.org/10.1038/s41567-020-0932-7}{Nat. Phys. {\bf 16}, 1050
  (2020)} (Preprint
  {\href{https://arxiv.org/abs/2002.08953}{arXiv:2002.08953}})

\bibitem{PhysRevLett.130.160401}
G~A~L White, K~Modi and C~D Hill,  \emph{{Filtering Crosstalk from Bath
  Non-Markovianity via Spacetime Classical Shadows}},
  \href{https://doi.org/10.1103/PhysRevLett.130.160401}{Phys. Rev. Lett. {\bf
  130}, 160401 (2023)} (Preprint
  {\href{https://arxiv.org/abs/2210.15333}{arXiv:2210.15333}})

\bibitem{whitethesis}
G~A~L White, {\href{https://arxiv.org/abs/2405.05416}{\textit{{Many-time
  physics in practice: characterising and controlling non-Markovian quantum
  stochastic processes}}}}, {PhD Thesis}, The University of Melbourne (2024)

\bibitem{Luchnikov_2020}
I~A Luchnikov, S~V Vintskevich, D~A Grigoriev and S~N Filippov,  \emph{{Machine
  Learning Non-Markovian Quantum Dynamics}},
  \href{https://doi.org/10.1103/PhysRevLett.124.140502}{Phys. Rev. Lett. {\bf
  124}, 140502 (2020)} (Preprint
  {\href{https://arxiv.org/abs/1902.07019}{arXiv:1902.07019}})

\bibitem{10.21468/SciPostPhys.13.2.028}
C~Guo,  \emph{{Quantifying Non-Markovianity in Open Quantum Dynamics}},
  \href{https://doi.org/10.21468/SciPostPhys.13.2.028}{SciPost Phys. {\bf 13},
  028 (2022)} (Preprint
  {\href{https://arxiv.org/abs/2205.05970}{arXiv:2205.05970}})

\bibitem{white2023unify}
G~A~L White, P~Jurcevic, C~D Hill and K~Modi,  \emph{{Unifying non-Markovian
  characterisation with an efficient and self-consistent framework}} (2023)
  (Preprint {\href{https://arxiv.org/abs/2312.08454}{arXiv:2312.08454}})

\bibitem{Baumgratz_2013}
T~Baumgratz, D~Gross, M~Cramer and M~B Plenio,  \emph{{Scalable Reconstruction
  of Density Matrices}},
  \href{https://doi.org/10.1103/PhysRevLett.111.020401}{Phys. Rev. Lett. {\bf
  111}, 020401 (2013)} (Preprint
  {\href{https://arxiv.org/abs/1207.0358}{arXiv:1207.0358}})

\bibitem{Holzapfel_2015}
M~Holz{\"a}pfel, T~Baumgratz, M~Cramer and M~B Plenio,  \emph{{Scalable
  reconstruction of unitary processes and Hamiltonians}},
  \href{https://doi.org/10.1103/PhysRevA.91.042129}{Phys. Rev. A {\bf 91},
  042129 (2015)} (Preprint
  {\href{https://arxiv.org/abs/1411.6379}{arXiv:1411.6379}})

\bibitem{Shrapnel_2018}
S~Shrapnel, F~Costa and G~Milburn,  \emph{{Quantum Markovianity as a supervised
  learning task}},  \href{https://doi.org/10.1142/S0219749918400105}{Int. J.
  Quantum Inf. {\bf 16}, 1840010 (2018)} (Preprint
  {\href{https://arxiv.org/abs/1901.05158}{arXiv:1901.05158}})

\bibitem{PhysRevE.98.042114}
C~Guo, Z~Jie, W~Lu and D~Poletti,  \emph{{Matrix product operators for
  sequence-to-sequence learning}},
  \href{https://doi.org/10.1103/PhysRevE.98.042114}{Phys. Rev. E {\bf 98},
  042114 (2018)} (Preprint
  {\href{https://arxiv.org/abs/1803.10908}{arXiv:1803.10908}})

\bibitem{Guo_2020}
C~Guo, K~Modi and D~Poletti,  \emph{{Tensor-network-based machine learning of
  non-Markovian quantum processes}},
  \href{https://doi.org/10.1103/PhysRevA.102.062414}{Phys. Rev. A {\bf 102},
  062414 (2020)} (Preprint
  {\href{https://arxiv.org/abs/2004.11038}{arXiv:2004.11038}})

\bibitem{torlai2020quantum}
G~Torlai, C~J Wood, A~Acharya, G~Carleo, J~Carrasquilla and L~Aolita,
  \emph{{Quantum process tomography with unsupervised learning and tensor
  networks}},  \href{https://doi.org/10.1038/s41467-023-38332-9}{Nat. Commun.
  {\bf 14}, 2858 (2023)} (Preprint
  {\href{https://arxiv.org/abs/2006.02424}{arXiv:2006.02424}})

\bibitem{PhysRevA.106.022411}
C~Guo,  \emph{{Reconstructing non-Markovian open quantum evolution from
  multiple time measurements}},
  \href{https://doi.org/10.1103/PhysRevA.106.022411}{Phys. Rev. A {\bf 106},
  022411 (2022)} (Preprint
  {\href{https://arxiv.org/abs/2205.06521}{arXiv:2205.06521}})

\bibitem{guo2023}
X~Zhang, Z~Wu, G~A~L White, Z~Xiang, S~Hu, Z~Peng, Y~Liu, D~Zheng, X~Fu,
  A~Huang, D~Poletti, K~Modi, J~Wu, M~Deng and C~Guo,  \emph{{Randomised
  benchmarking for characterising and forecasting correlated processes}} (2023)
  (Preprint {\href{https://arxiv.org/abs/2312.06062}{arXiv:2312.06062}})

\bibitem{Eisert_2020}
J~Eisert, D~Hangleiter, N~Walk, I~Roth, D~Markham, R~Parekh, U~Chabaud and
  E~Kashefi,  \emph{{Quantum certification and benchmarking}},
  \href{https://doi.org/10.1038/s42254-020-0186-4}{Nat. Rev. Phys. {\bf 2}, 382
  (2020)} (Preprint
  {\href{https://arxiv.org/abs/1910.06343}{arXiv:1910.06343}})

\bibitem{Kliesch_2021}
M~Kliesch and I~Roth,  \emph{{Theory of Quantum System Certification}},
  \href{https://doi.org/10.1103/PRXQuantum.2.010201}{PRX Quantum {\bf 2},
  010201 (2021)} (Preprint
  {\href{https://arxiv.org/abs/2010.05925}{arXiv:2010.05925}})

\bibitem{Figueroa-Romero_2024}
P~Figueroa-Romero, M~Papi{\v c}, A~Auer, M-H Hsieh, K~Modi and I~{de Vega},
  \emph{{Operational Markovianization in randomized benchmarking}},
  \href{https://doi.org/10.1088/2058-9565/ad3f44}{Quantum Sci. Technol. {\bf
  9}, 035020 (2024)} (Preprint
  {\href{https://arxiv.org/abs/2305.04704}{arXiv:2305.04704}})

\bibitem{liu2024}
Z~Liu, Y~Xiao and Z~Cai,  \emph{{Non-Markovian Noise Suppression Simplified
  through Channel Representation}} (2024) (Preprint
  {\href{https://arxiv.org/abs/2412.11220}{arXiv:2412.11220}})

\bibitem{Yang_2022}
S-X Yang, P~Figueroa-Romero and M-H Hsieh,  \emph{{Machine Learning of Average
  Non-Markovianity from Randomized Benchmarking}} (2022) (Preprint
  {\href{https://arxiv.org/abs/2207.01542}{arXiv:2207.01542}})

\bibitem{GST-quantum}
E~Nielsen, J~K Gamble, K~Rudinger, T~Scholten, K~Young and R~Blume-Kohout,
  \emph{{Gate Set Tomography}},
  \href{https://doi.org/10.22331/q-2021-10-05-557}{Quantum {\bf 5}, 557 (2021)}
  (Preprint {\href{https://arxiv.org/abs/2009.07301}{arXiv:2009.07301}})

\bibitem{Li_2024}
Z-T Li, C-C Zheng, F-X Meng, H~Zeng, T~Luan, Z-C Zhang and X-T Yu,
  \emph{{Non-Markovian quantum gate set tomography}},
  \href{https://doi.org/10.1088/2058-9565/ad3d80}{Quantum Sci. Technol. {\bf
  9}, 035027 (2024)} (Preprint
  {\href{https://arxiv.org/abs/2307.14696}{arXiv:2307.14696}})

\bibitem{Tanggara_2024}
A~Tanggara, M~Gu and K~Bharti,  \emph{{Strategic Code: A Unified
  Spatio-Temporal Framework for Quantum Error-Correction}} (2024) (Preprint
  {\href{https://arxiv.org/abs/2405.17567}{arXiv:2405.17567}})

\bibitem{kam_2024}
J~F Kam, S~Gicev, K~Modi, A~Southwell and M~Usman,  \emph{{Detrimental
  non-Markovian errors for surface code memory}} (2024) (Preprint
  {\href{https://arxiv.org/abs/2410.23779}{arXiv:2410.23779}})

\bibitem{Kobayashi_2024}
F~Kobayashi, H~Manabe, G~A~L White, T~Farrelly, K~Modi and T~Stace,
  \emph{{Tensor-network decoders for process tensor descriptions of
  non-Markovian noise}} (2024) (Preprint
  {\href{https://arxiv.org/abs/2412.13739}{arXiv:2412.13739}})

\bibitem{White_2024_Many}
G~A~L White, L~C~L Hollenberg, C~D Hill and K~Modi,  \emph{{Practical learning
  of multi-time statistics in open quantum systems}} (2024) (Preprint
  {\href{https://arxiv.org/abs/2412.17862}{arXiv:2412.17862}})

\bibitem{2105.03333}
L~Xiang, Z~Zong, Z~Zhan, Y~Fei, C~Run, Y~Wu, W~Jin, Z~Jia, P~Duan, J~Wu, Y~Yin
  and G~Guo,  \emph{{Quantify the Non-Markovian Process with Intervening
  Projections in a Superconducting Processor}} (2021) (Preprint
  {\href{https://arxiv.org/abs/2105.03333}{arXiv:2105.03333}})

\bibitem{Giarmatzi_2023}
C~Giarmatzi, T~Jones, A~Gilchrist, P~Pakkiam, A~Fedorov and F~Costa,
  \emph{{Multi-time quantum process tomography on a superconducting qubit}}
  (2023) (Preprint {\href{https://arxiv.org/abs/2308.00750}{arXiv:2308.00750}})

\bibitem{Palsson_2017}
M~S Palsson, M~Gu, J~Ho, H~M Wiseman and G~J Pryde,  \emph{{Experimentally
  modeling stochastic processes with less memory by the use of a quantum
  processor}},  \href{https://doi.org/10.1126/sciadv.1601302}{Sci. Adv. {\bf
  3}, e1601302 (2017)} (Preprint
  {\href{https://arxiv.org/abs/1602.05683}{arXiv:1602.05683}})

\bibitem{goswami_2020_increasing}
K~Goswami, Y~Cao, G~A Paz-Silva, J~Romero and A~G White,  \emph{{Increasing
  communication capacity via superposition of order}},
  \href{https://doi.org/10.1103/PhysRevResearch.2.033292}{Phys. Rev. Res. {\bf
  2}, 033292 (2020)} (Preprint
  {\href{https://arxiv.org/abs/1807.07383}{arXiv:1807.07383}})

\bibitem{Guo_2020a}
Y~Guo, X-M Hu, Z-B Hou, H~Cao, J-M Cui, B-H Liu, Y-F Huang, C-F Li, G-C Guo and
  G~Chiribella,  \emph{{Experimental Transmission of Quantum Information Using
  a Superposition of Causal Orders}},
  \href{https://doi.org/10.1103/PhysRevLett.124.030502}{Phys. Rev. Lett. {\bf
  124}, 030502 (2020)} (Preprint
  {\href{https://arxiv.org/abs/1811.07526}{arXiv:1811.07526}})

\bibitem{rubino2021experimental}
G~Rubino, L~A Rozema, D~Ebler, H~Kristj{\'a}nsson, S~Salek, P~A Gu{\'e}rin, A~A
  Abbott, C~Branciard, {\v C}~Brukner, G~Chiribella and P~Walther,
  \emph{{Experimental quantum communication enhancement by superposing
  trajectories}},
  \href{https://doi.org/10.1103/PhysRevResearch.3.013093}{Phys. Rev. Res. {\bf
  3}, 013093 (2021)} (Preprint
  {\href{https://arxiv.org/abs/2007.05005}{arXiv:2007.05005}})

\bibitem{Stromberg_2023}
T~Str{\"o}mberg, P~Schiansky, R~W Peterson, M~T Quintino and P~Walther,
  \emph{{Demonstration of a Quantum Switch in a Sagnac Configuration}},
  \href{https://doi.org/10.1103/PhysRevLett.131.060803}{Phys. Rev. Lett. {\bf
  131}, 060803 (2023)} (Preprint
  {\href{https://arxiv.org/abs/2211.12540}{arXiv:2211.12540}})

\bibitem{Yin_2023}
P~Yin, X~Zhao, Y~Yang, Y~Guo, W-H Zhang, G-C Li, Y-J Han, B-H Liu, J-S Xu,
  G~Chiribella, G~Chen, C-F Li and G-C Guo,  \emph{{Experimental
  super-Heisenberg quantum metrology with indefinite gate order}},
  \href{https://doi.org/10.1038/s41567-023-02046-y}{Nat. Phys. {\bf 19}, 1122
  (2023)} (Preprint
  {\href{https://arxiv.org/abs/2303.17223}{arXiv:2303.17223}})

\bibitem{Nie_2022}
X~Nie, X~Zhu, K~Huang, K~Tang, X~Long, Z~Lin, Y~Tian, C~Qiu, C~Xi, X~Yang,
  J~Li, Y~Dong, T~Xin and D~Lu,  \emph{{Experimental Realization of a Quantum
  Refrigerator Driven by Indefinite Causal Orders}},
  \href{https://doi.org/10.1103/PhysRevLett.129.100603}{Phys. Rev. Lett. {\bf
  129}, 100603 (2022)} (Preprint
  {\href{https://arxiv.org/abs/2011.12580}{arXiv:2011.12580}})

\bibitem{xi_experimental_2024}
C~Xi, X~Liu, H~Liu, K~Huang, X~Long, D~Ebler, X~Nie, O~Dahlsten and D~Lu,
  \emph{Experimental {Validation} of {Enhanced} {Information} {Capacity} by
  {Quantum} {Switch} in {Accordance} with {Thermodynamic} {Laws}},
  \href{https://doi.org/10.1103/PhysRevLett.133.040401}{Phys. Rev. Lett. {\bf
  133}, 040401 (2024)} (Preprint
  {\href{https://arxiv.org/abs/2406.01951}{arXiv:2406.01951}})

\bibitem{Zhu_2023}
G~Zhu, Y~Chen, Y~Hasegawa and P~Xue,  \emph{{Charging Quantum Batteries via
  Indefinite Causal Order: Theory and Experiment}},
  \href{https://doi.org/10.1103/PhysRevLett.131.240401}{Phys. Rev. Lett. {\bf
  131}, 240401 (2023)} (Preprint
  {\href{https://arxiv.org/abs/2105.12466}{arXiv:2105.12466}})

\bibitem{Ringbauer_2016}
M~Ringbauer, C~Giarmatzi, R~Chaves, F~Costa, A~G White and A~Fedrizzi,
  \emph{{Experimental test of nonlocal causality}},
  \href{https://doi.org/10.1126/sciadv.1600162}{Sci. Adv. {\bf 2}, 1600162
  (2016)} (Preprint
  {\href{https://arxiv.org/abs/1602.02767}{arXiv:1602.02767}})

\bibitem{cao2023semideviceindependent}
H~Cao, J~Bavaresco, N-N Wang, L~A Rozema, C~Zhang, Y-F Huang, B-H Liu, C-F Li,
  G-C Guo and P~Walther,  \emph{{Semi-device-independent certification of
  indefinite causal order in a photonic quantum switch}},
  \href{https://doi.org/10.1364/OPTICA.483876}{Optica {\bf 10}, 561 (2023)}
  (Preprint {\href{https://arxiv.org/abs/2202.05346}{arXiv:2202.05346}})

\bibitem{antesberger2024higherorder}
M~Antesberger, M~T Quintino, P~Walther and L~A Rozema,  \emph{{Higher-Order
  Process Matrix Tomography of a Passively-Stable Quantum Switch}},
  \href{https://doi.org/10.1103/PRXQuantum.5.010325}{PRX Quantum {\bf 5},
  010325 (2024)} (Preprint
  {\href{https://arxiv.org/abs/2305.19386}{arXiv:2305.19386}})

\end{thebibliography}

\providecommand{\newblock}{}

\end{document}